# Modern trends in Superconductivity and Superfluidity

Selected Chapters.

*M.Yu. Kagan*
*P.L. Kapitza Institute for Physical Problems*
kagan@kapitza.ras.ru

Chapter 1. Hydrodynamics of rotating superfluids with quantized vortices.

While many forthcoming Chapters of the present manuscript deal with the microscopical models for the description of Quantum Gasses, Fluids and Solids, as well as strongly correlated electron systems, the first four Chapters are devoted to purely phenomenological (macroscopic) description of hydrodynamic phenomena in anisotropic and inhomogeneous superfluids and superconductors.

The presentation in these Chapters is based on Landau ideas. Let us stress that Landau theory of hydrodynamics [1.1] based on the conservation laws in the differential form (together with Quantum Mechanics [1.2] and Statistical Physics [1.3]) is one of the masterpieces of Landau-Lifshitz course of the Theoretical Physics. While the essence of this theory is well known (serves as a mothermilk) to Russian researchers from the first steps of their scientific career, in the West many young physicists, specializing on macroscopic models, do not have a deep background in this subject. A good style meanwhile is to know fluently both microscopics and phenomenology from our point of view.

That is why from the pedagogical reasons we decided to start the present Chapter with brief description (in Section 1.1) of the Landau theory for hydrodynamics in classical [1.1] and superfluid [1.1, 1.4] liquids. We present Landau scheme of the conservation laws for the description of slowly varying in space and time Goldstone (gapless) collective modes.

We will discuss Landau [1.1] (or Landau-Tissa [1.1, 1.5], as often referred to in the West) two-velocity superfluid hydrodynamics for $^4$He [1.1, 1.4, 1.5] which contains normal and superfluid velocities $\vec{v}_n$ and $\vec{v}_s$, as well as normal and superfluid densities $\rho_n$ and $\rho_s$, thus describing both the normal and superfluid motion in helium [1.6, 1.28]. We will derive the spectrum and damping of first and second sound in superfluid helium (in He-II). We will get $\omega = c_{\mathrm{I}}k$ and $\omega = c_{\mathrm{II}}k$ (the spectrum is linear for both waves) and compare the velocities $c_{\mathrm{I}}$ and $c_{\mathrm{II}}$ of these sound waves. We will stress that while in first sound total mass-current $\vec{j} = \rho_s \vec{v}_s + \rho_n \vec{v}_n \neq 0$, in the second sound wave $\vec{j} = 0$ but the relative velocity $\vec{W} = \vec{v}_n - \vec{v}_s \neq 0$.

In Section 1.2 we proceed to the hydrodynamics of rotating superfluids with a large number of quantized vortices. We start this Section with famous Andronikashvilli experiments (on non-classical moment of inertia in rotating He-II) [1.7] and discuss Feynmann-Onsager quantization of the vortex lines in superfluid helium [1.8, 1.9, 1.29, 1.30]. We introduce the notion of first and second critical angular velocities of rotation $\Omega_{\mathrm{C1}}$ and $\Omega_{\mathrm{C2}}$ in helium and stress their correspondence with first and second critical magnetic fields $H_{\mathrm{C1}}$ and $H_{\mathrm{C2}}$ in type – II superconductors [1.10, 1.26, 1.27]. Than we will present the scheme of macroscopic averaging [1.11] for a large number of vortices and construct nonlinear elasticity theory for a 2D (triangular) vortex lattice in helium [1.12]. The linearization of this theory yields well-known Tkachenko modes [1.13] (which describe longitudinal and transverse oscillations of the vortex lattice), as well as Lord Kelvin (Thomson) bending oscillations of the vortex lines [1.14]. We include dissipation in the system and discuss Hall-Vinen friction coefficients $\beta$ and $\beta'$ [1.15, 1.24, 1.25] for the scattering of normal component on the vortex lattice. In the end of Section 1.2 we will construct the complete system of equations (which describe hydrodynamics of slow rotations [1.12, 1.22]) based on the Landau scheme of the conservation laws. We will discuss briefly another elegant method to derive a system of hydrodynamic equations based on Poisson brackets [1.19] and emphasize, nevertheless, the advantages of Landau method especially in nonlinear regime.

In the next Section (Section 1.3) we consider the different limit for hydrodynamics of fast rotating superfluid. In this case due to umklapp processes [1.10, 1.21] the normal excitations are bound to vortex lattice, and hence the normal and superfluid velocities perpendicular to the vortex lines coincide $\vec{v}_{n\perp} = \vec{v}_{s\perp}$. In the same time in the direction parallel to the vortex lines due to translational invariance $\vec{v}_{n\parallel} \neq \vec{v}_{s\parallel}$. Thus we have one-velocity motion in the plane of the vortex lattice $\vec{v}_\perp = \vec{j}_\perp / \rho$ and two-velocity motion parallel to the vortex lines. Hence we are in a strongly anisotropic situation where we have a crystal in the plane of the vortex lattice and a standard superfluid in the direction perpendicular to the lattice and parallel to the vortex lines. We will construct the full system of hydrodynamic equations for a regime of rapid rotations and analyze the spectrum and damping of collective excitations (first of all of a second sound mode) in this regime. We will get that the spectrum is linear and sound wave can freely propagate only along the vortex lines where $\vec{W}_\parallel = \vec{v}_{n\parallel} - \vec{v}_{s\parallel} \neq 0$, while the spectrum is overdamped in the perpendicular direction. More specifically for $k_\perp = 0$ the spectrum reads $\omega^2 = c_2^2 k_z^2$, while for $k_z = 0$: $\omega = -\dfrac{i\kappa_\perp k_\perp^2}{C_p}$ [1.12], where $\kappa_\perp$ is heat conductivity in the direction perpendicular to the vortex lines, $C_p$ is a specific heat and $k_z$, $k_\perp$ are the components of the $\vec{k}$-vector along and perpendicular to the vortex lines, respectively.

In the Section 1.4 we consider an opposite case of a single bended line. Here, as shown by Lord Kelvin, the spectrum for the bending oscillations is almost quadratic $\omega = \dfrac{k_z^2}{2m}\ln\dfrac{1}{dk_z}$ [1.10, 1.14] ($d$ is a normal vortex core, which in the case of superfluid $^4$He is of the order of the interatomic distance). Naive considerations show that such quasi 1D system (as a bended vortex line) should be completely destroyed by the thermal fluctuations and experience in analogy with the biophysical systems a phase-transition to the state of a globula [1.10]. We show, however, that in fact [1.12] the bending oscillations correspond to rotation: $\left[\vec{u}\,\dot{\vec{u}}\right] \neq 0$, where $\vec{u}$ and $\dot{\vec{u}}$ are respectively a local displacement of the bended vortex from a nondeformed position and its time derivative (local velocity). Thus the quanta of the bending oscillations in fact have the rotation moment (- ħ) (diamagnetic situation) and hence the gap ħΩ appears in their spectrum. This gap stabilizes the fluctuations of this 1D system providing a finite ratio $\sqrt{\left\langle \vec{u}^2 \right\rangle / R^2} \ll 1$ of a mean square displacement $\left\langle \vec{u}^2 \right\rangle$ to the radius squared $R^2$ of the rotating vessel with helium. It is a reason why a regular (triangular) vortex lattice can be directly visualized (the photograph of the lattice with small vortex displacements can be obtained) in the experiments of Packard [1.10].

We conclude the Chapter with the discussion in Section 1.5 how we can realize the hydrodynamics of fast rotations at not very high frequencies $\Omega \ll \Omega_{C2}$ (note that the second critical angular velocity $\Omega_{C2}$ is very high in superfluid $^4$He). We propose to use for that regime $^4$He-$^3$He mixtures, where $^4$He is superfluid and has a large number of vortices. In this case according to the Bernulli law [1.1] $p + \rho v_s^2 = const$ all the $^3$He impurities will be driven by the gradient of the pressure ($\nabla p$) to the vortex core and organize inside the core the quasi 1D normal component with a free motion only along the vortex lines. This is just a desirable regime of rapid rotations which we discuss in Section 1.2. Another possibility is to use an isotropic triplet phase (B-phase) of superfluid $^3$He [1.16, 1.31] where we can make the umklapp processes very effective already at moderate rotation frequencies [1.12]. We stress that a recent revival of interest to the rotating superfluids with large number of vortices is connected with the intensive experimental and theoretical studies of rotating Bose-condensates in the restricted geometry of magnetic traps in ultracold gasses and mixtures [1.17]. We will discuss the most important results in this field especially those connected with the possible melting of the vortex lattice [1.17, 1.11, 1.43-1.48] in Subsection 1.2.6.

1.1. The foundation of Landau theory for superfluid hydrodynamics.

We will start this Section with a brief description of Landau approach to the hydrodynamics of the classical liquid.

1.1.1. The essence of the hydrodynamics. Description of the Goldstone modes.

Generally speaking, hydrodynamics is a science which describes all <u>slowly-varying</u> in space and time processes in the liquid (the elasticity theory does the same in the solid [1.18]). For these processes in momentum space a frequency $\omega \to 0$ when the wave-vector $\vec{k} \to 0$. Thus hydrodynamics describes all low-lying gapless (<u>Goldstone</u>) modes.

Moreover, hydrodynamics assumes local thermodynamic equilibrium (equilibrium in small volume), so we can adopt reduced description of the system [1.3] and introduce finite number of local variables. For ordinary liquid (with one velocity $\vec{v}$) the canonical variables in Landau scheme are $\rho(\vec{r},t)$ – density; $\vec{j}(\vec{r},t)$ – density of linear momentum, and $S(\vec{r},t)$ – entropy of the unit volume.

### 1.1.2. Landau scheme of the conservation laws. Euler equation.

According to Landau the system of equations for classical liquid contains an equation for mass-conservation:

$$\frac{\partial \rho}{\partial t} + div(\vec{j}) = 0. \ \vec{j} = \rho \vec{v}, \qquad (1.1.1)$$

an equation for linear momentum conservation:

$$\frac{\partial j_i}{\partial t} + \frac{\partial}{\partial x_k}(\Pi_{ik}) = 0, \qquad (1.1.2)$$

$\Pi_{ik}$ is momentum flux (in ideal dissipationless liquid $\Pi_{ik} = \rho v_i v_k + p \delta_{ik}$; $p$ – is pressure), and an equation for entropy increase:

$$\frac{\partial S}{\partial t} + div\left(S\vec{v} + \frac{\vec{q}}{T}\right) = \frac{R}{T}, \qquad (1.1.3)$$

where $R \geq 0$ is positively defined dissipative function, $\vec{q}$ is dissipative entropy flux, $T$ is temperature (in ideal liquid $R = 0$ and $\vec{q} = 0$). Note that if we take into account only one source of dissipation connected with the heat flux, then $\vec{q} = -\kappa \vec{\nabla} T$ ($\kappa$ is heat conductivity) and correspondingly the dissipative function $R \sim \kappa \dfrac{\left(\vec{\nabla} T\right)^2}{T} > 0$ is quadratic in gradients. In general case in Eqs. (1.1.2) and (1.1.3) in the expression for $\Pi_{ik}$ and $R$ enter also viscous contributions $\pi_{ik} = \Pi_{ik}^{dis} = \eta\left(\dfrac{\partial v_i}{\partial x_k} + \dfrac{\partial v_k}{\partial x_i} - \dfrac{2}{3}\delta_{ik} div\vec{v}\right) + \xi \delta_{ik} div\vec{v}$, where $\eta$ and $\xi$ are viscosity coefficients ($\xi$ is often called second viscosity) and accordingly $R \sim \pi_{ik} \dfrac{\partial v_i}{\partial x_k}$.

The system of 3 equations for $\partial \rho / \partial t$, $\partial j_i / \partial t$ and $\partial S / \partial t$ (1.1.1) – (1.1.3) are consistent with an equation of total energy conservation:

$$\frac{\partial E}{\partial t} + div\vec{Q} = 0. \qquad (1.1.4)$$

While deriving (1.1.4) we can use the Galilean invariance and introduce the internal energy $E_0$ in the reference frame $K_0$ (where the liquid is at a rest) according to:

$$E = E_0 + \rho \frac{\vec{v}^2}{2} = E_0 + \frac{\vec{j}^2}{2\rho}. \qquad (1.1.5)$$

In Eq. (1.1.4) $\vec{Q}$ is an energy flux. In an ideal dissipationless liquid:

$$\vec{Q} = \left(W + \rho \frac{v^2}{2}\right)\vec{v}, \qquad (1.1.6)$$

where $W = E_0 + P$ is the density of the enthalpy (the enthalpy of the unit volume). To get (1.1.4) with $\vec{Q}$ from (1.1.6) we should utilize the thermodynamic identity for internal energy:

$$dE_0 = TdS + \mu d\rho \,, \qquad (1.1.7)$$

where $\mu$ is the chemical potential, and use the expression for pressure:

$$P = TS + \mu\rho - E_0 = W - E_0 \,. \qquad (1.1.8)$$

From the expression (1.1.8) we get:

$$dP = SdT + \rho d\mu \qquad (1.1.9)$$

### Euler equation

From Eq. (1.1.2) for linear momentum conservation and Eq. (1.1.1) for mass-conservation we can get for an ideal dissipationless liquid:

$$\frac{\partial \vec{v}}{\partial t} + \left(\vec{v}\vec{\nabla}\right)\vec{v} + \frac{1}{\rho}\vec{\nabla}p = 0 \;. \qquad (1.1.10)$$

It is a famous Euler equation for an ideal classical liquid. At zero temperature ($T = 0$) we can rewrite (1.1.10) as:

$$\frac{\partial \vec{v}}{\partial t} + \vec{\nabla}\left(\mu + \frac{v^2}{2}\right) = \left[\vec{v}\, rot\, \vec{v}\right], \qquad (1.1.11)$$

where $\mu$ is the chemical potential.

### Potential irrotational flows.

Note that for $rot\, \vec{v} = 0$ we have so-called potential (irrotational) flows. In this case:

$$\frac{\partial \vec{v}}{\partial t} + \vec{\nabla}\left(\mu + \frac{v^2}{2}\right) = 0 \,, \qquad (1.1.12)$$

Correspondingly we can represent velocity as a gradient of some scalar potential $\vec{v} = \vec{\nabla}\varphi$.

For stationary potential flows $\dfrac{\partial \vec{v}}{\partial t} = 0$ and $\vec{\nabla}\left(\mu + \dfrac{v^2}{2}\right) = 0$. Thus

$$\mu + \frac{v^2}{2} = const \,, \qquad (1.1.13)$$

This equation is often called Bernoulli equation (see [1.1]).

### 1.1.3. Sound waves in classical liquid. Damping of sound waves.

Linearizing the system of equations (1.1.1) – (1.1.3) for an ideal (dissipationless) liquid we get the following set of equations:

$$\frac{\partial \delta\rho}{\partial t} + \rho_0 div\, \vec{v} = 0 \,, \qquad (1.1.14)$$

$$\rho_0 \frac{\partial \vec{v}_i}{\partial t} + \vec{\nabla}_i \delta P = 0 \,, \qquad (1.1.15)$$

$$\frac{\partial \delta S}{\partial t} + S_0\, div\, \vec{v} = 0 \,, \qquad (1.1.16)$$

where $\delta\rho$, $\delta P$ and $\delta S$ are slowly varying and small (in amplitude) deviations of density, pressure and entropy from equilibrium values in the sound wave. Thus $\rho(\vec{r},t) = \rho_0 + \delta\rho(\vec{r},t)$ with $\left|\delta\rho\right| << \rho_0$ and so on. We also assume that the velocity $\vec{v}$ in the sound wave is small $\vec{v} = \delta\vec{v}(\vec{r},t)$. In fact it is a condition $v << c$, where $c$ is sound velocity, which is required.

In ideal liquid all the motions are adiabatic. It means that the entropy of unit mass $S_M = S/\rho$ = const (where $S$ is an entropy of unit volume). Thus $\delta S$ in Eq. (1.1.16) yields $\delta S = S_M \delta \rho$ and Eq. (1.1.16) for entropy becomes equivalent to Eq. (1.1.14) for density. Moreover the pressure $P = TS + \mu\rho - E_0 = T\rho S_M + \mu\rho - E_0$ (which in general case is the function of $S_M$ and $\rho$) for adiabatic sound wave is a function only of $\rho$. Thus for small derivations of pressure from equilibrium values $\delta P = \left(\dfrac{\partial P}{\partial \rho}\right)_{S_M} \delta \rho$, where by its definition we can introduce the sound velocity squared as:

$$c^2 = \left(\frac{\partial P}{\partial \rho}\right)_{S_M} \quad \text{and} \quad \delta P = c^2 \delta \rho \qquad (1.1.17)$$

Correspondingly we can rewrite Eq. (1.1.15) as:

$$\rho_0 \frac{\partial \vec{v}_i}{\partial t} + c^2 \vec{\nabla}_i \delta \rho = 0 \qquad (1.1.18)$$

If we take the time derivative $(\partial/\partial t)$ from Eq. (1.1.14) for density and space derivative ($div$) from Eq. (1.1.15) for linear momentum, we will get:

$$\frac{\partial^2 \delta \rho}{\partial t^2} + \rho_0 \, div \, \dot{\vec{v}} = 0, \qquad (1.1.19)$$

$$\rho_0 \, div \, \dot{\vec{v}} + c^2 \Delta \delta \rho = 0, \qquad (1.1.20)$$

where $\Delta$ is an operator of Laplacian.

Substitution of (1.1.20) into (1.1.19) finally yields:

$$\frac{\partial^2 \delta \rho}{\partial t^2} - c^2 \Delta \delta \rho = 0. \qquad (1.1.21)$$

If we assume that $\delta \rho$ varies in a monochromatic sound wave $\delta \rho \sim e^{-i\omega t + i\vec{k}\vec{r}}$ we will get for the spectrum:

$$\omega^2 = c^2 k^2. \qquad (1.1.22)$$

Thus the sound spectrum is linear as expected. It corresponds to compressible liquid where $\delta \rho \neq 0$ and $\vec{j} \neq 0$.

<u>Damping of the sound waves.</u>

The damping of sound in hydrodynamics theory is given by higher gradients connected with dissipative terms in the system of equations (1.1.1) – (1.1.3). The spectrum with an account of damping reads [1.1]:

$$\omega = ck + i\gamma, \qquad (1.1.23)$$

where

$$\gamma = \frac{\omega^2}{2\rho c^2}\left\{\left(\frac{4}{3}\eta + \xi\right) + \kappa\left(\frac{1}{C_V} - \frac{1}{C_P}\right)\right\}, \qquad (1.1.24)$$

$\eta$ and $\xi$ are coefficients of first and second viscosity, $\kappa$ is heat conductivity, $C_P$ and $C_V$ are specific heat at constant pressure and constant volume [1.3]. We can see that the damping $\gamma \sim \omega^2$ and thus it is small for $\omega \to 0$. Moreover in hydrodynamics (collisional) theory we can write $\gamma \approx \omega^2 \tau = c^2 k^2 \tau$ and

$$\gamma/\omega \approx \omega\tau \ll 1, \qquad (1.1.25)$$

where we introduced a characteristic (scattering) time $\tau$. Note that if we consider the damping due to the presence of the first viscosity $\eta$ and take into account a simple estimate from the kinetic theory [1.32] $\eta \sim \rho l \bar{v} \sim \rho \bar{v}^2 \tau$, than $\gamma \sim \dfrac{\omega^2 \eta}{\rho c^2} \sim \dfrac{\bar{v}^2}{c^2} \omega^2 \tau$ and indeed $\gamma / \omega \sim \dfrac{\bar{v}^2}{c^2} \omega \tau$, where $\tau = l / \bar{v}$ is a scattering time, $l$ is a length of the mean-free path and $\bar{v}$ is thermal (average) velocity (in Boltzman molecular gas, for example). Note that in Boltzman gas $c^2 \sim \bar{v}^2$ and hence $\gamma \sim \omega^2 \tau$.

We can conclude that hydrodynamic description (with the small damping and the propagating sound waves) is valid for small frequencies $\omega \tau \ll 1$ or, correspondingly, for small wave-vectors $kl \ll 1$. For larger frequencies $\omega \tau \gg 1$ we are in a ballistic (or Knudsen) regime (see for example Physical Kinetics [1.32]). In this regime we should start our theoretical analysis with a good kinetic equation in classical or degenerate case (see Chapter 16) and derive the equations in the collisionless (ballistic) regime.

_Equation for heat conductivity. Overdamped temperature waves._

Let us analyze now Eq. (1.1.3) for the entropy increase. In the absence of the drift velocity (for the liquid at a rest) $\bar{v} = 0$ and Eq. (1.1.3) reads:

$$\frac{\partial S}{\partial t} + div\left(\frac{\vec{q}}{T}\right) = \frac{R}{T}, \qquad (1.1.26)$$

where $\vec{q} = -\kappa \vec{\nabla} T$ is a heat flux, $\kappa$ is heat conductivity, $T$ is temperature, and $R$ is dissipative function.

If we are interested in small entropy and temperature deviations from the equilibrium values, having in mind temperature waves, we can linearize (1.1.22). After linearization we get:

$$\frac{\partial \delta S}{\partial t} - \frac{\kappa}{T_0} \Delta \delta T = 0. \qquad (1.1.27)$$

where the entropy $S$ and temperature $T$ are given by: $S = S_0 + \delta S(\vec{r}, t)$ and $T = T_0 + \delta T(\vec{r}, t)$. To find the spectrum of the temperature waves (which will be overdamped as we will see soon) we should express $\dfrac{\partial \delta S}{\partial t}$ in (1.1.23) via the time derivative of the temperature $\dfrac{\partial \delta T}{\partial t}$.

For almost incompressible fluid (which is a legitimate approximation in this case) $\delta S = \rho \delta S_M$, where $S_M$ is an entropy of a unit mass. Assuming that correct thermodynamic variables for mass entropy $S_M$ are $p$ – pressure and $T$ – temperature, and, moreover that we can put $p = const$ in temperature waves (see [3.1]), we can represent the time derivative:

$$\frac{\partial \delta S}{\partial t} \approx \rho \frac{\partial \delta S_M}{\partial t} \approx \rho \left(\frac{\partial S_M}{\partial T}\right)_P \frac{\partial \delta T}{\partial t} = \frac{\rho C_P}{T_0} \frac{\partial \delta T}{\partial t}, \qquad (1.1.28)$$

where $C_p = T_0 \left(\dfrac{\partial S_M}{\partial T}\right)_p$ is specific heat at constant pressure (see [1.3]).

Correspondingly Eq. (1.1.23) reads:

$$\frac{\rho C_P}{T_0} \frac{\partial \delta T}{\partial t} - \frac{\kappa}{T_0} \Delta \delta T = 0. \qquad (1.1.29)$$

Eq. (1.1.25) is a famous equation for the heat conductivity (or a Fourier equation as mathematicians often call it). For the monochromatic temperature wave $\delta T \sim e^{-i\omega t + i \vec{q} \vec{r}}$ we get:

$$i\omega = \frac{\kappa}{\rho C_P} q^2 \text{ or } \omega = -i \frac{\kappa}{\rho C_P} q^2. \qquad (1.1.30)$$

Thus we conclude that the spectrum of the temperature waves in a classical liquid is overdamped and quadratic in the wave-vector $q$.

1.1.4. Rotational fluid. Vorticity conservation. Inertial mode.

If $rot\,\vec{v} \neq 0$ we can take $rot$ from l.h.s and r.h.s. of the Euler equation (1.1.11). Then we get:

$$\frac{\partial rot\,\vec{v}}{\partial t} = rot\left[\vec{v}\,rot\,\vec{v}\right], \ \ \text{or} \ \ \frac{\partial \vec{\omega}}{\partial t} = rot\left[\vec{v}\,\vec{\omega}\right], \quad\quad (1.1.31)$$

where $\vec{\omega} = rot\,\vec{v}$. This equation is an equation for vorticity conservation. Note that for solid-state rotation $\vec{v} = \left[\vec{\Omega}\,\vec{r}\right]$ and hence $\vec{\omega} = rot\,\vec{v} = 2\vec{\Omega}$, where $\vec{\Omega}$ is an angular frequency of rotation.

<u>Inertial mode in classical fluid.</u>

Let us consider uniformly rotating liquid with an angular velocity $\vec{\Omega}$ and find the spectrum $\omega(k)$ in rotation frame for small variations of the velocity field $\vec{v} = \left[\vec{\Omega}\,\vec{r}\right] + \delta\vec{v}(\vec{r},t)$ on top of solid-state rotation. Let us consider an incompressible fluid. Then $div\,\vec{v} = div\,\delta\vec{v}(\vec{r},t) = 0$. Correspondingly $rot\,\vec{v} = 2\vec{\Omega} + rot\,\delta\vec{v}(\vec{r},t)$. The linearization of Euler equation with an account of the thermodynamic identity $\frac{1}{\rho}\vec{\nabla}P = \vec{\nabla}\mu$ reads:

$$\left\{\frac{\partial}{\partial t} + \left(\vec{v}_0\vec{\nabla}\right) - \vec{\Omega}x\right\}\delta\vec{v} + \frac{1}{\rho_0}\vec{\nabla}\left(P - \rho\frac{\left[\vec{\Omega}\,\vec{r}\right]^2}{2}\right) + 2\left[\vec{\Omega}\,\delta\vec{v}\right] = 0 \quad\quad (1.1.32)$$

where $\vec{v}_0 = \left[\vec{\Omega}\,\vec{r}\right]$ is a linear velocity of solid-state rotation. Note that in curly brackets in the l.h.s. of Eq. (1.1.32) stands the operator which transforms the time derivative of the vector $\delta\vec{v}$ to the rotation frame. Thus

$$\frac{\partial \delta\vec{v}}{\partial t'} + \frac{1}{\rho_0}\vec{\nabla}\left(P - \rho\frac{\left[\vec{\Omega}\,\vec{r}\right]^2}{2}\right) + 2\left[\vec{\Omega}\,\delta\vec{v}\right] = 0, \quad\quad (1.1.33)$$

where operator $\partial/\partial t'$ refers to the rotation frame. In Eq. (1.1.33) we recognize a Coriolis force $2\left[\vec{\Omega}\,\delta\vec{v}\right]$ and a centrifugal force $\vec{\nabla}\left(\frac{\left[\vec{\Omega}\,\vec{r}\right]^2}{2}\right)$ which are always present in the rotation frame. We can also introduce an effective pressure $P_{eff} = P - \rho\frac{\left[\vec{\Omega}\,\vec{r}\right]^2}{2}$. Then

$$\frac{\partial \delta\vec{v}}{\partial t'} + \frac{1}{\rho_0}\vec{\nabla}P_{eff} + 2\left[\vec{\Omega}\,\delta\vec{v}\right] = 0 \quad\quad (1.1.34)$$

Representing $P_{eff} = P_{0eff} + \delta P$ we get:

$$\frac{\partial \delta\vec{v}}{\partial t'} + \frac{1}{\rho_0}\vec{\nabla}\delta P + 2\left[\vec{\Omega}\,\delta\vec{v}\right] = 0. \quad\quad (1.1.35)$$

Applying operator of rot to l.h.s. of (1.1.34) we get:

$$\frac{\partial}{\partial t'}rot\,\delta\vec{v} + 2rot\left[\vec{\Omega}\,\delta\vec{v}\right] = 0. \quad\quad (1.1.36)$$

If $\vec{\Omega} = \Omega\vec{e}_z$ than for $div\,\delta\vec{v} = 0$ we get $2rot\left[\vec{\Omega}\,\delta\vec{v}\right] = -2\Omega\frac{\partial \delta\vec{v}}{\partial z}$ and

$$\frac{\partial}{\partial t'}rot\,\delta\vec{v} - 2\Omega\frac{\partial \delta\vec{v}}{\partial z} = 0 \quad\quad (1.1.37)$$

For the monochromatic wave $\delta\vec{v} \sim e^{-i\omega t + i\vec{k}\vec{r}}$

$$-i\omega\left[\vec{k}\,\delta\vec{v}\right] = ik_z 2\Omega\delta\vec{v}. \qquad (1.1.38)$$

If we take a vector product with $\vec{k}$ of the l.h.s and the r.h.s of (1.1.38) we get with an account of incompressibility condition $\vec{k}\,\delta\vec{v} = 0$

$$-i\omega k^2\delta\vec{v} = ik_z 2\Omega\left[\vec{k}\,\delta\vec{v}\right]. \qquad (1.1.39)$$

Comparing (1.1.38) and (1.1.39) we conclude that $-i\omega\left[\vec{k}\,\delta\vec{v}\right] = ik_z 2\Omega\dfrac{ik_z 2\Omega}{-i\omega k^2}\left[\vec{k}\,\delta\vec{v}\right]$ and correspondingly

$$\omega^2 k^2 = k_z^{\,2}4\Omega^2. \qquad (1.1.40)$$

Finally

$$\omega^2 = 4\Omega^2\frac{k_z^{\,2}}{k^2}, \text{ or } \omega = \frac{2(\vec{\Omega}\vec{k})}{k}. \qquad (1.1.41)$$

This is a well-known inertial mode in an uncompressible ideal fluid. Note that it is a Goldstone mode for $k_z/k \to 0$ where $k^2 = k_z^{\,2} + k_\perp^{\,2}$ and $\vec{k}_\perp = (k_x, k_y)$. Note also that the spectrum $\omega(k)$ in (1.1.41) is a spectrum in rotation frame.

Let us discuss compressible rotating fluid now. According to Sonin [1.24] the spectrum in this case reads:

$$\omega^2 = 4\Omega^2\frac{\omega^2 - c^2 k_z^{\,2}}{\omega^2 - c^2 k^2}. \qquad (1.1.42)$$

The solution of this equation yields two branches for the spectrum:

$$\omega^2 = \frac{1}{2}\left(4\Omega^2 + c^2 k^2\right) \pm \left[\frac{1}{4}\left(4\Omega^2 + c^2 k^2\right)^2 - \left(2\Omega c k_z\right)^2\right]^{1/2}. \qquad (1.1.43)$$

In a fluid at a rest (for $\Omega = 0$) $\omega_1 = ck$ and $\omega_2 = 0$. Thus we see that rotation adds a second mode with non-zero frequency to the sound mode. The reason for that according to Sonin is a Coriolis force: rotation makes the fluid rigid in the direction perpendicular to rotation axis.

We can distinguish two regimes in Eq. (1.1.43): the regime of large $k$-vectors $k > \dfrac{2\Omega}{c}$ (or almost incompressible liquid $c \to \infty$) and the regime of small $k$-vectors $k < \dfrac{2\Omega}{c}$.

In the regime of large $k$-vectors the solution of (1.1.43) yields $\omega_1^{\,2} = \dfrac{4\Omega^2 k_z^{\,2}}{k^2}$ for the inertial mode and

$$\omega_2^{\,2} = c^2 k^2 + \left(\frac{2\Omega k_\perp}{k}\right)^2 \qquad (1.1.44)$$

for the modified sound wave.

In the opposite limit of small $k$-vectors $k < \dfrac{2\Omega}{c}$ (compressible liquid, $c$ is finite):

$$\omega_1^{\,2} = 4\Omega^2 + c^2 k^2 \text{ and } \omega_2^{\,2} = c^2 k_z^{\,2}. \qquad (1.1.45)$$

Note that usually the space scale $c/\Omega$ in a liquid (like $^4$He) is extremely large (of the order of hundreds of meters) and is not relevant to any real laboratory experiment where typical size of the container with rotating liquid is 0.1 cm ÷ 1 cm. Thus the regime of small $k$-vectors where incompressibility condition fails is very difficult to realize experimentally.

1.1.5. Two-velocity hydrodynamics for superfluid helium. $\vec{v}_n$ and $\vec{v}_s$, $\rho_n$ and $\rho_s$.

Switching our considerations now to the hydrodynamics of a superfluid helium we should start from the well-known phase diagram of $^4$He (see [1.1, 1.4, 1.5, 1.8, 1.29? 1.33-1.35] and Fig.1.1).

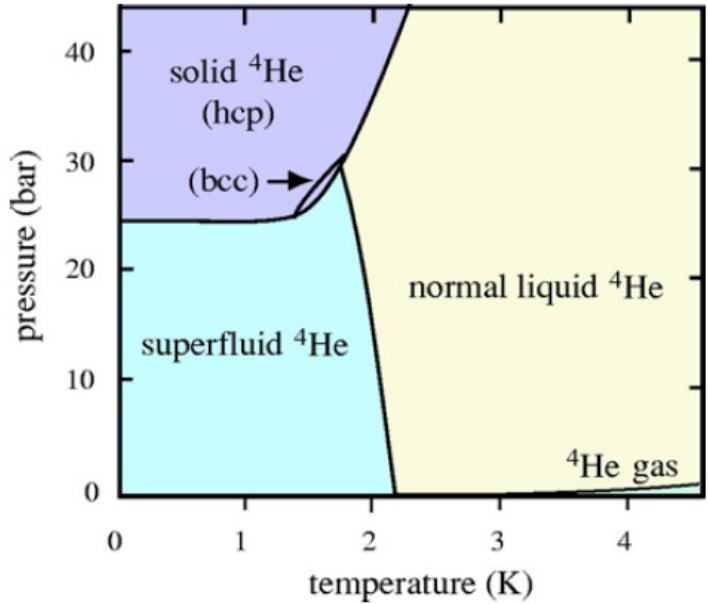

Fig.1.1. Phase diagram of $^4$He [1.29]. There is no triple point where the liquid, solid and gas phases would meet. At $P = 0$ $^4$He becomes superfluid for $T < 2.2$ K (see [1.1]). For pressures $P > 26$ Bar and low temperatures we have solid phase (with hcp and bcc crystalline structures). There is a shallow minimum on the $P$-$T$ curve for solid-superfluid phase-boundary at $T \sim (0,5-0,6)$ K.

According to Kapitza at zero (and small) pressures and temperatures $T \approx 2.2$ K $^4$He becomes superfluid. It has zero viscosity $\eta = 0$ and according to Landau [1.1] and Tissa [1.5] can be described by two-velocity hydrodynamics. Note that for pressures $P > 26$ Bar and low temperatures $^4$He becomes solid and is either in hcp or bcc crystalline phases (see Chapter 2 for more details). At low pressures and high temperatures we have normal $^4$He which is in a phase of a normal bosonic liquid or in a gas phase (see Fig.1.1).

For temperatures $0 < T < T_C$ Landau and Tissa proposed to describe superfluid $^4$He (or He-II in terminology of Kapitza) in terms of superfluid and normal densities $\rho_n$ and $\rho_s$, and superfluid and normal velocities $\vec{v}_n$ and $\vec{v}_s$ respectively, where

$$\rho = \rho_s + \rho_n \qquad (1.1.46)$$

is a total density, and a total mass current (total density of linear momentum) is given by:

$$\vec{j} = \rho_s \vec{v}_s + \rho_n \vec{v}_n = \vec{j}_0 + \rho \vec{v}_s. \qquad (1.1.47)$$

In Eq. (1.1.47)

$$\vec{j}_0 = \rho_n (\vec{v}_n - \vec{v}_s) = \rho_n \vec{W} \qquad (1.1.48)$$

(where $\vec{W} = \vec{v}_n - \vec{v}_s$ is relative velocity) describes the motion of normal component in the reference frame where a superfluid component is at a rest $(\vec{v}_s = 0)$.

According to Landau the normal density in 3D case is given by:

$$\rho_n = -\frac{1}{3} \int \frac{\partial n_B(\varepsilon / T)}{\partial \varepsilon} p^2 \frac{d^3 \vec{p}}{(2\pi)^3} \ , \qquad (1.1.49)$$

where $n_B(\varepsilon/T)$ is bosonic distribution function of the quasiparticles and correspondingly $n_B(\varepsilon/T) = \dfrac{1}{e^{\varepsilon/T}-1}$. Thus the normal density is closely connected with the spectrum of elementary excitations [1.1, 1.8] in superfluid helium (see Fig. 1.2).

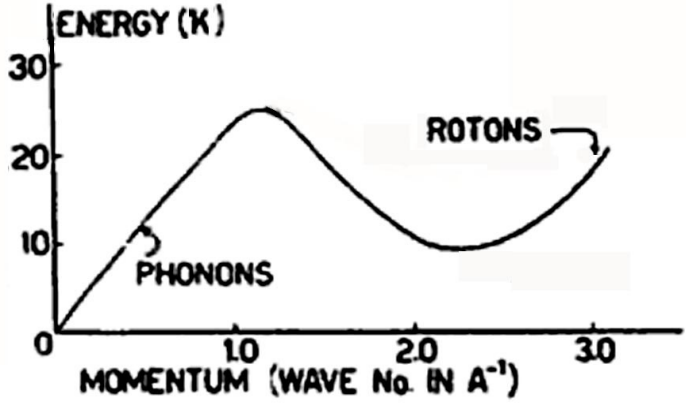

Fig. 1.2. The spectrum of elementary excitations in superfluid $^4$He [1.6, 1.10, 1.29, 1.36]. At small wave numbers the spectrum is almost linear and corresponds to phonons, while at larger wave numbers there is a minimum in $E(k)$ which corresponds to rotons.

There are two main branches of the elementary excitations according to Landau, Feynman theory - phonons with almost linear spectrum $\varepsilon = cp$ at small $p$ and rotons $\varepsilon = \Delta + \dfrac{(p-p_0)^2}{2m^*}$. The sound velocity in $^4$He at zero pressure $c = 2.4 \cdot 10^4$ cm/sec. The roton spectrum is described by the parameters $\Delta = 8.7$ K; $p_0/\hbar = 1.9 \cdot 10^8$ cm$^{-1}$ and $m^* = 0.16\, m$ of $^4$He. Correspondingly the phonon contribution to $\rho_n$ reads (see [1.10]):

$$\left(\rho_n\right)_{ph} = \frac{2\pi^2 T^2}{45 c^5}, \quad (\hbar = 1) \qquad (1.1.50)$$

while the roton contribution $\left(\rho_n\right)_r \sim e^{-\Delta/T}$.

The simple estimates show that the phonon contribution to $\rho_n$ is dominant for low temperatures $T \leq 0.6$ K, while for higher temperatures rotons are more important. Note that the condition $\left(\rho_n(T=T_C)\right)_r = \rho$ yields a rough estimate for $T_C \approx 2.2$ K in superfluid helium ($^4$He) (see Fig. 1.3).

Note also that the spectrum of elementary excitations on Fig. 1.2 according to Landau defines the critical velocity $v_C = \min\,(\varepsilon/p)$ for the destruction of the superfluid flow.

The system of hydrodynamic equations.

To derive the system of hydrodynamic equations for superfluid $^4$He we should use again the Galilean invariance for the total energy:

$$E = E_0 + \vec{j}_0 \vec{v}_s + \rho \frac{\vec{v}_s^2}{2}, \qquad (1.1.51)$$

where $E_0$ is an internal energy in the reference frame $K_0$ where $\vec{v}_s = 0$, $\vec{j} = \vec{j}_0 + \rho\vec{v}_s$. Moreover the internal energy $E_0$ satisfies the thermodynamic identity [1.1, 1.6]:

$$dE_0 = TdS + \mu d\rho + \vec{W}d\vec{j}_0, \qquad (1.1.52)$$

where $\vec{j}_0 = \rho_n \vec{W}$ and $\vec{W} = \vec{v}_n - \vec{v}_s$ is relative velocity.

Applying again Landau scheme of the conservation laws after some simple algebra we get the following system of equations:

$$\frac{\partial \rho}{\partial t} + div \vec{j} = 0, \qquad (1.1.53)$$

$$\frac{\partial \vec{v}_s}{\partial t} + \vec{\nabla}\left(\frac{v_s^2}{2} + \mu + \Psi\right) = \left[\vec{v}_s \, rot \, \vec{v}_s\right], \qquad (1.1.54)$$

$$\frac{\partial S}{\partial t} + div\left(S\vec{v}_n + \frac{\vec{q}}{T}\right) = \frac{R}{T}, \qquad (1.1.55)$$

$$\frac{\partial j_i}{\partial t} + \frac{\partial \Pi_{ik}}{\partial x_k} = 0. \qquad (1.1.56)$$

The first equation (1.1.53) yields the conservation of mass, the second equation (1.1.54) is Euler equation for superfluid velocity $\vec{v}_s$, the third equation (1.1.55) is an increase of entropy due to normal motion only (superfluid component does not carry an entropy and thus we have $S\vec{v}_n$ in (1.1.55)). Finally fourth equation (1.1.55) corresponds to the conservation of total mass current $\vec{j} = \rho_s \vec{v}_s + \rho_n \vec{v}_n$. In Eq. (1.1.55) the momentum flux:

$$\begin{aligned}\Pi_{ik} &= \rho_n v_{ni} v_{nk} + \rho_s v_{si} v_{sk} + P\delta_{ik} + \pi_{ik} = \\ &= v_{si} j_k + v_{nk} j_{0i} + P\delta_{ik} + \pi_{ik}\end{aligned} \qquad (1.1.57)$$

is a natural generalization of the expression for ordinary classical fluid, $\pi_{ik}$ is dissipative momentum flux corresponding to viscous contribution $P = TS + \mu\rho + \rho_n(\vec{v}_n - \vec{v}_s)^2 - E_0$ is the pressure and accordingly:

$$dP = SdT + \rho d\mu + \vec{j}_0 d\vec{W} \qquad (1.1.58)$$

These 4 equations (2 of them have 3 cartesian projections, so in total we have 8 equations) are consistent with the energy-conservation law (see [1.1]):

$$\frac{\partial E}{\partial t} + div \vec{Q} = 0, \qquad (1.1.59)$$

where the energy flux

$$\vec{Q} = TS\vec{v}_n + \left(\mu + \frac{v_s^2}{2}\right)\vec{j} + \vec{v}_n(\vec{v}_n \vec{j}_0) + \vec{Q}_{dis} \qquad (1.1.60)$$

and dissipative part of the energy flux:

$$Q_{dis}^i = q_i + \pi_{ik} v_{nk} + \Psi(j_i - \rho v_{ni}). \qquad (1.1.61)$$

Correspondingly the dissipative function $R$ in the equation for entropy increase (1.1.55) reads:

$$R = -\frac{\vec{q}\vec{\nabla}T}{T} - \pi_{ik}\frac{\partial v_{nk}}{\partial x_k} - \Psi div(\vec{j} - \rho\vec{v}_n) > 0. \qquad (1.1.62)$$

Note that $R$ should be positively defined, quadratic in gradients function.

### 1.1.6. First and second sound modes in superfluid liquid.

Two-fluid hydrodynamics describes not only a standard first sound wave, but also a second sound wave. First sound mode as usual demands finite compressibility of the system $\delta\rho \neq 0$; $\delta P \neq 0$ and is governed by the relation (see Subsection 1.1.3):

$$\omega^2 = c_I^2 k^2; \; c_I^2 = \left(\frac{\partial P}{\partial \rho}\right)_s \qquad (1.1.63)$$

For the second sound we can consider an incompressible superfluid where $\delta\rho \neq 0$. Thus from the continuity equation we get $div\,\delta\vec{j} = 0$ and can safely put:

$$\delta\vec{j} = \rho_s\delta\vec{v}_s + \rho_n\delta\vec{v}_n = 0 \qquad (1.1.64)$$

Hence

$$\delta\vec{v}_s = -\frac{\rho_n}{\rho_s}\delta\vec{v}_n \qquad (1.1.65)$$

in a second sound wave.

From the conservation of linear momentum we have $\dfrac{\partial\delta j_i}{\partial t} + \nabla_i\delta P = 0$ and correspondingly for $\delta j_i = 0$ we get $\delta P = 0$. On the other hand we know that

$$dP = SdT + \rho d\mu \qquad (1.1.66)$$

Hence if $\delta P = 0$ then

$$d\mu = -S\frac{dT}{\rho} \qquad (1.1.67)$$

in a second sound wave.

Now we can consider the linearized Euler equation for superfluid motion $\dfrac{\partial\delta v_{si}}{\partial t} + \nabla_i\delta\mu = 0$ and rewrite it via $\delta v_{ni}$ and $\delta T$ as follows:

$$-\frac{\rho_n}{\rho_s}\frac{\partial\delta v_{ni}}{\partial t} - \frac{S_0}{\rho}\nabla_i\delta T = 0 \quad . \qquad (1.1.68)$$

Note that the relative velocity in the second sound wave $\delta\vec{W} = \delta\vec{v}_n - \delta\vec{v}_s = \left(1 + \dfrac{\rho_n}{\rho_s}\right)\delta\vec{v}_n = \dfrac{\rho}{\rho_s}\delta\vec{v}_n$.

Finally we can use the equation for entropy increase. In the absence of dissipation it is given by:

$$\frac{\partial\delta S}{\partial t} + S_0\,div\,\delta\vec{v}_n = 0 \qquad (1.1.69)$$

Expressing again $\delta S$ via $\delta T$: $\delta S = \left(\dfrac{\partial S}{\partial T}\right)_P\delta T = \dfrac{C_p}{T_0}\delta T$ we get

$$\frac{\partial\delta T}{\partial t} + \frac{S_0 T}{C_P}\,div\,\delta\vec{v}_n = 0 \,, \qquad (1.1.70)$$

where $C_P$ is specific heat per unit mass at constant pressure, $S_0$ is entropy. The system of equations (1.1.68) and (1.1.70) allows us to find the spectrum of the second sound wave. Differentiation of Eq. (1.1.70) with respect to $\partial/\partial t$ and of Eq. (1.1.68) with respect to $\partial/\partial x_i$, and after that the substitution of (1.1.68) to (1.1.70) yields:

$$\frac{S_0}{\rho_n}\frac{\rho_s}{\rho}\Delta T - \frac{C_P}{S_0 T}\ddot{T} = 0 \quad . \qquad (1.1.71)$$

Thus for the spectrum we get

$$\omega^2 = c_{II}^2 k^2 \qquad (1.1.72)$$

where

$$c_{II}^2 = \frac{TS^2\rho_s}{C_P\rho_n\rho} \qquad (1.1.73)$$

is a velocity of a second sound squared and we skipped the subscript for the entropy ($S = S_0$). It is a second sound Goldstone mode which distinguishes superfluid liquid from a normal one where we only have overdamped temperature waves. We can say that a second sound is a sound in the subsystem of thermal normal excitations (phonons and rotons). For low temperatures $T \leq 0.6$ K in superfluid $^4$He the normal density $\rho_n$ as well as entropy $S$ and specific heat $C_P$ are mostly governed by phonons with linear spectrum. In this region in 3D system $c_{II} = c_I / \sqrt{3}$ (In 2D $c_{II} = c_I / \sqrt{2}$) (see Fig. 1.4). At higher temperatures $c_{II}$ is governed mostly by rotons. For $T = T_C$ ($\lambda$-point) $c_{II} = 0$ (see Fig. 1.4).

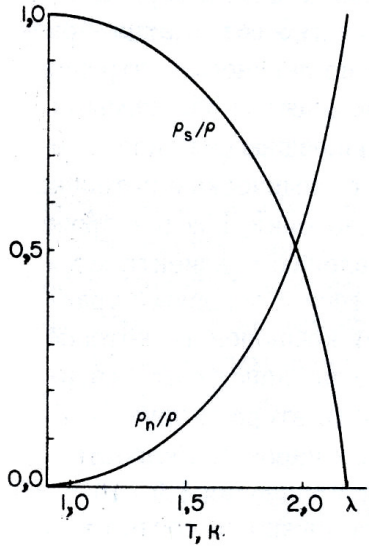

Fig. 1.3. The temperature dependence of $\rho_s/\rho$ and $\rho_n/\rho$ in superfluid $^4$He from Superfluid Hydrodynamics by S. Patterman [1.28]. For $T = 0$ $\rho_s/\rho = 1$ and $\rho_n/\rho = 0$ while in the $\lambda$-point (at $T_C = 2.2$ K) vice a versa $\rho_s/\rho = 0$ and $\rho_n/\rho = 1$.

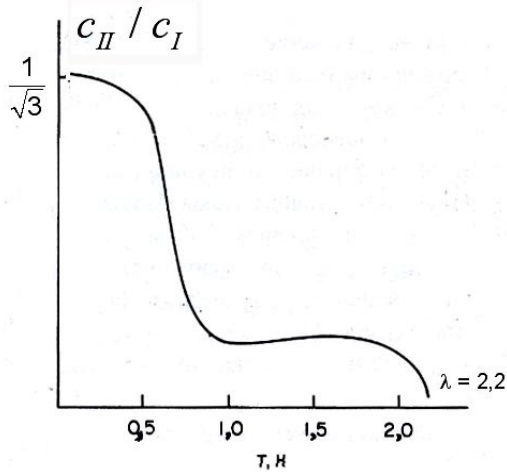

Fig. 1.4. The temperature dependence of the ratio $c_{II}/c_I$ in superfluid $^4$He from Patterman [1.28]. For $T \leq 0.5$ K $c_{II}/c_I \rightarrow 1/\sqrt{3}$. In the $\lambda$-point (at $T = 2.2$ K) $c_{II}/c_I = 0$.

1.1.7. Gross-Pitaevskii equation for dilute Bose-gas. Connection between superfluid hydrodynamics and microscopic theory at $T = 0$.

For weakly non-ideal repulsive (Bogolubov) Bose-gas we can establish the connection between the microscopic equations for superfluid hydrodynamics for irrotational liquid $\left( rot\, \vec{v}_s = 0 \right)$ at $T = 0$ and gradients and time derivatives of the order parameter:

$$\Psi(\vec{r},t)=\sqrt{n_s(\vec{r},t)}e^{i\varphi(\vec{r},t)} \qquad (1.1.74)$$

where $\rho_s(\vec{r},t)=mn_s(\vec{r},t)=m\,|\,\Psi\,|^2$.

To establish this connection we should use Gross-Pitaevskii (GP) equation for the order parameter $\Psi$ [1.50, 1.51]:

$$i\hbar\frac{\partial\Psi}{\partial t}=\left(-\frac{\hbar^2\Delta}{2m}+V_{ext}(\vec{r})+g\left|\Psi(\vec{r},t)\right|^2\right)\Psi(\vec{r},t), \qquad (1.1.75)$$

where $V_{ext}(\vec{r})$ - is an external potential (confinement potential of a magnetic trap, for example, see Chapter 4), $g=\dfrac{4\pi\hbar^2 a}{m}$ is a coupling constant and $a$ is an s-wave scattering length. GP equation is valid if the Bose-gas is dilute $na^3 << 1$ and the number of particles in the trap is much larger than 1. As usual we can introduce $\vec{v}_s=\dfrac{\hbar}{m}\vec{\nabla}\varphi$ for superfluid velocity and $\rho_s=m\left|\Psi(\vec{r},t)\right|^2$ for superfluid density. Substituting (1.1.74) into (1.1.75) and separating real and imaginary parts in (1.1.75) we obtain 2 equations for superfluid hydrodynamics at $T=0$ and in the absence of vortices $\left(rot\,\vec{v}_s=0\right)$:

$$\begin{aligned}
&\frac{\partial\rho}{\partial t}+div(\rho\,\vec{v}_s)=0,\\
&\frac{\partial\vec{v}_s}{\partial t}+\vec{\nabla}\left(\mu+\frac{v_S^2}{2}\right)=0
\end{aligned} \qquad (1.1.76)$$

In (1.1.76) we introduce a chemical potential:

$$m\mu=V_{ext}(\vec{r})+gn \qquad (1.1.77)$$

assuming that the phase of the order parameter $\varphi$ in (1.1.74) varies more rapidly in space and time than $\left|\Psi\right|=\sqrt{n_s}$ (quasiclassical or eikonal approximation [1.2]). Note that $\delta\mu=\dfrac{1}{\rho}\delta P$ in (1.1.77) and hence the pressure $P$ contains the term $gn^2/2$ [1.50].

## 1.2. Hydrodynamics of rotating superfluids.

In this Section we provide the basic notion for the standard hydrodynamics of a superfluid liquid with large number of quantized vortices, which includes the elasticity of the vortex lattice and the scattering of normal excitations on the vortices [1.6, 1.11, 1.12]. We will start the presentation with the short description of the famous Andronikashvilli experiments which measure the ratio of $\rho_n/\rho$ or, more precisely, non-classical moment of inertia in slowly rotating superfluid $^4$He [1.7].

### 1.2.1. Andronikashvilli experiments in rotating helium.

If we rotate a cylindrical vessel with an ordinary normal (classical) liquid (see Fig. 1.5), than due to the boundary conditions in the viscous liquid on the container wall the tangential component of the velocity $\vec{v}_\tau=\vec{V}_{wall}=\left[\vec{\Omega}\vec{R}\right]$ , where $\vec{\Omega}$ is an angular velocity of the rotation, $R$ is the radius of a cylindrical vessel which contains the liquid. Thus in this case all the liquid participates in solid-state rotation. In contrast to this in superfluid liquid only normal component follows the walls of the container $\vec{v}_n=\left[\vec{\Omega}\vec{R}\right]$, while frictionless superfluid component stays at rest at slow rotations.

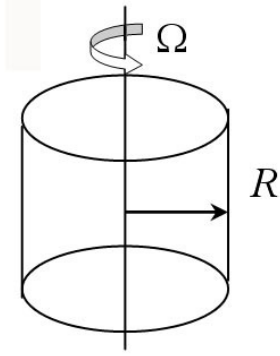

Fig. 1.5. Cylindrical vessel with superfluid $^4$He rotates with an angular velocity $\vec{\Omega}$. $R$ is the radius of the vessel.

As a result at small angular frequencies the response of a superfluid on rotation is governed not by the total moment of inertia $I \sim MR^2$ (M- is the total mass of the liquid in the container), but only by its normal fraction $\frac{\rho_n}{\rho} I \equiv \left(1 - \frac{\rho_s}{\rho}\right) I$. It is non-classical moment of inertia. This fact was used in Andronikashvilli experiments to measure $\rho_n/\rho$ [1.7]. In his first experiments he used a sequence of parallel disks with small distance (smaller than viscous penetration length $\delta_\eta = \left(\frac{2\eta}{\omega \rho_n}\right)^{1/2}$ see[1.1, 1.7] and Fig. 1.6) between them. Then he can consider that all the normal part of the volume of liquid $^4$He between the rotating disks participates in rotation.

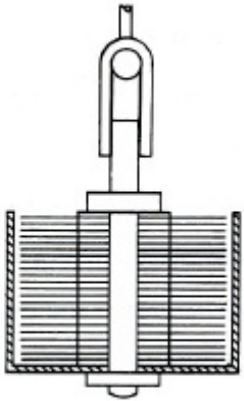

Fig. 1.6. First experiments of Andronikashvilli [1.7] with a sequence of parallel disks for measurements of the non-classical moment of inertia. The viscous penetration length $\delta_\eta = \left(\frac{2\eta}{\omega \rho_n}\right)^{1/2}$ is smaller than the distance between the disks [1.28].

The scheme of updated experiments on the measurements of non-classical moment of inertia is presented on Fig. 1.7a. In updated Andronikashvilli-type of experiments torsional oscillator is used. It is an ideal method for measuring of the non-classical moment of inertia (and thus of a transition to a superfluid state $\rho_s/\rho \neq 0$) especially in solid $^4$He (see Chapter 2 and [1.37]). In these experiments the time-period $\tau_0$ for the returning of the rode (of the string) in the initial position is measured using a torsional oscillator with a resonance frequency $\nu_0 = 1/\tau_0$.

More specifically experimentalists measure the period $\tau_0 = 2\pi\sqrt{I / K}$ for the container with Helium which is attached to a torsional rode. Note that $I$ is the moment of inertia and $K$ is rotating rigidity of the rode.

This scheme was used in recent experiments of Chan et al. [1.37] on the search of supersolidity [1.38, 1.39] in solid crystalline ⁴He. The typical resonant period in this type of experiments $\tau_0 \sim 1$ msec, stability in $\tau$ is given: $\delta\tau \sim \dfrac{\delta K}{K}\tau_0 \sim 5 \cdot 10^{-7}$, and the quality factor $Q = \nu_0/\Delta\nu \sim 2 \cdot 10^6$ (see Fig. 1.7b), where $\nu_0 = 1/\tau_0 \sim 1000$ Hz is the resonance frequency.

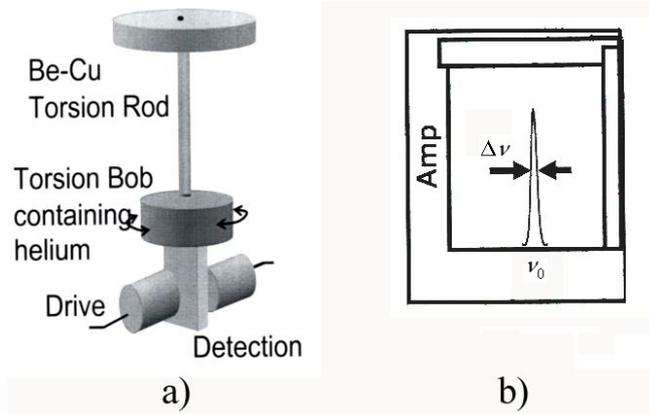

a)                    b)

Fig. 1.7. A qualitative scheme of updated Andronikashvilli-type of experiments on torsional oscillator by Chan et al. [1.37] for the search of supersolidity in solid ⁴He. On Fig. 1.7a we present a sketch of the experimental setup with torsional bob containing helium. On Fig. 1.7b we show the amplitude of the oscillations and the quality factor of the measurements $Q = \nu_0/\Delta\nu$ for the resonance frequency $\nu_0 = 1/\tau_0$, where $\tau_0 = 2\pi\sqrt{I / K}$ is the resonance period, $I$ is the moment of inertia and $K$ is rotating rigidity of the rode.

### 1.2.2. Feynman-Onsager quantized vortices. Critical angular velocities $\Omega_{C1}$ and $\Omega_{C2}$

The situation with non-classical moment of inertia occurs at low angular velocities $\Omega < \Omega_{C1}$. For $\Omega < \Omega_{C1}$ ($\Omega_{C1}$ is the first critical angular velocity) the first Feynman-Onsager [1.8,1.9] quantized vortex line appears in the center of cylindrical vessel. According to Feynman and Onsager the velocity field of single quantized vortex can be found from the quantization of the circulation of a superfluid velocity:

$$\Gamma_v = \oint_C \vec{v}_s d\vec{l} = 2\pi\frac{\hbar}{m}, \qquad (1.2.1)$$

where $\hbar$ is Planck constant, $m$ is the mass of ⁴He atom.

In fact $\Gamma_v = \vartriangle\varphi\hbar / m$ where $\vartriangle\varphi = 2\pi$ is the phase change on the contour. Note that in the absence of vortices we can introduce a scalar complex order-parameter $\Psi = \sqrt{n_s}e^{i\varphi}$ which describes a superfluid state of ⁴He [1.10]. Correspondingly $|\Psi|^2 = n_s$ and $mn_s = \rho_s$ - superfluid density, while $\vec{v}_s = \hbar / m \vec{\nabla}\varphi$ is superfluid velocity. Thus $2\pi\hbar / m$ is a natural unit to measure the circulation of a superfluid velocity in (1.2.1). It is often called as a circulation quanta. Note that, in principle, the vortices with larger amount of circulation quanta $\Gamma_v = 2\pi n\hbar / m$ $(n > 1)$ can be also stabilized in superfluid ⁴He. When the first vortice appears at $\vec{r} = 0$ (in the center of cylindrical vessel) the superfluid velocity still can be represented as $\vec{v}_s = \hbar / m \vec{\nabla}\varphi$ for all distances $\vec{r} \neq 0$. Hence it immediately follows that $\Gamma_v = \oint_C \vec{v}_s d\vec{l} = \dfrac{\hbar}{m}\Delta\varphi$. In the absence of vortices the phase change $\vartriangle\varphi = 0$. In the presence of a first quantized vortice $\vartriangle\varphi = 2\pi$ and we get (1.2.1).

Correspondingly from the Gauss theorem $\iint_S rot\vec{v}_s d\vec{s} = \oint_C \vec{v}_s d\vec{l} = 2\pi\dfrac{\hbar}{m}$ we easily obtain that:

$$rot\,\vec{v}_s = \frac{2\pi\hbar}{m}\delta(\vec{r})\vec{e}_z \qquad (1.2.2)$$

The solution for $\vec{v}_s$ which satisfies (1.2.1) (and also (1.2.2)) reads:

$$\vec{v}_s = \frac{\hbar}{m}\frac{\vec{e}_\varphi}{r} \qquad (1.2.3)$$

where $\vec{e}_\varphi$ is a tangential unit vector (remind that in cylindrical coordinates $(r, \varphi, z)$ we have the triade of mutually perpendicular unit vectors $\vec{e}_\varphi, \vec{e}_r, \vec{e}_z$ and thus $d\vec{l} = r\,d\varphi\,\vec{e}_\varphi$ in (1.2.1)).

Critical angular velocity $\Omega_{C1}$

The first critical angular velocity $\Omega_{C1}$ can be found according to [1.10] from the minimization of the Free-energy in the rotating frame:

$$\Delta F = E_v - \vec{M}_v\vec{\Omega} \leq 0 \quad \text{for } \Omega \geq \Omega_{C1} \qquad (1.2.4)$$

where $\Delta E = E_v$ is the kinetic energy of the vortex line

$$E_v = \int \rho_s \frac{v_s^2}{2}dV = \pi L\frac{\rho_s}{2}\frac{\hbar^2}{m^2}\ln\frac{R}{d}, \qquad (1.2.5)$$

$L$ is the height of the container and $d$ is the vortex-core, which is normal due to the violation of Landau criteria of superfluidity ( $v_s = \dfrac{\hbar}{md} \geq c_I$ - exceeds the sound velocity for $d$ of the order of interatomic distance). Note that from the spectrum of quasiparticle excitations in superfluid $^4$He (see Fig 1.2) we get $v_c = \min(\varepsilon / p) = c_I$ for the critical velocity of a destruction of the superfluid flow. In superfluid $^4$He $d \sim (3 \div 4)$ Å, $R \sim 0.1$ cm for a typical radius of the container with helium.

In the same time $M_v$ in (1.2.4) is an angular momentum associated with a vortex line:

$$M_v = \rho_s\int rv_s dV = \pi\rho_s L\frac{\hbar}{m}\frac{R^2}{2} \quad . \qquad (1.2.6)$$

Correspondingly the first critical angular velocity:

$$\Omega_{C1} = \frac{E_v}{M_v} = \frac{\hbar}{mR^2}\ln\frac{R}{d} \qquad (1.2.7)$$

can reach 0,1 rotations/sec in $^4$He.

Critical angular velocity $\Omega_{C2}$

At very large angular velocities the normal cores of the vortex lines start to overlap. As a result all the volume of a superfluid helium becomes normal. It is evident that

$$\Omega_{C2} = \frac{\hbar}{m\pi d^2} \sim 10^{11}\,\text{sec}^{-1}, \qquad (1.2.8)$$

where $S = \pi d^2$ is an area of one vortex core. Thus $\Omega_{C2}$ in dense superfluid helium is very high and practically unachievable. Note that in dilute weakly non-ideal Bose-gasses in a confined geometry of magnetic traps from the solution of Gross-Pitaevskii equation [1.10, 1.50, 1.51] the vortex core $\xi_0 = \dfrac{1}{\sqrt{na}}$, where $a$ is an s-wave scattering length (see [1.10, 1.39, 1.40] and Chapter 6). Thus in dilute case $na^3 < 1$: $\xi_0 >> a$ and the condition $\Omega_{C2} = \dfrac{\hbar}{m\pi\xi_0^2}$ is easier to fulfill (note that often $a \sim d$ in a dilute repulsive gas).

Of course from the definitions (1.2.7), (1.2.8) we have very direct correspondence between critical angular velocities $\Omega_{C1}$ and $\Omega_{C2}$ in superfluid $^4$He and critical magnetic fields in type-II superconductors.

<u>Macroscopic averaging for a large number of vortices</u>

In between $\Omega_{C1}$ and $\Omega_{C2}$ we have a system of quantized linear vortices. Their velocity circulations $\Gamma_v$ (see (1.2.1) and Fig. 1.8) cancel each other inside the vortex region and enhance each other outside the vortex region - thus the vortices of the same "charge" (same circulation) mimic solid-state rotation for superfluid component effectively performing macroscopic averaging (see [1.11, 1.6]) over an area containing a large number of vortices (but still much smaller than a container area $\pi R^2$). We can introduce an averaged vorticity $\vec{\bar{\omega}} = rot\vec{v}_s \approx 2\vec{\bar{\Omega}}$ (thus an averaged superfluid vorticity $\vec{v}_s \neq \dfrac{\hbar}{m}\vec{\nabla}\varphi$). The average vorticity is connected with the number of vortices in 1 cm$^2$ (with the 2D vortex density):

$$n_v \frac{\hbar}{m} = \left|\vec{\bar{\omega}}\right| \approx 2\Omega, \qquad (1.2.9)$$

where $n_v = 1/\pi b^2$ and $b$ is the mean distance between the vortices. The unit vector $\vec{v} = \dfrac{\vec{\bar{\omega}}}{|\vec{\bar{\omega}}|}$ defines the average orientation of vortex lines. We can say that the macroscopic averaging is valid in the long wavelength-limit $\hbar >> b$ or equivalently when $kb << 1$.

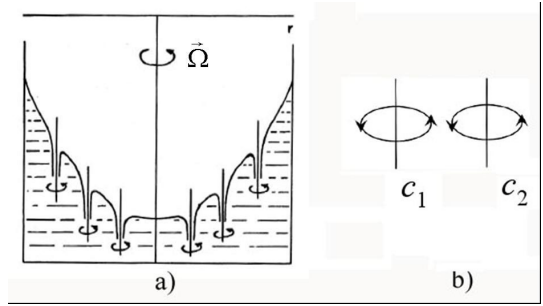

Fig. 1.8. a) Rotating superfluid helium with large number of vortices for angular velocities $\Omega_{C1} << \Omega << \Omega_{C2}$. b) The vortex circulations cancel each other inside the vortex region and enhance each other outside the vortex region. Thus the superfluid component mimics the solid-state rotation.

1.2.3. Vortex lattice. Nonlinear elasticity theory. Vorticity conservation law.

For $\Omega_{C1} << \Omega << \Omega_{C2}$ the developed structure of the vortices form a triangular lattice similar to Abrikosov lattice in type-II superconductors in a magnetic field (see [1.26] and Fig 1.9). The state of the vortex lattice, whose parameters vary slowly in space and time, in the non-linear elasticity theory [1.12] is convenient to describe by the vectors the $\vec{e}_a(\vec{r}, t)$, $a = 1, 2$ which are equal to the local values of the vectors of elementary translations of the lattice in the 2D plane perpendicular to the vortex lines.

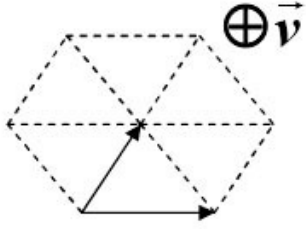

Fig 1.9. 2D triangular vortex lattice for $\Omega_{C1} \ll \Omega \ll \Omega_{C2}$ in the plane perpendicular to the vortex lines (to the unit vector $\vec{v} = \dfrac{\vec{\omega}}{|\vec{\omega}|}$, where $\vec{\omega} = rot\,\vec{v}_s$). The lattice is convenient to describe in terms of the vectors $\vec{e}_1(\vec{r},t)$, $\vec{e}_2(\vec{r},t)$, which are equal to the local values of the vectors of elementary translations of the 2D lattice.

We can also define the vectors of the reciprocal lattice: $\vec{e}^a = -\dfrac{1}{s}\varepsilon^{ab}\left[\vec{v}\vec{e}_b\right]$ where $s$ - is the area of elementary cell in real lattice, $\varepsilon^{12} = -\varepsilon^{21} = 1$; $\varepsilon_{11} = \varepsilon_{22} = 0$, $\vec{v} = \dfrac{1}{s}\left[\vec{e}_1\vec{e}_2\right]$ is a unit vector parallel to the vortex lines (we will see that $\vec{v} = \dfrac{\vec{\omega}}{|\vec{\omega}|}$, where $\vec{\omega} = rot\,\vec{v}_s$ is vorticity – see Fig 1.9). Note that $\left|\vec{e}^a\right| \neq 1$ - vectors of the reciprocal lattice $\vec{e}^1$ and $\vec{e}^2$ are not the unit vectors. There are evident equalities:

$$\vec{e}^a\vec{e}_b = \delta^a_b; \quad e_{ai}e^a_k = \delta_{ik} - v_i v_k \qquad (1.2.10)$$

between real and reciprocal lattices, where $\{i, k\} = \{x, y, z\}$ are Cartesian coordinates. Nonlinear elasticity is described similar to General Theory of Relativity by metric tensors [1.42]:

$$g_{ab} = \vec{e}_a\vec{e}_b; \quad g^{ab} = \vec{e}^a\vec{e}^b; \quad g_{ac}g^{cb} = \delta^b_a. \qquad (1.2.11)$$

We can also introduce a local velocity of the vortex lattice $\vec{v}_L(\vec{r},t)$ in the direction perpendicular to the vortex lines. Thus $(\vec{v}_L\vec{v}) = 0$ by definition. If $\vec{v}_L$ is known and the functions $\vec{e}^a(\vec{r})$ at the initial time moment are specified, we can determine $\vec{e}^a$ at the nearby time moment, i.e. the derivative $\dfrac{\partial \vec{e}^a}{\partial t}$ via $\vec{v}_L(\vec{r},t)$. To establish this connection we express the physically infinitely small differential of the coordinates $d\vec{r}$ (which is large compared with the lattice period, but small compared with the distance over which the vortex configuration varies) in the form (see [1.12] for more details):

$$d\vec{r} = \vec{e}_a dN^a \qquad (1.2.12)$$

The quantities $dN^a$ for two lattice points separated by a distance $d\vec{r}$ are the two projections of $d\vec{r}$ measured in units of the corresponding lattice periods. These quantities, obviously, are not altered by an arbitrary elastic deformation (in the absence of dislocations).

From (1.2.12) we get:

$$dN^a = \vec{e}^a d\vec{r} = \vec{e}^a_0 d\vec{r}_0, \qquad (1.2.13)$$

where $\vec{e}^a_0$ and $d\vec{r}_0$ are respectively the period of the reciprocal lattice and the difference between the coordinates of the undeformed state. From (1.2.13) we see that:

$$\vec{e}^a = \vec{\nabla}N^a. \qquad (1.2.14)$$

Correspondingly

$$rot\,\vec{e}^a = rot\left(\vec{\nabla}N^a\right) = 0. \qquad (1.2.15)$$

Thus the specified functions $N^a(\vec{r},t)$ determine completely the configuration of the vortex lines. We can say that $N^a(\vec{r},t)$ defines the number of sites (or nodes) of the vortex lattice in the direction of an elementary vector $\vec{e}^a$. Moreover if we consider the vortex lattice without vacancies or dislocations, than each site of the lattice is singly occupied and the number of sites (or nodes) is conserved. We can conclude that $N^a$ is a topological invariant (see Andreev, Kagan [1.12]) and we have a very convenient technique for all the problems of the non-linear theory of elasticity.

To establish the relation between $\dfrac{\partial \vec{e}^a}{\partial t}$ and $\vec{v}_L(\vec{r},t)$ we note that the unit vector $\vec{n} = \vec{e}^1 / e^1$ is the normal to the corresponding crystallographic plane (to the plane defined by the vectors $\vec{e}_2$ and $\vec{v}$) and $d = 1/e^1$ is proportional to the local value of the interplanar distance. From simple geometric considerations we get:

$$\dot{\vec{n}} + V(\vec{n}\vec{\nabla})\vec{n} = -\vec{\nabla}V + \vec{n}(\vec{n}\vec{\nabla}V);$$
$$\dot{d} + V(\vec{n}\vec{\nabla})d = d(\vec{n}\vec{\nabla})V, \qquad (1.2.16)$$

where $V = (\vec{v}_L\vec{n}) = (\vec{v}_L\vec{e}^1)/e^1$ is the projection of the velocity normal to the considered crystallographic plane.

From (1.2.16) we obtain:

$$\dot{\vec{e}}^1 + \vec{\nabla}(\vec{e}^1\vec{v}_L) = V[\vec{n}\,rot\,\vec{e}^1]. \qquad (1.2.17)$$

A similar relation holds also for $\vec{e}^2$.

Taking into account that $rot\,\vec{e}^a = 0$ we finally get:

$$\frac{\partial \vec{e}^a}{\partial t} + \vec{\nabla}(\vec{e}^a\vec{v}_L) = 0. \qquad (1.2.18)$$

Thus comparing (1.2.18) and (1.2.14) we obtain:

$$\vec{v}_L = -\vec{e}_a \dot{N}^a \qquad (1.2.19)$$

Note that the variables analogous to $N^a$ were considered for the vortex lattice by Volovik and Dotsenko [1.23].

In macroscopic hydrodynamics of a rotating superfluid liquid one introduce an averaged velocity $\vec{v}_s$ of the superfluid component, whose $rot$ is determined by the direction and the density of the vortex lines, as well as the circulation quantum $2\pi\hbar/m$. Since the unit area $s$ of the real lattice is equal to $g^{1/2}$, where $g$ is determinant of the metric tensor $g_{ab}$, we have:

$$rot\vec{v}_s = \frac{2\pi\hbar}{ms}\vec{v} = \frac{2\pi\hbar}{mg}\big[\vec{e}_1\vec{e}_2\big] = \frac{2\pi\hbar}{m}\big[\vec{e}^1\vec{e}^2\big]. \qquad (1.2.20)$$

By differentiating (1.2.20) with respect to time and using (1.2.18) we obtain the equation of the vorticity conservation:

$$\dot{\vec{\omega}} = rot\,\dot{\vec{v}}_s = rot\big[\vec{v}_L\vec{\omega}\big]. \qquad (1.2.21)$$

### 1.2.4. Hydrodynamics of slow rotations. Hall-Vinen friction coefficients $\beta$ and $\beta'$.

In accord with the general Landau method of deriving the hydrodynamics equations from the conservation laws [1.1], we introduce in the case of slow rotations two velocities $\vec{v}_s$ and $\vec{v}_n$ and search for the system of equations in the form of the conservation laws (similar to the system of equations (1.1.53) – (1.1.56)):

$$\dot{\rho} + div\,\vec{j} = 0; \quad \frac{\partial j_i}{\partial t} + \frac{\partial}{\partial x_k}\Pi_{ik} = 0$$

$$\dot{S} + div\left(S\vec{v}_n + \vec{q}/T\right) = \frac{R}{T}; \quad \dot{\vec{v}}_s = \left[\vec{v}_L\, rot\,\vec{v}_s\right] + \vec{\nabla}\,\varphi, \tag{1.2.22}$$

where $\rho$, $S$ and $\vec{j}$ are the mass, entropy and momentum per unit volume, while $\Pi_{ik}$, $\vec{q}$, $R > 0$ and $\varphi$ are the quantities to be determined. We must also find the connection between the velocities $\vec{v}_s, \vec{v}_n$ and $\vec{v}_L$. The criterion is the requirement that the energy conservation equation should be automatically obtained form the system of Eqs. (1.2.22).

The Galilean transformation formulas:

$$E = \rho\frac{v_s^2}{2} + \vec{j}_o\vec{v}_s + E_0; \quad \vec{j} = \vec{j}_o + \rho\vec{v}_s \tag{1.2.23}$$

connect again the energy $E$ per unit volume and the momentum $\vec{j}$ with their values $E_0$ and $\vec{j}_0$ in a system where $\vec{v}_s = 0$. The energy $E_0$ can be regarded as a function of $\rho$, $S$, $\vec{j}_0$, and the metric tensor $g^{ab}$, so that:

$$dE_0 = TdS + \mu d\rho + (\vec{v}_n - \vec{v}_s, d\vec{j}_o) + \frac{1}{2}h_{ab}dg^{ab}. \tag{1.2.24}$$

Equations (1.2.22), (1.2.23), (1.2.24) differ from the corresponding equations of Bekarevich, Khalatnikov [1.6] in that the vortex conservation condition (1.2.21) is taken into account in (1.2.22) in the equation of the superfluid motion, and also in that the dependence of the energy of the deformation of the vortex lattice is fully taken into account in the identity (1.2.24).

Differentiating with respect to time the first equation for energy in (1.2.23) we obtain by using (1.2.23) and (1.2.24):

$$\dot{E} + div\{\vec{Q}_0 + \vec{q} + v_{nk}\pi_{ik} + \Psi(\vec{j} - \rho\vec{v}_n) + v_{Lk}h_{ai}e_i^a e_k^b\} =$$

$$= R + \frac{\vec{q}\vec{\nabla}T}{T} + \pi_{ik}\frac{\partial v_{ni}}{\partial x_k} + \Psi div(\vec{j} - \rho\vec{v}_n) + \{\vec{v}_L - \vec{v}_n, [rot\,\vec{v}_s, \vec{j} - \rho\vec{v}_n] + \vec{e}^a div(h_{ab}e^b)\}. \tag{1.2.25}$$

While deriving Eq. (1.2.25) we used Eq. (1.2.18) for $\dfrac{\partial\vec{e}^1}{\partial t}$ and $\dfrac{\partial\vec{e}^2}{\partial t}$ which yield:

$$\frac{\partial g^{ab}}{\partial t} = \frac{\partial}{\partial t}(\vec{e}^a\vec{e}^b) = -\left(\vec{e}^b\vec{\nabla}\right)\left(\vec{v}_L\vec{e}^a\right) - \left(\vec{e}^a\vec{\nabla}\right)\left(\vec{v}_L\vec{e}^b\right). \tag{1.2.26}$$

In (1.2.25)

$$\vec{Q}_0 = \left(\mu + \frac{v_s^2}{2}\right)\vec{j} + ST\vec{v}_n + \vec{v}_n(\vec{j}_o\vec{v}_n),$$

$$\pi_{ik} = \Pi_{ik} - P\delta_{ik} - v_{si}j_k - v_{nk}j_{0i} - h_{ab}e_i^a e_k^b. \tag{1.2.27}$$

$$\Psi = -\left(\mu + \frac{v_s^2}{2} + \varphi\right); \quad P = -E_0 + TS + \mu\rho + (\vec{v}_n - \vec{v}_s, \vec{j}_o).$$

Note that from (1.2.26) it follows the equality

$$\frac{1}{2}h_{ab}\dot{g}^{ab} + \frac{1}{2}v_{ni}h_{ab}\frac{\partial g^{ab}}{\partial x_i} =$$

$$= -div\{h_{ab}\vec{e}^a(\vec{v}_L - \vec{v}_n, \vec{e}^b)\} - \frac{\partial v_{ni}}{\partial x_k}(h_{ab}e_i^a e_k^b) + \{\vec{v}_L - \vec{v}_n, \vec{e}^a div(h_{ab}\vec{e}^b)\} \tag{1.2.28}$$

The form of (1.2.25), (1.2.27) enables us to determine the energy-flux vector:

$$\vec{Q} = \vec{Q}_0 + \vec{q} + v_{nk}\pi_{ki} + \Psi(\vec{j} - \rho\vec{v}_n) + v_{Lk}h_{ab}e_i^a e_k^b \tag{1.2.29}$$

and the dissipative function

$$R = -\frac{\vec{q}\vec{\nabla}T}{T} - \pi_{ik}\frac{\partial v_{ni}}{\partial x_k} - \Psi div(\vec{j} - \rho\vec{v}_n) - \left\{\vec{v}_L - \vec{v}_n,\left[rot\,\vec{v}_s,\vec{j} - \rho\vec{v}_n\right] + \vec{e}^{\,a}div(h_{ab}\vec{e}^{\,b})\right\} \quad (1.2.30)$$

From the condition that $R$ is positive it follows that the unknown quantities $\vec{q}$, $\pi_{ik}$, $\Psi$ and $\vec{v}_L - \vec{v}_{s\perp}$ - (the symbol $\perp$ means that we are dealing with the projection of the corresponding vector on a plane perpendicular to $\vec{v}$ ) can be written in the general case as linear combinations of all conjugated variables $\vec{\nabla}T$, $\frac{\partial v_{ni}}{\partial x_k}$, etc., contained in (1.2.30). We shall not write out the unwieldy general formulas and confine ourselves, as usual, to the mutual friction effects described by the last term in (1.2.30). We have:

$$\vec{v}_L - \vec{v}_{n\perp} = -\alpha\left\{\vec{j}_\perp - \rho\vec{v}_{n\perp} + \frac{mg^{1/2}}{2\pi\hbar}\left[\vec{e}^{\,a}\vec{v}\right]div(h_{ab}\vec{e}^{\,b})\right\} -$$

$$-\beta\left\{\left[\vec{v},\vec{j} - \rho\vec{v}_n\right] + \frac{mg^{1/2}}{2\pi\hbar}\vec{e}^{\,a}div(h_{ab}\vec{e}^{\,b})\right\}, \qquad (1.2.31)$$

where $\alpha$ and $\beta$ are certain coefficients with $\beta > 0$. Since at $T = 0$ there is no normal part, so that $\vec{v}_L$ should be independent of $\vec{v}_n$, it follows that the constant $\alpha$ should be equal to - $1/\rho$. If we put $\alpha = -1/\rho_s + \beta'$, the constants $\beta$ and $\beta'$ will vanish at $T \to 0$, that coincide with the friction coefficients introduced by Bekarevich and Khalatnikov [1.6] and differ from the constants of Hall and Vinen [1.15] by a factor $\rho_n/2\rho\rho_s$.

At zero temperature we obtain from (1.2.31):

$$\vec{v}_L - \vec{v}_{s\perp} = \frac{mg^{1/2}}{2\pi\hbar\rho}\left[\vec{e}^{\,a}\vec{v}\right]div(h_{ab}\vec{e}^{\,b}) \ , \quad (1.2.32)$$

which is the generalization of a known relation [1.23, 1.43] to the case of arbitrary and not small deformations of the vortex lattice.

<u>The physical meaning of $\beta$, $\beta'$ - coefficients. Elementary estimates.</u>

The coefficient $\beta$ is a dissipative coefficient, while $\beta'$ is a Hall-like dissipationless coefficient. According to Iordansky [1.25] to get the feeling about coefficients $\beta$-s we should consider the elementary excitations with momentum $\vec{p}$ and energy $\varepsilon_0(\vec{p})$ approaching the vortex and thus experiencing the velocity field of the vortex $\vec{v}_s$ (see Fig. 1.10).

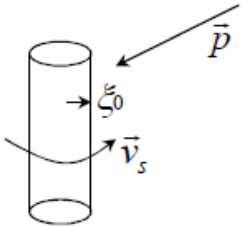

Fig. 1.10. The scattering of the normal excitations with momentum $\vec{p}$ on the vortex line, which creates the velocity field $\vec{v}_s$ (outside the vortex core $\xi_0 \sim d$).

According to the requirements of the Galilean invariance $\varepsilon(\vec{p}) = \varepsilon_0(\vec{p}) - \vec{p}\vec{v}_s$ or more rigorously

$$\varepsilon(\vec{p}) = \varepsilon_0(\vec{p}) - \vec{p}(\vec{v}_s - \vec{v}_n) \qquad (1.2.33)$$

for the spectrum of quasiparticles (if normal excitations have a drift velocity $\vec{v}_n$). Thus according to Iordansky and Sonin [1.24, 1.25], an effective interaction:

$$H_{int} = \vec{p}(\vec{v}_s - \vec{v}_n) \qquad (1.2.34)$$

arises in our system. According to the elementary kinetic theory

$$\beta \sim \frac{1}{n_V \sigma l} \sim \frac{1}{n_V \sigma v \tau_U} \quad , \qquad (1.2.35)$$

where $v = \dfrac{\partial \varepsilon_0}{\partial p}$ is group velocity for elementary excitations, $\sigma$ is the cross-section, $n_V$ is the density of vortices, $\tau_U$ is the scattering time due to umklapp processes $1/\tau_U \sim \beta \rho_s \Omega$ [1.15]. Note that we need umklapp processes [1.32] connected with the vortex lattice to have the relaxation of the quasiparticle linear momentum. For a rough estimate we can find an impact parameter $a$ ($\sigma = \pi a^2$) from the quasiclassical turning point, that is $\varepsilon_0(\vec{p}) - \vec{p}\vec{v}_s(a) = 0$ (for $\vec{v}_n = 0$). For a single vortex line $\vec{v}_s = \dfrac{\hbar}{mr}\vec{e}_\varphi$ and if we have, for example, a roton with an energy

$$\varepsilon_0(\vec{p}) = \Delta_0 + \frac{(p - p_0)^2}{2m_*} \text{ for } p \approx p_0, \text{ then } \Delta_0 = \frac{\hbar p_0}{ma} \text{ or}$$

$$a = \frac{\hbar p_0}{m\Delta_0} \geq (\xi_0 \sim d) \qquad (1.2.36)$$

for the impact parameter. If $a \sim \xi_0 \sim d$ we can say that the normal component scatters on the vortex cores, which are also normal.

1.2.5. Linearization of the elasticity theory. Connection between $\vec{v}_s$ and $\vec{u}$ in linearized theory.

When we expand the hydrodynamic energy and hydrodynamic equations in powers of deformation it is convenient to put

$$N^a = N_0^a - u^a \qquad (1.2.37)$$

for the number of nodes $N^a$, where $N_0^a = \vec{e}_o^{\,a}\vec{r}$ and $\vec{e}_o^{\,a}$ is a non-deformed unit vector of a reciprocal lattice. We can also introduce the displacement

$$\vec{u} = \vec{e}_{oa}u^a \qquad (1.2.38)$$

which is a 2D vector perpendicular to the axes of undeformed vortices.

Shear, compression and bending elastic modulus.

We represent the elastic energy (the energy of vortex repulsion due to the lattice rigidity) $E_{el}$ per unit lattice volume in the form:

$$E_{el} = E_1 + E_2 \qquad (1.2.39)$$

where

$$E_1 = \frac{\pi \rho_s}{2}\left(\frac{\hbar}{m}\right)^2 n_V \ln \frac{1}{n_V d^2} \qquad (1.2.40)$$

is a compression energy, $n_V = \dfrac{1}{s} = \dfrac{1}{\sqrt{g}}$ is the density of vortices on unit area, $E_2$ is shear energy (which depends on the shape of a deformed unit cell) (see Elasticity theory [1.18] for the details).

Accordingly

$$h_{ab} = \frac{2\partial E_{eL}}{\partial g^{ab}} = h_{ab}^{(1)} + h_{ab}^{(2)} . \quad (1.2.41)$$

Differentiating $E_1$ with respect to $g^{ab}$ with the allowance for the identity $-\frac{1}{g}dg = g_{ab}dg^{ab}$ and linearizing the result with respect to the deviations $\delta g^{ab}$ of the metric tensor from its value $g^{ab}(0)$ in an undeformed triangular lattice, we obtain:

$$h_{ab}^{(1)} = \rho_s \frac{\hbar\Omega}{m} \left\{ \ln\frac{b}{d}\left( g_{ab}^{(0)} - \delta g_{ab} + \frac{1}{2}g_{ab}^{(0)}\delta g_c^c \right) - \frac{1}{4}g_{ab}^{(0)}\delta g_c^c \right\}, \quad (1.2.42)$$

where $\pi b^2 \sim \frac{\hbar}{m\Omega} \sim \frac{1}{n_V}$ is the mean distance between the vortex lines, $\Omega = (\pi\hbar/m)g_o^{-1/2}$ is the angular velocity of the rotation, and

$$\delta g_{ab} = g_{ac}^{(0)}g_{bd}^{(0)}\delta g^{cd}; \quad \delta g_c^c = g_{cd}^{(0)}\delta g^{cd}. \quad (1.2.43)$$

The constant term appears in (1.2.42) because at equilibrium the energy that has a minimum is the energy in the rotating coordinate frame. We express the shear part $h_{ab}^{(2)}$ of the full $h_{ab}$ in the form:

$$h_{ab}^{(2)} = \mu_s\left( \delta g_{ab} - \frac{1}{2}g_{ab}^{(0)}\delta g_c^c \right), \quad (1.2.44)$$

where $\mu_s = \rho_s\frac{\hbar\Omega}{4m}$ is the shear modulus calculated by Tkachenko [1.13] for the triangular lattice.

The quantities $\delta g_{ab}$ and $\vec{e}^a$ can be easily expanded in the displacement $\vec{u}$ by using Eqs. (1.2.11) and (1.2.14). As a result we get the following expression for the elastic terms that enter in Eq. (1.2.25) for $\vec{v}_L$:

$$\vec{F} = \vec{e}^a div(h_{ab}\vec{e}^b) = 2\Omega\lambda\vec{\nabla}_\perp\rho_s + \rho_s\frac{\hbar\Omega}{4m}\left( 2\nabla_\perp div\vec{u} - \Delta_\perp\vec{u} \right) -$$
$$-\rho_s 2\Omega\lambda\frac{\partial^2\vec{u}}{\partial z^2}, \quad (1.2.45)$$

where

$$\lambda = \frac{\hbar}{2m}\ln\frac{b}{d} . \quad (1.2.46)$$

If according to Tkachenko we introduce the compression modulus $\mu_c = -\mu_s = -\frac{\hbar\Omega}{4m}\rho_s$ (where $\mu_s$ is the shear modulus), and a bending modulus $\mu_b = \rho_s 2\Omega\lambda$ (see [1.14]) we can rewrite Eq. (1.2.35) as:

$$\vec{F} = 2\Omega\lambda\nabla_\perp\rho_s + \mu_c\nabla_\perp div\vec{u} - \mu_s\Delta_\perp\vec{u} - \mu_b\frac{\partial^2\vec{u}}{\partial z^2} . \quad (1.2.47)$$

Note that it is convenient to introduce the longitudinal ($c_l$) and the transverse sound velocities ($c_t$) in the vortex lattice according to the standard relations of Elasticity theory [1.18]:

$$\mu_c = 2\rho_s(c_l^2 - 2c_t^2) \text{ and } \mu_s = 2\rho_s c_t^2 . \quad (1.2.48)$$

In our case

$$c_t^2 = c_l^2 = \frac{\hbar\Omega}{8m} = \frac{\mu_s}{2\rho_s}. \qquad (1.2.49)$$

As the result for the elastic terms which enter the system of equation for hydrodynamics of slow rotations:

$$\vec{F} = 2\Omega\lambda\vec{\nabla}_\perp\rho_s - 2\rho_s c_t^2\vec{\nabla}_\perp div\,\vec{u} - 2\rho_s c_t^2\Delta_\perp\vec{u} - \mu_b\frac{\partial^2\vec{u}}{\partial z^2}, \qquad (1.2.50)$$

$$\frac{\partial}{\partial x_k}(h_{ab}e_i^a e_k^b) = -2\rho_s\Omega\lambda\vec{\nabla}div\,\vec{u} + \vec{F} \qquad (1.2.51)$$

Let us emphasize the first term in the right-hand side of Eq. (1.2.45) $-2\rho_s\Omega\lambda\nabla_i div\,\vec{u}$ can be left out upon the normalization of the pressure in (1.2.38)

$$P \to P - 2\rho_s\Omega\lambda div\,\vec{u} \qquad (1.2.52)$$

and a simultaneous replacement of the chemical potential $\mu$ in Eq. (1.2.22) for the superfluid motion (for $\partial\vec{v}_s/\partial t$) by the chemical potential of the liquid without allowance for elasticity. Indeed, in the linear approximation we have:

$$d\mu = -\frac{S}{\rho}dT + \frac{1}{\rho}dP - \frac{1}{\rho}\vec{j}_0 d(\vec{v}_n - \vec{v}_s) + \frac{1}{2\rho}h_{ab}dg^{ab} =$$
$$= -\frac{S}{\rho}dT - \frac{1}{\rho}\vec{j}_0 d(\vec{v}_n - \vec{v}_s) + \frac{1}{\rho}d(P - \rho_s 2\Omega\lambda div\,\vec{u}) \qquad (1.2.53)$$

Let us point out that when Eq. (1.2.35) for $\vec{F}$ is substituted in Eq. (1.2.26) for $\vec{v}_L$ at $T = 0$, we get the customary employed equation (see [1.24, 1.43])

$$\vec{v}_L - \vec{v}_{s\perp} = \frac{m}{2\pi\hbar\rho_s n_V}\left[\vec{F}\vec{V}\right] \approx \frac{1}{2\rho_s\Omega}\left[\vec{F}\vec{e}_z\right], \qquad (1.2.54)$$

where $\vec{e}_z = \vec{V}^{(0)}$ is a unit vector in the direction of the undeformed vortex lines and $\rho_s(T = 0) = \rho$.

<u>Connection between $\vec{v}_s$ and $\vec{u}$ in linearized theory for $T = 0$.</u>

Let us find first the relation between $\vec{v}_s$ and $\vec{u}$. To get this relation in linearized theory we use Eq. (1.2.19) $\vec{v}_L = -\vec{e}_a\dot{N}^a$ and the Eq. (1.2.27) $N_0^a = N^a - u^a$, where $\vec{u} = \vec{e}_{0a}u^a$. Note also that $\vec{e}_0^a = \vec{\nabla}N_0^a$, where $\vec{e}_0^a$ are the vectors of an undeformed but uniformly rotating lattice. Therefore $\dot{\vec{e}}_0^a = \left[\vec{\Omega}\vec{e}_0^a\right]$ and

$$\dot{N}_0^a = \left[\vec{\Omega}\vec{e}_0^a\right]\vec{r} = -\vec{v}_0\vec{e}_0^a \qquad (1.2.55)$$

where $\vec{v}_0 = \left[\vec{\Omega}\vec{r}\right]$ is a linear velocity of a solid state rotation. We should also use the condition $\vec{v}_L\vec{V} = 0$ where $\vec{v}_L = \vec{v}_0 + \delta\vec{v}_L$ and $\vec{V} = \vec{e}_z + \delta\vec{V}$ in linearized theory. Having in mind that $\delta\vec{V} \sim \frac{\partial\vec{u}}{\partial z}$ we finally get:

$$\vec{v}_L = \vec{v}_0 + \dot{\vec{u}} + \left(\vec{v}_0\vec{\nabla}\right)\vec{u} - \vec{e}_z\left(\vec{v}_0\frac{\partial\vec{u}}{\partial z}\right) - \left[\vec{\Omega}\vec{u}\right]. \qquad (1.2.56)$$

Now we can find the relation between $\vec{v}_s$ and $\vec{u}$ using Eq. (1.2.46). To find this relation it is important to note that in linearized theory $\vec{v}_s = \vec{v}_0 + \delta\vec{v}_s$ and correspondingly $\delta\vec{v}_{s\perp}$ in (1.2.54) reads:

$$\vec{v}_{s\perp} = \delta\vec{v}_s - \delta v_{sz}\vec{e}_z - \vec{e}_z\left(\vec{v}_0\frac{\partial\vec{u}}{\partial z}\right). \quad (1.2.57)$$

Substituting (1.2.57) in (1.2.56) we get the cancellation of the term $\vec{e}_z\left(\vec{v}_0\frac{\partial\vec{u}}{\partial z}\right)$ in the relative velocity $\vec{v}_L - \vec{v}_{s\perp}$ and moreover:

$$\vec{v}_L - \vec{v}_{s\perp} = \dot{\vec{u}} + \left(\vec{v}_0\vec{\nabla}\right)\vec{u} - [\vec{\Omega}\vec{u}] - (\delta\vec{v}_s - \delta v_{sz}\vec{e}_z) = \frac{1}{2\rho_s\Omega}\left[\vec{F}\,\vec{e}_z\right]. \quad (1.2.58)$$

Introducing the two-dimensional projection of $\delta\vec{v}_s$ on the $(x,y)$ plane:

$$\delta\vec{v}_{s2D} = \delta\vec{v}_s - \delta v_{sz}\vec{e}_z \quad (1.2.59)$$

we can rewrite (1.2.63) in the form:

$$\left\{\frac{\partial}{\partial t} + \left(\vec{v}_0\vec{\nabla}\right) - \vec{\Omega}\times\right\}\vec{u} - \delta\vec{v}_{s2D} = \frac{1}{2\rho_s\Omega}\left[\vec{F}\,\vec{e}_z\right]. \quad (1.2.60)$$

Now we can use that in curly brackets in the l.h.s. of (1.2.60) just stands an operator which is responsible for the transformation of the vector quantity to the rotation frame. Thus we derive an important relation between the time derivative of $\vec{u}$ in the rotation frame $\frac{\partial\vec{u}}{\partial t'}$ and the two dimensional projection of the superfluid velocity $\delta\vec{v}_{s2D}$:

$$\frac{\partial\vec{u}}{\partial t'} = \delta\vec{v}_{s2D} + \frac{1}{2\rho_s\Omega}\left[\vec{F}\,\vec{e}_z\right]. \quad (1.2.61)$$

Linearized Euler equation at $T = 0$.

Returning back to Euler equation for superfluid velocity $\vec{v}_s$ at $T = 0$ we get in linearized theory:

$$\frac{\partial\vec{v}_s}{\partial t} + \vec{\nabla}\left(\mu + \frac{v_s^2}{2}\right) = \left[\vec{v}_L\,rot\,\vec{v}_s\right] \approx \left[\vec{v}_s\,rot\,\vec{v}_s\right] + \frac{1}{2\rho\Omega}\left[\left[\vec{F}\vec{e}_z\right]rot\,\vec{v}_s\right] \quad (1.2.62)$$

In the rotation frame in direct analogy with the situation in an ideal rotating fluid (see Subsection 1.1.4) we get:

$$\frac{\partial\delta\vec{v}_s}{\partial t'} + \vec{\nabla}\frac{\delta P}{\rho_0} + \left[2\vec{\Omega}\,\delta\vec{v}_s\right] = \frac{1}{\rho_0}\vec{F}. \quad (1.2.63)$$

Applying an operator of rot to the l.h.s. and to the r.h.s. of this equation we finish with:

$$\frac{\partial}{\partial t'}(rot\,\delta\vec{v}_s) + rot\left[2\vec{\Omega}\,\delta\vec{v}_s\right] = \frac{1}{\rho_0}rot\,\vec{F}. \quad (1.2.64)$$

This equation together with the relation (1.2.61) (which establishes the connection between $\delta\vec{v}_{s2D}$ and $\frac{\partial\vec{u}}{\partial t'}$) and the continuity equation

$$\frac{\partial\delta\rho}{\partial t} + \rho_0\,div\,\delta\vec{v} + (\vec{v}_0\vec{\nabla})\delta\rho = \frac{\partial\delta\rho}{\partial t'} + \rho_0\,div\,\delta\vec{v} = 0 \quad (1.2.65)$$

(where we used the transformation $\dfrac{\partial}{\partial t} \to \dfrac{\partial}{\partial t} + \left(\vec{v}_0 \vec{\nabla}\right)$ to the rotation frame for a scalar) helps us to find the spectrum of collective excitations in a rotating macroscopically averaged superfluid at long wave-vectors $kb << 1$.

   1.2.6. Collective modes of the lattice. Tkachenko waves and Lord Kelvin waves. Melting of the vortex lattice.

   According to Sonin in the general case of compressive rotating superfluid with a triangular vortex lattice the spectrum of collective excitations in the long wavelength limit reads (see [1.24]):

$$\omega^2 = \left(2\Omega + \lambda k_z^2\right)\left[2\Omega \frac{\omega^2 - c_t^2 k_z^2}{\omega^2 - c_I^2 k^2} + \lambda k_z^2 + \frac{c_t^2 k_\perp^4}{2\Omega}\right], \qquad (1.2.66)$$

where $c_t^2 = \dfrac{\mu_s}{2\rho_s} = \dfrac{\hbar \Omega}{8m}$ is Tkachenko sound velocity squared, $c_I^2$ is first sound velocity squared,

$\lambda = \dfrac{\hbar}{2m}\ln\dfrac{b}{d}$ and $k_z$ and $k_\perp$ are the projections of the wave-vector $\vec{k}$ ($k^2 = k_\perp^2 + k_z^2$) on the undeformed vortex axis (on the $z$-axis) and on the $xy$ plane perpendicular to undeformed vortex axis. For $k_\perp = 0$ the spectrum (1.2.66) contains 2 modes – a sound mode with a spectrum $\omega_1^2 = c_I^2 k_z^2$ and Lord Kelvin (Thomson) mode for bending oscillations of the vortex lines $\omega_2^2 = \left(2\Omega + \lambda k_z^2\right)^2$.

   In the same time for $k_z = 0$ (and thus $k = k_\perp$) there are 2 modes again $\omega_1^2 = 4\Omega^2 + c_I^2 k_\perp^2$ and $\omega_2^2 = \omega_T^2 \approx \dfrac{c_t^2 c_I^2 k_\perp^4}{4\Omega^2 + c_I^2 k_\perp^2}$ for $c_I >> c_t$. First mode is usually called an inertial mode. It has a gap. Note that the second mode has a nontrivial dispersion in the denominator. Moreover, for $k_\perp > \dfrac{2\Omega}{c_I}: \omega_T^2 = c_t^2 k_\perp^2$ and we have a linear spectrum for Tkachenko mode, while for very small $k_\perp < \dfrac{2\Omega}{c_I}: \omega_T^2 = \dfrac{c_t^2 c_I^2 k_\perp^4}{4\Omega^2}$ and correspondingly $\omega_T = \dfrac{c_t c_I k_\perp^2}{2\Omega}$ – the spectrum of Tkachenko mode becomes quadratic.

   According to Baym [1.11] the quadratic character of the spectrum at very small $k$-vectors leads to dramatic consequences for purely 2D flows with $k_z = 0$. Namely the mean displacement squared of a single vortex line from equilibrium due to the excitations of very long wavelength Tkachenko modes is logarithmically divergent:

$$\frac{\left\langle \vec{u}^2 \right\rangle}{b^2} \sim \frac{T}{\Omega} \frac{n_v}{nL} \ln \frac{k_{max}}{k_{min}} \quad (1.2.67)$$

where in dense $^4$He $k_{max} \sim \dfrac{2\Omega}{c_I}, k_{min} \sim \dfrac{2\pi}{R}$, $n$ and $n_v$ are 3D density of particles and 2D density of vortices, and $L$ is the height of the container (or effective third dimension of a quasi-2D magnetic trap). Thus $\dfrac{n_v}{nL}$ is dimensionless density ratio. The linear in T dependence of $\left\langle \vec{u}^2 \right\rangle \big/ b^2$ requires that $T >> \omega_T^{max}$ since in this regime bosonic distribution function for Tkachenko waves $n_B\left(\dfrac{\omega_T}{T}\right) \sim \dfrac{T}{\omega_T}$. From (1.2.67) it follows that $\left\langle \vec{u}^2 \right\rangle \big/ b^2$ is logarithmically divergent in infrared region of small $k \sim 1/R$. The strong effect of compressibility on the Tkachenko mode in the long wavelength limit $\omega_T = \dfrac{c_t c_l k_\perp^2}{2\Omega}$ was studied by Reato [1.52].

Note, however, that in practice it is very difficult to fulfill the condition

$$\frac{2\pi}{R} < k_\perp < \frac{2\Omega}{c_l} \qquad (1.2.68)$$

in dense $^4$He where the typical size of the container $R \sim 0.1$ cm. Usually the transition to quadratic regime requires unachievably high angular velocities in dense liquid. In helium, for example, $c_l \sim 2.4 \cdot 10^4$ cm/sec, $\Omega_{C1} = \dfrac{\hbar}{mR^2}\ln\dfrac{R}{d} \sim 1$ rot/sec and we need to demand $\Omega \geq 10^7 \Omega_{C1}$ to get the quadratic regime, which is practically impossible. We notice again that $n$L is an effective two-dimensional density of particles (number of particles in the unit area). Thus

$$\frac{n_v}{nL} = \frac{N_{vortices}}{N_{particles}} = \frac{1}{p}, \qquad (1.2.69)$$

where $p$ is a dimensionless filling factor. In terms of $p$ Eq. (1.2.67) reads:

$$\frac{\left\langle \vec{u}^2 \right\rangle}{b^2} \sim \frac{T}{\Omega}\frac{1}{p}\ln(Rk_{max}), \qquad (1.2.70)$$

where $R$ is the size of the container.

In dense $^4$He $mc_1^2 \sim (20 \div 30)$ K, $T \sim 1$ K, $\hbar\Omega/k_B \sim 10^{-11}$ K for $\Omega \sim \Omega_{C1}$. Thus $mc_1^2 \gg T \gg \Omega$ and the requirement for macroscopic averaging $k_\perp b \ll 1$ is automatically fulfilled for $k_\perp < 2\Omega/c_l$ since $\Omega/mc_1^2 \ll 1$. Correspondingly, $k_\perp < 2\Omega/c_l < 1/b$. The requirement $\omega_T^{max} \ll T$ is also fulfilled since for $k_\perp^{max} \sim \dfrac{2\Omega}{c_l}$: $\omega_T^{max} \sim c_t k_\perp^{max} \sim \dfrac{c_t\Omega}{c_l} \ll \Omega \ll T$ for $c_t \ll c_l$. Finally in dense $^4$He we always have $p \gg 1$ for the filling factor since only at $\Omega = \Omega_{C2}$ $N_{particles} \sim N_{vortices}$ (for $\Omega = \Omega_{C2}$ the normal cores which in $^4$He have interatomic size start to overlap). For $\Omega < \Omega_{C2}$: $N_{particles} \gg N_{vortices}$ and thus $p \gg 1$.

In dilute Bose-gasses $mc_1^2 = \dfrac{4\pi a}{m}n$, where $a$ is the s-wave scattering length. To get the "inverse ratio" $\Omega/mc_1^2 \gg 1$ we have to consider very small densities $na^3 \ll 1$ and very large angular velocities. Note that in this limit $k_\perp < 1/b < 2\Omega/c_l$ and the spectrum of Tkachenko mode will be always quadratic in $k_\perp$. The filling factor $\nu$ in dilute Bose-gasses in magnetic traps can be also much smaller than in helium since the vortex cores $\xi_0 \sim 1/\sqrt{na} \gg d$ are much larger than in dense $^4$He.

Note also that in real $^4$He the situation is always three-dimensional $k_z \neq 0$. The spectrum still has two branches $\omega_1^2(k_z, k_\perp)$ and $\omega_2^2(k_z, k_\perp)$. Moreover for Tkachenko mode according to Williams and Fetter [1.43 1977] we can use an approximate form $\omega_T^2 = \omega_2^2 \approx 4\Omega^2\dfrac{k_z^2}{k^2} + \dfrac{\hbar\Omega}{4m}k^2\dfrac{k_\perp^4}{k^4} + 2\Omega\lambda\dfrac{k_z^2}{k^2}\left(1 + \dfrac{k_z^2}{k^2}\right)$, or introducing $\cos\theta = k_z/k$ ($\sin\theta = k_\perp/k$) we get $\omega_T^2 = 4\Omega^2\cos^2\theta + \dfrac{\hbar\Omega}{4m}k^2\sin^4\theta + 2\Omega\lambda\cos^2\theta\left(1 + \cos^2\theta\right)$, where $\lambda = \dfrac{\hbar}{2m}\ln\dfrac{b}{d}$.

Than according to Baym [1.11] the mean displacement squared

$$\langle \vec{u}^2 \rangle = \frac{T}{\rho} \int\limits_0^{} \frac{d^3 \vec{k}}{(2\pi)^3} \frac{1+\cos^2 \theta}{\omega_r^2} \sim \frac{T}{\rho} \int\limits_{-1}^{1} \frac{d \cos \theta}{4\pi^2} \int\limits_0^{} k^2 dk \frac{1+\cos^2 \theta}{4\Omega^2 \cos^2 \theta + \dfrac{\hbar k^2 \Omega}{4m}\sin^4 \theta + 2\Omega\lambda \cos^2 \theta (1+\cos^2 \theta)}$$

becomes finite in the infrared limit $k \to 0$ in 3D case. However even in this case for the quasi two-dimensional thin film or a slab geometry in $z$-direction (for the system restricted by the two planes separated by the distance $L$ in $z$-direction and with a discrete set of $k_z = 2\pi n/L$) Gifford and Baym [1.11] predict the logarithmic divergence of the correlator of the two displacements:

$$\frac{\langle |\vec{u}(\vec{r}) - \vec{u}(\vec{r}')|^2 \rangle}{b^2} \sim T \ln R_\perp \qquad (1.2.71)$$

at finite temperatures and very large distances $R_\perp^2 >> Lb$, where $\vec{R} = \vec{r} - \vec{r}' = \{\vec{R}_\perp, R_z\}$.

### An approach to collective modes based on Gross-Pitaevskii-equation.

Note that at large filling factors $\nu >> 1$ there is another more microscopic (but still mean-field) approach to study the spectrum and even the damping of collective modes (see Matvienko et al. [1.44]). It is based on the mean-field solution for the GP equation for rotating dilute Bose-gas. Note that for a stationary problem $\Psi(\vec{r}, t) = \Psi(\vec{r})e^{i\mu t}$ and GP equation can be written as a nonlinear Schroedinger equation for the charged particles in strong magnetic field. Namely

$$\hat{H}\Psi(r) = \left\{ \left( \hat{p} - \frac{e}{c}\vec{A} \right)^2 + V_{ext}(\vec{r}) - \frac{m\Omega^2 r^2}{2} + g \,|\,\Psi(\vec{r})\,|^2 \right\} \Psi(\vec{r}) = \mu\Psi(\vec{r}); \qquad (1.2.72)$$

where $\vec{A} = m\left[\vec{\Omega}\vec{r}\right] = \left[\vec{B}\vec{r}\right]$ is an effective vector potential in radial (cylindrical) gauge [1.3] $\vec{B}$ is a magnetic field.

The chemical potential $\mu$ in (1.2.72) plays a role of an averaged energy on one particle $E$ in the standard Schroedinger equation. If an external trapping potential in (1.2.72) is also quadratic (confinement potential – see Chapter 4) $V_{ext}(\vec{r}) = \dfrac{m\omega^2 r^2}{2}$, than neglecting the nonlinear term $g\,|\Psi|^2\,\Psi$ we can get from (1.2.72) an equation for Landau levels which is well known from Quantum mechanics. The spectrum of this equation reads for close values of angular velocity and trapping frequency $\Omega \sim \omega$ (more precisely for $|\Omega - \omega| / \Omega << 1$)

$$E_{n,l} = \hbar\omega\left( n + \frac{1}{2} \right) + \hbar(\Omega - \omega)l + \frac{p_z^2}{2m}, \qquad (1.2.73)$$

where $n = n_r + l$ is a principal quantum number and $l$ is an orbital momentum ($n_r$ is radial quantum number).
For the lowest Landau level (LLL) $n = 0$ and for purely 2D motion ($p_z = 0$) we get:

$$E_{0l} = \frac{\hbar\omega}{2} + \hbar(\Omega - \omega)l \,. \qquad (1.2.74)$$

The corresponding $\Psi$-function of the LLL

$$\Psi_{0l} = f_l(z)e^{-\frac{\omega|z|^2}{2}} \qquad (1.2.75)$$

where $z = x + iy, \overline{z} = x - iy,$ and $|z|^2 = z\overline{z}$

To have vortices we should demand that $f_l(z)$ is analytic function of $z$, which does not have any poles. Moreover $f_l(z = z_i) = 0$ for the vortex solutions centered at the points $z = z_i$. In the most simple case according to Matvienko et al., $f_l$ should behave linearly (proportional to ($z - z_i$)) close to each vortex core. Such mean-field solution corresponding to the triangular lattice can be described by some θ-function of Yakobi (see [1.44]) similar to Abrikosov solution for type-II superconductors [1.26]. Note that we can safely neglect the nonlinear term in (1.2.72) if $gn \triangleleft |\Omega - \omega|$. The authors of [1.44] also managed to derive the spectrum and damping of collective modes in the same type of formalism.

<u>Melting of the vortex lattice.</u>

If $\frac{\langle \vec{u}^2 \rangle}{b^2}$ exceeds some finite number (which is less that 1), then according to Lindeman criterium [1.53] the 2D vortex lattice starts to melt. For finite temperatures we have classical melting while for $T = 0$ we can still have quantum melting. The last case is very interesting both for vortex lattices in rotating gasses and for the search of supersolidity in Quantum crystals (see chapter 2). The quantum melting for the 2D lattice requires $\langle \vec{u}^2 \rangle / b^2 \geq 0.07$ (see Cooper et al., [1.45]).

Note that according to Baym at zero temperature the mean displacement squared for purely 2D flows ($k_z \equiv 0$) reads:

$$\frac{\langle \vec{u}^2 \rangle}{b^2} \sim \frac{n_v}{n_s L} \left( \frac{mc_l^2}{\Omega} \right)^{1/2} \sim \frac{1}{p} \left( \frac{mc_l^2}{\Omega} \right)^{1/2}, \qquad (1.2.76)$$

where $n_S$ is a superfluid particle density. In dense $^4$He the ratio $\left( \frac{mc_l^2}{\Omega} \right)^{1/2}$ as we already discussed is very large. For $\Omega \geq \Omega_{C1} \sim 1$ rot / sec : $\left( \frac{mc_l^2}{\Omega} \right)^{1/2} \sim 10^6$. However $\nu$ is also very large and thus at $T = 0$ $\frac{\langle \vec{u}^2 \rangle}{b^2} << 1$ in $^4$He.

In weakly non-ideal Bogolubov Bose gas

$$\left( \frac{mc_l^2}{\Omega} \right)^{1/2} = \left( \frac{4\pi a n_s}{m\Omega} \right)^{1/2} \sim \left( \frac{4}{m\Omega \xi_0^2} \right)^{1/2} \sim \frac{b}{\xi_0} \sim \left( \frac{\Omega_{C2}}{\Omega} \right)^{1/2}, \qquad (1.2.77)$$

where $m\Omega = b^2$ and $\xi_0 = 1/\sqrt{na}$. Note that for $\Omega = \Omega_{C2} : n_v = na$ and $p = \frac{nL}{n_v} = \frac{L}{a}$. Hence at $T = 0$ in a dilute Bose gas

$$\frac{\langle \vec{u}^2 \rangle}{b^2} \sim \frac{1}{p} \frac{b}{\xi_0}. \qquad (1.2.78)$$

Thus if we have small enough $p \geq 1$ (for that we should have almost 2D trap with $L \geq a$ for $z$-dimension) we can reach the regime of quantum melting $\langle \vec{u}^2 \rangle / b^2 \geq 0.07$. In this case according to Cooper et al., [1.45] we will have a phase transition form a vortex crystal to a strongly correlated phase of a vortex liquid. The strongly correlated vortex liquid (in contrast with the vortex crystal) is closely related to incompressible liquid states which according to Laughlin [1.49] and Haldane et al., [1.50]arise in the physics of the Fractional Quantum hall Effect (FQHE). Note that in practice $\nu_C \sim (5 \div 6)$ for a phase-transition from vortex crystal to vortex-liquid [1.45] in dilute rotating Bose gasses.

Note again that in this Chapter we mostly used Landau scheme of the conservation laws to derive the nonlinear hydrodynamics of slow rotations and to get the spectrum of the collective modes. There are other methods to derive these equations based on Poisson brackets (PB) [1.19, 1.20] or on Gross-Pitaevskii (GP) equation [1.45]. These methods are also very elegant ones. However they are not purely phenomenological and use some additional microscopic arguments.

The method of PB, for example, utilizes some additional microscopic equation for vortex lines. We can say that the other approaches do not provide the nonlinear fluxes in the system of hydrodynamic equations for slow rotations and do not describe the nonlinear elasticity theory of the vortex lattice in such a straightforward and a single-valued fashion as Landau scheme of the conservation laws. Note that Baym and Chandler [1.11] in their classical paper also considered equatins for slow-rotations but only in a form linearized in the lattice deformations.

### 1.3. Hydrodynamics of fast rotations.

In connection with dilute quantum gasses in rotation the researchers usually understand rapid rotations as a quantum limit when $\frac{n_v}{nL} = \frac{N_{vortices}}{N_{particles}} = \frac{1}{p}$ and we are in the regime of vortex liquid (the vortex lattice is melted). In this Section we will always work in a mean-field hydrodynamic regime (classical limit $\nu \gg 1$) mostly considering dense superfluid $^4$He. Thus the vortex lattice is always present in our considerations. Nevertheless even here we can distinguish between slow and rapid rotations, having in mind absolutely different regimes and phenomena in comparison with dilute quantum gasses. Namely we will construct strongly anisotropic hydrodynamics with two different velocities of normal and superfluid component $\vec{v}_{n\parallel} \neq \vec{v}_{s\parallel}$ in the direction parallel to the vortex lines and with only one velocity $\vec{v}_{n\perp} = \vec{v}_{s\perp} = \vec{j}_\perp / \rho$ in the direction perpendicular to the vortex lines. This hydrodynamics, describing the crystal in perpendicular to the vortex lines direction and a free superfluid in a parallel direction, can be realized at large rotation frequencies with the help of intensive umklapp processes [1.32] for the scattering of normal excitations on the 2D vortex lattice. Throughout this Section we will often use a term of fast rotations to distinguish this regime from rapid rotations in quantum gasses considered in the end of the Section 1.2.

### 1.3.1. The foundation of the hydrodynamics of fast rotations. The role of umklapp processes.

According to Andreev, Kagan [1.12] the two different approaches are possible to a hydrodynamic description of rotating systems, i.e. to a description in which the quantities are assumed to vary slowly in space and time and the expansions are in terms of gradients. In the first one the initial state of the system is assumed to be at a rest and the expansion is in all the gradients including the components of the velocity curl connected with the uniform rotation. This is hydrodynamics of slow rotations (in terminology of Andreev, Kagan considered in the previous Section).

In the same time since uniform rotation is always in thermodynamic equilibrium, another approach is possible in which the velocity curl that corresponds to uniform rotation is not assumed small, and the expansion is only in terms of nonequilibrium gradients, in the gradients on top of the uniform rotation. This expansion corresponds to hydrodynamics of fast rotations in the terminology of Andreev, Kagan.

Note that hydrodynamics of slow rotations, considered in Section 1.2, contains two independent velocities of normal and superfluid motions. The interaction of the normal excitations with the vortex lines are taken into account as a mutual friction force proportional to the difference between normal and superfluid velocities. For such a description to be valid it is necessary in any case that the excitations mean free path time $\tau_N$ which is connected with their scattering on each other, should be considerably smaller than the analogous time $\tau_U$ connected with the scattering of normal excitations on the vortex lattice. Note that otherwise an introduction of the velocity of normal component as an independent thermodynamic variable is meaningless. But even if the condition $\tau_U \gg \tau_N$ is satisfied, the usual equations are valid only for not very low angular velocities of rotation.

Let us clarify this situation, considering for simplicity on the basis of usual equations for hydrodynamics of slow rotations, the temperature oscillations and the related with them oscillations of the relative velocity $\vec{W}_\perp = \vec{v}_{n\perp} - \vec{v}_{s\perp}$ perpendicular to the vortex lines. In this case such oscillations are analogous in many respect to temperature oscillations in crystal under conditions of phonon hydrodynamics (see Gurevich [1.21]). In this case $\tau_U$ and $\tau_N$ play the role of the times of phonon relaxation due to normal and umklapp processes respectively. In both cases there are two oscillation modes, whose frequencies can be expressed as the functions of the wave vectors in the form:

$$\omega_{1,2} = -i\gamma \pm \left( c_{II}^2 k^2 - \gamma^2 \right)^{1/2}, \qquad (1.3.1)$$

where $c_{II}$ is a second sound velocity and

$$\gamma = 1/\tau_U + c_{II}^2 k^2 \tau_N. \qquad (1.3.2)$$

In the case of a rotating superfluid liquid we have (see [1.15] and Section 1.2)

$$1/\tau_U \sim (B\Omega), \qquad (1.3.3)$$

where $\Omega$ is an angular frequency of rotation and $B = \dfrac{2\rho\rho_s}{\rho_n}\beta$ is one of the two dimensionless parameters introduced by Hall and Vinen [1.15] to define the mutual friction force. From the view-point of a hydrodynamics of slow rotations, both modes are hydrodynamic, since both frequencies $\omega_{1,2}$ tend to zero when $k$ and $\Omega$ tend to zero simultaneously (for $\Omega \to 0$ the inverse umklapp time $1/\tau_U \to 0$ and $\gamma$ also tends to zero for $\Omega \to 0$ and $k \to 0$).

In fast rotation hydrodynamics, however, i.e. as $k \to 0$ and at constant $\Omega$, $\gamma \neq 0$ and only one (heat-conduction) mode is hydrodynamic (gapless) mode.

The usual equations are thus hydrodynamic in the sense of slow rotations. For given $\Omega$, however, their validity is restricted by the condition $\tau_U \gg \tau_N$. If, however, this condition is satisfied and the motion frequency $\omega$ satisfies the inequality $\omega\tau_U \ll 1$, we can replace the ordinary equations by the much simpler equations of fast rotations which will be derived in this Section. Note that in the crystals (see [1.21]) for $\tau_U \gg \tau_N$ and $\omega\tau_U \ll 1$ we can exactly in the same way replace the equations of phonon-hydrodynamics by the usual equations of the elasticity theory [1.18].

In fast rotation hydrodynamics we introduce for a superfluid liquid with a vortex lattice only one independent velocity of macroscopic motion in the direction perpendicular to the vortex lines, i.e. the system behaves for these directions as an ordinary crystal. Since the longitudinal total momentum of the excitations (parallel to the vortex lines) is preserved by the demands of the homogeneity of the system in the lines direction, we introduce in this direction two velocities and a system behaves as a standard superfluid liquid. Note that hydrodynamics of fast rotations is valid for given $\Omega$ at sufficiently low frequencies $\omega \ll \tau_U^{-1}$ and $\omega \ll \tau_N^{-1}$ no matter what is the ratio of $\tau_U$ and $\tau_N$. At temperatures of the order of 1 K in liquid He-II the dimensionless constant $B$ of Hall and Vinen is of the order of unity and the validity of fast rotations hydrodynamics is restricted to frequencies $\omega \ll (\tau_U^{-1} \sim \Omega)$ (see (1.3.3)).

### 1.3.2. The system of the nonlinear equations for the hydrodynamics of fast rotations.

In fast rotations hydrodynamics we must introduce one velocity in the direction perpendicular to the vortices and two independent velocities in the longitudinal direction. Under these considerations it is not convenient to use as the hydrodynamic variable the true superfluid velocity which we define as $\vec{V}_s$ in this Section. Instead of it we introduce a single perpendicular velocity $\vec{v}_\perp$ ( $\vec{v}_\perp \vec{\nu} = 0$ )defined by the equality:

$$\vec{v}_\perp = \frac{\vec{j}_\perp}{\rho}, \qquad\qquad (1.3.4)$$

where $\vec{j}_\perp$ is the exact value of the perpendicular momentum. The motion in the longitudinal direction will be described by two velocities $v_{n\parallel}$ and $v_{s\parallel}$, with $v_{s\parallel} = \vec{V}_s\vec{\nu}$. Thus we put:

$$\vec{v}_s = \vec{v}_\perp + v_{s\parallel}\vec{\nu}; \quad \vec{v}_n = \vec{v}_\perp + v_{n\parallel}\vec{\nu} \qquad (1.3.5)$$

We emphasize once more that the velocity $\vec{v}_s$ introduced by us does not coincide, generally speaking, with the true superfluid velocity $\vec{V}_s$. Nevertheless Eq. (1.2.23) for Galilean transformation of the energy and the linear momentum, as well as Eq. (1.2.24) for the differential of an internal energy $E_0$ (in the frame where $\vec{v}_s = 0$) are still valid in terms of new velocities $\vec{v}_n$ and $\vec{v}_s$ from (1.3.5). Indeed from the definition of $\vec{v}_n$ and $\vec{v}_s$ in (1.3.5) it follows that the relative velocity $\vec{W} = \vec{v}_n - \vec{v}_s = (v_{n\parallel} - v_{s\parallel})\vec{\nu}$ and $\vec{j}_0 = (\vec{j} - \rho\vec{v}_s) = \vec{j}_\parallel - \rho\vec{v}_{s\parallel} = j_{0\parallel}\vec{\nu}$ have only longitudinal components. Thus the term $\vec{W}d\vec{j}_0$ in $dE_0$ can be written in the form $(v_{n\parallel} - v_{s\parallel})dj_{0\parallel}$, which corresponds precisely to the correct expression for a system that is superfluid only in the longitudinal direction. We emphasize that the one-dimensional densities of the normal ($\rho_n$) and superfluid components ($\rho_s = \rho - \rho_n$), defined by the formula $j_{0\parallel} = \rho_n(v_{n\parallel} - v_{s\parallel})$, differ substantially, generally speaking, from the corresponding "microscopic" three-dimensional quantities which enter in the expressions (1.2.42), (1.2.45) and (1.2.48) for the elastic moduli (for $\frac{1}{2}h_{ab} = \frac{\partial E_{el}}{\partial g^{ab}}$) of a vortex lattice. Note that all the definitions connected with the kinematics of the vortex lattice, particularly expression (1.2.13) $\vec{v}_L = -\vec{e}_a\dot{N}^a$, remain the same as before. Since the number of the independent velocity components is now smaller by two than in slow rotation hydrodynamics, the number of equations in hydrodynamics of fast rotations should be correspondingly less than in (1.2.22). Specifically the last equation of the system (1.2.22) (the three component equation for the superfluid motion) should be replaced by one component scalar equation for the fast rotation. We derive it by using the formula:

$$\vec{v}_s\vec{\nu} = \vec{V}_s\vec{\nu} \qquad (1.3.6)$$

and the relation (1.2.21) for the vorticity conservation which in the notation of the present Section can be written in the form:

$$\dot{\vec{V}}_s + \vec{\nabla}\varphi = \left[\vec{v}_L \, rot\vec{V}_s\right], \qquad (1.3.7)$$

where $\varphi$ is a certain scalar.

Differentiating (1.3.6) with respect to time, and taking into account (1.3.7) as well as the relation $\vec{\nu} = \frac{rot\vec{V}_s}{\left|rot\vec{V}_s\right|}$, we obtain:

$$\vec{\nu}\dot{\vec{v}}_s = -\vec{\nu}\vec{\nabla}\varphi - \dot{\vec{\nu}}\left(\vec{V}_s - \vec{v}_s\right) \qquad (1.3.8)$$

An expression for $\dot{\vec{\nu}}$ can be easily derived from (1.3.7):

$$\dot{\vec{\nu}} + (\vec{v}_L\vec{\nabla})\vec{\nu} = (\vec{\nu}\vec{\nabla})\vec{v}_L - \vec{\nu}(\vec{\nu},(\vec{\nu}\vec{\nabla})\vec{v}_L) \qquad (1.3.9)$$

Substituting it in (1.3.8) we obtain after simple transformations:

$$\vec{\nu}\left(\dot{\vec{v}}_s + \vec{\nabla}\left(\mu + \frac{v_s^2}{2} + \Psi\right) - \left[\vec{v}_L \, rot\vec{v}_s\right]\right) = 0 \qquad (1.3.10)$$

where

$$\Psi = \varphi - \mu - \frac{v_s^2}{2} - \vec{v}_L\left(\vec{V}_s - \vec{v}_s\right) \qquad (1.3.11)$$

It is convenient to choose Eq. (1.3.10) with yet undetermined scalar $\Psi$ as a required scalar equation which together with the first three equations of the system (1.2.22) constitute the complete system of equations for the hydrodynamics of fast rotations [1.12].

Differentiating as usual the first equation of (1.2.23) for total energy with respect to time and using the aforementioned complete system of equations, we reduce the equation for $\dot{E}$ to the form:

$$\dot{E} = -div\{\vec{Q}_0 + \vec{q} + v_{nk}\pi_{ik} + v_{Lk}h_{ab}e_i^a e_k^b + \Psi(\vec{j} - \rho\vec{v}_n)\} + R + \frac{\vec{q}\vec{\nabla}T}{T} + \pi_{ik}\frac{\partial v_{ni}}{\partial x_k} +$$
$$+ \Psi div(\vec{j} - \rho\vec{v}_n) + \{\vec{v}_L - \vec{v}_\perp, \vec{F} + (\vec{j} - \rho\vec{v}_n, \vec{v}) \cdot [rot\,\vec{v}_s, \vec{v}]\} \qquad (1.3.12)$$

where the expressions for $\vec{Q}_0$ and $\pi_{ik}$ formally coincide with those given in Eq. (1.2.27) of the preceding Section. From (1.3.12) we find the dissipation function:

$$R = -\frac{\vec{q}\vec{\nabla}T}{T} - \pi_{ik}\frac{\partial v_{ni}}{\partial x_k} - \Psi div(\vec{j} - \rho\vec{v}_n) - \{\vec{v}_L - \vec{v}_\perp, \vec{F} + (\vec{j} - \rho\vec{v}_n, \vec{v}) \cdot [rot\,\vec{v}_s, \vec{v}]\}, \qquad (1.3.13)$$

Confining ourselves, as in the preceding Section, to consideration of only the last term in (1.3.13), we write down the expression for the relative velocity of the vortices and of the matter in the following general form:

$$\left(\vec{v}_L - \vec{v}_\perp\right)_\alpha = -\hat{B}_{\alpha\beta}G_\beta \qquad (1.3.14)$$

where

$$\vec{G} = \vec{F} + (\vec{j} - \rho\vec{v}_n, \vec{v}) \cdot [rot\,\vec{v}_s, \vec{v}], \qquad (1.3.15)$$

and $\alpha$ and $\beta$ are two-dimensional spatial indices in a plane perpendicular to $\vec{v}$. The matrix of the coefficients $\hat{B}_{\alpha\beta}$ satisfies the Onsager relations [1.32]:

$$\hat{B}_{\alpha\beta}(\vec{v}) = \hat{B}_{\beta\alpha}(-\vec{v}). \qquad (1.3.16)$$

Therefore

$$\left(\vec{v}_L - \vec{v}_\perp\right)_\alpha = \left(\frac{mg^{1/2}}{2\pi\rho\hbar} - B'\right)\left[\vec{G}, \vec{v}\right]_\alpha - \hat{B}_{\alpha\beta}G_\beta, \qquad (1.3.17)$$

where $B_{\alpha\beta}$ is the symmetric part of $\hat{B}_{\alpha\beta}$, $B_{\alpha\beta}(T=0) = B'(T=0) = 0$. For an arbitrary deformed lattice $B_{\alpha\beta}$ is an arbitrary symmetric tensor. We point out that the second term in the expression for $\vec{G}$, in contrast to the analogous term in the slow-rotation hydrodynamics, is of the second order in the deviations from the state of the uniform rotation. Therefore $B_{\alpha\beta}$ has in fast-rotation hydrodynamics the meaning of the diffusion coefficient of the vortices. The coefficient $B'$ describes an effect of the Hall type in the diffusion. In an undeformed triangular lattice the tensor $B$ reduces to a scalar.

We recall that the complete system of the equations for the hydrodynamics of fast rotations consists of the first three equations of (1.2.22) for $\dot{\rho}$, $\dot{j}$ and $\dot{S}$, and of Eq. (1.3.10) for the superfluid motion.

### 1.3.3. Linearized system of equations of fast rotations. The spectrum and the damping of the second sound mode.

Linearizing the equations for $\dot{S}$, $\dot{\rho}$, $\dot{j}$ and $v_i\dot{v}_{si}$ near the uniform rotation we obtain the following set of the equations that describes the oscillations of the temperature and the associated oscillations of the relative velocity $\delta W_{\parallel} = \delta v_{n\parallel} - \delta v_{s\parallel}$ along the vortices (compare with Eqs. (1.1.65) and (1.1.67) for a second sound wave):

$$
\begin{cases}
\delta \dot{T} + \left( \vec{v}_o \vec{\nabla} \right) \delta T + \dfrac{TS\rho_s}{C_p \rho} \nabla_z \delta W_{\parallel} + \dfrac{div\,\vec{q}}{\rho C_p} = 0, \\
\delta \dot{W}_{\parallel} + \left( \vec{v}_o \vec{\nabla} \right) \delta W_{\parallel} + \dfrac{S}{\rho_n} \nabla_z \delta T = 0,
\end{cases}
\tag{1.3.18}
$$

where $S$ and $C_p$ are the entropy and heat capacity per unit mass at a constant pressure, and the liquid is assumed incompressible $\delta P = \delta \rho = 0$, $\rho_n \delta v_{n\parallel} + \rho_s \delta v_{s\parallel} = 0$. In (1.3.18) $\vec{v}_0 = \left[ \vec{\Omega} \vec{r} \right]$. Equations (1.3.18) are reduced to the equations with constant coefficients by transforming to a rotation frame. This corresponds to the substitution $\dfrac{\partial}{\partial t} \to \dfrac{\partial}{\partial t'} - \left( \vec{v}_o \vec{\nabla} \right) - \Omega \times$ for vectors and $\dfrac{\partial}{\partial t} \to \dfrac{\partial}{\partial t'} - \left( \vec{v}_o \vec{\nabla} \right)$ for scalars (where $\partial/\partial t'$ is the time derivative in the rotation frame). The heat flux in (1.3.18) can be set equal to $\vec{q} = -\kappa_{\perp} \vec{\nabla}_{\perp} T$ ($\kappa_{\perp}$ is the heat conductivity in the direction perpendicular to the vortex axis), since the equations contain other considerably larger terms with $\partial/\partial z$. As the result we get:

$$
\begin{cases}
\dfrac{\partial \delta T}{\partial t'} + \dfrac{c_{II}^2 \rho_n}{S} \dfrac{\partial \delta W_{\parallel}}{\partial z} - \dfrac{\kappa_{\perp}}{\rho C_p} \Delta_{\perp} \delta T = 0 \\
\dfrac{\partial \delta W_{\parallel}}{\partial t'} + \dfrac{S}{\rho_n} \dfrac{\partial \delta T}{\partial z} = 0,
\end{cases}
\tag{1.3.19}
$$

where $c_{II}^2 = \dfrac{TS^2 \rho_s}{C_p \rho^2 \rho_n}$ is the second sound velocity.

The differentiation of the first equation in (1.3.19) with respect to $\partial/\partial t'$ and of a second one with respect to $\partial/\partial z$, and after that the substitution of the second equation in the first one yields:

$$
\dfrac{\partial^2 \delta T}{\partial t'^2} - c_{II}^2 \dfrac{\partial^2 \delta T}{\partial z^2} - \dfrac{\kappa_{\perp}}{\rho C_p} \Delta_{\perp} \dfrac{\partial T}{\partial t'} = 0 \tag{1.3.20}
$$

Correspondingly $\omega^2 - c_{II}^2 k_z^2 + \dfrac{\kappa_{\perp}}{C_p} k_{\perp}^2 i\omega = 0$ and the spectrum in rotation frame reads:

$$
\omega_{1,2} = -i \dfrac{\kappa_{\perp} k_{\perp}^2}{2\rho C_p} \pm \sqrt{ c_{II}^2 k_z^2 - \left( \dfrac{\kappa_{\perp} k_{\perp}^2}{2\rho C_p} \right)^2 } \tag{1.3.21}
$$

For $k_{\perp} = 0$ $\omega_{1,2} = \pm c_{II} \left| k_z \right|$, for $k_z = 0$ $\omega_{1,2} = i(-1 \pm 1) \dfrac{\kappa_{\perp} k_{\perp}^2}{2\rho C_p}$. It is interesting to compare this spectrum with the overdamped temperature waves in ordinary classical liquid with the spectrum $\omega = -i \dfrac{\kappa k^2}{\rho C_p}$ considered in Section 1.1 (see Eq. (1.1.30)), and the propagating second sound waves with the spectrum $\omega^2 = c_{II}^2 k^2$ (see Eq. (1.1.69)) for irrotational superfluid (the same propagating spectrum of the second sound $\omega^2 = c_{II}^2 k^2$ we get in the hydrodynamics of slow rotations). From the spectrum (1.3.21) follows, that in the hydrodynamics of fast rotations the oscillations of the temperature propagate in the form of the second sound only along the axis of the vortices, while in perpendicular direction they are ordinary damped thermal waves (the second root corresponds at small $k_z$ to $\delta T \to 0$ and $\delta W_{\parallel} = const$).

Another collective modes in the hydrodynamics of fast rotations.

To establish another application of the derived equations we consider the oscillations of the transverse velocity $\delta \vec{v}_\perp$ and of the displacement $\vec{u}$ in a state with simultaneous uniform rotation and a uniform heat flux $Q = TS v_{n\parallel}$ along the vortex lines. This problem is of interest because a substantial role in it is played by the second term of the expression (1.3.15) for $\vec{G}$. Confining ourselves for simplicity to the case of low temperatures $\rho_n << \rho$ and neglecting in (1.3.17) the term with $B_{\alpha\beta}$ and $B'$, we rewrite this equation in the rotation frame in the form:

$$\ddot{\vec{u}} - \dot{\vec{v}}_\perp = \frac{1}{2\Omega\rho}\left[\vec{F}\vec{e}_z\right] - iv_{n\parallel}k_z\dot{\vec{u}} - \frac{i}{2\Omega}v_{n\parallel}\left\{k_z\left[\vec{e}_z\delta\vec{v}_\perp\right] + \left[\vec{e}_z\vec{k}_\perp\right]\frac{\vec{k}_\perp\delta\vec{v}_\perp}{k_z}\right\}, \qquad (1.3.22)$$

where $\vec{F}$ is defined by (1.2.50) and $v_{n\parallel} = Q/TS$. The second equation that connects $\vec{u}$ with $\vec{v}_\perp$ is obtained by projecting on the $\{xy\}$ plane of the second equation for $\partial j_i / \partial t$ of the system (1.2.22) and excluding from it the pressure using the incompressibility condition at low temperatures $div\,\vec{v}_s = 0$. We have in the rotation frame:

$$\ddot{\delta\vec{v}}_\perp + \left[2\vec{\Omega}\delta\vec{v}_\perp\right] - \frac{\vec{k}_\perp}{k^2}\left(\vec{k}_\perp\left[2\vec{\Omega}\delta\vec{v}_\perp\right]\right) + \frac{1}{\rho}\vec{F} - \frac{\vec{k}_\perp}{\rho k^2}\left(\vec{k}_\perp\vec{F}\right) = 0 \qquad (1.3.23)$$

If $\vec{k}$ is small enough, Eqs. (1.3.22), (1.3.23) describe in the principle approximation the independent oscillations of $\delta\vec{v}_\perp$ and $\vec{u}$. The former have a frequency:

$$\omega = \frac{2\left(\vec{\Omega}\vec{k}\right)}{k} \qquad (1.3.24)$$

and are inertial waves in an ordinary (classical) hydrodynamics of the incompressible rotational fluid considered in Section 1.1.

The second mode constitutes the oscillations of the displacement $\vec{u}$ and are peculiar to the fast rotations hydrodynamics. The frequency of this mode in the rotation frame:

$$\omega = v_{n\parallel}k_z; \quad v_{n\parallel} = Q/TS \qquad (1.3.25)$$

It must be emphasized that this mode exists only at finite temperature. It vanishes at $T = 0$, i.e. in the total absence of the normal component. Here lies an essential difference between hydrodynamics of slow-rotations and fast-rotations at finite temperatures.

In the hydrodynamics of slow rotations the velocity $\vec{v}_s$ and the displacement $\vec{u}$ are not independent variables, but are connected by the additional condition (1.2.20) for $rot\,\vec{v}_s$ and vectors of elementary translations $\vec{e}_a$. In fast-rotations hydrodynamics the same takes place at $T = 0$, when the difference between $\vec{v}_s$ and $\vec{V}_s$ vanishes. The presence of the root with $\omega = 0$ means here simply compatibility of hydrodynamic equations with the supplementary condition (1.2.20).

We should also make a remark concerning the spectrum (1.3.24). Although the mode (1.3.24) has zero gap, its frequency, generally speaking, does not tend to zero as $\vec{k} \to 0$. This highlights the distinction of the fast-rotations hydrodynamics, for the validity of which the condition $\vec{k} \to 0$ is generally speaking not sufficient and one more condition is required. To determine this condition we note that the frequency in a rotating coordinate frame can be regarded as an eigenvalue of the operator:

$$i\left\{\frac{\partial}{\partial t} + \left(\left[\vec{\Omega}\vec{r}\right]\vec{\nabla}\right) - \vec{\Omega}\times\right\} = i\frac{\partial}{\partial t} - \Omega J_z, \qquad (1.3.26)$$

where $J_z = S_z + L_z$, $L_z = -i \left[ \vec{r} \vec{\nabla}_{\perp} \right]$ in the orbital angular momentum and $S_z = -i\varepsilon_{\alpha\beta}$ is the spin of the vector field $\delta\vec{v}_{\perp}$. The derivative $\partial\omega/\partial\Omega$, as can be seen from (1.3.26) is equal to $-<J_z>$. For the frequencies of all the modes in the fast-rotations hydrodynamics to tend to zero we must satisfy besides the condition $\vec{k} \to 0$ also $<J_z> \to 0$. The latter is equivalent for the spectrum (1.3.24) to the condition $k_z/k \to 0$, where $k^2 = k_z^2 + k_{\perp}^2$.

## 1.4. Opposite case of a single bended vortex line for extremely slow rotations ($\Omega \sim \Omega_{C1}$).

In the last Section of the present Chapter we will consider an opposite case of a single bended vortex line aligned with the axis of a cylindrical vessel of radius $R$. It is known that such a state corresponds to the thermodynamic equilibrium at $\Omega \sim \Omega_{C1} = \dfrac{\hbar}{mR^2}\ln\dfrac{R}{d}$. As it was shown by Hall [1.14], the equation that describes the bending oscillations of the vortex lines is reduced to the Schroedinger equation with an effective mass:

$$m^* = \frac{m}{\ln(\hbar / p_z d)} \qquad (1.4.1)$$

($p_z$ is the momentum of the oscillations parallel to the vortex line) by introducing the wave-function:

$$\Psi = const(u_x + iu_y) . \qquad (1.4.2)$$

In Eq. (1.3.28) $\vec{u}$ is the two-dimensional vector of the displacement of the vortex line in a plane perpendicular to the rotation axis (see Fig. 1.11). If we choose $const = \left(\pi\rho_s / m\right)^{1/2}$ in Eq. (1.4.2) the energy of the oscillations also becomes identical with the energy given by Schroedinger expression:

$$E = \frac{\hbar^2}{2m^*}\int\left|\frac{\partial\Psi}{\partial z}\right|^2 dz \qquad (1.4.3)$$

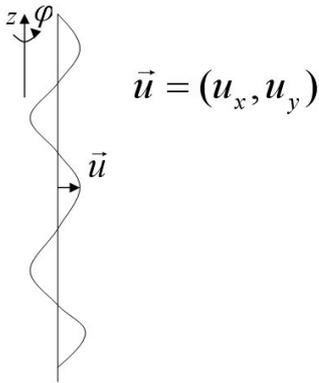

$$\vec{u} = (u_x, u_y)$$

Fig. 1.11. The bended vortex line situated in the center of the cylindrical vessel. $\vec{u}$ is the 2D displacement perpendicular to the vortex line. $z$ is the axis of the vessel, $\varphi$ is rotation angle.

### 1.4.1. Stabilization of the bending oscillations by rotation.

Note that in our case rotation around $z$-axis on an angle $\varphi$ is simultaneously the gauge transformation of the $\Psi$-function: $\Psi \to \Psi e^{i\varphi}$. The generator of this gauge transformation is the operator $\hbar\hat{N}$, where $\hat{N}$ - is the number of particles operator. In the second quantization technique the operator of the number of particles $\hat{N} = \sum_{p_z} b_{p_z}^+ b_{p_z}$ commutes with the Hamiltonian

$\hat{H} = \sum_{p_z} \dfrac{p_z^2}{2m^*} b_{p_z}^+ b_{p_z}$ of the 1D Bose-gas of the oscillating quanta carrying the momentum $p_z$.

From the other hand the operator of the $z$-component of the angular momentum in our case equals to $\hat{J}_z = \hbar \hat{N}$ and hence $\hat{J}_z$ also commutes with the Hamiltonian $\hat{H}$. Thus the quanta of the vortex line oscillations besides a linear momentum $p_z$ posses also an independent quantum number – an intrinsic angular momentum – $\hbar$ ("diamagnetic situation" according to Andreev, Kagan [1.12]). That is why in our case the bended vortex line in reality does not vibrate but rotates around $z$-axis (we can prove that in second quantization technique $\left[ \hat{\vec{u}} \, \hat{\vec{u}} \right] \neq 0$).

Thus the real thermodynamic equilibrium takes place only in the rotating frame and hence the spectrum in this frame reads:

$$\varepsilon(p_z) = \varepsilon_0(p_z) - \Omega \frac{J_z}{N} = \hbar\Omega + \varepsilon_0(p_z) = \hbar\Omega + \frac{p_z^2}{2m}\ln\frac{\hbar}{d\,p_z}. \qquad (1.4.4)$$

Thus the spectrum of the bended vortex line acquires a gap $\hbar\Omega$. Note that in the macroscopic hydrodynamics of slow rotations for $\Omega_{C1} << \Omega << \Omega_{C2}$: $\hbar\omega = 2\hbar\Omega + \dfrac{\hbar^2 k_z^2}{2m}\ln\dfrac{b}{a}$ and we have a gap $2\hbar\Omega$ for Lord Kelvin waves (see Section 1.2). The difference between the macroscopic gap $2\Omega$ and the "microscopic" gap $\Omega = \Omega_{C1}$ is connected with the fact that for a single bended vortex line we cannot introduce a macroscopic superfluid velocity of the solid state rotations $\vec{v}_0 = \left[ \vec{\Omega}\vec{r} \right]$ and hence the orbital angular momentum $\hat{L}_z = 0$ (see the discussion in the end of the Section 1.3). Correspondingly $\hat{J}_z = \hat{S}_z$ (is purely intrinsic angular momentum) and $\partial\omega/\partial\Omega = \hbar$ (and not $2\hbar$).

Thermodynamics of a bended vortex line.

The presence of the energy gap $\hbar\Omega$ causes the spectrum (1.3.30) to satisfy the Landau criterion for superfluidity. The critical velocity is:

$$v_C = \min\frac{\varepsilon(p_z)}{p_z} = \left( \frac{\hbar\Omega}{m}\ln\frac{\hbar}{2md^2\Omega} \right)^{1/2}, \qquad (1.4.5)$$

where $d$ is the vortex core. For $\Omega \sim \Omega_{C1}$ $v_C \sim \dfrac{\hbar}{mR}$.

For the same reason, the divergences at small momenta, which are customary for one-dimensional systems, are absent in this case. Indeed, let us calculate again the mean displacement from the equilibrium position of the bended vortex line squared $\langle \vec{u}^2 \rangle$ (see Section 1.2) assuming that the condition $T >> \Omega$ (more rigorously $k_B T >> \hbar\Omega$) is satisfied. In this case the bosonic distribution function $n_B\left( \dfrac{\varepsilon(p_z)}{T} \right) \approx \dfrac{T}{\varepsilon(p_z)}$ and

$$\langle \vec{u}^2 \rangle = \frac{m}{\pi\rho_s}\left\langle \left|\Psi\right|^2 \right\rangle = \frac{m}{\pi\rho_s}\int\frac{dp_z}{2\pi\hbar} n_B\left( \frac{\varepsilon(p_z)}{T} \right) = \frac{mT}{\pi\hbar\rho_s v_C}, \qquad (1.4.6)$$

where the critical velocity $v_C$ is given by (1.4.5).

### 1.4.2. Visualization of the vortex lattice in rotating superfluid. Packard experiments.

The ratio of this quantity to the square of the vessel radius is of the order of

$$\frac{\left\langle \vec{u}^2 \right\rangle}{R^2} \sim \frac{d}{R} \frac{Tmd^2}{\hbar^2} \frac{1}{\ln \dfrac{\hbar}{2md^2\Omega}} \sim \frac{d}{R} \frac{Tmd^2}{\hbar^2} \frac{1}{\ln(R^2/d^2)} \qquad (1.4.7)$$

for $\Omega \sim \Omega_{C1}$.

In superfluid $^4$He for $\Omega = \Omega_{C1} = \dfrac{\hbar}{mR^2}\ln\dfrac{R}{d} \sim 1\dfrac{rot}{\sec}$, $d \sim 3\overset{\circ}{\mathrm{A}}$ and $R \sim 0.1$ cm: $\dfrac{\left\langle \vec{u}^2 \right\rangle}{R^2} \ll 1$.

Note that without the gap $\left\langle \vec{u}^2 \right\rangle \sim T \int \dfrac{dp_z}{p_z^2}$ and we have a strong infrared divergence at $p_z \to 0$. Hence without the gap 1D bended vortex will form a globula as in many 1D systems (like in polymers for example). The small value of $\left\langle \vec{u}^2 \right\rangle / R^2 \ll 1$ can provide an explanation of Packard experiments [1.20]. He injected electrons in the vortex core, applied the voltage in the direction parallel to the vortex lines and got the photograph of the vortex lattice in $^4$He on the screen. In his experiments on visualization of the vortex lattice he observed small vortex displacements from equilibrium positions in the triangular lattice (see Fig 1.12).

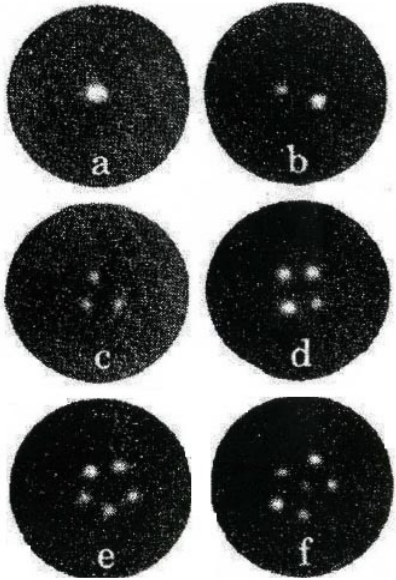

Fig 1.12. Vortex arrays in superfluid $^4$He from Yarmchuk, Gordon, Packard [1.20] and Vinen [1.15]. On Figs. a-f the number of vortices changes from 1 to 6.

Note that in alkali gases in magnetic traps a low-temperature regime of Bose Einstein condensation (BEC) was achieved in 1995 (see Chapter 4) [1.54]. Quantum vortices, which are the main signature of the superfluidity, are found in rotating condensate [1.47, 1.55] similar to that in liquid helium (see Section 1.2). The snapshot of the vortex lattice in dilute Bose-gasses is presented on Fig. 1.13.

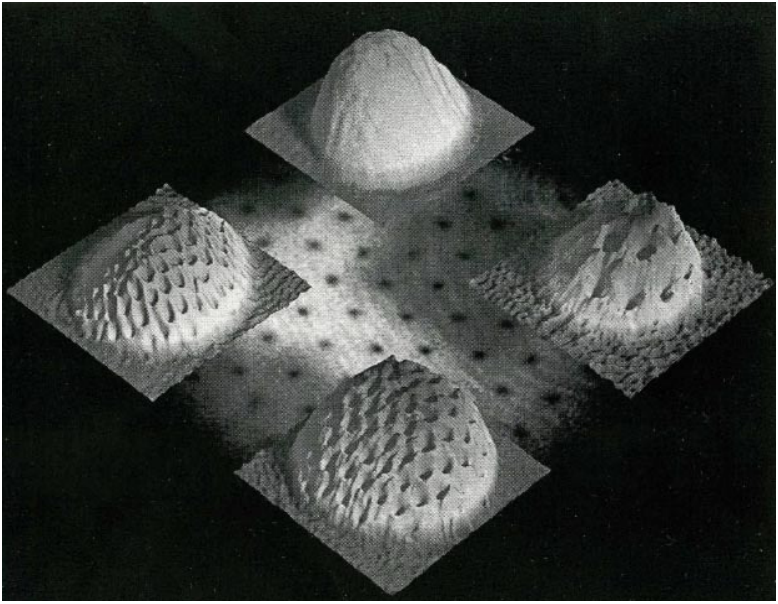

Fig. 1.13 Vortex lattice in rotating dilute Bose gasses of alkali elements ($^7$Li, $^{23}$Na, $^{87}$Rb)

Later on the vortex lattice was discovered also in BCS-phase of fermionic gasses [1.17] ($^6$Li and $^{40}$K) in the regime of Feshbach resonance (see Chapter 4). From both Figures we can see almost regular distribution of vortices forming triangular lattice both in dense superfluid helium and in dilute gasses. Hence the mean displacements squared of the vortex lines are finite and small in comparison with the intervortex distance $b$ in both experimental pictures.

### 1.4.3. Contribution to normal density and specific heat from bended vortex lines.

Returning back to the thermodynamic contribution from bended vortex lines we can calculate also the one-dimensional density of the normal component $\rho_h$ (see Section 1.1):

$$\rho_n = -\int p_z^2 \frac{\partial n_B(\varepsilon(p_z)/T)}{\partial \varepsilon} \frac{dp_z}{2\pi\hbar} = \frac{2T\Omega}{v_c^3} \qquad (1.4.8)$$

With decreasing $\Omega$ the density $\rho_n$ varies proportionally to $\Omega^{-1/2}$ (note that anyway $\Omega \geq \Omega_{C1}$). We can also write down an expression for the specific heat $C_\Omega$ per unit length of the vortex line at constant angular velocity $\Omega$:

$$C_\Omega = C_0(T) + \frac{2\Omega}{v_C}, \qquad (1.4.9)$$

where

$$C_0(T) = \frac{3\xi(3/2)}{4\hbar} \left( \frac{mT}{\pi \ln \dfrac{\pi}{mTd^2}} \right)^{1/2}, \qquad (1.4.10)$$

and $\xi(3/2)$ is Rieman function.

In contrast to Eq. (1.4.7) for $\langle \vec{u}^2 \rangle$ and Eq. (1.4.8) for $\rho_n$, in the case of the specific heat only the second correction term in (1.4.9) depends on the rotation frequency for $T >> \hbar\Omega$.

The thermal oscillations delocalize the vortex line. As a result, the average velocity curl differs from a $\delta$-function (compare with Eq. (1.2.2) in Section 1.2) and is determined by the probability distribution of the values of the distance $r$ of the vortex line from the vessel axis. Since the distribution is obviously Gaussian at $T >> \hbar\Omega$, we have in accord with (1.4.7):

$$\left\langle rot\,\vec{v}_s(\vec{r})\right\rangle = \frac{2\hbar}{m\left\langle\vec{u}^2\right\rangle}\exp\left\{-\frac{r^2}{\left\langle\vec{u}^2\right\rangle}\right\} = \frac{2\pi\hbar^2\rho_s v_C}{m^2 T}\exp\left\{-\frac{r^2 mT}{\pi\hbar\rho_s v_C}\right\} \qquad (1.4.11)$$

Note that the bending oscillations of the vortex lines as we already mentioned have an intrinsic angular momentum and thus produce an interesting effect of the angular-momentum transport by the heat flux in the absence of the matter flux. The angular momentum flux is:

$$L = -\hbar\left|\Psi\right|^2 v_n = -\frac{Tv_n}{v_C} = -\frac{Q}{Sv_C}, \qquad (1.4.12)$$

where $Q = TSv_n$ is the heat flux.

The torque carried by the heat flux between solid surfaces perpendicular to the rotation axis at $v_n \sim v_C$ is of the order of $N_v T$, where $N_v$ is the total number of vortices. At $T \sim 1$ K the torque measured in dyn·cm is of the order of $10^{-12}\,\Omega$ [sec$^{-1}$] $S_0$ [cm$^2$], where $S_0$ is the area of the solid surface. Although this is a small quantity, it seems to be experimentally observable.

1.5. Experimental situation and discussion. How to achieve the limit of the fast rotations at not very high frequencies in He II- $^3$He mixtures and in superfluid $^3$He-B.

Concluding the present Chapter let us stress once more that:
1) We present a hydrodynamic description of the two-velocity hydrodynamics of a superfluid $^4$He and dilute Bose gasses following Landau scheme of the conservation laws.
2) We generalize the hydrodynamic scheme on the presence of the quantized linear vortices and construct nonlinear hydrodynamics of slow and fast rotations with an account of the vortex lattice and the friction forces between normal excitations and vortices.
3) We analyze collective modes of the vortex lattice and in particular the shear (Tkachenko mode) as well as the bending mode of Lord Kelvin. We find out their contribution to the vortex displacements and to the thermodynamics. We also analyze the possibility to melt the vortex lattice in quasi two-dimensional rotating Bose gasses due to the contribution of the quadratic in $k$-vektor Tkachenko modes at very low frequencies.

The last topic which should be considered in this Chapter is how to achieve the limit of the hydrodynamics of fast rotations at not very high frequencies.

In the end of Section 1.3 we mentioned that when the Hall-Vinen coefficient $\beta$ becomes of the order of one, the umklapp relaxation time for the scattering of the normal excitations on the vortices $\tau_U \sim 1/\beta\Omega$ is of the order of $1/\Omega$ and the condition $\omega\tau_U \ll 1$ for the validity of the hydrodynamics of fast rotations require the condition $\omega < \Omega$ for the frequencies. With decreasing temperature, however, the situation changes since the bending oscillations of the vortex lines begin to produce a substantial contribution to the density of the normal component. This contribution $\rho_{n\Omega} = 2T\Omega/v_C^3$ exceeds the phonon contribution $\rho_n^{ph} = T^4/c_t^5$ (see Eq. (1.147)) for $T \le 0.1$ K in very clean superfluid $^4$He thus creating the possibility for the hydrodynamics of rapid rotations for $\Omega \sim \Omega_{C1}$. Note that the quasiparticles which correspond to the quanta of the bended oscillations, are localized on the vortex lines and can move easily only along them. Correspondingly we have only one component of the normal velocity $v_{n\parallel}$ and can speak only about highly anisotropic hydrodynamics of fast rotations.

A similar but even more clearly pronounced situation arises at low temperatures in the solutions of $^3$He in He-II (see Chapter 15), owing to the absorption of the impurities by the vortex cores (see [1.56]). In other words in the rotating solution all the $^3$He-impurities (which serve as the normal excitations in this system) are localized in the vortex core due to the gradient of Bernulli integral $\vec{\nabla}\left(\mu + \dfrac{v_s^2}{2}\right)$ (see Eq. (1.1.13)). Note that $\vec{v}_s$ increases when we approach the vortex core thus the pressure gradient causes  the localization of the impurities inside the core.

From the other hand they can again move freely along the vortex lines. We again have only one $v_{n\parallel}$ and can speak only about the hydrodynamics of fast rotations.

Finally in superfluid $^3$He-B, in which as in He-II, the orbital hydrodynamics, is isotopic the condition $\tau_U \sim \tau_N$ starts to be satisfied at temperatures $T \leq (T_C \sim 1$ mK) for $\Omega \leq 1$ sec$^{-1}$ because of the rather large $\tau_N$ for the scattering of (fermionic) quasiparticles on each other in $^3$He-B. The higher angular velocities in $^3$He-B should be described by the fast-rotations hydrodynamics.

We can repeat once more that for the dilute Bose gasses we should distinguish between the slow rotations with vortex lattice and rapid rotations when the vortex lattice is melted. In the same time in dense superfluids (He-II, $^3$He-B, $^3$He- He-II solutions) we should distinguish between the isotropic hydrodynamics of the slow rotations and the anisotropic hydrodynamics of the fast rotations. However the vortex lattice is present in both types of the hydrodynamics descriptions in dense superfluids.

We should like also to clarify more detaily the similarity between the hydrodynamics of rotating superfluids in the presence of the vortex lattice and the second sound regime in alkali crystals briefly mentioned in the Introduction to the Section 1.3.

Note that in the alkali crystals the second sound as a propagating mode with a drift velocity $\vec{v} \neq \dot{\vec{u}}$ ( $\dot{\vec{u}}$ is the lattice velocity) exists only in the frequency window $\gamma_U = \dfrac{1}{\tau_U} < \omega < \gamma_N = \dfrac{1}{\tau_N}$ for weak umklapp processes with $\gamma_U < \gamma_N$ (see Gurevich [1.21] for a review). In the same time for small frequencies $\omega < \gamma_U < \gamma_N$ the second sound is overdamped.

The same estimates distinguish hydrodynamics of slow and fast rotations. Namely, the total damping of a second sound $\dfrac{\text{Im}\,\omega}{\omega} = \dfrac{\gamma_U}{\omega} + \dfrac{\omega}{\gamma_N}$ is a sum of the damping due to umklapp process and $\text{Im}\,\omega \sim \gamma_U$ and a standard hydrodynamic damping $\text{Im}\,\omega \sim \omega^2 \tau_N \sim \dfrac{\omega^2}{\gamma_N}$ (see Eq. (1.1.25)). For $\gamma_U < \omega < \gamma_N$, $\dfrac{\text{Im}\,\omega}{\omega} << 1$ - the second sound wave is propagating and we have the hydrodynamics of the slow rotations. In the same time for $\omega < \gamma_U < \gamma_N$, $\dfrac{\text{Im}\,\omega}{\omega} >> 1$ (due to the contribution $\gamma_U / \omega$) and we have an overdamped second sound in the direction perpendicular to the vortex lines. Thus we restore the hydrodynamics of the fast rotations.

References to Chapter 1

Chapter 4. Quantum hydrodynamics of the p-wave superfluids with the symmetry of $^3$He-A.

Content.

4.1. Orbital hydrodynamics of bosonic and fermionic superfluids with the symmetry of A-phase of $^3$He.
4.1.1 Orbital hydrodynamics and collective modes in bosonic regime.
4.1.2. Orbital waves, the paradox of the intrinsic angular momentum and anomalous current in fermionic superfluids.
4.1.3. The spectrum of orbital waves in fermionic and bosonic superfluids with the symmetry of A-phase.
4.2 Two approaches to a complicated problem of chiral anomaly, anomalous current in fermionic (BCS) A-phase.
4.2.1. Supersymmetric hydrodynamics of the fermionic A-phase.
4.2.2 The different approach based on the formal analogy with quantum electrodynamics. Dirac equation in a magnetic field.
4.2.3 How to reach the hydrodynamic regime $\omega\tau \ll 1$.
4.2.4 The concept of the spectral flow, the cancellation of anomalies and the role of damping.
4.2.5 Experimental situation and discussion. Anomalous spin currents in 2D axial phase.

References to Chapter 4.

In the present chapter we derive the equations of orbital hydrodynamics and analyze the spectrum of collective excitations for bosonic and fermionic p-wave triplet superfluids with the symmetry of A-phase.

We discuss the spectrum of orbital waves, the paradox of the intrinsic angular momentum and the complicated problem of chiral anomaly (mass current nonconcervation) in the sperfluid hydrodynamics of the fermionic A-phase at $T = 0$.

We present two different approaches to the chiral anomaly, one based on supersymmetric hydrodynamics [4.1; 4.2] and another one on the formal analogy between Bogolubov – de Gennes equations for $^3$He-A [4.23] and the Dirac equation in quantum electrodynamics (QED, [4.17; 4.3; 4.4]). We are motivated by the experimental discovery of superfluid and superconductive fermionic systems with nodal (Dirac) points and lines, which exist in the complex order parameter or in the energy spectrum of the superfluid $^3$He-A (see Chapter 19), organic and heavy-fermion superconductors, ruthenates (Sr$_2$RuO$_4$) (see Chapter 9) and p-wave Fermi gasses in the regime of Feshbach resonance (see Chapter 6). Note that both competing approaches, which we discuss in this Chapter, are very general. An approach, connected with the construction of the supersymmetric  hydrodynamics, is based on the inclusion of the fermionic Goldstone mode in the low-frequency hydrodynamic action [4.5; 4.1; 4.2]. It can be useful for all nodal fermionic superfluids and superconductors with zeroes of the superconductive gap such as $^3$He-A, Sr$_2$RuO$_4$, UPt$_3$, UNi$_2$Al$_3$ and U$_{1-x}$Th$_x$Be$_{13}$ [4.5]. The second approach is also very nice and general. It is connected with the appearance of the Dirac-like spectrum of fermions with a zero mode [4.3; 4.4], which also arises in many condensed-matter systems such as $^3$He-A, chiral superconductor Sr$_2$RuO$_4$, organic conductor α-(BEDT-TTF)$_2$I$_3$, 2D semiconductors, or recently discovered graphene [4.6 - 4.9].

## 4.1 Orbital hydrodynamics of bosonic and fermionic superfluids with the symmetry of A-phase of $^3$He.

In the previous chapters we considered mostly hydrodynamics of superfluid $^4$He, which is a hydrodynamics of the isotropic bosonic superfluid. At zero temperatures hydrodynamics of a superfluid 4He is trivial. It is described in terms of the two equations, first one for superfluid velocity $\vec{v}_s$ (or mass current $\vec{j}_s = \rho_s(T=0)\vec{v}_s$, where superfluid density $\rho_s(T=0) = \rho$ - equals to the total density) [4.14; 4.16; 4.25]. The second equation is a conservation of mass (see Chapter 1). Note that the order parameter in the superfluid 4He is a scalar complex function $\Delta = \sqrt{\rho_s}e^{i\chi}$, where $\vec{v}_s = \dfrac{\hbar}{m_4}\vec{\nabla}\chi$ and $\rho_s = |\Delta|^2$ (see [4.13]). The order parameter in a bosonic or a fermionic superfluid with the symmetry of A-phase (where the role of the order parameter plays a superfluid gap, see Chapter 6) has a more complicated (tensor) structure. It's orbital part is characterized by three mutually perpendicular unit vectors $\vec{e}_1$, $\vec{e}_2$  and $\vec{l}$ , where $\vec{l} = [\vec{e}_1 \times \vec{e}_2]$ and $\vec{l}^2 = \vec{e}_1^2 = \vec{e}_2^2 = 1$ (see [4.10; 4.18; 4.56] and Chapter 6). In the homogeneous case $\vec{l}$ , $\vec{e}_1$, $\vec{e}_2$ coincide with the Cartesian unit vectors $\vec{e}_z$, $\vec{e}_y$, $\vec{e}_x$. However in general case they are slowly varying functions of $\vec{r}$ and $t$ . The orbital part of the order parameter in the A-phase is a complex vector

$$\vec{\Delta} = \Delta_0 e^{i\chi}(\vec{e}_1 + i\vec{e}_2), \qquad\qquad (4.1.1)$$

where $\Delta_0$ is the amplitude of the order parameter (of the superfluid gap in case of Fermi-liquids and Fermi-gasses). Note that $\vec{\Delta}$ in (4.1.1) corresponds to the spherical function $Y_{11}$ and thus $l = l_z = 1$ for the orbital momentum and its projection in the A-phase.

The superfluid velocity $\vec{v}_s$ in case of Fermi-liquid (3He-A) or Fermi gas is given by [4.10]:

$$\vec{v}_s = \frac{\hbar}{2m_3}\left(e_{1i}\vec{\nabla}e_{2i} + \vec{\nabla}\chi\right), \qquad (4.1.2)$$

where a factor $2m_3$ reflects the pairing of two fermions [4.10; 4.13] and their subsequent bose-condensation in superfluid $^3$He [4.12]. The additional (with respect to superfluid $^4$He) variable $\vec{l}$ corresponds to the quantization axis of the angular momentum of the p-wave pairs in the superfluid $^3$He.

Note that in the next chapters we will consider superfluidity (or superconductivity) in Fermi systems of the two types: strong coupling superfluidity, where we have tightly bound pairs (or difermionic molecules) well separated from each other. In this case the pairing takes place in real space. We will often call the Bose-Einstein condensation of local pairs in this case as a BEC limit of the superfluidity [4.19; 4.20]. Another type of superfluidity in Fermi systems corresponds to the creation and simultaneous bose-condensation of the extended Cooper pairs, which strongly overlap with each other in real space. The phenomenon of pairing takes place in momentum space in this limit. This is a standard BCS-type of superfluidity or superconductivity [4.13; 4.14]. Thus when we are speaking about bosonic superfluid we have in mind either elementary bosons (atoms of $^4$He), or composed bosons (molecules of $^6$Li$_2$ and $^{40}$K$_2$ formed in BEC-limit for p-wave superfluid Fermi-gasses, which we will consider in Chapter 6 [4.21]). Correspondingly when we are speaking about fermionic superfluids we have in mind BCS-type of pairing (as in $^3$He-A and $^3$He-B for example).

### 4.1.1 Orbital hydrodynamics and collective modes in bosonic regime.

In this subsection we will consider bosonic (BEC) regime having in mind first of all diatomic molecules with p-wave symmetry, which arise in ultracold Fermi-gasses of $^6$Li and $^{40}$K in the regime of Feshbach resonance (see Chapter 4). In BEC regime at $T = 0$ we could safely define the density of the orbital momentum of p-wave molecules (local pairs) as $\vec{L} = \frac{\hbar}{2m}\rho\vec{l}$. Correspondingly the total mass-current at $T = 0$ reads:

$$\vec{j}_B = \rho\vec{v}_s + \frac{\hbar}{2m}rot(\rho\vec{l}) = \rho\vec{v}_s + rot\vec{L}, \qquad (4.1.3)$$

where $rot\vec{L}$ - term in the r.h.s. of (4.1.3) is analogous to a diamagnetic displacement current well known in the Electrodynamics of Continuous Media [4.11]. Hydrodynamic energy for molecular A-phase reads:

$$E_B = E_0(\rho, \vec{l}, \partial_i\vec{l}) + \rho\frac{v_s^2}{2}, \qquad (4.1.4)$$

where we used Galilean transformation for $E_B$ to the coordinate frame where the superfluid velocity $\vec{v}_s = 0$. In this frame bosonic energy is $E_0$. The differential $dE_0$ in general case reads:

$$dE_0 = \mu d\rho + F_{ik}d\nabla_i l_k + M_k dL_k, \qquad (4.1.5)$$

where

$$F_{ik} = \frac{\partial E_0}{\partial\nabla_i l_k}, \qquad (4.1.6)$$

and $\vec{L}_B = \frac{\hbar\rho}{2m}\vec{l}$ is the density of orbital momentum.

The internal energy $E_0$ is connected with thermodynamic and liquid crystal like orbital energy. The term $F_{ik}d\nabla_i l_k$ provides quadratic in gradients contribution to $E_0$. It corresponds to the energy of the orbital deformation, which is similar to the deformation energy in the liquid crystals [4.22]. Finally $\mu$ in (4.1.5) is a chemical potential and

$\vec{M} = \dfrac{1}{2} rot\vec{v}_s$ .        (4.1.7)

The differential of the total energy $dE_B$ in (4.1.4) can be rewritten as:

$$dE_B = \left(\mu + \frac{\hbar}{4m}\left(rot\vec{v}_s \cdot \vec{l}\right) + \frac{v_s^2}{2}\right)d\rho + F_{ik}d\nabla_i l_k + \frac{\hbar\rho}{4m}\left(rot\vec{v}_s\right)_k dl_k + \rho v_{sk} dv_{sk} ,$$        (4.1.8)

where $dE_0 = \left(\mu + \dfrac{\hbar}{4m}\vec{l}\,rot\vec{v}_s + \dfrac{v_s^2}{2}\right)d\rho + F_{ik}d\nabla_i l_k$ .

We use again Landau approach for the superfluid hydrodynamics (see Chapter 1). Then collecting all the terms for time derivative of the energy $\dfrac{\partial E_B}{\partial t}$ under the divergence (collecting $div\vec{Q}$), we get the energy conservation law in the form:

$$\frac{\partial E_B}{\partial t} + \nabla_i \left\{\left(\mu + \frac{v_s^2}{2} + \frac{\hbar}{4m}\vec{l}\,rot\vec{v}_s\right)\rho v_{si} - F_{ik}\frac{\partial l_k}{\partial t}\right\} = 0 ,$$        (4.1.9)

where we can define again an energy flux $Q_i = \left(\mu + \dfrac{v_s^2}{2} + \dfrac{\hbar}{4m}\vec{l}\,rot\vec{v}_s\right)\rho v_{si} - F_{ik}\dfrac{\partial l_k}{\partial t}$ . Thus $\dfrac{\partial E_B}{\partial t} + div\vec{Q} = 0$ , and the energy conservation law (4.1.9) is consistent with a system of three hydrodynamic equations for $\dfrac{\partial \vec{l}}{\partial t}$ , $\dfrac{\partial \vec{v}_s}{\partial t}$ and $\dfrac{\partial \rho}{\partial t}$ . This system of equations yields [4.39]:

$$\frac{\partial \rho}{\partial t} + div(\rho\vec{v}_s) = 0 ,$$        (4.1.10)

$$\frac{\partial v_{si}}{\partial t} + \nabla_i\left(\mu + \frac{v_s^2}{2} + \frac{\hbar}{4m}\vec{l}\,rot\vec{v}_s\right) = \frac{\hbar}{2m}\partial_i\vec{l}\left[\vec{l}\times\dot{\vec{l}}\right] ,$$        (4.1.11)

$$\frac{\partial \vec{l}}{\partial t} + \left(\vec{v}_s\vec{\nabla}\right)\vec{l} + \left[\vec{l}, \frac{1}{2}rot\vec{v}_s - \frac{2m}{\hbar\rho}\nabla_i\vec{F}_i\right] = 0 ,$$        (4.1.12)

where we introduced $\vec{F}_i = \dfrac{\partial E_0}{\partial \nabla_i \vec{l}}$ (see 4.1.6).

The first equation (4.1.10) in the system (4.1.10)-(4.1.12) is a standard continuity equation at $T = 0$ . Note that in (4.1.10) we used that $div\left(rot\dfrac{\rho\vec{l}}{2}\right) = 0$ and thus we can also represent it in the form $\dfrac{\partial \rho}{\partial t} + div\vec{j}_B = 0$ . In other words, the displacement current as usual does not contribute to $div\vec{j}_B$ .

Second equation (4.1.11) is an equation for superfluid velocity with a non-trivial right-hand side. It corresponds to the well known (in the physics of $^3$He-A) Mermin-Ho [4.24] identity for $rot\vec{v}_s$ :

$$\nabla_i v_{sj} - \nabla_j v_{si} = \frac{\hbar}{2m}\vec{l}\left[\partial_i\vec{l}, \partial_j\vec{l}\right] .$$        (4.1.13)

The condition (4.1.13) is based on the definition (4.1.2) and plays an important role in the derivation of the system of equations (4.1.10 − 4.1.12). Note that we can also define the superfluid velocity in different way. To do that we should introduce three angles $\varphi^a(\vec{r}, t)$ defining the orientation of the trio of the mutually orthogonal unit vectors $\vec{e}_1$ , $\vec{e}_2$ and $\vec{l} = [\vec{e}_1, \vec{e}_2]$ . Let

$$\delta \vec{\theta} = \vec{\lambda}_a(\varphi)d\varphi^a \qquad (4.1.14)$$

be the Cartan form defining the infinitesimal rotation $\delta\theta$ corresponding to two neighboring points $\varphi^a$ and $\varphi^a + \delta\varphi^a$ in the group space $\{\varphi^a\}$ of the three-dimensional rotation group. Setting $\{\varphi^a\} = \vec{\varphi}$ where the vector $\vec{\varphi}$ is directed along the rotation axis and its magnitude is $tg\,\theta/2$ (where $\theta$ is the rotation angle) we get (see Andreev, Marchenko [4.30]):

$$\lambda_{ai} = \frac{2}{1+\varphi^2}\left(\delta_{ai} + \varepsilon_{iba}\varphi^b\right), \qquad (4.1.15)$$

where $\varepsilon_{iba}$ is the Levi-Civita tensor. We can easily verify that the quantities $\vec{\lambda}_a$ in (4.1.15) satisfy the "flatness" conditions:

$$\frac{\partial \vec{\lambda}_a}{\partial \varphi^b} - \frac{\partial \vec{\lambda}_b}{\partial \varphi^a} + \left[\vec{\lambda}_a, \vec{\lambda}_b\right] = 0. \qquad (4.1.16)$$

Since the conditions (4.1.16) are covariant with respect to a change of the coordinates $\varphi^a$, they retain their form also for any other parameterization of the rotations.

The superfluid velocity $\vec{v}_s$ in this approach has the components:

$$v_{si} = -\frac{\hbar}{2m}\vec{l}\,\frac{\delta\vec{\theta}}{\delta x_i} = -\frac{\hbar}{2m}\vec{l}\,\vec{\lambda}_a\,\frac{\partial\varphi^a}{\partial x_i}. \qquad (4.1.17)$$

We can easily check that this definition coincides with the standard definition (4.1.2) for the superfluid velocity and moreover the "flatness" condition (4.1.16) automatically guaranties the fulfillment of the Mermin – Ho identity (4.1.13). The Mermin-Ho identity yields non zero vorticity in ${}^3$He-A even in the absence of singular vortices (see Chapter 2) and, only with an account of continuous textures of the $\vec{l}$ - vector.

Third equation (4.1.12) is an equation for vector $\vec{l}$ or for the density of orbital momentum

$$\vec{L}_B = \frac{\hbar\rho}{2m}\vec{l}.$$

First two equations on $\partial\rho/\partial t$ and $\partial\vec{v}_s/\partial t$ after linearization yield the sound spectrum $\omega = c_s q$. Note that physically the sound Goldstone mode corresponds to the gauge transformation $\vec{\Delta} \to \vec{\Delta}e^{i\chi}$ of the order parameter (similar to the origin of a sound mode in superfluid ${}^4$He). On the level of $\vec{l}$ -vector it corresponds to the rotation on angle $\varphi$ around vector $\vec{l}$.

The equation on $\partial\vec{l}/\partial t$ after linearization yields the spectrum of orbital waves. In bosonic (molecular) superfluid it is quadratic in $q$: $\omega \sim q^2/m$ at small $q$. We can say that the A-phase is an orbital Ferromagnet [4.18].

Physically the second (orbital) Goldstone mode is connected with the rotations of the $\vec{l}$ -vector on angle $\gamma$ around the perpendicular (to vector $\vec{l}$) axis (see [4.26]).

Note that, as usual, the system of hydrodynamic equations (4.1.10-4.1.12) is compatible not only with the energy conservation (4.1.9) but also with the linear momentum conservation laws (see Chapter 1):

$$\frac{\partial j_B^i}{\partial t} + \frac{\partial}{\partial x_k}\left(\Pi_{ik}\right) = 0, \qquad (4.1.18)$$

where $\Pi_{ik} = \rho v_{si}v_{sk} + (P\delta_{ik} - F_{mn}\nabla_m l_n)\delta_{ik} + F_{km}\nabla_i l_m - \varepsilon_{ikj}\dot{L}_j$ is a momentum tensor and $P = -E_0 + \mu\rho + F_{ik}\nabla_i l_k + \frac{1}{2}rot\vec{v}_s\vec{L}$ is the pressure.

### 4.1.2 Orbital waves. The paradox of the intrinsic angular momentum and anomalous current in fermionic superfluids.

In fermionic (BCS-type) superfluids with a symmetry of A-phase we deal with extended Cooper pairs. The pairing takes place in momentum space. The fermionic quasiparticle spectrum in momentum space reads (see [4.10] and Chapter 6):

$$E_p = \sqrt{\left(\frac{p^2}{2m} - \mu\right)^2 + \frac{\left|\vec{\Delta p}\right|^2}{p_F^2}}\,, \qquad (4.1.19)$$

where $\vec{\Delta} = \Delta_0(\vec{e}_x + i\vec{e}_y)$ is the complex order parameter and $\Delta_0$ is the magnitude of the superfluid gap.

In fact, $\left|\vec{\Delta p}\right|^2 = \Delta_0^2 p^2 \sin^2\theta = \Delta_0^2 \left[\vec{p} \times \vec{l}\right]^2$, where $\vec{l} = \left[\vec{e}_x \times \vec{e}_y\right]$ is the unit vector of the orbital momentum (see Fig. 4.1). We note that Fermi momentum $p_F$ is fixed by the fixed density $n = \frac{p_F^3}{3\pi^2}$. The angle $\theta$ is between the momentum $\vec{p}$ and the axis $\vec{l} = \vec{e}_z$ of the quantization of the orbital momentum.

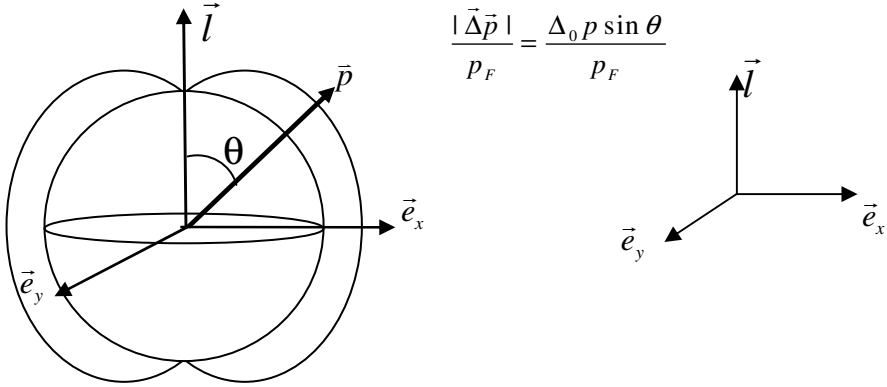

Fig. 4.1. The topology of the superfluid gap in the BCS A-phase. There are two nodes in the quasiparticle spectrum corresponding to the south and north poles [4.2].

Note that for a standard s-wave pairing, the quasiparticle spectrum in BCS fermionic superfluid is given by [4.12; 4.14] $E_p = \sqrt{\left(\frac{p^2}{2m} - \mu\right)^2 + \Delta^2}$. It has no zeroes (no nodes), and therefore the topology of the s-wave superfluid is trivial in momentum space. But for the triplet A-phase the quasiparticle spectrum in BCS superconductor (or superfluid) has two nodes for $\frac{p^2}{2m} = \mu$ and $\theta = 0, \pi$. Note that in BCS-superfluid $\mu \simeq \varepsilon_F$ and thus for the nodal points $p \simeq p_F$. Hence there are fermionic quasiparticles with practically zero energy in BCS A-phase. They play the role of a fermionic Goldstone mode (additional to Goldstone bosonic modes associated with sound and orbital waves). We will include the fermionic Goldstone mode in the hydrodynamics and construct the supersymmetric hydrodynamic action in Section 4.2, which describes both fermionic and bosonic Goldstone modes (see [4.1; 4.2; 4.5]) on the equal grounds. In this subsection we would like to stress that the topological effects, connected with the presence of the nodes in $E_p$, also are important for the spectrum of the orbital waves in the BCS domain at low temperatures $T \to 0$. Their spectrum is different from the spectrum in bosonic A-phase for small $\omega$ and $\vec{q}$. Note that for bosonic superfluid (with p-wave molecules or local pairs) $\mu \simeq -\dfrac{|E_b|}{2} < 0$ (where $E_b$ is a binding energy of a molecule) and hence the quasiparticle energy

$E_p = \sqrt{\left(\dfrac{p^2}{2m} + |\mu|\right)^2 + \dfrac{\Delta_0^2 p^2}{p_F^2} \sin^2\theta}$ has no nodes. Thus its topology in momentum space is trivial just as for the s-wave BCS pairing.

In BCS superfluid A-phase the symmetry requirements also allow us to write an additional anomalous term in the total mass current at $T = 0$:

$\vec{j}_{tot} = \vec{j}_B + \vec{j}_{an}$,        (4.1.20)

where

$\vec{j}_{an} = -\dfrac{\hbar}{4m} C_0 (\vec{l} \, rot \vec{l} \,)\vec{l}$                (4.1.21)

is an anomalous current.

In the BEC superfluid the important coefficient $C_0 = 0$ and the anomalous current is absent. Formally it is connected with the integral

$N_{3D}(0) \int d\varsigma_p \left(1 - \dfrac{\varsigma_p}{|\varsigma_p|}\right)$,        (4.1.22)

where $\varsigma_p = \dfrac{p^2}{2m} - \mu$ is a quasiparticle spectrum in the normal state. In BEC superfluid the chemical potential $\mu < 0$, $\varsigma_p = \dfrac{p^2}{2m} + |\mu| > 0$ and thus $C_0 = 0$. (Note that $N_{3D}(0) = \dfrac{m p_F}{2\pi^2}$ is the density of states in 3D Fermi gas).

In BCS superfluid $\mu \simeq \varepsilon_F > 0$ and the integral in (4.1.18) is nonzero. Moreover in naive estimates it defines the total density in the BCS superfluid ($C_0 \approx \rho$) at least in the weak-coupling case $\Delta_0 \ll \varepsilon_F$.

Thus it is a difficult question whether $C_0 = 0$ in BCS phase. If $C_0 \neq 0$ the spectrum of orbital waves is strongly modified. Moreover for nonzero $C_0$ we cannot get rid of anomalous current in (4.1.16) and (4.1.17). That is very unpleasant since the anomalous current $\vec{j}_{an}$ violates the conservation law (4.1.14) for the total mass current (for the total linear momentum) $\vec{j}_{tot}$. Namely the time derivative of $\vec{j}_{tot}$ cannot be expressed as a divergence of any momentum tensor $\Pi_{ik}$ (in contrast with a bosonic phase see (4.1.18)):

$\dfrac{\partial j_{tot}^i}{\partial t} \neq -\dfrac{\partial}{\partial x_k} \Pi_{ik}$.        (4.1.23)

Thus the presence of the anomalous current destroys the superfluid hydrodynamics of the A phase as $T \to 0$. Its contribution to the equation for the total linear momentum ($\dfrac{\partial j_{tot}^i}{\partial t}$) can be compensated only by adding a term with the relative normal velocity $\vec{W} = (\vec{v}_n - \vec{v}_s)$ and normal density $\rho_n(T = 0)(\vec{v}_n - \vec{v}_s)$ to the total current $\vec{j}_{tot}$ already at $T = 0$ (see [4.3; 4.4]). We would like to stress that it is preferable to construct a closed set of the hydrodynamic equations at $T = 0$ only in terms of a superfluid density $\rho_s$ and superfluid velocity $\vec{v}_s$ by putting, as usual, $\rho_s(T = 0) = \rho$. This scenario perfectly works for bosonic A-phase. But it is not clear whether we can describe fermionic (BCS) A-phase only in terms of the condensate oscillations.

Note that, as we already mentioned, the presence of the anomalous current (with $C_0 \neq 0$) also significantly modifies the spectrum of orbital waves.

In subsection 4.1.1 we mentioned that $\omega \sim q^2$ in bosonic A-phase. More precisely in BEC A-phase for small $\omega$ and $q$,

$$\rho\omega \sim \rho\frac{q_z^2}{m}, \qquad (4.1.24)$$

or equivalently $\omega \sim \frac{q_z^2}{m}$, where $\vec{l} \parallel \vec{e}_z$ (see Fig. 4.1) is a quantization axis of the orbital momentum. In the same time we will show in Chapter 7 (see also [4.2]) that

$$(\rho - C_0)\omega \sim \rho\frac{q_z^2}{m}ln\frac{\Delta_0}{v_F|q_z|}. \qquad (4.1.25)$$

We will also show diagrammatically that $C_0 \simeq \rho$ and hence $(\rho - C_0) \ll \rho$ in (4.1.25) in the weak-coupling case $\Delta_0 \ll \varepsilon_F$. The most straightforward way to obtain (4.1.25) is to use the diagrammatic technique of Galitskii, Vaks, Larkin [4.26] for the spectrum of collective excitations in p-wave and d-wave superfluids. The technique is based on the solution of the Bethe-Salpeter integral equation in the superfluid state (see Chapters 5 and 7). The details of the derivation of (4.1.25) will be presented in Chapter 7 (Section 7.4). Here we would like to emphasize that the spectrum (4.1.24) in bosonic A-phase corresponds to the density of angular momentum

$$\vec{L}_B = \frac{\hbar\rho}{2m}\vec{l}, \qquad (4.1.26)$$

while the spectrum (4.1.25) in the fermionic A-phase corresponds to the intrinsic angular momentum

$$\vec{L}_F = \frac{\hbar}{2m}(\rho - C_0)\vec{l}, \qquad (4.1.27)$$

which is different from (4.1.26) for $C_0 \neq 0$ and moreover $(\rho - C_0) \ll \rho$ for $C_0 \simeq \rho$ in the weak-coupling case. We note that there are several competing evaluations of $\vec{L}_F$ which are based not on the spectrum of orbital waves, but on the exact microscopic representation of the static ground-state Hamiltonian of the BCS A-phase. Here different groups provide conflicting results for $\vec{L}_F$. In [4.27 ;4.28] for $\vec{l} = const$ the evaluation of the intrinsic angular momentum yields $\vec{L}_F = \frac{\hbar\rho}{2m}\vec{l}$ in agreement with the bosonic phase, while the inclusion of the inhomogeneous textures of the $\vec{l}$-vector restores the expression (4.1.27).

We note that according to Legget [4.29], the total N-particle microscopic Hamiltonian $\hat{H}$ exactly commutes with the z-projection of the angular momentum $\hat{L}_z = \hbar\hat{N}/2$. This fact is in favor of the result $\vec{L}_F = \frac{\hbar\rho}{2m}\vec{l}$ in the BCS A-phase.

Note that rigorously speaking for $^3$He-A the question about the existence of the superfluid hydrodynamics at $T = 0$ has to some extent a purely academic interest since on the phase-diagram of superfluid $^3$He at $T = 0$ A-phase does not correspond to a global minimum of the Ginzburg-Landau Free-energy [4.10], and hence only an isotropic B-phase is realized here (see Fig. 4.2a). To get A-phase as a global minimum of the Free-energy we should switch on a large magnetic field $B_{par} \geq \frac{T_{C1}}{\mu_B}$, where $\mu_B$ is Bohr magneton (see Chapter 19). In this case B-phase will be completely paramagnetically suppressed and the global minima of the Free-energy will correspond to the A1-phase with $S_z^{tot} = 1$ (for z-projection of a total spin $S_{tot} = 1$ of the Cooper pair) and A2-phase with two bose-condensates and correspondingly with two projections $S_z^{tot} = \pm 1$.

For magnetic fields being smaller than paramagnetic limit $B < B_{par}$, but still sufficiently large the phase-diagram of a superfluid $^3$He will be given by Fig. 4.2b, and we see that the superfluid A2-phase (which is an analog of A-phase in case of nonzero magnetic field) can exist at $T = 0$ in some interval of pressures $P$.

For superfluid p-wave Fermi-gasses in the regime of Feshbach resonance (see Chapters 5 and 7) usually only one projection of spin (say $\sigma = \uparrow$) is captured by magnetic trap and thus we have triplet Cooper pairs only with $S_z^{tot} = 1$. We can say that we are dealing with 100% - polarized A1-phase here. However while spin sectors of A1 and A2 phases are very different, the orbital (or gauge-orbital sectors [4.10]) are very similar with respect to the spectrum of collective excitations (sound waves and orbital waves) as well as orbital superfluid hydrodynamics. Thus in this Chapter we will mainly discuss A-phase while in Chapter 7 we will discuss 100%-polarized A1-phase. Note that the orbital structure of the order parameter both in A1 and A2-phases is governed by the spherical function $Y_{11}$ and thus corresponds to $l = l_z = 1$ for the relative orbital momentum of a triplet pair.

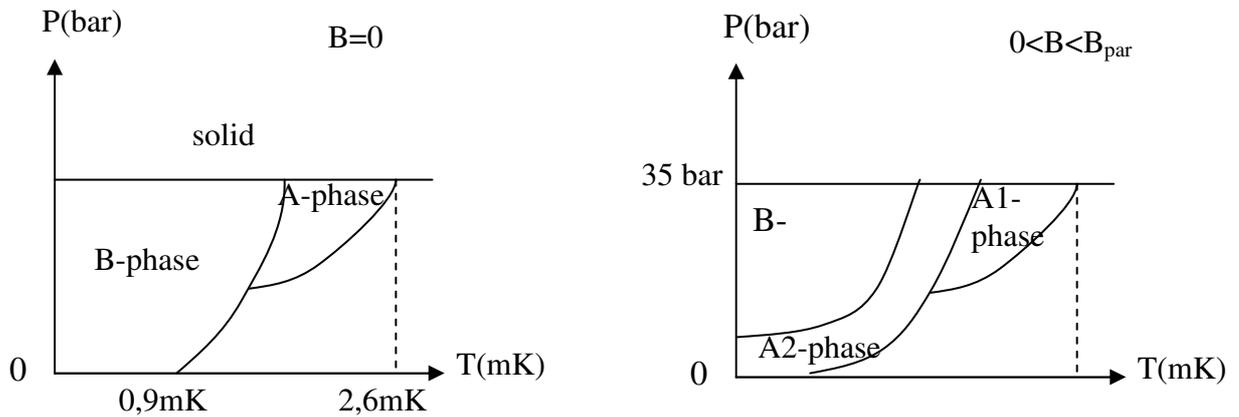

Fig. 4.2 Phase diagram (pressure versus temperature) in the superfluid $^3$He in the absence of magnetic field (a) and in the presence of magnetic field (b), which is sufficient enough but smaller than a paramagnetic limit $B < B_{par}$ (required for total suppression of B-phase). For zero temperatures only isotropic B-phase corresponds to global minima of Ginzburg-Landau Free-energy [4.10] at $T = 0$ on Fig. (a). However on Fig. (b) there are regions of pressure where anisotropic A2-phase corresponds to a global minimum of the Free-energy at $T = 0$.

## 4.2 Two approaches to a complicated problem of anomalous current in fermionic (BCS) A-phase.

In this Section we will reconsider two different approaches to the complicated problem of the anomalous current (which is often called the problem of chiral anomaly – see this Section and Section 4.4). These approaches (see [4.1; 4.3; 4.4]) were worked out in the late 1980-s, but are actively discussed till the present time.

### 4.2.1 Supersymmetric Hydrodynamics of the A-phase.

The main idea of [4.1] (see also [4.2]) was to check whether an anomalous current $\vec{j}_{an}$ (more precisely, the term $\vec{j}_{an}\vec{v}_s$ in the total energy) is directly related to the zeroes of the superfluid gap (see (4.1.5) and Fig. 4.1). Andreev, Kagan assumed in [4.1] that in a condensed matter system at low frequencies, the only physical reason for an anomaly (which can produce an anomalous current) can be related to the infrared singularity. We note that the ultraviolet singularities are absent in condensed matter systems [4.12; 4.13], in contrast to quantum electrodynamics [4.17]. Strong (critical) fluctuations are also suppressed in three-dimensional systems. The main idea in [4.1] was therefore to check the dangerous infrared regions where the gap is practically zero. For that, the authors of [4.1] considered the total hydrodynamic action of the fermionic (BCS) A phase for low frequencies and small $q$ vectors as a sum of bosonic and fermionic contributions:

$$S_{tot} = S_B + S_F, \qquad (4.2.1)$$

where $S_B(\rho, \vec{l}, \vec{v}_s)$ is the bosonic action related to the zeroes of the superfluid gap (see Fig. 4.3). Generally speaking, the idea in [4.1] was to use super-symmetric hydrodynamics to describe all the zero-energy Goldstone modes, including the fermionic Goldstone mode tha comes from the zeroes of the gap.

The authors of [4.1] were motivated by the nice paper of Volkov and Akulov [4.5], where the massless fermionic neutrino was for the first time included in the effective infrared Lagrangian for the electroweak interactions.

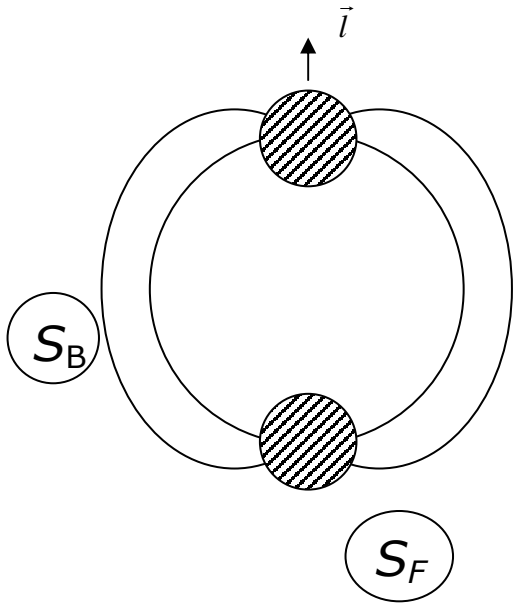

Fig. 4.3 A qualitative illustration of the fermionic ($S_F$) and bosonic ($S_B$) contributions to the total hydrodynamic action $S_{tot}$ of the A phase at $T \to 0$.

Bosonic part of the total action.

Bosonic part of the total action $S_B$ in (4.2.1) describes sound waves and orbital waves in purely bosonic (molecular) limit and does not contain an anomalous term $\vec{j}_{an} \vec{v}_s$ in the bosonic energy. Formally $S_B = \int L_B d^4 x$, where $x = (\vec{r}, t)$ and bosonic Lagrangian reads

$$L_B = \vec{L} \frac{\delta \vec{\theta}}{\delta t} - E_B, \qquad (4.2.2)$$

where $\frac{\delta \vec{\theta}}{\delta t}$ is a variational derivative, which defines an angular frequency, connected with the rotation angle $\theta$ (see Eq.'s (4.1.14), (4.1.15) and (4.1.16) and [4.1], [4.30] for more details on the parametrization of the three-dimensional rotation group which governs $\delta \vec{\theta}$).

The bosonic Lagrangian $L_B$ corresponds to bosonic energy $E_B$ in (4.1.4). Thus a bosonic part of the total action describes 3 equations (4.10), (4.11), (4.12) for bosonic hydrodynamics at $T = 0$. They contain the sound waves $\omega = c_s q$ and the orbital waves $\omega \sim \alpha q^2$.

Fermionic part of the total action.

A characteristic peculiarity of $^3$He-A and fermionic (BCS) phase of triplet superfluid Fermi gas is the fact that in addition to the usual ground-state degeneracy related to spontaneous breaking of continuous symmetries, there exists an additional degeneracy related to the vanishing of the quasiparticle energy $E_p$ (see Fig. 4.1 and 4.2 and eq. (4.1.15)) at the two points of the Fermi surface. It turns out that the states

$$a_1^+ |0\rangle \text{ and } a_2^+ |0\rangle, \qquad (4.2.3)$$

(where $|0\rangle$ is the quasiparticle vacuum, and $a_1^+$ and $a_2^+$ are the creation operators for quasiparticles with momenta $p_F \vec{l}$ and $-p_F \vec{l}$, respectively) have ground state energy. Similary to the way in which the usual degeneracy leads in the hydrodynamic description to the appearance of Bose fields, which vary slowly in space and time, this additional degeneracy leads to the appearance of "Fermionic goldstones", i.e. slowly varing anticommuting (Grassman) fields [4.31-4.36] $a_1(x)$, $a_2(x)$, $a_1^*(x)$ and $a_2^*(x)$, where $x = (\vec{r}, t)$. In fact, it is more convenient in this case to make use of certain linear combinations of these fields. The reason for this is that in the systems with Cooper pairing the quantities $a_1, a_2 ...$ are the subject to complicated gauge transformation laws. We introduce their linear combinations $\varphi_1(x)$, $\varphi_2(x)$, $\varphi_1^*(x), \varphi_2^*(x)$, so that they satisfy the same anticommutation relations:

$$\{\varphi_1^*, \varphi_1\} = \{\varphi_1, \varphi_2\} = \{\varphi_2^*, \varphi_1\} = ... = 0 \qquad (4.2.4)$$

as before, but under the gauge transformation $\vec{\Delta} \to \vec{\Delta} e^{i\chi}$ they transform as

$$\varphi_{1,2} \to \varphi_{1,2} e^{i\chi/2} ; \; \varphi_{1,2}^* \to \varphi_{1,2}^* e^{-i\chi/2} . \qquad (4.2.5)$$

We note that on account of the known properties of the mentioned linear (Bogolubov or u-v [4.13; 4.14]) transformations for spatially homogeneous systems, the subscripts 1,2 refer, as before, to the states with momenta $p_F \vec{l}$ and $-p_F \vec{l}$ .

The presence of the additional degeneracy of the ground state and the related Goldstone character of the fields $\varphi$ is due to the symmetry properties of the A-phase. We shall convince ourselves below of this independently, by determining the general form of the fermionic part of a Lagrangian $L_F$ satisfying all the necessary symmetry requirements.

We emphasize the following important point: hydrodynamics deals with the slowly varying quantities corresponding to the small statistical volume near the certain points of the momentum space. For the fermionic variables (in contradiction to the bosonic ones) this automatically leads to a small spatial fermionic density. In the Lagrangian $L_F$ ($\int L_F d^4 x = S_F$ in (4.2.1)) we can therefore limit ourselves to the consideration of the terms which are quadratic in $\varphi$ and $\varphi^*$.

The Lagrangian $L_F$ of the Fermi subsystem, which together with $L_B$ forms the total Lagrangian of superhydrodynamics, must be hermitian, invariant to rotations and gauge transformations, as well as with respect to the reflections $z \to -z$, $t \to -t$, where the $z$ axis is directed along the vector $\vec{l}$ . Moreover, on account of the momentum conservation, the Lagrangian $L_F$ must contain the products $\varphi_1^* \varphi_1$, $\varphi_2^* \varphi_2$, $\varphi_1 \varphi_2$, .... but not $\varphi_1^* \varphi_2$, $\varphi_2^* \varphi_1$ ... The fields $\varphi$, $\varphi^*$ behave as scalars under rotations. Under reflections they have the transformation properties:

$$\varphi_1^z = \varphi_2 ; \; \varphi_2^z = \varphi_1 ; \; \varphi_{1,2}^T = \varphi_{1,2}^* ; \; (\varphi_{1,2}^*)^T = \varphi_{1,2} . \qquad (4.2.6)$$

Here the subscripts z and T denote respectively the operations $z \to -z$ and $t \to -t$. The operation T is accompanied, as always, by a reversal of the order of factors.

There is a unique expression not containing derivatives and satisfying the abovementioned requirements:

$$g(\varphi_1^* \varphi_1 + \varphi_2^* \varphi_2). \qquad (4.2.7)$$

The coefficient $g$ appearing here is in reality a function of the magnitude of the momentum. This function should vanish at the point $p = p_F$. One may assume that the term (4.2.7) is absent from the Lagrangian $L_F$, since the equation $g(p_F) = 0$ is in fact a definition of the excitation momentum $p_F$. The existence of a zero in the function $g(p)$ is that "topological" property of $^3$He-A (and other BCS A-phases) which, together with the vector character of the order parameter $\vec{\Delta}$, is responsible for the gapless nature of the fields $\varphi$.

There exists a unique hermitian invariant involving the time derivatives:

$$\frac{i}{2}\left(\varphi_1^* \dot{\varphi}_1 + \varphi_2^* \dot{\varphi}_2 - \dot{\varphi}_1^* \varphi_1 - \dot{\varphi}_2^* \varphi_2\right). \qquad (4.2.8)$$

In the same time there are two invariants which are linear in the spatial derivatives. One of them contains the vector $\vec{l}$. Owing to the conditions $l^z = l$, $l^T = -l$ it has the form:

$$i\vec{l}\,(\varphi_1^* \vec{\nabla}\varphi_1 - \varphi_2^* \vec{\nabla}\varphi_2) - i\vec{l}\,(\vec{\nabla}\varphi_1^* \cdot \varphi_1 - \vec{\nabla}\varphi_2^* \cdot \varphi_2). \qquad (4.2.9)$$

The second invariant contains order-parameter $\vec{\Delta}$ and on account of the transformation properties $\vec{\Delta}^z = \vec{\Delta}$, $\vec{\Delta}^T = -\vec{\Delta}^*$ it yields:

$$i\vec{\Delta}(\varphi_1^* \vec{\nabla}\varphi_2^* + \varphi_2^* \vec{\nabla}\varphi_1^*) - i\vec{\Delta}^*(\vec{\nabla}\varphi_1 \cdot \varphi_2 + \vec{\nabla}\varphi_2 \cdot \varphi_1). \qquad (4.2.10)$$

Thus for $L_F$ we get:

$$L_F = \frac{i}{2}\left(\varphi_1^* \dot{\varphi}_1 + \varphi_2^* \dot{\varphi}_2 - \dot{\varphi}_1^* \varphi_1 - \dot{\varphi}_2^* \varphi_2\right) + i\frac{v_l}{2}\vec{l}\,(\varphi_1^* \vec{\nabla}\varphi_1 - \varphi_2^* \vec{\nabla}\varphi_2 - \vec{\nabla}\varphi_1^* \cdot \varphi_1 + \vec{\nabla}\varphi_2^* \cdot \varphi_2) +$$
$$+ i\frac{v_t}{2}\frac{\vec{\Delta}}{\Delta_0}(\varphi_1^* \vec{\nabla}\varphi_2^* + \varphi_2^* \vec{\nabla}\varphi_1^*) - i\frac{v_t}{2}\frac{\vec{\Delta}^*}{\Delta_0}(\vec{\nabla}\varphi_1 \cdot \varphi_2 + \vec{\nabla}\varphi_2 \cdot \varphi_1) \qquad , \qquad (4.2.11)$$

where the longitudinal and the transverse velocities $v_l$ and $v_t$ are the functions of the density. In $^3$He-A, for example, $v_t \sim \left(\dfrac{T_c}{\varepsilon_F}\right)v_l \ll (v_l \sim v_F)$ since $\dfrac{T_c}{\varepsilon_F} \sim 10^{-3}$.

The Lagrangian (4.2.11) refers to the spatially homogeneous case, when $\vec{l} = const$ and $\vec{\Delta} = const$. To treat the spatially inhomogeneous case it is necessary to note the following. Since the states 1 and 2 have a finite momentum $\pm p_F \vec{l}$, the "genuine" fields $\Psi_{1,2}$ are related to the slowly varying fields $\varphi_{1,2}$ by the equations:

$$\Psi_1 = \varphi_1 e^{ip_F \vec{l}\vec{r}}; \ \Psi_2 = \varphi_2 e^{-ip_F \vec{l}\vec{r}} \qquad (4.2.12)$$

and correspondingly for $\Psi_1^*$ and $\Psi_2^*$.

For $rot\,\vec{l} \neq 0$ the transformations of the form (4.2.12) do not exist. It is necessary to make use of the fields $\Psi$ and expand not only in terms of gradients, but in terms of the combinations $\vec{\nabla} \pm ip_F \vec{l}$. The corresponding Lagrangian $L_F$ is obtained from (4.2.11) by means of the substitution $\varphi \to \Psi$ and $\vec{\nabla} \to \vec{\nabla} \pm ip_F \vec{l}$.

Up to the total derivatives (which appear when the differentiations are transposed either completely to $\Psi_1$ or completely to $\Psi_2^*$) we have then:

$$L_F = \Psi_1^* \left( i\frac{\partial}{\partial t} + i\frac{v_l}{2}\left\{ \vec{l}, \left( \vec{\nabla} - ip_F\vec{l} \right) \right\} \right)\Psi_1 + \Psi_2 \left( i\frac{\partial}{\partial t} - i\frac{v_l}{2}\left\{ \vec{l}, \left( \vec{\nabla} - ip_F\vec{l} \right) \right\} \right)\Psi_2^* +$$

$$+ i\frac{v_t}{2}\left( \Psi_1^* \left\{ \frac{\vec{\Delta}}{\Delta_0}, \left( \vec{\nabla} - ip_F\vec{l} \right) \right\}\Psi_2^* + \Psi_2 \left\{ \frac{\vec{\Delta}^*}{\Delta_0}, \left( \vec{\nabla} - ip_F\vec{l} \right) \right\}\Psi_1 \right) \qquad (4.2.13)$$

where the curly brackets stand, as before, for anticommutators.

In addition, it is necessary to add to the Lagrangian independent invariant terms which contain explicitly the spatial derivatives of $\Delta$. The time derivatives may be omitted, since according to the system of Equations (4.2.10), (4.2.11), (4.2.12) they are quadratic in the spatial derivatives.

The invariants which are linear in the spatial derivatives $\frac{\partial \Delta_i}{\partial x_k}$ and are at the same time of zero order in the derivatives of the fields $\Psi$ can be of the two types. They may contain expressions obtained by means of the contractions with $\delta_{ik}$ or $\varepsilon_{ikl}$, or the terms of the form

$$\underbrace{\Delta_{l_1}^* \Delta_{l_2}^* \cdots}_{n} \underbrace{\Delta_{m_1} \Delta_{m_2} \cdots}_{n} \frac{\partial \Delta_i}{\partial x_k} \qquad (4.2.14)$$

and their complex conjugates, or the terms of the form

$$\underbrace{\Delta_{l_1}^* \Delta_{l_2}^* \cdots}_{n} \underbrace{\Delta_{m_1} \Delta_{m_2} \cdots}_{n-1} \frac{\partial \Delta_i}{\partial x_k} . \qquad (4.2.15)$$

All the terms of the first type are obviously genuine scalars. They are therefore invariant with respect to the transformations $z \to -z$ and consequently they must be multiplyed by the combinations of the $\Psi$ fields $\Psi_1^*\Psi_2^* + \Psi_2^*\Psi_1^*$ and $\Psi_1\Psi_2 + \Psi_2\Psi_1$, which vanish on the account of the anticommutation relations. There exist only three independent expressions of the second type which are invariant with respect to the rotation. Namely, these are $div\vec{l}$, $(\vec{l} rot\vec{l})$ and $(\vec{v}_s\vec{l})$. They are all pseudoscalars, and therefore must be multiplied by $(\Psi_1^*\Psi_1 - \Psi_2^*\Psi_2)$. Moreover only the last two of them are invariant with respect to the transformation $t \to -t$. Taking all this into account we have:

$$L_F = \Psi^* \Lambda \Psi , \qquad (4.2.16)$$

where

$$\Psi = \begin{pmatrix} \Psi_1 \\ \Psi_2^* \end{pmatrix} ; \quad \Psi^* = \begin{pmatrix} \Psi_1^* & \Psi_2 \end{pmatrix} , \qquad (4.2.17)$$

$$\Lambda = i\frac{\partial}{\partial t} + \begin{pmatrix} iv_l\vec{l}\left( \vec{\nabla} - ip_F\vec{l} \right) & -iv_t\dfrac{\vec{\Delta}}{\Delta_0}\vec{\nabla} \\[2ex] iv_t\dfrac{\vec{\Delta}^*}{\Delta_0}\vec{\nabla} & -iv_l\vec{l}\left( \vec{\nabla} - ip_F\vec{l} \right) \end{pmatrix} + \begin{pmatrix} i\dfrac{v_l}{2}div\vec{l} & i\dfrac{v_t}{2}div\dfrac{\vec{\Delta}}{\Delta_0} \\[2ex] i\dfrac{v_t}{2}div\dfrac{\vec{\Delta}^*}{\Delta_0} & -i\dfrac{v_l}{2}div\vec{l} \end{pmatrix} + a(\vec{l} rot\vec{l}) + b(\vec{v}_s\vec{l}) ,$$

$$(4.2.18)$$

and $a$ and $b$ are functions of the density. The function $b(\rho)$ is determined from the requirement of the Galilean invariance of the Lagrangian $L_F$. Under a Galilean transformation $\vec{r} = \vec{r}' + \vec{V}t$, we have $\vec{v}_s = \vec{v}_s' + \vec{V}$ and according to Quantum mechanics [4.37]:

$$\Psi_{1,2} = \Psi_{1,2}' \exp\left\{ im\vec{V}\left( \vec{r}' + \frac{\vec{V}t}{2} \right) \right\} \qquad (4.2.19)$$

$$\vec{\Delta} = \vec{\Delta}' \exp\left\{ i2m\vec{V}\left( \vec{r}' + \frac{\vec{V}t}{2} \right) \right\} . \qquad (4.2.20)$$

From (4.2.19), (4.2.20) it is clear, that, acting on quantities which transform like $\Psi_{1,2}$, the invariant operators are

$$i\frac{\partial}{\partial t} + i\vec{v}_s\vec{\nabla} + \frac{mv_s^2}{2}; \quad \vec{\nabla} - im\vec{v}_s \qquad (4.2.21)$$

and, correspondingly, thei complex conjugated operators acting on the quantities which transform like $\Psi_{1,2}^*$. Since in the adopted approximation one should neglect terms involving $v_s^2$, the Gallilean-invariant Lagrangian is obtained from the expression for $\vec{v}_s = 0$ by means of the substitution

$$\frac{\partial}{\partial t} \rightarrow \frac{\partial}{\partial t} + \vec{v}_s\vec{\nabla}; \quad \vec{\nabla} \rightarrow \vec{\nabla} \mp im\vec{v}_s, \qquad (4.2.22)$$

where the upper sign refers to the operators acting on the quantities which transform like $\Psi_{1,2}$, and the lower sign – to the quantities transforming like $\Psi_{1,2}^*$. As a result of this we find that $b = mv_l - p_F$. In the weak-coupling approximation $v_l = v_F$ and $b = 0$. In this approximation the Lagrangian $L_F$ corresponds to the well-known (see Volovik et al [4.3]) Bogolubov equations [4.57] for the BCS A-phase, linearized in $\vec{\nabla} \pm ip_F\vec{l}$.

The effective bosonic action.

We apply the results obtained above to the calculation of the effective action of the bosonic subsystem. For this it is necessary to eliminate the fermionic subsystem by evaluating the functional integral over the fermionic (Grassman) fields. To faciliate the calculations we proceed to the Euclidean formuladion by substituting $-\frac{\partial}{\partial \tau}$ for $i\frac{\partial}{\partial t}$ and setting $\tau = it$. The effective action is

$$S_{eff} = S_B + \Delta S_B, \qquad (4.2.23)$$

where $S_B = \int L_B d^4x$, $L_B$ is given by (4.2.2) and $x = (\vec{r}, \tau)$.
According to general rules (see for example [4.1] and [4.35]):

$$\Delta S_B = \ln\int D\Psi D\Psi^* \exp\left(\int d^4x \Psi^* \Lambda \Psi\right) = \ln\left(Det\Lambda\Lambda_0^{-1}\right) = Tr\ln\left(\Lambda\Lambda_0^{-1}\right) = Tr\left(F_1 - \frac{1}{2}F_2 + ...\right), \quad (4.2.24)$$

where $\Lambda$ is defined by (4.2.18), $\Lambda_0$ is a normalizing operator,

$$F_1 = \delta\Lambda\Lambda_0^{-1}, \quad F_2 = \left(\delta\Lambda\Lambda_0^{-1}\right)^2, \quad \delta\Lambda = \Lambda - \Lambda_0. \qquad (4.2.25)$$

The operator $\Lambda_0$ is usually chosen equal to the operator $\Lambda$ in the unperturbed equlibrium state. In an accord with the hydrodynamic character of the theory, which we are developing, we chose $\Lambda_0$ in the following manner. In the spatially homogeneous case the operator $\Lambda^{-1} \equiv G$ is an operator, whoose matrix elements $G(x_1, x_2)$ coincide with the fermionic Green's function. It can be easily determined by solving the equation $\Lambda G = \delta(x_1 - x_2)$. Thus we get for $G$:

$$G(x) = -\frac{e^{ip_F\vec{l}\vec{r}}}{2\pi^2v_lv_t^2}\left[\tau^2 + v_t^{-2}\frac{|\vec{\Delta}\vec{r}|^2}{\Delta_0^2} + v_l^{-2}\left(\vec{l}\vec{r}\right)^2\right]^{-2}\begin{pmatrix} \tau + i\dfrac{\vec{l}\vec{r}}{v_l} & \dfrac{i}{v_t}\dfrac{\vec{\Delta}\vec{r}}{\Delta_0} \\ \dfrac{i}{v_t}\dfrac{\vec{\Delta}^*\vec{r}}{\Delta_0} & \tau - i\dfrac{\vec{l}\vec{r}}{v_l} \end{pmatrix}, \qquad (4.2.26)$$

where $x = x_1 - x_2$.

In the general case, when the quantities $\vec{l}$, $\vec{\Delta}$, $p_F(\rho)$, ... are slowly varying functions of the coordinates and time, we introduce instead of $x_1$ and $x_2$ the "center of mass" variable $X = \dfrac{x_1 + x_2}{2}$ and relative variable $x = x_1 - x_2$. Then we will get the Green's function $G(X, x)$ of the "local-equilibrium", which is obtained from (4.2.26) by setting $\vec{l} = \vec{l}(X)$, $\vec{\Delta} = \vec{\Delta}(X)$, $p_F = p_F(X)$ and so on. We consider as a defenition of the operator $\Lambda_0$ the requirement that the matrix elements $\left(\Lambda_0^{-1}\right)_{x_1, x_2}$ should be equal to the functions $G(X, x)$.

The product $\Lambda G$ of $\Lambda$ and any other operator $G$ (defined by its matrix elements $G(x_1, x_2)$ ) has the matrix elements which are obtained from $G(x_1, x_2)$ by applying the operator (4.2.18), where all the differential operators act on the first argument $x_1$ and the arguments in $\vec{l}(x_1)$, $\vec{\Delta}(x_1)$, $p_F(x_1)$ etc contain also $x_1$. The action of the operator $\Lambda_0$ inverse to $G(X, x)$ is obviously defined by the first two terms in (4.2.18), where the differentiations must be fulfiled with respect to $x_1$, while $X$ must appear in the arguments $\vec{l}$, $\vec{\Delta}$, $p_F$ etc. Making use of the equalities $x_1 = X + (x/2)$, $\dfrac{1}{2}\partial/\partial x_1 = \partial/\partial x + \dfrac{1}{2}\partial/\partial X$ and expanding in terms of the gradients of the slowly varying functions, it is easy to calculate the operator $\delta\Lambda$ in (4.2.24) and (4.2.25). In doing this we should keep in mind that in our theory only the hydrodynamic asymptotic behavior is meaningful, i.e. the asymptotics for large $|x|$. Accordingly one has to retain only the leading terms for $|x| \to \infty$. Moreover, since the original bosonic action $S_B$ contains the density itself, while the quantities $\vec{\Delta}$ and $\vec{l}$ enter in $S_B$ only via derivatives, in the action $\Delta S_B$, as well, the zero-order terms in the derivatives will be absent by defenition, and the density may be considered constant. Thus in $\Delta S_B$ one must take into account only the spatial derivatives of $\vec{\Delta}$ and $\vec{l}$. (As noted above, the consideration of the time derivatives would lead to the terms of the higher order of smallness).

We write the formula (4.2.24) for $\Delta S_B$ in terms of matrix elements $F_1(X, x)$ of the operator $F_1$ in the $(X, x)$ representation.

$$\Delta S_B = \int d^4 X \lim_{x \to 0} tr\left\{ F_1(X, x) + \int d^4 x' F_1(X, x - x') F_1(X, x') + ... \right\}, \qquad (4.2.27)$$

where an operator $tr$ (in distinction from $Tr$ ) should be understood as a matrix rather than a complete operator.

We are interested in the part of $\Delta S_B$ which contains the terms of the lowest (second) order in the spatial derivatives $\partial/\partial X$. Such terms arise from the first term in the curly brackets in (4.2.27) and in this case they are proportional to $1/|x|$. In addition, the matrix elements $F_1$ contain also terms which are linear in the derivatives $\partial/\partial X$ and are proportional to $1/|x|^2$. On account of the second term in (4.2.27), which is given by the integral in the curly brackets, they contribute to the expression in the brackets proportional to $(\partial/\partial X)^2$ and $\ln \dfrac{\lambda}{|x|}$, i.e. it is exactly these terms which determine the hydrodynamic asymptotics. Thus we can restrict our attention only to the integral term in (4.2.27). The resulting expression for $F_1$, linear in the derivatives $\partial/\partial X$ (in a reference frame in which the coordinate axes for a given $X$ are respectively along $\vec{e}_1$, $\vec{e}_2$ and $\vec{l}$ ) has the form:

$$F_1 = -\frac{p_F}{4\pi^2 v_l v_t} \frac{e^{ip_F Z}}{(\tau^2 + \tau_n^2)^2} \sigma_i^{Tr} (\tau \tau_k \omega_{ik} - \varepsilon_{imk} \tau_k \tau_l \omega_{ml}) , \qquad (4.2.28)$$

where $\sigma_i^{Tr}$ are the transposed Pauli Matrices, $\tau_i = \left(\frac{x}{v_t}, \frac{y}{v_t}, \frac{z}{v_l}\right)$, $T_i = \left(\frac{X}{v_t}, \frac{Y}{v_t}, \frac{Z}{v_l}\right)$ and

$$\omega_{ik} = \left(\frac{\partial l_i}{\partial T_k} - \frac{\partial l_k}{\partial T_i}\right). \qquad (4.2.29)$$

Substituting $F_1$ from (4.2.28) into (4.2.27) we get

$$\Delta S_B = -\int d^4 X \, \frac{p_F^2}{32\pi^2 v_l^2} \, \omega_{ik} \int \frac{d|x'|}{|x'|} . \qquad (4.2.30)$$

The logarithmically divergent integral must be cut off in the weak-coupling case $\Delta_0 < \varepsilon_F$ at the upper limit (which is the wavelength of the notion $\lambda$) and at the lower limit at the mean free-path length $l_{MF}$ (in superfluid Fermi-liquid like $^3$He-A $l_{MF} \sim d\left(\frac{\varepsilon_F}{T_C}\right)^2$, where $d$ is an interatomic distance. We can also represent $l_{MF}$ as $v_F/\omega_c$ where the hydrodynamic frequencies $\omega$ are limited by $\omega_c$). In the chosen coordinate system, in view of $dl_z = 0$, one can represent $\omega_{ik}^2$ in the form:

$$\omega_{ik}^2 = 2v_l^2 \left(\frac{\partial l_\alpha}{\partial Z}\right)^2 + v_t^2 \left(\frac{\partial l_\alpha}{\partial X_\beta} - \frac{\partial l_\beta}{\partial X_\alpha}\right)^2 , \qquad (4.2.31)$$

where $\alpha, \beta = 1, 2$.

Therefore (4.2.30) corresponds to the following invariant expression for the fermionic contribution $\Delta L_B$ to the effective Lagrangian $L_{eff} = L_B + \Delta L_B$:

$$\Delta L_B = -\frac{p_F^2 v_l}{16\pi^2} \left\{ \left[\vec{l} \, rot \, \vec{l}\right]^2 + \frac{v_t^2}{v_l^2} \left(\vec{l} \, rot \, \vec{l}\right)^2 \right\} \ln \frac{\lambda}{l_{MF}} . \qquad (4.2.32)$$

Thus the elimination of the Fermionic degrees of freedom leads to the appearance in $L_{eff}$ of nonlocal (albeit weakly logarithmically divergent) terms, similar to the well-known (see [4.3; 4.38]) terms in the energy of $^3$He-A in the region $\left(\varsigma_0 \sim \frac{v_F}{T_C}\right) \ll \lambda \ll l$.

More specifically these terms lead only to a strong renormalization of the liquid crystal (De Gennes [4.22]) type coefficients $K_2$ and $K_3$ in $E_{0B}$ [4.38] (and after Fourier transform - to a logarithmic renormalization of the spectrum of orbital waves $\omega \sim \frac{q_z^2}{m} \ln \frac{\Delta_0}{v_F |q_z|}$ ).

Let us repeat that if we hope to obtain an anomalous term $\vec{j}_{an} = -\frac{\hbar}{4m} C_0 (\vec{l} \, rot \, \vec{l}) \vec{l}$ in total current with a large coefficient $C_0 \sim \rho$, we must get the term $\vec{j}_{an} \vec{v}_s$ in $\Delta L_B$. But due to small fermionic density $\rho_F \sim \Psi^* \Psi$ near the south and north poles (small statistical weight of the fermionic pockets on Fig. 4.4 in comparison with the total density $\rho$) we have to get $C_0 \sim \delta(\vec{p} \pm p_F \vec{l})$ in momentum space or accordingly $C_0 = const$ in real space. That is why in order to obtain an anomalous term in the current we must find very strong delta-functional infra-red divergencies in $\Delta L_B$. In our approach we found only weak (logarithmic) infra-red singularities in $\Delta L_B$, but we did not find a strong $\delta$-functional singularity when we accurately evaluated an integral over fermionic (Grassman) variables. Hence even if anomalous current exists in the BCS A-phase, it is not directly connected with the dangerous regions of the momentum space near zeroes of the gap (even if the chiral anomaly exists, in $^3$He-A it does not have an infra-red character).

### 4.2.2 The different approach based on the formal analogy with Quantum Electrodynamics.

The authors of [4.3; 4.4] proposed a different, and also rather nice approach based on a formal analogy between an anomalous current in $^3$He-A (and other 3D BCS A-phases) and the chiral anomaly in Quantum Electrodynamics (QED). They assume that the anomalous current in the fermionic A-phase is not directly related to the zeroes of the gap (and hence is not contained even in the super-symmetric hydrodynamics). They believe that it is related to the global topological considerations, and therefore a topological term should be added to supersymmetric hydrodynamics. To illustrate this point, they solve the microscopic Bogolubov- de Gennes (BdG) [4.23; 4.57] equations for fermionic quasiparticles in a given inhomogeneous twisted texture ($\vec{l} \parallel rot\vec{l}$) of the $\vec{l}$-vector. To be more specific they consider the case

$$\vec{l} = \vec{l}_0 + \delta\vec{l} \qquad (4.2.33)$$

with

$$l_z = l_{0z} = e_z,\ l_y = \delta l_y = Bx,\ l_x = 0, \quad (4.2.34)$$

where $\vec{e}_z$ is the direction of a nonperturbed $\vec{l}$ vector. In this case

$$\vec{l} rot\vec{l} = l_x \frac{\partial l_y}{\partial x} = B = const \qquad (4.2.35)$$

and, accordingly,

$$\vec{j}_{an} = -\frac{\hbar}{4m} C_0 B \vec{e}_z . \qquad (4.2.36)$$

Solution of BdG equation. Analogy with Dirac equation in magnetic field.

After linearization BdG equations become equivalent to Dirac equation in homogeneous magnetic field $B = (\vec{l} rot\vec{l})$. Namely after linearization BdG equations read $\hat{H}\chi_F = E\chi_F$, where the doublet $\chi_F$ has the form $\chi_F = \begin{pmatrix} u(x) \\ v(x) \end{pmatrix} \exp(ip_z z + ip_y y)$ and for the Hamiltonian we have:

$$\hat{H} = \varsigma(p_z)\hat{\sigma}_3 + v_t(\hat{\sigma}_1 \frac{1}{i}\partial_x - \hat{\sigma}_2(p_y - eBx)), \qquad (4.2.37)$$

where $\{\hat{\sigma}_1, \hat{\sigma}_2, \hat{\sigma}_3\}$ - are Pauli matrices, $\varsigma(p_z) = \frac{p_z^2 - p_F^2}{2m}$, $v_t = v_F \frac{\Delta_0}{\varepsilon_F}$ - corresponds to weak-coupling limit ($v_t \ll v_F$), and near the nodes $e = \frac{p_z}{p_F} = \pm 1$ is an electric charge. The solution of BdG equations yields for the doublet $\chi_F(x)$ (see Volovik, Balatskii, Konyshev [4.3]):

$$\chi_{n_L}(x) = \theta(-eB)\begin{pmatrix} \alpha_{n_L} f_{n_L}(\tilde{x}) \\ i\beta_{n_L} f_{n_L-1}(\tilde{x}) \end{pmatrix} + \theta(eB)\begin{pmatrix} \alpha_{n_L} f_{n_L-1}(\tilde{x}) \\ i\beta_{n_L} f_{n_L}(\tilde{x}) \end{pmatrix}, \qquad (4.2.38)$$

where $\theta$ is a step-function, $f_{n_L}(\tilde{x}) = f_{n_L}\left(x - \dfrac{p_y}{eB}\right)$ - is the ortho-normalized wave-function of the harmonic oscillator [4.32], $f_{-1} = 0$ ; $\left|\alpha_{n_L}\right|^2 = \dfrac{E_{n_L} + \varsigma(p_z)}{2E_{n_L}}$ and $\left|\beta_{n_L}\right|^2 = \dfrac{E_{n_L} - \varsigma(p_z)}{2E_{n_L}}$ are Bogolubov coefficients ( $\left|\alpha_{n_L}\right|^2 + \left|\beta_{n_L}\right|^2 = 1$ ).

Accordingly for the spectrum

$$E_{n_L}(p_z) = \pm\sqrt{\varsigma^2(p_z) + \tilde{\Delta}_{n_L}^2} \ , \qquad (4.2.39)$$

where $\tilde{\Delta}_{n_L}^2 = 2n_L v_f^2 p_F \left|eB\right|$ is a gap squared and $n_L$ is a quantum number for the Landau level [4.32]. The solution for $\chi_{n_L}(x)$ in (4.2.38) contains the level asymmetry.

Namely, for $n_L \neq 0$ (see Fig. 4.4) all the levels are gapped $\tilde{\Delta}_{n_L}^2 \neq 0$ and doubly degenerate with respect to $p_z \to -p_z$ . Their contribution to the total mass-current is zero for $T \to 0$ .

Zero mode and anomalous current at $T = 0$ .

However for $n_L = 0$ there is no gap $\Delta_0 = 0$ and we have an asymmetric chiral branch which exists only for $p_z < 0$ (only for one sign of $eB$ ). The energy spectrum for $n_L = 0$ yields:

$$E_0 = \varsigma(p_z) . \qquad (4.2.40)$$

We can say that there is no gap for zeroth Landau level. Moreover in BCS (fermionic) A-phase $E_0 = 0$ for $\left|p_z\right| = p_F$ - the chiral level crosses the origin in Fig 4.4, so we have a zero mode.

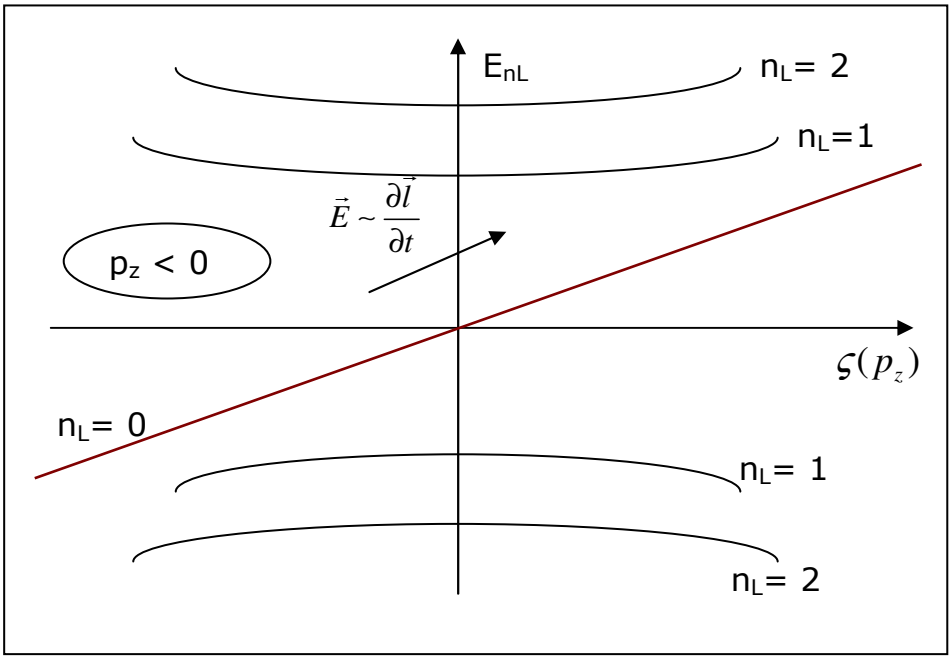

Fig. 4.4 Level structure of the Dirac equation in the magnetic field $B = (\vec{l}\,rot\,\vec{l})$ from [4.2]. All the levels with $n_L \neq 0$ are doubly degenerate. The zeroth level is chiral. It crosses the origin at $\left|p_z\right| = p_F$ in the BCS (fermionic) A-phase. We also illustrate the concept of the spectral flow, which will be discussed in Subsection 4.2.5.

We note that in the bosonic A-phase the chemical potential $\mu \simeq -\dfrac{|E_b|}{2} < 0$ ($|E_b|$ is a molecular binding energy). Thus $E_0 = \varsigma(p_z) = \dfrac{p_z^2}{2m} - \mu \ge |\mu|$ and the zeroth Landau level does not cross the origin. The absence of a zero mode in molecular A-phase is the physical reason why the coefficient $C_0 = 0$ there.

The zeroth Landau level gives an anomalous contribution to the total current in fermionic A-phase:

$$\vec{j}_{an}(\vec{r} = 0) = -\vec{e}_z(\vec{l}\,rot\vec{l}) \int_{p_z < 0} \frac{p_z}{2\pi^2}\,d\varsigma(p_z) = -\frac{\hbar C_0}{4m}(\vec{l}\,rot\vec{l})\vec{l} \ , \qquad (4.2.41)$$

$$\frac{(\vec{l}\,rot\vec{l})\,p_z}{2\pi^2\,p_F} = \frac{eB}{2\pi^2} = \int |f_0|^2 \frac{dp_y}{2\pi} \ , \qquad\qquad (4.2.42)$$

and hence

$$C_0 \simeq \frac{mp_F^3}{3\pi^2} \simeq \rho \qquad\qquad (4.2.43)$$

in the fermionic phase (in 100%-polarized A1 phase in magnetic traps, which we will consider in Capter 7, $C_0 \simeq \rho_\uparrow = \dfrac{mp_{F\uparrow}^3}{6\pi^2}$ ).

We note that $f_0(x - p_y\,|eB|)$ in (4.2.42) is an eigen-function of a zeroth Landau level. It is easy to see that the integral for $C_0$ in (4.2.41) and (4.2.42) is governed by the narrow cylindrical tube inside the Fermi sphere (see Fig. 4.5) with the length $p_F$ parallel to the $\vec{l}$ vector and with the radius of the cylinder squared given by:

$$\left\langle p_y^2 \right\rangle \sim p_F\,|eB| \ . \qquad\qquad (4.2.44)$$

According to the ideas of [4.3; 4.18], this tube plays the role of a vortex in the momentum space, thus providing a normal core and an anomalous current at $T = 0$.

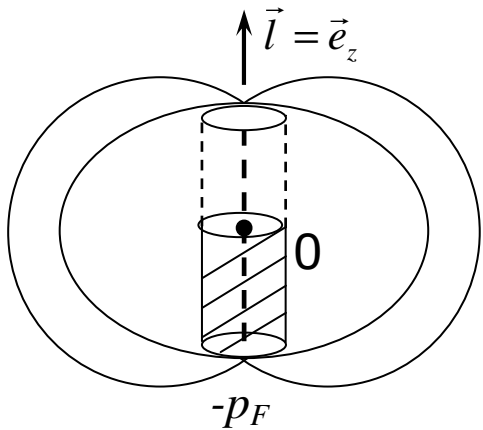

Fig. 4.5 The contribution to the coefficient $C_0$ is governed by a narrow cylindrical tube of the length $p_F$ and the width $\left\langle p_y^2 \right\rangle \sim p_F\,|eB|$ inside the Fermi sphere [4.2]

We note that a key result in [4.3; 4.4] related to the absence of the gap for the energy of the zeroth Landau level (see (4.2.40)) is pretty stable with respect to small modifications of the texture of the $\vec{l}$ -vector in (4.2.33), (4.2.34). Our careful analysis shows that the account of small bending corrections with $\left[\vec{l}rot\vec{l}\right]^2$ to the twisted texture (small tilting of the magnetic field with respect to the $(x, y)$ plane $\vec{B} = B_0\vec{e}_z + B_1\vec{e}_x$) as well as an account of small inhomogenities of a magnetic field $B = B_0 + B_1 x$, which lead to a double-well effective potential, does not suppress the zero mode in the spectrum of the BdG equations. In other words, an account of these corrections does not lead to the appearance of a gap $\tilde{\Delta}_{n_L} = 0$ for the zeroth Landau level.

### 4.2.3 How to reach the Hydrodynamic regime $\omega\tau \ll 1$.

Inspite of the zero mode stability, the authors of [4.1] expressed their doubts regarding the calculation of $C_0$ based on Dirac equation in the homogeneous magnetic field $B = \vec{l}rot\vec{l}$. From their standpoint, the calculation of $C_0$ in (4.2.41), (4.2.42) is an oversimplification of a complicated many-particle problem. In particular, they emphasized the role of the finite damping $\gamma = 1/\tau$ and of the other residual interactions in destroying the chiral anomaly (which is connected with the states inside the Fermi sphere on Fig. 4.5). Thus they hope to restore the superfluid hydrodynamics at $T = 0$ without the normal velocity $\vec{v}_n$ and the normal density $\rho_n$. Indeed, if the damping $\gamma$ is larger than the level spacing of the Dirac equation, we have

$$\omega_0 = v_t p_F \sqrt{\frac{|\vec{l}rot\vec{l}|}{p_F}} \qquad (4.2.45)$$

in the case where $\varsigma(p_z) = 0$, and then the contribution from the zeroth Landau level should be washed out by the damping (see Fig. 4.6). As a result the hydrodynamic regime will be established.

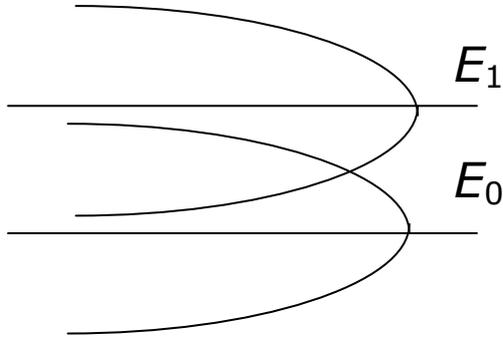

Fig 4.6 The possible role of damping in reaching the hydrodynamic limit for low frequencies and small $\vec{k}$ vectors for $\gamma > \omega_0$ ($\omega_0 = E_1 - E_0$ is the level spacing) [4.2].

The damping $\gamma$ for the chiral fermions (for the fermions living close to the nodes) in a very clean fermionic (BCS) A-phase without impurities is defined at $T = 0$ by the different decay processes (see [4.13]).

It is natural to assume that the only parameter that determines $\gamma$ at $T = 0$ for chiral fermions is the gap $\Delta_0\langle\theta\rangle = \Delta_0\langle p_\perp\rangle / p_F$ (where $\sin\theta \approx \theta$ close to the nodes). The leading term in decay processes is given by the emission of an orbital wave (see Fig. 4.7). It is given by (see [4.2])

$$\gamma \sim \left[\frac{\Delta_0^2 p_\perp^2 / p_F^2 + v_F^2(p_z - p_F)^2}{\varepsilon_F}\right]. \qquad (4.2.46)$$

For $p_z = p_F$ ($\varsigma(p_z) = 0$), we have

$$\gamma \sim \frac{\Delta_0^2}{\varepsilon_F} \frac{p_\perp^2}{p_F^2}. \qquad (4.2.47)$$

We note that for the chiral fermions on the zeroth Landau level, we have

$$\frac{\langle p_\perp \rangle}{p_F} = \left( \frac{|\vec{l}\, rot\vec{l}\,|}{p_F} \right)^{1/2} \qquad (4.2.48)$$

and the level spacing for $\varsigma(p_z) = 0$ is

$$\omega_0 \sim \Delta_0 \frac{\langle p_\perp \rangle}{p_F}. \qquad (4.2.49)$$

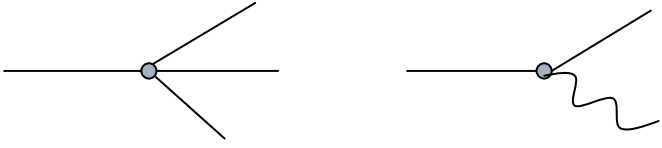

Fig. 4.7 Different decay processes for the damping of chiral fermions at $T = 0$: the standard three-fermion decay process and the decay process with an emission of the orbital wave [4.2]

Hence, $\gamma / \omega_0 \ll 1$ close to the zero mode for these two decay processes (the second decay process is a standard three-fermion decay, which also yields $\gamma / \omega_0 \ll 1$). Thus a ballistic regime is established. It is therefore difficult to wash out the contribution from the zeroth Landau level by the different decay processes in superclean $^3$He-A phase at $T = 0$. We note that the hydrodynamic regime $\omega \tau \ll 1$ could be easily reached in the presence of nonmagnetic impurities or in the presence of aerogel [4.40; 4.41; 4.42] (see Section 4.2.6).

### 4.2.4 The concept of the Spectral Flow and the Exact Anomaly Cancellation

If the anomalous current exists in a superclean fermionic A-phase at $T = 0$, it should be compensated somehow. According to Volovik et al [4.3], the deficit in the equation for the conservation of the total linear momentum due to the presence of the anomalous current $\vec{j}_{an}$:

$$\frac{\partial j_{an}^i}{\partial t} + \frac{\partial}{\partial x_k} \pi_{ik} = I_l \qquad (4.2.50)$$

with a source term

$$\vec{I} = \frac{3\hbar}{4m} C_0 \vec{l} \left( rot\vec{l} \cdot \frac{\partial \vec{l}}{\partial t} \right) \qquad (4.2.51)$$

is exactly compensated by the quasiparticle contribution $\vec{P}_{quasi}$:

$$\frac{\partial P_{quasi}^i}{\partial t} + \frac{\partial \Phi_{ik}}{\partial x_k} = -I_i, \qquad (4.2.52)$$

where $\vec{P}_{quasi} = \rho_n(T=0)(\vec{v}_n - \vec{v}_s)$ in the hydrodynamic regime. Correspondingly the total current in fermionic A-phase

$$\vec{j}_{tot} = \vec{j}_B + \vec{j}_{an} + \vec{P}_{quasi} \qquad (4.2.53)$$

is still conserved.

We note that a normal density $\rho_n(T=0) \sim \left|\left[\vec{l} \, rot\vec{l}\right]\right| / \Delta_0$ is a non-analytic function in $^3$He-A and is related to the nonzero bending. The arguments in [4.3] in favor of exact anomaly cancellation are connected with the nonconservation of the axial current $j_5$ in QED (see [4.17]), where the source term $\vec{I}$ is compensated via the Schwinger term $\vec{E}\vec{B} \sim \dfrac{\partial \vec{l}}{\partial t} \, rot\vec{l}$ ($\vec{E} = \dfrac{\partial \vec{l}}{\partial t}$ is an electric field and $\vec{B} = rot\vec{l}$ is a magnetic field). Physically, according to [4.3] and [4.43], this cancellation is due to the spectral flow from the negative to the positive energy values along the anomalous branch with $n_L = 0$ in Fig. 4.4 and then to the quasiparticle bath in the presence of an electric field $\vec{E} \sim \dfrac{\partial \vec{l}}{\partial t}$ (of a time-dependent texture of the $\vec{l}$ vector). Considering the tube, which produces $\vec{j}_{an}$ on Fig. 4.5, as a vortex in the momentum space, Volovik et al [4.3] and Stone et al [4.27] use the analogies between the physics of bulk $^3$He-A and the physics of the vortex core. They also consider the role of damping exactly opposite to our considerations. Note that in the physics of a vortex-core in the case of cylindrical symmetry there is again one anomalous level which crosses the zero energy (see Fig. 4.8).

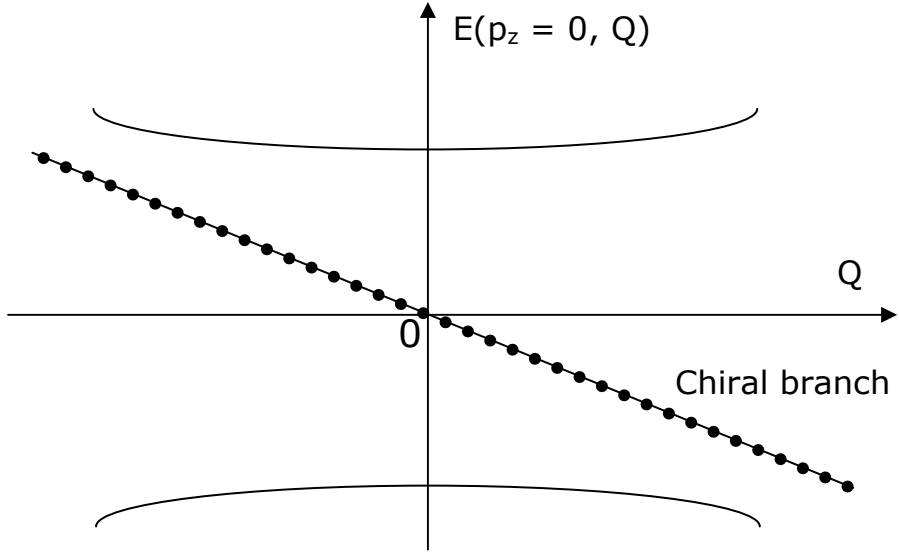

Fig. 4.8 The level structure in the vortex core of $^3$He-A [4.2]. All the branches are even in the generalized angular momentum $Q$, but one branch $E(p_z = 0, Q) = -\omega_0 Q$, which crosses zero energy at $Q = 0$, is chiral (odd in $Q$). It participates in the momentum exchange between the fermions in the vortex core and the fermions of the heat bath in the hydrodynamic limit $\omega\tau \ll 1$ according to [4.18].

At $T = 0$, as a function of the generalized angular momentum $Q$, it represents the set of discrete points separated by a minigap $\omega_0 \sim \Delta_0^2 / \varepsilon_F$. Therefore, at $T = 0$ and in the superclean case $\gamma = 1/\tau \to 0$, the spectral flow from negative to positive energies is totally suppressed. Thus in the ballistic regime $\omega\tau \gg 1$ according to Volovik and Stone it is very difficult to transfer momentum to the quasiparticles and in this way to guarantee the conservation of the total current $\vec{j}_{tot}$ in (4.2.53). In the same time in the hydrodynamic regime $\omega\tau \ll 1$ it is easy to transfer momentum along an anomalous branch to quasiparticles and thus to restore a conservation of the total current.

The authors of [4.1] vice versa think that for $\omega\tau \ll 1$ the coefficient $C_0$ in front of the anomalous current becomes small $C_0(\omega\tau \ll 1) \to 0$. Thus $\vec{j}_{an} \to 0$ and it is not necessary to add the quasiparticle current in the expression for $\vec{j}_{tot}$ in (4.2.53). In other words, for $\omega\tau \ll 1$ :

$$\vec{j}_{tot} = \vec{j}_B = \rho\vec{v}_s + rot\left(\frac{\hbar\rho\vec{l}}{2m}\right)$$ - the total current $\vec{j}_{tot}$ coincides with the bosonic current $\vec{j}_B$. Hence

the superfluid hydrodynamics is restored (without $\rho_n(T=0)$ and $\vec{v}_n$ ) in BCS A-phase at low frequencies and wave-vectors according to the philosophy of [4.1]. Note that in the physics of the vortex core the analogous discussion about an existence of chiral anomaly and it's possible contribution to one of the Hall-Vinen friction coefficients (see Chapter 1) was started in [4.44] by Thouless (who believes in a Berry phase [4.55] without an anomaly) and was thoroughly investigated by Kopnin et al in [4.45] for different temperature regimes in clean and dirty limits. The authors of [4.45] derived an anomalous contribution to the friction coefficients in the dirty limit $\omega\tau \ll 1$ (large number of impurities) and at finite temperatures. Returning back to the bulk $^3$He A-phase, we can say that an ideology of Volovik and Stone on exact anomalies cancellation does not work in ballistic (superclean limit) at $T=0$ (both in bulk and for the vortices). Hence the question of how the total current (total linear momentum) is conserved in this case remains open for an infinite system (without the walls). We note that at very small but finite temperatures $T \ll T_C$ there is a finite number of normal quasiparticles in the system. Hence the damping $\gamma = 1/\tau \sim T^n$ becomes finite and the possibility of the spectral flow and the momentum exchange with the thermal bath restores at low frequencies $\omega < \gamma$ according to Volovik and Stone. However we would like to stress that at $T \neq 0$ the relative normal velocity $\vec{v}_n - \vec{v}_s = \dfrac{\partial E_0}{\partial \vec{P}_{quasi}}$ becomes an additional hydrodynamic variable (see Chapter 1 and [4.16]) and hence the cancellation of the linear momentum deficit in (4.2.50), (4.2.52) will occur automatically.

Thus, the problem of the exact anomaly compensation exists only at $T=0$. We think that in this case the exact cancellation between the time derivatives of the anomalous and quasiparticle currents should be demonstrated explicitly by deriving and solving the kinetic equations for the nodal quasiparticles both in the ballistic and the hydrodynamic regime. Note that an approach based on the kinetic equation for quasiparticles at different temperatures and the impurity concentrations in a vortex core of the s-wave superconductors and the superfluid $^3$He was worked out by Kopnin et al [4.45] in the case of a singular vortex.

In the case of the nonsingular vortex structures in $^3$He-A (which can be produced by the textures of $\vec{l}$ vector via Mermin-Ho identity, for example) we should mention also papers [4.43] where the authors consider the scattering of quasiparticles on the walls of the container for a finite system to obtain a finite damping $\gamma$ at $T=0$. The importance of the prehistory of the orbital texture for the spectral flow concept was also stressed by Volovik in these papers.

### 4.2.5 Expreimental situation and discussion.

Concluding this Section we would like to emphasize once more that we discuss a complicated problem of chiral anomaly and mass-current non-conservation in BCS A-phase at $T=0$. We presented two different approaches to this problem – one based on sypersymmetric hydrodynamics, another one – on the formal analogy with Dirac equation in QED-theory. We evaluate the damping $\gamma = 1/\tau$ due to different decay processes in superclean BCS A-phase at $T=0$ and find that $\gamma$ is small in comparison with the level spacing $\omega_0$ of the BdG-equation. To reach the hydrodynamic regime $\omega\tau \ll 1$ we need a sufficient amount of aerogel or nonmagnetic impurities at $T=0$. We assume that both in a hydrodynamic and in a ballistic regime at $T=0$ we have to derive a reliable kinetic equation to demonstrate explicitly an exact cancellation between time-derivatives of anomalous current $\vec{j}_{an} = -\dfrac{\hbar}{4m}C_0(\vec{l}\,rot\vec{l}\,)\vec{l}$ and quasiparticle contribution $\vec{P}_{quasi}$ in the equation for the conservation of the total linear momentum $\vec{j}_{tot}$. Note that for the full theoretical analysis of the problem the other residual interactions different from damping are aslo important for nodal fermions. To check whether a chiral anomaly has an infra-red manifestation (which was not caught in the approach based on supersymmetric hydrodynamics of [4.1]) it will be useful to derive a complete set of Ward identities (see [4.13; 4.17]) between self-energies of chiral fermions $\Sigma$ and the corresponding vertices $\Gamma$. The idea is to find in this approach either a strong infra-red singularity or a powerful reexpansion of the quasiparticle spectrum for $\omega, k \to 0$.

Note that the importance of the residual Fermi-liquid like interactions for the analysis of half-integer vortex in $^3$He-A was recently emphasized by Leggett et al [4.46].

We invite experimentalists to enter this very interesting problem. It will be important to measure a spectrum and damping of orbital waves in superfluid A-phase of $^3$He at low temperatures $T \ll T_C$. The spectrum is quadratic for low frequencies $(\rho - C_0)\omega \sim \rho \dfrac{q_z^2}{m}\ln\dfrac{\Delta_0}{v_F \mid q_z \mid}$ and contains a density of intrinsic angular momentum $L_F = \dfrac{\hbar}{2m}(\rho - C_0)$ near the linear in frequency term (see Chapter 7 for more details). Moreover it is possible to show that in the weak-coupling limit $\Delta_0 \ll \varepsilon_F$: $\dfrac{(\rho - C_0)}{\rho} \sim \dfrac{\Delta_0^2}{\varepsilon_F^2} \ll 1$ and thus $L_F$ is very small in comparison with a standard (bosonic) angular momentum $L_B = \dfrac{\hbar\rho}{2m}$. (In $^3$He-A we are in a weak-coupling limit $\dfrac{\Delta_0}{\varepsilon_F} \sim 10^{-3}$ and $\dfrac{\rho - C_0}{\rho} \sim 10^{-6}$). Note that at higher frequencies $\omega > \Delta_0^2 / \varepsilon_F$ the spectrum of orbital waves is almost linear $\omega^2 \ln\dfrac{\Delta_0}{\mid\omega\mid} \sim q_z^2 v_F^2 \ln\dfrac{\Delta_0}{v_F\mid q_z\mid}$, where $\vec{e}_z \parallel \vec{l}$ (see [4.26]). Note also that in a strong-coupling case $\Delta_0 \geq \varepsilon_F$: $C_0 \ll \rho$ and we restore the hydrodynamics without an anomalous term.

The damping of the orbital waves provides an evaluation of the orbital viscosity in $^3$He-A at low temperatures $T \ll T_C$. Note that even in this case it is an interesting possibility to get an overdamped (diffusive) character of the spectrum at low frequencies. This possibility is supported theoretically in [4.47], where Brusov et al obtained several overdamped modes in the partially polarized A1-phase via the functional integral technique in the hydrodynamic limit of small $\omega$ and $\vec{q}$.

Basically we propose to extend the measurements of orbital inertia and orbital viscosity (performed by Bevan et al [4.47] in non-singular vortex textures in A-phase) to low temperatures $T \ll T_C$. Of couse to have this possibility we need to create a spin-polarization (which according to Fig. 4.1 extends A2-phase to low temperatures on phase-diagram of a superfluid $^3$He).

Note that another possibility to get an overdamped diffusive spectrum was considered in [4.48] in the impurity diagrammatic technique [4.49; 4.50] for the hydrodynamic regime $\omega\tau \ll 1$ of spin waves in a frustrated two-dimensional AFM, which models strongly underdoped cuprates. We note that in the opposite high-frequency regime, the spectrum of spin waves is linear.

We would like to emphasize that according to the ideology of [4.1] the overdamped spectrum could serve as a precursor for anomaly-free (bosonic-like) spectrum of orbital waves at very low frequencies (where the superfluid hydrodynamics might be restored).

We also note that a crossover from the ballistic to the hydrodynamic regime $\omega\tau \ll 1$ could occur due to both the aerogel (the nonmagnetic impurities) or at finite temperature $T \neq 0$, which is always present in a real experiment. In the last case, the damping $\gamma \sim T^n$ is temperature dependent.

In aerogel we can definitly fulfil the inequality $\omega_0 < \gamma$ (where $\omega_0$ is the level spacing in Dirac equation and $\gamma$ is damping) already at $T = 0$. To fulfil this inequality we should be in a moderately clean $^3$He-A phase, that is in the presence of a sufficient amount of aerogel. Note that aerogel serves like a non-magnetic impurity for p-wave fermionic superconductor like $^3$He-A [4.49; 4.50]. We can say that effectively the damping $\gamma$ is an external parameter which depends only upon the concentration of aerogel $x$. Moderately clean case means that $\omega_0 < \gamma < \Delta_0$, where $\omega_0$ is given by (4.2.45). It can be achieved experimentally since it is possible to have $\gamma$ as large as $0,1T_C$ in A-phase of $^3$He [4.40; 4.41].

Thus a very interestin experimental proposal is to check our conjecture (that $C_0(\omega_0 \ll \gamma)$ is small) by creating a twisted texture $\vec{l} \parallel rot\vec{l}$ and varying the aerogel concentration. Then it is interesting to decrease the aerogel concentration drastically and to answer experimentally the question whether $\gamma(x \to 0)$ is larger or smaller than $\omega_0$ in superclean $^3$He-A.

The similar project with the impurities can be also proposed for the magnetic traps if it will be possible to increase the lifetime of p-wave fermionic A1-phase (see Chapter 7 for more details). Experimentally both in $^3$He-A and in magnetic traps we can measure either the anomalous current directly or the spectrum of orbital waves, which is usually easier. Anyway, these measurements will allow us to compare $C_0$ or $(\rho - C_0)$ in moderately clean case and in extremely clean case and thus to check the conjecture of [4.1; 4.2] about the destruction of the chiral anomaly in the hydrodynamic regime.

Note that in the presence of the sufficient amount of aerogel the phase-diagram of superfluid $^3$He changes considerably. The global minima now are B-like and A-like phases which differ considerably from B and A-phases in superclean case. Their precise symmetry is a subject of a hot debate today (see discussion in [4.40; 4.41; 4.42]).

Anomalous spin currents in 2D axial phase.

Let us also consider briefly the spin and orbital hydrodynamics in the axial phase, which is the 2D analog of A-phase (see [4.10]). The quasiparticle energy in the fermionic 2D phase reads:

$$E_p = \sqrt{\left(\frac{p^2}{2m} - \mu\right)^2 + \frac{\Delta_0^2 p^2}{p_F^2}} \,. \qquad (4.2.54)$$

It has only one nodal point $E_p = 0$ for $\mu = 0$ and $p = 0$.

Note that in this phase the order parameter is given by $\vec{\Delta} = \Delta_0(\vec{e}_x + i\vec{e}_y)$ in similarity with 3D, but $\left|\vec{\Delta}\vec{p}\right|^2 = \Delta_0^2 p^2$ in two dimensions. An axial phase can be realized in 2D magnetic traps or in thin films or submonolayers of $^3$He (see also Chapter 16). The most interesting effect for the physics of the $^3$He 2D films [4.10; 4.11; 4.18; 4.53] is connected with the existence of the topological invariant $Q$, which comes to the physics of helium from the Quantum Hall Effect (QHE). This invariant is written as

$$Q = \frac{\pi}{2}\varepsilon_{\alpha\beta}\int \frac{d^2\vec{p}}{(2\pi)^2}\vec{n}\left[\partial_\alpha\vec{n} \times \partial_\beta\vec{n}\right], \qquad (4.2.55)$$

where $\varepsilon_{\alpha\beta} = -\varepsilon_{\beta\alpha}$ is asymmetric tensor, the components of the unit vector $\vec{n}$ in the momentum space are given by [4.51; 4.53]:

$$\vec{n} = \frac{1}{E_p}\left(-\Delta_0 p_x, \Delta_0 p_y, \varsigma_p\right), \qquad (4.2.56)$$

and $\varsigma_p = \dfrac{p^2}{2m} - \mu$, while $E_p$ is given by (4.2.54). In the theory of Quantum Hall Effect the topological invariant $Q$ governs the quantization of the Hall conductivity $\sigma_{xy}$. It is also important in the 2D space-time continuum $(x, \tau)$ for Haldane effective action, which defines an important difference in the spectrum of collective excitations (gapped or gappless) between the spin chains with integer and half-integer spins $S$.

It is easy to check (see [4.51;4.53]) that for 2D axial phase:

$$Q = \int\limits_0^\Lambda dp^2 \frac{d}{dp^2}\left[n_s(p^2)\right] = \frac{1}{2}\left(1 + \frac{\mu}{|\mu|}\right), \qquad (4.2.57)$$

where $n_s(p^2) = \dfrac{1}{2}\left(1 - \dfrac{\varsigma_p}{E_p}\right)$ is a superfluid density.

Thus $Q = 0$ in the BEC phase (where the chemical potential is negative $\mu = -|\mu|$) and $Q = 1$ in the BCS phase. We can say that the BCS phase has a nontrivial topology in contrast to the BEC phase (see [4.51] and [4.53]).

Correspondingly for $\mu \to +0$ $Q = 1$ while for $\mu \to -0$ $Q = 0$. Thus there is a jump in $Q$ ($\Delta Q = 1$) for a point $\mu = 0$. It is reasonable to assume that a point $\mu = 0$ is a singular point at $T = 0$ [4.18; 4.52; 4.53]. In Chapter 7 we will prove that it is a point of a quantum phase-transition (or even a topological phase-transition). To measure the nontrivial topological effects in two dimenssions, we propose to performe experiments with a Josephson current (see [4.2; 4.13]) between two thin films of $^3$He or between two magnetic traps containing superfluid Fermi-gasses: one with a two-dimensional axial BCS phase and the topological charge $Q = 1$ and another one with the planar 2D phase with $Q = 0$ (see [4.10; 4.51]). We hope that it will be possible to measure directly $\Delta Q = 1$ in this type of experiments.

We note that in the 2D axial phase, the $\vec{l}$ vector $\vec{l} = \left[\vec{e}_x \vec{e}_y\right] = \vec{e}_z$ is perpendicular to the plane of the 2D film. Hence, the orbital waves, connected, as we discussed in Section 4.1, with the rotation of the $\vec{l}$ vector around a perpendicular axis, are gapped. The sound wave is the only Goldstone mode in the gauge-orbital sector. Moreover $\vec{l} \perp rot\,\vec{l}$ and it is impossible to create a twisted texture in two dimensions. Therefore the anomalous current is absent: $\vec{j}_{an} = -\dfrac{\hbar}{4m}C_0(\vec{l}\,rot\,\vec{l})\,\vec{l} = 0$. Hence there is no problem with the mass current nonconservation at $T = 0$ in 2D axial phase [4.2].

Nontrivial topological effects possibly exist in the spin sector [4.51] in 2D. Here, the anomalous spin current was predicted by Volovik, Solov'ev and Yakovenko in the presence of an inhomogeneous magnetic field $\vec{B}(\vec{r})$ for 2D $^3$He-A film (for the axial BCS phase). It reads:

$$j_{\alpha,i}^{spin} \sim Q\varepsilon_{izk}l_z\partial_k B_\alpha^\perp, \qquad (4.2.58)$$

where $\vec{B}_\perp \vec{d} = 0$ and $\vec{d}$ is the spin vector in the 2D $^3$He film (see [4.10]).

Another possibility is to measure the contribution of the massless Majorana fermions (see [4.17]) for the edge states on the surface of superfluid $^3$He-B and a rough wall (or on the surface of a vibrating wire in the Lancaster experiments) [4.54].

Chapter 6. Composed particles, trios and quartets in resonance quantum gases and mixtures.

6.1. Two-particle pairing and phase-separation in Bose-gas with one or two sorts of bosons.

    6.1.1. Lattice model with van der Waals interaction between bosons.

    6.1.2. Two-particle T-matrix problem.

    6.1.3. Thresholds for extended s-wave, p-wave and d-wave two-bosons pairings.

    6.1.4. Bethe-Salpeter integral equation for an s-wave pairing of the two bosons.

    6.1.5. Possibility of p-wave and d-wave pairing of the two-bosons.

    6.1.6. Total phase separation.

    6.1.7. Phase diagram of the system.

    6.1.8. Two-band Hubbard model for the two sorts of bosons.

    6.1.9. Slave-boson formulation of the t-J model. Application to high-$T_C$ systems.

6.2. Composed fermions in the Fermi-Bose mixture with attractive interaction between fermions and bosons.

    6.2.1. The theoretical model.

    6.2.2. Intermediate coupling case in 2D.

    6.2.3. Bethe-Salpeter integral equation.

    6.2.4. Crossover (Saha) temperature.

    6.2.5. Three and four particles bound states in the Fermi-Bose mixture.

6.3. Bound states of three and four resonantly interacting particles.

    6.3.1. Atom-molecule scattering length for three resonantly interacting fermions in 3D. Skorniakov-Ter-Martirosian integral equation.

    6.3.2. Three resonantly interacting bosons in 3D. Efimov effect.

    6.3.3. Three resonantly interacting bosons in 2D.

    6.3.4. The three-particle complex $f_\sigma b, b$ in 2D case.

    6.3.5. Dimer-dimer scattering for four resonantly interacting fermions in 3D. Exact integral equation for four-fermion problem.

    6.3.6. Four particles bound states.

    6.3.7. Phase diagram of a Fermi-Bose mixture in 2D.

    6.3.8. Phase diagram of 2D Bose-gas.

    6.3.9. The role of the dimer-fermion and dimer-dimer scattering lengths for the lifetime of the resonance Fermi-gas.

Reference list to Chapter 6.

In the beginning of this Chapter we consider two-boson pairing for bosons of the same ($bb$) or different ($b_1b_2$) sorts. The two-boson pairing was first proposed by Nozieres and Saint James in their famous paper [6.31]. Here we consider Bose-gas with van der Waals interacting potential between particles and the two-band Hubbard model with attraction between bosons of different sorts [6.18] and repulsion between bosons of same sort. We also discuss a competing (to two-boson pairing) phenomena of phase-separation in one band and two-band [6.18, 6.12] bosonic models. In the end of the first Section we consider briefly the possibility of two-holon pairing which arises in the 2D underdoped t-J model if we assume the scenario of spin-charge separation between spinons and holons advocated by Anderson [6.41] and Lee [6.42].

The second Section will be devoted to the model of Fermi-Bose mixture with attractive interaction between fermions and bosons [6.5, 6.6, 6.13, 6.14] and the possibility to get composed fermions $f_\sigma b$ in this model. Note that the attraction between fermions and bosons is opposite to the Fermi-Bose mixtures of $^3$He and $^4$He [6.20, -6.22] or $^6$Li and $^7$Li [6.21] (which will be considered in more details in Chapters 11 and 12). In these systems usually the interaction between fermions and bosons is repulsive. The Feshbach resonance effect helps to change the sign of fermion-boson interaction [6.15].

In the next Sections we will consider 3 and 4-particle complexes which can appear in resonance Fermi-Bose-gasses and mixtures, where the scattering length $a$ is much larger than the range of the potential $r_0$. At first we will consider the scattering amplitude $a_{2-1}$ for the scattering of elementary fermion $f_\sigma$ on the composed boson $f\uparrow f\downarrow$. Here we will present exact solution of Skorniakov-Ter-Martirosian integral equation for three particles in resonance approximation and get $a_{2-1} = 1.18|a| > 0$ [6.23], which corresponds to repulsion between fermion and molecule (dimer). We also consider so-called Efimov effect [6.22, 6.23] which predicts a lot of bound states for three bosons $bbb$ in a 3D case with the energies ranging from $|E_3| \approx 1/ma^2$ for shallow levels till $|E_3| \approx 1/mr_0^2$ for deep three-particle levels. The number of levels in the resonance approximation $a >> r_0$ is governed by $N \sim \dfrac{1}{\pi} \ln \dfrac{a}{r_0}$ in 3D [6.7]. We will show also that Efimov effect is absent in 2D case [6.3] and thus the number of bound states for three bosons $bbb$ [6.11] or two bosons and one fermion $bbf_\sigma$ is finite. Moreover all of them have the energies of the order of $|E_b| = 1/ma^2$. Thus all 3-particle levels in 2D case are shallow (or quasiresonance) in origin.

We proceed than to the 4-particle problem. At first we solve Skorniakov-Ter-Martirosian equations 4 fermions [6.23, 6.10] and find that the scattering length $a_{2-2}$ [6.24-6.26] for dimer-dimer scattering (for scattering of one molecule $f\uparrow f\downarrow$ on the other). Here in the resonance approximation we get an exact result $a_{2-2} = 0.6|a| > 0$ which is different from mean-field result $a_{2-2} = 2|a|$ and from the result of the ladder approximation of Pieri and Strinati [6.1] which predicts $a_{2-2} = 0.75|a|$ if we neglect the dynamics (the possibility for two molecules to form virtual states with 3 and 1 particle [6.24-6.26]).

We also discuss the shallow bound states for 4 bosons [6.2], 3 bosons and 1 fermion and 2 bosons and 2 fermions in 2D case where Efimov effect is absent. Note that all the binding energies of the 4-particle complexes can be expressed in terms of $|E_b| = 1/ma^2$ only in 2D [6.3, 6.4].

In the end of the Chapter we discuss the importance of the obtained results for the phase diagram and life-time of ultracold Fermi-Bose-gases and mixtures. In particular we use the value of $a_{2-1}$ to estimate the inelastic scattering time in the resonant Fermi-gas [6.8] and 4-particle binding energies $E_4$ to complete phase diagram of the resonance Fermi-Bose mixture with attraction between fermions and bosons. Here we pay a special attention on the complexes ($f\uparrow b$, $f\downarrow b$) formed by two composed fermions $f_\sigma b$ and stress the analogy between ultracold Fermi-Bose mixtures in magnetic traps and strongly-interacting mixture of spinons and holons in underdoped high-$T_C$ compounds in the framework of Laughlin ideas [6.34, 6.35] on spin-charge confinement. According to Laughlin, the spinons and holons experience the phenomenon of spin-charge confinement in analogy with the confinement [6.47] in quark-gluon plasma (in quark bags) in QCD-physics [6.32, 6.34, 6.46]. We think [6.40] that for strongly-correlated quasi-2D (layered) cuprates the philosophy of Laughlin is more adequate that the philosophy of Anderson and Lee [6.41, 6.42] on spin-charge separation, which is based on the analogy with 1D-physics. We emphasize that due to linear (string-like) confinement potential between spinons $f_{i\sigma}$ and holons $b_i$ [6.36 – 6.38], the composite hole $h_{i\sigma} = f_{i\sigma}b_i$ is formed on the lattice. It represents a compact object (a bag or spin-polaron [6.39]). The Cooper pairs in this system are formed by the residual (dipole-dipole) interaction [4.40] between two composite holes and effectively represent 4-particle complex [6.19, 6.23, 6.24, 6.26, 6.27], consisting of two spinons and two holons. Thus the superconductive gap reads $\Delta_{ij} = <h_{i\sigma}h_{j-\sigma}> = <f_{i\sigma}b_i, f_{j-\sigma}b_j>$. It is formed by two composite holes on the lattice with a total spin $S_{tot} = 0$ of a pair. We will consider these ideas more detaily in Chapter 13 on the basic of 2D t-J model and advocate the scenario of BCS-BEC crossover in the d-wave channel for pairing of two composite hole (two strings or two spin polarons) in underdoped cuprates.

We stress also the importance of dimer-dimer amplitude $a_{2-2}$ for the phase-diagram of the BCS-BEC crossover and the spectrum of collective excitations in resonance Fermi-gas. We will consider these properties in detail in Chapter 7.

Note that the creation of three and four particle complexes, as well as the evaluation of different scattering amplitudes, for the two-partical potentials $V(r) = \alpha r^\delta + \dfrac{\beta}{r^\gamma}$; $1 \le \{\delta, \gamma\} \le 2$, which are the sum of confainement and Coulomb (or dipolar) parts, play an important role in the non-relativistic problem for barions [6.81] and in Quantum Chromodynamics [6.82], especially when the problems of inhomogeneous superconductivity are studied in the QCD theory at zero temperature. Note that the potentials $V(r) = \alpha r^2 + \dfrac{\beta}{r^2}$ ($\delta = \gamma = 2$) allow also for exact solutions at least on the level of the two-particle problem. The creation of three and four particle complexes was also predicted for the quark-gluon plasma in the high-temperature limit. This limit can be realized experimentally in heavy ion collisions. Here three and four particle complexes can be formed on the way from a gas to a liquid in the hydrodynamic regime for the quark-gluon plasma (see [5.85] and references therein).

### 6.1. Two-particles pairing and phase-separation in Bose-gas with one or two sorts of bosons.

As we already discussed in Chapter 5, in contrast with two-particle Cooper pairing in Fermi-systems (see Chapters 11 - 15), the essence of a superfluidity is Bose-systems is one-particle Bose-Einstein condensation (BEC). This asymmetry between Fermi and Bose (two-particle versus one-particle condensation) was challenged in a pioneering paper by Valatin and Butler [6.48]. They proposed a BCS-like variational function for the description of an attractive Bose-gas. The most difficult problem with the validity of their description is connected with the tendency towards phase separation which arises in attractive Bose systems. Later on Nozieres and Saint-James [6.31] conjectured that in a Bose-system with a short-range, hard-core repulsion and a van der Waals attractive tail, in principle, it is possible to create a two-particle bosonic bound state and to escape collapse. Unfortunately their calculations in three-dimensional (3D) systems showed that, at least for one sort of structureless boson, either standard one-particle BEC is more energetically beneficial, or that a phase separation takes place earlier than the two-particle condensation. Note that the same result was obtained earlier by Jordanski [6.49] for the case of weak van der Waals attraction.

The important development of the ideas of Nozieres and Saint James belongs to Rice and Wang [6.50]. These authors claimed that in two dimensions (where already an infinitely small attraction leads to the bound state in a symmetrical potential well) it is possible to realize a two-particle boson pairing. Moreover, this two-particle pairing results, for small momenta $q\xi_0 < 1$ ($\xi_0$ is a coherence length) in a linear, soundlike, dispersion law of quasiparticles at $T = 0$ in an analogy with a standard one-particle Bose-condensation for weakly repulsive Bogolubov Bose-gas.

To escape a collapse in a 2D attractive Bose-gas, the authors of [6.50] introduced in their model a Hartree-Fock shift of the chemical potential $\mu_B \sim Un$, connected with the short-range repulsion $U$. This shift in the case of $U > 2V$, where $V$ is the magnitude of the van der Waals tail, leads to a positive compressibility in the system $\kappa^{-1} = d\mu_B/dn = U - 2V > 0$.

The main goal of this Section is to construct a phase diagram of a 2D dilute Bose-gas with the van der Waals interaction between particles, by taking into account on equal grounds the full contribution of a hard-core repulsion $U$ and a van der Waals tail $V$ (see [6.18]). Throughout the paper we will consider the lattice model, and will base our results on the exact solution of the two-particle T-matrix problem presented in [6.51, 6.52] in connection with a fermionic t-J model (see Chapter 13). Note that effectively a lattice model with van der Waals interaction between bosons is a bosonic analog of a famous fermionic t-J model considered in Chapter 13 for high-$T_C$ systems. We will study the possibility of different two-boson pairings, as well as the possibility of a total phase separation in the system. We will also consider the two sorts of structureless bosons described by the two-band bosonic Hubbard model [6.53-6.56] (note that fernionic one-band and two-band Hubbard models are considered in Chapters 8, 9, 10). In the case of attraction between bosons of two different sorts, we will find a possibility of an s-wave two-boson pairing $< b_1 b_2 > \neq 0$.

### 6.1.1. Lattice model with van der Waals interaction between bosons.

The model under consideration is described by the following Hamiltonian on the 2D square lattice [6.18]:

$$\hat{H} = -t\sum_{<ij>} b_i^+ b_j + \frac{U}{2}\sum_i n_i^2 - \frac{V}{2}\sum_{<ij>} n_i n_j , \qquad (6.1.1)$$

where $n_i = b_i^+ b_i$ is a 2D boson density on site $i$. We will work in the limit of strong hard-core repulsion $U >> \{V; t\}$, and restrict ourselves mostly to a low-density limit which in the lattice case yields $n_B d^2 << 1$, $d$ being the interatomic distance. In (6.1.1) $t$ is hopping integral for bosons, $<ij>$ are nearest lattice sites on the square lattice, $V$ is van der Waals attraction on neighboring sites, $U$ is onsite repulsion, $b_i^+$ and $b_i$ are bosonic creation and annihilation operators on site $i$ of the lattice. Note that in the case of $V = 0$, the model (6.1.1) is just the Bose-Hubbard model, extensively studied in the literature for the case of 2D $^4$He submonolayers, as well as for the flux lattices and Josephson arrays in the type-II superconductors (see [6.53-6.56]). As we already mentioned in the introduction to this Section, a model (6.1.1) is, to some extent, a Bose analog of the fermionic t-J model considered by Kagan and Rice in [6.52]. After Fourier transformation from (6.1.1) we obtain:

$$\hat{H} = \sum_p \varepsilon_p b_p^+ b_p + \frac{U}{2}\sum_{k_1 k_2 q} b_{k_1}^+ b_{k_2}^+ b_{k_2-q} b_{k_1+q} - \sum_{k_1 k_2 q} V(q) b_{k_1+q}^+ b_{k_2}^+ b_{k_2-q}, \qquad (6.1.2)$$

where $\varepsilon(p) = -2t(\cos p_x d + \cos p_y d)$ is an uncorrelated bosonic spectrum on the square lattice, and $V(q) = V[\cos q_x d + \cos q_y d]$ is a Fourier transform of the van der Waals tail. As a result, a total interaction in the momentum space is given by the formula:

$$V_{eff}(q) = \frac{U}{2} - V(q) \qquad (6.1.3)$$

### 6.1.2. Two-particle T-matrix problem.

An instability toward a two-particle boson pairing manifests itself (just as in the case of Cooper pairing of two fermions in Chapters 5 and 9) in the appearance of a pole at a temperature $T = T_C$ in the solution of the Bethe-Salpeter equation for the two-particle vertex $\Gamma$ for zero total momentum of the two bosons ($\vec{p}$ and $-\vec{p}$) (see [6.18, 6.57]). To proceed to the solution of this equation, we must solve at first the T-matrix problem [6.58] for the two bosonic particles in vacuum. Here we can use the results of [6.52] (for the T-matrix for two fermions), since the solution of the two-particle problem does not depend upon statistics of colliding particles. For the T-matrix problem it is convenient to expand $V_{eff}(q)$ in (6.1.3) with the eigenfunctions of the irreducible representation of the lattice symmetry group $D_4$ (see also Chapters 9 and 13). This yields:

$$V_{eff}(\text{extended s}-\text{wave}) = \frac{U}{2} - \frac{V}{2}(\cos p_x d + \cos p_y d)(\cos p_x 'd + \cos p_y 'd);$$

$$V_{eff}(\text{p}-\text{wave}) = -\frac{V}{2}(\sin p_x d \sin p_x 'd + \sin p_y d \sin p_y 'd); \qquad (6.1.4)$$

$$V_{eff}(\text{d}-\text{wave}) = -\frac{V}{2}(\cos p_x d - \cos p_y d)(\cos p_x 'd - \cos p_y 'd),$$

where we use the functions $\varphi_S = (\cos p_x d + \cos p_y d)$, $\varphi_p = (\sin p_x d + i \sin p_y d)$ and $\varphi_d = (\cos p_x d - \cos p_y d)$ respectively for extended s-wave, p-wave and d-wave channels on the square lattice.

Note that, for spinless bosons, which we formally consider in (6.1.1), the total spin of the Bose pair is zero ($S_{tot} = 0$). Hence only s-wave ($l = 0$ in the absence of lattice) and d-wave ($l = 2$ in the absence of lattice) pairings are allowed by the symmetry of the pair $\Psi$-function. A p-wave pairing ($l = 1$ in the absence of lattice) is allowed only for an odd total spin ($S_{tot} = 1, 3 \ldots$) of the two bosons (the total pair $\Psi$-function which is product of orbital part and spin part is symmetric for two bosons). Nevertheless we will conserve the results for the p-wave pairing in our paper because the generalization of (6.1.1) for the case of bosons with internal degrees of freedom is straightforward. The T-matrix problems for p- and d-wave channels are very simple. Solutions of these problems for the two particles with a total momentum zero and a total energy $E$ yield:

$$T_{d,p}(E) = -\frac{\frac{V}{2}}{1 + \frac{V}{2}I_{d,p}}, \qquad (6.1.5)$$

where

$$I_{d,p} = \int_0^{2\pi}\int_0^{2\pi}\frac{dp_x}{2\pi}\frac{dp_y}{2\pi}\frac{\left|\varphi_{d,p}\right|^2}{E + 4t(\cos p_x d + \cos p_y d)} =$$

$$= \int\frac{d\omega}{2\pi}\frac{d^2\vec{p}}{(2\pi)^2}G_0(\omega + E, \vec{p})G_0(-\omega, -\vec{p})\left|\varphi_{d,p}\right|^2, \qquad (6.1.6)$$

$G_0(\omega, \vec{p}) = \dfrac{1}{\omega - \varepsilon(p) + io}$ is vacuum Green function and $\varphi_{d,p}$ are the functions for d-wave and p-wave channels.

### 6.1.3. Thresholds for extended s-wave, p-wave and d-wave two-bosons pairings.

Let us find the thresholds for the bound states in the extended s-wave, d-wave and p-wave channels. The appearance of a bound state means that $E = -W - \tilde{E}$, where $W = 2zt = 8t$ is a bandwidth for the 2D square lattice ($z = 4$ is a number of nearest neighbors on the square lattice). For the threshold $\tilde{E} = 0$. An exact solution of (6.1.5) and (6.1.6), which involves the calculation of elliptic integrals of first and second order (see [6.18, 6.52]), yields for p-wave and d-wave thresholds:

$$\left(\frac{V_C}{4t}\right)_{p-wave} \approx \frac{1}{1-\frac{2}{\pi}} \approx 2.8$$

$$\left(\frac{V_C}{4t}\right)_{d-wave} \approx \frac{1}{\frac{4}{\pi}-1} \approx 3.7$$

(6.1.7)

Note that a threshold for a p-wave pairing is lower. Now let us proceed to an s-wave channel. Here an ordinary s-wave pairing is suppressed by large hard-core repulsion $U$, however an extended s-wave pairing with a symmetry of the order parameter $\Delta_S^{ex} = \Delta_0(\cos p_x d + \cos p_y d)$ is allowed (for ordinary s-wave pairing $\Delta_S$ is just $\Delta_0$ and does not depend upon momentum $\vec{p}$). In real space this pairing corresponds to the particles on the neighboring sites. Moreover the pair $\Psi$-function is zero in the region of a hard core ($r < r_0$), and is centered (has a maximum) in the region of a van der Waals attraction for $r_0 < r < r_1$ (see Fig.6.1). On the lattice $r_0 \sim d/2$ and $r_1 \sim d$.

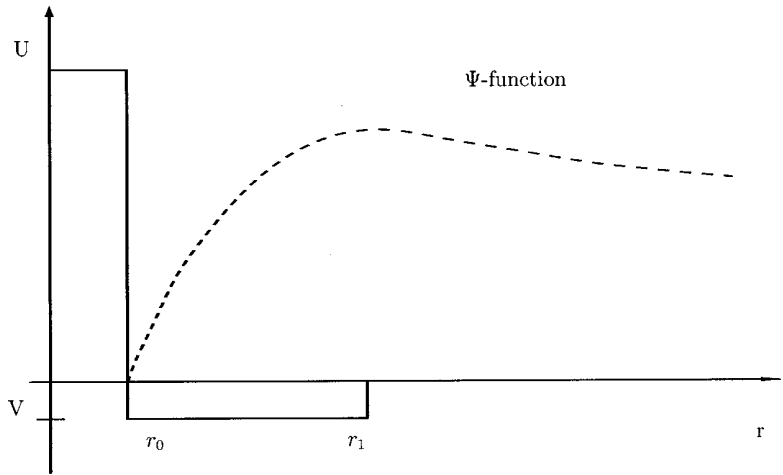

Fig.6.1. The $\Psi$-function of an extended s-wave pairing. $r_0$ is the radius of the hard core repulsion, and $r_1$ is the radius of the van der Waals attraction. On the lattice, $r_0 \sim d/2$ and $r_1 \sim d$.

One can see that the $\Psi$-function has a region of zero values (for $r < r_0$). But it has no nodes because it does not change its sign for all values of $r$ ($\Psi \geq 0$). The rigorous calculation of the threshold for an extended s-wave pairing yields [6.51, 6.52, 6.18]:

$$\left(\frac{V_C}{4t}\right)_{s-wave} = 1 \qquad (6.1.8)$$

Thus the threshold for s-wave pairing is the lowest (compare with (6.1.7)). Moreover for $V > V_{CS} = W/2$ an energy of the bound state has the form:

$$\left|\tilde{E}_b^{\ S}\right| = \left|\tilde{E}\right| = 8We^{-\frac{\pi V}{(V-V_{CS})}} \qquad (6.1.9)$$

Of course, in a strong coupling case (for $V >> W$) $\left|\tilde{E}_b^{\ S}\right| \approx V$. Correspondingly, the bound states for p-wave and d-wave pairings yields [6.18, 6.52]:

$$\left|E_b^p\right| = W\frac{(V-V_{Cp})}{V}\frac{1}{\ln\frac{V}{\left|V-V_{Cp}\right|}} \quad \text{for } V > V_{Cp} = 1.4W;$$

(6.1.10)

$$\left|E_b^d\right| = W\frac{(V-V_{Cd})}{V} \quad \text{for } V > V_{Cd} = 1.85W.$$

We can see that for $V >> W$: $\left|\tilde{E}_b^{\ S}\right| \approx V$ while $\left|\tilde{E}_b^{\ p}\right| \sim \left|\tilde{E}_b^{\ d}\right| \sim W$ and extended s-wave bound state correspond to a global minima in agreement with general theorems of quantum mechanics ($\left|\tilde{E}_b^{\ s}\right| > \left|\tilde{E}_b^{\ p}\right| > \left|\tilde{E}_b^{\ d}\right|$ for fixed $V > V_{Cd} > V_{Cp} > V_{Cs}$).

The T-matrix in an s-wave channel for small and intermediate values of $V$ is given by [6.18, 6.52]:

$$T_s(\tilde{E}) = \frac{W(1 - V/4t)}{\dfrac{1}{\pi}(1 - V/4t)\ln\dfrac{8W}{\tilde{E}} - \dfrac{V}{4t}}. \qquad (6.1.11)$$

The most important is that a strong Hubbard repulsion $U$ acts only as an excluded volume (for $r < r_0 \sim d/2$), and effectively drops out from (6.1.11) at low energies. It manifests itself only as an additional pole (at very large energies $E = U > 0$) in total analogy with antibound state [6.59] in fermionic Hubbard model (it will be considered in Chapter 14). For $V << 4t$ the T-matrix:

$$T_s(\tilde{E}) \approx \frac{\pi W}{\ln \dfrac{8W}{\left|\tilde{E}\right|}}. \qquad (6.1.12)$$

corresponds to repulsion and coincides with the T-matrix for the 2D Bose-Hubbard model at low density. The same T-matrix was obtained by Fukuyama et al., for 2D fermionic Hubbard model at low density. For $V = 4t$: $T_s(\tilde{E}) = 0$ and there is no interaction at all. Finally, for $V > 4t$, $T_s(\tilde{E}) < 0$ corresponds to an attraction and reflects the appearance of the bound state [6.18, 6.52].

### 6.1.4. Bethe-Salpeter integral equation for an s-wave pairing of the two bosons.

Let us consider at first the most interesting case of $V > 4t$ and find the critical temperature for an extended s-wave pairing of the two bosons. The solution of the Bethe-Salpeter equation [4.60] for bosonic systems reads [6.61, 6.18, 4.20]:

$$\Gamma_s = \frac{T_s}{1 + T_s \iint \dfrac{dp_x dp_y}{(2\pi)^2} \dfrac{cth\dfrac{\varepsilon_p - \mu}{T}}{2(\varepsilon_p - \mu)}}, \qquad (6.1.13)$$

where $T_s$ is a T-matrix for s-wave pairing and $\Gamma_s$ is an s-wave harmonic of the total two-particle vertex $\Gamma$ in the Cooper channel (for zero total momentum $\vec{P} = \vec{p}_1 + \vec{p}_2 = 0$ and zero total Matsubara frequency $\Omega = \Omega_1 + \Omega_2 = 0$ of the two incoming particles – see Chapter 5 and 9 also for the case of two fermions). For low density of bosons $n_B d^2 << 1$ one has $\mu = -4t + \tilde{\mu}$ and $\xi_p = \varepsilon_p - \mu = \dfrac{p^2}{2m} - \tilde{\mu}$ for the uncorrelated quasiparticle spectrum counted from the chemical potential level.

The most substantial difference of (6.1.13) from an analogous equation for two fermions is the replacement of $th\dfrac{\xi_p}{2T}$ by $cth\dfrac{\xi_p}{2T}$ in its kernel. Moreover, as shown by Miyake in [6.62] for the 2D attractive Fermi gas $\tilde{\mu} = \varepsilon_F - \dfrac{\left|E_b\right|}{2}$ (where $|E_b|$ is a binding energy of a pair in vacuum). So, in a weak-coupling case, when $\varepsilon_F >> |E_b|$, the chemical potential $\tilde{\mu} \approx \varepsilon_F > 0$ is positive. In contrast to this, we shall see below that a bosonic chemical potential $\tilde{\mu}$ is always negative even in the weak-coupling case, when a binding energy is much smaller than a degeneracy temperature $|E_b| < T_0 = \dfrac{2\pi n}{m}$.

Another very important point (see also Chapters 9 and 13) is that the T-matrix, which enters into the Bethe-Salpeter equation, must be calculated for a total energy $\tilde{E} = 2\tilde{\mu}$ by [6.18, 6.52] of colliding bosons. The chemical potential $\tilde{\mu}$ can be determined from the requirement of the number of particle conservation. This requirement yields:

$$n_B = \iint \frac{d^2\vec{p}}{(2\pi)^2} \frac{1}{\exp\left\{\dfrac{p^2/2m - \tilde{\mu}}{T}\right\} - 1}. \qquad (6.1.14)$$

From (6.1.14) for the temperatures $|E_b| < T < T_0 < W$ we obtain:

$$\tilde{\mu} = -T \exp\left(\frac{T_0}{T}\right) < 0. \qquad (6.1.15)$$

Note that a standard Hartree-Fock shift $nU$ drops out from the expression for bosonic quasiparticle spectrum $\xi_p = \varepsilon_p - \mu$ both in the Bethe-Salpeter equation (6.1.13) and in the equation for the number of particle conservation (6.1.14) in similarity with a fermionic problem. Now we are ready to solve the Bethe-Salpeter equation (6.1.14). The critical temperature $T_C$ corresponds to the pole in (6.1.13),

$$1 + \frac{md^2 T_s(2\tilde{\mu})}{2\pi} I = 0, \qquad (6.1.16)$$

where

$$I = \int\limits_0^{\sim W/T_C} \frac{dy \, cth\left(y + \dfrac{|\tilde{\mu}|}{2T_C}\right)}{y + \dfrac{|\tilde{\mu}|}{2T_C}} \qquad (6.1.17)$$

and $y = \dfrac{p^2}{4mT_C}$.

An analysis of (6.1.17) shows that the main contribution to the integral comes from the lower limit of integration.

Hence providing $|\tilde{\mu}|/T_C \ll 1$ we have:

$$I \approx \int\limits_0^{W/T_C} \frac{dy}{\left(y + \dfrac{|\tilde{\mu}|}{2T_C}\right)^2} \approx \frac{2T_C}{|\tilde{\mu}|}. \qquad (6.1.18)$$

As a result (6.1.18) can be represented in the following form:

$$\frac{T_C}{|\tilde{\mu}|} = -\frac{\pi}{md^2 T_s(2\tilde{\mu})}. \qquad (6.1.19)$$

It is useful now to represent $T_s(2\tilde{\mu})$ in terms of the binding energy $E_b$. Utilizing (6.1.9) and (6.1.11) we can write:

$$T_s(2\tilde{\mu}) = -\frac{\pi W}{\ln\dfrac{2|\tilde{\mu}|}{|E_b|}} = -\frac{4\pi}{md^2 \ln\dfrac{2|\tilde{\mu}|}{|E_b|}}. \qquad (6.1.20)$$

It is important to mention here that $\tilde{\mu} < 0$, and hence the T-matrix in (6.1.20) does not contain an imaginary part. In the fermionic case $\tilde{\mu} \approx \varepsilon_F > 0$, and the T-matrix contains an imaginary part corresponding to the resonant scattering. As a result, from (6.1.20) we obtain:

$$4ln\frac{2|\tilde{\mu}|}{|E_b|} = \frac{\tilde{\mu}}{T_S}. \qquad (6.1.21)$$

Assuming that $|E_b| \ll T_C \ll T_0$, we get (see also Chapter 8): $\tilde{\mu}(T_C) = -T_C \exp\left(\dfrac{T_0}{T_C}\right)$ and

$\dfrac{\tilde{\mu}}{T_C} = \exp\left(-\dfrac{T_0}{T_C}\right)$.

Later on we will justify this assumption.

As a result from (6.1.21) we will obtain:

$$T_C = \frac{T_0}{\ln\left(\dfrac{1}{4}\ln\dfrac{2|\tilde{\mu}|}{|E_b|}\right)}. \qquad (6.1.22)$$

Recall that in the case of the fermionic s-wave pairing in two dimensions a critical temperature according to Miyake [6.62] (see also Chapter 8) reads: $T_C = \sqrt{2\varepsilon_F|E_b|}$.

Let us analyze expression (6.1.22). As we already know $|E_b| = 8We^{-\frac{1}{\lambda}}$, where

$$\lambda = \frac{(V - V_{Cs})}{\pi V} \qquad (6.1.23)$$

Then a condition $|E_b| \ll T_0$ means:

$$\lambda \ll \frac{1}{\ln\dfrac{W}{T_0}} \ll 1. \qquad (6.1.24)$$

Hence $\ln\left(T_0/|E_b|\right) = 1/\lambda - \ln\left(W/T_0\right) \approx 1/\lambda$, and

$$T_C \approx \frac{T_0}{\ln\left(\dfrac{1}{4}\ln\dfrac{T_0}{|E_b|}\right)} \approx \frac{T_0}{\ln\left(\dfrac{1}{4\lambda}\right)}, \qquad (6.1.25)$$

which is in an agreement with [6.50]. Note that $T_C$ from (6.1.25) satisfies the conditions $|E_b| \ll T_C \ll T_0$, so an assumption used for the derivation of $T_C$ is justified.

For $T < T_C$ the spectrum of the quasiparticles acquires a gap:

$$E_p = \sqrt{\left(\frac{p^2}{2m} + |\tilde{\mu}|\right)^2 - \Delta^2} \qquad (6.1.26)$$

Note that at low densities of bosons a gap $\Delta$ becomes isotropic in the principal approximation.

The gap $\Delta$ together with the chemical potential $\tilde{\mu}$ must be defined self-consistently from the two coupled equations:

$$1 = \frac{\lambda}{4}\int\limits_{|\tilde{\mu}|/2T}^{\sim W/2T} \frac{dz\, cth\sqrt{z - \Delta^2/4T^2}}{\sqrt{z - \Delta^2/4T^2}}, \qquad (6.1.27)$$

and

$$n_B = \frac{\lambda}{4}\int\limits_{|\tilde{\mu}|}^{\sim W} d\xi \frac{1}{\exp\left\{\dfrac{\sqrt{\xi^2 - \Delta^2}}{T}\right\} - 1}, \qquad (6.1.28)$$

where $\xi_p = \dfrac{p^2}{2m} + |\tilde{\mu}|$ and $z = \xi/2T$.

Of course, the solution of the system of equations (6.1.27) and (6.1.28) exists only if $|\tilde{\mu}| > \Delta$ , or, in other words, only if $E_p^2 = |\tilde{\mu}|^2 - \Delta^2 > 0$ . The exact solution of these equations yields for zero temperature in an agreement with [6.50]:

$$|\tilde{\mu}(T=0)| = \Delta = \frac{|E_b|}{2}. \qquad (6.1.29)$$

This result is very important. It justifies our scenario, leading to a linear, soundlike spectrum of the quasiparticles for a small momenta $p$. Indeed,

$$E_p = \sqrt{\left(\frac{p^2}{2m}\right)^2 + \frac{p^2}{2m}|\tilde{\mu}|} = \sqrt{\left(\frac{p^2}{2m}\right)^2 + \frac{p^2}{2m}|E_b|}. \qquad (6.1.30)$$

From (6.1.30) for the case $p\xi_0 << 1$, where $\xi_0 = \dfrac{1}{\sqrt{2m|E_b|}}$ is the coherence length of the boson pair, we immediately obtain a linear dispersion law:

$$E_p = cp. \qquad (6.1.31)$$

In (6.1.31) $c^2 = \dfrac{|E_b|}{2m}$ is a sound velocity squared. This means that an inverse compressibility of the system $\kappa^{-1} \sim c^2$ is positive. This fact proves the stability of a superfluid paired state and excludes the possibility of the collapse of the pairs in the system. Note also that close to $T_C$ one has:

$$\Delta(T) = \Delta(0)\sqrt{\frac{T_C - T}{T_C}}, \qquad (6.1.32)$$

which is similar to the BCS theory. We would like to mention that bosonic pairs in the limit $|E_b| << T_0$ are extended in full analogy with the BCS theory. That is, the coherence length in this limit,

$$\xi_0 >> \frac{1}{\sqrt{n}} >> 1 \qquad (6.1.33)$$

is larger than the mean distance between the bosons. The Bose pairs are strongly overlapping in this limit. The pairing takes place in the momentum space in an analogy with the Cooper pairing in the BCS picture of superconductivity.

In the opposite limit $|E_b| >> T_0$ the pairs are local and the situation closely resembles BEC (or bipolaronic) limit for the fermionic systems [4.12. 4.13, 6.63] (see Chapters 5 and 8). That is, the creation of the bosonic bound pairs is associated with the crossover temperature [4.14, 4.17, 6.18] (see Chapter 8 also):

$$T_* = \frac{|E_b|}{\ln 1/nd^2} \qquad (6.1.34)$$

The Bose condensation of the pairs occurs at lower temperature [6.64, 6.65, 4.17]:

$$T_C = \frac{T_0}{\ln\ln(1/nd^2)}. \qquad (6.1.35)$$

Note that this temperature is obtained in Fisher-Hohenberg theory [6.64] from the ansatz $\tilde{\mu}(T_C) = -T_C \exp(T_0/T_C) + f_0 T_0 = 0$, where $f_0 = 1/\ln(1/nd^2)$ is a repulsive interaction between the local pairs and $f_0 T_0$ is a Hartree-Fock contribution to the chemical potential $\mu_B$ in 2D repulsive Bose-gas (see also Chapter 8). Thus the superfluid transition takes place only in case of a residual repulsion between the pairs. Also note that in a dilute Bose-gas in 2D the Berezinski-Kosterlitz-Thouless (BKT) contribution of vortices [6.66, 6.67 is important only very close to $T_C$, so the mean field expression (6.1.35) gives a very good estimate for the exact BKT critical temperature: $\dfrac{T_C - T_{BKT}}{T_C} \sim \dfrac{1}{\ln\ln(1/nd^2)} \ll 1$.

In the case of the local pairs the coherence length is small:

$$\xi_0 \ll \frac{1}{\sqrt{n}} \qquad (6.1.36)$$

The pairs are compact, and the pairing takes place in the real space.

### 6.1.5. Possibility of p-wave and d-wave pairing of the two-bosons.

Now let us analyze the solution of the Bethe-Salpeter equation for p- and d-wave two-boson pairings. Here the critical temperatures should be found from the conditions (see also Chapter 9 and [6.18, 6.52]):

$$1 + T_{p,d}(2\tilde{\mu})\tilde{I}_{p,d} = 0, \qquad (6.1.37)$$

where

$$\tilde{I}_{p,d} = \int\limits_{0}^{2\pi/d}\int\limits_{0}^{2\pi/d} \frac{dp_x dp_y}{(2\pi)^2}\, cth\frac{\varepsilon - \mu}{2T}\frac{1}{2(\varepsilon - \mu)}\left|\varphi_{p,d}\right|^2. \qquad (6.1.38)$$

In a low-density limit the $\varphi$-functions can be approximated by the following expressions [6.68, 6.18, 6.52]:

$\varphi_p = (p_x + ip_y)d = pde^{i\varphi}$;

$\varphi_d = \dfrac{1}{2}(p_x^2 - p_y^2)d^2 = \dfrac{1}{2}p^2d^2\cos 2\varphi.$

Hence after an angular integration we obtain:

$$\tilde{I}_p = \frac{m}{2\pi}\int p\,dp\,\frac{cth\dfrac{\xi_p}{2T_C}}{2\xi_p}p^2d^2;$$

$$\qquad (6.1.39)$$

$$\tilde{I}_d = \frac{m}{16\pi}\int p\,dp\,\frac{cth\dfrac{\xi_p}{2T_C}}{2\xi_p}p^4d^4,$$

where again $\xi_p = \dfrac{p^2}{2m} + |\tilde{\mu}|$.

Additional factors $p^2d^2$ and $p^4d^4$ in the integral expressions for $\tilde{I}_p$ and $\tilde{I}_d$ reflect a well-known fact, that for slow 2D particles in vacuum an s-wave harmonics of the scattering amplitude behaves as $f_0 \sim \ln(1/p^2d^2)$, whereas for a magnetic number $m \neq 0$, the scattering amplitude vanishes for $p$ goes to zero as $f_m \sim (pd)^{2m}$ (see Quantum Mechanics [4.19]). The additional factor $p^4d^4$ leads to the absence of an infra-red singularity for $\varepsilon \to 0$ in $\tilde{I}_d$:

$$\tilde{I}_d \sim \int \frac{d\varepsilon \cdot \varepsilon^2}{\varepsilon^2} \sim \varepsilon \to 0 \quad \left(\varepsilon \sim \frac{p^2}{2m}\right) \qquad (6.1.40)$$

For the p-wave channel the infra-red singularity becomes logarithmically weak:

$$\tilde{I}_p \sim \int \frac{d\varepsilon \cdot \varepsilon}{\varepsilon^2} \sim \ln\varepsilon \qquad (6.1.41)$$

This means that the Bethe-Salpeter equation has no solutions in the p- and d-wave channels for $|V|/t < 1$.

Hence the boson pairing with a large coherence length $\xi_0 > \dfrac{1}{\sqrt{n}}$ is absent in a p-wave channel as well as in a d-wave channel. Here only the limit of the local pairs is possible. For p- and d-wave channels local pairs are created at the crossover (Saha [4.21, 4.14, 4.17]) temperature $T_*$ given by (6.1.34), where binding energies $|E_b^p|$ and $|E_b^d|$ are given by (6.1.10) for $V > V_{Cp}$ and $V > V_{Cd}$ correspondingly. Remind that for a fixed $V > V_{Cd} > V_{Cp}$: $|E_b^p| > |E_b^d|$ and thus $T_*^p > T_*^d$. Providing that the interaction between the local pairs is repulsive in p-wave and d-wave channels, the mean-field temperature of the Bose condensation of the local pairs should be determined again from the ansatz for the chemical potential $|\tilde{\mu}(T_C)|$ with an account of the repulsive Hartree-Fock shifts in p-wave and d-wave channels (see [6.64]). In the next Subsection we will show, however, that in the van der Waals model (6.1.2) there is a competing phenomenon of the total phase separation which takes place earlier (at smaller values of $V/t$) than binding in different (s-wave, p-wave, d-wave channels). Thus it is difficult to get two-boson pairing in the present model.

### 6.1.6. Total phase separation.

As we discussed in Subsection 6.1.1, the real collapse is prohibited in our system by large Hubbard repulsion $U$. However the phase separation on the two large clusters is allowed. The first cluster corresponds to the Mott-Hubbard [4.38, 6.69] Bose solid. In this cluster $n_b d^2 \to 1$, that is, each site on the quadratic lattice is practically occupied by one boson. Such a cluster is localized due to Mott-Hubbard consideration for large $U >> W$ (where $W$ is a bandwidth). It has no kinetic energy. However, it has a potential energy of the order of -2$V$ for one particle. A second cluster has a very small boson density $n_b d^2 \to 1$. In this cluster for $V < 4t$ the energy per particle is $\varepsilon = -\dfrac{W}{2} + \dfrac{4\pi}{m} f_0 n$, where $f_0 = 1/\ln(1/nd^2)$ in the absence of a bound state. Rigorously speaking (see also Chapters 13 and 15), at a given bosonic density $n$ the phase separation (according to Maxwell construction) results in the formation of the two clusters with the densities $n_1 > n$ and $n_2 < n$ ([6.51, 6.52]), where $n_1 d^2$ is close to or identically equal to 1 (one boson per site). The phase separation for $V < 4t$ takes place if the energy per particle in the cluster with the density $n_1$ becomes smaller than the energy per particle in the cluster with the density $n_2$,

$$-2\eta V \le -\frac{W}{2}, \qquad (6.1.42)$$

where $\eta$ is an unknown numerical coefficient of the order of 1. Note that in the fermionic t-J model, considered in [6.51, 6.52] (see Chapter 13), the Mott-Hubbard cluster with $nd^2 = 1$ has an antiferromagnetic order for electrons with spins $S = \frac{1}{2}$ on the square lattice. Hence instead of $2\eta V$ in (6.1.42) one should write $1.18J$ – the energy per bond on 2D AFM square lattice. As a result (see Chapter 13) in a fermionic case $J_{ps} = 3.8t$. In our system $V_{ps} \approx 2t$, due to the absence of kinetic energy and zero point energy in the case of structureless bosons. In the same time in our case for $nd^2 \to 0$ the phase separation between the Bose solid (see Chapter 2) and the one-particle BEC takes place. According to Dagotto and Riera [6.70] for $n_B d^2 \to 1$, the phase separation takes place already for small values of $|V|/t$.

In principle, another scenario of the phase separation connected with the creation of quartets [6.71] is also possible in our system. It requires an evaluation of the four-particle vertex which is often impossible to do analytically (except for the resonance approximation when the scattering length is much larger than intersite distance $d$ – see the last part of the Chapter). However, we think that our scenario of total phase separation takes place earlier, for smaller values of $V/t$ than the quartet formation. This is in agreement with numerical calculations [6.70] for the 2D fermionic t-J model on the square lattice.

### 6.1.7. Phase diagram of the system.

In this Subsection we will complete the phase diagram of the system. At first, note that for $V < V_{ps}$ (when the T-matrix for an s-wave channel is repulsive) we have at low density a standard Bogolubov Bose-gas with a hard-core repulsion. It will be unstable toward a standard one-particle BEC at a critical temperature given again by Fisher-Hohenberg type of formula (see (6.1.35)). For $V > 2t$ a total phase separation on two large clusters takes place in our system. One of these clusters contains a Mott-Hubbard Bose-solid, another one contains a Bose-gas with one-particle condensation (see Fig. 6.2).

For large densities $n = n_C \leq 1$ ($n_C \equiv 1$ in [6.56] for structureless bosons) the system will undergo a transition to the Mott-Hubbard Bose solid. As a result, on a qualitative level the phase diagram for our system has the form, presented in Fig. 6.3. Note that our model could be important for the study of the biexcitonic pairing in semiconductors [6.72] (see also the end of the Chapter). In this context we should mention the important results of Lozovik et al. [6.73]. It could be also important for the understanding of the physics of the gas of kinks and steps on a solid interface of $^4$He (see Chapter 2). Note that if we change the sigh of the nearest neighbors interaction $V$ from attractive in the Hamiltonian (6.1.1) on repulsive, we will get a bosonic model with on-site and intersite repulsions, which gives a disproportionation (a density wave) in some range of parameters $U$, $|V|$, $W$ and density $n_B d^2$ and can possibly serve as a simple toy model for bosonic supersolidity on the lattice (see Chapter 2) in case of moderate values of on-site repulsion $U$ (when double occupation of the site is not totally prohibited) [6.80].

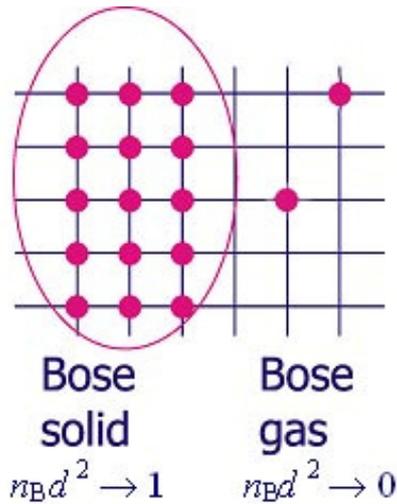

**Bose solid** $\quad n_B d^2 \rightarrow 1 \qquad$ **Bose gas** $\quad n_B d'^2 \rightarrow 0$

Fig. 6.2. Phase separation on two large clusters. First one corresponds to Bose solid with one boson per site $n_B d^2 \rightarrow 1$, the second one to dilute Bose-gas $n_B d^2 \rightarrow 0$ with repulsion between bosons.

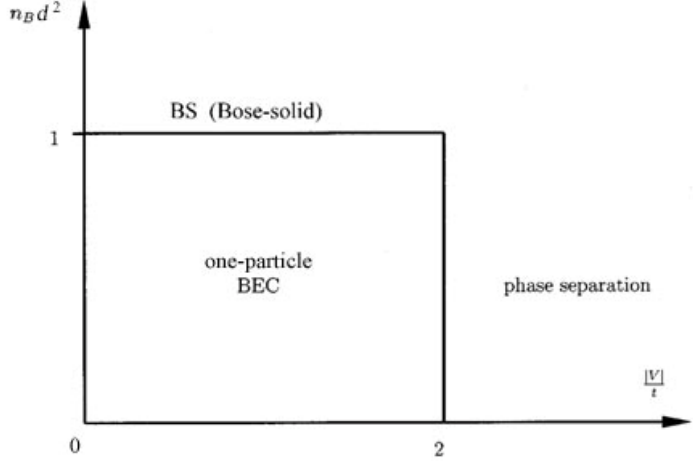

Fig. 6.3. Qualitative phase diagram of the 2D Bose-gas with the van der Waals interaction on the square lattice.

### 6.1.8. Two-band Hubbard model for the two sorts of bosons.

Let us consider the two-band Hubbard model for the two sorts of structureless bosons. The Hamiltonian of the system has the form [6.18]:

$$\hat{H} = -t_1 \sum_{<ij>} b_{1i}^+ b_{1j} - t_2 \sum_{<ij>} b_{2i}^+ b_{2j} + \frac{U_{11}}{2} \sum_i n_{1i}^2 + \frac{U_{22}}{2} \sum_i n_{2i}^2 - \frac{U_{12}}{2} \sum_i n_{1i} n_{2i} \qquad (6.1.43)$$

where $t_1$ and $t_2$ and $n_1$ and $n_2$ are, respectively, the hopping matrix elements and densities for bosons of sorts 1 and 2. For simplicity, we will consider the case $t_1 = t_2$, which corresponds to the equal masses $m_1 = 1/(2t_1 d^2) = m_2$ of bosons. We also assume that the bottoms of the bands coincide. In the Hamiltonian (6.1.43) $U_{11}$ and $U_{22}$ are Hubbard onsite repulsions for bosons of sorts 1 and 2 respectively. Finally $U_{12}$ is an onsite attraction between bosons of two different sorts (two-band Hubbard model will be also considered in Chapters 10, 14 and 15 for the case of intraband and interband repulsions).

In the present Subsection we will consider the low-density limit of the model (6.1.43), when both $n_1 d^2 << 1$ and $n_2 d^2 << 1$ on the square 2D lattice. In this limit we must replace the Hubbard interaction $U_{12}$ by the corresponding T-matrix. The relevant expression for the T-matrix $T_{12}$ is given by:

$$T_{12}(\tilde{E}) = \frac{U_{12}}{1 - U_{12} \int \frac{d^2 \vec{p}}{(2\pi)^2} \frac{1}{p^2/2m + |\tilde{E}|}}, \qquad (6.1.44)$$

where $\tilde{E}$ is given again by $\tilde{E} = E + W$. The T-matrix has the pole for the energy:

$$|\tilde{E}| = |E_b| = W \exp\left(-\frac{4\pi}{md^2 U_{12}}\right) \qquad (6.1.45)$$

in the intermediate coupling case ($|E_b| < W$ ).In the extremely strong coupling case $U_{12} > W$ the pole corresponds to the energy $|E| = |E_b| = U_{12}$. The pole in the T-matrix reflects the appearance of the bound state of the two bosons of different sorts ($b_1 b_2$).

Now we can solve the two-particle problem in the presence of the bosonic background. A simple analysis shows that only local bosonic pairs are possible in our case. They are formed at a high crossover temperature $T_* = \dfrac{|E_b|}{\ln W / T_0}$ , where $T_0 = \min\{T_{01}, T_{02}\}$, and $T_{01}$ and $T_{02}$ are degeneracy temperatures for bosons of the two sorts. Correspondingly the pairs are bose condensed at lower temperature $T_{C12} = \dfrac{T_0}{\ln(1/\lambda)}$, where $\lambda = \dfrac{mU_{12}}{4\pi}$. Our results are valid is the case $|E_b| > \{T_{01}, T_{02}\}$. In the opposite case of higher densities, when at least one of the temperatures $T_{01}$ or $T_{02}$ is larger then $|E_b|$, we have at first a standard one-particle condensation for bosons with higher density. As a result the two-particle pairing between bosons of different sort can take place only as a second superfluid transition inside the superfluid phase with one-particle BEC.

The obtained results for $T_*$ and $T_{C12}$ in low-density case $|E_b| > \{T_{01}, T_{02}\}$ can be applied for anisotropic magnetic traps [6.18] where one of the oscillator frequencies is large: $\{\omega_x; \omega_y\} << T_C < (T_0 \sim \omega_z)$. Then the system occupies only the lowest energy level in $z$-direction and becomes effectively two-dimensional. In this case formulas for $T_C$ are qualitatively correct, because it is possible to make the coherence length of the Bose-gas $\xi_0$ smaller than an effective size of the trap $R(\varepsilon)$ for $\varepsilon \sim T_C$. Another limitation on the two-particle pairing is connected with the energy release of the order of $Q \sim |E_b|$ when the pairs are created. Due to this energy release the most energetical particles can overcome the potential barrier and evaporate from the trap (a process which is analogous to an evaporative cooling technique considered in Chapter 5).

Let us now analyze the stability of our system with respect to quartet formation. For simplicity let us consider first an extremely strong-coupling case $\{U_{11}, U_{22}, U_{12}\} > W$ (remind that $m_1 = m_2$ and thus $W_1 = W_2 = W$). In this case the local pairs of two bosons of different sorts have onsite character $<b_1 b_2> \neq 0$ on the lattice. To escape local quartets $(<b_1 b_2 b_1 b_2>)$ creation we must satisfy the inequality $U_{11} + U_{22} - 4U_{12} > 0$.

The situation is less trivial in the intermediate coupling case when for the binding energy $|E_b|$ in (6.1.45) we have $T_0 < |E_b| < W$. In this case the bose pairs $b_1 b_2$ has the radius $a >> d$ (though $a < 1/\sqrt{n}$ for equal densities $n_1 = n_2$ and masses $m_1 = m_2$ when degeneracy temperatures $T_{01} = T_{02} = T_0 = 2\pi n/m$). Effectively this situation corresponds to the resonance case for shallow bound states $|E_b| = \dfrac{1}{ma^2} << \dfrac{1}{mr_0^2}$ (where $r_0 = d$ on the lattice).

In Section 6.3 we will define the binding energies $|E_3|$ of the three-particle $b_1 b_2 b_1$ and $|E_4|$ of the four-particle $b_1 b_2 b_1 b_2$ complexes in this case. We will also present variational calculations for the binding energy and radius of larger bosonic droplets which contain $N > 4$ particles in the intermediate coupling case in 2D. Note that the phase-diagram of the two-component bosons was also investigated on the optical lattices by Kuklov, Prokof'ev and Svistunov [6.83] and by Demler's group [6.84].

Note that the results on two-boson pairing are also important for SU(2) - slave-boson theories of high-$T_C$ superconductivity [6.42-6.44] and Schwinger-boson theories of 2D magnets [6.74-6.76].

### 6.1.9. Slave-boson formulation of the t-J model. Application to high-$T_C$ systems.

The superconductive pairing in the 2D fermionic t-J model will be considered in details in Chapter 13. Here we will briefly discuss the problem which arises in the approaches to the underdoped t-J model based on the scenarios of spin-charge separation advocated by Anderson et al. and Lee et al. The Hamiltonian of the canonical 2D t-J model reads:

$$\hat{H} = -t \sum_{<ij>\sigma} c_{i\sigma}^+ (1 - n_{i-\sigma}) c_{j\sigma} (1 - n_{j-\sigma}) + J \sum_{<ij>} \left( \vec{S}_i \vec{S}_j - \frac{1}{4} n_i n_j \right), \quad (6.1.47)$$

where $c_{i\sigma}^{+}$ and $c_{j\sigma}$ are creation and annihilation operators for electrons on neighboring ($<ij>$) sites $i$ and $j$ with spin projection $\sigma$, $n_{i\sigma} = c_{i\sigma}^{+} c_{i\sigma}$ is onsite electron density with spin projection $\sigma$ ( $n_i = \sum\limits_{\sigma} n_{i\sigma}$ ), and $\vec{S}_i = \frac{1}{2} c_{i\mu}^{+} \vec{\sigma}_{\mu\nu} c_{i\nu}$ is electron spin ( $\vec{\sigma}_{\mu\nu}$ are Pauli matrices). In slave-boson formulation of the t-J model (based on the scenario of spin-charge separation) close to half-filling ($nd^2 \to 1$ on the 2D square lattice) an electron according to Anderson and Lee [6.41-6.44] can be represented as a product of spinon (fermion with charge 0 and spin ½) and holon (boson with charge $|e|$ and spin 0):

$$c_{i\sigma}^{+} = f_{i\sigma}^{+} b_i . \qquad (6.1.48)$$

A superconductive d-wave gap (fro electrons):

$$\Delta_d = < c_{i\uparrow} c_{j\downarrow} - c_{i\downarrow} c_{j\uparrow} > \qquad (6.1.49)$$

is a direct product

$$\Delta_d = \Delta_{sp} \Delta_h \qquad (6.1.50)$$

of a spinon d-wave gap

$$\Delta_{sp} = < f_{i\uparrow} f_{j\downarrow} - f_{i\downarrow} f_{j\uparrow} > \qquad (6.1.51)$$

and a holon s-wave gap

$$\Delta_h = < b_i b_j > . \qquad (6.1.52)$$

Then a natural question arises whether $<b_i> \neq 0$ and, accordingly, $\Delta_h = <b_i><b_j>$, or $<b_i> = 0$ but $<b_i b_j> \neq 0$. In other words, whether a one particle or two-particle condensation of the holons takes place in our system [6.42, 6.77].

This problem is a very difficult one and, surely deserves a very careful analysis. Our preliminary considerations show, however, that the more beneficial conditions for the two-particle condensation may arise in the SU(2) formulation of the t-J model [6.43], which assumes the appearance of two sorts of holons $b_1$ and $b_2$. Note that in the standard U(1) formulation of the model [6.44] with one sort of holons an effective potential of the two-holon interaction on neighboring sites appearing after the Hubbard-Stratonovich transformation has a form:

$$\left( \frac{8t^2}{J} - \frac{J}{4} \right) \sum\limits_{<ij>} b_i^{+} b_j^{+} b_i b_j . \qquad (6.1.53)$$

and thus corresponds to the repulsion for $t > J$ (in 2D cuprates usually $J \sim (\frac{1}{2} \div \frac{1}{3}) t < t$). This observation excludes the possibility of the two-holon pairing in the U(1) formulation of the t-J model.

In the SU(2) case it will be desirable to derive conditions when $<b_1> = <b_2> = 0$ but $<b_1 b_2> \neq 0$ for two sorts of holons. For such a nondiagonal pairing, as already discussed above, it is easier to satisfy the stability criteria [6.18]. Note also that the same situation with two sorts of bosons and a possible attraction between them can be realized for 2D magnetic systems. The corresponding bosonic Hamiltonian can be obtained here after a Schwinger transformation of spins [6.74-6.76] in extended Heisenberg models.

Concluding this section we would like to emphasize that we analyzed the possibility of the formation of boson pairs with s-wave symmetry and an appearance of total phase separation in a 2D Bose-gas. In addition to that we considered the case of boson pairs with the symmetries of p- and d-wave type. We also considered the qualitative phase diagram for the 2D Bose-gas with the van der Waals interaction between the particles, which, besides a standard one-particle BEC, contains the regions of the Mott-Hubbard Bose solid and a total phase separation. We also considered the situation for two sorts of bosons described by the two-band Hubbard model, and found the conditions for the two-particle pairing between bosons of different sorts. We discussed the applicability of our results for the different physical systems ranging from 2D magnetic traps or optical lattices, submonolayers of $^4$He and excitons in the semiconductors till Schwinger bosons in magnetic systems and holons in the slave-boson theories of high-$T_C$ superconductors.

Note that for high-$T_C$ systems we considered in this Section slave-boson formulation of the t-J model based on the ideas of spin-charge separation. These ideas were transferred to high-$T_C$ materials from the 1D physics of doped spin chains. We think, however, that for quasi-2D high-$T_C$ systems more suitable are the ideas of spin-charge confinement, which are based on the formation of the AFM-string (of the linear trace of the frustrated spins) which accompany the hole motion on 2D (or 3D) AFM-background of spins ½. In the next Section of the present Chapter and in Chapter 13 we will consider the scenario of spin-charge confinement (introduced by Laughlin et al. [6.34, 6.35]), more detaily with a special emphasis on the formation of composite hole (or composed fermions) there.

### 6.2. Composed fermions in the Fermi-Bose mixture with attractive interaction between fermions and bosons.

In Section 5.2 of the Chapter 5 we considered composed bosons $f_\uparrow f_\downarrow$ which arise in the Fermi-gas with attraction or in a broad Feshbach resonance in the framework of the one-channel resonance approximation.

In Section 6.1 of the present Chapter we considered the possibility of two-boson pairing $bb$ or $b_1 b_2$ in the Bose-gas with one ore two sorts of bosons.

For the sake of completeness (to restore the full "supersymmetry" between fermions and bosons, treating them on equal grounds) we will analyze in this Section a possibility to form composed fermions $f_\sigma b$ in the Fermi-Bose mixture with attractive interaction between fermions and bosons. Note that in an optical dipole trap it is possible to get an attractive scattering length of the fermion-boson interaction with the help of Feshbach resonance [4.29]. Note also that even in the absence of Feshbach resonance it is experimentally possible now to create a Fermi-Bose mixture with attractive interaction between fermions and bosons. For example in [6.5, 6.6] such a mixture of $^{87}$Rb (bosons) and $^{40}$K (fermions) was experimentally studied. Moreover, the authors of [6.5, 6.6] experimentally observed the collapse of a Fermi gas with the sudden disappearance of fermionic $^{40}$K atoms when the system enters into the degenerate regime. We cannot exclude in principle that it is just a manifestation of the creation of quartets $f_\uparrow b f_\downarrow b$ in the system. Note that in the regime of a strong attraction between fermions and bosons, a phenomenon of phase separation with the creation of larger clusters or droplets is also possible. Note also that for a large mismatch between fermionic and bosonic densities $n_F << n_B$, a much slower collapse in the Bose subsystem of $^{87}$Rb was experimentally observed. Here, after the formation of composed fermions, a lot of residual (unpaired) bosons are still present. This fact probably can explain a slow collapse in bosonic subsystem.

### 6.2.1. The theoretical model.

The model of a Fermi-Bose mixture has the following form on a lattice:

$$\hat{H}' = \hat{H}'_F + \hat{H}'_B + \hat{H}'_{FB},$$

$$\hat{H}'_F = -t_F \sum_{<ij>\sigma} f_{i\sigma}^+ f_{j\sigma} + U_{FF} \sum_i n_{i\uparrow}^F n_{i\downarrow}^F - \mu_F \sum_{i\sigma} n_{i\sigma}^F,$$

$$\hat{H}'_B = -t_B \sum_{<ij>} b_i^+ b_j + \frac{U_{BB}}{2} \sum_i n_i^B n_i^B - \mu_B \sum_i n_i^B,$$

$$\hat{H}'_{FB} = -U_{BF} \sum_i n_i^B n_{i\sigma}^F.$$

(6.2.1)

This is a lattice analog of the standard Hamiltonian considered for example by Efremov and Viverit [6.21] for the case of repulsive interaction between fermions and bosons in $^7$Li-$^6$Li mixture (see Chapter 12). Note that in the Fermi-Bose mixture of $^3$He and $^4$He (which will be considered in Chapter 11) the fermion-boson interaction also corresponds to repulsion (see a classical paper [6.20] by Bardeen, Baym, Pines). In the Hamiltonian (6.2.1) $t_F$ and $t_B$ are fermionic and bosonic hopping amplitudes, $f^+_{i\sigma}$, $f_{i\sigma}$ and $b^+_i$, $b_i$ are fermionic and bosonic creation and annihilation operators. The Hubbard interactions [6.58] $U_{FF}$ and $U_{BB}$ correspond to hard-core repulsions between particles of the same sort. The interaction $U_{BF}$ corresponds to the attraction between fermions and bosons. $W_F = 8t_F$ and $W_B = 8t_B$ are the bandwidths in 2D. Finally, $\mu_F$ and $\mu_B$ are fermionic and bosonic chemical potentials. For the square lattice the uncorrelated spectra of fermions and bosons after Fourier transformation read: $\xi_{p\sigma} = -2t_F(\cos p_x d + \cos p_y d) - \mu_F$ for fermions and $\eta_p = -2t_B(\cos p_x d + \cos p_y d) - \mu_B$ for bosons, where $d$ is a lattice constant.

### 6.2.2. Intermediate coupling case in 2D.

In the intermediate coupling case in 2D (see Chapter 8 and [4.17] for more details) $\dfrac{W_{BF}}{\ln W_{BF}/T_{0BF}} < U_{BF} < W_{BF}$ the energy of the bound state (for a formation of a composed fermion $f_{i\sigma}b_j$) reads [6.19]:

$$|E_b| = \frac{1}{2m_{BF}d^2} \frac{1}{\exp\left\{\dfrac{2\pi}{m_{BF}U_{BF}}\right\} - 1}, \qquad (6.2.2)$$

where $m_{BF} = \dfrac{m_B m_F}{m_B + m_F}$ is an effective mass, $W_{BF} = \dfrac{4}{m_{BF}d^2}$ is an effective bandwidth, $m_B = \dfrac{1}{2t_B d^2}$ and $m_F = \dfrac{1}{2t_F d^2}$ are the band masses of elementary bosons and fermions. Finally $T_{0BF} = \dfrac{2\pi n}{m_{BF}}$ is an effective degeneracy temperature. For simplicity we consider the case of equal densities $n_B = n_F = n$ which is more relevant for the physics of holons and spinons in underdoped high-$T_C$ materials.

Note that in the intermediate coupling case in 2D the binding energy for pairing between fermions and bosons $|E_b|$ is larger than bosonic and fermionic degeneracy temperatures:

$$|E_b| > \left\{ T_{0B} = \frac{2\pi n_B}{m_B} ; T_{0F} = \frac{2\pi n_F}{m_F} = \varepsilon_F \right\}, \qquad (6.2.3)$$

but smaller than the bandwidths $|E_b| < \{W_B, W_F\}$. In this case the pairing of fermions and bosons $\langle f_\sigma b\rangle$ takes place earlier (at higher temperatures) than both Bose-Einstein condensation of bosons ($\langle b\rangle \neq 0$ or $\langle bb\rangle \neq 0$) and superconductive pairing of fermions $\langle f_\sigma f_{-\sigma}\rangle \neq 0$. Note that a matrix element $\langle c_\sigma\rangle = \langle f_\sigma b\rangle$ is nonzero only for the transitions between the states with $|N_B;N_F\rangle$ and $\langle N_B - 1; N_F - 1|$, where $N_B$ and $N_F$ are the numbers of particles of elementary bosons and fermions, respectively [6.19]. For superconductive state a matrix element for a quartet $\langle c_\sigma c_{-\sigma}\rangle = \langle f_\sigma b, f_\sigma b\rangle \neq 0$ only for the transitions between the states with $|N_B;N_F\rangle$ and $\langle N_B - 2; N_F - 2|$. Note also that in the case of a very strong attraction $U_{BF} > W_{BF}$ we have a natural result $|E_b| = U_{BF}$, and the effective mass $m_{BF}^* = m_{BF}\dfrac{U_{BF}}{W_{BF}} >> m_{BF}$ is additionally enhanced on the lattice (see Nozieres, Schmitt-Rink [4.12] and Chapter 8). Finally let us emphasize that the diagonal Hubbard interactions $U_{FF}$ and $U_{BB}$ satisfy the inequalities $U_{FF} > \dfrac{W_F}{\ln W_F/|E_b|}$ and $U_{BB} > \dfrac{W_B}{\ln W_B/|E_b|}$ in the intermediate coupling case. Now let us consider the temperature evolution of the system.

### 6.2.3. Bethe-Salpeter integral equation.

The temperature evolution is governed again (as in the case of two-fermion or two-boson pairing) by the corresponding Bethe-Salpeter (BS) equation. After analytical continuation $i\omega_n \to \omega + io$ (see [4.60]) the solution of this equation for the two-particle total vertex $\Gamma$ acquires a form:

$$\Gamma(\vec{q},\omega) = \frac{-U_{BF}d^2}{1 - U_{BF}d^2 \int \frac{d^2\vec{p}}{(2\pi)^2} \frac{1 - n_F(\xi_p) + n_B(\eta_{q-p})}{\xi(\vec{p}) + \eta(\vec{q}-\vec{p}) - \omega - io}}, \qquad (6.2.4)$$

where $\xi(p) = \dfrac{p^2}{2m_F} - \tilde{\mu}_F$ and $\eta(\vec{q}-\vec{p}) = \dfrac{p^2}{2m_B} - \tilde{\mu}_B$ are spectra of fermions and bosons at low densities $n_F d^2 << 1$ and $n_B d^2 << 1$, the chemical potentials $\tilde{\mu}_F = \dfrac{W_F}{2} + \mu_F$ and $\tilde{\mu}_B = \dfrac{W_B}{2} + \mu_B$ are counted from the bottoms of the bands. Note that in the pole of BS equation enters the temperature factor $1 - n_F(\xi(\vec{p})) + n_B(\eta(\vec{q}-\vec{p}))$ in contrast with the factor $1 - n_F(\xi(\vec{p})) - n_F(\xi(\vec{q}-\vec{p}))$ for the standard two-fermion superconductive pairing and $1 + n_B(\eta(\vec{p})) + n_B(\eta(\vec{q}-\vec{p}))$ for the two-boson pairing (considered in Section 6.1). The pole of the Bethe-Salpeter equation corresponds to the spectrum of the composed fermions:

$$\omega = \xi_p^* = \frac{p^2}{2(m_B + m_F)} - \mu_{comp}. \qquad (6.2.5)$$

Note that in (6.2.5)

$$\mu_{comp} = \tilde{\mu}_B + \tilde{\mu}_F + |E_b| \qquad (6.2.6)$$

is a chemical potential of composed fermions. Similary to (6.2.6) in Chapter 8 for two-fermion pairing we will get $\mu_{comp} = \mu_B = 2\tilde{\mu}_F + |E_b|$ for a chemical potential [4.14, 4.17] of a composed boson (molecule or dimer) $f_\sigma f_{-\sigma}$, while instead of (6.2.5) we will obtain $\omega = \dfrac{p^2}{4m_F} - \mu_B$ for a pole of the fermionic BS-equation. Note also that composed fermions are well-defined quasiparticles, since the damping of quasiparticles equals to zero in the case of the bound state ($E_b < 0$), but it becomes nonzero and is proportional to $E_b$ in the case of the virtual state ($E_b > 0$). The process of a dynamical equilibrium (boson + fermion $\leftrightarrow$ composed fermion) is again governed by the standard Saha formula (see Sections 5.2, 6.1 and [4.20]).

### 6.2.4. Crossover (Saha) temperature.

In the 2D case Saha temperature reads [6.19]:

$$\frac{n_B n_F}{n_{comp}} = \frac{m_{BF}T}{2\pi} \exp\left\{-\frac{|E_b|}{T}\right\}. \qquad (6.2.7)$$

The crossover temperature $T_*$ is defined, as usual, from the requirement that the number of composed fermions equals the number of unbound fermions and bosons: $n_{comp} = n_B = n_F = n_{tot}/2$. (where $n_{tot} = n_B + n_F$ is a total density). This conditions yields:

$$T_* = \frac{|E_b|}{\ln(|E_b|/2T_{0BF})} >> \{T_{0B}, T_{0F}\}. \qquad (6.2.8)$$

Note that in the Boltzmann regime $|E_b| > \{T_{0B}, T_{0F}\}$. In fact we deal here with the pairing of two Boltzmann particles.

That is why this pairing does not differ drastically from the pairing of two particles of the same type of statistics. Indeed, if we substitute $\tilde{\mu}_B + \tilde{\mu}_F$ in (6.2.6) on $2\tilde{\mu}_B$ or $2\tilde{\mu}_F$ we will get the familiar expressions for the chemical potentials of composed bosons consisting either of two elementary bosons [6.18] or of two elementary fermions [4.12, 4.17, 4.14].

For lower temperatures $T_0 < T < T_*$ (where $T_0 = \dfrac{2\pi n}{m_B + m_F}$ is the degeneracy temperature for composed fermions) the numbers of elementary fermions and bosons (for the case of equal densities $n_B = n_F$) are exponentially small. The chemical potential of composed fermions reads $\mu_{comp} = -T \ln(T / T_0)$. Hence $\left| \mu_{comp} \right| << \left| E_b \right|$ for $T << T_*$.

### 6.2.5. Three and four particles bound states in the Fermi-Bose mixture.

Note that in a general case to complete the phase-diagram of the Fermi-Bose mixture model with attraction between fermions and bosons we should determine also the binding energies of trios and quartets $|E_3|$ and $|E_4|$ for the complexes $f_\sigma b, b$ and $f_\sigma b, f_{-\sigma} b$ (the complex $f_\sigma b, f_{-\sigma} b$ is not formed due to repulsive interaction between a composed fermion $f_\sigma b$ and elementary fermion $f_{-\sigma}$). The knowledge of $|E_3|$ and $|E_4|$ will help us to find the hierarchy of inequalities between the different Saha temperatures, which for equal masses $m_B = m_F$ and equal densities $n_B = n_F$ read:

$$T_*^{(2)} = \frac{|E_b|}{\ln(|E_b| / T_0)}, \; T_*^{(3)} = \frac{|E_3|}{2\ln(|E_3| / T_0)}, \; T_*^{(4)} = \frac{|E_4|}{3\ln(|E_4| / T_0)}, \qquad (6.2.9)$$

where $T_0 = \dfrac{2\pi n}{m}$ is degeneracy temperature, $T_*^{(3)}$ and $T_*^{(4)}$ are Saha temperatures for the formation of composed trios and quartets. If, as we will prove in the next Section, $T_*^{(4)} > \{ T_*^{(2)}; T_*^{(3)} \}$, then the phase-diagram of the Fermi-Bose mixture is trivialized. Namely for high temperatures $T > T_*^{(4)}$ the elementary fermions and bosons prevail in the system, while for lower temperatures $T < T_*^{(4)}$ the quartets $f_\uparrow b, f_\downarrow b$ prevail in the system.

The quartets are already composed bosons. They are bose-condensed below the temperature governed again by the Fisher-Hohenberg type of the formula [6.64]:

$$T_C \sim \frac{T_0}{16 \ln \ln \left( \alpha \dfrac{|E_4|}{T_0} \right)}, \qquad (6.2.10)$$

where the coefficient $\alpha$ will be also defined in the next Section for scattering of molecule on molecule in 2D.

### 6.3. Bound states of three and four resonantly interacting particles.

In this Section we will complete the phase-diagram of 2D Fermi-Bose mixture with attraction between fermions and bosons, as well as a phase-diagram of 2D Bose-gas with one or two sorts of bosons by calculating exactly the bound states energies $|E_3|$ for the three-particle complexes $bbb$, $b_1 b_2, b_1$, $f_\sigma b, b$ and $|E_4|$ for the four-particle complexes $bbbb$, $b_1 b_2 b_1 b_2$, $f_\sigma bbb$ and $f_\sigma b, f_{-\sigma} b$ in the resonance (one-channel) approximation [6.25, 6.26]. We will also present variational calculations of the binding energy of the larger droplet $|E_N|$ with the number of particles in the droplet $N > 4$.

Finally to complete the phase-diagram of the attractive Fermi-gas in the regime of BCS-BEC crossover we will determine the scattering amplitudes $a_{2-1}$ for the scattering of molecules $f_\uparrow f_\downarrow$ on elementary fermions (atoms) $f_\sigma$, and $a_{2-2}$ for the scattering of molecules $f_\uparrow f_\downarrow$ on other molecules $f_\uparrow f_\downarrow$ in the resonance approximation for BEC-domain $a >> r_0$ (see Chapter 5). Note that the scattering length $a_{2-1}$ governs also the inelastic scattering time or a lifetime of the 3D resonance Fermi-gas in BEC-domain, where $|E_b| = 1/ma^2 < 1/mr_0^2$, and $a > 0$.

In the resonance approximation we will get diagrammatically the results for the binding energies $|E_3|$ and $|E_4|$ only in terms of the two-particle binding energy $|E_b|$, while for the scattering amplitudes we will get $a_{2-1}$ and $a_{2-2}$ only in terms of the two-particle s-wave scattering length $a$ [6.25, 6.26].

### 6.3.1. Atom-molecule scattering length for three resonantly interacting fermions in 3D. Skorniakov-Ter-Martirosian integral equation.

In this Subsection we will present diagrammatic method [6.11] to rederive the famous Skorniakov and Ter-Martirosian result (firstly obtained for scattering of neutrons on deutrons in nuclear physics) [6.23] for dimer-fermion ($f_\uparrow f_\downarrow$; $f_\uparrow$) scattering length $a_{2-1}$ in the case of three resonantly interacting fermions in 3D.

Following Skorniakov and Ter-Martirosian in the presence of the weakly bound resonance level $-|E_b|$ in a two-particle cross-section, we can limit ourselves to the zero-range interaction potential ($|a| >> r_0$) between fermions. A two-fermion vertex (two-particle T-matrix in vacuum) can be approximated by a simple one-pole structure, which reflects the presence of the s-wave resonance level in a spin-singlet state:

$$T_{2\alpha\beta,\gamma\delta}(E, \vec{p}) = T_2(E, \vec{p})(\delta_{\alpha\gamma}\delta_{\beta\delta} - \delta_{\alpha\delta}\delta_{\beta\gamma}) = T_2(E, \vec{p})\chi(\alpha, \beta)\chi(\gamma, \delta), \qquad (6.3.1)$$

where in the 3D case

$$T_2(E, \vec{p}) = \frac{4\pi}{m^{3/2}} \frac{\sqrt{|E_b|} + \sqrt{p^2/2m - E}}{E - p^2/4m + |E_b|} \qquad (6.3.2)$$

is given by the ladder sequence of vacuum diagrams (see Section 6.2 and Fig. 5.13). $E$ is the total energy and $\vec{p}$ is the total momentum of the incoming particles, $m$ is the fermionic mass, and $|E_b|$ = $1/ma^2$ is the binding energy ($a = a_F$ – s-wave scattering length of two fermions). Indices $\alpha$, $\beta$ and $\gamma$, $\delta$ denote spin state of incoming and outgoing particles. The function $\chi(\alpha, \beta)$ stands for the spin-singlet state $\chi(\alpha, \beta) = \delta_{\alpha\uparrow}\delta_{\beta\downarrow} - \delta_{\alpha\downarrow}\delta_{\beta\uparrow}$. Note that the pole of the vacuum T-matrix (6.3.2) coincides with the pole of two-particle vertex $\Gamma$ obtained for two-fermion pairing in Section 5.2 for $\mu_F = 0$ (and $\mu_B = 2\mu_F + |E_b| = |E_b|$). The simplest process which contributes to the dimer-fermion interaction is the exchange of a fermion (Fig. 6.4).

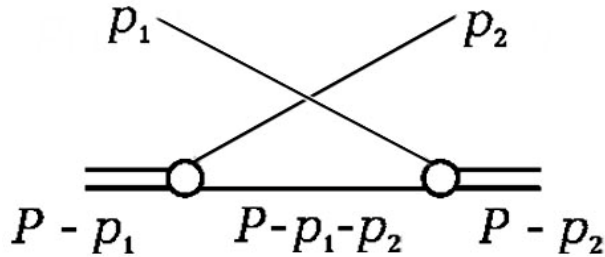

Fig. 6.4. The simplest exchange diagram for the three-particle interaction. The double line corresponds to dimer, the single lines – to elementary fermions [6.19]. Intermediate single line stands for the vacuum Green's function $G(P-p_1-p_2)$, where $P$ is the total 4-momentum of dimer and elementary fermion.

We will denote it as $\Delta_3$ (see [6.26]). Its analytical expression is:

$$\Delta_{3\alpha,\beta}(p_1, p_2, P) = -\delta_{\alpha\beta}G(P - p_1 - p_2), \qquad (6.3.3)$$

where we introduce four-momenta $P = (E, \vec{p})$, $p_1$ and $p_2$. In (6.3.3) $G(P) = \dfrac{1}{\omega - \dfrac{p^2}{2m} + io}$ is a bare fermion Green's function in vacuum. The minus sign on the right-hand side of (6.3.3) comes from the permutation of the two fermions and corresponds to the bare repulsive interaction between dimer and fermion. In order to obtain a full dimer-fermion scattering vertex $T_3$ we need to build a ladder again from $\Delta_3$ blocks. One can easily verify that the spin projection is conserved in every order of $T_3$, and thus $T_{3\alpha\beta} = \delta_{\alpha\beta}T_3$. An equation for $T_3$ will have the diagrammatic representation shown in Fig. 6.5, and in analytical form it is written as:

$$T_3(p_1, p_2, P) = -G(P - p_1 - p_2) - i\sum_q G(P - p_1 - q)G(q)T_2(P - q)T_3(q, p_2, P), \qquad (6.3.4)$$

where $\sum_q = \int d^3\vec{q}\,d\Omega / (2\pi)^4$.

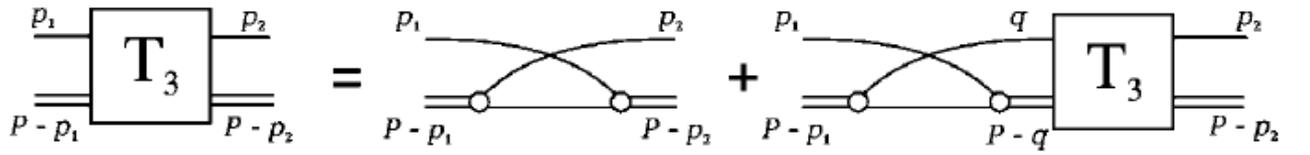

Fig. 6.5. The graphic representation of the equation for the full dimer-fermion scattering vertex $T_3$. Intermediate double line corresponds to two-particle scattering vertex $T_2(p\text{-}q)$ (which describes the bound state of a dimer). Intermediate single lines stand for the vacuum one-particle Green-functions $G(q)$ and $G(P\text{-}p_1\text{-}q)$.

Note that for three resonantly interacting bosons $bb,b$ the sign near the first term in right-hand side of (6.3.4) should be changed on "plus". This means that we have a bare attraction between a molecule $bb$ and elementary boson $b$. Correspondingly the bare interaction is also attractive for a molecule $f_\sigma b$ interacting with elementary boson $b$, but it is repulsive (due to Pauli principle) for the interaction of the molecule $f_\sigma b$ with an elementary fermion $f_{-\sigma}$ having an opposite spin projection.

Returning back to three resonantly interacting fermions ($f\uparrow f\downarrow$ and $f_\sigma$) we can integrate out the frequency $\Omega$ in (6.3.4) by closing the integration contour in the lower half-plane, since both $T_2(P - q)$ and $T_3(q, p_2; P)$ are analytical functions in this region. Moreover, if we are interested in the low-energy s-wave dimer-fermion scattering length $a_3 = a_{2\text{-}1}$, we can safely put $P = (E, \vec{p}) = (-|E_b|, 0)$ and $p_2 = 0$. The full vertex $T_3$ is connected with $a_3 = a_{2\text{-}1}$ by the following relation [6.25, 6.26]:

$$\left(\frac{8\pi}{m^2 a}\right)T_3(0, 0, -|E_b|) = \frac{3\pi}{m}a_{2\text{-}1}. \qquad (6.3.5)$$

Introducing a new function $a_{2\text{-}1}(k)$ according to the formula:

$$\left(\frac{8\pi}{m^2 a}\right)T_3\left(\left\{\frac{k^2}{2m}, \vec{k}\right\}, 0, -|E_b|\right) = \frac{3\pi}{m}a_{2\text{-}1}(k), \qquad (6.3.6)$$

and substituting it into equation (6.3.4), we obtain the Skorniakov-Ter-Martirosian (STM) equation for the scattering amplitude:

$$\frac{\frac{3}{4}a_{2\text{-}1}(k)}{\sqrt{m|E_b|} + \sqrt{\frac{3k^2}{4} + m|E_b|}} = \frac{1}{k^2 + m|E_b|} - 4\pi\int \frac{a_{2\text{-}1}(q)}{q^2(k^2 + q^2 + \vec{k}\vec{q} + m|E_b|)}\frac{d^3\vec{q}}{(2\pi)^3}. \qquad (6.3.7)$$

Solving this equation numerically one obtains the well-known result for the dimer-fermion scattering length:

$$a_{2-1} = a_{2-1}(0) = 1.18|a| > 0. \qquad (6.3.8)$$

This result is quite nice since dimer-fermion scattering length in it depends only upon numerical coefficient (1.18) and two-particle s-wave scattering length $a$ [6.23, 6.8].

### 6.3.2. Three resonantly interacting bosons in 3D. Efimov effect.

As was first shown by Danilov [6.9] (see also the paper by Minlos and Fadeev [6.7]) in the 3D case, the problem of three resonantly interacting bosons cannot be solved in the resonance approximation. This statement stems from the fact that in the case of identical bosons (or in the case of the complexes with two bosons and one fermion $f_\sigma b, b$) the homogeneous part of STM-equation (6.3.7) has a nonzero solution at any negative energies. The physical meaning of this mathematical artifact was elucidated by Efimov [6.52], who showed that a Hamiltonian with only two-particle interaction leads to the appearance of an attractive $1/r^2$ interaction in a three-body system. Since in the attractive $1/r^2$ potential in 3D a particle falls to the center, the short-range physics is important and one cannot replace the exact pair interaction by its resonance approximation. In the excellent review article by Jensen et al. [6.3] it was nicely illustrated that Efimov effect is present in the dimensions 2, 3 < D < 3.8. In 3D case it creates the attractive (centripetal) effective potential $V_{eff}(\rho) \approx -\dfrac{1.26}{\rho^2}$ (see Fig. 6.6) if we introduce convenient relative coordinates ($m_1 = m_2 = m_3$):

$$\begin{aligned} \vec{x} &= \vec{r}_2 - \vec{r}_3, \\ \vec{y} &= \vec{r}_1 - \frac{\vec{r}_2 + \vec{r}_3}{2}, \end{aligned} \qquad (6.3.9)$$

which correspond to the relative radius-vector ($\vec{x}$) inside the dimer (23) and the radius-vector ($\vec{y}$) between the elementary particle 1 (falling on dimer) and the dimer's center of mass.

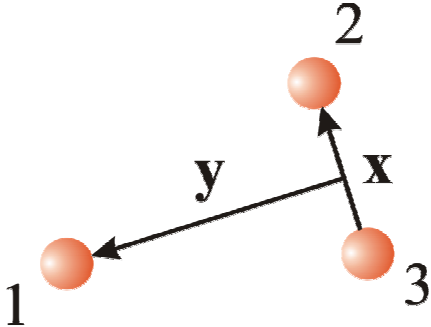

Fig. 6.6. Convenient coordinates $\vec{x} = \vec{r}_2 - \vec{r}_3$ and $\vec{y} = \vec{r}_1 - \dfrac{\vec{r}_2 + \vec{r}_3}{2}$ for the description of fermion (1)-dimer (23) interaction (see Ref. [6.3]).

Vectors $\vec{x}$ and $\vec{y}$ define the plane. In this plane we can introduce a hyperradius $\rho$ and angles $\alpha_1$, $\alpha_2$ such as:

$$\begin{aligned} x_i &= \rho \sin \alpha_i; \quad i = 1, 2 \\ y_i &= \rho \cos \alpha_i; \quad i = 1, 2 \end{aligned} \qquad (6.3.10)$$

If we rewrite the three-particle Hamiltonian:

$$\hat{H} = \sum_{j=1}^{3} \frac{1}{2m_j} \nabla_j^2 + \sum_{j,k=1}^{3} V\left(\left|\vec{r}_j - \vec{r}_k\right|\right) = \hat{T} + \hat{V} \qquad (6.3.11)$$

in terms of $\vec{x}$, $\vec{y}$ and $\vec{R}_{c.m.} = \frac{\vec{r}_1 + \vec{r}_2 + \vec{r}_3}{3}$ (for equal masses $m_1 = m_2 = m_3$) and then introduce a hyperradius $\rho$ and angles $\alpha_i$, then in the new coordinates the kinetic part of the Hamiltonian $\hat{T}$ in (6.3.11) will contain the centripetal term according to [6.3].

Note that in 2D case $V_{eff}(\rho) \approx \frac{1.5}{\rho^2} > 0$ - the potential is repulsive and Efimov effect is absent. Thus the STM equation in 2D has a finite number of the solutions for the binding energies $E_3$ of the three-particle complexes $bbb$ and $fbb$.

In 3D case Efimov effect leads to the appearance of the strongly-bound three-particle levels with an energy [6.7]:

$$|E_{3n}| = \frac{1}{mr_0^2} \exp\left\{ -\frac{2\pi n}{\varsigma_0} \right\}, \qquad (6.3.12)$$

where $\zeta_0 = 1.006$. The total number of levels is

$$N = \frac{1}{\pi} \ln \frac{a}{r_0}, \qquad (6.3.13)$$

where $r_0$ is the range of the potential.

They lie in the interval:

$$\frac{1}{ma^2} < |E_{3n}| < \frac{1}{mr_0^2}. \qquad (6.3.14)$$

Mathematically Efimov's effect is connected with the properties of the kernel of the integral STM equation. If we perform analytically the angular integration in (6.3.7) of the type $\int_{-1}^{1} \frac{d\cos\theta}{a + b\cos\theta} = \frac{1}{b} \ln \frac{a+b}{a-b}$ and introduce dimensionless intermediate and incoming momenta $y = qa$ and $x = ka$ (measured in terms of the scattering length $a$) as well as dimensionless three-particle energy $W = \frac{|E_3|}{|E_b|} = ma^2|E_3|$ (measured in terms of the two-particle binding energies $|E_b|$) and dimensionless 3-particle vertex $\tau_3 = \frac{T_3}{|E_b|} = ma^2 T_3$ then in terms of $x$, $y$ and $W$ the homogeneous part of STM-equation for $\tau_3$ reads:

$$\tau_3(x, W) = \frac{2}{\pi} \int_0^{a/r_0} \frac{y^2 dy}{2xy} \ln\left( \frac{W + x^2 + y^2 + xy}{W + x^2 + y^2 - xy} \right) \frac{1}{\sqrt{W + 3/4\, y^2 - 1}} \cdot \tau_3(y, W), \qquad (6.3.15)$$

where sign "plus" in front of the integral corresponds for three bosons to the attractive dimer-monomer interactions (Three-particle binding energies $E_3$ correspond to the poles of $\tau_3$, and consequently for $E = E_3$ homogeneous part of STM-equation should have nonzero solutions). It is possible to show that the kernel in the right-hand side of integral equation (6.3.15) is not limited. The integral of the kernel is of the order of 1 on the upper limit so the solutions with large three-particle binding energies $W = \frac{|E_3|}{|E_b|} \gg 1$ are possible. These solutions are "sitting" on the upper limit $y_{max} = \frac{a}{r_0} \to \infty$ of the integral in (6.3.15). For them we can get $x \sim x_{max} = \frac{a}{r_0} \gg 1$ and $W \sim W_{max} = \frac{a^2}{r_0^2} \gg 1$. Effectively they describe the situation when the three-particle system goes far away from the resonance (when we add a third particle to resonantly interacting pair of particles). That is why we have deep three-particle levels (see (6.3.14)) in 3D.

6.3.3. Three resonantly interacting bosons in 2D.

As we already mentioned, Efimov effect is absent in 2D. Therefore it is possible to describe the binding energies $|E_3|$ of the three-particle complexes $bbb$ and $f_\sigma b,b$ in terms of the two-particle binding energies $|E_b|$ only.

As in the 3D case, the cornerstone in the diagrammatic technique is the two-particle resonance scattering vertex $T_2$. For two resonantly interacting particles with total mass $2m$ (we assume that all the particles under considerations have the same mass $m$, which in case of Fermi-Bose mixture means that $m_B = m_F$) it can be written in 2D as:

$$\left(\frac{8\pi}{m^2 a}\right) T_2(P) = -\frac{4\pi}{m} \frac{\alpha}{\ln\left(\left(\frac{p^2}{4m} - E\right)\middle/|E_b|\right)}, \qquad (6.3.16)$$

where $P = (E, \vec{p})$ is 4 momentum and we introduce the factor $\alpha = \{1, 2\}$ in order to take into account whether or not two particles are indistinguishable. It is $\alpha = 2$ for the case of a resonance interaction between identical bosons and $\alpha = 1$ for the case of a resonance interaction between fermion and boson or for the case of two distinguishable bosons. Note that $T_2(P)$ contains a typical for 2D systems logarithm in denominator of (6.3.16). We start with a system of three resonantly interacting identical bosons-$bbb$ in 2D. An equation for the dimer- (elementary) boson scattering vertex $T_3$ which describes interaction between three bosons has the same diagrammatic form as shown in Fig. 6.5; however, the rules for its analytical notation changed. It can be written as:

$$T_3(p_1, p_2, P) = G(P - p_1 - p_2) + i \sum_q G(P - p_1 - q) G(q) T_2(P - q) T_3(q, p_2, P), \qquad (6.3.17)$$

where $\displaystyle\sum_q = \int d^3\vec{q}\, d\Omega / (2\pi)^4$, $P = (E, 0)$ and one should put $\alpha = 2$ for the two-particle vertex $T_2$ in (6.3.16). As we already mentioned the opposite signs in (6.3.4) for fermions and (6.3.17) for bosons are due to the permutation properties of the particles involved: an exchange of fermions results in a minus sign, while an analogous exchange of bosons brings no extra minus. Finally, we note that the three-particle s-wave (s-wave channel of a boson-dimer scattering) binding energies $E_3$ correspond to the poles of $T_3(0, 0; - |E_3|)$ and, consequently, at energies $E = E_3$ the homogeneous part of (6.3.17) has a nontrivial solution. Introducing the same dimensionless variables $x$, $y$ for initial and intermediate momenta, $W$ for dimensionless three-particle energy $|E_3|$ and $\tau_3$ for dimensionless three-particle T-matrix $T_3$ we will get in a 2D case (after the corresponding angular integration $\displaystyle\int_0^{2\pi} \frac{d\varphi}{a + b\cos\varphi} = \frac{\pi}{\sqrt{a^2 - b^2}}$):

$$\tau_3(x, W) = \frac{2}{\pi} \int_0^\infty \frac{y\, dy}{\sqrt{\left(W + x^2 + y^2\right)^2 - (xy)^2}} \frac{1}{\ln(W + 3/4\, y^2)} \cdot \tau_3(y, W). \quad (6.3.18)$$

It is possible to show that the kernel in 2D STM-equation is limited (in contrast to the 3D case). For large $y$ the integral of the kernel behaves as $\dfrac{1}{\ln y} \ll 1$.

Hence (in spite of the fact that the integral in (6.3.18) again (as in a 3D case) is governed by large $y$) there are no solutions for which $x \sim pa$ and $W$ are much larger than 1. Correspondingly deep three-particle levels are absent in the 2D case and the system does not go far away from the resonance when we add the third particle to the two resonantly interacting particles.

Numerical solutions of (6.3.18) are obtained by finding the eigenvalues $\lambda(E)$ of the kernel $\hat{K}$ of the homogeneous part of the integral equation: $T_3 = \hat{K} T_3$. Than $\lambda = 1$ is the condition for the appearance of a three-particle bound state. More precisely: $\lambda(E = E_3) = 1$. The numerical solutions for binding energies of three identical bosons in 2D are presented on Fig. 6.7. We can see that $\lambda(E/|E_b|)$ crosses the horizontal line $\lambda = 1$ for two three-particle levels at $\left|E_3^{(1)}\right| = 1.27\left|E_b\right|$ and $\left|E_3^{(2)}\right| = 16.52\left|E_b\right|$. The third level on Fig. 6.7 for all energies corresponds to $\lambda(E) < 1$, and hence does not represent a bound state. Thus we obtained two s-wave three-particle bound states $\left|E_3^{(1)}\right| \approx 1.27\left|E_b\right|$ and $\left|E_3^{(2)}\right| \approx 16.52\left|E_b\right|$ in agreement with the results of Bruch and Tjon [6.57, 6.3].

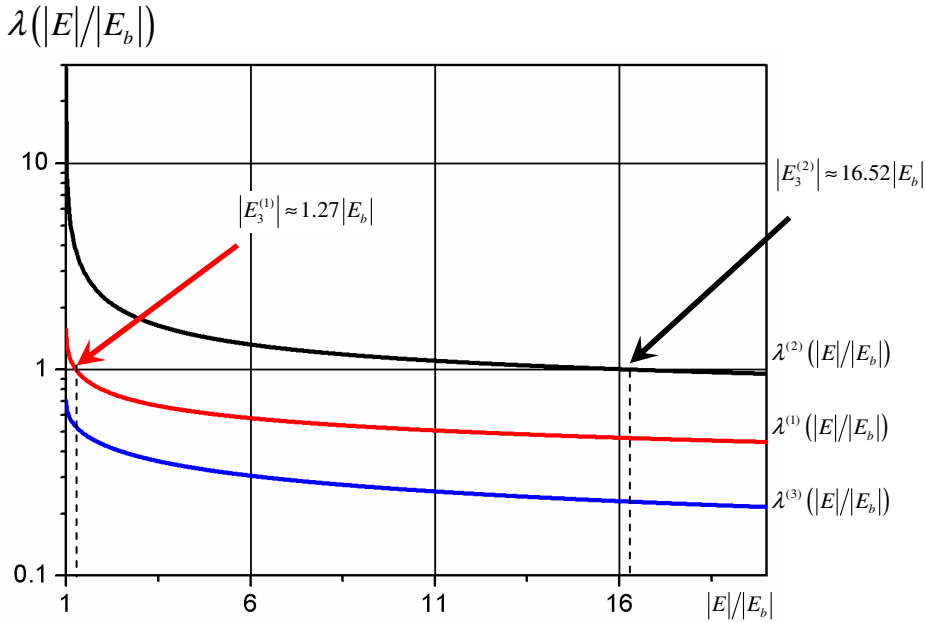

Fig. 6.7. The typical level structure $\lambda\left(\left|E\right|/\left|E_b\right|\right)$ for numerical solution of a homogeneous part of STM-equation. In the case of three identical bosons in 2D two levels $\lambda^{(1)}\left(\left|E\right|/\left|E_b\right|\right)$ and $\lambda^{(2)}\left(\left|E\right|/\left|E_b\right|\right)$ cross the horizontal line $\lambda = 1$ correspondingly for three-particle binding energies $\left|E_3^{(1)}\right| = 1.27\left|E_b\right|$ and $\left|E_3^{(2)}\right| = 16.52\left|E_b\right|$, while for the third level $\lambda^{(3)}\left(\left|E\right|/\left|E_b\right|\right) < 1$ for all energies $\left|E\right|/\left|E_b\right|$ and hence it does not represent the three-particle bound state.

### 6.3.4. The three-particle complex $f_\sigma b,b$ in 2D case.

Let us now consider a complex $f_\sigma b,b$ consisting of one fermion and two bosons. As noted above we consider bosons and fermions with equal masses $m_B = m_F = m$. In agreement with the results of Section 6.2 we assume that an attractive fermion-boson interaction $U_{FB}$, characterized by the radius of the interaction $r_{FB}$ (for the lattice models $r_{FB} = d$), yields a resonant two-body bound state with an energy $E = -\left|E_b\right|$, where $\left|E_b\right| \ll 1/mr_{FB}^2$ is a shallow bound state (note that for the lattice models the requirement $\left|E_b\right| \ll 1/mr_{FB}^2$ is satisfied in the intermediate coupling case for 2D Fermi-Bose mixture with attraction between fermions and bosons considered in Section 6.2). At the same time a boson-boson interaction $U_{BB}$, characterized by the interaction radius $r_{BB}$, does not yield a resonance. This condition is also fulfilled in our model for Fermi-Bose mixture, since both $U_{FF} > 0$ and $U_{FB} > 0$ correspond to repulsion. Hence, if we are interested in the low-energy physics, the only relevant interaction is $U_{FB}$, and we can ignore the boson-boson interaction $U_{BB}$ (the latter will give only small corrections of the order $\left|E_b\right|mr_{FB}^2 \sim \left|E_b\right|md^2 \ll 1$ at low energies). In order to determine three-particle bound states one has to find the poles in the composed fermion $f_\sigma b$ - elementary boson $b$ scattering vertex $T_3$. Since we neglect the boson-boson interaction $U_{BB}$ the vertex $T_3$ is described by the same diagrammatic equation (see Fig. 6.5 and (6.3.17)) as in the problems of three resonantly interacting bosons. The analytical form of this equation also coincides with (6.3.17) with the minor correction that the resonance scattering vertex $T_2$ now corresponds to the interaction ($U_{FB}$) between a boson and a fermion, and therefore we should put $\alpha = 1$ in equation (6.3.17) for $T_2$. Solving the STM-equation for $T_3$ we find that the $f_\sigma b, b$ complex has only s-wave bound state with energy $|E_3| \approx 2.39 |E_b|$ [6.25, 6.26]. The same result holds for the complex $b_1 b_2 b_1$ with two bosons of one sort and one boson of different sort. Note that a complex $bff$ - consisting of a boson and two spinless identical fermions (or a complex with a boson and spin "up" and spin "down" fermion $bf\uparrow f\downarrow$) with resonance interaction $U_{FB}$ does not have any three-particle bound states in the 2D case.

### 6.3.5. Dimer-dimer scattering for four resonantly interacting fermions in 3D. Exact integral equation for four-fermion problem.

Now we can proceed to the problem of dimer-dimer scattering for two molecules $f\uparrow f\downarrow$ in 3D case. This problem was firstly solved by Petrov et al. [6.8] by studying the Schrödinger equation for a 4-fermion wave function.

Inspired by the work of Petrov et al. [6.8] we are looking for a special vertex, which describes an interaction of two fermions constituting a first dimer with a second dimer (considered as a single object). An obvious candidate for this vertex would be the sum of all the diagrams with two fermionic and one dimer incoming line. It would be natural to suppose that these diagrams should have the same set of outgoing lines - two fermionic and one dimer. However, in this case there will be a whole set of disconnected diagrams contributing to our sum that describes the interaction of a dimer with only one fermion. As was pointed out by Weinberg [6.10], one can construct a good integral equation of Lippmann-Schwinger type only for connected class of diagrams. Thus we are forced to give attention to the asymmetric vertex $\Phi_{\alpha\beta}(q_1, q_2; p_2, P)$ corresponding to the sum of all diagrams with one incoming dimer, two incoming fermionic lines and two outgoing dimer lines (see Fig. 6.8). This vertex $\Phi_{\alpha\beta}(q_1, q_2; p_2, P)$ is rather straightforwardly related to the standard dimer-dimer scattering vertex $T_4(p_1, p_2; P)$:

$$T_4(p_1, p_2, P) = \frac{i}{2} \sum_{k,\alpha,\beta} \chi(\alpha, \beta) G(P + p_1 - k) G(k) \Phi_{\alpha\beta}(P + p_1 - k, k; p_2, P). \quad (6.3.19)$$

Note that, by definition, in any order of interaction $\Phi$ contains only connected diagrams.

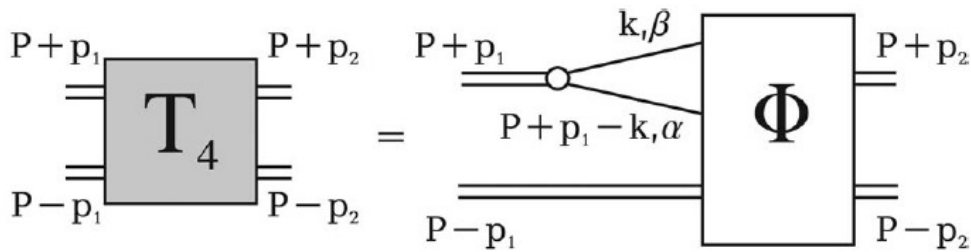

Fig. 6.8. The vertex $\Phi$ represents the full dimer-dimer scattering matrix $T_4$ with one dimer line being cut.

The spin part of the vertex $\Phi_{\alpha\beta}$ has the simple form $\Phi_{\alpha\beta}(q_1, q_2; P, p_2) = \chi(\alpha, \beta) \Phi(q_1, q_2; P, p_2)$. A diagrammatic representation of the equation on $\Phi$ is given in Fig. 6.9. One can assign some physical meaning to the processes described by these diagrams. The diagram of Fig. 6.9a represents the simplest exchange process in a dimer-dimer interaction. The diagram of Fig. 6.9b accounts for a more complicated nature of a "bare" (irreducible by two dimer lines) dimer-dimer interaction.

Finally the diagram of Fig. 6.9c allows for a multiple dimer-dimer scattering via a bare interaction. The last term in Fig. 6.9 means that we should add another three diagrams analogous to Figs. 6.9a, 6.9b and 6.9c, but with the two incoming fermions ($q_1$ and $q_2$) exchanged. The analytical equation for the vertex $\Phi$ can be written as:

$$\Phi(q_1, q_2; p_2, P) = -G(P - q_1 + p_2)G(P - q_2 - p_2) -$$

$$-i\sum_k G(k)G(2P - q_1 - q_2 - k)T_2(2P - q_1 - k)\Phi(q_1, k; p_2, P) +$$

$$+\frac{1}{2}\sum_{Q,k} G(Q - k_1)G(2P - Q - q_2)T_2(2P - Q)T_2(Q)G(k)G(Q - k)\Phi(k, Q - k; p_2, P) + (q_1 \leftrightarrow q_2)$$

$$(6.3.20)$$

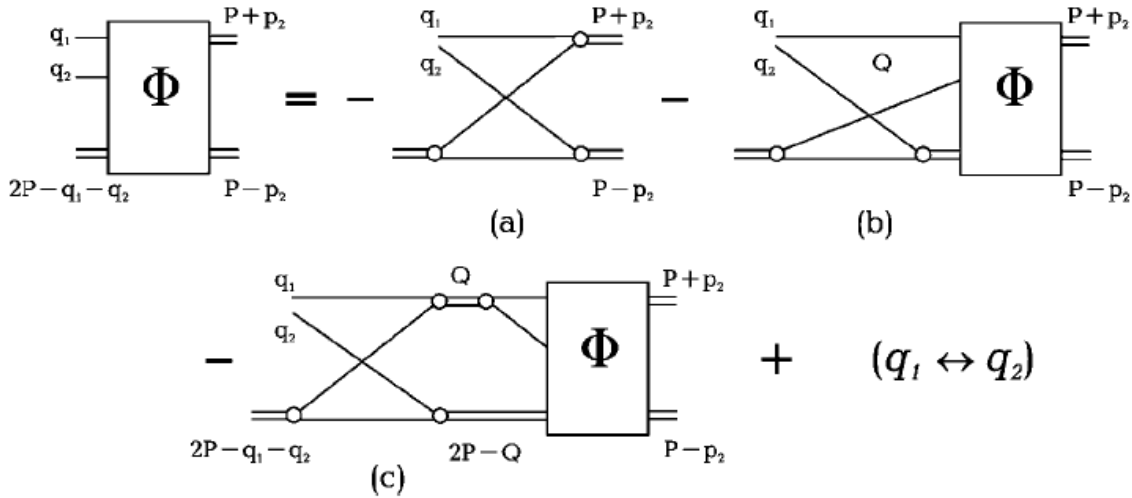

Fig. 6.9. The graphic representation of the equation on function $\Phi$ describing dimer-dimer scattering.

Since we are looking for an s-wave scattering length we can put $p_2 = 0$ and $P = \{0, -|E_b|\}$. At this point we have a single closed equation for the vertex $\Phi$ in momentum representation, which, we believe, is analogous to the equation of Petrov et al. in coordinate representation. To make this analogy more prominent we have to exclude frequencies from the equation (6.3.19). This exclusion requires some more mathematical efforts, but we succeeded in doing that in our second (more extended) article on this subject (see Appendix A in Ref. [6.25]).

The dimer-dimer scattering length is proportional to the full symmetrized vertex $T_4(p_1, p_2; P)$:

$$\left(\frac{8\pi}{m^2 a}\right)^2 T_4(0, 0, -2|E_b|, 0) = \frac{2\pi(2a_{2-2})}{m}. \qquad (6.3.21)$$

In the Born approximation we can consider only the contribution of the first term (Fig. 6.9a) in $\Phi$. Then in this approximation (for the simplest exchange process in dimer-dimer interaction) $\Phi \sim GG$ and $T_4 \sim \Sigma GG\Phi \sim \Sigma GGGG$ (see Fig. 6.10) where the symbol $\Sigma$ stands for the sum in these estimates.

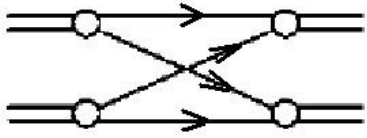

Fig. 6.10. Simplest exchange process which defines the Born approximation for the dimer-dimer T-matrix $T_4$.

In the Born approximation $a_{2-2}(f{\uparrow}f{\downarrow},f{\uparrow}f{\downarrow}.) = 2|a|$. This result in is intuitively transparent since only the interaction between fermions with different spin projections has a resonance character according to Pauli principle. We will get it in the mean-field description of BCS-BEC crossover and the spectrum of collective excitations in Chapter 7.

If one skips the second term in (6.2.50), i.e., one omits the diagram in Fig. 6.9b, one arrives at the ladder approximation of Pieri and Strinati [6.1]. In this approximation we take into account the multiple scattering of two dimmers on each other without the loss of their identity. In the ladder approximation we get $a_{2-2} = 0.78|a|$ in agreement with the results of Pieri and Strinati. The exact equation (6.3.20) corresponds to the summation of all diagrams. Moreover the diagrams on the Fig. 6.9b (omitted in [6.1]) with an account of crossing ($q_1 \leftrightarrow q_2$) describe two types of processes:

1) there are two incoming dimmers in the beginning, than they are virtually decaying and exchange one particle with each other;

2) there are two incoming dimmers in the beginning, than they are virtually decaying forming a virtual 3-particle complex and one elementary fermion. After that 2 dimers are again created.

Note that mathematical structure of the second term in (6.3.20) $\Sigma GGT_2\Phi$ resembles to some extent a mathematical structure of the integral term in STM equation for the trios $\Sigma GGT_2T_3$. Thus it takes into account dynamics and moreover allows to write a closed equation on $\Phi$ in (6.3.20) only in terms of two-particle T-matrix $T_2$ and one-particle Green functions $G$ (without explicit utilization of the three-particle T-matrix $T_3$). Exact solution of (6.3.20) with an account of all these terms (Figs. 6.9a. 6.9b and 6.9c) together with crossing yields:

$$a_{2-2} = 0.6|a| \qquad (6.3.22)$$

in agreement with Petrov et al.

Note also that our approach allows one to find the dimer-dimer scattering amplitude in the 2D case also. Here

$$f_{2-2}(\varepsilon) = \frac{1}{\ln(1.6|E_b|/\varepsilon)} \qquad (6.3.23)$$

in agreement with Petrov, Baranov and Shlyapnikov [6.24]. We will use this result to find coefficient $\alpha$ in formula (6.2.10) for $T_C$ of the bose condensed quartets $f{\uparrow}b,f{\downarrow}b$ in the 2D Fermi-Bose mixture with attractive interaction between fermions and bosons. This formula of Fisher-Hohenberg type $T_C \sim \dfrac{T_0}{16\ln\ln\left(\alpha\dfrac{|E_4|}{T_0}\right)}$ contains in denominator the coefficient $\alpha$ which describes the scattering of quartet on quartet (in the repulse 2D Bose-gas of quartets $f{\uparrow}b,f{\downarrow}b$). Thus it is easy to show that for equal masses of fermions and bosons $m_B = m_F$ $\alpha = 4 \cdot 1.6 = 6.4$, where 1.6 is a coefficient under logarithm for scattering of dimer on dimer in 2D in (6.3.23). The binding energy $|E_4|$ for the quartet $f{\uparrow}b,f{\downarrow}b$ we will define in the next Subsection.

Finally we would like to mention that our results allow one to find the dressed fermionic Green's function, chemical potential, and sound velocity as a function of two-particle scattering length $a$ in the case of the dilute superfluid Bose-gas of weakly interacting dimers $f{\uparrow}f{\downarrow}$ at low temperatures. The problem of dilute superfluid Bose-gas of di-fermionic molecules was solved by Popov [6.79] and later deeply investigated by Keldysh and Kozlov in connection with the problem of excitonic (or bi excitonic) superfluidity [6.78]. Those authors managed to reduce the gas problem to a dimer-dimer scattering problem in vacuum, but were unable to express the dimer-dimer scattering amplitude in a single two-fermion parameter. A direct combination of our results with those ones of Popov, Keldysh and Kozlov allows one to get all the thermodynamic values of a dilute superfluid resonance gas of composed bosons. This strategy will be fulfilled in Chapter 7.

### 6.3.6. Four particles bound states.

In the 3D case a homogeneous part of the four-particle integral equation (6.3.20) has a nontrivial solution for any negative values of the energy for the 4-particle complexes $bbbb$, $f_\uparrow b f_\downarrow b$ and $fbbb$. Thus Efimov effect manifests itself again in a four-particle problem. However for the same complexes in 2D Efimov effect is absent and the homogeneous part of the four-particles equation has the finite number of solutions for the negative energy $E < 0$, which can be represented again only in terms of the two-particle binding energy $|E_b|$.

First we will consider four identical resonantly interacting bosons $bbbb$. Any two bosons would form a stable dimer with binding energy $E = - |E_b|$. We are going to find a four-particle binding energy of an s-wave bound state of two dimers. Generally speaking (see Section 6.1 for example), a bound state could emerge in channels with larger orbital moments; however, this question will be considered in the next Chapters. To find a binding energy we should examine the analytical structure of the dimer-dimer scattering vertex $T_4$ and find its poles. The set of equations for $T_4$ has the same diagrammatic structure as those shown in Figs. 6.8 and 6.9. The analytical expression for the first equation can be written as:

$$T_4(p_1, p_2, P) = \frac{i}{\alpha} \sum_k G(P + p_1 - k) G(k) \Phi_{\alpha\beta}(P + p_1 - k, k; p_2, P), \quad (6.3.24)$$

and the equation for the vertex $\Phi$:

$$\Phi(q_1, q_2; p_2, P) = G(P - q_1 + p_2) G(P - q_2 - p_2) +$$
$$+ i \sum_k G(k) G(2P - q_1 - q_2 - k) T_2(2P - q_1 - k) \Phi(q_1, k; p_2, P) -$$
$$- \frac{1}{2\alpha} \sum_{Q,k} G(Q - q_1) G(2P - Q - q_2) T_2(2P - Q) T_2(Q) G(k) G(Q - k) \Phi(k, Q - k; p_2, P) +$$
$$+ (q_1 \leftrightarrow q_2),$$

(6.3.25)

where two-particle T-matrix $T_2$ in 2D should be taken from (6.3.16) and one should put $\alpha = 2$ for the case of identical resonantly interacting bosons. Solving the above equations for the poles of $T_4$ as a function of the variable $P = \{0, E\}$, we found two bound states for the $bbbb$ complex (see Table 6.1). Certainly, for the validity of our approximation we should have $|E_4| \ll 1/mr_0^2$. For the case of four bosons $bbbb$ it means that $197|E_b| \ll 1/mr_0^2$ for the deepest level and hence $a/r_0 \gg \sqrt{197}$. This case can still be realized in the Feshbach resonance scheme.

Table 6.1. Bound states of three and four resonantly interacting particles in 2D [6.25, 6.26].

| System | Relative[a] interaction | Number of bound states | Energy (in $\|E_B\|$)[b] | $\alpha$[c] |
|--------|--------|--------|--------|--------|
| $bbb$ | $U_{bb}$ | 2 | 1.27, 16.52 | 2 |
| $fbb$ | $U_{fb}$ | 1 | 2.39 | 1 |
| $fbbb$ | $U_{fb}$ | 1 | 4.10 | 1 |
| $bf_\uparrow bf_\downarrow$ | $U_{fb}$ | 2 | 2.84, 10.64 | 1 |
| $bbbb$ | $U_{bb}$ | 2 | 22, 197 | 2 |

[a] Interaction that yields resonance scattering. All other interactions are negligible.
[b] $m = m_B = m_F$.
[c] The indistinguishability parameter in (6.3.16).

The case of a four-particles complex $f_\uparrow b f_\downarrow b$, consisting of resonantly interacting fermions and bosons is described by the same equations (6.3.24), (6.3.25), but with the parameter $\alpha = 1$. In this case we found two bound states $|E_4| \approx 2.84 |E_b|$ and $|E_4| \approx 10.64 |E_4|$. They are also listed in Table 6.1. The same result is valid for two pairs of bosons of different sort $b_1 b_2 b_1 b_2$.

In order to obtain the bound states of the $fbbb$ complex one has to find the energies $P = \{0, E\}$ corresponding to nontrivial solutions of the following homogeneous equation:

$$\Phi(q_1, q_2; p_2, P) = i \sum_k G(k) G(2P - q_1 - q_2 - k) T_2(2P - q_1 - k) \Phi(q_1, k; p_2, P) + (q_1 \leftrightarrow q_2), \quad (6.3.26)$$

This equation corresponds to the diagrams of Fig. 6.9b with an account of crossing (and does not contain the diagrams of Fig. 6.9c). We found a bound state for the $fbbb$ complex with an energy $|E_4| \approx 4.10 |E_b|$.

We summarize the results concerning binding energies of three and four resonantly interacting particles in 2D in the Table 6.1. Note that all our calculations corresponds to the case of particles with equal masses $m = m_B = m_F$, though they can be easily generalized to the case of different masses.

### 6.3.7. Phase diagram of the Fermi-Bose mixture in 2D.

In Subsection (6.2.5) we wrote the chain of the Saha crossover temperatures $T_*^{(2)} = \dfrac{|E_b|}{\ln(|E_b|/T_0)}$, $T_*^{(3)} = \dfrac{|E_3|}{2\ln(|E_3|/T_0)}$, and $T_*^{(4)} = \dfrac{|E_4|}{3\ln(|E_4|/T_0)}$ for two-particle $(f_\sigma b)$, three-particle $(f_\sigma b, b)$, and four-particle $(f_\uparrow b, f_\downarrow b)$ complexes in the 2D Fermi-Bose mixture with attractive (and resonance in the intermediate coupling case) interaction between fermions and bosons. The knowledge of the binding energies of three-particle $(|E_3|)$ and four-particle $(|E_4|)$ complexes allows us to complete the phase diagram of a Fermi-Bose mixture. Namely the deepest levels for $m = m_B = m_F$ correspond to $|E_3| \approx 2.4 |E_b|$ and $|E_4| \approx 10.6 |E_b|$ for $fbb$ and $f_\uparrow b, f_\downarrow b$ complexes, respectively. Thus for equal densities of fermions and bosons $(n_B = n_F = n = n_{tot}/2)$ and, hence, for equal degeneracy temperatures $T_{0B} = T_{0F} = T_0 = 2\pi n/m$ we have the following hierarchy of Saha temperatures:

$$T_*^{(4)} > \{T_*^{(3)}; T_*^{(2)}\}. \quad (6.3.27)$$

Then, as we already assumed in Subsection 6.3.5, the phase diagram becomes rather simple. Namely: for $T > T_*^{(4)}$ elementary fermions and bosons prevail in the system, while for $T < T_*^{(4)}$ – the quartets $f_\uparrow b, f_\downarrow b$ prevail in the system in case of equal densities $n_B = n_F$. The quartets are bose-condensed below the temperature: $T_C = T_0/(16 \ln(6.4 |E_4|/T_0))$, where $|E_4| \approx 10.64 |E_b|$.

### 6.3.8. Phase diagram of 2D Bose-gas.

In the 2D Bose-gas with resonance interaction between bosons or in strongly disbalanced Fermi-Bose mixture for $n_B \gg n_F$ (see [6.5, 6.6]) the formation of large droplets containing $N > 4$ particles is possible. The limitation on the number of particles in the droplet is connected with the repulsive hard-core with the radius $r_0$:

$$R_N > r_0, \quad |E_N| < \frac{1}{mr_0^2}, \qquad (6.3.28)$$

where $R_N$ and $E_N$ are the radius and the binding energy of $N$-particle droplet.

For $N > 4$ the exact solution of the STM-equations is practically impossible (it requires too much of the computer time). Thus we have to restrict ourselves with the variational calculations of $R_N$ and $E_N$ (see Hummer, Son [6.4]). For 2D case they yield $R_N \sim ae^{-N}$ and $E_N = 1/mR_N^2 \sim |E_b|e^{2N}$ under the condition: $N < N_{max} \sim 0.9 \ln a/r_0$ ([6.4, 6.40]). The large droplets were experimentally observed in the disbalanced Fermi-Bose mixture of $^{87}$Rb (bosons) and $^{40}$K (fermions) for $n_B > n_F$ by Modugno et al. [6.6].

### 6.3.9. The role of the dimer-fermion and dimer-dimer scattering lengths for the lifetime of the resonance Fermi-gas.

In case of the resonance Fermi-gas in 3D the dimer-fermion scattering length $a_{2-1} = 1.18 \, |a| > 0$ and the dimer-dimer scattering length $a_{2-2} = 0.6 \, |a| > 0$ define the relaxation rates of inelastic dimer-fermion and dimer-dimer collisions in the BEC-domain for $a > 0$.
Namely according to Petrov, Salomon and Shlyapnikov [6.8] the relaxation rates read:

$$\alpha_{2-1} = C_1 \frac{\hbar r_0}{m}\left(\frac{r_0}{a}\right)^{3,33},$$

$$\alpha_{2-2} = C_2 \frac{\hbar r_0}{m}\left(\frac{r_0}{a}\right)^{2,55} \qquad (6.3.28)$$

for dimer-fermion ($\alpha_{2-1}$) and dimer-dimer ($\alpha_{2-2}$) inelastic scatterings. In (6.3.28) $C_1$ and $C_2$ are numerical coefficients.

Correspondingly the inverse inelastic scattering times $1/\tau_{2-1} \sim n_{atoms}\alpha_{2-1}$ and $1/\tau_{2-2} \sim n_{atoms}\alpha_{2-2}$ define the molecular transitions from the shallow to deep vibrational levels in the potential well. In course of this process, according to the discussion in Chapter 5 the molecules lose their identity and do not participate in the symmetrization of the dimer wave function. Thus practically $\frac{1}{\tau_{loss}} \sim \frac{1}{\tau_{2-1}} + \frac{1}{\tau_{2-2}}$, where $\frac{dn_{mol}}{dt} \sim \alpha_{2-1}n_{mol}n_{atom} + \alpha_{2-2}n_{mol}^2 \sim \frac{n_{mol}}{\tau_{loss}}$ and the inverse time of the losses of the resonance dimers $1/\tau_{loss}$ coincides by order of magnitude with the inverse lifetime $1/\tau_{lifetime}$ of the molecular BEC-condensate.

In the regime of Feshbach resonance for $B \to B_0$: $1/a \to 0$ and the lifetime of Bose-condensate strongly increases. Note that if the numbers of atoms and molecules coincide by the order of magnitude (for $n_{atom} \sim n_{mol}$) the inverse inelastic scattering time $\frac{1}{\tau_{2-1}} \sim \left(\frac{r_0}{a}\right)^{3,33}\frac{T_0}{\hbar}(r_0 p_F)$ (where $T_0$ is degeneracy temperature) while $\frac{1}{\tau_{2-2}} \sim \left(\frac{r_0}{a}\right)^{2,55}\frac{T_0}{\hbar}(r_0 p_F)$ and hence for $1/a \to 0$ the inverse loss time $\frac{1}{\tau_{loss}} \approx \frac{1}{\tau_{2-2}}$ is defined by dimer-dimer inelastic scatterings mostly (see [6.8]). Note that usually for the resonance Fermi-gas in BEC-regime $\tau_{lifetime} \leq 10$ sec.

Concluding this Section let us emphasize that in the resonance approximation we derive and solve exactly the integral equations for trios and quartets in 3D and 2D.

We evaluate exact scattering amplitudes of molecule on atom ($a_{2-1}$) and molecule on molecule in the 3D and the 2D resonance Fermi-gas.

We calculate the binding energies of all the possible complexes, consisting of three *bbb*, $b_1b_2b_1$, *fbb*, and four *bbbb*, $b_1b_2b_1b_2$, *fbbb*, and $f_\uparrow bf_\downarrow b$ particles as the functions only of the two-particle binding energy $|E_b|$ in the 2D case.

We construct the phase-diagram for the resonance Fermi-Bose mixture in 2D with equal densities of fermions and bosons $n_B = n_F$. We discuss also the possibility of the formation of the large droplets containing 5 and more particles in the 2D Bose-gas and in the disbalanced Fermi-Bose mixture for $n_B > n_F$.

Note that a binding energy of a four-particle complex $b_1b_2b_1b_2$ is important to complete the phase diagram of the two-band Bose-Hubbard model with repulsion between bosons of one sort and attraction between bosons of different sorts, considered in Section 6.1. In the intermediate coupling case $T_0 < |E_b| < W$ in 2D this model as we already mentioned in Section 6.1 corresponds to the resonance interaction between bosons of different sort. Moreover for $|E_4| < W$ the quartets $b_1b_2b_1b_2$ are extended $1/\sqrt{n} > a >> d$ (not sitting on one site of the lattice) and thus even strong Hubbard repulsions between bosons of the same sort $U_{11} > 0$ and $U_{22} > 0$ cannot prevent the formation of the quartets.

Reference list to Chapter 6.

Chapter 7 BCS-BEC crossover and the spectrum of collective excitations in s-wave and p-wave resonance superfluids.

7.1 Phase-diagram of the resonance Fermi-gas in 3D and 2D.
7.1.1 Self-consistent T-matrix approximation.
7.1.2 Equation for $T_C$. Dilute BCS and BEC limits. Cooper pairs and local pairs.
7.1.3 Saha temperature in dilute BEC limit
7.1.4 Unitary limit of the BCS-BEC crossover. Fermi-Bose mixture in strong coupling case $|a|p_F \geq 1$
7.2 Self-consistent Leggett theory for T = 0 in 3D and 2D case
7.2.1 Chemical potential and superfluid gap
7.2.2 Sound velocity in BCS and BEC limits
7.2.3 BCS-BEC crossover and Leggett theory in 2D case
7.2.4 Experimental measurement of the superfluid gap (infrared spectroscopy)
7.3 Anderson-Bogolubov theory for collective excitations
7.3.1 Diagrammatic approach
7.3.2 The spectrum of collective excitations
7.3.3 Landau critical velocities
7.4 Feshbach resonance and phase-diagram for p-wave superfluid Fermi-gas
7.4.1 Quasiparticle spectrum and nodal points in p-wave superfluid with a symmetry of A1-phase
7.4.2 Leggett equations for p-wave superfluid. Sound velocity in BCS and BEC limits
7.4.3 Specific heat and normal density in 3D A1-phase
7.4.4 Specific heat and normal density in 2D case
7.4.5 Indications of quantum phase-transition. Quantum critical point
7.4.6 The spectrum of orbital waves in three-dimensional p-wave superfluid with a symmetry of A1-phase

In this Chapter, using the knowledge of dimer-dimer scattering length $a_{2\text{-}2} = 0.6|a| > 0$, we complete the phase-diagram of the resonance Fermi-gas in 3D. We present more detaily the scheme of BCS-BEC crossover developed by Nozieres and Schmitt-Rink [4.12] and construct the phase-diagram of the resonance Fermi-gas in the self-consistent T-matrix approximation [4.14]. We define the crossover line $\mu(T) = 0$ which effectively separates BEC-domain (for which $\mu < 0$ and $a > 0$) from the BCS-domain (for which vise versa $\mu > 0$ and $a < 0$). We discuss the critical temperatures of extended Cooper pairing $T_C^{\text{BCS}}$ and bose-condensation for local pairing of two fermions $T_C^{\text{BEC}}$ in dilute BCS and BEC limits $|a| p_F \ll 1$, where $p_F$ is the Fermi-momentum and $a$ is an s-wave two-particle scattering length. We also provide qualitative considerations for the phase-diagram in the unitary limit $1/a \to 0$ and close to unitarity (for large values of the gas parameter $|a| p_F \gg 1$). We observe that in the substance a unitary limit is effectively inside the BCS-domain. We extend our results at finite temperatures on 2D resonance Fermi-gas.

In the second part of this Chapter we provide a brief description of the self-consistent Leggett scheme [4.13] for the BCS-BEC crossover at zero temperature ($T = 0$). We solve the Leggett equations and find the behavior of the superfluid gap $\Delta$ and chemical potential $\mu$ as the functions of the gas parameter $a p_F$ in the dilute BCS and BEC limits, as well as close to unitarity [5.16]. The knowledge of the gap $\Delta$ and chemical potential allows us to solve Bogolubov-Anderson equations [6.1, 6.2] and determine the spectrum of collective excitations both in BCS and BEC domains. For small $\omega$ and $q$ the spectrum is linear and we can find the behavior of the sound velocity (and Landau critical hydrodynamic velocity [6.17]) as a function of $a p_F$. Note that the sound velocity can be defined already from the static Leggett equations (from the knowledge of compressibility or the chemical potential of the system).

In the third part of this Chapter we will extend our results on BCS-BEC crossover on p-wave resonance Fermi-gas [6.3]. We present experimental results on p-wave Feshbach resonance [6.4-6.6] in BEC-domain for 100%-polarized Fermi-gas where the triplet molecules $^{40}K_2$ and $^6Li_2$ with $S_{\text{tot}} = S_{\text{tot}}^z = l = 1$ for total spin of the pair $S_{\text{tot}}$ and relative orbital momentum of the pair $l$ are created. The p-wave superfluid has a symmetry of triplet A1-phase here.

We will construct the phase-diagram of the BCS-BEC crossover in p-wave resonance superfluid and define again a crossover line $\mu(T) = 0$ which separates BEC-domain of the A1-phase from the BCS-domain [6.3]. We will also solve the self-consistent Leggett equations at $T = 0$ and find the behavior of the superfluid gap and sound velocity. For low temperatures $T \ll T_C$ we will determine also the temperature behavior of the specific heat $C_v(T)$ and normal density $\rho_n(T)$ both in 3D and 2D triplet superfluid Fermi-gas. We will show that globally phase-diagram of the BCS-BEC crossover for p-wave resonance superfluid resembles that of s-wave resonance superfluid (considered in Section 6.1). However, in p-wave case there is a special point on the phase-diagram, namely $\mu(T = 0) = 0$ which corresponds to quantum phase-transition [6.7, 6.8]. Close to this point the temperature behavior of the normal density $\rho_n(T)$ and specific heat $C_v(T)$ is different in classical and quantum limits $|\mu| \to 0$, $T \to 0$ but either $\dfrac{|\mu|}{T} \to 0$ or vice versa $\dfrac{T}{|\mu|} \to 0$ [6.9]. Note that the point $\mu(T = 0) = 0$ separates in 3D the BCS state with the nodes in the quasiparticle energy (and zeroes in the superconductive gap for triplet A1-phase [6.9, 6.10]) form the gapped BEC-domain (where the quasiparticle energy has no zeroes).

### 7.1. Phase-diagram of the resonance Fermi-gas in 3D and 2D cases.

In Chapter 5 we presented a basic knowledge connected with the Bethe-Salpeter integral equation [5.61], which describes the temperature evolution of the 3D resonance Fermi-gas. We also considered the dilute molecular (BEC) limit at high temperatures $T > T_C^{\text{BEC}}$. In Chapter 6, from the solution of Skorniakov-ter-Martirosian equation, we define mean-field ($a_{2-2} = 2|a|$) and exact ($a_{2-2} = 0.6|a|$) dimer-dimer (or molecule-molecule) scattering length which describes a weak repulsion between the composed bosons $f↑f↓$ in the dilute BEC limit and for the temperatures $T < T_*$ (where $T_*$ is the crossover Saha temperature which corresponds to the formation of local pairs or molecules $f↑f↓$ for positive scattering length $a>0$). In this Section we will complete the phase-diagram of the resonance Fermi-gas in 3D and in 2D case.

### 7.1.1. Self-consistent T-matrix approximation.

In Chapter 5 we wrote two equations for the T-matrix in attractive 3D Fermi-gas. The first one is for the T-matrix (or scattering length) in vacuum (see Eq. (5.2.6) and Fig. 5.14):

$$T_{vac}(\omega = 0, q = 0) = \frac{4\pi a}{m} = -\frac{|U_0|}{1 - |U_0|K_{vac}(0,0)}, \qquad (7.1.1)$$

where

$$K_{vac}(0,0) = \int \frac{d^3\vec{p}}{(2\pi)^3} \frac{d\Omega}{(2\pi)} G_{vac}(\Omega, \vec{p})G_{vac}(-\Omega, -\vec{p}) = \int \frac{d^3\vec{p}}{2\varepsilon_p} \qquad (7.1.2)$$

is a Cooper loop in vacuum (a product of two vacuum Green-functions $G_{vac}(\Omega, \vec{p}) = \dfrac{1}{\Omega - \dfrac{p^2}{2m} + io}$

in the Cooper channel for zero total frequency $\omega = 0$ and total momentum $\vec{q} = 0$) and $U_0 = -|U_0|$ is Fourier-harmonic of the attractive two-particle potential, $U(q)$ for $q = 0$.

The second one (see Eq. (5.2.13)) is for the T-matrix in substance:

$$T(i\omega_n, \vec{q}) = -\frac{|U_0|}{1 - |U_0|K(i\omega_n, \vec{q})}, \qquad (7.1.3)$$

where $K(i\omega_n, \vec{q})$ is a Cooper loop in substance for total Matsubara frequency of two (incoming in the Cooper channel) particles $\omega_h$ and total momentum $\vec{q}$. For fermions $\omega_h = \pi T(2n + 1)$ – we also emphasized in Chapter 5 that in the T-matrix approximation (when we neglect the difference between irreducible bare vertex $U_{eff}(q)$ and the Fourier-harmonic of the bare vacuum interaction $U(q)$) the equation (7.1.3) for the T-matrix in substance coincides with the solution of the Bethe-Salpeter integral equation for a total two-particle vertex $\Gamma(\vec{q}, \omega)$ in the Cooper channel.

According to general quantum-mechanical prescriptions [5.19, 5.20], we should make a renormalization procedure [5.23, 5.20, 5.60] and rewrite (7.1.3) only in terms of observables. Effectively we should replace $U_0$ in (7.1.3) by scattering amplitude $a$ from (7.1.1). This yields:

$$-\frac{1}{|U_0|} + K_{vac}(0,0) = \frac{m}{4\pi a} \qquad (7.1.4)$$

and correspondingly [7.53]:

$$T(i\omega_n, \vec{q}) = \frac{1}{-\dfrac{1}{|U_0|} + K(i\omega_n, \vec{q})} = \frac{1}{\dfrac{m}{4\pi a} + K(i\omega_n, \vec{q}) - K_{vac}(0,0)} = \frac{4\pi a/m}{1 + \dfrac{4\pi a}{m}\big(K(i\omega_n, \vec{q}) - K_{vac}(0,0)\big)} \qquad (7.1.5)$$

That is precisely a renormalization procedure which comes in condensed matter physics from quantum electrodynamics [7.11, 7.20]. Note that a Cooper loop $K(i\omega_n, \vec{q})$ in (7.1.5) reads:

$$K(i\omega_n, \vec{q}) = \sum_{\Omega_n} \int \frac{d^3\vec{p}}{(2\pi)^3} G_M(i\Omega_n, \vec{p})G_M(i\Omega_n - i\omega_n, \vec{p} - \vec{q}) \qquad (7.1.6)$$

In graphical form we sum up in the T-matrix approximation the dressed ladder diagrams in particle-particle channel presented on Fig. 7.1 [7.53, 5.17, 7.12 – 7.15].

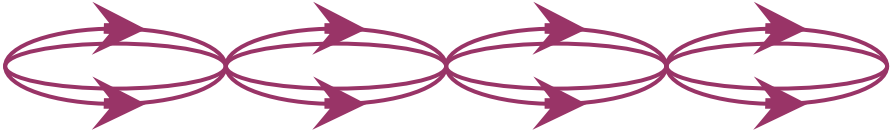

Fig. 7.1. The dressed ladder diagrams in particle-particle channel for the T-matrix given by Eq.'s (7.1.5) (7.1.6).

The Dyson equation [5.20, 5.60] for dressed Matsubara Green's function (which enter in the expression (7.1.6) for a Cooper loop) yields:

$$G_M(i\omega_n, \vec{q}) = \frac{1}{i\omega_n - \varepsilon(q) + \mu - \Sigma_M(i\omega_n, \vec{q})}. \qquad (7.1.7)$$

Finally the Matsubara self-energy $\Sigma_M$ in the self-consistent T-matrix approximation reads (see Fig. 7.2):

$$\Sigma_M(i\omega_n, \vec{q}) = \sum_{\Omega_n} \int \frac{d^3\vec{p}}{(2\pi)^3} G_M(i\Omega_n - i\omega_n, \vec{p} - \vec{q}) T(i\Omega_n, \vec{p}) \qquad (7.1.8)$$

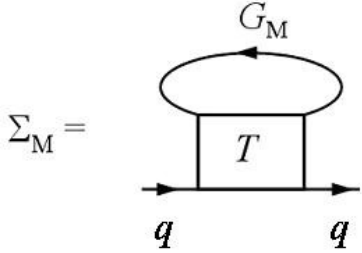

Fig. 7.2. The self-energy $\Sigma_M(i\omega_n, \vec{q})$ in the self-consistent T-matrix approximation. $q$ is a four momentum $\{\omega_n; \vec{q}\}$.

The dressed one-particle Green's function should be normalized on total density for one spin projection $n_{tot}/2$:

$$\sum_{\omega_n} \int \frac{d^3\vec{q}}{(2\pi)^3} G_M(i\omega_n, \vec{q}) = \sum_{\omega_n} \int \frac{d^3\vec{q}}{(2\pi)^3} \frac{1}{G_0^{-1}(i\omega_n, \vec{q}) - \Sigma_M(i\omega_n, \vec{q})} = \frac{n_{tot}}{2} = \frac{p_F^3}{6\pi^2}, \qquad (7.1.9)$$

where

$$G_0(i\omega_n, \vec{q}) = \frac{1}{i\omega_n - \varepsilon(\vec{q}) + \mu} \qquad (7.1.10)$$

is a bare Matsubara Green's function.

The system of equations (7.1.5)-(7.1.9) constitute the self-consistent T-matrix approximation. This approximation is energy and momentum conserving [7.12-7.15, 7.53, 5.19], and is exact in leading order in the gas parameter $ap_F$ for dilute Fermi-systems (see for example Galitskii-Bloom results [5.46, 7.16] for 3D and 2D repulsive Fermi-gas).

We solve the system of equations (7.1.5)-(7.1.9) perturbatively. As a result in the first iteration we should calculate the T-matrix in (7.1.5) with the zeroth order Matsubara Green's functions $G_0(i\omega_n, \vec{q})$ see (7.1.10).

Then the renormalized Cooper loop $K(i\omega_n, \vec{q}) - K_{vac}(0,0)$ (which enters in the denominator of (7.1.5)) reads:

$$K_{ren} = K(i\omega_n, \vec{q}) - K_{vac}(0,0) = \int \frac{d^3\vec{p}}{(2\pi)^3} \left[ -\frac{1 - n_F(\xi_p) - n_F(\xi_{-p+q})}{i\omega_n - \xi_p - \xi_{-p+q}} - \frac{1}{2\varepsilon_p} \right]. \qquad (7.1.11)$$

Correspondingly:

$$T_0(i\omega_n, \vec{q}) = \frac{4\pi a/m}{1 + \dfrac{4\pi a}{m} K_{ren}(i\omega_n, \vec{q})} \,. \qquad (7.1.12)$$

The pole of the T-matrix in (7.1.5) for $\omega_n = \vec{q} = 0$ determines the equation for $T_C$:

$$1 + \frac{4\pi a}{m} K_{ren}(0,0) = 0 \,. \qquad (7.1.13)$$

The self-energy in the first iteration reads:

$$\Sigma_1(i\omega_n, \vec{q}) = \sum_{\Omega_n} \int \frac{d^3\vec{p}}{(2\pi)^3} T_0(i\Omega_n, \vec{p}) G_0(i\Omega_n - i\omega_n, \vec{p} - \vec{q}) \,. \qquad (7.1.14)$$

It enters in the equation of the conservation of the number of particles:

$$\sum_{\omega_n} \int \frac{d^3\vec{q}}{(2\pi)^3} \frac{1}{G_0^{-1}(i\omega_n, \vec{q}) - \Sigma_1(i\omega_n, \vec{q})} = \frac{n_{tot}}{2} = \frac{p_F^3}{6\pi^2} \,. \qquad (7.1.15)$$

### 7.1.2. Equation for $T_C$

The pole of the T-matrix in the 3D case reads:

$$1 + \frac{4\pi a}{m} \int \left[ \frac{th\dfrac{\varepsilon_p - \mu}{2T}}{2(\varepsilon_p - \mu)} - \frac{1}{2\varepsilon_p} \right] \frac{d^3\vec{p}}{(2\pi)^3} \,. \qquad (7.1.16)$$

As we already discussed in Chapter 5, in the BCS-domain for $a < 0$ and $|a|p_F < 1$ (dilute Fermi-gas) the chemical potential $\mu \approx \varepsilon_F > 0$ ($\varepsilon_F = p_F^2/2m$) and the critical temperature: $T_C^{BCS} \sim \varepsilon_F \exp\left\{ -\dfrac{\pi}{2|a|p_F} \right\}$.

In the BEC-domain for $a > 0$ and $ap_F < 1$: the fermionic chemical potential $\mu < 0$ and bosonic chemical potential $\mu_B(T_C) = 2\mu + |E_b| = 0$ ($\mu(T_C) = -|E_b|/2$). A temperature of the Bose-Einstein condensation $T_C^{BEC} \sim 3.31 \dfrac{(n_{tot}/2)^{2/3}}{2m} \approx 0.2\varepsilon_F$ in an ideal Bose-gas of molecules (composed bosons) with the mass $m_B = 2m$ and density $n_B = \dfrac{n_{tot}}{2} = \dfrac{p_F^3}{6\pi^2}$.

### 7.1.3. Self-energy in dilute BEC limit

In the BEC-domain in the dilute case (for $ap_F < 1$) there is one more characteristic temperature (see Chapter 5) namely: $T_* = \dfrac{|E_b|}{3/2\ln\dfrac{|E_b|}{\varepsilon_F}} >> T_C^{BEC}$ - Saha temperature, which describes smooth crossover ($|E_b| = 1/ma^2$ is a molecule binding energy). For $T = T_*$ the number of molecules $(n_B = n_{F\sigma} = n_{tot}/4)$ equals to the number of unpaired fermions with one spin projection $\sigma$. For $T_C < T << T_*$ we have a slightly non-ideal Bose-gas of composed molecules in a normal (non-superfliud) state. The self-energy $\Sigma_1(i\omega_n, \vec{q})$ in the first iteration to the self-consistent T-matrix approximation (7.1.14) reads (see [7.53]):

$$\Sigma_1(i\omega_n, \vec{q}) = \frac{16\pi a |\mu| \dfrac{n_{tot}}{m} \dfrac{2|\mu|}{|E_b|}}{i\omega_n + q^2/2m - \mu + \mu_B} \qquad (7.1.17)$$

Note that $\Sigma_1(i\omega_n, \vec{q})$ has a "holelike" dispersion $i\omega_n + \xi_q + \mu_B$ in contrast to the "particlelike" dispersion $i\omega_n - \xi_q$  ($\xi_q = q^2/2m - \mu$) in the Matsubara Green's function of the zeroth approximation $G_0$ (see 7.1.10). For $T_C < T << T_*$: $n_{\text{tot}} \approx 2n_B$ (number of unpaired fermions is small) and thus:

$$\frac{n_{tot}}{2} = \frac{1}{2\pi^2} \int_0^\infty \frac{k^2 dk}{\exp\left\{ \dfrac{k^2/4m - |E_b| - 2\mu}{T} \right\} - 1}, \qquad (7.1.18)$$

where we took into account that $\mu_B = 2\mu + |E_b|$. The same result can be restored from the form of the dressed Green's function $G^{-1}(i\omega_n, \vec{q}) = G_0^{-1}(i\omega_n, \vec{q}) - \Sigma_1(i\omega_n, \vec{q})$ in the first iteration with $\Sigma_1$ given by Eq. (7.1.17).

We remind that the T-matrix $T(i\omega_n, \vec{q}) = \dfrac{4|\mu| \, 4\pi a/m}{i\omega_n - q^2/4m + \mu_B}$ for small $\omega_n$ and $\vec{q}$. Correspondingly the dressed Green's function can be rewritten as:

$$G(i\omega_n, \vec{q}) \approx \frac{1}{i\omega_n - \xi_q - \dfrac{16\pi a |\mu| n/m}{i\omega_n + \xi_q + \mu_B}} = \frac{1}{i\omega_n - \xi_q - \dfrac{\Delta_{PG}^2}{i\omega_n + \xi_q + \mu_B}}, \qquad (7.1.19)$$

where we put $2|\mu|/|E_b| \approx 1$  ($|\mu| \approx |E_b|/2$ for $T << T_*$) and introduced the pseudogap $\Delta_{PG}^2 = 16\pi a |\mu| n/m$ according to [7.30]. Note that $\Delta_{PG}^2 \sim |E_b| \varepsilon_F a p_F$. Intensive discussion of the pseudogap state we can also find in [7.49, 7.50, 7.51]. We will return to this interesting problem in the next Section.

The spectral function $A(\omega, \vec{q}) = -\dfrac{1}{\pi} \text{Im} G(\omega + io, \vec{q})$ reads:

$$A(\omega, \vec{q}) \approx \left( 1 - \frac{4\pi a |\mu| n/m}{\xi_q^2} \right) \delta(\omega - \xi_q) + \frac{4\pi a |\mu| n/m}{\xi_q^2} \delta(\omega + \xi_q + \mu_B). \qquad (7.1.20)$$

(Effectively in (7.1.20) we made analytical continuation $i\omega_n \to \omega + io$). For $T << T_*$ the form of $A(\omega, \vec{q})$ in (7.1.20) reflects the existence of two bands (see Chapter 8 and [7.53, 5.17]): the filled bosonic band and, separated by large correlation gap $\Delta = |E_b|$ ($\Delta >> \Delta_{PG}$ for dilute BEC regime) the almost empty band of unbound fermions. Integrating the spectral weight it is easy to check that in this regime:

$$\frac{4\pi a |\mu|}{m} \frac{1}{2\pi^2} \int_0^\infty \frac{k^2 dk}{\xi_k^2} = \frac{n_{tot}}{2} \approx n_B, \qquad (7.1.21)$$

where we used that $|\mu| \approx |E_b|/2 = 1/2ma^2$ in $\xi_k = k^2/2m + |\mu|$. Note that the specific heat of the system

$$C_v = \frac{\partial E}{\partial T} = \frac{\partial}{\partial T} \left[ \int \frac{k^2}{4m} \frac{k^2 dk}{2\pi^2} \exp\left( -\frac{k^2}{4mT} \right) \right] \exp\frac{\mu_B}{T} \sim \frac{n_{tot}}{2} = const \qquad (7.1.22)$$

is temperature independent in agreement with general thermodynamic requirements.

### 7.1.4. Phase-diagram of the resonance Fermi-gas in 3D

In the first iteration to the self-consistent T-matrix approximation the numerical calculations yield the following qualitative phase-diagram of the BCS-BEC crossover (see Fig 7.3 and [7.53]).

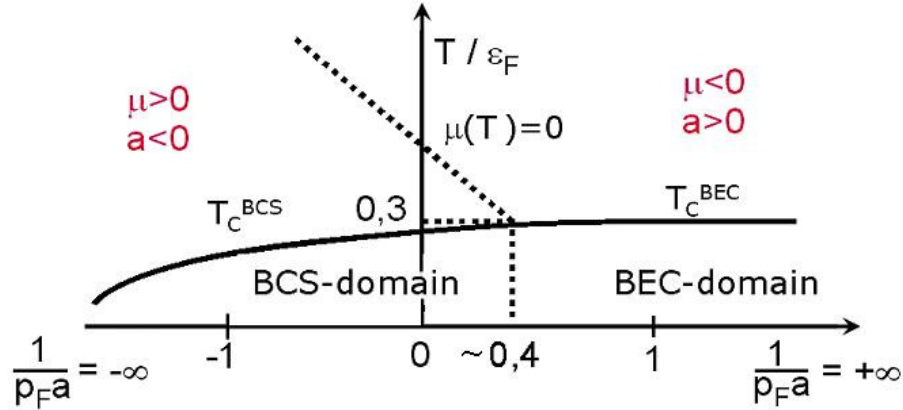

Fig 7.3. Phase-diagram of the BCS-BEC crossover in the resonance Fermi-gas in 3D (numerical calculations for $T_C$ and $\mu$ versus $1/p_F a$ for the first iteration to the self-consistent T-matrix approximation) [7.53].

On Fig 7.3 we present a critical temperature $T_C$ versus inverse gas parameter $1/p_F a$. The dashed line $\mu(T) = 0$ effectively separates BCS-domain (with $\mu > 0$) from BEC-domain (with $\mu < 0$). Note that according to [7.53] $\mu(T_C) = 0$ for $T_C \approx 0.29\varepsilon_F$ and $1/p_F a \approx 0.37$. Thus in substance the border between BCS and BEC-domain effectively lies in the BEC region of positive scattering length $a > 0$. In the same time a unitary limit in vacuum $1/p_F a = 0$ is effectively in the BCS-domain of positive chemical potential $\mu > 0$. Note that on Fig 7.3 the region of dilute Fermi-gas corresponds to the interval [-∞, -1] for $1/p_F a$, while the region of dilute Bose-gas (of composed bosons or molecules) corresponds to the interval [1, ∞]. Close to the unitarity (in the interval [-1, 1]) we have strongly interacting gas and cannot apply exact diagrammatic expansions of Galitskii [5.56] and Beliaev [7.18]. It is interesting to note that while in dilute limit for BEC-domain according to Gor'kov and Melik-Barchudarov the critical temperature $T_C^{BCS} = 0.28\,\varepsilon_F \exp\left\{-\dfrac{\pi}{2|a|\,p_F}\right\}$ has a preexponential factor $0.28\,\varepsilon_F$, in dilute limit for BEC-domain we have very interesting correction to Einstein formula obtained by Prokof'ev and Svistunov [7.19]:

$$T_C^{BEC} = 0.2\,\varepsilon_F\left[1 + 1.3a_{2-2}n_{tot}^{1/3}\right]. \qquad (7.1.23)$$

These corrections just take into account the repulsive dimer-dimer scattering length $a_{2-2} = 0.6|a| > 0$ and correspond to 3D weakly non-ideal Bose-gas with repulsion between composed bosons.

### 7.1.5. Unitary limit.

In the unitary limit $1/p_F a = 0$ there is only one scale, namely, Fermi-energy $\varepsilon_F$ both for kinetic and potential energy. Thus in this limit the total energy of the system reads:

$$E = \frac{3}{5}\varepsilon_F N\beta. \qquad (7.1.24)$$

It does not depend upon the gas parameter $1/p_F a$ and depends only on the universal coefficient $\beta$. This coefficient depends only upon the number of components in the Fermi-gas. For the gas with the spin $S = \frac{1}{2}$ the number of components is 2 ($S_z = \pm \frac{1}{2}$) and $\beta \approx 0.44 > 0$. Thus it is a gas phase ($E > 0$) according to Monte Carlo simulations by Astrakharchik et al. [7.20] and Carlson et al. [7.21]. In a gas phase the chemical potential $\mu = dE/dN = \beta \varepsilon_F = 0.44 \varepsilon_F$ at temperature $T = 0$.

For the mixture of protons and neutrons the number of components is 4 and $\beta < 0$. We have a liquid phase here ($E < 0$) according to Heiselberg et al. [7.22]. To some extent it is an answer on the question formulated some time ago by Zel'dovich in Moscow: whether we can get a dilute liquid in the system of resonantly interacting neutrons? The answer of Heiselberg is negative: without protons we are in the gas-phase for $|a| p_F \gg 1$ (see also [7.32]).

### 7.1.6. Qualitative interpretation of the intermediate region of large values of $|a| p_F \gg 1$ ($-1 < 1/a p_F < 1$).

In the strong coupling limit $|a| p_F > 1$ for BCS-domain ($\mu > 0$ and $a < 0$) nothing drammatic happens. We can qualitatively represent Gor'kov-Melik-Barchudarov result as:

$$T_C = A\mu \exp\left\{ -\frac{\pi}{2|a|\sqrt{2m\mu}} \right\} \text{ for } \mu > 0 \qquad (7.1.25)$$

(where we replaced $\varepsilon_F$ by $\mu$ and $p_F$ by $\sqrt{2m\mu}$).

In the unitary limit $1/a \to -0$, the chemical potential $\mu > 0$ and $T_C = A\mu \approx 0.15 \varepsilon_F$ according to numerical calculations of Prokof'ev, Troyer [7.23]. In the same time $\mu \approx 0.44 \varepsilon_F$ according to [7.20, 7.21] and thus $A \approx 0.35$ in (7.1.25), which is close to preexponential factor 0.28 (in front of $\varepsilon_F$) of Gor'kov, Melik-Barchudarov result.

The situation is more complicated in the strong-coupling BEC limit. Here for $\mu < 0$ and $1 \le a p_F \le 3$ ($a > 0$) the molecules with the small binding energies $|E_b| \le 2\varepsilon_F$ are formed. In the same time $n_{tot} a^3 = \frac{p_F^3 a^3}{3\pi^2} \le 1$ for $a p_F \le 3$ and thus local pairs do not overlap but only touch each other. Hence we have an intermediate situation between the tightly bound molecules and extended Cooper pairs here and can speculate about a formation of Fermi-Bose mixture of molecules and unpaired fermions here (see also Chapter 8). Nevertheless the theory of BCS-BEC crossover near the unitary limit is still far from being completed.

### 7.2. Self-consistent Leggett theory for $T = 0$.

In this Section we will derive and solve the system of Leggett equations [5.13] in 3D resonance Fermi-gas. This system contains the equation for the chemical potential and the equation of the self-consistency for the superfluid gap and describe BCS-BEC crossover in a resonance Fermi-gas at zero temperatures.

### 7.2.1. Leggett equations for chemical potential and superfluid gap.

The Leggett's equation for the superfluid gap is represented as follows (see Fig. 7.4):

$$\Delta = U_0 \int F_s(\omega, \vec{q}) \frac{d^3 \vec{q}}{(2\pi)^3} \frac{d\omega}{2\pi}, \qquad (7.2.1)$$

where

$$F_s(\omega, \vec{q}) = \frac{\Delta}{\omega^2 + E_q^2} \qquad (7.2.2)$$

is the anomalous Green's function (see [5.20, 5.60]) allowing the Wick shift $\omega \to i\omega$, $U_0$ is the zeroth Fourier component of the two-particle interaction $U(q)$.

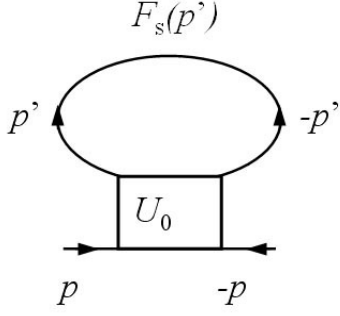

Fig. 7.4. Self-consistent equation for the gap $\Delta$. $U_0$ is the zeroth Fourier component of the two-particle interaction $U(q)$. $F(p')$ is the anomalous Green's function [7.24]. $p$ and $p'$ are four-momenta $\{\omega, \vec{p}\}$ and $\{\omega', \vec{p}'\}$.

Note that in a superfluid state at $T = 0$ according to Abrikosov, Gor'kov, Dzyaloshinskii [5.60] (see also [5.20]) we should introduce two Green's functions: normal Green's function $G_s$ of a superfluid state and anomalous Green's function $F_s$ (instead of one Green's function $G_N$ in a normal state for $T > T_C$). In Euclidean form (after the Wick transformation $\omega \to i\omega$) the Green's functions of a superfluid state read:

$$G_s(i\omega, \vec{q}) = -\frac{i\omega + \xi_q}{\omega^2 + E_q^2}, \qquad (7.2.3)$$

$$F_s(i\omega, \vec{q}) = -\frac{\Delta}{\omega^2 + E_q^2}, \qquad (7.2.4)$$

$$F_s^+(i\omega, \vec{q}) = -\frac{\Delta^+}{\omega^2 + E_q^2}, \qquad (7.2.5)$$

where $E_q^2 = \Delta^2 + \xi_q^2$ is uncorrelated spectrum squared of one-particle excitations in superconductor, $\xi_q = q^2/2m - \mu$ and $\Delta$ is a superfluid gap. $F_s^+$ and $\Delta^+$ in (7.2.5) are hermitian conjugated from $F$ and $\Delta$. In the graphical form the normal $G_s$ and anomalous Green's functions $F_s$ and hermitian conjugated $F_s^+$ are represented on Fig. 7.5.

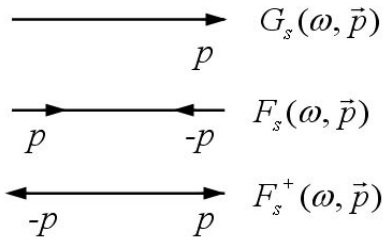

Fig. 7.5. Graphical representation for normal ($G_s$) and anomalous ($F_s$ and $F_s^+$) Green's functions in the superconductor at $T = 0$. $p$ is four-momentum $\{\omega, \vec{p}\}$.

Note that in the absence of an external field, a superfluid gap is real $\Delta = \Delta^+$ and thus $F_s^+(\omega, \vec{p}) = F_s^*(-\omega, -\vec{p}) = F_s(-\omega, -\vec{p})$ for anomalous Green's function $F_s^*$.

Returning back to the equation (7.2.1) for the superfluid gap we can rewrite it as follows:

$$\Delta = |U_0| \int \frac{d^3\overline{p}\, d\omega}{(2\pi)^4} \frac{\Delta}{(\omega^2 + E_p^2)}, \qquad (7.2.6)$$

or for non-zero superfluid gap $\Delta \neq 0$:

$$1 = |U_0| \int \frac{d^3\overline{p}\, d\omega}{(2\pi)^4} \frac{1}{(\omega^2 + E_p^2)} = |U_0| \int \frac{d^3\overline{p}}{(2\pi)^3} \frac{1}{2E_p}, \qquad (7.2.7)$$

where we made frequency integration on the complex half-plane for $\omega$ (see Fig. 7.6). This is a standard self-consistency equation familiar for the BCS-theory [5.20, 5.60, 5.11, 5.13] with $E_p = \sqrt{\Delta^2 + \xi_p^2}$ for the spectrum. Taking into account renormalization condition (7.1.4):

$$\frac{m}{4\pi a} = -\frac{1}{|U_0|} + \int \frac{d^3\overline{p}}{(2\pi)^3} \frac{1}{2\varepsilon_p}$$ we can represent (7.2.7) as follows:

$$1 + \frac{4\pi a}{m} \int \frac{d^3\overline{p}}{(2\pi)^3} \left( \frac{1}{2E_p} - \frac{1}{2\varepsilon_p} \right) = 0, \qquad (7.2.8)$$

where $E_p = \sqrt{(\varepsilon_p - \mu)^2 + \Delta^2}$ and $a$ is an s-wave scattering length in vacuum. Equation (7.2.8) is a first Leggett's equation.

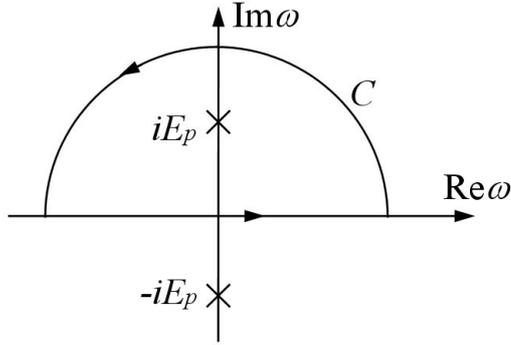

Fig. 7.6. Complex half-plane for frequency integration in the first Leggett's equation (7.2.7) for a superfluid gap $\Delta$. We have one pole $\omega = iE_p$ inside the closed contour $C$ in the upper half-plane.

To derive the second Leggett's equation we should use two facts. The first one is connected with the definition of superfluid density $n_s$ via the integral of a normal Green's function $G_s$ in superconductor:

$$n_s = \int \frac{d\omega}{2\pi} \frac{d^3\overline{p}}{(2\pi)^3} G_S(\omega, \overline{p}). \qquad (7.2.9)$$

The second one is connected with the absence of normal excitations in superconductor at $T = 0$. Hence normal density equals to zero and superfluid density $n_S(T=0) = \frac{n_{tot}}{2} = \frac{p_F^3}{6\pi^2}$. Correspondingly after frequency integration over the contour $C$ (see Fig. 7.6) we will get from (7.2.9):

$$n_S(T=0) = \int \frac{d^3\overline{p}}{(2\pi)^3} \frac{E_p - \xi_p}{2E_p} = \int \frac{d^3\overline{p}}{(2\pi)^3} \frac{1}{2}\left(1 - \frac{\xi_p}{E_p}\right) = \frac{n_{tot}}{2} = \frac{p_F^3}{6\pi^2} \qquad (7.2.10)$$

where $\xi_p = p^2/2m - \mu$. It is a second Leggett's equation for chemical potential $\mu$. Note that the equations (7.2.8), (7.2.10) are valid both in BCS ($\mu > 0$ and $a < 0$) and BEC ($\mu < 0$ and $a > 0$) domains. They can be also applied to describe unitary limit $1/p_F a = 0$. Note also that both Nozieres-Schmitt-Rink scheme at $T \neq 0$ and Leggett's scheme at $T = 0$ provide smooth crossover between BCS and BEC regions, at least for singlet s-wave pairing. In the next section we will show that for triplet p-wave pairing (and the symmetry of A1-phase) we can have the point of quantum phase-transition [7.7, 7.8] (or even topological phase-transition [7.9, 7.25]) for a special point $\mu(T=0)=0$ separating BCS and BEC domains at $T=0$.

In dilute BCS-limit for $|a|p_F \ll 1$ the solution of Leggett's equations yield: $\mu \approx \varepsilon_F > 0$ for the chemical potential and

$$\Delta \approx 1.75 T_C^{BCS} \approx 1.75 \cdot 0.28 \varepsilon_F \exp\left\{-\frac{\pi}{2|a|p_F}\right\} \approx 0.5 \varepsilon_F \exp\left\{-\frac{\pi}{2|a|p_F}\right\} \qquad (7.2.11)$$

for the superfluid gap.

For dilute BEC-limit $ap_F \ll 1$ the superfluid gap reads:

$$\Delta = \sqrt{2\mu_B |E_b|}, \qquad (7.2.12)$$

where $|E_b| = 1/ma^2$ is a binding energy and $\mu_B = \frac{4\pi a_{2-2}}{m_B} n_B > 0$ is a bosonic chemical potential for weakly repulsive Bose-gas of molecules (dimers) with the mass $m_B = 2m$ and density $n_B = n_{tot}/2$. In (7.2.12) the dimer-dimer scattering length $a_{2-2} = 0.6|a| > 0$. Note that bosonic chemical potential $\mu_B$ is of Hartree-Fock origin and $\mu = -\frac{|E_b|}{2} + \frac{\mu_B}{2} < 0$ for $ap_F \ll 1$. Of course, $\mu_B \ll |E_b|$ in dilute BEC-domain. Thus we have a standard BCS-gap and a standard ratio $\frac{2\Delta}{T_C} \approx 3.5$ in dilute BCS-limit while in dilute BEC-limit the gap is pretty unconventional (see (7.2.12)). Physically the gap $\Delta$ in BEC-domain reflects a creation of bound pairs (molecules). To get the filling of $\Delta^2 = 2\mu_B |E_b|$ in BEC limit let us consider the normal Green's function $G_s(i\omega, \vec{q}) = -\frac{i\omega + \xi_q}{\omega^2 + E_q^2}$. It can be rewritten as:

$$G_s(i\omega, \vec{q}) = -\frac{i\omega + \xi_q}{\omega^2 + E_q^2} = -\frac{1}{i\omega - \xi_q - \frac{\Delta^2}{i\omega + \xi_q}} = \frac{1}{G_0^{-1} - \Sigma}. \qquad (7.2.13)$$

We can see from (7.2.13) that a normal Green's function of a superfluid state has a two-pole structure, where a resonance self-energy $\Sigma$ in (7.2.13) reads:

$$\Sigma(\omega, \vec{p}) = \frac{\Delta^2}{i\omega + \xi_q}. \qquad (7.2.14)$$

It is interesting to note that at $T = 0$ a superconductive gap $\Delta$ in (7.2.14) coincides with a pseudogap $\Delta_{PG}$ for a normal state in BEC-domain for $T \ll T^*$ in (7.1.19). Moreover if we put $\mu_B = 0$ in the expression for the self-energy of a normal state in (7.1.23) we will recover (7.2.14) for self-energy of a superfluid state.

In the same time at $T = T_C^{BEC}$ a superfluid gap $\Delta(T_C^{BEC}) = \sqrt{2|E_b|\mu_B(T_C^{BEC})} = 0$ since $\mu_B(T_C^{BEC}) = 0$, while a pseudogap $\Delta_{PG}$ does not drastically changes at $T_C$ and is nonzero. Note that we can rewrite the pseudogap in (7.1.23) as $\Delta_{PG}{}^2 \approx 2|E_b|\mu_B(T=0)$, where $\mu_B(T=0) = \frac{4\pi a_{2-2}}{m_B} n_B = \frac{\pi a_{2-2} n_{tot}}{m}$ is a Hartree-Fock contribution to the chemical potential in the repulsive Bose-gas of composed bosons. Moreover let us stress that on the level of the T-matrix approximation in the expression for dimer-dimer scattering length $a_{2-2}$ in (7.1.19) enters mean-field result $a_{2-2} = 2|a|$, while in a more elaborate approach beyond the simple T-matrix scheme it should be $a_{2-2} = 0.6|a| > 0$ (see [7.34]). Nevertheless the similarity between $\Delta(T = 0)$ and $\Delta_{PG}(T=0)$ is striking and not accidental. More careful analysis shows, however, that for non-zero temperatures $0 < T < T_C$ inside the superfluid phase, the superfluid gap reads:

$$\Delta^2(T) = 2|E_b|\mu_B(T), \text{ where } \mu_B(T) = \frac{\pi a_{2-2} n_s(T)}{m} \text{ (see [5.60])}.$$ Thus in the superfluid gap squared

enters the superfluid density $n_s(T)$, while in the pseudogap enters the total density $n_{tot}$. This fact clarifies the situation. Indeed, for $T = 0$ the superfluid density $n_s(T = 0) = n_{tot}$, and hence $\Delta(T = 0) = \Delta_{PG}$. In the same time for $T = T_C^{BEC}$ the superfluid density $n_s(T_C^{BEC}) = 0$ and thus $\Delta(T_C^{BEC}) = 0$ while $\Delta_{PG}(T_C^{BEC})$ is still governed by $n_{tot}$ and does not differ much from $\Delta_{PG}(T = 0)$. At finite temperatures $0 < T < T_C^{BEC}$ the superfluid density $n_s(T) < n_{tot}$, and correspondingly $\Delta(T) < \Delta_{PG}$.

In the unitary limit at $T=0$ the superfluid gap $\Delta \approx 0.5\varepsilon_F$ and the chemical potential $\mu \approx 0.44\varepsilon_F > 0$ are found by Carlson et al. [7.21] in the framework of Monte Carlo simulations.

Note that in Leggett's self-consistent scheme $\mu(T = 0) = 0$ for the value of the gas parameter (see [5.16]):

$$\frac{1}{p_F a_0} \approx 0.553 \qquad (7.2.16)$$

Thus the border between BCS- ($\mu > 0$) and BEC-domains ($\mu < 0$) lies in the region of positive values of the scattering length $a_0 > 0$ (for the values of gas parameter $a_0 p_F \approx 1.8$ ). The superfluid gap at $a_0 p_F$ satisfies the relation:

$$\left(\frac{\Delta}{\varepsilon_F}\right)_0^2 = 2\frac{1}{p_F a_0}, \qquad (7.2.17)$$

where the index 0 indicates that the relevant physical quantities are taken for $\mu(T = 0) = 0$. Thus from Eq. (7.2.17) we get $\Delta_0 \approx 1.05\varepsilon_F$ for $a_0 p_F \approx 1.8$. Both the chemical potential $\mu$ and the gap $\Delta$ vary linearly as function of $a - a_0$ near the point $\mu = 0$ and $a = a_0$. Namely (see [6.16]):

$$\frac{\Delta}{\varepsilon_F} = \left(\frac{\Delta}{\varepsilon_F}\right)_0\left[1 - \frac{\pi}{4}\frac{1}{p_F a_0}\frac{\mu}{\varepsilon_F}\right]; \qquad (7.2.18)$$

$$\frac{\mu}{\varepsilon_F} = \left(\frac{1}{p_F a_0} - \frac{1}{p_F a}\right)\frac{1}{\left(\frac{\pi}{8}\frac{1}{p_F a_0} + \frac{1}{\pi}p_F^2 a_0^2\right)} \approx \frac{a - a_0}{a_0}\frac{1}{\left(\frac{\pi}{8} + \frac{1}{\pi}p_F^3 a_0^3\right)}. (7.2.19)$$

In would be interesting to compare these results with those of Monte Carlo calculations and with experiments to check how good is the quantitative description of BCS-BEC crossover at $T = 0$ by the self-consistent Leggett's theory.

### 7.2.2. Sound velocity in BCS and BEC limits.

Finally note that, as we mentioned in the Introduction, the sound velocity in the resonance Fermi-gas can be obtained not only from the solution of the dynamical problem for the spectrum of collective excitations (see the next Section), but also from the thermodynamic identity for compressibility in a static case at $T=0$:

$$\kappa^{-1} \sim c_s^2 = \frac{n}{m}\frac{\partial\mu}{\partial n} \qquad, \qquad (7.2.20)$$

where $n$ is total density and $\mu$ is chemical potential. In dilute BCS-limit for $|a|p_F << 1$: the chemical potential $\mu \approx \varepsilon_F$ and the superfluid gap $\Delta \sim T_C^{BCS} << \mu$. Here effectively we can calculate sound velocity neglecting the derivative $\partial\Delta/\partial\mu$ and putting $n = p_F^3/3\pi^2$, while neglecting the small difference between $\mu$ and $\varepsilon_F$ connected with the superfluid gap squared:

$\mu = \varepsilon_F - \alpha \dfrac{\Delta^2}{\varepsilon_F} \ln \dfrac{\varepsilon_F}{\Delta} \approx \varepsilon_F$ (see the foundations of the BCS-theory [5.11]) . Thus $\dfrac{\partial \mu}{\partial n} = \dfrac{2}{3} \dfrac{\mu}{n}$ and

$\dfrac{n}{m} \dfrac{\partial \mu}{\partial n} = \dfrac{2}{3} \dfrac{\mu}{m} \approx \dfrac{p_F^2}{3m^2}$ . Correspondingly $c_S^2 = \dfrac{v_F^2}{3}$ and $c_S^{BCS} = \dfrac{v_F}{\sqrt{3}}$ (where $v_F = p_F / m$ is Fermi velocity). We get the well-known result for Bogolubov-Anderson sound velocity in neutral (non-charged) superfluid Fermi-gas. Note that in normal Fermi-gas the sound waves will be overdamped (see [5.20]) and only zero-sound mode will be propagating at $T \to 0$. The superfluid gap causes the final damping of the first sound mode at small frequencies: $\mathrm{Im}\,\omega \sim \omega^2 \tau$, where $\gamma = 1/\tau \sim \Delta^2 / \varepsilon_F$ is an inverse scattering time.

To get the sound velocity in the dilute BEC-limit we should recollect that $\mu = -\dfrac{|E_b|}{2} + \dfrac{\mu_B}{2}$ and $\mu_B(T) = \dfrac{4\pi a_{2-2} n_B}{m_B} = \dfrac{\pi a_{2-2} n}{m}$ . Thus $\dfrac{\partial \mu}{\partial n} = \dfrac{1}{2} \dfrac{\partial \mu_B}{\partial n} = \dfrac{\pi a_{2-2}}{2m}$ and

$\dfrac{n}{m} \dfrac{\partial \mu}{\partial n} = \dfrac{\pi a_{2-2} n}{2m^2} = \dfrac{\mu_B}{m_B}$ . Correspondingly $c_S^{BEC} = \sqrt{\dfrac{\mu_B}{m_B}}$ and we recover the result for Bogolubov sound velocity in 3D gas of (composed) bosons with weal repulsion (between bosons): $\mu_B = m_B (c_s^{BEC})^2$ .

Note that in the dilute limit $|a| p_F \ll 1$   $c_S^{BEC} \ll c_S^{BCS}$ . The sound velocities become equal in the intermediate region for large values of the gas parameter $|a| p_F \gg 1$ .

In the unitary limit $c_S \approx 0.4 \varepsilon_F$. Note that for $\mu(T = 0) = 0$ and $a_0 p_F \approx 1.8$ the sound velocity reads [6.16]:

$$c_S^2 = \dfrac{p_F^2}{3m^2} \dfrac{\dfrac{\pi}{2} \dfrac{1}{p_F^2 a_0^2}}{\left(1 + \dfrac{\pi^2}{8} \dfrac{1}{p_F^3 a_0^3}\right)} \approx 0.132 v_F^2 . \qquad (7.2.21)$$

### 7.2.3. BCS-BEC crossover for the 2D resonance Fermi-gas.

Let us discuss briefly the BCS-BEC crossover and Leggett's equations for the 2D resonance Fermi-gas.

For symmetric attractive potential well in 2D according to quantum mechanics we have a bound-state of two-particles even for infinitely small attraction between them [5.19]. This is in contrast with the 3D case where the bound-state even in symmetric potential well is formed only for deep enough potentials: $|U| \geq \gamma / m r_0^2$, where $\gamma$ is numerical coefficient, $|U|$ is the depth of the attractive potential and $r_0$ is the width of the well. Thus there is a threshold for a bound state in 3D. On the level of the two-particle T-matrix in vacuum (see Chapter 5) it corresponds to $|\beta| = 1$ for the value of the Born parameter $\beta$. Correspondingly in 3D we have a shallow bound-state $|E_b| \ll 1 / m r_0^2$ for $0 < \dfrac{|\beta| - 1}{|\beta|} \ll 1$ .

In 2D the threshold for the formation of the bound-state is absent. Thus composed bosons (molecules $f \uparrow f \downarrow$) are every time present in 2D attractive Fermi-gas. However the important parameter, namely the ratio of $|E_b| / \varepsilon_F$ is still present in 2D Fermi-gas. Namely, for $\varepsilon_F > |E_b|$ we are still in the BCS-domain according to Miyake [6.62]. Here the crossover temperature $T_*$ (which corresponds to the molecule formation) and the critical temperature $T_C^{\mathrm{BCS}}$ coincide and are given by the famous Miyake formula: $T_* = T_C^{BCS} \sim \sqrt{2 \varepsilon_F |E_b|}$ (see Chapter 6). Moreover as it was shown by Beasley, Mooij and Orlando [7.36] the difference between $T_C^{BCS}$ (obtained in the mean-field theory) and exact (for 2D) Berezinskii-Kosterlitz-Thouless (BKT) critical temperature ([6.66, 6.67]) is small:

$$\frac{\left| T_C^{BCS} - T_C^{BKT} \right|}{T_C^{BCS}} \sim \frac{T_C^{BCS}}{\varepsilon_F} << 1 \quad for\ |E_b| << \varepsilon_F. \qquad (7.2.22)$$

Thus the mean-field formula for $T_C$ obtained by Miyake is a very good estimate for a critical temperature, which is very close to the exact $T_C^{BKT}$.

The Cooper pairs are extended for $|E_b| < \varepsilon_F$. Thus in the BCS-domain we have simultaneously collective Cooper pairing in momentum space in substance and the two-particle pairing in real space in vacuum. In the same time for $\varepsilon_F < |E_b|$ we are in BEC-domain. Here we have two characteristic temperatures instead of one: Saha crossover temperature $T_* = |E_b| / \ln(|E_b|/\varepsilon_F)$ for the formation of molecules and critical temperature $T_C^{BEC}$ of the bose-condensation. Note that for $T << T_*$ in 2D case we have again a slightly non-ideal Bose-gas of composed molecules with repulsion between them. As we derived in Chapter 6, repulsive interaction between molecules (dimers) in 2D is described by the coupling constant $f_{2-2} = 1/\ln(1.6|E_b|/\varepsilon_F)$. Hence according to Fisher-Hohenberg theory (see Chapter 6 and [6.64, 6.66]) the mean-field critical temperature is given by:

$$T_C^{BEC} = \frac{\varepsilon_F}{4\ln(1/f_{2-2})} \qquad (7.2.23)$$

It is again very close to exact BKT critical temperature since

$$\frac{\left| T_C^{BEC} - T_C^{BKT} \right|}{T_C^{BEC}} \sim f_{2-2} << 1 \qquad (7.2.24)$$

for $|E_b| >> \varepsilon_F$.

Leggett's equations in the 2D case.

The mean-field Leggett's scheme for BCS-BEC crossover in 2D resonance Fermi-gas at $T = 0$ works even better then the mean-field approaches at $T = T_C$ since at low temperatures all the vortex-antivortex pairs are well bound in molecules. Thus Leggett's equations well describe the density of superfluid component $n_S$, the velocity of sound $c_S$, the behavior of the superfluid gap $\Delta$, and the chemical potential $\mu$ at $T = 0$ ([7.24]).

The Leggett's equations in 2D resonance Fermi-gas with s-wave pairing are the evident generalization of the Leggett's equations for the 3D case and yield:

$$\frac{n_{tot}}{2} = \frac{p_F^2}{4\pi} = \int\limits_0^{\sim -1/r_0} \frac{pdp}{4\pi} \left( 1 - \frac{\xi_p}{E_p} \right); \qquad (7.2.25)$$

$$\frac{m}{4\pi} \ln(mr_0^2\,|\,E_b\,|) = \int\limits_0^{\sim -1/r_0} \frac{pdp}{4\pi} \frac{1}{E_p}, \qquad (7.2.26)$$

where $r_0$ is the range of the attractive potential $U(r) = -|U|e^{-r/r_0}$, $\xi_p = p^2/2m - \mu$, $E_p = \sqrt{\xi_p^2 + \Delta^2}$ is the quasiparticle spectrum in the superfluid state, $E_b$ is the two-particle binding energy and in the l.h.s. of Eq. (7.2.26) we have a typical 2D logarithm. For shallow level the binding energy $|E_b| << 1/mr_0^2$ and is given by

$$|E_b| = \frac{1}{mr_0^2} \exp\left\{ -\frac{4\pi}{m|U_0|} \right\}, \qquad (7.2.27)$$

where $|U_0| \sim r_0^2 |U|$ is the zeroth Fourier component ($q = 0$) of the vacuum potential. Thus Eq. (7.2.26) for the superfluid gap can be represented in the familiar form

$$1 - |U_0| \int\limits_0^{\sim 1/r_0} \frac{p\, dp}{4\pi} \frac{1}{E_p} = 0. \qquad (7.2.28)$$

We also assume that the Fermi energy $\varepsilon_F \ll 1/mr_0^2$ or $p_F r_0 \ll 1$ – we have dilute Fermi-gas in 2D. The solution of Leggett's equations yield:

$$\ln \frac{|\mu| - \mu + \dfrac{\Delta^2}{2|\mu|}}{|E_b|} = 0, \qquad (7.2.29)$$

and

$$2\varepsilon_F = |\mu| + \mu + \frac{\Delta^2}{2|\mu|}. \qquad (7.2.30)$$

In the BCS-case $\mu = |\mu| > 0$ we get from (7.2.29)

$$\frac{\Delta^2}{2|\mu||E_b|} = 1. \qquad (7.2.31)$$

Accordingly from (7.2.30) we obtain

$$2\varepsilon_F = 2|\mu| + \frac{\Delta^2}{2|\mu|}. \qquad (7.2.32)$$

If we substitute an expression for $\Delta^2$ in (7.2.31) into eq. (7.2.32) we get:

$$2\varepsilon_F = 2|\mu| + |E_b| \text{ or } \mu = |\mu| = \varepsilon_F - \frac{|E_b|}{2} > 0. \qquad (7.2.33)$$

In the same time deep in BCS-domain (for $\varepsilon_F \gg |E_b|/2$):

$$\mu \approx \varepsilon_F \text{ and } \Delta^2 = 2\varepsilon_F |E_b|. \qquad (7.2.34)$$

We get a famous result of Miyake [6.62] (see also Varma et al. [7.47]) and Randeria et al. [7.48].

In the BEC-limit $\mu = -|\mu| < 0$. Thus from (7.2.29) we obtain

$$\frac{\Delta^2}{2|\mu|} + 2|\mu| = |E_b| \quad . \qquad (7.2.35)$$

Correspondingly eq. (7.2.30) reads

$$2\varepsilon_F = \frac{\Delta^2}{2|\mu|} \text{ or } \Delta^2 = 4\varepsilon_F |\mu|. \qquad (7.2.36)$$

Substitution of (7.2.36) into (7.2.35) yields: $2\varepsilon_F + 2|\mu| = |E_b|$. Accordingly

$$|\mu| = \frac{|E_b|}{2} - \varepsilon_F \text{ or } \mu = -\frac{|E_b|}{2} + \varepsilon_F. \qquad (7.2.37)$$

Hence deep in BEC-domain (for $|E_b|/2 \gg \varepsilon_F$): $\mu \approx -\dfrac{|E_b|}{2} < 0$, and $\Delta^2 \approx 2\varepsilon_F |E_b|$ again in agreement with Miyake et al. Note that the result of Miyake et al. are valid not only in deep BCS and BEC regions, but also in the intermediate case, close to the point $\mu = 0$. Here $\varepsilon_F \approx |E_b|/2$ and for the gap we get: $\Delta \approx 2\varepsilon_F \approx |E_b|$.

### 7.2.4. Gap spectroscopy in 3D.

The superfluid gap $\Delta$ was measured from the threshold of the absorption $\omega = 2\Delta$ of the radiofrequency waves in the experiments of Grimm's group [7.26, 7.27] (see also [7.28, 7.29, 7.31]). The gap $\Delta$ varies in a magnetic field. Far from Feshbach resonance field (for $|a|p_F < 1$):

$$\Delta_{BCS} \sim \varepsilon_F \exp\left\{-\frac{1}{|a(B)|p_F}\right\} \text{ where } a(B) \approx a_{bg}\left(1+\frac{\Delta}{B-B_0}\right) \text{ in dilute BCS-domain.}$$

In dilute BEC-domain $\Delta^2_{BEC} = 2\mu_B|E_b| \sim \frac{a_{2-2}}{a^2} \sim \frac{1}{a}$ for $a > 0$ and $ap_F < 1$. Thus with the help of Feshbach resonance we can measure the gap $\Delta$ for all the values of the gas parameter $ap_F$ (for all the values of $\Delta B = B - B_0$) in the BCS and BEC domains.

### 7.3. Anderson-Bogolubov theory for collective excitations.

In this section we will present the diagrammatic approach for studying the spectrum of collective excitations in the BCS-BEC crossover at $T = 0$. To do that we will derive a set of the Bethe-Salpeter integral equations for the total vertices $\Gamma(\vec{q}, \omega)$ in the superfluid state of the resonance Fermi-gas at $T = 0$.

#### 7.3.1. Diagrammatic approach

In order to obtain a system of Bethe-Salpeter equations [5.61], we have to introduce two vertices $\Gamma_{11}$ and $\Gamma_{12}$ (see Fig. 7.7) in the Cooper channel for the superfluid state (instead of one vertex $\Gamma$ in the normal state considered in Chapter 5).

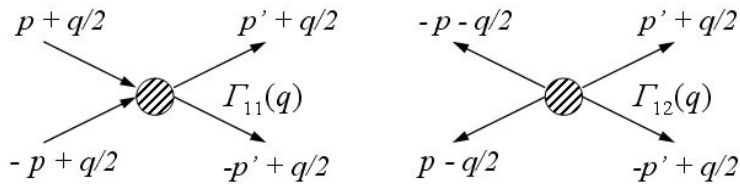

Fig. 7.7. Two vertices $\Gamma_{11}(q)$ and $\Gamma_{12}(q)$ in superconductive state. $p = (\omega, \vec{p})$, $q = (\Omega, \vec{q})$ are four-momenta.

The first one $\Gamma_{11}$ corresponds to the scattering of two atoms with opposite spins and with two incoming lines and two outgoing ones, while the second one $\Gamma_{12}$ has four outgoing lines. The second vertex ($\Gamma_{12}$) is nonzero only in the superfluid state, while $\Gamma_{11}$ exists also in the normal state.

The Bethe-Salpeter integral equation for the vertex $\Gamma_{11}(q)$ reads:

$$\Gamma_{11}(q) = U(\vec{q}) - U(\vec{q})\Gamma_{11}(q)\int\frac{d^3\vec{p}''}{(2\pi)^3}\frac{d\Omega''}{2\pi}G_s\left(p''+\frac{q}{2}\right)G_s\left(-p''+\frac{q}{2}\right) -$$
$$- U(\vec{q})\Gamma_{12}(q)\int\frac{d^3\vec{p}''}{(2\pi)^3}\frac{d\Omega''}{2\pi}F_s\left(p''+\frac{q}{2}\right)F_s\left(-p''+\frac{q}{2}\right), \tag{7.3.1}$$

where we introduced again the normal and anomalous Green's functions $G_s(i\omega, \vec{q}) = -\frac{i\omega+\xi_q}{\omega^2+\xi_q^2+\Delta^2}$, $F_s(i\omega, \vec{q}) = -\frac{\Delta}{\omega^2+\xi_q^2+\Delta^2}$ in the Euclidean formulation according to (7.2.4) and (7.2.5) (see Fig 7.5). Note that in (7.3.1) $q = (\Omega, \vec{q})$ is four-momentum and we assume the superfluid gap $\Delta$ to be real. Graphically the Bethe-Salpeter equation for $\Gamma_{11}(q)$ is presented on Fig. 7.8.

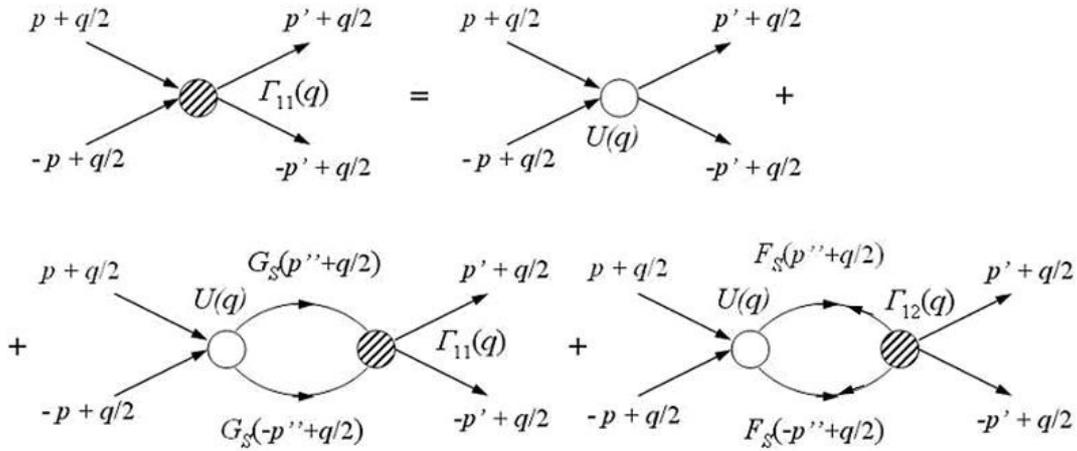

Fig. 7.8. The Bethe-Salpeter equation for the vertex $\Gamma_{11}(q)$.

If we would like to perform the normalization procedure and replace the vacuum interaction $U(\vec{q})$ by the scattering length $a$ (by $4\pi a/m$) (in (7.3.1)), we should simultaneously replace $G_S G_S$ in (7.3.1) by $(G_S G_S - G_0 G_0)$ where $G_0(\vec{q}, \omega) = (i\omega - \varepsilon_q)^{-1}$ is the vacuum Green's function in Euclidian formulation (see (7.1.2) and (7.1.4)). Analyzing the structure of eq. (7.3.1), we see that it is natural to introduce now the elementary response functions $\chi_{ij}$ (see Bogolubov [7.1] and Anderson [7.2]) corresponding to the various bubles appearing in Fig. 7.8. Specifically we define:

$$-\chi_{11}(q) = \frac{1}{g} + \int \frac{d^4 p}{(2\pi)^4} \left[ G_S\left(p + \frac{q}{2}\right) G_S\left(-p + \frac{q}{2}\right) - G_0(p) G_0(-p) \right],$$
$$\chi_{12}(q) = \int \frac{d^4 p}{(2\pi)^4} F_S\left(p + \frac{q}{2}\right) F_S\left(-p + \frac{q}{2}\right),$$
(7.3.2)

where $g = 4\pi a/m$ is a coupling constant.

Then Eq. (7.3.1) takes the form:

$$\Gamma_{12}(q)\chi_{12}(q) = 1 + \Gamma_{11}(q)\chi_{11}(q) \qquad (7.3.3)$$

Now we have to derive a second Bethe-Salpeter equation for $\Gamma_{12}(q)$. Graphically it has the form shown on Fig. 7.9.

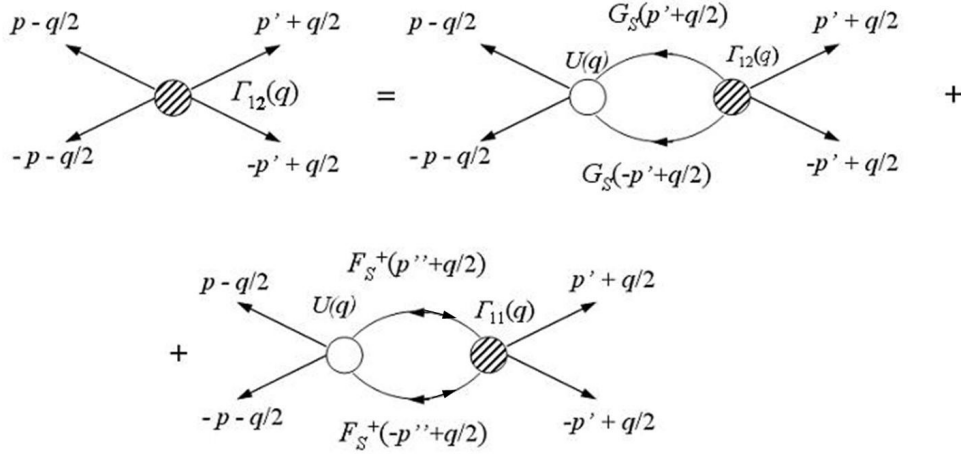

Fig. 7.9. The Bethe-Salpeter equation for $\Gamma_{12}(q)$.

A small difference between Fig. 7.8 and Fig 7.9 is that the anomalous Green's functions $F^+(q)$ appearing in this equation have their arrows outside (instead of inside as in Fig. 7.8). However for real $\Delta$, they are simply related to the preceding ones by $F^+(p) = F(- p)$. In algebraic form BS-equation for $\Gamma_{12}$ reads:

$$\Gamma_{12}(q) = -U(\vec{q})\Gamma_{12}(q)\int \frac{d^4 p''}{(2\pi)^4} G_S\left(p''-\frac{q}{2}\right)G_S\left(-p''-\frac{q}{2}\right)-$$
$$-U(\vec{q})\Gamma_{11}(q)\int \frac{d^4 p''}{(2\pi)^4} F_S^+\left(p''+\frac{q}{2}\right)F_S^+\left(-p''+\frac{q}{2}\right). \qquad (7.3.4)$$

Note the absence of a free term $U(\vec{q})$ in (7.3.4). Renormalization requires again the substitution of $G_S G_S$ by $(G_S G_S - G_0 G_0)$ and the replacement of $U(\vec{q})$ by $g = 4\pi a/m$. We see that in Eq. (7.3.4) naturally appear the same response-functions $\chi_{11}$ and $\chi_{12}$ as in (7.3.3), but for the four-momentum $- q$. This leads to the following form of the second BS-equation:

$$\Gamma_{12}(q)\chi_{11}(-q) = \Gamma_{11}(q)\chi_{12}(-q) \qquad (7.3.5)$$

From (7.3.3), (7.3.5) the vertices $\Gamma_{11}$ and $\Gamma_{12}$ are immediately obtained. However, we are only interested in the pole of these vertices, for which they diverge. This allows us to solve only the homogeneous equations corresponding to (7.3.3) and (7.3.5), leading to the following equation for the collective mode:

$$\chi_{11}(q)\chi_{11}(-q) = \chi_{12}^{\ 2}(q), \qquad (7.3.6)$$

where we have used the fact that taking the explicit form of $F_S(p)$ in (7.3.2) into account, one has $\chi_{12}(- q) = \chi_{12}(q)$. This result has already been derived in [7.33].

In order to obtain a more convenient equation for the collective mode, we can perform the integration over the frequency variable in (7.3.2). This is easily done by closing the contour in the upper half-plane (see Fig 7.6) where the quantity to be integrated has two poles, located at $\omega_1 = iE_+$ and $\omega_2 = iE_-$. Here we have introduced the convenient notation $E_\pm = E(\vec{p} \pm \vec{q}/2)$, and we will use similarly $\xi_\pm = \xi(\vec{p} \pm \vec{q}/2)$. The results are:

$$-\chi_{11}(\omega,\vec{q}) = \frac{1}{2}\int \frac{d^3\overline{p}}{(2\pi)^3}\left\{\frac{(E_+ + E_-)(E_+E_- + \xi_+\xi_-) + i\omega(E_+\xi_- + E_-\xi_+)}{E_+E_-[(E_+ + E_-)^2 + \omega^2]} - \frac{1}{E_p}\right\}, \qquad (7.3.7)$$

$$\chi_{12}(\omega,\vec{q}) = \frac{1}{2}\int \frac{d^3\overline{p}}{(2\pi)^3}\frac{\Delta^2}{E_+E_-}\frac{E_+ + E_-}{[(E_+ + E_-)^2 + \omega^2]}, \qquad (7.3.8)$$

where in (7.3.7) we have used the Leggett's self-consistency equation for the gap (7.2.8) (which reads: $\frac{1}{g} = \int \frac{d^3\bar{p}}{(2\pi)^3}\left(\frac{1}{2\varepsilon_p} - \frac{1}{2E_p}\right) = 0$ ) to get rid of the term $1/g$ in Eq. (7.3.2). Finally to get the spectrum of collective excitations $\omega = \omega(q)$ in the standard form, we should go back to ordinary frequencies in (7.3.7), (7.3.8) by the inverse Wick transformation $i\omega \to \omega$.

From (7.3.7) and (7.3.8) we can check that $\chi_{12}(\omega, \bar{q}) = \chi_{12}(-\omega, -\bar{q})$. In the same time $\chi_{11}(\omega, \bar{q}) \neq \chi_{11}(-\omega, -\bar{q})$ due to the presence of a linear in frequency anomalous term in (7.3.7). It is possible to show that this term is small and can be often neglected in BCS-limit, but it is more substantial in BEC-limit.

### 7.3.2. The spectrum of collective excitations.

The details of the expansions of $E_{\pm} = \sqrt{\xi_{\pm}^2 + \Delta^2}$ and $\xi_{\pm}$ at small $\bar{q}$, as well as the expansions of the expressions under integrals in (7.3.7) and (7.3.8) in small $\omega$ are presented in [5.16]. Here we will only give the final result for the spectrum. For $a < 0$ and $\mu \approx \varepsilon_F$ (BCS-domain) we get:

$$\omega^2 = c_s^2 q^2 \text{ and } c_s^{BCS} = \frac{v_F}{\sqrt{3}} \qquad (7.3.9)$$

for Bogolubov-Anderson sound mode.

In the same time for $a > 0$ and $\mu < 0$ (BEC-domain) (see Section 7.2.2):

$$\omega^2 = c_s^2 q^2 + \left(\frac{q^2}{2m_B}\right)^2 \text{ and } c_s^{BEC} = \sqrt{\frac{\mu_B}{m_B}} \qquad (7.3.10)$$

Thus we get the standard Bogolubov spectrum for slightly non-ideal Bose-gas with repulsion (see [5.20, 5.60]). The correction to the Bogolubov spectrum (7.3.10) due to the composed character of the molecules (bosons) are important only for large wave-vectors $qa \gg 1$ [6.16].

If we introduce a healing length (or a coherence length) $\xi_0 = \frac{1}{\sqrt{n_B a_{2-2}}}$ [5.20, 5.60, 7.35], where $n_B = n_{tot}/2$ and $a_{2-2} = 0.6a$, then there is a small parameter in the theory $\frac{a}{\xi_0} \sim (n_B a^3)^{1/2} \ll 1$. For $q \leq 1/\xi_0$ we get the sound-like linear regime, for $q > 1/\xi_0$ quadratic regime. In the same time the corrections due to the composed character of the bosons start to be important for $q \geq 1/a \gg 1/\xi_0$ (see Fig. 7.10).

Analytically one obtains [6.16]:

$$\omega^2 = \left(c_s q\right)^2 \frac{16 - (qa)^2}{16 + (qa)^2} + \frac{q^2}{4m} \ . \qquad (7.3.11)$$

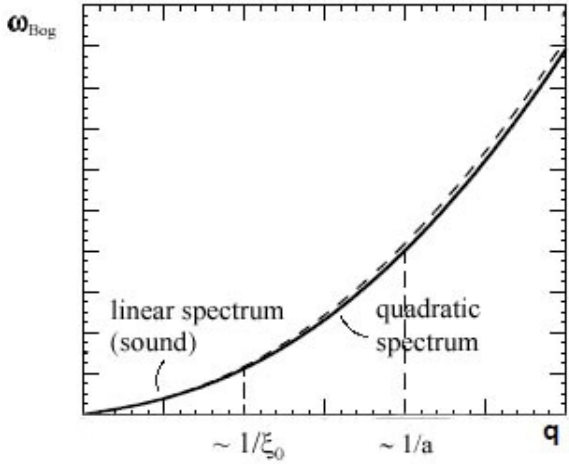

Fig. 7.10. The spectrum of collective excitations in dilute BEC-regime. The solid line is Bogolubov spectrum for pointlike bosons. The dashed line indicates the deviations from Bogolubov spectrum at large wave-vectors $q \geq 1/a \gg 1/\xi_0$.

Note that in a simple version of the theory for the collective mode given by (7.3.7) and (7.3.8), we will get a mean-field result (see Chapter 6) for the dimer-dimer scattering length $a_{2-2} = 2a$ and correspondingly for bosonic chemical potential $\mu_B = \pi n_{tot} a_{2-2}/m$. However in a more rigorous theory constructed by Combescot and Leironais [7.34] the exact result for $a_{2-2} = 0.6a$ in the expression for $\mu_B$ and sound velocity in BEC-regime can be recovered.

### 7.3.3. Landau critical velocity.

We have now basically all the information required to calculate the critical velocity $v_C$ (which corresponds to the destruction of the superfluid flow [7.17]) in the BCS-BEC crossover. Indeed, according to Landau criterion (see Chapter 1), it is given by $v_C = \min[\omega(q)/q]$ where $\omega(q)$ is the energy of the elementary excitation. In our case we have two types of the excitations. First we have bosonic excitations corresponding to the collective mode. If the disperdion relation has an upward curvature (as in our case or in superfluid $^4$He), then the minimum of $[\omega(q)/q]$ is obtained for $q \to 0$, which gives for $v_C$ the sound velocity $c_S$. However, we have also take into account fermionic single-particle excitations $\omega(q) = \sqrt{\xi_p^2 + \Delta^2}$ and we have to find again the minimum of $[\omega(q)/q]$. Thus we get according to [6.16] that:

$$v_C = \min\left\{ c_S ; \left( \frac{\sqrt{\Delta^2 + \mu^2} - \mu}{m} \right)^{1/2} \right\}. \qquad (7.3.12)$$

Note that deep inside BCS-region for $|a| p_F \ll 1$: $c_S = v_F / \sqrt{3}$, but $v_C = \Delta/p_F \ll c_S$ due to the contribution of the one-particle fermionic excitations (due to the unbinding of the Cooper pairs). The resulting curve for the critical velocity $v_C$ and sound velocity $c_S$ as the functions of the inverse gas parameter $1/p_F a$ are presented on Fig. 7.11. The result displays a kink in the maximum for $v_C$, occurring when one switches from bosonic to fermionic excitations. It occurs very near unitarity on the BEC-side. It is worth to note also that in terms of critical velocity, the "strength" of the superfluid is at its highest around unitarity and not on the BEC-side, as one might naively assume.

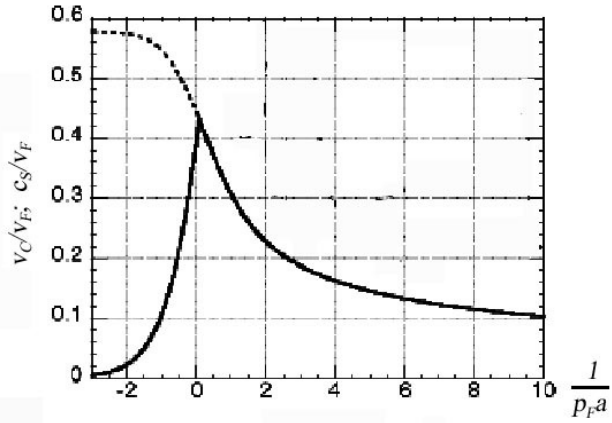

Fig. 7.11. Critical velocity $v_C$ (full line) and sound velocity (dashed line) as functions of $1/p_F a$ [6.16].

### 7.4. Feshbach resonance and phase-diagram for p-wave superfluid Fermi-gas

The first experimental results on p-wave Feshbach resonance [7.4-7.6] in ultracold fermionic gasses $^{40}$K and $^6$Li make the field of quantum gasses closer to the interesting physics of superfluid $^3$He (see Chapters 11 and 12) and the physics of unconventional superconductors, such as $Sr_2RuO_4$, for example. In this context, it is important to bridge the physics of ultracold gasses and the low-temperature physics of quantum liquids and anomalous superconductors, and thus to enrich both communities with the experience and knowledge accumulated in each of these fields. In this Section we will describe the transition from the weakly bound Cooper pairs with p-wave symmetry (with a relative orbital momentum of a pair $l = 1$) to strongly bound local p-wave pairs (triplet molecules - see also Chapter 6). We will try to reveal the nontrivial topological effects related to the presence of the nodes in the superfluid gap of the 100%-polarized p-wave A1-phase in three dimensions. We note that the symmetry of the A1-phase (where we have a pairing of two fermions with total spin $S = 1$ and its projection $S_z = 1$) is relevant both to ultracold Fermi gasses in the regime of p-wave Feshbach resonance and to superfluid $^3$He in the presence of large spin-polarization (see Chapters 11,12).

### 7.4.1. Feshbach resonance for 100%-polarized p-wave resonance superfluids

In the first experiments on p-wave Feshbach resonance, experimentalists measured the molecule formation in the ultracold fermionic gas of $^6$Li atoms close to the resonance magnetic field $B_0$ [7.4, 7.5].

In the last years, analogous experiments on p-wave molecules formation in spin-polarized fermionic gas of $^{40}$K-atoms were started [7.6]. The lifetime of p-wave triplet molecules is still rather short [7.4-7.6]. Nevertheless the physicists working in ultracold quantum gasses began to study intensively the huge bulk of experimental and theoretical wisdom accumulated in the physics of superfluid $^3$He [7.10] and anomalous complex superconductors [7.37].

To understand the essence of p-wave Feshbach resonance, we recall the basic quantum mechanical formula for the p-wave scattering amplitude in vacuum [5.19, 7.38-7.40]:

$$f_{l=1}(E) = \frac{p\,p'}{\dfrac{1}{V_p} + \dfrac{2mE}{\pi r_0} + i(2mE)^{3/2}}, \qquad (7.4.1)$$

where $l = 1$ is the relative orbital momentum for the two-particle problem in the p-wave channel, $E$ is the two-particle energy, $V_p = r_0^2 a_p$ is the scattering volume, $a_p$ is the p-wave scattering length, $r_0$ is the interaction range, and $p$ and $p'$ are incoming and outgoing momenta for the scattering amplitude $f_{pp'}(E)$. For Feshbach resonance in the fermionic systems, $p \sim p' \sim p_F$ and usually $p_F r_0 < 1$. The p-wave scattering length $a_p$, and hence the scattering volume $V_p$, diverge in the resonance magnetic field $B_0$ (see Fig. 7.12 and Chapter 5). Thus $1/V_p = 1/a_p = 0$ for $B = B_0$. The imaginary part of the scattering amplitude $f_p$ is small and nonzero only for positive energies $E > 0$. Hence the p-wave Feshbach resonance is intrinsically narrow. We note that for negative energies $E < 0$, there is a triplet molecular bound state:

$$|E_b| = \frac{\pi r_0}{2mV_p} = \frac{\pi}{2m \, r_0 \, a_p}. \qquad (7.4.2)$$

In the unitary limit, the molecular binding energy $|E_b| \to 0$ for triplet p-wave molecules.

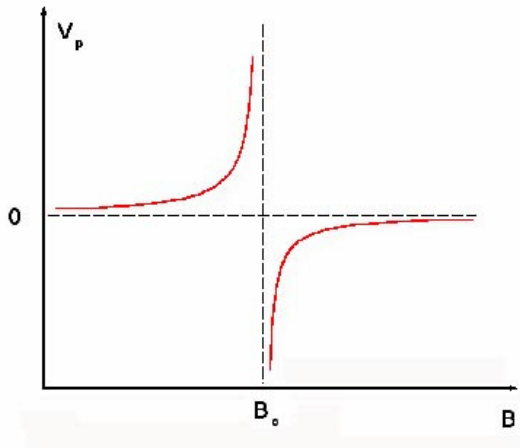

Fig. 7.12. Sketch of the p-wave Feshbach resonance. The scattering volume $V_p$ diverges at $B = B_0$ [7.3].

The first theoretical articles on p-wave Feshbach resonance often deal with mean-field two-channel description of the resonance [7.38-7.40] (see Chapter 5). However, in this Chapter we will use again an analog of a resonance approximation and will study the p-wave Feshbach resonance in the framework of one-channel description, which is more close to the physics of superfluid $^3$He and captures the essential physics of the BCS-BEC crossover in p-wave superfluids rather well.

In magnetic traps (in the absence of the so-called dipolar splitting (see [7.4-7.7])) fully (100%) polarized gas or, more precisely, one hyperfine component of the gas is usually studied. In the language of $^3$He the fermionic pairs with $S_{tot} = S_z^{tot} = 1$, or $|\uparrow\uparrow>$ - pairs are considered for p-wave triplet A1-phase in 3D.

### 7.4.2. The global phase diagram of the BCS-BEC crossover in 100%-polarized A1-phase

A qualitative picture of the global phase-diagram of the BCS-BEC crossover in 100%-polarized A1-phase is presented in Fig. 7.13. In its gross features, it resembles the phase-diagram of the BCS-BEC crossover for s-wave pairing described in Section 7.1 (see Fig. 7.3). However, there is a very interesting question about the origin of the point $\mu(T = 0) = 0$ for 3D A1-phase. We will show in what follows that at the point $\mu(T = 0) = 0$, we probably deal with a quantum phase-transition [7.7, 7.8] or even topological phase-transition [7.9, 7.25].

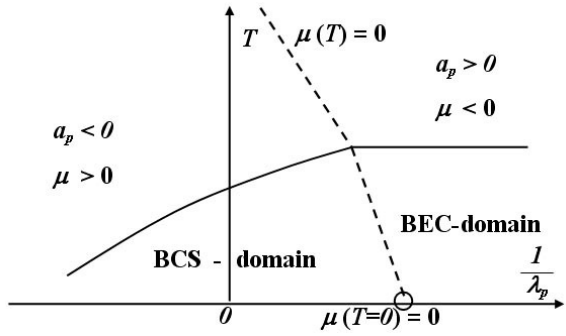

Fig. 7.13. Qualitative picture of the BCS-BEC crossover in the 100%-polarized A1-phase for p-wave superfluid. $\lambda_p = V_p \, p_F^3$ is the gas parameter. We indicate the line where $\mu \, (T) = 0$ and the point of the quantum phase-transition $\mu \, (T = 0) = 0$ [7.3].

On the global phase diagram, the BCS-domain with the chemical potential $\mu > 0$, occupies the region of the negative values of the gas parameter $\lambda_p = V_p \, p_F^3 < 0$ (or the negative values of the p-wave scattering length $a_p$). It also stretches to the small positive values of the inverse gas parameter $1/\lambda_p \leq 1$ and is separated from the BEC-domain (where $\mu < 0$ and the inverse gas parameter is large and positive $1/\lambda_p \geq 1$) by the line $\mu(T) = 0$. In the Feshbach resonance regime, the density of "up" spins $n_\uparrow = p_F^3/6\pi^2$ is usually fixed. Deep inside the BCS-domain (for small absolute values of the gas parameter $|\lambda_p| << 1$) we have the standard BCS-like formula for the critical temperature of the A1-phase:

$$T_{Cp} \simeq 0.1 \varepsilon_F \, e^{-\frac{\pi}{2|\lambda_p|}}. \qquad (7.4.3)$$

Note that the prefactor in (7.4.3) for the 100% polarized A1-phase is defined by the second order diagrams of Gor'kov and Melik-Barchudarov type [5.42] and is approximately equal to $0.1\varepsilon_F$ ( see the analysis of Kagan, Chubukov [7.41, 7.42]).

Deep in BEC-domain (for $\lambda_p << 1$) the Einstein formula is again applicable in the leading approximation for Bose-condensation of triplet p-wave molecules with the density $n_\uparrow/2$ and mass $2m$ yielding for a critical temperature

$$T_{Cp} = 3.31 \frac{\left(n_\uparrow / 2\right)^{2/3}}{2m}. \; (7.4.4)$$

Note that the triplet molecules repel each other again. In the unitary limit, $1/\lambda_p = 0$. Hence from (7.4.3) we get that $T_{Cp} \approx 0.1\varepsilon_F$ here, and we are still in the BCS-regime (as for the case of s-wave pairing considered in Section 7.1). In the rest of this Section, we consider low temperatures $T << T_C$, i.e. we will work deep in the superfluid parts of the phase diagram for the BCS and BEC-domains of the A1-phase.

### 7.4.3. Quasiparticle energy and nodal points in the A1-phase

For $\mu > 0$ (BCS-domain) there are two nodes in the spectrum for $p^2/2m = \mu$ and for angles $\theta = 0$ or $\pi$ (see the discussion in Chapter 4). For $\mu < 0$ (BEC-domain) $p^2/2m - \mu = p^2/2m + |\mu|$ and there are no nodes. The important point $\mu = 0$ is a boundary between the totally gapped BEC-domain and the BCS-domain with two nodes of the quasiparticle spectrum, which correspond to the south and north poles in Fig. 7.14. The point $\mu = 0$ for $T = 0$ is often called the point of the topological quantum phase transition [7.9, 7.25, 7.43].

### 7.4.4. Leggett equations for A1-phase

The Leggett equations for the 100%-polarized A1-phase in three dimensions are the evident generalization of the standard Leggett equations for the s-wave BCS-BEC crossover, derived in Section 7.2. The equation for the chemical potential reads:

$$n_\uparrow = \frac{p_F^3}{6\pi^2} = \int_0^{1/r_0} \frac{p^2 dp}{2\pi^2} \int_{-1}^{1} \frac{dx}{2} \left(1 - \frac{\xi_p}{E_p}\right) \frac{1}{2}, \qquad (7.4.5)$$

where $\xi_p = \left(\frac{p^2}{2m} - \mu\right)$, $E_p = \sqrt{\left(\frac{p^2}{2m} - \mu\right)^2 + \frac{\Delta_0^2 p^2}{p_F^2}\sin^2\theta}$ - is a quasiparticle spectrum, $r_0$ is the range of the potential and $x = cos\theta$. This equation defines the chemical potential $\mu$ for a fixed density $n$. The momentum distribution for the function $\frac{1}{2}\left(1 - \frac{\xi_p}{E_p}\right)$ in (7.4.5) is depicted in Fig.

7.14 for the different values of $\mu$ corresponding to the BCS and BEC domains.

The second self-consistency equation defines the magnitude of the superfluid gap $\Delta_0$. It is given by:

$$-\pi\, m\, \text{Re} \frac{1}{f_{l=1}(2\mu)} = \int_{-1}^{1} \frac{dx}{2} \int_0^{1/r_0} p^4 dp \left\{\frac{1}{E_p} - \frac{1}{\xi_p}\right\}, \qquad (7.4.6)$$

where

$$Re \frac{1}{f_{l=1}(2\mu)} = \left(\frac{1}{V_p} + \frac{4m\mu}{\pi\, r_0}\right) \qquad (7.4.7)$$

is the real part of the inverse scattering amplitude in p-wave channel for total energy $E = 2\mu$ of colliding particles.

Deep in the BCS-domain, the solution of the Leggett equations yields:

$$\Delta_0 \sim \varepsilon_F\, e^{-\frac{\pi}{2|\lambda_p|}} \sim T_{Cp};\quad \mu \approx \varepsilon_F > 0. \qquad (7.4.8)$$

In three dimensions the sound velocity coincides with the result for s-wave pairing and reads: $c_S = \left(\frac{n_\uparrow}{m}\frac{d\mu}{dn_\uparrow}\right)^{1/2} = \frac{v_F}{\sqrt{3}}$.

Deep in the BEC-domain the superfluid gap yields: $\Delta_0 \approx 2\varepsilon_F\sqrt{p_F r_0} << \varepsilon_F$ for

$p_F r_0 << 1$, and the chemical potential $\mu = -\frac{|E_b|}{2} + \frac{\mu_B}{2} < 0$, just as in the s-wave case. Note that the

binding energy of a triplet pair (molecule) $|E_b| = \frac{\pi}{2m\, r_0\, a_p}$ . Accordingly:

$$\mu_B \approx \frac{4\varepsilon_F}{3}\sqrt{p_F r_0} \qquad (7.4.9)$$

is a bosonic chemical potential which governs the repulsive interaction between two p-wave molecules on a mean-field level.

The sound velocity deep in the BEC-domain is given by:

$$c_S = \left(\frac{n_B}{2m}\frac{d\mu_B}{dn_B}\right)^{1/2} \approx \frac{v_F}{\sqrt{3}}\sqrt{p_F r_0} << v_F \qquad (7.4.10)$$

for $p_F r_0 \ll 1$, where $n_B = n_\uparrow/2$ is the bosonic density.

At $\mu \to 0$ (more rigorously, for $|\mu| < \Delta_0^2/\varepsilon_F$) we have:

$$\Delta_0(\mu = 0) = 2\varepsilon_F \sqrt{p_F r_0} \qquad (7.4.11)$$

for the magnitude of the superfluid gap.

For the gas parameter $\lambda_p$ in the point $\mu = 0$, we obtain:

$$\lambda_p(\mu = 0) = \frac{3\pi}{4} > 0. \qquad (7.4.12)$$

Hence, the interesting point $\mu = 0$ is effectively in the BEC-domain (in the domain of the positive p-wave scattering length $a_p > 0$). Accordingly, for $\mu = 0$ the binding energy is:

$$|E_b| = \frac{4}{3}\varepsilon_F p_F r_0. \qquad (7.4.13)$$

The sound velocity squared for $\mu = 0$ is given by:

$$c_S^2 = \frac{v_F^2}{3} p_F r_0 \qquad (7.4.14)$$

and coincides with (7.4.10) obtained deep in the BEC-domain. A careful analysis of the Leggett equations close to $\mu = 0$ shows that the derivative $\partial\Delta/\partial\mu$ also have no singularities at this point. The second derivative $\partial n_\uparrow^2/\partial\mu^2$ is also continuous at $\mu = 0$, and hence the anomaly appears only in higher derivatives in qualitative agreement with the numerical calculations [7.44] in three dimensions.

### The gap and compressibility close to $\mu = 0$ in 2D axial phase

Let us briefly consider the situation with the gap and compressibility close to $\mu = 0$ in the 2D case. Note that the quasiparticle energy in the 2D case reads (see Chapter 3 and [7.9, 7.24, 7.38]): $E_p = \sqrt{\left(\frac{p^2}{2m} - \mu\right)^2 + \frac{\Delta_0^2 p^2}{p_F^2}}$. It has only one nodal point $E_p = 0$ for $\mu = 0$ and $p = 0$. The anomalies in compressibility close to $\mu = 0$ in the 2D case are also stronger then in three dimensions. The careful analysis of the compressibility in two dimensions shows [7.3, 7.24] its continuous behavior, but with a kink, which is developed in $\partial n_\uparrow/\partial\mu$ in 100%-polarized axial phase for $\mu = 0$. This kink can be obtained both on the level of analytic [7.3] and numerical [7.44] calculations. To be more specific:

$$\partial n_\uparrow/\partial\mu \sim 1 + \frac{\mu\varepsilon_F}{\Delta_0^2}[1 - sign\mu], \qquad (7.4.15)$$

where $sign\mu = 1$ for $\mu \geq 0$ and -1 for $\mu < 0$.

Hence from (7.4.15) we get:

$$\partial n_\uparrow/\partial\mu \sim 1 \text{ for } \mu \to +0 \text{ and } \partial n_\uparrow/\partial\mu \to 1 + \frac{2\mu\varepsilon_F}{\Delta_0^2} \text{ for } \mu \to -0. \qquad (7.4.16)$$

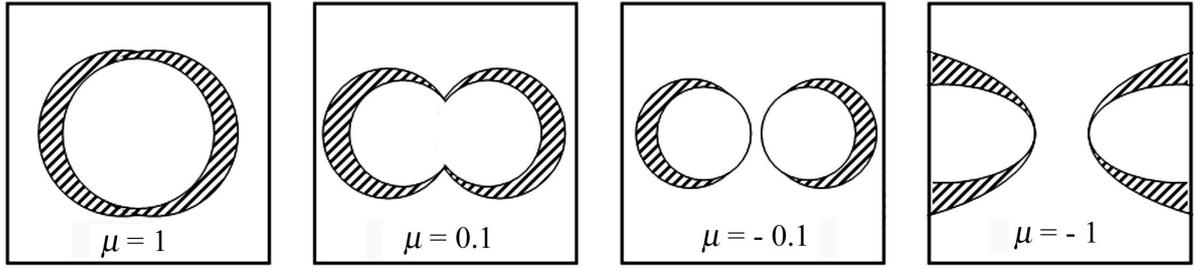

Fig. 7.15. Schematic momentum distribution of the function $\frac{1}{2}\left(1 - \frac{\xi_p}{E_p}\right)$ entering (7.4.5) in the $(p_x, p_z)$ plane for $p_y = 0$, $\Delta_0 = 1$, and $\varepsilon_F = p_F^2/2m$ in the BCS-BEC crossover for the 3D A1 phase. The different values of $\mu$ correspond to the situation deep in the BCS domain ($\mu = 1$), deep in the BEC domain ($\mu = -1$), and in the important region close to $\mu = 0$ ($\mu = +0.1$ and $\mu = -0.1$). $\mu$ is measured in terms of $\varepsilon_F$ [7.3].

7.4.5. Specific heat at low temperatures $T \ll T_C$ in the A1 phase. Classical and quantum limits. Quantum critical point $\mu(T = 0) = 0$

In this Subsection we study the thermodynamic functions, namely the specific heat $C_v$ in three-dimensional p-wave superfluids with the A1 symmetry at low temperatures $T \ll T_C$. Our goal is to find nontrivial contributions to $\rho_n$ and $C_v$ from the nodal points on the mean-field level.

### Specific heat in the three-dimensional A1-phase

The fermionic (quasiparticle) contribution to $C_v$ at the mean-field level in three dimensions reads (see [5.20]) for 100%-polarized A1-phase:

$$C_v^F = \frac{\partial}{\partial T}\int n_F\left(E_p/T\right)E_p\frac{d^3\vec{p}}{(2\pi)^3} \approx N_{3D}(0)\int_{-\infty}^{\infty}d\xi_p\int_{-1}^{1}\frac{d\cos\theta}{2}\frac{E_p^2}{T^2}\frac{e^{E_p/T}}{\left(e^{E_p/T}+1\right)^2}, \qquad (7.4.17)$$

where $N_{3D}^{\sigma}(0) = \frac{mp_F}{4\pi^2}$ is the density of states at the Fermi surface for one spin-projection $\sigma$, $n_F\left(E_p/T\right) = \frac{1}{\left(e^{E_p/T}+1\right)}$ is the quasiparticle (Fermi-Dirac) distribution function, and $E_p$ is the quasiparticle energy given by (7.4.4). For $T \ll \Delta_0$ the fermionic distribution function $n_F \approx e^{-E_p/T}$. In the same time close to the south and north poles $\Delta = \Delta_0|\theta|$ and $d\cos\theta = |\theta| |d|\theta|$. Correspondingly the results of the calculations deep in the BCS-domain (for $\mu \approx \varepsilon_F > 0$) yields (see [7.10]):

$$C_v^F \sim N_{3D}^{\sigma}(0)\frac{T^3}{\Delta_0^2}. \qquad (7.4.18)$$

Thus due to the nodal contribution we have a power-law dependence of the specific heat instead of an exponential one (typical for s-wave pairing) in the BCS A1-phase.

Deep in the BEC-domain (for $\mu \approx -|E_b|/2 < 0$) $C_v$ behaves in the exponential fashion:

$$C_v{}^F \sim \frac{(2mT)^{3/2}}{4\pi^2} \frac{E_b^2}{4T^2} e^{-\frac{|E_b|}{2T}}, \qquad (7.4.19)$$

with $|E_b|$ given by (7.4.2).

Note that bosonic (phonon) contribution from the sound waves yields:

$$C_v{}^B \sim \frac{\partial}{\partial T} \int \frac{p^2 dp}{2\pi^2} c_S p \frac{1}{\exp\{c_S p / T\} - 1} \sim \frac{T^3}{c_S{}^3} \frac{1}{2\pi^2} \qquad (7.4.20)$$

with a sound velocity $c_S \approx \frac{v_F}{\sqrt{3}}$ deep in BCS-domain and $c_S \approx \frac{v_F}{\sqrt{3}} \sqrt{p_F r_0}$ deep in the BEC-domain. Note that for an important point $\mu = 0$ the expression (7.4.20) is still valid. Moreover $c_S \approx \frac{v_F}{\sqrt{3}} \sqrt{p_F r_0}$ again.

### Specific heat of fermionic quasiparticles in classical limit close to $\mu \to 0$

Finally in the interesting region of small $\mu$ and low temperatures $T \to 0$, $|\mu| \to 0$, but $|\mu|/T \to 0$ (according to the terminogoly of Hertz, Millis [7.7, 7.8]) we have a nontrivial temperature dependence for $C_v{}^F$ (see [7.3]):

$$C_v{}^F \sim \frac{(2mT)^{3/2}}{2\pi^2} \frac{\varepsilon_F T}{\Delta_0^2}. \qquad (7.4.21)$$

This formula is valid both for $\mu \to -0$ (BEC-domain) and for $\mu \to +0$ (BCS-domain). In the same time $C_v{}^F \sim T^{5/2}$ for $(|\mu| \ll T \ll \Delta_0^2/\varepsilon_F$. Note that for small $|\mu|$ but intermediate temperatures $|\mu| \ll \Delta_0^2/\varepsilon_F \ll T \ll \Delta_0$ (when of course we are still in the classical limit) we get more expected result:

$$C_v \sim \frac{(2mT)^{3/2}}{4\pi^2}. \qquad (7.4.22)$$

Note that at low temperatures $|\mu| \ll T \ll \Delta_0^2/\varepsilon_F$ the bosonic contribution $C_v{}^B$ in (7.4.20) becomes dominant (over $C_v{}^F$ in (7.4.21)) for the total specific heat $C_v = C_v{}^F + C_v{}^B$. Note also that the condition $T \ll \Delta_0^2/\varepsilon_F$ is equivalent to the condition $T \gg mc_S{}^2$ when we analyze the bosonic contribution $C_v{}^B$ to the specific heat.

### Specific heat of fermionic quasiparticles in quantum limit

We note that in the opposite quantum limit $T \ll |\mu| \ll \Delta_0^2/\varepsilon_F$ (or equivalently $T \to 0$, $|\mu| \to 0$, but $T/|\mu| \to 0$) we get:

$$C_v{}^F \sim \sqrt{\frac{T}{|\mu|}} \frac{\varepsilon_F T}{\Delta_0^2} \frac{(2mT)^{3/2}}{4\pi^2} \text{ for } \mu \to +0 \text{ (BCS-domain)}, \qquad (7.4.23)$$

and

$$C_v{}^F \sim \frac{|\mu|^3}{T^3} e^{-\frac{|\mu|}{T}} \frac{\varepsilon_F T}{\Delta_0^2} \frac{(2mT)^{3/2}}{4\pi^2} \text{ for } \mu \to -0 \text{ (BEC-domain)}. \qquad (7.4.24)$$

Hence in the quantum limit fermionic contribution $C_v{}^F$ behaves very differently for $\mu \to -0$ and $\mu \to +0$ in contrast with the classical limit. It is important to note that for $|\mu|/T \sim 1$ results (7.4.22) and (7.4.23) coincide by order of magnitude and moreover here $C_v^F \sim \dfrac{(2mT)^{3/2}}{4\pi^2}\dfrac{\varepsilon_F T}{\Delta_0^2}$ coincides with the classical limit (7.4.21).

Summarizing we see that a power-law fermionic contribution $C_v^F \sim T^{5/2}$ at low temperatures and $C_v^F \sim T^{3/2}$ at intermediate temperatures can be separated from bosonic contribution $C_v^B \sim T^3$ close to the important point $\mu = 0$. We also see very different behavior of $C_v^F$ in BCS and BEC-domains in the quantum limit $T/|\mu| \to 0$.

Note that for $T \neq 0$ we are effectively always in the classical limit $|\mu|/T \to 0$, because the chemical potential $\mu$ is continuous close to $\mu = 0$. Hence the real phase transition occurs only at $T = 0$ (in the important point $\mu(T = 0) = 0$ (see [7.9, 7.43, 7.3, 7.25])). That is why the point $\mu(T = 0) = 0$ corresponds to the point of quantum phase transition.

### Specific heat for 100%-polarized axial phase in 2D

In the two-dimensional 100%-polarized axial phase the fermionic contribution to the specific heat deep in the BCS-domain (for $\mu \approx \varepsilon_F > 0$) reads [7.24]:

$$C_v^F \sim \frac{m}{\pi}\sqrt{\Delta_0 T}\left(\frac{\Delta_0}{T}\right)^2 e^{-\Delta_0/T} \qquad (7.4.24)$$

and we get the standard exponential behavior of the specific heat similar to the case of s-wave superconductors.

In the same time deep in the BEC-domain (for $\mu \approx -|E_b|/2 < 0$) it is given by

$$C_v^F \sim \frac{mT}{\pi}\frac{|\mu|^2}{T^2}e^{-|\mu|/T} \sim \frac{mT}{\pi}\frac{E_b^2}{4T^2}e^{-\frac{|E_b|}{2T}}. \qquad (7.4.25)$$

Close to the important point $\mu = 0$ in the classical limit $|\mu| < T < \Delta_0^2/\varepsilon_F$ (or $|\mu|/T \to 0$) the fermionic contribution to the specific heat is given by:

$$C_v^F \sim \int_0^\infty \frac{p\,dp}{2\pi}\frac{E_p^2}{T^2}\frac{e^{E_p/T}}{(e^{E_p/T}+1)^2}. \qquad (7.4.26)$$

The quasiparticle spectrum for $|\mu| < \Delta_0^2/\varepsilon_F$ can be represented as follows:

$$E_p = \sqrt{\left(\frac{p^2}{2m} - \mu\right)^2 + \frac{\Delta_0^2 p^2}{p_F^2}} \approx \sqrt{\left(\frac{p^2}{2m}\right)^2 + \frac{\Delta_0^2 p^2}{p_F^2}} \qquad (7.4.27)$$

Formally for small $|\mu|$ it is similar to Bogolubov weakly non-ideal Bose-gas with the sound velocity $(c_S^2)_{\text{Bose-gas}} = \Delta_0^2/p_F^2$. Therefore in the equation of $C_v^F$ in (7.4.26) we can utilize the well known method introduced for Bogolubov Bose-gas by Hugenholtz and Pines [7.46]. Namely we can approximate $E_p$ by the linear spectrum $E_p \approx \Delta_0 p/p_F$ and simultaneously substitute the upper limit of integration in (7.4.26) by $p_{max} = \Delta_0 2m/p_F$, where for $p_{max}$ we get $\left(\dfrac{p_{max}^2}{2m}\right)^2 = \dfrac{\Delta_0^2 p_{max}^2}{p_F^2}$ in the expression for $E_p$. Than (7.4.26) can be rewritten as:

$$C_v^{\,F} \sim \int_0^{p_{max}} \frac{p\,dp}{2\pi} \frac{\Delta_0^2 p^2}{T^2 p_F^2} \frac{e^{\frac{\Delta_0}{T}\frac{p}{p_F}}}{\left(e^{\frac{\Delta_0}{T}\frac{p}{p_F}}+1\right)^2}. \qquad (7.4.28)$$

Introducing now the new dimensionless variable $y = \dfrac{\Delta_0}{T}\dfrac{p}{p_F}$ and having in mind that

$$y_{max} = \frac{\Delta_0}{T}\frac{p_{max}}{p_F} = \frac{\Delta_0^{\,2}}{T}\frac{2m}{p_F^{\,2}} = \frac{\Delta_0^{\,2}}{\varepsilon_F T} >> 1 \text{ for } T < \Delta_0^2/\varepsilon_F \text{ we finally get:}$$

$$C_v^{\,F} \sim \int_0^\infty \frac{y\,dy}{2\pi}\, y^2\, \frac{e^y}{(e^y+1)^2}\, \frac{T^2 p_F^2}{\Delta_0^2} \sim \frac{mT}{\pi}\frac{T\varepsilon_F}{\Delta_0^2}. \qquad (7.4.29)$$

At the same time $C_v^{\,F}$ behaves very differently in quantum limit $T << |\mu| << \Delta_0^2/\varepsilon_F$ (or $T/|\mu| \to 0$). Here both in the BCS-domain (for $\mu \to +0$) and in the BEC-domain (for $\mu \to -0$):

$$C_v^{\,F} \sim \frac{mT}{\pi}\frac{\varepsilon_F T}{\Delta_0^2}\frac{|\mu|^3}{T^3}e^{-|\mu|/T}. \qquad (7.4.30)$$

Thus the difference between 3D and 2D case is the following: in both cases $C_v^{\,F}$ in quantum limit is different from $C_v^{\,F}$ in classical limit. However in 3D, on top of that, BCS and BEC results ($\mu \to +0$ and $\mu \to -0$) in quantum limit are different, while in 2D they coincide.

The fermionic contribution in quantum limit interpolates again for $|\mu|/T \sim 1$ with the classical limit $C_v^{\,F} \sim \dfrac{mT}{\pi}\dfrac{\varepsilon_F T}{\Delta_0^2}$. Finally for higher temperatures $T \sim \Delta_0^2/\varepsilon_F$ we have $C_v^{\,F} \sim \dfrac{mT}{4\pi}$.

Note that bosonic (phonon) contribution to the specific heat $C_v^{\,B} \sim \dfrac{T^2}{c_S^{\,2}}\dfrac{1}{2\pi}$ for low temperatures $T << \Delta_0^2/\varepsilon_F$ has the same order of magnitude as $C_v^{\,F}$ in (7.4.28) close to the point $\mu = 0$, since $c_S^{\,2} \sim v_F^{\,2} \geq \Delta_0^2/p_F^{\,2}$ for $\Delta_0 \leq \varepsilon_F$ (see [7.24]).

### 7.4.6. Normal density in the three-dimensional A1-phase

The fermionic (quasiparticle) contribution to the normal density in the 3D A1-phase reads (see [5.20]):

$$\rho_n^{\,F} = -\frac{1}{3}\int p^2 \frac{\partial n_F(E_p/T)}{\partial E_p}\frac{d^3\vec{p}}{(2\pi)^3}. \qquad (7.4.31)$$

Deep in the BCS-domain, the evaluation of $\rho_n^{\,F}$ yields:

$$\rho_n^{\,F} \sim \rho \frac{T^2}{\Delta_0^2}, \qquad (7.4.32)$$

where $\rho = mn\!\uparrow$ is a mass-density for up-spins. We note that rigorously speaking, Eq. (7.4.31) yields longitudinal component of the normal density tensor $\rho_{nl}$. There is a small transverse contribution also $\rho_{ntr} \sim T^4$ firstly obtained by Volovik in [7.9, 7.43].

Deep in the BEC-domain the normal density is exponential:

$$\rho_n^F \sim \frac{m}{\pi^2}(2mT)^{3/2}e^{-\frac{|E_b|}{2T}}. \qquad (7.4.33)$$

Bosonic (phonon) contribution from the sound waves is given by:

$$\rho_n^B \sim \frac{T^4}{c_S^5}. \qquad (7.4.34)$$

Eq. (7.4.33) is valid for BCS and BEC domains, and also close to the important point $\mu = 0$.

### Normal density of fermionic quasiparticles in classical and quantum limits

Close to the important point $\mu = 0$ at low temperatures $|\mu| << T << \Delta_0^2/\varepsilon_F$ (and hence in the classical limit $|\mu|/T \to 0$) we have:

$$\rho_n^F \sim \frac{m}{\pi^2}\frac{\varepsilon_F T}{\Delta_0^2}(2mT)^{3/2}. \qquad (7.4.35)$$

Eq. (7.4.35) is valid both for $\mu \to +0$ (BCS-domain) and for $\mu \to -0$ (BEC-domain). In the opposite quantum limit $T/|\mu| \to 0$ ($T < |\mu| < \Delta_0^2/\varepsilon_F$) we obtain in the BCS-domain ($\mu \to +0$):

$$\rho_n^F \sim \frac{m}{\pi^2}\frac{\varepsilon_F T}{\Delta_0^2}2mT\sqrt{2m|\mu|}. \qquad (7.4.36)$$

In the same time in BEC-domain ($\mu \to -0$):

$$\rho_n^F \sim \frac{m}{\pi^2}\frac{\varepsilon_F T}{\Delta_0^2}e^{-|\mu|/T}2m|\mu|\sqrt{2mT}. \qquad (7.4.37)$$

Therefore the behavior of $\rho_n^F$ is again very different in the BCS and BEC domains in the quantum limit.

For $|\mu|/T \sim 1$ the results (7.4.36) and (7.4.37) coincide with (7.4.35) by order of magnitude.

At intermediate temperatures $|\mu| << \Delta_0^2/\varepsilon_F << T << \Delta_0$ the quasiparticle contribution to the normal density yields:

$$\rho_n^F \sim \frac{m}{\pi^2}(2mT)^{3/2} \qquad (7.4.38)$$

as expected. Note that the bosonic contribution $\rho_n^B$ from (7.4.34) prevails at these temperatures. Thus close to $\mu = 0$ we can again separate the fermionic (quasiparticle) contribution to $\rho_n$ ($\rho_n^F \sim T^{5/2}$ at low temperatures and $\rho_n^F \sim T^{3/2}$ at intermediate temperatures) from the bosonic contribution (($\rho_n^B \sim T^4$) close to $\mu = 0$. We also see very different behavior in the BCS and BEC domains in the quantum limit $T/|\mu| \to 0$.

### Normal density in the two-dimensional A1 phase

In 100%-polarized axial phase deep in the BCS regime at $\mu \sim \varepsilon_F > \Delta_0$ the fermionic contribution to the normal density reads:

$$\rho_n^F \sim \frac{m}{\pi}m\Delta_0 e^{-\Delta_0/T}. \qquad (7.4.39)$$

Note that here the result for the BCS phase is exponential, since there is no cusps in the superfluid gap where practically gapless fermionic quasiparticles with $\Delta = \Delta_0|\theta|$ live in the 3D case.

Deep in the BEC-regime for $\mu \approx -|E_b|/2 < 0$:

$$\rho_n^F \sim \frac{m}{\pi}mT e^{-\frac{|E_b|}{2T}}. \qquad (7.4.40)$$

In the same time bosonic (phonon) contribution from the sound waves:

$$\rho_n{}^B \sim \frac{T^3}{c_S{}^4},\qquad\qquad (7.4.41)$$

and the Eq. (7.4.41) is valid not only in dilute (deep) BCS and BEC regimes, but also close to the important point $\mu = 0$.

In the classical limit $|\mu| < T < \Delta_0{}^2/\varepsilon_F$ the fermionic contribution:

$$\rho_n{}^F \sim \frac{m^2 T}{\pi}\left(\frac{T\varepsilon_F}{\Delta_0{}^2}\right)^2.\qquad\qquad (7.4.42)$$

coincides again with the bosonic contribution given by (7.4.40) by order of magnitude. At higher temperatures $T \geq \Delta_0{}^2/\varepsilon_F$: $\rho_n{}^F \sim \frac{m^2 T}{\pi}$ as expected.

In the quantum limit $T << |\mu| << \Delta_0{}^2/\varepsilon_F$ the fermionic contribution is exponential:

$$\rho_n{}^F \sim \left(\frac{|\mu|}{T}\right)^2\left(\frac{\varepsilon_F T}{\Delta_0{}^2}\right)^2 \frac{m^2 T}{\pi} e^{-|\mu|/T}.\qquad\qquad (7.4.43)$$

Eq. (7.4.43) is valid both in the BEC-domain (for $\mu \to -0$) and in the BCS-domain (for $\mu \to +0$). It interpolates with a classical limit for $|\mu|/T \sim 1$.

### 7.4.7. The spectrum of orbital waves in the three-dimensional p-wave superfluids with the symmetry of A1-phase

In Chapter 3 we briefly considered the spectrum of orbital waves in the hydrodynamic (low-frequency regime) of the bosonic (BEC) and fermionic (BCS) A-phase, having in mind first of all superfluid $^3$He-A and p-wave superfluid Fermi-gasses in the regime of Feshbach resonance. Note that in the last case we usually have 100%-polarized A1-phase with Cooper pairs having $S_z{}^{tot} = 1$ for $z$-projection of total spin, while in $^3$He-A $S_z{}^{tot} = \pm 1$ for the triplet Cooper pairs. However with respect to orbital hydrodynamics and orbital sector of collective excitations (sound waves and orbital waves), the situation in A-phase and A1-phase is similar.

With this remark we can start our considerations having in mind (see Chapter 3) that for small $\omega$ and $\vec{q}$, $\rho\omega \sim \rho q_z{}^2/m$, or equivalently

$$\omega \sim q_z{}^2/m\qquad\qquad (7.4.44)$$

for orbital waves in BEC-domain. We will show that in the weak-coupling BCS-domain

$$(\rho - C_0)\omega \sim \rho \frac{q_z{}^2}{m}\ln\frac{\Delta_0}{v_F|q_z|},\qquad\qquad (7.4.45)$$

where $C_0$ is the coefficient near anomalous current $\vec{j}_{an} = -\frac{\hbar}{4m}C_0(\vec{l}, rot\,\vec{l})\vec{l}$ in BCS A1-phase which we carefully discussed in connection with chiral anomaly and mass-current non-conservation in Chapter 3.

As we already mentioned in Chapter 3, the most straightforward way to get the spectrum is to use Galitskii, Vaks, Larkin [3.26] diagrammatic technique for collective excitations in p-wave and d-wave superfluids. The solution of the Bethe-Salpeter integral equation for the Goldstone spectrum ($\omega \to 0$ when $q \to 0$) of orbital waves in [3.26] involves the Ward identity [3.17] between the total vertex $\Gamma$ and the self-energy $\Sigma$, which is based on the generator of rotations of the $\vec{l}$ vector around perpendicular axis. In the general form, for small $\omega$ and $\vec{q} = q_z \vec{e}_z$ it is given by:

$$\int_{-1}^{1} \frac{d\cos\theta}{2}\cos^2\theta \int \frac{p^2 dp}{2\pi^2}\left[\frac{\omega^2}{8E_p^3}+\frac{\omega\xi_p}{4E_p^3}-\frac{p_F^2}{m^2}q_z^2\frac{1}{4E_p^3}\right]=0 \qquad (7.4.46)$$

Deep in the BCS domain (for $\mu \sim \varepsilon_F > 0$), we can replace $\frac{p^2 dp}{2\pi^2}$ with $N_{3D}(0)d\xi_p$ (where $N_{3D}(0) = mp_F^2/2\pi^2$) and $p_z^2/m^2$ with $v_F^2\cos^2\theta$.

This yields:

$$N_{3D}(0)\int_{-1}^{1}\frac{d\cos\theta}{2}\cos^2\theta\int d\xi_p\left[\frac{\omega^2}{8E_p^3}+\frac{\omega\xi_p}{4E_p^3}-\frac{v_F^2\cos^2\theta q_z^2}{4E_p^3}\right]=0. \qquad (7.4.47)$$

Using the estimates

$$\int_{-\varepsilon_F}^{\infty}\frac{d\xi_p}{E_p^3}=\frac{1}{\Delta_0^2\sin^2\theta} \qquad (7.4.48)$$

and

$$\int_{-\varepsilon_F}^{\infty}\frac{\xi_p d\xi_p}{E_p^3}\approx\frac{1}{\varepsilon_F}, \qquad (7.4.49)$$

we obtain

$$N_{3D}(0)\left\{\frac{\omega^2}{\Delta_0^2}\ln\frac{\Delta_0}{\omega}+\frac{\omega}{\varepsilon_F}-\frac{v_F^2 q_z^2}{\Delta_0^2}\ln\frac{\Delta_0}{v_F|q_z|}\right\}=0. \qquad (7.4.50)$$

More rigorously, the equation for the spectrum is biquadratic due to the rotation of the $\vec{l}$ vector, as it should be for bosonic excitations:

$$\left(\frac{\omega^2}{\Delta_0^2}\ln\frac{\Delta_0}{\omega}+\frac{\omega}{\varepsilon_F}\right)^2 \sim \left(\frac{v_F^2 q_z^2}{\Delta_0^2}\ln\frac{\Delta_0}{v_F|q_z|}\right)^2. \qquad (7.4.51)$$

For small frequencies $\omega < \Delta_0^2/\varepsilon_F$ the spectrum is quadratic:

$$\omega\frac{\Delta_0^2}{\varepsilon_F} \sim \frac{v_F^2 q_z^2}{\Delta_0^2}\ln\frac{\Delta_0}{v_F|q_z|}, \qquad (7.4.52)$$

or equivalently,

$$\omega\frac{\Delta_0^2}{\varepsilon_F^2} \sim \frac{q_z^2}{m}\ln\frac{\Delta_0}{v_F|q_z|}. \qquad (7.4.53)$$

Hence, comparing (7.4.53) and (7.4.45), we obtain

$$\frac{\rho-C_0}{\rho} \sim \frac{\Delta_0^2}{\varepsilon_F^2} << 1, \qquad (7.4.54)$$

and therefore $C_0 \approx \rho$ deep in BCS domain.

In superfluid $^3$He-A, for example, $\Delta_0/\varepsilon_F \sim T_C/\varepsilon_F \sim 10^{-3}$ [7.10], and hence $\frac{\rho-C_0}{\rho} \sim 10^{-6}$ as we mentioned already in Chapter 3. Note that in the strong-coupling case $\Delta_0 \geq \varepsilon_F$: $C_0 << \rho$ and we restore the bosonic spectrum (7.4.44). At the same time, for larger frequencies $\Delta_0^2/\varepsilon_F < \omega < \Delta_0$ the spectrum is almost linear:

$$\omega^2\ln\frac{\Delta_0}{\omega} \sim v_F^2 q_z^2\ln\frac{\Delta_0}{v_F|q_z|}. \qquad (7.4.55)$$

Deep in the BEC-domain for $\mu \approx -|E_b|/2 < 0$, it follows from (7.4.44) that

$$\omega^2 + |\mu|\omega \sim |\mu|\frac{q_z^2}{m}. \qquad (7.4.56)$$

Of course, the exact equation is again biquadratic due to the rotation:

$$\left(\omega^2 + |\mu|\omega\right)^2 \sim \left(|\mu|\frac{q_z^2}{m}\right)^2. \qquad (7.4.57)$$

Hence, $\omega \sim q_z^2 / m$ for $\omega < |\mu|$ in agreement with (7.4.44). Moreover this means that $\dfrac{\rho - C_0}{\rho} = 1$

deep in the BEC-domain, and therefore $C_0 = 0$.

Summarizing Section 7.4 we should emphasize, that:

(i) Motivated by the recent experiments on p-wave Fechbach resonance in ultracold Fermi-gasses, we solve the Leggett equations and construct the phase-diagram of the BCS-BEC crossover in 100%-polarized A1-phase.

(ii) We evaluate compressibility, low temperature specific heat and normal density and find the indications of quantum phase transition close to the important point $\mu(T = 0) = 0$. These indications are connected with a cusp in compressibility as well as with different behavior of $C_v$ and $\rho_n$ in the classical ($|\mu|/T \to 0$) and the quantum ($T/|\mu| \to 0$) limits.

Volovik, Green, Read call this point – the point of topological quantum phase transition. It separates gapless from the gapped regions on the phase diagram.

(iii) Deep in BCS and BEC domains the crossover ideas of Leggett and Nozieres-Schmitt-Rink work pretty well. The phase diagram in these regions resembles in gross features the phase diagram of BCS-BEC crossover for s-wave pairing in resonance superfluids considered in Section 7.1 and 7.2.

(iv) We derived the spectrum of collective excitations (including sound waves and orbital waves) in p-wave superfluids with the symmetry of A1-phase.

Reference list to Chapter 7

Chapter 9. Superconductivity in the low-density electron systems with repulsion.

Content

9.1. Kohn-Luttinger mechanism of SC in purely repulsive Fermi-systems.

The modern physics of superconductivity (SC) is a very rapidly progressing field of condensed-matter physics where the colossal intellectual efforts of the researchers are concentrated and the new knowledge is accumulated very intensively giving rise to the development of neighboring areas of physics, chemistry, material science and engineering.

The latest progress in this area during last 25 years of experimental and theoretical research in connected with the physics of high-$T_C$ superconductivity in cuprates and other related materials such as plumbates-bismithates, magnesium diborides, superconductors based on FeAs and so on.

One of the most essential and unresolved question in this area connected with the mechanism of superconducting pairing: whether it is of electron-phonon origin as in standard BCS-like [30] superconductors such as Hg, Pb, Nb, Al, or is due to electron-electron interaction as in new unconventional superconductive systems such as ruthenates, organic superconductors, heavy-fermion compounds and so on.

In this chapter we will advocate non- phonon mechanism of superconductivity based on electron-electron interaction. Here according to Prof. P.W. Anderson [9.1] there are two basic questions:

1)    to convent the sign of the Coulomb interaction;

2)    to understand the properties of the normal state in high-$T_C$ materials and other unconventional SC-systems.

We agree with these statements. We will address them in Galitskii-Bloom [9.2, 9.3] Fermi-gas approach for low density electron systems. We will prove the existence of SC at low-density limit, in purely repulsive Fermi-systems where we are far from antiferromagnetic (AFM) and structural instabilities. Moreover in this limit we can develop a regular perturbation theory.

The small parameter of the problem is gas-parameter:

$$|a|p_F \ll 1; \hbar = 1, \qquad (9.1.1)$$

where $a$ – is the s-wave scattering length, $p_F$ – is Fermi-momentum. Critical temperatures ($T_C$-s) which we obtain are not very low. Our theory often "works" even for rather high densities because of intrinsic nature of SC-instability. Our philosophy throughout this chapter is to solve exactly low-density limit and then go to higher densities. The basic mechanism which we address here is enhanced Kohn-Luttinger mechanism of SC. We will also check the normal state of low-density electron systems with respect to marginality [9.5].

### 9.2. Unconventional superconductive systems.

During last 30 years there is a huge progress in experimental and theoretical investigation of unconventional (anomalous) superconductive systems. Among the new materials with anomalous superconductive pairing there are examples of triplet p-wave ($S_{tot} = l = 1$) and singlet d-wave ($S_{tot} = 0$; $l = 2$) superconductors as well as multiband s-wave SC-systems ($S = l = 0$).

The p-wave SC with orbital momenta of the Cooper pair $l = 1$ and total spin of the pair $S_{tot} = 1$ is realized in:

- fermionic isotopes of alkali elements $^6$Li and $^{40}$K in magnetic traps in the regime of Feshbach resonance at ultralow temperatures $T_C \sim (10^{-6} \div 10^{-7})$ K [9.8, 9.9]

- superfluid A and B phases of $^3$He at very low critical temperatures $T_C \sim 1$ mK [9.6, 9.7]

- heavy-fermion superconductors $U_{1-x}Th_xBe_{13}$, $UNi_2Al_3$, $T_C \sim (0.5 \div 1)$ K, heavy electron mass m* $\sim (100 \div 200)m_e$ due to strong correlations [9.10]

- ruthenates $Sr_2RuO_4$, $T_C \sim 1$ K (part of the community assumes d-wave pairing in these materials) [9.12]

- organic superconductor $\alpha$-(BEDT-TTF)$_2$I$_3$, $T_C \sim 5$ K [9.11].

Singlet d-wave pairing is realized in:
- heavy-fermion SC UPt$_3$, $T_C \sim 0.5$ K, large effective mass m* $\sim 200$ m$_e$
- high-$T_C$ superconductors [9.13] $La_{2-x}Sr_xCuO_4$, $YBa_2Cu_3O_{7-\delta}$, $Bi_2Sr_2Ca_2Cu_3O_{10}$, $Tl_2Sr_2Ca_2Cu_3O_{10}$, $Hg_2Ba_2Ca_2Cu_3O_{10}$.

In all the families of high-$T_C$ materials the elementary block CuO is present. They are called cuprates. $T_C$-s are in the range from 36 K for lanthanum-based family to 160 K for Hg-based family under pressure (the record established $T_C$ in cuprates for today). Note that part of the community still assumes standard s-wave pairing in cuprates.

The highest $T_C$ in unconventional SC corresponds to neutron stars which consist of 98% of bineutrons and 2% of biprotons. For bineutrons $S_{tot} = l = 1$, but there is a strong spin-orbit coupling. So the total rotating moment $J = \left| \vec{S}_{tot} + \vec{l} \right| = 2$. According to theoretical predictions $T_C \sim (10^8 \div 10^{10})$ K for neutron stars.

Finally in the end of this section we would like to mention several unconventional SC systems of the multiband character with s-wave pairing, namely MgB$_2$ [9.14] (very promising for technical applications in electronics and energetics), and recently discovered family on the basus of FeAs such as BaFe$_2$(As$_{1-x}$P$_x$)$_2$ (with the coexistence of ferromagnetic fluctuations and SC) [9.15]. Note that part of the experimental community still hopes to demonstrate p-wave or d-wave SC in FeAs-based compounds [9.14c].

### 9.3. 3D and 2D Fermi-gas with repulsion. Triplet p-wave pairing.

The basic model to study non-phonon mechanism of SC in low density electron systems is a Fermi-gas model. In Fermi-gas with attraction s-wave scattering length $a < 0$ and we have unconventional s-wave pairing ($S = l = 0$) with a critical temperature:

$$T_{CO} \approx 0.28\, \varepsilon_F\ e^{-\frac{\pi}{2|a|p_F}} \qquad (9.3.1)$$

This result was obtained by Gor'kov and Melik-Barhudarov [9.16] soon after the appearance of a famous BCS-theory [9.0] and differs from the BCS-result in preexponential factor $0.28\varepsilon_F$ (instead of Debye-frequency $\omega_D$ in the phonon models typical for unconventional SC systems).

### 9.3.1. 3D Fermi-gas with repulsion.

In Fermi-gas with repulsion $a > 0$ – repulsive interaction of two particles in vacuum. We will show that for the effective interaction of two particles in substance (via polarization of a fermionic background) we can convert the sign of the interaction in the triplet p-wave channel (for $S_{\text{tot}} = l = 1$) and get:

$$T_{C1} \sim \varepsilon_F e^{-\frac{1}{(ap_F)^2}}, \qquad (9.3.2)$$

where $\varepsilon_F = p_F^2/2m$ is Fermi-energy. This highly nontrivial result was obtained by Fay, Laser [9.17] and Kagan, Chubukov [9.18]. The most important is to understand what is effective interaction $U_{eff}$? We will show that in momentum space in first two orders of the gas-parameter (9.1.1):

$$U_{eff}(\vec{p}, \vec{k}) = \frac{4\pi a}{m} + \left(\frac{4\pi a}{m}\right)^2 \Pi(\vec{p} + \vec{k}) \qquad (9.3.3)$$

where $\Pi(\vec{p} + \vec{k})$ is a static polarization operator. It is given by a standard formula [9.19, 9.20]:

$$\Pi(q) = \int \frac{d^3\vec{p}}{(2\pi)^3} \frac{n_F(\varepsilon_{p+q}) - n_F(\varepsilon_p)}{(\varepsilon_p - \varepsilon_{p+q})} \qquad (9.3.4)$$

where $\varepsilon_p = \dfrac{p^2}{2m}$; $\varepsilon_{p+q} = \dfrac{(\vec{p} + \vec{q})^2}{2m}$ are the energy spectra and $n_F(\varepsilon_p) = \dfrac{1}{(e^{\varepsilon_p/T} + 1)}, n_F(\varepsilon_{p+q})$ are Fermi-Dirac distribution functions. At low temperatures $T \ll \varepsilon_F$ polarization operator besides a regular part contains a Kohn's anomaly [9.21] of the form:

$$\Pi_{\text{sing}}(q) \sim (q - 2p_F)\ln|q - 2p_F| \quad \text{in 3D} \qquad (9.3.5)$$

In real space Kohn's anomaly leads to Friedel oscillations (RKKY interaction) [9.22]. For dimensionless product of $U_{eff}$ and 3D density of states:

$$N_{3D}(0) = \frac{mp_F}{2\pi^2} \qquad (9.3.6)$$

we have in real space (see Fig. 9.1.):

$$N_{3D}(0)U_{eff}(r) \sim (ap_F)^2 \frac{\cos(2p_F r)}{(2p_F r)^3} \qquad (9.3.7)$$

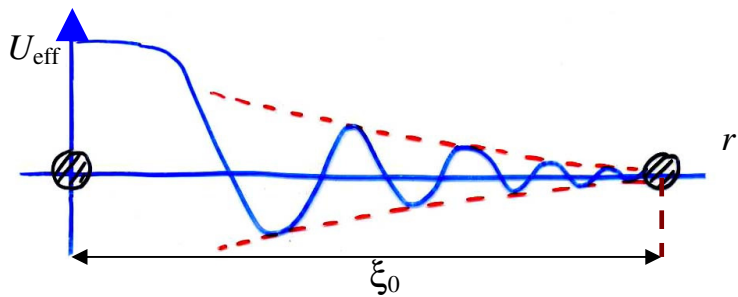

Fig. 9.1. Friedel oscillations in the effective interaction of two particles via polarization of a fermionic background. $\xi_0$ is a coherence length of a Cooper pair.

Thus we start from pure hard-core repulsion in vacuum and get the competition between repulsion and attraction in substance.

It is possible to show [9.4] that a singular part of $U_{\text{eff}}$ "plays" in favor of attraction for large orbital momenta $l \gg 1$, while a regular part of $U_{\text{eff}}$ – in favor of repulsion. Standard s-wave pairing is suppressed by hard-core repulsion. However as it was shown in [9.17, 9.18] the Kohn-Luttinger effect can be extended from the large momenta $l \gg 1$ to $l = 1$ and the attractive contribution is dominant even for p-wave channel. The exact solution [9.17, 9.18] yields for the critical temperature:

$$T_{C1} \sim \varepsilon_F e^{-\frac{5\pi^2}{4(2\ln 2 - 1)(ap_F)^2}} \qquad (9.3.8)$$

### 9.3.2. Triplet p-wave pairing.

Note that diagrammatically effective interaction $U_{\text{eff}}$ corresponds to the irreducible bare vertex and is given by the set of diagrams which cannot be separated by two lines running in the same direction [9.19, 9.20] (thus $U_{\text{eff}}$ does not contain the Cooper loop). Correspondingly the formula (9.3.3) can be represented as:

$$U_{\text{eff}}(\vec{p}, \vec{k}) = \frac{4\pi a}{m} + \left(\frac{4\pi a}{m}\right)^2 \Phi(\vec{p}, \vec{k}), \qquad (9.3.9)$$

where $\Phi(\vec{p}, \vec{k})$ is given by 4 diagrams of Kohn-Luttinger [9.4] (see Fig. 9.2.)

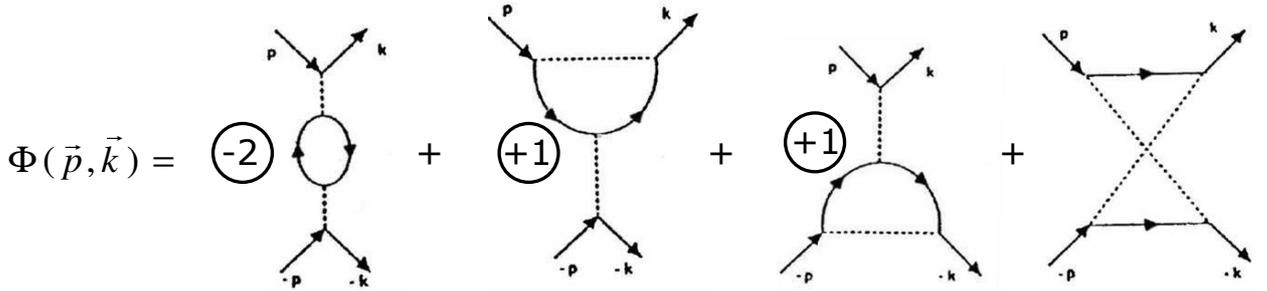

Fig. 9.2. Four diagrams of Kohn-Luttinger which give the contribution to the irreducible bare vertex for the Cooper channel.

First three diagrams cancel each other exactly for contact interaction and as a result in accordance with (9.3.3):

$$\Phi(\vec{p}, \vec{k}) = \Pi(\vec{p} + \vec{k}) \qquad (9.3.10)$$

is given by fourth diagram on Fig. 9.2. Algebraically $\Pi(\vec{p} + \vec{k})$ is a static polarization operator (9.3.4) in a "crossed" channel (for $\tilde{q} = \left|\vec{p} + \vec{k}\right|$ instead of $q = \left|\vec{p} - \vec{k}\right|$). For incoming and outgoing momenta $\vec{p}$ and are $\vec{k}$ lying on the Fermi-surface $\left|\vec{p}\right| = \left|\vec{k}\right| = p_F$ the "crossed" momentum:

$$\tilde{q}^2 = 2p_F^2(1 + \cos\theta), \qquad (9.3.11)$$

where an angle $\theta = \angle \vec{p}\vec{k}$.

The polarization operator $\Pi(\tilde{q})$ is given by static Lindhard function [9.23]:

$$\Pi(\tilde{q}) = \frac{N_{3D}(0)}{2}\left[1 + \frac{4p_F^2 - \tilde{q}^2}{4p_F\tilde{q}}\ln\frac{2p_F + \tilde{q}}{\left|2p_F - \tilde{q}\right|}\right] \qquad (9.3.12)$$

Integration of $\Pi(\tilde{q})$ with first Legendre polynomial $P_1(\cos\theta) = \cos\theta$ yields the desired result for p-wave harmonic [9.17, 9.18]:

$$\Pi_1 = \int_{-1}^{1} P_1(\cos\theta)\frac{d\cos\theta}{2}\Pi(\tilde{q}(\cos\theta)) = \frac{N_{3D}(0)}{5}(1-2\ln 2) < 0. \qquad (9.3.13)$$

The p-wave critical temperature reads:

$$T_{C1} \sim \frac{2e^C \varepsilon_F}{\pi} e^{-\frac{13}{\lambda^2}}, \qquad (9.3.14)$$

where $\lambda = 2a\, p_F/\pi$ in 3D is effective gas-parameter of Galitskii [9.2], $C$ is Euler constant.

### 9.3.3. Model-independent considerations of Prof. P.Nozieres.

The model-independent proof of the possibility of the p-wave pairing in Fermi-systems with repulsion belongs to Prof. P.Nozieres [9.24]. His way of reasoning is the following: usually for static effective interaction in fermionic substance we have [9.23, 9.25]:

$$U_{eff}(q) = \frac{U_0(q)}{\varepsilon(q)}, \qquad (9.3.15)$$

where $\varepsilon(q) = 1 + U_0(q)\Pi(q)$ is static dielectric function and $\Pi(q)$ is polarization operator given by (9.3.4). It is known from solid-state physics [9.21, 9.23] that $\Pi(q)$ is decreasing function of $q$ which behaves as follows (see Fig. 9.3.).

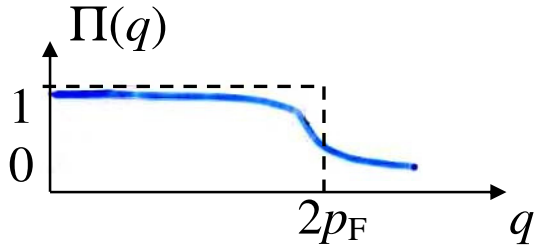

Fig. 9.3. Behavior of static polarization operator $\Pi(q)$ as a function of $q$ with the decrease for $0 < q \leq 2\, p_F$.

That is why effective interaction $U_{eff}$ in (9.3.15) decreases in the interval $0 \leq q \leq 2\, p_F$ which is important for superconductivity (see Fig. 9.4.).

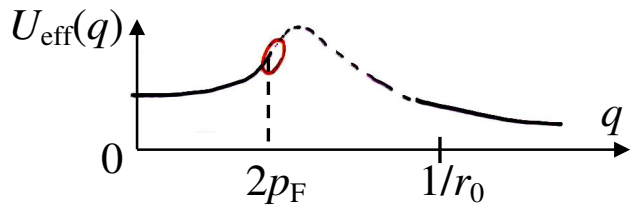

Fig. 9.4. Behavior of effective interaction $U_{eff}$ as a function of $q$.

Note that $U_{eff}(q)$ should strongly decrease for $q > 1/r_0$, where $r_0$ is the range of the potential. But in the Fermi-gas $1/r_0 \gg 2\, p_F$ according to (9.3.1) and hence an interval $[0, 2\, p_F]$ where $U_{eff}(q)$ increases is an intermediate asymptotics (see Fig. 9.4).

Finally for first Legendre polynomial (which corresponds to p-wave pairing) we have:

$$q^2 = 2\, p_F^2 (1-\cos\theta), \qquad (9.3.16)$$

and accordingly

$$P_1(\cos\theta) = \cos\theta = 1 - \frac{q^2}{2\, p_F^2} \qquad (9.3.17)$$

Moreover $\cos\theta = 1$ for $q = 0$ and -1 for $q = 2\, p_F$. Thus $U_{eff}(\cos\theta)$ and $\cos\theta$ behave as follows (see Fig. 9.5.) as functions of $\cos\theta$.

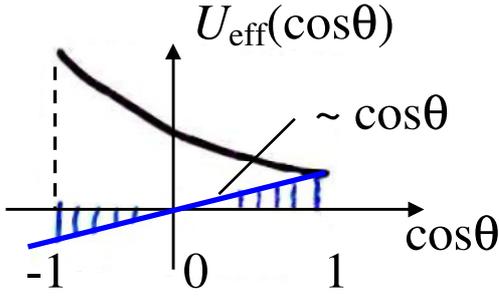

Fig. 9.5. Behavior of $U_{\text{eff}}(\cos\theta)$ and $\cos\theta$ as a function of $\cos\theta$ on the relevant interval [-1,1].

As a result the p-wave harmonic of $U_{\text{eff}}$:

$$U_{eff}^{l=0} = \int\limits_{-1}^{1} U_{eff}(\cos\theta)\cos\theta\,\frac{d\cos\theta}{2} < 0 \qquad (9.3.18)$$

if $U_{\text{eff}}(\cos\theta)$ for $\cos\theta \in$ [-1, 0] is larger then $U_{\text{eff}}(\cos\theta)$ for $\cos\theta \in$ [0, 1] as in Fig. 9.5. Hence for all effective potential increasing for $q$ changing form 0 to $2p_F$ we have p-wave SC in the isotropic case and the Kohn's anomaly does not play a decisive role here (in contrast with the case of SC pairing with large orbital momenta $l >> 1$). Note that in 3D Fermi-gas model $U_{\text{eff}}(\cos\theta)$ is given by (9.3.3) and decreases on the interval [-1,1] due to crossing $q \rightarrow \tilde{q}$ (see (9.3.9)).

### 9.3.4. Two-dimensional case.

In 2D Fermi-gas with repulsion dimensionless effective interaction in momentum space reads:

$$N_{2D}(0)U_{eff}(\tilde{q}) = f_0 + f_0^2\Pi(\tilde{q})\frac{4\pi}{m}, \qquad (9.3.19)$$

where $N_{2D}(0) = \dfrac{m}{2\pi}$ is 2D density of states,

$$f_0 = \frac{1}{2\ln(1/p_F r_0)} \qquad (9.3.20)$$

is 2D gas-parameter of Bloom [9.3]. Correspondingly in real space effective interaction

$$N_{2D}(0)U_{eff}(r) \sim f_0^2\frac{\cos(2p_F r)}{(2p_F r)^2} \qquad (9.3.21)$$

contains much more stronger 2D Friedel oscillations [9.22].

However the 2D Kohn's anomaly [9.21] in polarization operator $\Pi(q)$ has one-sided character

$$N_{2D}(0)U_{eff}(\tilde{q}) \sim f_0^2\,\text{Re}\sqrt{\tilde{q}-2p_F} = 0 \qquad (9.3.22)$$

for the important interval for SC $0 \le \tilde{q} \le 2p_F$. Hence strong 2D Kohn's anomaly is ineffective for SC in second order of perturbation theory with respect to gas-parameter $f_0$ (9.3.20). As it was shown by Chubukov [9.26] SC appears only in the third order of perturbation theory where the Kohn's anomaly changes its character first diagram on (Fig. 9.6) and reads:

$$N_{2D}(0)U_{eff}(\tilde{q}) \sim f_0^3\,\text{Re}\sqrt{2p_F - \tilde{q}} \ne 0 \qquad (9.3.23)$$

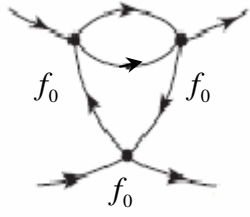

Fig. 9.6. Third order diagrams which contribute to effective interaction and SC in 2D Fermi-gas according to [9.26, 9.28]. First diagram on this figure was originally evaluated in [9.26] by Chubukov.

Accordingly critical temperature of the p-wave pairing reads in [9.26]:

$$T_{C1} \sim \varepsilon_F \exp\left\{-\frac{1}{4.1 f_0^3}\right\} \qquad (9.3.24)$$

Note that the first third order diagram on Fig. 9.6 considered in [9.26] is still irreducible with respect to Cooper channel.

Later on the authors of [9.27] and [9.28] considered all irreducible skeleton diagrams of the third order in $f_0$ on Fig. 9.6 on equal grounds and obtained $-6.1 f_0^3$ (instead of $-4.1 f_0^3$) in the exponent of (9.3.24) for p-wave pairing critical temperature in 2D Fermi-gas.

## 9.4. SC in 3D and 2D Hubbard model with repulsion at low electron density.

One of the basic models to describe unconventional normal and superconductive states of strongly-correlated electron systems is a famous Hubbard model originally introduced by Hubbard [9.29] for the explanation of metal-insulator transition in half-filled narrow band metals. In real space the Hamiltonian of the Hubbard model reads:

$$\hat{H}' = \hat{H} - \mu\hat{N} = -t\sum_{<i\,j>\sigma}c_{i\sigma}^+ c_{j\sigma} + U\sum_i n_{i\uparrow}n_{i\downarrow} - \mu\sum_i n_{i\sigma}, \qquad (9.4.1)$$

where $n_{i\sigma} = c_{i\sigma}^+ c_{i\sigma}$ is electron density with spin-projection $\sigma = |\uparrow\downarrow>$ on site $i$, $t$ – is hopping integral, $U$ is on site Hubbard repulsion between two electrons with opposite spin-projections, $\mu$ is chemical potential.

### 9.4.1. 3D Hubbard model at low density.

After Fourier transformation the Hamiltonian (9.4.1) reads:

$$\hat{H}' = \sum_{\vec{p}\sigma}\left[\varepsilon(p) - \mu\right]c_{p\sigma}^+ c_{p\sigma} + U\sum_{\vec{p}\vec{p}'\vec{q}}c_{\vec{p}\uparrow}^+ c_{\vec{p}'+\vec{q}\downarrow}^+ c_{\vec{p}+\vec{q}\downarrow}c_{\vec{p}'\uparrow}, \qquad (9.4.2)$$

where $\varepsilon(p) = -2t\left(\cos p_x d + \cos p_y d + \cos p_z d\right)$ is uncorrelated electron spectrum in 3D simple cubic lattice, $d$ is intersite distance [9.29, 9.30, 9.31]. For small electron densities $nd^3 < 1$ (where $n = p_F^3/3\pi^2$ – electron density in 3D) the spectrum reads:

$$\varepsilon(p) \approx -\frac{W}{2} + tp^2 d^2, \qquad (9.4.3)$$

where $W = 12t$ is a bandwidth in 3D for the simple cubic lattice. If we introduce the uncorrelated band mass according to [9.32]:

$$m = \frac{1}{2td^2}, \qquad (9.4.4)$$

then for the spectrum we have:

$$\varepsilon(p) = -\frac{W}{2} + \frac{p^2}{2m}. \qquad (9.4.5)$$

Correspondingly for the chemical potential at low densities:

$$\mu = -\frac{W}{2} + \varepsilon_F. \qquad (9.4.6)$$

Thus according to [9.30, 9.31, 9.32] the 3D Hubbard model at low electron densities becomes equivalent to 3D Fermi-gas with δ-functional (hard-core) repulsive interaction between particles. The s-wave scattering length $a$ in the renormalization scheme of Kanamori for Hubbard interaction [9.33, 9.34] reads:

$$\frac{4\pi a}{m} = T = \frac{Ud^3}{1 - Ud^3 K_{vac}(0,0)}, \qquad (9.4.7)$$

where $T$ – is a T-matrix, $K_{vac}(0,0) = -\int \frac{d^3\vec{p}}{(2\pi)^3} \frac{m}{p^2}$ - is a Coper loop for two particles in vacuum for total frequency $\Omega = 0$ and total momentum of two particles $\vec{P} = \vec{p}_1 + \vec{p}_2 = 0$.

Diagrammatically $K_{vac}$ is a product of two vacuum Green-functions, while Lipman-Shwinger equation [9.35] for the T-matrix is illustrated on Fig. 9.7

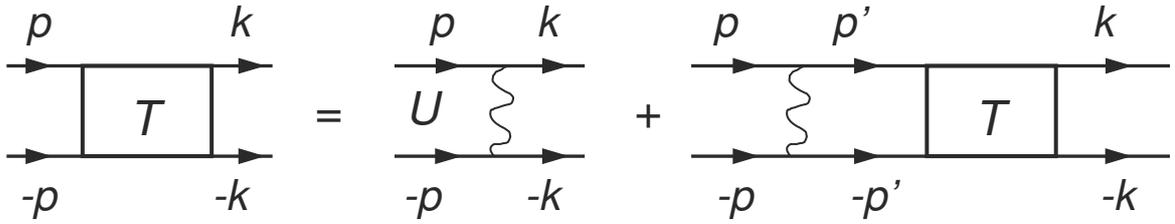

Fig. 9.7 Lipman-Shwinger equation (Bethe-Solpeter equation in vacuum) for the T-matrix [9.19, 9.35] in Kanamori renormalization scheme for Hubbard interaction.

In the strong-coupling limit $U >> W$ of the Hubbard model and at low density we have according to [9.33] (see also [9.36]):

$$a \sim d \qquad (9.4.8)$$

and thus effective gas-parameter for the Hubbard model reads:

$$\lambda = \frac{2ap_F}{\pi} = \frac{2dp_F}{\pi} \qquad (9.4.9)$$

Correspondingly, in analogy with 3D Fermi-gas model, the Hubbard model at low electron density is unstable towards triplet p-wave pairing below the critical temperature [9.30-9.32]

$$T_{C1} \sim \varepsilon_F \exp\left\{-\frac{13}{\lambda^2}\right\}, \qquad (9.4.10)$$

where λ is given by (9.4.9). Note that in the absence of the lattice for strong short-range interaction $a \sim r_0$, where $r_0$ is the range of the potential.

### 9.4.2. 2D Hubbard model.

The uncorrelated electron spectrum for 2D square lattice reads:

$$\varepsilon(p) = -2t\left(\cos p_x d + \cos p_y d\right) \qquad (9.4.11)$$

Correspondingly, for low electron density $nd^2 << 1$ (where $n = p_F^2/2\pi$ is 2D electron density) the spectrum reads again (see (9.4.4) and (9.4.5)):

$$\varepsilon(p) \approx -\frac{W}{2} + \frac{p^2}{2m}, \qquad (9.4.12)$$

the band mass is still $m = \dfrac{1}{2td^2}$, but now $W = -8t$ - a bandwidth for 2D square lattice.

Correspondingly, the chemical potential $\mu \approx -\dfrac{W}{2} + \varepsilon_F$ is given again by (9.4.6).

It is convenient to introduce dimensionless 2D density of electrons:

$$n_{el} = \frac{2\varepsilon_F}{W} \qquad (9.4.13)$$

which measures the filling of the band in terms of the half-filled band ($n_{el} = 1$ for $\varepsilon_F = \dfrac{W}{2}$ in the half-filled case actual for high-$T_C$ superconductors).

The gas-parameter of 2D Hubbard model at the strong-coupling limit $U >> W$ and low density was derived in [9.37] and reads:

$$f_0 = \frac{1}{\ln\left(4W / \varepsilon_F\right)} = \frac{1}{\ln\left(8 / n_{el}\right)} \qquad (9.4.14)$$

Correspondingly, in similarity with 2D Fermi-gas model the normal state of 2D Hubbard model at low density is unstable towards triplet p-wave pairing below the critical temperature [9.26-9.28, 9.31, 9.32].

$$T_{C1} \sim \varepsilon_F \exp\left\{-\frac{1}{6.1 f_0^3}\right\} \qquad (9.4.15)$$

with gas-parameter $f_0$ given by (9.4.14).

### 9.4.3. Qualitative phase-diagram at low density in 2D.

It is interesting to construct qualitative phase-diagram for different SC-instabilities in the 2D Hubbard model at low electron densities. This project was realized in [9.32]. To do that it is important to note (see [9.32]) that p-wave pairing with critical temperature given by (9.4.15) is the most energetically beneficial (corresponds to the highest $T_C$) for $U >> t$ and dimensionless electron density $0 < n_{el} = \dfrac{2\varepsilon_F}{W} \leq 2/3$. In the same time in the weak-coupling case $U \leq 0.3\ t$ and $0 < n_{el} \leq 2/3$ the highest $T_C$ corresponds to d$_{xy}$-pairing according to [9.32] and [9.38]. For d$_{xy}$-pairing it is important to take into account the quadratic corrections to parabolic spectrum:

$$\varepsilon(p) - \mu = -2t\left(\cos p_x d + \cos p_y d\right) - \mu = \frac{p^2 - p_F^2}{2m} - \frac{(p_x^4 + p_y^4)d^2}{24m} \qquad (9.4.16)$$

We should also remember that the superconductive gap for d$_{xy}$-pairing is given by [9.38]:

$$\Delta_{d_{xy}} \sim \Delta_0^d \sin(p_x d) \sin(p_y d) \sim \Delta_0^d p_x p_y d^2 \sim \Delta_0^d \sin 2\varphi \qquad (9.4.17)$$

for low electron density, where $\varphi$ is the angle between momentum $\vec{p}$ and $x$-axis of the square lattice. In the same time more traditional d$_{x2-y2}$-pairing actual for optimally doped high-$T_C$ materials [9.39] correspond to the gap [9.38, 9.40]:

$$\Delta_{d_{x^2-y^2}} \sim \Delta_0^d (\cos p_x d - \cos p_y d) \sim \Delta_0^d (p_x^2 - p_y^2) d^2 \sim \Delta_0^d \cos 2\varphi \qquad (9.4.18)$$

The critical temperature of d$_{xy}$-pairing is given by (see [9.38]):

$$T_C^{d_{xy}} \sim \varepsilon_F \exp\left\{-\frac{20}{f_0^2 n_{el}^2}\right\} \qquad (9.4.19)$$

In contrast to p-wave pairing it is described by the second-order (in the gas-parameter $f_0$) contribution to the effective interaction $U_{eff}$. This result [9.38] for d$_{xy}$-pairing was confirmed later on by Zanchi and Schulz [9.41] in the framework of renormalization group (RG) approach. We can also mention in this respect the papers of Hlubina et al. in which the authors get $d_{xy}$-pairing in weak-coupling case for $n_{el} \leq 0.62$ [9.64]. Finally for larger electron densities $n_{el} \geq 2/3$ both in weak-coupling ($U \ll W$) and strong-coupling case $d_{x2-y2}$-pairing (which is more conventional for optimally doped high-$T_C$ materials) is realized [9.42-9.47]. As a result the qualitative phase-diagram of the repulsive-$U$ Hubbard model at low and moderate electron densities in 2D case looks like as follows (see Fig. 9.8).

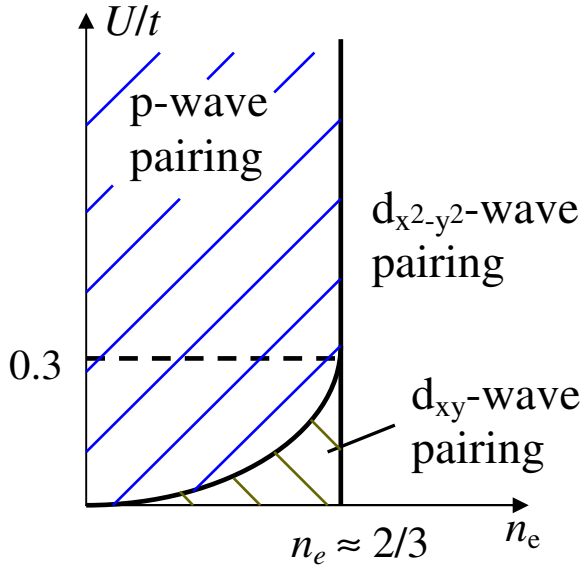

Fig. 9.8. Qualitative phase-diagram for SC- pairing in the 2D Hubbard model at low and moderate electron densities.

### 9.4.4. SC in 2D Hubbard model at larger electron densities $n_{el} \leq 1$.

At larger densities $n_{el} \leq 1$ close to half-filling $\left( \varepsilon_F \approx \dfrac{W}{2} \right)$ the spectrum of electrons becomes almost hyperbolic [9.44, 9.45]:

$$\varepsilon(p) = \pm \frac{(p_x^2 - p_y^2)}{2m} \qquad (9.4.20)$$

close to the corner points where the Fermi-surface almost touches the Brillouin zone $(0,\pi)$; $(0,-\pi)$ and $(\pi,0)$; $(-\pi,0)$ (see Fig. 9.9).

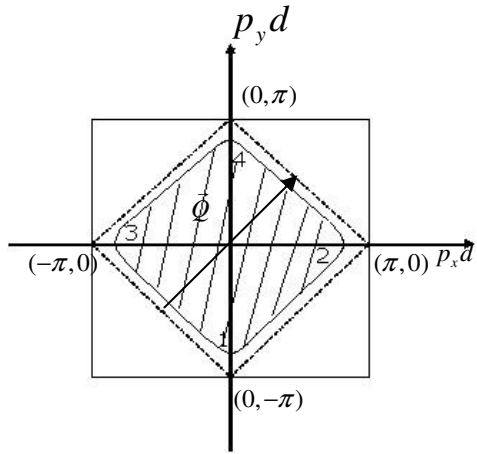

Fig. 9.9. Almost half-filling situation ($n_{el} \to 1$) for the 2D Hubbard model on the square lattice. At the corner points $(0,\pi)$; $(0,-\pi)$ and $(\pi,0)$; $(-\pi,0)$ the Fermi-surface almost touches the Brillouin zone. We also show the nesting-vector $\vec{Q} = \left( \dfrac{\pi}{d}, \dfrac{\pi}{d} \right)$ on the figure.

There are here the extended almost flat parts of the Fermi-surface which satisfy the perfect nesting criteria for exactly half-filled case ($n_{el} = 1$).

$$\varepsilon(\vec{P} + \vec{Q}) = -\varepsilon(\vec{P}),\qquad(9.4.21)$$

where $\vec{Q} = (\pi/d, \pi/d)$ is a nesting-vector for the 2D square lattice.

As a result the Kohn's anomaly becomes logarithmically strong here as in 1D-case (see [9.21,9.45]). Additional increase for $T_C$ is due to Van-Howe singularity in the density of states $N_{2D}(0)$ which is also logarithmically strong [9.23,9.45]. As a result both polarization operator $\Pi(q)$ in (9.3.4) and the Cooper loop (in substance) $K(0,0) = \int \dfrac{1 - 2n_F(\varepsilon_p)}{2(\varepsilon_p - \mu)} \dfrac{d^2\vec{p}}{(2\pi)^2}$ contains $\ln^2$-contribution. Thus in the weak-coupling case for the coupling constant

$$f_0 = \frac{U}{8\pi\, t} \ll 1 \qquad(9.4.22)$$

and in the second order of perturbation theory (in $f_0$) for the effective interaction we have:

$$U_{eff} \sim f_0 + f_0^2 \ln^2\frac{\mu}{t},\qquad(9.4.23)$$

where $\mu \ll t$ is chemical potential close to half-filling and we still assume that $\mu > T_C^d$. Note that in the expansion (9.4.23) an effective parameter of the perturbation theory is [9.44]:

$$f_0 \ln^2\frac{\mu}{t},\qquad(9.4.24)$$

Correspondingly as it was shown by Kozlov [9.44] the equation for the $d_{x2-y2}$-wave critical temperature $T_C^d$ reads:

$$f_0^2 \ln^3\frac{t}{\mu} \ln^2\frac{\mu}{T_C^d} \sim 1 \qquad(9.4.25)$$

or equivalently

$$T_C^{\ d} \sim \mu \exp\left\{ -\frac{1}{f_0^2 \ln^3\dfrac{t}{\mu}} \right\}.\qquad(9.4.26)$$

Thus we see that small coupling constant $f_0 \ll 1$ (9.4.22) is enhanced by large value of $\ln^3\dfrac{t}{\mu} \gg 1$ in (9.4.26) – nice result of Kozlov. For the sake of completeness let us mention here the other important articles where $d_{x2-y2}$-wave pairing with rather high $T_C$ was obtained by a variety of computational approaches for 2D Hubbard model at optimal doping $n_{el} \sim (0.8 \div 0.9)$ [9.67-9.73, 9.78].

### 9.4.5. Parquette solution at weak-coupling and close to half-filling.

Very close to half-filling when $\mu \sim T_C$ we have so-called doubly-logarithmic parquette solution of Dzyaloshinslii, Yaroshenko [9.45] with the competition between SC and SDW-instability in particle-particle (SC) and particle-hole (SDW – spin density wave) channels. Here for $\mu \sim T_C$ from (9.4.26) we get:

$$f_0^2 \ln^4\frac{t}{\mu} \sim f_0^2 \ln^4\frac{t}{T_C^d} \sim 1 \qquad(9.4.27)$$

Hence for the $d_{x2-y2}$-wave critical temperature:

$$T_C^{\ d} \sim t \exp\left\{-\frac{const}{\sqrt{f_0}}\right\} - \qquad (9.4.28)$$

- an elegant result of [9.45]. We should mention here also the results of [9.74-9.78]. The maximal critical temperature in the 2D Hubbard model according to qualitative considerations of Kivelson et al. [9.48,9.48,9.65] corresponds to intermediate coupling case U/W~1 and optimal concentrations $n_{el} \sim (0.8 \div 0.9)$. $T_C$ here can reach the desired values of $10^2$K realistic for optimally doped cuprates [9.66].

The border between AFM (or SDW) phase and superconductive phase of the high-$T_C$ superconductor (described by 2D Hubbard model) in weak-coupling case $U \ll W$ and very close to half-filling (for doping concentrations $x = (1 - n_{el}) \ll 1$) according to Kirelson [9.48] at the temperature $T \rightarrow 0$ is given by:

$$x_C = (1 - n_{el}) \sim \exp\left\{-2\pi\sqrt{\frac{t}{U}}\right\} \qquad (9.4.29)$$

and we have the same expression for $\frac{1}{\sqrt{f_0}} \sim \sqrt{\frac{t}{U}}$ in the exponent of (9.4.29) for $x_C$.

Another interesting observation belongs to Kopaev and Belyavsky [9.48] namely that the spectrum (9.4.11) on the square lattice at half-filling satisfies also "mirror nesting" property:

$$\varepsilon(P) = -\varepsilon(-P + Q) \qquad (9.4.30)$$

This property is in favor of Cooper pairing with large total moment of a pair $\vec{Q}$ in a clean case (no impurities).

## 9.5. SC transitions in the jelly model for Coulomb electron plasma.

Very recently in connection with the high-$T_C$ physics Alexandrov and Kabanov [9.49] raised the very important question of the role of full Coulomb interaction (which is not reduced to onsite Hubbard repulsion but is extended over several coordinate spheres) for non-phonon mechanisms of superconductivity. They claimed that in the 3D jelly model for reasonable electron densities $r_S \leq 20$, where

$$r_S = \frac{1.92}{p_F a_B} \qquad (9.5.1)$$

is correlative radius and $a_B = \frac{1}{me^2}$ ($\hbar = 1$) is Bohr radius for electron, the superconductive critical temperatures correspond to the pairing with large orbital momenta ($l \gg 1$) and are very low.

Indeed both the perturbative analysis of Chubukov and Kagan [9.50] (see also [9.31]) as well as numerical calculations of Alexandrov and Kabanov [9.49] provide small critical temperatures in 3D dense electron plasma.

### 9.5.1. Cascade of SC-transitions in the dense electron plasma.

To be more specific the authors of [9.50] predicted a cascade of SC-transitions with orbital moment of the Cooper pair $l$ being dependent upon electron density $r_S$:

$$l > l_C = \frac{|\ln r_S|}{\sqrt{r_S}}\left(1 + \frac{7}{2}\frac{\ln|\ln r_S|}{|\ln r_S|} + ...\right) \qquad (9.5.2)$$

The critical temperature of the superconductive transition:

$$T_C \sim \varepsilon_F \exp\left\{-\left(\frac{\pi}{4}\right)^2\frac{1}{l^4}\right\} \qquad (9.5.3)$$

corresponds for given density $r_S$ to $l = l_C(r_S)$ from (9.5.2)and is very low. The authors of [9.79] get qualitatively the same result numerically with maximal but still very low $T_C$ corresponding to f-wave pairing ($l = 3$) for $0 < r_S \leq 18$. We can honestly say that the 3D jelly model is not very promising for superconductivity with reasonable high-$T_C$ in case of dense Coulomb plasma.

### 9.5.2. The dilute electron plasma.

The possibility of the p-wave SC with critical temperatures $T_{C1}$ in the range ~$(10^{-3} \div 10^{-2})$ K for 3D electron plasma of intermediate and small electron densities ($r_S >> 1$) (which are relevant for simple and noble metals like Na, K, Ag, Au or for semimetals) was predicted in the papers [9.52-9.54]. Here $T_{C1} \sim \varepsilon_F \exp\left\{-1/|\lambda_1|\right\}$, where $|\lambda_1| = 0.07$ in [9.52] and $|\lambda_1| = 0.06$ in [9.54].

The most realistic region of densities for p-wave pairing in 3D dilute plasma is possibly $20 \leq r_S \leq 35$ according to [9.52, 9.54]. In the same time some groups [9.55, 9.56] belive more in Khodel-Shaginyan [9.56] type of Fermi-surface reconstruction (and not SC-transition) at these densities or more close to Wigner crystallization instability. It is a very difficult question in particular in 2D where rigorously speaking we should sum up an infinite parquette class of diagrams for the effective interaction (irreducible bare vertex) $U_{eff}$ in the Cooper channel.

### 9.6. SC and phase separation in Shubin-Vonsovsky model.

The situation for superconductivity with reasonable $T_C$ becomes much more favorable on the lattice if we consider so-called Shubin-Vonsovsky model [9.55] with onsite Hubbard repulsion $U$ and additional Coulomb repulsion $V$ on the neighboring sites [9.56]. In real space the Hamiltonian of the Shubin-Vonsovsky model reads:

$$\hat{H}' = \hat{H} - \mu\hat{N} = -t\sum_{<ij>\sigma} c_{i\sigma}^+ c_{j\sigma} + U\sum_i n_{i\uparrow} n_{i\downarrow} + \frac{V}{2}\sum_{<ij>} n_i n_j - \mu\sum_{i\sigma} n_{i\sigma}, \qquad (9.6.1)$$

where $n_i = \sum_\sigma n_{i\sigma}$ - total density on site $i$ (for both $\uparrow$ and $\downarrow$ projection of electron spin. It is reasonable to assume (see [9.56]) that in (9.6.1) we have the following estimates for the parameters of the model:

$$U \sim \frac{e^2}{\varepsilon a_B}; \quad V \sim \frac{e^2}{\varepsilon d}; \quad W \sim \frac{1}{md^2}, \qquad (9.6.2)$$

where $W$ is a bandwidth, $\varepsilon$ is the effective dielectric permittivity and $a_B \sim \frac{\varepsilon}{me^2}$ ($\hbar = 1$) is Bohr radius in ionic media. Thus for $\varepsilon \sim 1$, $a_B \sim 0.5$ Å and $d \sim (3 \div 4)$ Å [9.56] and in the limit $a_B/d << 1$ we come to the following hierarchy of parameters:

$$U >> V >> W. \qquad (9.6.3)$$

The effective vacuum interaction in Shubin-Vonsovsky model in real space behaves as follows (see Fig. 9.10) in the strong-coupling case (9.6.3).

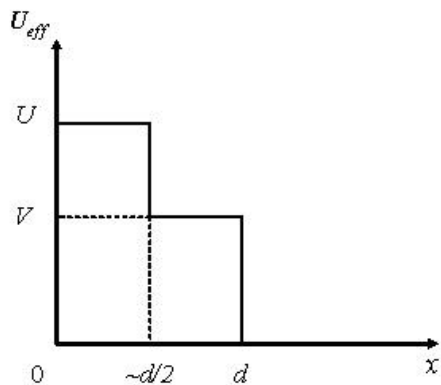

Fig. 9.10. Effective vacuum interaction in Shubin-Vonsovsky model in the strong-coupling case (9.6.3).

### 9.6.1 p-wave SC in Shubin-Vonsovsky model at low density.

Note that Shubin-Vonsovsky model [9.55] is the most repulsive and the most unbeneficial model for SC. In the same time it is useful toy-model to study the effects of intersite Coulomb repulsion on SC and Mott-Hubbard metal-dielectric transition [9.29, 9.57] as well as the physics of the nanoscale phase separation [9.58, 9.59].

After Fourier transformation the Hamiltonian of the Shubin-Vonsovsky model reads:

$$\hat{H}' = \sum_{\vec{p}\sigma}[\varepsilon(p)-\mu]c^+_{p\sigma}c_{p\sigma} + U\sum_{\vec{p}\vec{p}'\vec{q}}c^+_{\vec{p}\uparrow}c^+_{\vec{p}'+\vec{q}\downarrow}c_{\vec{p}+\vec{q}\downarrow}c_{\vec{p}'\uparrow} + \sum_{\substack{\vec{p}\vec{p}'\vec{q}\\ \sigma\sigma'}}V(\vec{p},\vec{p}')c^+_{\vec{p}\sigma}c^+_{\vec{p}'+\vec{q}\sigma'}c_{\vec{p}+\vec{q}\sigma'}c_{\vec{p}'\sigma}, \quad (9.6.4)$$

where in 3D case:

$$V(\vec{p},\vec{p}') = V\big(\cos(p_x - p_x')d + \cos(p_y - p_y')d + \cos(p_z - p_z')d\big) \quad (9.6.5)$$

At the low density $pd \ll 1$ the expansion up to quadratic terms gives effective vacuum interaction in 3D for s-wave and p-wave harmonics respectively:

$$U^S_{eff\,vac} = U + 6V + o(p^2d^2), \quad (9.6.6)$$

and

$$U^P_{eff\,vac} = 2V\vec{p}\vec{p}'d^2 \quad (9.6.7)$$

The renormalization of effective vacuum interaction in the framework of Kanamori T-matrix approximation [9.33] yields for s-wave and p-wave scattering lengths [9.58] in the strong-coupling $U \gg V \gg W$ and low density $nd^3 < 1$ case:

$$a_S \sim d, \quad (9.6.8)$$

and

$$a_P \sim d, \quad (9.6.9)$$

where $T_S = \dfrac{4\pi a_S}{m}$ and $T_P = \dfrac{4\pi}{m}2a_P(\vec{p}\vec{p}')d^2$ are T-matrices in s-wave and p-wave channels.

Thus the dimensionless s-wave gas-parameter $\lambda_S = \lambda = 2d\ p_F/\pi$ as in the repulsive-$U$ Hubbard model (see (9.9.4)) while the dimensionless p-wave gas-parameter:

$$\lambda_p \sim (p_F d)^3 \quad (9.6.10)$$

in agreement with general quantum-mechanical consideration for slow particles ($p_F\ d\ < 1$) in vacuum (see [9.35]).

Thus in 3D a normal state of the Shubin-Vonsovsky model at strong-coupling and low densities is again unstable towards triplet p-wave pairing. The irreducible bare vertex in substance reads:

$$N_{2D}(0)U^{l=1}_{eff} = \lambda^2_S\Pi_{l=1} + \lambda_P \quad (9.6.11)$$

Thus critical temperature in the main order of the s-wave gas-parameter is given by (9.4.10) again.

The presence of the additional Coulomb repulsion $V$ on the neighboring sites in the model changes only the preexponential factor in (9.4.10). Thus situation in the 3D Shubin-Vonsovsky model and low electron density is much more favorable for SC with reasonable $T_C$ even in the most repulsive strong-coupling case (in contrast with the situation in 3D jelly model for dense electron plasma).

Analogously in strong-coupling and low density 2D case the s-wave dimensionless gas-parameter is again $f_S = f_0 = \dfrac{1}{\ln(4W/\varepsilon_F)} = \dfrac{1}{\ln(8/n_{el})}$ (see (9.4.14)) as in the 2D Hubbard model, while the dimensionless p-wave gas-parameter:

$$f_P \sim p^2_F d^2 \quad (9.6.13)$$

in 2D again in agreement with general quantum-mechanical considerations [9.35] for slow particles in vacuum. The irreducible bare vertex in substance reads in 2D:

$$U_{eff}^{l=1} N_{2D}(0) = -6.1 f_S^3 + 2 p_F^2 d^2 \qquad (9.6.14)$$

Hence in the main order of 2D s-wave gas-parameter the p-wave critical temperature is given again by (9.4.15) in exact similarity with 2D Hubbard model. The presence of the additional Coulomb repulsion $V$ again changes only the preexponential factor.

### 9.6.2. Localization and phase separation in Shubin-Vonsovsky model at larger densities.

At larger dimensionless densities $n_{el} \geq 0.5$ ($n_{el} = 0.5$ corresponds to quarter-filling of the band) there are, however, extended regions of phase-separation in Shubin-Vonsovsky model in the strong-coupling limit $U >> V >> W$.

To be more specific in the model [9.55] there are two types of localization: Mott-Hubbard localization with an appearance of AFM-state at half-filling ($n_{el} \to 1$) [9.29, 9.57] and Verwey localization with an appearance of checkerboard charge-ordered (CO) state at quarter-filling ($n_{el} \to \frac{1}{2}$) [9.59].

Close to $n_{el} = \frac{1}{2}$ and $n_{el} = 1$ we have extended regions of nano-scale phase-separation [9.56-9.58]. The qualitative phase-diagram of the Shubin-Vonsovsky model in the strong-coupling case is presented on Fig.9.11.

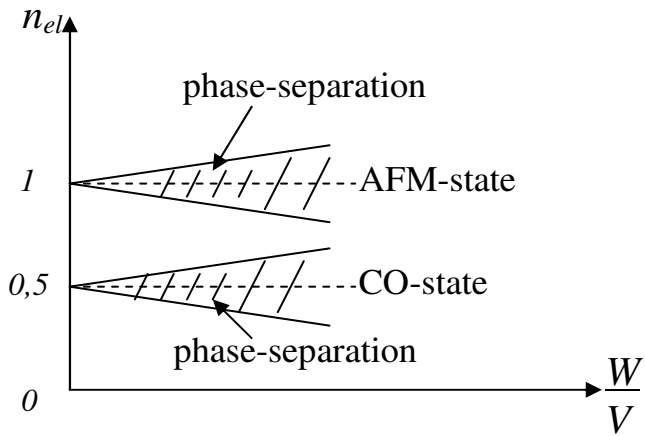

Fig.9.11. Qualitative phase-diagram of the Shubin-Vonsovsky model in the strong-coupling case. At $n_{el} = 1$ AFM-state appears in the model, while at $n_{el} = \frac{1}{2}$ we have the checkerboard CO-state.

If we increase the density from $n_{el} \to 0$ to $n_{el} = \frac{1}{2} - x$, then at the critical concentrations:

$$x_C \sim \left(\frac{W}{V}\right)^{3/5} \text{ in 3D} \quad (9.6.16)$$

and

$$x_C \sim \left(\frac{W}{V}\right)^{1/2} \text{ in 2D} \quad (9.6.17)$$

the system undergoes the first-order phase-transition into a phase-separated state with nano-scale metallic clusters inside charge-ordered checkerboard insulating matrix (see Fig.9.12). At critical concentration $n_{el} = \frac{1}{2} - x_C$ the metallic clusters start to touch each other. As a result all the sample volume becomes metallic for $n_{el} < \frac{1}{2} - x_C$. The more detailed analysis of Verwey localization at $n_{el} = \frac{1}{2}$ and nano-scale phase-separated state for $\frac{1}{2} - x_C < n_{el} < \frac{1}{2}$ will be a subject of Chapter 17.

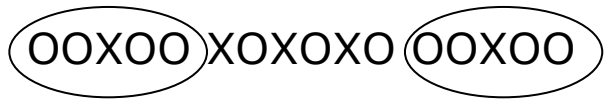

Fig.9.12. Phase-separated state at the densities $\frac{1}{2} - x_C < n_{el} < \frac{1}{2}$ with nano-scale metallic clusters inside CO checkerboard insulating matrix. At critical concentration $n_{el} = \frac{1}{2} - x_C$ the metallic clusters start to touch each other and all the volume of the sample becomes metallic.

Thus we can not extend our calculations for $T_C$ in homogeneous case to densities larger than $n_{el} = \frac{1}{2} - x_C$ in the strong-coupling limit of Shubin-Vonsovsky model. However it is interesting to construct SC phase-diagram of the model for $n_{el} < 0.5 - x_C$ and to find the regions of p-wave, $d_{xy}$ and $d_{x2-y2}$-pairing [9.56]. The work along these lines is in progress [9.61].

Another interesting problem is to consider the opposite weak-coupling Born case $W > U > V$ which can be realized for large Bohr radius $a_B > d$ or correspondingly for large dielectric permittivity $\varepsilon >> 1$ (note that real high-$T_C$ materials are in the difficult intermediate coupling regime $\varepsilon \geq 1$ and $a_B \sim d$). In the weak-coupling case Verwey localization [9.59] is absent for $n_{el} \to \frac{1}{2}$ and we can construct superconductive phase-diagram of the Shubin-Vonsovsky model for p-wave, $d_{xy}$ and $d_{x2-y2}$-pairing for all densities including almost half-filled case [9.61].

The first numerical results in this respect were obtained in [9.65], where the authors found (besides p-wave, $d_{xy}$ and $d_{x2-y2}$-pairing) also the regions of extended s-wave pairing and g-wave pairing for $0 < n_{el} < 0,95$.

Chapter 10. Strong $T_C$ enhancement in spin-polarized Fermi-gas and in two-band superconductors.

Content

In previous chapter (Chapter 9) we discuss the possible unconventional SC of p-wave and d-wave character in strongly correlated fermion systems with repulsive interaction between particles. The basic mechanism of SC which we discussed at low electron densities was Kohn-Luttinger mechanism [9.4] and its extension on triplet p-wave pairing by Fay, Laser [9.17] and Kagan, Chubukov [9.18]. However the critical temperatures of 3D and 2D p-wave pairing given by (9.4.10) and (9.4.14) are usually rather low, especially in 3D.

The legitimate question then appears whether it is possible to increase $T_C$ already at low density and to get experimentally feasible critical temperatures?

The answer to this question is positive and is connected with 2 possibilities:

1)        to apply external magnetic field (or strong spin-polarization) to triplet p-wave SC [10.1, 10.2];

2)        to consider the two-band situation especially for multiband SC with one narrow band [10.3, 10.4].

In both cases the most important idea – is to separate the channels for the formation of Cooper pair and for the preparation of the effective interaction [10.1, 10.3].

10.1. $T_C$ enhancement in spin-polarized neutral Fermi-gas.

In magnetic field (or in the presence of the strong spin-polarization) the Cooper pair according to [10.1] is formed by two fermions with spins "up", while an effective interaction is prepared by two fermions with spins "down" (see Fig.10.1).

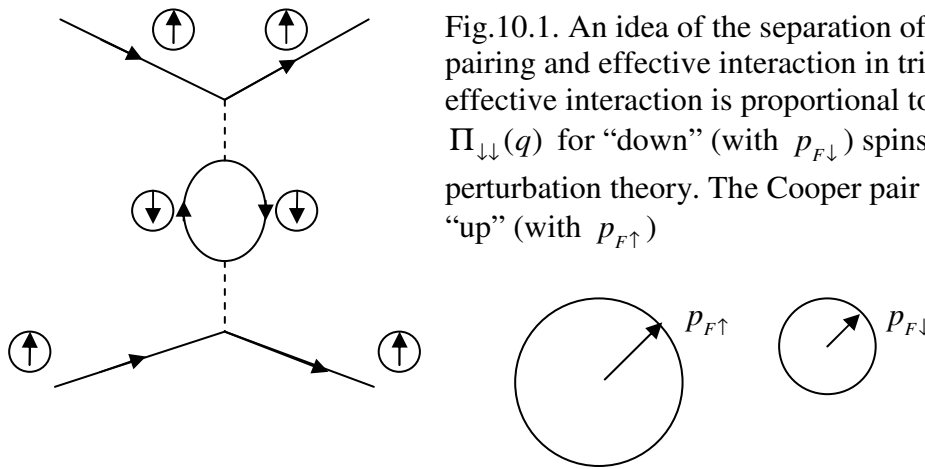

Fig.10.1. An idea of the separation of the channels for Cooper pairing and effective interaction in triplet neutral SC. An effective interaction is proportional to polarization operator $\Pi_{\downarrow\downarrow}(q)$ for "down" (with $p_{F\downarrow}$) spins in the second order of perturbation theory. The Cooper pair is formed by two spins "up" (with $p_{F\uparrow}$)

10.1.1. 3D spin-polarized Fermi-gas.

As a result the Kohn's anomaly is strongly increased. In the absence of magnetic field (for $H = 0$) the Kohn's anomaly in 3D reads: $(q - 2p_F) \ln|q - 2p_F|$ (see (9.3.5)) where $q^2 = 2p_F^2(1 - \cos\theta)$, (see (9.3.11)). Physically Kohn's anomaly is important for $q \rightarrow 2p_F$ (diameter of Fermi-sphere) or for angles $\theta \rightarrow \pi$ - backward scattering between incoming and outgoing momenta $\vec{p}$ and $\vec{k}$ for the Cooper channel ($q = |\vec{p} - \vec{k}|$; $|\vec{p}| = |\vec{k}| = p_F$). In terms of the angle $\theta$ between $\vec{p}$ and $\vec{k}$ ($\theta = \angle \vec{p}\vec{k}$) the Kohn's anomaly for $H = 0$ reads:

$$\Pi_{sing}(q) \sim (q - 2p_F) \ln|q - 2p_F| \sim (\pi - \theta)^2 \ln(\pi - \theta) \qquad (10.1.1)$$

and only second derivative of $\Pi_{sing}(\theta)$ with respect to $(\pi - \theta)$ diverges. In the same time for $H \neq 0$ the Fermi-momenta for the particles with spins "up" and "down" are different ($p_{F\uparrow} \neq p_{F\downarrow}$) and thus:

$$\Pi_{sing}(\theta) \sim (q_\uparrow - 2p_{F\downarrow}) \ln|q_\uparrow - 2p_{F\downarrow}| \sim (\theta - \theta_c) \ln(\theta - \theta_c), \qquad (10.1.2)$$

where $\theta_C$ differs form $\pi$ proportionally to ($p_{F\uparrow} / p_{F\downarrow}$ -1). Correspondingly already first derivative of $\Pi_{sing}(\theta)$ with respect to ($\theta - \theta_C$) diverges. Unfortunately fot $H \neq 0$ there is a competing process namely the decrease of the density of states of the fermions with "down" spins in $\Pi_{\downarrow\downarrow}(\theta)$ on Fig.10.1:

$$N_\downarrow^{3D}(0) = \frac{mp_{F\downarrow}}{2\pi^2} \qquad (10.1.3)$$

As a result of this competition the critical temperature $T_C^{\uparrow\uparrow}$ (for Cooper pair formed by two "up" spins) has strongly non-monotonous (reentrant) behavior with large maximum (see Fig.10.2).

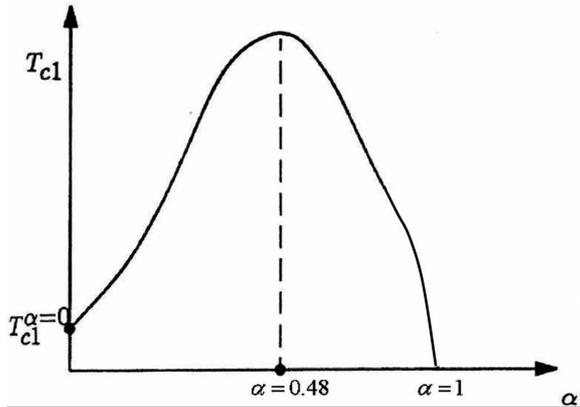

Fig.10.2. Dependence of the p-wave critical temperature $T_C^{\uparrow\uparrow}$ (for pairing of two "up" spins) from the degree of spin-polarization in 3D polarized Fermi-gas with repulsion.

In terms of the polarization degree:

$$\alpha = \frac{n_\uparrow - n_\downarrow}{n_\uparrow + n_\downarrow}, \qquad (10.1.4)$$

where $n_\uparrow = p_{F\uparrow}^3 / 6\pi^2$ and $n_\downarrow = p_{F\downarrow}^3 / 6\pi^2$ are the densities of "up" and "down" fermions in 3D. the critical temperature has a pronounced maximum for $\alpha = 0.48$ (48% of the polarization). Correspondingly $\delta = p_{F\uparrow} / p_{F\downarrow} = 1.43$ at maximum [9.31, 10.1], where in 3D:

$$\alpha = \frac{\delta^3 - 1}{\delta^3 + 1} \qquad (10.1.5)$$

Note that polarization operator $\Pi_{\downarrow\downarrow}(q)$ in Fig.10.1 reads in 3D:

$$\Pi_{\downarrow\downarrow}(q_\uparrow) = \frac{N_\downarrow^{3D}(0)}{2}\left[1 + \frac{4p_F^2 - q_\downarrow^2}{4p_{F\downarrow}q_\uparrow}\ln\frac{2p_{F\downarrow} + q_\uparrow}{|2p_{F\downarrow} - q_\downarrow|}\right] \qquad (10.1.6)$$

Correspondingly in terms of $\delta = p_{F\uparrow}/p_{F\downarrow}$ from (10.1.5) and angle $\theta$ the transferred momentum squared $q_\uparrow^2 = 2p_{F\uparrow}^2(1 - \cos\theta)$ the polarization operator is given by:

$$\Pi_{\downarrow\downarrow}(\theta) = \frac{N_\downarrow^{3D}(0)}{2}\left[1 + \frac{1 - \delta^2\sin^2\frac{\theta}{2}}{2\delta\sin\frac{\theta}{2}}\ln\frac{1 + \delta\sin\frac{\theta}{2}}{\left|1 - \delta\sin\frac{\theta}{2}\right|}\right] \qquad (10.1.7)$$

The evaluation of p-wave harmonics $\Pi_{\downarrow\downarrow}^{l=1}$ is again elementary and can be done analytically as in unpolarized case (for $\delta = 1$) (see [10.1]). For us it is important to represent

$$T_{C1}^{\uparrow\uparrow} = T_{C1}(\alpha = 0)e^{\frac{f(\alpha)}{\lambda^2}}, \qquad (10.1.8)$$

(where $\lambda = 2a\,p_F/\pi$ is 3D gas-parameter) and show that for small $\alpha \ll 1$

$$f(\alpha) = \frac{5}{9}\alpha\frac{(7 - 4\ln 2)}{(2\ln 2 - 1)^2} > 0 \qquad (10.1.9)$$

positive and linear in $\alpha$. Note that for not very strong external magnetic field $\mu_B H < \varepsilon_{F0}$ ($\varepsilon_{F0} = p_{F0}^2/2m$ – Fermi-energy for $H = 0$) we have the linear relation between $\alpha$ and $H$:

$$\alpha \approx \frac{3}{2}\frac{\mu_B H}{\varepsilon_{F0}} \ll 1 \qquad (10.1.10)$$

in 3D case. Thus $T_{C1}^{\uparrow\uparrow}$ increases for $\alpha \ll 1$. In the opposite case of practically 100% polarization ($\alpha \to 1$ or $1 - \alpha \to 0$) we have:

$$f(\alpha) = -\frac{9}{2^{2/3}}\frac{1}{(1 - \alpha)} \to -\infty \qquad (10.1.11)$$

Hence $T_{C1}^{\uparrow\uparrow} \to 0$ for $\alpha \to 1$. The asymptotic behavior of $f(\alpha)$ for $\alpha \ll 1$ and $(1-\alpha) \ll 1$ justifies the appearance of the maximum of $T_{C1}^{\uparrow\uparrow}$ in between. In maximum:

$$\max T_{C1}^{\uparrow\uparrow} = T_{C1}^{\uparrow\uparrow}(\alpha = 0.48) \sim \varepsilon_F e^{-\frac{7}{\lambda^2}}, \qquad (10.1.12)$$

and we see that in the second order of perturbation theory the expression in the main exponent for $T_C$ increases practically in 2 times (from $\lambda^2/13$ to $\lambda^2/7$) giving a large $T_C$-enhancement in 3D spin-polarized Fermi-gas. Note that the behavior of the critical temperature at $\alpha \to 1$ is not accidental in this approximation, namely $T_{C1}^{\uparrow\uparrow}(\alpha = 1) = 0$ because of the absence of "down" spins in 100% polarized gas. $T_{C1}^{\uparrow\uparrow}(\alpha = 1)$ is non-zero only in the third order of the gas-parameter $\lambda = 2a\,p_F/\pi$ which is in accordance with Pauli principle for slow particles in vacuum [9.35] (p-wave harmonic of the scattering amplitude $f_l \sim (a\,p_F)^3$).

### 10.1.2. 2D spin-polarized Fermi-gas.

This effect is even more pronounced in 2D spin-polarized Fermi-gas. Here, as we discussed in Chapter 11) in unpolarized case the strong Kohn's anomaly $\operatorname{Re}\sqrt{q - 2p_F}$ is ineffective for SC in the second order of perturbation theory for effective interaction $U_{\text{eff}}(q)$. However the situation is dramatically changed in spin-polarized case.

For $\alpha \neq 0$ in 2D:

$$\Pi_{\sin g}(q) \sim \operatorname{Re}\sqrt{q_\uparrow - 2p_{F\downarrow}} \neq 0 \qquad (10.1.13)$$

and thus Kohn's anomaly is now "switched on" for SC already in the second order of the gas-parameter.

The exact calculation of $\Pi_{\downarrow\downarrow}(q)$ in 2D yields in p-wave channel (for magnetic number $m = 1$):

$$N_\downarrow^{2D} U_{eff}^{m=1} = 8 f_0 \frac{\delta - 1}{\delta^2}, \qquad (10.1.14)$$

where in 2D

$$\alpha = \frac{\delta^2 - 1}{\delta^2 + 1} \qquad (10.1.15)$$

Correspondingly the critical temperature for p-wave pairing of two "up" spin reads in second order of 2D gas-parameter $f_0$:

$$T_{C1}^{\uparrow\uparrow} \sim \varepsilon_F \exp\left\{ -\frac{\delta^2}{8 f_0 (\delta - 1)} \right\}, \qquad (10.1.16)$$

From (10.1.16) we see that for $\delta = 0$ (unpolarized case) $T_{C1}^{\uparrow\uparrow} \to 0$ and superconductive transition arises only in third order of the gas-parameter for the effective interaction.
For $\alpha \to 1$ ($\delta \to \infty$) we again have $T_{C1}^{\uparrow\uparrow} \to 0$ in second order in $f_0$.
Thus we have a very large maximum in between again. It corresponds to $\delta = p_{F\uparrow} / p_{F\downarrow} = 2$ or, $\alpha = 0.6$. Hence for $T_{C1}^{\uparrow\uparrow}$ in maximum we get:

$$\max T_{C1}^{\uparrow\uparrow} = T_{C1}^{\uparrow\uparrow} (\alpha = 0.6) \sim \varepsilon_F \exp\left\{ -\frac{1}{2 f_0^2} \right\}, \qquad (10.1.17)$$

The dependence of $T_{C1}^{\uparrow\uparrow}$ from $\alpha$ in 2D spin-polarization Fermi-gas is shown on Fig.10.3.

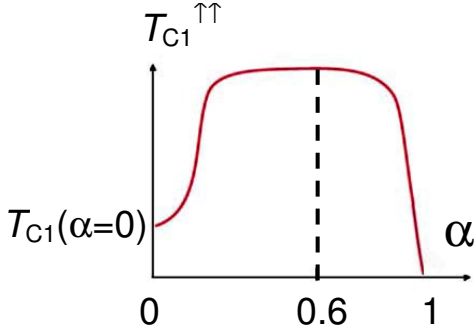

Fig.10.3. Dependence of the critical temperature $T_{C1}^{\uparrow\uparrow}$ for two "up" spins from the polarization degree $\alpha$ in 2D spin-polarized Fermi-gas.

The maximum is very broad in 2D – it stretches form $\alpha \sim 0.1$ till $\alpha \sim 0.9$.
Note that in this approximation (in second order in gas parameter evaluation of $T_{C1}^{\uparrow\uparrow}$) $T_{C1}^{\uparrow\uparrow}(\alpha = 1) = 0$ again due to the absence of "down" spins in 100% polarized case.

### 10.1.3. Spin-polarized superfluid $^3$He.

The application of the theory for dilute spin-polarized mixture will be a subject of the next Chapter (Chapter 13). Here we would like to emphasize that even for dense superfluid $^3$He (where $\lambda \sim 1.3$) we get reasonable estimates for $T_{C1}^{\uparrow\uparrow}$ - increase. These estimates are confirmed by the experiments of Frossati et al., [9.5] in Kamerlingh-Ohnes Laboratory in Leiden. Namely in unpolarized $^3$He-A:

$$T_{C1}(\alpha = 0) = 2.7\ mK . \qquad (10.1.18)$$

In the same time for $\alpha = 6\%$ of the polarization (in magnetic fields $\sim 15$ T) $T_{C1}^{\uparrow\uparrow}(\alpha = 6\%) \approx 3.2$ mK and we have 20% increase of $T_C$.
In maximum for A1-phase of superfluid $^3$He we predict

$$\max T_{C1}^{\uparrow\uparrow} = 6.4 T_{C1}^{\uparrow\uparrow} (\alpha = 0.6) . \qquad (10.1.19)$$

The similar estimate for reentrant behavior of $T_{C1}^{\uparrow\uparrow}(\alpha)$ and for $T_{C1}^{\uparrow\uparrow}$ in maximum was proposed in [10.6] on the basis of more phenomenological approach based on Landau Fermi-liquid theory for superfluid $^3$He.

## 6.2 $T_C$ enhancement in quasi-2D charged SC in parallel magnetic field.

In 2D electron gas in parallel magnetic field: $\vec{H} = H\vec{e}_x$ (see Fig.10.4) the vector-potential

$$\vec{A} = Hz\vec{e}_y \qquad (10.2.1)$$

does not change the motion of electrons and Cooper pairs in $(xy)$ plane (the $\Psi$-function for this problem $\Psi(\vec{r}) = e^{ip_x x + ip_y y} \varphi(z)$).

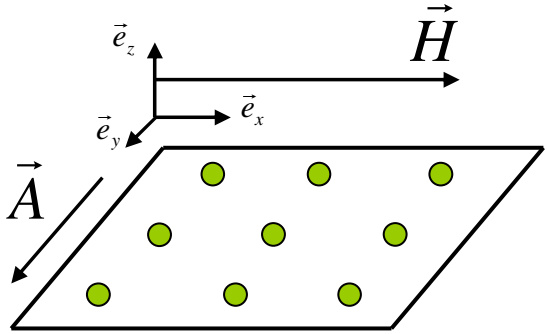

Fig.10.4. 2D electron gas in parallel magnetic field $\vec{H} = H\vec{e}_x$. Vector-potential $\vec{A} = Hz\vec{e}_y$ does not change the motion of electrons and Cooper pairs in $(xy)$ plane.

Hence in this geometry diamagnetic Meissner effect [10.7] (which changes the motion of the Cooper pair in plane if magnetic field is perpendicular to the plane) is totally suppressed [10.9-10.11]. Hence 2D SC in parallel magnetic field becomes equivalent to neutral (uncharged) $^3$He-monolayer and only Pauli paramagnetic effect is present [9.6]. According to this effect even if the singlet s-wave pairing organized by two electrons with spins "up" and "down" was present in SC at $H = 0$ it will be totally suppressed in magnetic fields exceeding Pauli paramagnetic limit [10.10,10.11]

$$H > H_P = \frac{T_{CO}}{g\mu_B}, \qquad (10.2.2)$$

where $\mu_B$ is Bohr magneton and $g$ is hyromagnetic ratio (or Lande factor) (note that for $T_{CO} \sim 1$ K and $g \sim 1$: $H_P \sim 1$ T). As a result only p-wave pairing with the symmetry of A1-phase (pairing of two electrons with spins "up") that is with $S_{tot} = S_z^{tot} = 1$ survives in large magnetic field (the two electrons should lie on the same Fermi-surface with the radius $p_{F\uparrow}$).

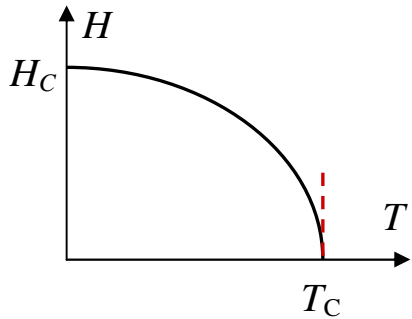

Fig.10.5. Standard behavior of a critical magnetic field as a function of temperature in 3D superconductors, which are subject of Meissner effect.

Hence instead of a standard behavior of a critical magnetic field $H_C$ versus temperature [10.7] presented on Fig 10.5 we have very unusual (reentrant behavior) of p-wave critical temperatures in 2D superconductor in parallel magnetic field (see Fig.10.6).

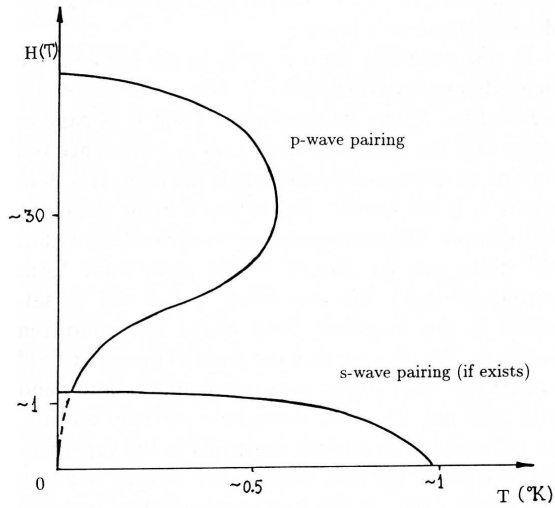

Fig.10.6. Suppression of s-wave pairing and reentrant behavior of p-wave critical temperatures in 2D SC in parallel magnetic field H(T). p-wave pairing is formed by two electrons living on larger Fermi-surface (with $p_{F\uparrow}$) while effective interaction is formed by two electrons with down spins (with $p_{F\downarrow}$)

The critical temperature of p-wave pairing depends upon polarization degree according to (10.1.16). At small polarization degrees:

$$\alpha = \frac{g\mu_B H}{\varepsilon_{F0}} << 1 \qquad (10.2.3)$$

in 2D case. For low density electron systems with $\varepsilon_F \sim 30$ K in high magnetic fields $H \sim 15$ T and $g \sim 1$ polarization degree $\alpha \sim 0.25$ and we can get $T_{C1}{}^{\uparrow\uparrow} \sim 0.5$ K in accordance with (10.1.16), (10.1.17) which is quite nice.

It is important to note that the results (10.1.16) and (10.1.17) are valid for 2D Fermi-gas with short-range repulsion or for 2D Hubbard model. The question is how these results will change in 2D Coulomb plasma (or 2D electron gas with screened Coulomb interaction). Here the coupling constant $f_0$ in (10.1.16) and (10.1.17) can be calculated only within accuracy of Random Phase Approximation (RPA) for metallic electron densities with correlation radius $r_s \sim \dfrac{1}{p_F a_B} >> 1$

( $a_B = \dfrac{\varepsilon}{m^* e^2}$ is Bohr radius, $m^*$ is effective mass and $\varepsilon$ is effective dielectric permittivity)

[10.11, 10.13]. It is shown in [10.13] that in 2D electron gas (2DEG):

$$U_{\text{eff}}(q_{||} = 0) = \frac{\pi e^2}{\kappa_{TF}} \qquad (10.2.4)$$

is 2D-projection of 3D screened Coulomb interaction. $\kappa_{TF} = 2\pi N_{2D}(0)e^2$ is Thomas-Fermi wave vector [10.11,10.13], $q_{||}$ is projection of vector $\vec{q}$ on $(xy)$ plane and $N_{2D}(0) = m/2\pi$ is a density of states in 2D.

The dimensionless coupling constant for screened Coulomb interaction reads:

$$f_0 = N_{2D}(0)U_{eff}(q_{||}=0) = \frac{m}{2\pi}\frac{\pi e^2}{\kappa_{TF}} = \frac{m}{2\pi}\frac{\pi e^2}{2\pi e^2 m/2\pi} = \frac{1}{2} \qquad (10.2.5)$$

- remarkable result which is valid also for 3D plasma where $N_{3D}(0) = m\, p_F/2\pi^2$ and $U_{eff}(q=0) = \dfrac{4\pi e^2}{\kappa_{TF}^2}$. Thus effective gas-parameter for screened Coulomb plasma $f_0 = \frac{1}{2}$ does not depend upon density both in 3D and 2D case.

Note that the RPA is exact in 3D and 2D for large densities and small correlation radius $r_S \ll 1$ (dense electron plasma considered in paragraph 9.5.1), but also it is not bad (at least in 3D) for standard metallic densities ($2 \le r_S \le 6$) because the effective parameter of expansion in the energy-functional of 3D electron plasma is actually $\sim r_S/6$ (see [10.11, 9.49, 9.50]). We can hope that (10.1.16) and (10.1.17) with $f_0 \sim \frac{1}{2}$ provides qualitative estimates for 2D dense electron plasma.

Note that the most important experimental limitation to this scenario is connected with the demands on the purity of the sample. In accordance with Abrikosov-Gor'kov results for paramagnetic impurities (more exactly Gor'kov, Larkin results for non-magnetic impurities in p-wave SC) it means that for inverse scattering time:

$$\gamma = 1/\tau < T_{C1} \quad (\hbar = k_B = 1). \qquad (10.2.6)$$

For $T_C \sim 0.5$ K we get then $\tau \ge 2 \cdot 10^{-11}$ see for scattering time of electrons on impurities. Accordingly in 2D case the sheet conductivity for $\varepsilon_F \sim 30$ K and $T_C \sim 0.5$ K:

$$\sigma_\square > \frac{e^2}{2\pi\hbar}\frac{2\varepsilon_F}{T_C} \sim 120\frac{e^2}{2\pi\hbar}. \qquad (10.2.7)$$

Hence plaquette resistance:

$$R_\square < \frac{1}{120}\frac{2\pi\hbar}{e^2} \sim 0.2\, k\Omega \qquad (10.2.8)$$

and mobility:

$$\mu_{el} = \frac{e\tau}{m^*} = 3 \cdot 10^4 \left(\frac{m_e}{m_e^*}\right) cm^2 V^{-1} s^{-1} \qquad (10.2.9)$$

Thus for semimetals or degenerate 2D semiconductive layers with light effective mass $m^* \sim 0.1$ $m_e$ (heterostructures, inverse layers):

$$\mu_{el} = 10^5\ cm^2 V^{-1} s^{-1}, \qquad (10.2.10)$$

which requires very clean samples.

Note that the highest mobility for GaAs-AlAs heterostructures utilized for the measurements of Fractional Quantum Hall effect $\mu_{el} \sim 10^7$. In these systems $\varepsilon_F \sim 30$ K for densities $n_{2D} \sim 10^{12}$ cm$^{-2}$, Bohr radius $a_B = \varepsilon / m^* e^2$ with effective dielectric permittivity $\varepsilon \sim 4$. Thus the correlation radius $r_S \sim 4$, so we are far from Wiegner crystal case ($r_S \sim 20$ in 2D) [10.11, 10.13] and can hope that RPA theory works quite well also in 2D for these densities. Thus we can justify the estimates for $T_{C1}^{\uparrow\uparrow}$ with $f_0 = \frac{1}{2}$ (10.1.16) and (10.1.17) for heterostructures.

Note that if we consider 2D organic superconductors like α-(BEDT-TTF)$_2$I$_3$ or intercalated systems (dichalcogenides) TaS$_2$, TaSe$_2$ then their mobilities are also very high, but they have Fermi-energies $\varepsilon_F \sim (10^3 \div 10^4)$ K which are very difficult to polarize even with very high magnetic fields (maximal possible stationary magnetic fields are equal to 30 T, maximal fields in pulsed regime are 100 T today, hence polarization degree $\alpha = \dfrac{g\mu_B H}{\varepsilon_{F0}}$ will be very low and $T_C$-s will be low).

If we start to think about 2D layer of p-wave heavy-fermion SC like UBe$_{13}$ with $m^* \sim 200\ m_e$ and $\varepsilon_F \sim (30 \div 50)$ K, than we should realize that it is very difficult now to grow very thin films of high quality and it is necessary also to take into account strong spin-orbital coupling and the influence of the crystalline fields in these substances.

It is a very nice challenge for experimentalists to grow very good quality 2D-samples and to apply very parallel magnetic fields or to change gradually tilting angle for a field from $\pi/2$ to $0$ and measure an appearance of p-wave SC at very small angles (note that already very small perpendicular component of the magnetic field will destroy SC via diamagnetic Meissner effect). In the end of this section let us mention that in principle a reentrant behavior of $T_C$ from magnetic field (similar to the one shown on Fig.10.6) can be realized even in 3D superconductors in so-called superquantun limit [10.14, 10.15]. This limit corresponds to very high magnetic fields and very small Fermi-energies when Larmor frequency $\omega_L \sim \mu_B H > \varepsilon_F$ (note that $\varepsilon_F \gg T_C$). In this situation we should take into account Landau quantization of electron levels in magnetic field (diamagnetic one-particle effect). Moreover for $\omega_L > \varepsilon_F$ only one Landau level is filled, so the spectrum of electrons becomes quasi one-dimensional $\varepsilon(p) = \dfrac{\hbar\omega_L}{2} + \dfrac{p_z^{\,2}}{2m}$ and we again have parquette situation with the necessity to sum up an infinite number of diagrams for effective interaction $U_{\text{eff}}$ in particle-particle and particle-hole channels. Halperin and Tesanovic [10.15] showed that as a result a reentrant behavior takes place (the SC appears again in very high fields).

### 10.3 Strong $T_C$ enhancement in the two-band SC.

Another possibility to enhance the critical temperature of p-wave pairing is to consider the two-band model with two sorts of fermions. The main idea again is that to separate the channels: the two particles of sort 1 form the Cooper channel (for $p_{F1} > p_{F2}$), then the effective interaction is formed by two particles of sort 2 (see Fig.10.7).

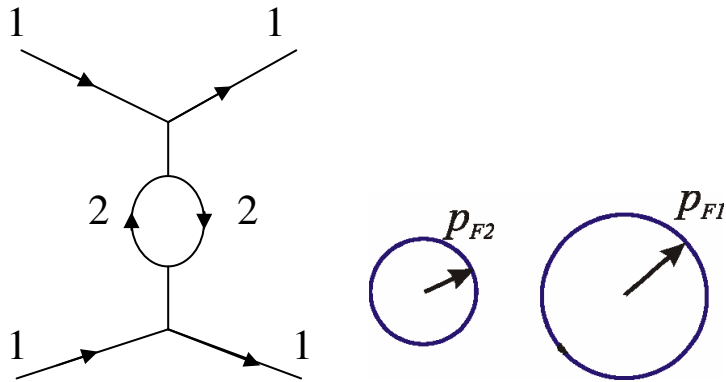

Fig.10.7. Effective interaction for two sorts of particles with $p_{F1} > p_{F2}$.

In this case all the results of section (10.1) are valid with the change of $\delta = p_{F\uparrow}/p_{F\downarrow}$ on $\delta = p_{F1}/p_{F2}$. The gas-parameters $\lambda$ and $f_0$ now depend upon the interaction between the particles of sort 1 and sort 2. In the simplest case it reads $\dfrac{U_{12}}{2}\sum_i n_{1i} n_{2i}$, where $U_{12}$ is interband interaction, $n_{1i}$ and $n_{2i}$ are the densities of the particles with sort 1 and 2 on site $i$.
Let us consider the physics which appears in the two-band model for the most interesting case of very different masses of the two sorts of fermions.

### 10.3.1. The two-band Hubbard model with one narrow band.

On the lattice the most general and simple model which describes this situation is the two-band Hubbard model with one narrow band. This model is very rich. It describes adequately mixed valence systems [10.16, 10.17] such as uranium based heavy-fermions and also some other novel

SC and transition metal systems with orbital degeneracy such as complex magnetic oxides (CMR systems) in optimally doped case. Moreover it contains such highly nontrivial effect as electron-polaron effect [10.18, 10.19] in the homogeneous case. It also shows the tendency towards phase separation [10.20] for a large mismatch between the densities of heavy and light bands as well as anomalous resistivity characteristics (we will study them in the next chapters). In addition to that it describes anomalous p-wave SC of enhanced Kohn-Luttinger type which we will study detaily in this paragraph. The Hamiltonian of the two-band Hubbard model in real space reads:

$$\hat{H}' = -t_h \sum_{<ij>\sigma} a_{i\sigma}^+ a_{j\sigma} - t_L \sum_{<ij>\sigma} b_{i\sigma}^+ b_{j\sigma} - \varepsilon_0 \sum_{i\sigma} n_{ih}^\sigma - \mu \sum_{i\sigma} (n_{i\sigma}^L + n_{i\sigma}^h) +$$

$$+ U_{hh} \sum_i n_{ih}^\uparrow n_{ih}^\downarrow + U_{LL} \sum_i n_{iL}^\uparrow n_{iL}^\downarrow + \frac{U_{hL}}{2} \sum_i n_{iL} n_{ih} \qquad (10.3.1)$$

where $t_h$ and $t_L$ are hopping integrals in heavy and light bands, $-\varepsilon_0$ is the center of gravity of the heavy band, and the difference $\Delta$ between the bottoms of the bands is given by:

$$\Delta = -\varepsilon_0 + \frac{W_L - W_h}{2} = E_{min}^h - E_{min}^L \qquad (10.3.2)$$

$U_{hh}$ and $U_{LL}$ are intraband Hubbard interactions for heavy and light electrons, $U_{hL}$ is interband Hubbard interaction for heavy and light electrons, $n_{ih}^\sigma = a_{i\sigma}^+ a_{i\sigma}$, $n_{LL}^\sigma = b_{i\sigma}^+ b_{i\sigma}$ are the densities of heavy and light electrons on site $i$ with spin-projection $\sigma$, $\mu$ is chemical potential.

After Fourier transformation we obtain:

$$\hat{H}' = \sum_{\vec{p}\sigma} [\varepsilon_h(p) - \mu] a_{p\sigma}^+ a_{p\sigma} + \sum_{\vec{p}\sigma} [\varepsilon_L(p) - \mu] b_{p\sigma}^+ b_{p\sigma} + U_{hh} \sum_{\vec{p}\vec{p'}\vec{q}} a_{\vec{p}\uparrow}^+ a_{\vec{p'}\downarrow}^+ a_{\vec{p}-\vec{q}\downarrow} a_{\vec{p'}+\vec{q}\uparrow} +$$

$$+ U_{LL} \sum_{\vec{p}\vec{p'}\vec{q}} b_{\vec{p}\uparrow}^+ b_{\vec{p'}\downarrow}^+ b_{\vec{p}-\vec{q}\downarrow} b_{\vec{p'}+\vec{q}\uparrow} + \frac{U_{hL}}{2} \sum_{\substack{\vec{p}\vec{p'}\vec{q} \\ \sigma\sigma'}} a_{\vec{p}\sigma}^+ (b_{\vec{p'}\sigma'}^+ b_{\vec{p}-\vec{q}\sigma'}) a_{\vec{p'}+\vec{q}\sigma}, \qquad (10.3.3)$$

where in $D$ dimensions ($D = 2, 3$) for the hypercubic lattice:

$$\varepsilon_h(p) - \mu = -2t_h \sum_{a=1}^D \cos(p_a d) - \varepsilon_0 - \mu, \qquad (10.3.4)$$

and

$$\varepsilon_L(p) - \mu = -2t_L \sum_{a=1}^D \cos(p_a d) - \mu, \qquad (10.3.4)$$

are the uncorrelated quasiparticle energies for heavy and light bands (see Fig.10.8) and $p_a = \{p_x, p_y, \ldots\}$ are Cartesian projections of the momentum.

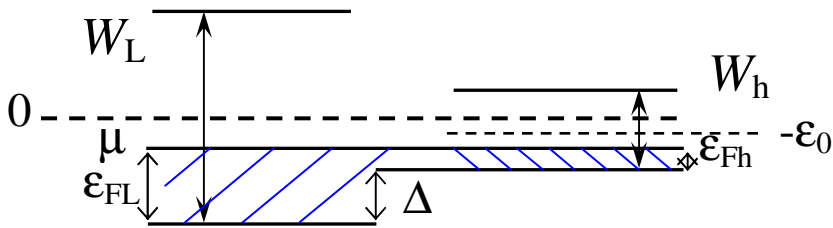

Fig.10.8. Band structure in the two-band model with one narrow band. $W_h$ and $W_L$ are the bandwidths of heavy and light electrons, $\varepsilon_{Fh}$ and $\varepsilon_{Fh}$ are the Fermi-energies, $\Delta = -\varepsilon_0 + \frac{W_L - W_h}{2} = E_{min}^h - E_{min}^L$ is the energy difference between the bottoms of the heavy and light bands, with ($-\varepsilon_0$) being the center of gravity of the heavy band. The center of gravity of the light band is zero, $\mu$ is chemical potential.

For low densities of heavy and light components $n_{\text{tot}}d^D = (n_h + n_L)\,d^D \ll 1$, the quasiparticle spectra are:

$$\varepsilon_h(p) - \mu = -\frac{W_h}{2} + t_h(p^2 d^2) - \varepsilon_0 - \mu,$$

$$\varepsilon_L(p) - \mu = -\frac{W_L}{2} + t_L(p^2 d^2) - \mu \qquad (10.3.5)$$

where $W_h = 4Dt_h$ and $W_L = 4Dt_L$ are bandwidths of heavy and light electrons for $D$ - dimensional hypercubic lattice (for $D = 3$ we have $W_h = 12t_h$ and $W_L = 12t_L$, for $D = 2$ the bandwidths read $W_h = 8t_h$ and $W_L = 8t_L$).

Introducing again (as in one-band Hubbard model in section 9.4) the bare masses of heavy and light components:

$$m_h = \frac{1}{2t_h d^2}; m_L = \frac{1}{2t_L d^2} \qquad (10.3.6)$$

and Fermi-energies:

$$\varepsilon_{Fh} = \frac{p_{Fh}^2}{2m_h} = \frac{W_h}{2} + \mu + \varepsilon_0, \quad \varepsilon_{FL} = \frac{p_{FL}^2}{2m_L} = \frac{W_L}{2} + \mu \qquad (10.3.7)$$

we finally obtain uncorrelated quasiparticle spectra (for temperatures $T \to 0$) as:

$$\varepsilon_h(p) - \mu = \frac{p^2}{2m_h} - \varepsilon_{Fh}, \quad \varepsilon_L(p) - \mu = \frac{p^2}{2m_L} - \varepsilon_{FL} \qquad (10.3.8)$$

In deriving (10.3.5) - (10.3.8) we implicitly assumed that the difference between the bottoms of the heavy and light bands $\Delta$ on Fig.10.8 is not too large, and hence the parabolic approximation for the spectra of both bands is still valid. We note that there is no one-particle hybridization in Hamiltonians (10.3.1), (10.3.3) but there is a strong two particle hybridization $\frac{U_{hL}}{2}\sum_i n_i^h n_i^L$ connected with interband Hubbard interaction $U_{12} = U_{hL}$.

We assume that $m_h \gg m_L$ and therefore:

$$W_h/W_L = m_L/\,m_h \ll 1 \qquad (10.3.9)$$

We also assume that the strong-coupling situation $U_{hh} \sim U_{LL} \sim U_{hL} \gg W_L \gg W_h$ takes place (note that $U_{hL}$ is large because in reality light particles experience strong scattering on the heavy ones as if on a quasiresonance level).

Finally throughout this Chapret we will consider the simplest case where the densities of the bands are of the same order: $n_h \sim n_L \sim n$ (where in 3D $n = \frac{p_F^{\,3}}{2\pi^2}$, and in 2D $n = \frac{p_F^{\,2}}{2\pi}$). In the end of this section note that the two-band Hubbard model with one narrow band is natural generalization of the well-known Falikov-Kimball model with one finite mass and one infinitely large mass[10.21] but contains much more rich physics due to a finite width of a heavy band (instead of a localized level in [10.21]) which allows an interesting dynamics of the heavy component.

## 10.3.2. The Kanamori T-matrix approximation.

In Chapter 9 (section 9.4) when we discussed SC in 3D and 2D one-band Hubbard model at low density we already acquainted ourselves with Kanamori T-matrix approximation [9.33]. Let us apply the same scheme for a two-band Hubbard model. In the 3D case the solution of the corresponding Lipman-Shwinger equations [9.35] yield (see Fig.10.9):

$$T_{hh} = \frac{U_{hh}d^3}{\left(1 - U_{hh}d^3 K_{hh}^{vac}(0,0)\right)} \approx \frac{U_{hh}d^3}{\left(1 + \dfrac{U_{hh}}{8\pi t_h}\right)};$$

$$T_{hL} \approx \frac{U_{hL}d^3}{\left(1 + \dfrac{U_{hL}}{8\pi t_{hL}^*}\right)}; \quad T_{LL} = \frac{U_{LL}d^3}{\left(1 + \dfrac{U_{LL}}{8\pi t_L}\right)}, \tag{10.3.10}$$

where $K_{hh}^{vac}(0,0) = -\int\limits_0^{\sim 1/d} \dfrac{d^3\vec{p}}{(2\pi)^3}\dfrac{m_h}{p^2}$ is a Cooper loop for heavy particles in vacuum, the effective hopping integrals

$$t_{hL}^* = \frac{1}{2d^2 m_{hL}^*} \text{ and } m_{hL}^* = \frac{1}{2t_{hL}^* d^2} = \frac{m_h m_L}{(m_h + m_L)} \approx m_L \tag{10.3.11}$$

is an effective mass for T-matrix $T_{hL}$ for $m_h \gg m_L$ (note that T-matrix $T_{hL}$ describes scattering of heavy electrons on the light ones).

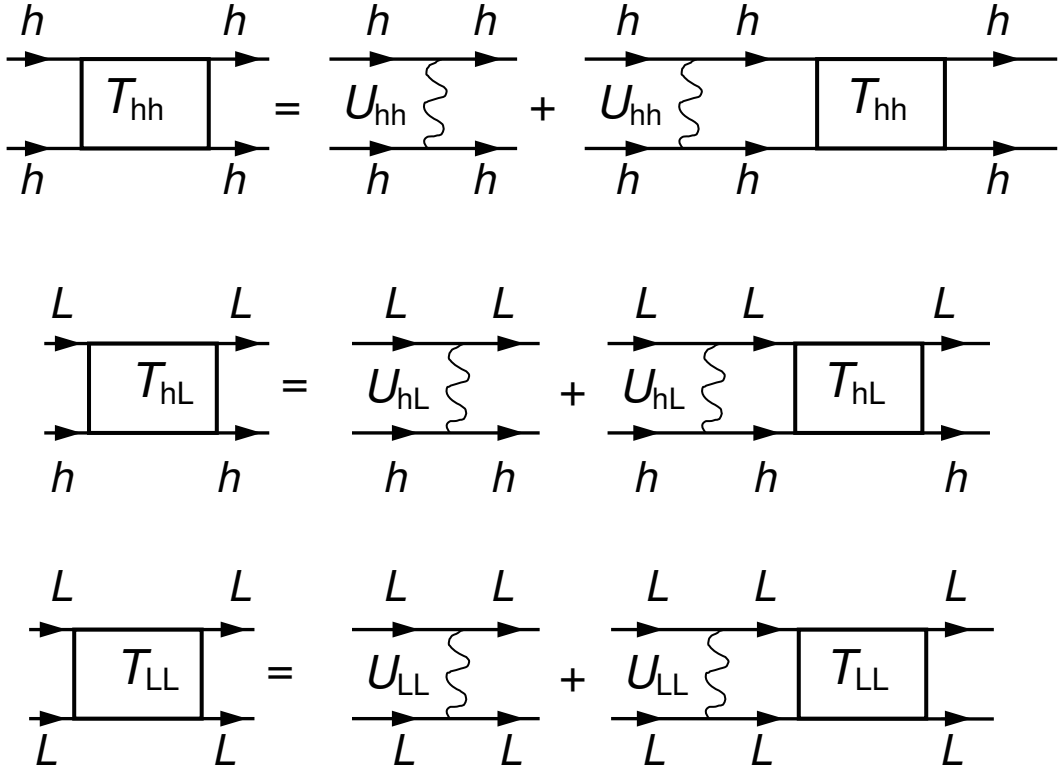

Fig.10.9. Lipman-Shwinger equations for T-matrices $T_{hh}$, $T_{LL}$ and $T_{hL}$ for the two-band model with heavy (h) and light (l) electrons, $U_{hh}$ and $U_{LL}$ are the intraband Hubbard interactions, $U_{hL}$ is the interband Hubbard interaction between heavy and light particles.

Accordingly $t_{hL}^* \sim t_L$ is an effective hopping integral for this T-matrix. The quantities $U_{hh}d^3$, $U_{hL}d^3$ and $U_{LL}d^3$ play the role of zeroth Fourier components for the intraband and interband Hubbard interactions $U_{hh}$, $U_{hL}$ and $U_{LL}$ in 3D. As a result in the strong-coupling case for $U_{hh} \sim U_{LL} \sim U_{hL} \gg W_L \gg W_h$ we have:

$$T_{hh} \approx 8\pi t_h d^3; \quad T_{hL} \approx T_{LL} \approx 8\pi t_L d^3 \tag{10.3.12}$$

The s-wave scattering lengths for the two-band Hubbard model $a = \dfrac{mT}{4\pi} = \dfrac{T}{8\pi t d^2}$ then read in the strong-coupling case:

$$a_{hh} = a_{hL} = a_{LL} = d \tag{10.3.13}$$

.

Correspondingly, the gas-parameter of Galitskii $\lambda = 2a\, p_F/\pi$ in the case of equal densities of heavy and light bands $n_h = n_L$ is given by:

$$\lambda = \left(\lambda_L \approx \frac{2dp_{FL}}{\pi}\right) = \left(\lambda_h \approx \frac{2dp_{Fh}}{\pi}\right) = \frac{2dp_F}{\pi} \qquad (10.3.14)$$

In the 2D case for strong Hubbard interactions and low densities the vacuum T-matrices for $n_h = n_L$ with logarithmic accuracy are given by:

$$T_{hh} \approx \frac{U_{hh}d^2}{\left(1 + \dfrac{U_{hh}d^2}{8\pi t_h}\displaystyle\int\limits_{\sim 1/p_F}^{\sim 1/d^2}\dfrac{dp^2}{p^2}\right)} \approx \frac{U_{hh}d^2}{1 + \dfrac{U_{hh}}{8\pi t_h}\ln\dfrac{1}{p_F^2 d^2}};$$

$$T_{LL} = \frac{U_{LL}d^2}{1 + \dfrac{U_{LL}}{8\pi t_L}\ln\dfrac{1}{p_F^2 d^2}}; \quad T_{hL} = \frac{U_{hL}d^2}{1 + \dfrac{U_{hL}}{8\pi t_{hL}^*}\ln\dfrac{1}{p_F^2 d^2}},$$

$$(10.3.15)$$

where $Ud^2$ plays the role of zeroth Fourier component of the Hubbard potential in 2D. As a result, in the strong-coupling case, the 2D gas-parameter of Bloom [9.3] for equal densities $n_h = n_L$ reads:

$$f_0 = f_{0L} = f_{0h} = \frac{1}{2\ln\dfrac{1}{p_F d}}. \qquad (10.3.16)$$

### 10.3.3. Evaluation of the self-energies of heavy and light bands

Let us now evaluate the self-energies of heavy and light bands. In the two-band model the self-energies of heavy and light particles read (see Fig.10.10):

$$\Sigma_h = \Sigma_{hh} + \Sigma_{hL} \text{ and } \Sigma_L = \Sigma_{LL} + \Sigma_{Lh}. \qquad (10.3.17)$$

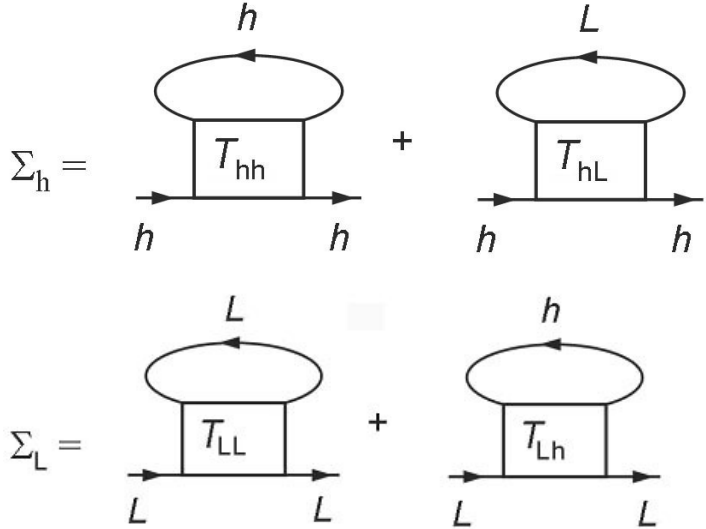

Fig.10.10. The self-energies of heavy and light particles in the T-matrix approximation. $T_{hh}$, $T_{LL}$, $T_{Lh}$, and $T_{hL}$ are full T-matrices of heavy and light particles in substance.

In 3D case the full T-matrices $T_{hh}$ in substance which enter in the first diagram for $\Sigma_h$ in Fig. 10.10 has the form:

$$T_{hh}(\Omega, \vec{p}) = \frac{U_{hh}d^3}{\left(1 - U_{hh}d^3 K_{hh}(\Omega, \vec{p})\right)}, \qquad (10.3.18)$$

where according to [10.4, 9.34]:

$$K_{hh}(\Omega, \vec{p}) = \int \frac{d^3 \vec{p}\,'}{(2\pi)^3} \frac{(1 - n_h^F(\varepsilon_{p'+p} - \mu) - n_h^F(\varepsilon_{-p'} - \mu))}{(\Omega - \varepsilon_h(p'+p) - \varepsilon_h(-p') + 2\mu + i0)} \quad (10.3.19)$$

is a Cooper loop in substance (the product of the two Green's functions in the Cooper (particle-particle) channel), $n_h{}^F(\varepsilon - \mu)$ is the Fermi – Dirac distribution function for heavy particles, and similarly for the full T-matrices $T_{hL}$, $T_{Lh}$ and $T_{LL}$ and Cooper loops $K_{hL}$, $K_{Lh}$ and $K_{LL}$

If we expand the T-matrix for heavy particles in first two orders in the gas-parameter, than according to Galitskii [9.2] we obtain:

$$T_{hh}(\Omega, \vec{p}) = \frac{4\pi a_h}{m_h} + \left(\frac{4\pi a_h}{m_h}\right)^2 (K_{hh} - K_{hh}^{vac}) + o[(\frac{4\pi a_h}{m_h})^3 (K_{hh} - K_{hh}^{vac})^2], \quad (10.3.20)$$

where

$$\frac{4\pi a_h}{m_h} \approx \frac{U_{hh} d^3}{(1 - U_{hh} d^3 K_{hh}^{vac})} \quad (10.3.21)$$

coincides with the Kanamori approximation for the vacuum T-matrix and

$$K_{hh}^{vac}(\Omega, \vec{p}) = \int \frac{d^3 \vec{p}\,'/(2\pi)^3}{(\Omega - \frac{(\vec{p}\,' + \vec{p})^2}{2m_h} - \frac{p'^2}{2m_h})} \quad (10.3.22)$$

is the Cooper loop in vacuum (rigorously speaking the scattering length is defined by $K_{hh}^{vac}(0,0)$, but the difference between $K_{hh}^{vac}(\Omega, \vec{p})$ and $K_{hh}^{vac}(0,0)$ is proportional to the gas-parameter $a_h p_{Fh}$ and is small). $K_{hh}$ in (10.3.20) is the full Cooper loop (cooperon) in substance for heavy particles given by (10.3.19). If we consider the low densities and the energies close to $\varepsilon_F$ we can show that the terms neglected in $T_{hh}$ are small with respect to the gas-parameter

$$\frac{4\pi a_h}{m_h}\left(K_{hh} - K_{hh}^{vac}\right) \sim a_h p_{Fh}. \quad (10.3.23)$$

The self-energy of heavy particles $\Sigma_{hh}$ in the first two orders of the gas-parameter is given by:

$$\Sigma_{hh}(p) = \sum_k T_{hh}(k + p) G_h(k) \approx \frac{4\pi a_h}{m_h} \sum_k G_h(k) - \left(\frac{4\pi a_h}{m_h}\right)^2 \sum_k (K_{hh} - K_{hh}^{vac}) G_h(k) + o(a_h p_{Fh})^3 \cdot \quad (10.3.24)$$

The first term becomes $\frac{4\pi a_h}{m_h} n_h$ which is just the Hartree-Fock contribution [9.2, 9.19]. In the second term we can make an analytic continuation $i\omega_n \to \omega + io$ for bosonic propagator $K_{hh}$ and fermionic propagator $G_h$ [10.22, 10.23]. As a result (bearing in mind that $\text{Im} K_{hh}^{vac} = 0$) we obtain the imaginary part of $\Sigma^{(2)}{}_{hh}$ as:

$$\text{Im}\Sigma_{hh}^{(2)}(\varepsilon, \vec{p}) = \left(\frac{4\pi a_h}{m_h}\right)^2 \sum_k \text{Im} K_{hh}(\varepsilon_k + \varepsilon - \mu, \vec{k} + \vec{p})[n_B(\varepsilon_k + \varepsilon - \mu) + n_F(\varepsilon_k - \mu)] =$$

$$= -\left(\frac{4\pi a_h}{m_h}\right)^2 \pi \int \frac{d^3 \vec{k}}{(2\pi)^3} \int \frac{d^3 \vec{p}\,'}{(2\pi)^3} \left[1 - n_h^F(\vec{p} + \vec{p}\,' + \vec{k}) - n_h^F(-\vec{p}\,')\right][n_B(\varepsilon_k + \varepsilon - \mu) + n_F(\varepsilon_k - \mu)] \cdot$$

$$\cdot \delta\left[\varepsilon + \varepsilon_h(\vec{k}) - \varepsilon_h(\vec{p} + \vec{p}\,' + \vec{k}) - \varepsilon_h(-\vec{p}\,') + \mu\right] \quad (10.3.25)$$

and similarly for the real part of $\Sigma_{hh}^{(2)}$:

$$\text{Re}\Sigma_{hh}^{(2)}(\varepsilon, \vec{p}) = \left(\frac{4\pi a_h}{m_h}\right)^2 \sum_k \left[\text{Re} K_{hh}(\varepsilon_k + \varepsilon - \mu, \vec{k} + \vec{p}) - \text{Re} K_{hh}^{vac}(\varepsilon_k + \varepsilon_p - 2\mu, \vec{k} + \vec{p})\right][n_B(\varepsilon_k + \varepsilon - \mu) + n_F(\varepsilon_k - \mu)] \quad (10.3.26)$$

where for the real part of a Cooper loop in vacuum we have:

$$\operatorname{Re} K_{hh}^{vac}(\varepsilon_k + \varepsilon_p, \vec{k} + \vec{p}) = \int \frac{d^3 \vec{p}'}{(2\pi)^3} P \frac{2m_h}{\vec{k}^2 + \vec{p}^2 - (\vec{p}' + \vec{k} + \vec{p})^2 - \vec{p}'^2} \qquad (10.3.27)$$

Thus $\operatorname{Re} K_{hh}^{vac}$ is calculated in resonance for $\Omega = \varepsilon_k + \varepsilon_p$ (or for $\varepsilon = \varepsilon_p$), where $P$ is the principal value. In (10.3.25) and (10.3.26) $n_B(\Omega) = \frac{1}{(e^{\Omega/T} - 1)}$ and $n_F(\Omega) = \frac{1}{(e^{\Omega/T} + 1)}$ are the bosonic and fermionic distribution functions and hence:

$$n_B(\varepsilon_k + \varepsilon - \mu) + n_F(\varepsilon_k - \mu) = \frac{1}{2}\left[ cth \frac{(\varepsilon_k + \varepsilon) - \mu}{2T} + th \frac{\varepsilon_k - \mu}{2T} \right] \qquad (10.3.28)$$

The real part of a Cooper loop in substance for heavy particles is given by:

$$\operatorname{Re} K_{hh}(\varepsilon_k + \varepsilon - \mu, \vec{k} + \vec{p}) = \int \frac{d^3 \vec{p}'}{(2\pi)^3} \frac{[1 - n_h^F(\vec{p} + \vec{p}' + \vec{k}) - n_h^F(-\vec{p}')]}{[\varepsilon + \varepsilon_h(\vec{k}) - \varepsilon_h(\vec{p} + \vec{p}' + \vec{k}) - \varepsilon_h(-\vec{p}') + \mu]} \qquad (10.3.29)$$

The analytic continuation for $\Sigma^{(2)}_{hh}$ in a 2D case is similar to the one in the 3D case.

We note that for $\Omega/T \gg 1$, the bosonic distribution function $n_B(\Omega) \to 0$ and the fermionic distribution function $n_F(\Omega) \to \theta(\Omega)$ - to the step-function. Hence at low temperatures $\operatorname{Im}\Sigma_{hh}$ and $\operatorname{Re}\Sigma_{hh}$ acquire the standard form [9.2, 9.19, 9.20]. We will analyze their behavior at finite temperatures more detaily in Chapters 16 and 17.

Note that for higher temperatures we should keep in mind that $n_B(\Omega) \to T/\Omega$ for $T \gg \Omega$. The fermionic distribution function is "washed out" by temperature. Accordingly, $n_B(\Omega) = \frac{1}{2}\left(1 - \frac{\Omega}{2T}\right)$. These approximations are important when we evaluate $\operatorname{Im}\Sigma$ for higher temperatures $T > W_h$ [9.32] in chapter 17.

We note that in contrast with the model of slightly non-ideal Fermi-gas (see [9.2, 9.19, 9.20]) the Hubbard model does not contain an exchange-type diagram for $\Sigma_{hh}$ (see Fig. 10.11) because the T-matrix in this diagram corresponds to incoming and outgoing heavy particles with the same spin-projection $a^+_\sigma a^+_\sigma a_\sigma a_\sigma$ while the Hubbard model contains only the matrix elements $a^+_\uparrow a^+_\downarrow a_\downarrow a_\uparrow$ with different spin-projections.

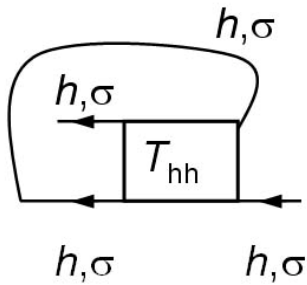

Fig. 10.11. An exchange – type diagram for the self-energy $\Sigma_{hh}{}^\sigma$ which contains the matrix element $a^+_\sigma a^+_\sigma a_\sigma a_\sigma$ and thus is absent in the Hubbard model.

We also note that when we expand the T-matrix up to second order in the gas-parameter, we implicitly assume that the T-matrix itself does not have a simple pole structure of the type of a bosonic propagator. This is a case for a partially filled heavy band $n_h d^D \ll 1$ and the low energy sector where $0 < \varepsilon < W_h \ll U_{hh}$. Effectively we neglect the lattice in this expansion.

However if we take the lattice into account, then we will get two poles for the full (unexpanded) T-matrix of heavy particles in (10.3.18). The first one is connected with the so-called antibound state predicted by Hubbard [9.29] and Anderson [10.24] and corresponds to large positive energy:

$$\varepsilon \sim U_{hh} > 0 \qquad (10.3.30)$$

Physically it describes an antibound pair of two heavy particles with energy of $U_{hh}$ on the same lattice site. It therefore reflects the presence of the upper Hubbard band already at low densities $n_h d^D \ll 1$ [10.25]. But we will show that the intensity of the upper Hubbard band (UHB) is small at low densities and for low energy sector. The more detailed discussion of the antibound state will be a subject of Chapter 16.

A second pole in the full T-matrix found by [10.26] corresponds to negative energy and in the 2D case yields:

$$\varepsilon \approx -2\varepsilon_{Fh} - \frac{2\varepsilon_{Fh}^2}{W_h} < 0 \qquad (10.3.31)$$

It describes the bound state of the two holes below the bottom of the heavy band ($\varepsilon < -2\varepsilon_{Fh}$). Therefore, it has zero imaginary part and does not contribute to Im$T$. (This mode produces non-analytic corrections to Re$\Sigma_{hh} \sim |\varepsilon|^{5/2}$ in 2D). We will also consider this mode more detailly in Chapter 16. In the forthcoming sections we can neglect both these two contributions to the T-matrices and self-energies.

### 10.3.4. Electron polaron effect.

For temperatures $T \to 0$ the one-particle Green's functions for heavy and light electrons are given by [9.19, 9.20]:

$$G_h(\omega, \vec{q}) = \frac{1}{(\omega - \varepsilon_h(q) + \mu - \Sigma_h(\omega, \vec{q}))} \approx \frac{Z_h}{(\omega - \varepsilon_h^*(q) + \mu + io)} \qquad (10.3.32)$$

and respectively

$$G_L(\omega, \vec{q}) \approx \frac{Z_L}{(\omega - \varepsilon_L^*(q) + \mu + io)}, \qquad (10.3.33)$$

where according to Galitskii [9.2]:

$$\varepsilon_h^*(q) - \mu = \frac{(q^2 - p_{Fh}^2)}{2m_h^*} \text{ and } \varepsilon_L^*(q) - \mu = \frac{(q^2 - p_{FL}^2)}{2m_L^*} \qquad (10.3.34)$$

are renormalized quasiparticle spectra, and

$$Z_h^{-1} = \left(1 - \frac{\partial \mathrm{Re}\Sigma_h^{(2)}(\omega, \vec{q})}{\partial \omega}\bigg|_{\substack{\omega \to 0 \\ q \to p_{Fh}}}\right); Z_L^{-1} = \left(1 - \frac{\partial \mathrm{Re}\Sigma_L^{(2)}(\omega, \vec{q})}{\partial \omega}\bigg|_{\substack{\omega \to 0 \\ q \to p_{FL}}}\right) - \qquad (10.3.35)$$

are inverse $Z$-factors [9.20, 9.19, 10.27, 10.28] of heavy and light electrons. The leading contribution to Im$\Sigma_h^{(2)}$ in (10.3.35) comes from the substitution of Re$\Sigma_{hL}^{(2)}(\omega, \vec{q})$ to $Z_h^{-1}$, which is described by a formula similar to (10.3.26), and yields:

$$1 - Z_h^{-1} = \lim_{\substack{\omega \to 0 \\ q \to p_{Fh}}} \frac{\partial \mathrm{Re}\Sigma_{hL}^{(2)}(\omega, \vec{q})}{\partial \omega} \sim -\left(\frac{4\pi a_{hL}}{m_{hL}^*}\right)^2 \iint \frac{d^D\vec{p}}{(2\pi)^D} \frac{d^D\vec{p}'}{(2\pi)^D} \cdot$$
$$\cdot \frac{[1 - n_L^F(\vec{p}'+\vec{p}) - n_h^F(-\vec{p}')]n_L^F(\vec{p}-\vec{q})}{[\varepsilon_L(\vec{p}-\vec{q}) - \varepsilon_L(\vec{p}'+\vec{p}) - \varepsilon_h(-\vec{p}') + \mu]^2}, \qquad (10.3.36)$$

where $n_B(\Omega) \to 0$, $n_F(\Omega)$ is a step function for $\Omega/T \gg 1$; $a_{hL} \approx d$ in 3D is connected with the vacuum T-matrix $T_{hL}$, and $m_{hL}^* \approx m_L$. Replacing in (10.3.36) $\frac{d^D\vec{p}}{(2\pi)^D} \frac{d^D\vec{p}'}{(2\pi)^D}$ by $N_L^2(0)d\varepsilon_L(\vec{p})d\varepsilon_L(\vec{p}')$ (where $N_L(0)$ is a density of states of light particles, $D = 2, 3$), and taking into account that in (10.3.36) $\varepsilon_L(\vec{p}-\vec{q}) - \mu < 0$ while $\varepsilon_L(\vec{p}'+\vec{p}) - \mu > 0$ we can easily verify that for $m_h \gg m_L$ or equivalently for $\varepsilon_{FL} \gg \varepsilon_{Fh}$ this expression contains a large logarithm both in 3D and 2D cases (see [10.18]). Hence, the $Z$-factor of the heavy particles in 3D in the leading approximation in the gas-parameter $\lambda = 2a\,p_F/\pi$ is given by:

$$Z_h^{-1} \approx 1 + 2\lambda^2 \ln\frac{m_h}{m_L} \qquad (10.3.37)$$

Correspondingly, in 2D:

$$Z_h^{-1} = 1 + 2f_0^2 \ln\frac{m_h}{m_L} \qquad (10.3.38)$$

where $f_0 = \dfrac{1}{2\ln(1/\,p_{FL}d)}$ - 2D gas-parameter of Bloom.

We note that the contribution to $Z_h^{-1}$ from $\mathrm{Re}\Sigma_{hh}^{(2)}$ does not contain a large logarithm and thus is smaller than the contribution from $\mathrm{Re}\Sigma_{hL}^{(2)}$ at low electron density. The analogous calculation for $Z_L$ with $\mathrm{Re}\Sigma_{Lh}^{(2)}$ and $\mathrm{Re}\Sigma_{LL}^{(2)}$ yields only $Z_L^{-1} \sim 1 + \lambda^2$ in 3D and $Z_L^{-1} \sim 1 + f_0^2$ in 2D (does not contain a large logarithm).

Correspondingly for the effective mass of a heavy particle according to [9.20, 10.27, 10.28] we have:

$$\frac{m_h}{m_h^*} = Z_h\left(1 + \frac{\partial\,\mathrm{Re}\Sigma_{hL}^{(2)}(\varepsilon_h(\vec{q}) - \mu, \vec{q})}{\partial\varepsilon_h(\vec{q})}\bigg|_{\varepsilon_h(q)\to\mu}\right), \qquad (10.3.39)$$

where the second contribution in brackets corresponds to the momentum dependence of the self-energy $\Sigma_{hL}$.

Therefore as usual the $Z$-factor contributes to the enhancement of a heavy mass:

$$\frac{m_h^*}{m_h} \sim Z_h^{-1} \sim \left(1 + 2\lambda^2 \ln\frac{m_h}{m_L}\right) \text{ in 3D} \qquad (10.3.40)$$

Analogous calculations for $Z$-factor contribution to the light mass yields only

$$\frac{m_L^*}{m_L} \sim Z_L^{-1} \sim 1 + \lambda^2 \text{ in 3D} \qquad (10.3.41)$$

If the effective parameter $2\lambda^2\ln(m_h/m_L) > 1$ we are in the situation of strong electron polaron effect (strong EPE). To obtain the correct polaron exponent in this region of parameters diagrammatically, we should sum up at least the so-called maximally crossed diagrams for $\mathrm{Re}\Sigma_{hL}$ [10.30]. But this exponent can be also evaluated in a different manner, based on the non-adiabatic part of the many particle wave-function [10.18] which describes the heavy particle dressed in a cloud of virtual electron – hole pairs of light particles. This yields:

$$\frac{m_h^*}{m_h} \sim Z_h^{-1} = \left(\frac{m_h}{m_L}\right)^{\frac{b}{(1-b)}}, \qquad (10.3.42)$$

where $b = 2\lambda^2$ in 3D and $b = 2f_0^2$ in 2D.

For $b = \frac{1}{2}$ or equivalently for $\lambda = \frac{1}{2}$ in 3D or $f_0 = \frac{1}{2}$ in 2D (as for the coupling constant for screened Coulomb interaction in 3D and 2D electron plasma in the RPA-theory - see preceding section 10.2 of this Chapter), we are in the so-called unitary limit. In this limit according to [10.18] the polaron exponent:

$$\frac{b}{(1-b)} = 1, \qquad (10.3.43)$$

and hence:

$$\frac{m_h^*}{m_h} = \frac{m_h}{m_L} \qquad (10.3.44)$$

or equivalently:

$$\frac{m_h^*}{m_L} = \left(\frac{m_h}{m_L}\right)^2. \qquad (10.3.45)$$

Thus starting from the ratio between the bare masses $m_h/m_L \sim 10$ (obtained, for instance, in LDA-approximation [10.31]) we finish in the unitary limit with $m_h^*/m_L \sim 100$ (due to many-body electron-polaron effect), which is a typical ratio for uranium-based heavy-fermion systems [10.32-10.36].

### 10.3.5. Other mechanisms of heavy mass enhancement

We note that rigorously speaking (see (10.3.39)) the momentum dependence of $\mathrm{Re}\,\Sigma_{hL}^{(2)}(\varepsilon_h(\vec{q}) - \mu, \vec{q})$ is also very important for the evaluation of the effective mass. Preliminary estimates of Prokof'ev, Kagan (see also [10.4] and [9.34]) show that in the zeroth approximation in $m_L/m_h$ in 3D case close to the Fermi-surface (for $\varepsilon_h(\vec{q}) - \mu = (q^2 - p_{Fh}^2)/2m_h \to 0$ and $q \to p_{Fh}$):

$$\mathrm{Re}\,\Sigma_{hL}^{(2)}(\varepsilon_h(\vec{q}) - \mu, \vec{q}) \approx 2\left(\frac{4\pi a_{hL}}{m_L}\right)^2 \int \frac{d^3\vec{p}}{(2\pi)^3}\Pi_{LL}(0,\vec{p})n_h^F(\vec{p}-\vec{q}), \quad (10.3.46)$$

where (see also (9.3.4))

$$\Pi_{LL}(0,\vec{p}) = \int \frac{d^3\vec{p}'}{(2\pi)^3}\frac{[n_L^F(\varepsilon_{p'+p} - \mu) - n_L^F(\varepsilon_{p'} - \mu)]}{\varepsilon_L(\vec{p}') - \varepsilon_L(\vec{p}' + \vec{p})} \qquad - \qquad (10.3.47)$$

is a static polarization operator for light particles. Having in mind that $|\vec{p} - \vec{q}| < p_{Fh}$ and $q \approx p_{Fh}$ we can see that $\vec{p} \to 0$ and use the asymptotic form for $\Pi_{LL}(0,\vec{p})$ at small $p << p_{FL}$ (if the densities of heavy and light bands are not very different and $p_{FL} \sim p_{Fh}$):

$$\lim_{p>0}\Pi_{LL}(0,\vec{p}) = N_L^{3D}(0)\left[1 - \frac{p^2}{12\,p_{FL}^2}\right], \qquad (10.3.48)$$

where $N_L^{3D}(0) = m_L p_{FL}/2\pi^2$ is the density of states for light electrons in 3D. The substitution of $\lim_{p\to 0}\Pi_{LL}(0,\vec{p})$ from (10.3.48) to (10.3.46) yields:

$$\mathrm{Re}\,\Sigma_{hL}^{(2)}(\varepsilon_h(\vec{q}) - \mu, \vec{q}) \approx \mathrm{Re}\,\Sigma_{hL}^{(2)}(0, p_{Fh}) - \frac{(q^2 - p_{Fh}^2)}{2m_h}\frac{\lambda^2}{9}\frac{m_h n_h}{m_L n_L}, \quad (10.3.49)$$

where $\lambda = 2a\,p_F/\pi$ is a 3D gas parameter, $n_h = p_{Fh}^3/3\pi^2$, $n_L = p_{FL}^3/3\pi^2$ are the densities of heavy and light bands.

The first term in (10.3.49) describes $\mathrm{Re}\,\Sigma_{hL}^{(2)}(\varepsilon_h(\vec{q}) - \mu, \vec{q})$ on the Fermi-surface (for $\varepsilon_h(q) - \mu = 0$ and $q = p_{Fh}$). It reads:

$$\mathrm{Re}\,\Sigma_{hL}^{(2)}(0, p_F) \approx \frac{4\lambda^2}{3}\frac{n_h}{n_L}\varepsilon_{FL}\left(1 - \frac{2p_{Fh}^2}{15\,p_{FL}^2}\right) > 0 \text{ for } p_{FL} \sim p_{Fh} \qquad (10.3.50)$$

It is renormalization of the effective chemical potential of the heavy band in the second order of the gas parameter due to the interaction of light and heavy particles.

We note that according to [9.19, 9.20] the renormalized heavy-particle spectrum is given by:

$$\varepsilon_h^*(q) - \mu = \left(\frac{q^2}{2m_h} - \mu_h^{eff}\right) + \frac{2\pi}{m_L} n_L(\mu) a_{hL} + \mathrm{Re}\,\Sigma_{hL}^{(2)}\left(\varepsilon_h(\vec{q}) - \mu, \vec{q}\right) = \frac{\left(q^2 - p_{Fh}^2\right)}{2m_h^*}, \qquad (10.3.51)$$

where the scattering length $a_{hL} \approx d$, an effective chemical potential $\mu_h^{eff} = \mu_h + W_h/2 + \varepsilon_0$ is counted from the bottom of a heavy band, and the Hartree-Fock term $\frac{2\pi}{m_L} n_L(\mu) a_{hL}$ represents the contribution to the self-energy $\mathrm{Re}\,\Sigma_{hL}^{(1)}$ in the first-order in the gas parameter. From (10.3.51) collecting the terms proportional to $\varepsilon_h(\vec{q}) - \mu = (q^2 - p_{Fh}^2)/2m_h$ we obtain:

$$\frac{(q^2 - p_{Fh}^2)}{2m_h^*} = \left[\varepsilon_h(q) - \mu\right]\left(1 - \frac{\lambda^2}{9}\frac{m_h n_h}{m_L n_L}\right). \qquad (10.3.52)$$

Correspondingly, the effective mass of a heavy particle is given by:

$$\frac{m_h}{m_h^*} = 1 + \frac{\partial\,\mathrm{Re}\,\Sigma_{hL}^{(2)}\left(\varepsilon_h(q) - \mu, q\right)}{\partial(\varepsilon_h(q) - \mu)}\Bigg|_{\varepsilon_h(q)\to\mu} = 1 - \frac{\lambda^2}{9}\frac{m_h n_h}{m_L n_L}. \qquad (10.3.53)$$

As a result we obtain much more dramatic (linear in $m_h/m_L$) enhancement of $m_h^*$ than in EPE (which yields only logarithmic $m_h/m_L$ enhancement $m_h/m_L^* \approx 1 - 2\lambda^2 \ln(m_h/m_L)$ due to the $Z$-factor of a heavy particle). For $m_h/m_L \sim 10$ the contribution to $m_h^*$ in (10.3.53) becomes larger then the contribution to $Z$-factor in (10.3.40) for a large density mismatch $n_h \geq 5n_L$ between the heavy and light band. In general in the second order of the gas parameter in 3D:

$$\frac{m_h^*}{m_h} = 1 + \frac{\lambda^2}{9}\frac{m_h n_h}{m_L n_L} + 2\lambda^2 \ln\frac{m_h}{m_L} \qquad (10.3.54)$$

We note that the contribution to $m_h^*/m_h$ from $\mathrm{Re}\,\Sigma_{hh}^{(2)}(\varepsilon_h(q) - \mu, \vec{q})$ associated with the "heavy-heavy" interaction is small in comparison with the contribution to $m_h^*$ from $\mathrm{Re}\,\Sigma_{hL}^{(2)}$ (which is associated with "heavy-light" interaction) due to the smallness of the ratio between the bare masses: $m_L/m_h << 1$. We can now collect the terms which do not depend upon $\varepsilon_h(q)$ - $\mu$ in (10.3.51). This gives the effective chemical potential of heavy electrons:

$$\mu_h^{eff} = \frac{p_{Fh}^2}{2m_h} + \frac{2\pi}{m_L} n_L(\mu) a_{hL} + \mathrm{Re}\,\Sigma_{hL}^{(2)}(0, p_{Fh}) \qquad (10.3.55)$$

We note that the contributions to $\mu_h^{eff}$ from the Hartree-Fock term $\frac{2\pi}{m_L} n_L(\mu) a_{hL}$ of heavy electrons and from $\mathrm{Re}\,\Sigma_{hh}^{(2)}(0, p_{Fh})$ (which is connected with "heavy-heavy" interactions) are small in comparison with "heavy-light" contributions due to the smallness of the ratio between the bare masses: $m_L/m_h << 1$.

## 2D situation

In 2D the static polarization operator for light particles is (see [10.4, 9.34, 10.55]):

$$\Pi_{LL}(0,\vec{p}) = \frac{m_L}{2\pi}\left[1 - \mathrm{Re}\sqrt{1 - \frac{4p_{FL}^2}{p^2}}\right], \qquad (10.3.56)$$

and hence for $p < 2p_{\mathrm{FL}}$: $\Pi_{LL}(0,\vec{p}) = \frac{m_L}{2\pi}$ - does not contain any dependence on $p^2$ in contrast to the 3D case. Thus EPE in 2D is a dominant mechanism of the heavy mass enhancement and in general in the second order of the gas parameter $\frac{m_h{}^*}{m_h} = 1 + 2f_0^2 \ln\frac{m_h}{m_L}$; $\frac{m_L{}^*}{m_L} \sim 1 + f_0^2$ in 2D.

In the end of this subsection we would like to note that the important role of the interband ("heavy"-"light") Hubbard repulsion $U_{hL}$ for the formation of a heavy mass $m^* \sim 100 m_e$ in a two-band Hubbard model was also emphasized in [8.56] for the LiV$_2$O$_4$ HF compound.

## The tendency towards phase-separation

We also note that for large density mismatch $n_h \gg n_L$ we could have a tendency towards phase-separation in a two-band model [10.4] in 3D. Namely if we evaluate the partial compressibility of the heavy component (the sound velocity of heavy particles squared):

$$\kappa_{hh}^{-1} \sim c_h^2 = \frac{n_h}{m_h}\left(\frac{\partial \mu_h^{eff}}{\partial n_h}\right) \qquad (10.3.57)$$

we already see the tendency towards phase-separation $\kappa_{hh}^{-1} < 0$ (towards negative compressibility) in the strong coupling limit and low densities $\lambda^2 \frac{m_h p_{Fh}}{m_L p_{FL}} \geq 1$ in qualitative agreement with more phenomenological (mean-field type) variational approach [10.20].

The tendency towards phase-separation at low electron fillings also manifests itself for the asymmetric Hubbard model (where only interband Hubbard repulsion $U_{hL}$ between heavy and light electrons is present and intraband Hubbard repulsions $U_{hh}$ and $U_{LL}$ are absent) in the limit of strong asymmetry $t_h \ll t_L$ [8.57] between heavy and light bandwidths.

In the end of this section we emphasize that the physics of EPE and evaluation of $Z_h$ in [10.18] are to some extent connected with the well-known results of Nozieres et al., [10.19, 10.38] on infrared divergences in the description of the Brownian motion of a heavy particle in a Fermi-liquid and on the infrared divergences for the problem of X-ray photoemission from the deep electron levels, as well as with the famous results of Anderson [10.39] on the orthogonality catastrophe for the 1D chain of $N$ electrons under the addition of one impurity to the system.

Finally we mention a competing mechanism [10.40] proposed by Fulde for the explanation of the effective mass in praseodymium (Pr) and in some uranium-based molecules like U(C$_8$H$_8$)$_2$. Later on Fulde et al., [10.40] generalized this mechanism on some other uranium-based heavy-fermion (HF) compounds with localized and delocalized orbitals. This mechanism has a quantum-chemical nature and is based on the scattering of conductive electrons on localized orbitals as if on the two-level systems. The mass-enhancement is here governed by non-diagonal matrix elements of the Coulomb interaction which are not contained in the simple version of a two-band model (10.3.1). In this context we also mention [10.41] where the authors considered the mass-enhancement of conductivity electrons due to their scattering on local $f$-levels splitted by the crystalline field.

We note that de Haas van Alphen (dHvA) experiments [10.42] together with ARPES (angle-resolved photoemission spectroscopy) experiments [10.43] and thermodynamic measurements [10.33, 10.34, 10.44] are the main instruments to reconstruct the Fermi-surface for HF-compounds and to determine the effective mass (thus verifying the predictions of different theories regarding the mass enhancement in uranium-based HF-compounds).

### 10.3.6. Anomalous superconductivity in the two-band model with one narrow band.

For the sake of completeness let us consider briefly superconductivity problem in the same type of models [10.3] and namely in the two-band model with narrow band [10.4]. Let us concentrate on a 2D case where critical temperatures are higher already at low densities and consider the most typical case (see Fig. 10.8) $m_h > m_L$ and $p_{Fh} > p_{FL}$. We assume however that still the mismatch between the densities is not large enough to produce phase-separation. Note that in 2D case where only EPE is present the restrictions on a homogeneous state could be more mild than in a 3D case. At low densities $n_L d^2 < n_h d^2 \ll 1$ the maximal $T_C$ corresponds to p-wave pairing and is governed by the enhanced Kohn – Luttinger mechanism of SC [9.4, 10.1, 10.3]. The general expression for the effective interaction $U_{\text{eff}}$ of the heavy particles (for the irreducible bare vertex for the Cooper channel) in the first two orders in the gas-parameter reads:

$$U_{eff}(\vec{p}_h, \vec{p}_h') = T_{hh} + T_{hh}^2 \Pi_{hh}(0, \vec{\tilde{q}}_h = \vec{p}_h + \vec{p}_h') - 2T_{hL}^2 \Pi_{LL}(0, \vec{q}_h = \vec{p}_h - \vec{p}_h'), \quad (10.3.58)$$

where $\vec{p}_h$ and $\vec{p}'_h$ are incoming and outgoing momenta for the heavy particles in the Cooper channel, $|\vec{p}_h| = |\vec{p}_h'| = p_{Fh}$ and:

$$q_h^2 = 2p_{Fh}^2(1 - \cos\varphi); \; \tilde{q}_h^2 = 2p_{Fh}^2(1 + \cos\varphi) \quad (10.3.59)$$

are transferred momentum squared (for $q_h^2$) and transferred momentum with an account of crossing squared (for $\tilde{q}_h^2$). These formulas are exactly analogous to (9.3.11), (9.3.16) but correspond to 2D case, φ is an angle between $\vec{p}_h$ and $\vec{p}'_h$. Note that, as we discussed in Chapter 7, both transferred momenta $q_h \le 2p_{Fh}$ and $\tilde{q}_h \le 2p_{Fh}$ for superconductivity problem. The second term in (10.3.58) is connected with an exchange diagram (see Fig.9.2 and (9.3.10)) for heavy electrons while the third term is a static polarization operator (10.3.47).

In (10.3.58) for $\Pi_{hh}$ and $\Pi_{LL}$ we get:

$$\Pi_{hh}(0, \vec{\tilde{q}}_h) = Z_h^2 \frac{m_h^*}{2\pi}\left[1 - \text{Re}\sqrt{1 - \frac{4p_{Fh}^2}{\tilde{q}_h^2}}\right],$$

$$\Pi_{LL}(0, \vec{q}_h) = Z_L^2 \frac{m_L^*}{2\pi}\left[1 - \text{Re}\sqrt{1 - \frac{4p_{FL}^2}{q_h^2}}\right] \quad (10.3.60)$$

where $Z_h$ and $m_h^*$ are Z- factor and effective mass of heavy particle, $Z_L$ and $m_L^*$ are Z- factor and effective mass of light particle, $p_{Fh}$ and $p_{FL}$ – are Fermi-momenta for heavy and light particles. Having in mind that $Z_L \sim m_L / m_L^* \sim (1 - f_0^2)$, we can put $Z_L \sim 1$ and $m_L^* \sim m_L$ in all the forthcoming estimates. Finally in (10.3.58) for $p_{Fh} > p_{FL}$ the Kanamori T-matrices read in the strong coupling case in 2D:

$$T_{hh} = \frac{4\pi}{m_h^*}\frac{1}{\ln\left(\frac{1}{p_{Fh}^2 d^2}\right)} > 0, \; T_{hL} = \frac{4\pi}{m_L^*}\frac{1}{\ln\left(\frac{1}{p_{Fh}^2 d^2}\right)} > 0 \quad (10.3.61)$$

Having in mind that $\tilde{q}_h \le 2p_{Fh}$ we get: $\Pi_{hh}(0, \vec{\tilde{q}}_h) = Z_h^2 m_h^*/2\pi$ – does not contain any dependence upon transferred momentum with crossing $\tilde{q}_h$.

In the same time $\Pi_{LL}(0, \vec{q}_h)$ contains nontrivial dependence upon $q_h$ for $p_{Fh} > p_{FL}$. We can say [10.1, 10.3, 10.4] that large 2D Kohn's anomaly becomes effective for SC – problem already in the second order of the gas-parameter and we have the pairing of heavy electrons via polarization of light electrons (see Fig. 10.12).

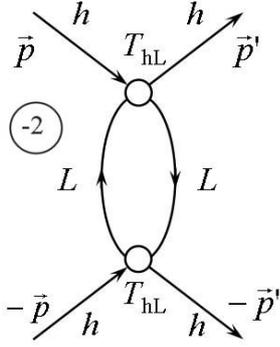

Fig. 10.12. The leading contribution to the effective interaction $U_{eff}$ for the p-wave pairing of heavy particles via polarization of light particles. The open circles stand for the vacuum T-matrix $T_{hL}$.

Note that a standard s-wave pairing is suppressed in a two-band Hubbard model by short-range Hubbard repulsion which yields $T_{hh} > 0$ in the first-order contribution to $U_{eff}$ in (10.3.58).

According to Landau-Thouless criterion [9.19] the maximal critical temperature in our model corresponds to triplet p-wave pairing (to pairing with magnetic quantum number $m = 1$ in 2D) and reads:

$$-U_{eff}^{m=1} N_{2D}^{h\,*}(0) Z_h^2 \ln \frac{\varepsilon_{Fh}^*}{T_{C1}} = 1, \qquad (10.3.62)$$

where $N_{2D}^{h\,*}(0) = m_h^*/2\pi$ is an effective 2D density of states for heavy electrons with effective heavy mass $m_h^*$; $\varepsilon_{Fh}^* = p_{Fh}^2/2m_h^*$ - is renormalized Fermi-energy for heavy electrons; $U_{eff}^{m=1}$ is a p-wave harmonic of the effective interaction. It is given by:

$$U_{eff}^{m=1} = \int_0^{2\pi} U_{eff}[q(\cos\varphi)]\cos\varphi \frac{d\varphi}{2\pi} \qquad (10.3.63)$$

It is shown in [10.1, 10.3, 10.4] that $U_{eff}^{m=1}$ depends upon relative occupation of the two bands $p_{Fh}/p_{FL}$ and yields:

$$U_{eff}^{m=1} = N_{2D}^{L\,*}(0) \frac{(p_{Fh}/p_{FL} - 1)}{(p_{Fh}^2/p_{FL}^2)} T_{hL}^2(-2) < 0. \qquad (10.3.64)$$

Moreover it corresponds to attraction. In (10.3.64) $N_{2D}^{L\,*}(0) = m_L^*/2\pi$ is effective 2D density of states for light electrons. We can see that $U_{eff}^{m=1} \to 0$ for $p_{Fh}/p_{FL} \to 1$ and $p_{Fh}/p_{FL} \to \infty$. It is easy to show that $U_{eff}^{m=1}$ has rather large and broad maximum [10.1, 10.3, 10.4] for $p_{Fh} = 2p_{FL}$ or equivalently for $n_h = 4n_L$ (see Fig. 10.13). In maximum an effective interaction reads:

$$U_{eff}^{m=1} = -\frac{1}{2} N_{2D}^{L\,*}(0) \left( \frac{4\pi}{m_L^* \ln(1/p_{Fh}^2 d^2)} \right)^2. \qquad (10.3.65)$$

Correspondingly Landau-Thouless criterion for superconducting temperature $T_{C1}$ yields:

$$\frac{m_h^*}{m_L^*} Z_h^2 2 f_0^2 \ln \frac{\varepsilon_{Fh}^*}{T_{C1}} = 1, \qquad (10.3.66)$$

where $f_0 = \dfrac{1}{\ln(1/p_{Fh}^2 d^2)}$ is a 2D gas parameter. For $f_0^2 \ln \dfrac{m_h}{m_L} \leq 1$ EPE is weak and $m_h^* \approx m_h$.

Thus $Z_h \approx 1$, $\varepsilon_{Fh}^* \approx \varepsilon_{Fh}$ and Landau-Thouless criterion reads: $\left(\dfrac{m_h}{m_L}\right) 2f_0^2 \ln \dfrac{\varepsilon_{Fh}}{T_{c1}} = 1$. An effective gas-parameter which enters the formula for $T_{C1}$ for weak EPE is thus $f_0 (\dfrac{m_h}{m_L})^{1/2}$. In weak-coupling Born case for $U_{hh} \sim U_{hL} \sim U_{LL} < W_h < W_L$ the critical temperature $T_{C1} \sim \varepsilon_{Fh} \exp\left\{-\dfrac{1}{2\tilde{f}_0^2}\right\}$

where $\tilde{f}_0 = \dfrac{\sqrt{m_L m_h} U_{hL}}{4\pi}$ is connected with interband Hubbard interaction $U_{hL}$ and proportional to geometric average of heavy and light masses $\sqrt{m_L m_h}$ (see [9.32, 10.3]). In opposite strong-coupling case $U_{hL} > W_h > W_L$ the critical temperature $T_{C1} \sim \varepsilon_{Fh} \exp\left\{-\dfrac{1}{2\tilde{f}_0^2}\right\}$ with effective gas-

parameter:

$$\tilde{f}_0 = \sqrt{\dfrac{m_h}{m_L}} \dfrac{1}{\ln\left(\dfrac{1}{p_{Fh}^2 d^2}\right)} \qquad (10.3.67)$$

It is interesting to emphasize that in the unitary limit $f_0 \to \frac{1}{2}$ the EPE yields for the heavy-mass enhancement $m_h^*/m_L \sim (m_h/m_L)^2$ and for $Z$- factor of heavy particle $Z_h \sim m_h/m_h^* \sim m_L/m_h$.

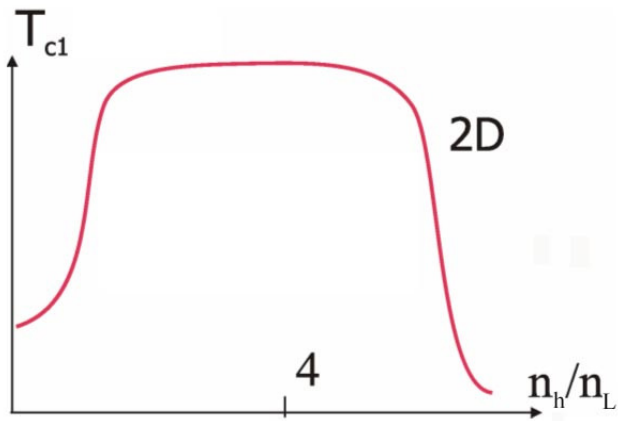

Fig. 10.13. Dependence of the critical temperature $T_{C1}$ from the relative filling of heavy and light bands $n_h/n_L$ in the two-band model with one narrow band. The maximum for $T_{C1}$ corresponds to $n_h/n_L = 4$ in 2D.

If we assume that still $m_L^* \approx m_L$ even in unitary limit (for $f_0^2 = 1/4$) then we get for the

combination $\dfrac{m_h^*}{m_L^*} Z_h^2$ in (10.3.66):

$$\dfrac{m_h^*}{m_L^*} Z_h^2 \sim \dfrac{m_h^*}{m_L} \left(\dfrac{m_h}{m_h^*}\right)^2 \sim \dfrac{m_h^2}{m_L^2} \dfrac{m_L^2}{m_h^2} \sim 1 \qquad (10.3.68)$$

Thus for the critical temperature $T_{C1}$ in the unitary limit $f_0 \rightarrow 1/2$ we can get:

$$T_{C1} \sim \varepsilon_{Fh}^* \exp\left\{-\frac{1}{2f_0^2}\right\} \sim \varepsilon_{Fh}^* e^{-2}. \qquad (10.3.69)$$

It means that for typical (for uranium-based HF compounds) values of $\varepsilon_{Fh}^* \sim (30 \div 50)$ K the critical temperatures are in the range of 5K already at low density which is quite reasonable.

Note that in a phase-separated state we have the droplets (clusters) with the density-ratio $n_h/n_L$ larger or smaller than the density ratio in a homogeneous state [10.4]. For example in a fully phase-separated state we have two large clusters (1,2) with $n_{h1} > n_h > n_{h2}$ (that is so-called Maxwell construction [10.46] typical for phase-separated systems and first-order phase transitons). Thus the expression (10.3.64) for the critical temperature $T_{C1}$ as a function of relative occupation is valid for both clusters but with local values of $(p_{Fh}/p_{FL})_1$ and $(p_{Fh}/p_{FL})_2$. Correspondingly the critical temperature will be different for these two clusters at least in zeroth approximation when we neglect the Josephson coupling [10.7, 10.19] or proximity effect [10.7, 10.19] between the neighboring clusters or droplets.

To finish the consideration of SC in the two-band model let us discuss briefly an effective interaction (irreducible bare vertex) for light electrons in the Cooper channel. It reads:

$$U_{eff}(\vec{p}_L, \vec{p}_L') = T_{LL} + T_{LL}^2 \Pi_{LL}(0, \tilde{\vec{q}}_L = \vec{p}_L + \vec{p}_L') - 2T_{hL}^2 \Pi_{hh}(0, \vec{q}_L = \vec{p}_L - \vec{p}_L'), \qquad (10.3.70)$$

where incoming and outgoing momenta $|\vec{p}_L| = |\vec{p}_L'| = p_{FL}$ and transferred momenta $\tilde{q}_L \leq 2p_{FL}$; $q_L \leq 2p_{FL}$ for SC-problem. The Kanamori T-matrix for light electrons $T_{LL}$ in strong coupling case reads $T_{LL} = \dfrac{4\pi}{m_L^*} \dfrac{1}{\ln\left(1/p_{Fh}^2 d^2\right)} > 0$. Using the expression for $\Pi_{LL}$ and $\Pi_{hh}$:

$$\Pi_{LL}(0, \tilde{\vec{q}}_L) = Z_L^2 \frac{m_L^*}{2\pi}\left[1 - \text{Re}\sqrt{1 - \frac{4p_{FL}^2}{\tilde{q}_L^2}}\right],$$

$$\Pi_{hh}(0, \vec{q}_L) = Z_h^2 \frac{m_h^*}{2\pi}\left[1 - \text{Re}\sqrt{1 - \frac{4p_{Fh}^2}{q_L^2}}\right], \qquad (10.3.70)$$

and having in mind that $p_{Fh} > p_{FL}$ we get:

$$U_{eff}(\vec{p}_L, \vec{p}_L') \approx T_{LL} + T_{LL}^2 \frac{m_L^* Z_L^2}{2\pi} - 2 \frac{m_h^*}{2\pi} Z_h^2, \qquad (10.3.71)$$

where we can put $Z_L \sim m_L/m_L^* \sim 1$.

Thus an effective interaction for light electrons does not contain any nontrivial dependence from $\vec{q}_L$ and $\tilde{\vec{q}}_L$ and hence anomalous superconductivity with magnetic quantum number $m \neq 0$ is absent for light electrons in this approximation. Note that the standard s-wave pairing for light electrons is also suppressed by first order repulsive term $T_{LL} > 0$ in $U_{eff}$ (10.3.71). However an inclusion of Moskalenko-Suhl-Geilihman term [10.47, 10.48], which describes rescattering of a Cooper pair between two bands and is given by:

$$K \sum_{pp'} (a_p^+ a_{-p}^+ b_p b_{-p'} + h.c.) \qquad (10.3.72)$$

in the Hamiltonian of the two-band model (10.3.3) already in the case of infinitely small $K$ makes the light band superconductive at the same temperature as the heavy one. This interesting fact was illustrated for standard s-wave pairing in [10.47] and for p-wave pairing in [10.45]. Thus $T_{C1}$ in (10.3.62) is a mutual SC temperature in the two-band model with one narrow band [10.45]. Of course superconductive gaps for heavy and light bands open simultaneously at $T = T_{C1}$, but then "live their separate lifes" for $T < T_{C1}$ (see [10.47]).

To conclude this section let us note that we discuss briefly the SC-instability which arises in the two-band model low electron density. The leading instability of the enhanced Kohn – Luttinger type [9.4, 10.1, 10.3] corresponds to triplet p-wave pairing of heavy electrons via polarization of light electrons. In 2D or quasi-2D case $T_C$ can reach experimentally realistic values already at low densities for layered dichalcogenides $CuS_2$, $CuSe_2$ and semimetallic superlattices InAs-GaSb, PbTe-SnTe with geometrically separated bands belonging to neighboring layers [10.3, 9.31]. Note that p-wave SC is widely discussed in 3D heavy-fermion systems like $U_{1-x}Th_xBe_{13}$ [9.10] and in layered ruthenates $Sr_2RuO_4$ [9.12, 10.49] with several pockets (bands) for conducting electrons [10.50]. Note also that when we increase the density of a heavy-band and go closer to half-filling ($n_h \rightarrow 1$) the d-wave superconductive pairing (as in $UPt_3$) becomes more beneficial in the framework of the spin-fluctuation theory in the heavy band [9.42, 9.43]. Different mechanisms of SC in HF-compounds including odd-frequency pairing introduced by Coleman, Miranda, Tsvelik are discussed in [10.51].

Note also that the multiband physics is important for some superconducting systems with conventional s-wave pairing including Nb [10.7]. $MgB_2$ [9.14] and new superconductors based on FeAs-compounds like $BaFe_2(As_{1-x}P_x)_2$ [9.15]. For these compounds superconductive gaps in different bands are also open simultaneously at the same critical temperature due to Suhl-Moskalenko-Geilihman theory. All of them are very important for technical applications and energetics.

Note that the two-band Hubbard model discussed in this section is applicable also for degenerate case when there are two orbitals belonging to the same atom (when one atom is a source of electrons of two sorts) In the two-band degenerate Hubbard model $U = U_{hh} = U_{LL} = U_{hL} + 2J_H$ (where $J_H$ is Hund's coupling) [10.52]. Close to half-filling this model becomes equivalent to the $t$-$J$ orbital model [10.53] and contains for $J < t$ and at optimal doping the SC d-wave pairing [10.53] governed by superexchange interaction between the different orbitals of AFM-type ($J > 0$) with $J \sim t^2/U \sim 300$ K for not very different values of $t_h$ and $t_L$. The physics of the $t$-$J$ model will be described in detail in Chapter 13.

## Reference to Chapter 10

Chapter 15. Nanoscale phase separation in complex magnetic oxides.

Nanoscale phase-separation together with the anomalous (marginal) behavior of resistivity are among the most interesting normal properties of many strongly-correlated electron systems.

In this Chapter we will consider another very interesting family of strongly-correlated electron systems, which is called manganites [15.1, 15.2] or the systems with colossal magnetoresistance (CMR-systems) [15.3]. These systems have a lot of striking similarities with high-$T_C$ superconductors (or cuprates) with respect to their crystalline and electronic structure especially in a layered class of manganites. They also exhibit reach phase-diagrams [15.4 - 15.10] with extended regions of nanoscale phase separation. In these regions we get ferromagnetic metallic droplets of the size of $10 - 1000$ Å embedded into the insulating matrices of antiferromagnetic or paramagnetic type in close similarity with the physics of stripes in high-$T_C$ superconductors [15.11 - 15.13]. The CMR-family is very promising for applications in magnetorecording technologies due to their anomalous transport properties and first of all due to the phenomena of colossal negative magnetoresistance (colossal up to $10^2 \div 10^3$ times decrease of resistivity in moderately strong magnetic fields $2 \div 6$ T).

15.1 Inhomogeneous states and nanoscale phase separation in complex magnetic oxides. Similarities with cuprates.

Manganites, the Mn-based magnetic oxide materials typified by $LaMnO_3$, have been under investigation for more than 50 years [15.1, 15.2] but attracted particular attention after the discovery in 1994 of the colossal magnetoresistance (CMR) effect first obtained for Ca-doped $LaMnO_3$ film [15.3]. There is currently considerable review literature on these materials (see e.g., [15.4 - 15.10]) and it is worthwhile noting that in [15.9] a bibliography of more than 600 references is given. This large body of the original and review literature is due to in part to the potential technological applications of colossal magnetoresistance, but also reflects the manganites suitability for studying the physics of strongly-correlated systems. In particular, the interaction of spin, charge and orbital degrees of freedom in these materials as well as their phase diagrams are of interest. On the other hand, the possibility of various types of inhomogeneous charge and spin states in manganites-lattice and magnetic polarons, droplets and stripes structures etc., - is currently receiving special attention.

Similar phenomena occur in many strongly correlated systems where the potential energy of the interaction of electrons exceeds their kinetic energy. In particular (as we mentioned in the introduction 15.1.1), such phenomena are being widely discussed in connection with high-$T_C$ superconductors [15.11 - 15.13]. Of earlier examples, ferromagnetic (FM) droplets (ferrons) in antiferromagnetic (AFM) state at low doping levels [15.14 - 15.16] and ferromagnetic spin-polarons in a paramagnetic (PM) [15.17] are particularly notable. A string (linear track of frustrated spins) created by a hole passing through an AFM insulator [15.18, 15.19] as well as paramagnetic polarons (spin-bags [15.20]) should also be mentioned in connection with mechanisms of SC in underdoped cuprates [15.21, 15.22 and Chapter 13]. All these phenomena are examples of the so-called electron nanoscale phase separation effect, which results from individual charge carriers changing their local electron environment (it is favorable for such regions to be as far apart as possible to minimize the Coulomb energy). Along with this small-scale (nano-scale) phase separation, manganites, as many other materials showing first-order transitions (between the AFM and FM-phases, for example), display yet another type of phase separation, related to the fact that there is a large region of coexistence for various phases in the material. One example of such large-scale separation is the formation of relatively large FM-droplets ($100 - 1000$ Å in size) in an AFM-matrix [15.23-15.24].

A noteworthy feature of manganites is the strong interaction between the electronic and lattice subsystems due to the fact that $Mn^{3+}$ is a Jahn-Teller ion [15.25] and therefore any phase-separation gives rise to elastic lattice distortions which can be detected experimentally. Another characteristic feature is charge ordering (CO), i.e., a regular (often checkerboard) arrangement of $Mn^{3+}$ and $Mn^{4+}$ ions, when the material in fact obtains an additional lattice period and hence acquires a superstructure. Note that charge ordering usually appears at quarter-filling that is at densities close to ½. Along with a superstructure, nontrivial spin and orbital ordering may result from charge ordering. An example is the well-known zigzag magnetic structure (referred to usually as CE) in compounds of the type $Pr_{0.5}Ca_{0.5}MnO_3$ [15.26, 15.27], in which charge ordering is accompanied by the formation of zigzag magnetic chains. The interaction of spin, charge, and orbital degrees of freedom can also lead to stripe [15.28] (rather than droplet) structures at high concentrations of an alkaline-earth element (Ca, Ba, Sr). Because of the strong interaction with the lattice, it turns out that in manganites (as opposed to HTSC systems) such structures are not dynamic but static and are observable with electron diffraction and X-ray small-angle scattering techniques.

In this Chapter, we will focus on the small-scale phase separation, in which the microscopic nature of charge redistribution manifests itself most clearly. We will consider free [15.29] and bound [15.30, 15.31] FM-polarons inside AFM, PM and CO-matrices [15.32] (including FM-polarons on frustrated AFM-lattices [15.33]) as well as so-called orbital ferrons [15.34] inside orbitally ordered matrix [15.34]. The basic models which we use for free and bound FM-polarons is ferromagnetic Kondo lattice model or double exchange model of de Gennes [15.35] (with or without Coulomb interactions). While for orbital ferrons we use the degenerate two-band Hubbard model (considered with respect to anomalous KL-superconductivity in Chapter 12) which is reduced to the orbital t-J model [15.37].

Note that the effects of phase-separation on the transport properties are considered in Chapter 16.

## 15.2. Crystal structure. Electronic and transport properties of manganites.

In this section we will study crystal structure, phase diagram and basic electronic and transport properties of manganites including the CMR-effect in this family of materials.

Ideal crystal structure of 3D manganites $Ln_{1-x}A_xMnO_3$ (Ln = La, Pr, Nd, Sm; A = Ca, Sr, Ba) corresponds to cubic cell (perovskite structure) see Fig. 15.1. Real structure is slightly orthorombically distorted due to Jahn-Teller (JT) effect [15.25].

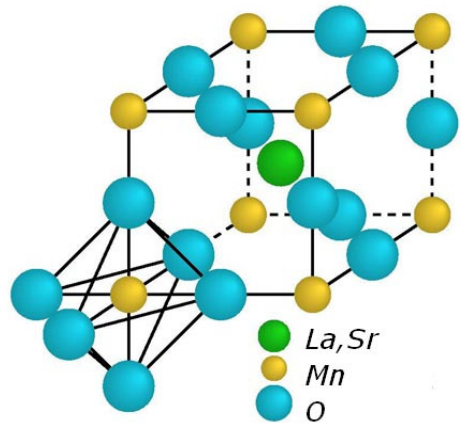

*La,Sr*
*Mn*
*O*

Fig. 15.1. Crystal structure of ideal 3D manganites $Ln_{1-x}A_xMnO_3$ (Ln = La, Pr, Nd, Sm; A = Ca, Sr, Ba).

### 15.2.1 Gross phase-diagram of manganites.

Gross phase-diagram of typical manganites $(La_{1-x}Ca_xMnO_3)$ is presented on Fig. 15.2. At low doping levels $0.02 < x < 0.16$ and low temperatures we have phase-separation (PS) on metallic FM droplets inside AFM insulating matrix [15.6, 15.14-15.16, 15.29]. At optimal doping concentration $x_{opt} = 0.25$ we have FM-metal for $T < T_C$ – Curie temperature, which exhibits phenomena of colossal negative magnetoresistance (CMR). For these concentrations we have two-band mixed-valence situation (see Chapter 16). At $x = 0.5$ (quarter-filling) and low temperatures we have checkerboard charge ordering (CO). Around $x = 0.5$ we have extended region of PS II on metallic FM droplets inside CO insulating matrix [15.32]. For $x \leq x_C = 0.16$ we have also the beautiful physics of orbital ordering [15.35] described by degenerate two-band Hubbard model with metallic orbital polarons inside insulating AFM orbital matrix [15.34]. Finally at very small doping levels $x \sim (1 \div 2)\%$ and low temperatures we have magnetic polarons bound by Coulomb interaction on Ca accepting centers with extended coat of slowly decreasing tails of spin-distortions. These bound magnetic polarons were first predicted by de Gennes [15.36] and obtained in a simple model in [15.30]. Effectively they behave as a magnetic impurity.

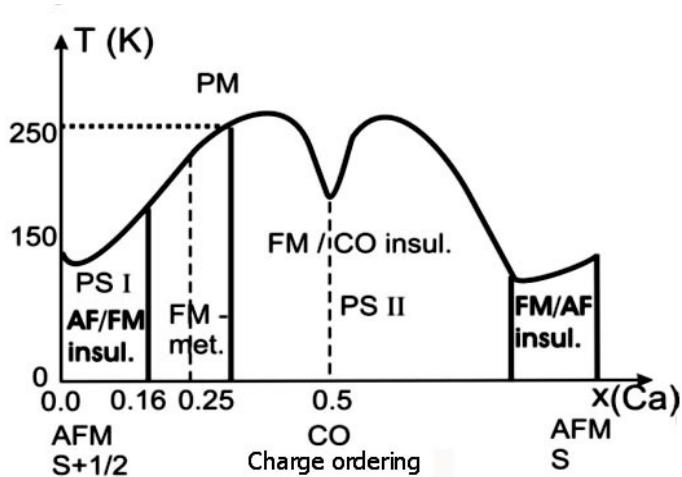

Fig. 15.2. Gross phase-diagram of typical 3D perovskite manganite La$_{1-x}$Ca$_x$MnO$_3$, $x$ is a hole concentration.

For $x = 0$ and $x = 1$ we have ideal AFM ordering for spins $S + \frac{1}{2}$ and spins $S$ ($S = 3/2$), respectively. Close to $x = 1$ we have again PS on FM droplets inside AFM insulating matrix at low temperatures. At high temperatures and low doping levels ($x \ll 1$ and $T > T_N$ – Neel temperature) and at $x \sim x_{opt}$ and $T > T_C$ temperature ferrons (FM-droplets) inside insulating PM matrix [15.17] are formed.

15.2.2. Resistivity at optimal concentrations.

At optimal concentration $x_{opt} = 0.25$ and $T_C = 250$ K we have simultaneously FM → PM transition coinciding with metal-insulator transition (see Fig. 15.3.).

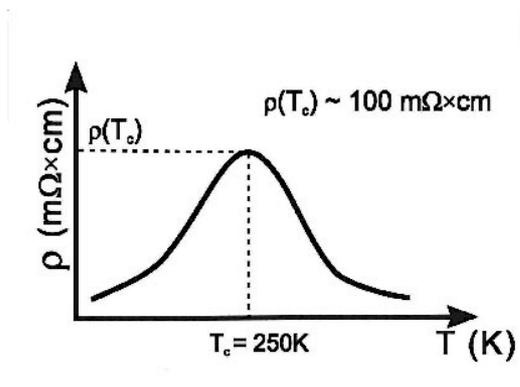

Fig. 15.3. Resistivity at optimal doping concentrations. At maximum (for $T_C = 250$ K) resistivity $\rho(T_C) \sim 100$ mΩ·cm.

For low temperatures $T < T_C$:

$$\rho \sim \rho_0 + AT^{\beta}, \qquad (15.2.1)$$

where $\beta \sim 2.5$ and $\dfrac{d\rho}{dT} > 0$ corresponds to metallic behavior.

For high temperatures $T > T_C$:

$$\rho \sim Be^{A/T}, \qquad (15.2.2)$$

and $\dfrac{d\rho}{dT} < 0$ corresponds to insulating thermoactivative behavior.

This curve can be explained in two ways: by electron tunneling from one FM-cluster (droplet) to a neighboring one via insulating barrier (see Chapter 18) or possibly in the framework of the two-band model, especially for layered quasi 2D manganites (see Chapter 16).

15.2.3. Colossal magnetoresistance.

At optimal concentration $x_{opt} = 0.25$ we have phenomena of colossal magnetoresistance [15.3], namely resistivity strongly decreases in the presence of a moderately strong magnetic fields $H \sim (2 \div 4)$ T (see Fig. 15.4).

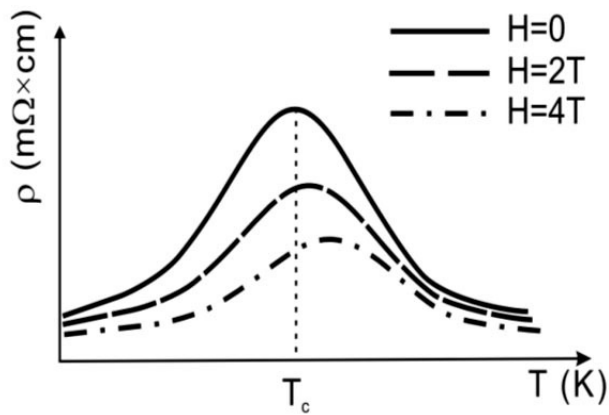

Fig. 15.4. Strong decrease of resistivity in the presence of magnetic field in optimally doped manganites. Solid line corresponds to the absence of magnetic field ($H = 0$), dashed line for $H = 2$ T and dashed-dotted line for $H = 4$ T.

Magnetoresistance is defined as (see [15.6] for example):

$$(MR) = \frac{\Delta R}{R(H)} = \frac{R(H) - R(0)}{R(H)} \approx \frac{R(0)}{R(H)} \sim -(10^2 \div 10^3) \quad - \quad (15.2.3)$$

colossal negative magnetoresistance.

It is interesting that for optimal concentration $x_{opt} = 0.25$ magnetization behaves as follows (see Fig. 15.5).

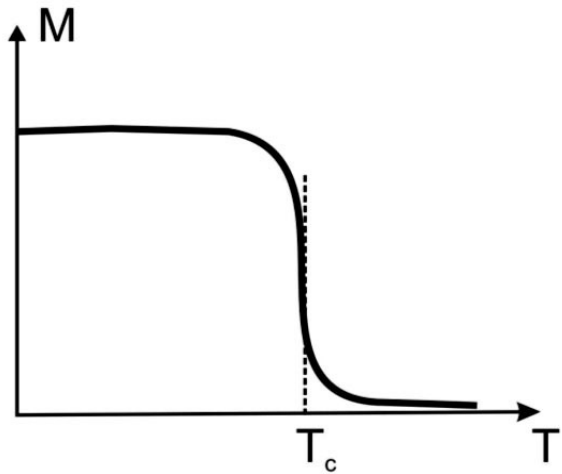

Fig. 15.5. Temperature dependence of magnetization at optimal concentration $x_{opt} = 0.25$.

### 15.2.4. Electronic structure of manganites.

Essential for electronic structure of manganites are 5 $d$-orbitals of Mn-ions splited by the crystalline field on 3 localized $t_{2g}$-orbitals $d_{xy}$, $d_{xz}$, $d_{yz}$ and 2 conductive $e_g$-orbitals $d_{3z^2-r^2}$, $d_{x^2-y^2}$. On localized $t_{2g}$-orbitals according to Hund's rule a local spin $S_{loc} = 3/2$ is formed.

Manganites are in a mixed valence situation, so $Mn^{3+}$ and $Mn^{4+}$-ions are present. For $Mn^{4+}$ ion both conductive orbitals $e_g$ are empty (see Fig. 15.6).

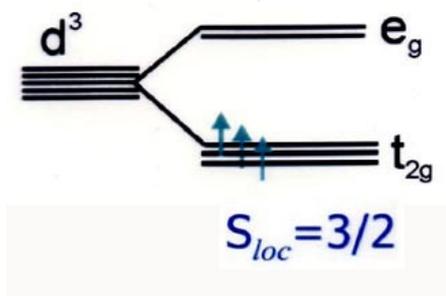

Fig. 15.6. Electronic structure of $d$-orbitals for $Mn^{4+}$ ion.

In the same time for $Mn^{3+}$ ion one conductive $e_g$-orbital is filled and one is empty (see Fig. 15.7). Note that $e_g$-conductive orbitals are additionally splitted by Jahn-Teller effect in $Mn^{3+}$ ion.

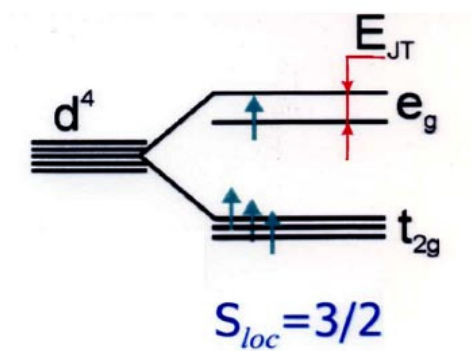

Fig. 15.7. Electronic structure of Jahn-Teller $Mn^{3+}$ ion.

This splitting is due to the different distortion of $d_{x^2-y^2}$-octahedra in *xy*-plane and $d_{3z^2-r^2}$-octahedra along *z*-axis (see Fig. 15.8). As a result Jahn-Teller gap ($E_{JT}$) appears. We assume that $E_{JT}$ is large, so for low doping concentration $x << x_C = 0.16$ only one conduction band is occupied. Note that close to $x_C = 0.16$ we have two occupied conductive bands and the beautiful physics of orbital ordering. Note also that for low doping a spin of conductivity electron $\vec{\sigma}$ is coupled ferromagnetically (oriented parallel) to the local spin $S_{loc} = 3/2$.

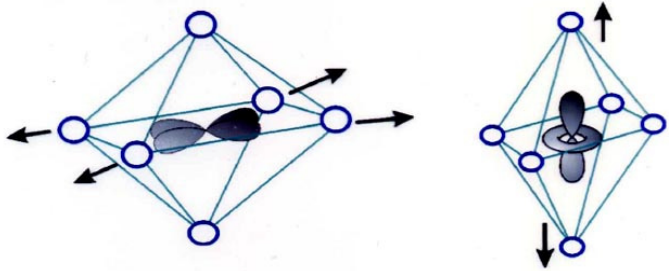

Fig. 15.8. Different distortion of $d_{x^2-y^2}$-octahedra in *xy*-plane and $d_{3z^2-r^2}$-octahedra along *z*-axis.

### 15.3. The minimal theoretical model for manganites.

The basic theoretical model for manganites is FM Kondo-lattice model or double-exchange model firstly introduced by de Gennes in 1960 [15.36]. The Hamiltonian of the double-exchange model reads:

$$\hat{H} = -J_H \sum_i \vec{S}_i \vec{\sigma}_i - t \sum_{<ij>\sigma} c^+_{i\sigma} c_{j\sigma} + J \sum_{<ij>} \vec{S}_i \vec{S}_j \qquad (15.3.1)$$

where $J_H$ is large Hund's type (FM) interaction between local spin $\vec{S}$ and spin of conductivity electron $\vec{\sigma}$ on site *i*, *t* – is hopping integral for conductivity electron and *J* is AFM (Heisenberg) exchange between local spins, $c^+_{i\sigma}$, $c_{j\sigma}$ - are creation and annihilation operators for conductivity electron with spin-projection $\sigma$ on sites *i* and *j*, respectively. For one conductive band we have the following hierarchy of parameters:

$$J_H S >> t >> JS^2. \qquad (15.3.2)$$

For real manganites the typical values of the parameters: $J_H S \approx 1$ eV; $t \approx 0.3$ eV (the bandwidth is narrowed by electron-phonon polaron effect); $JS^2 \approx 0.001$ eV. Double occupancy is prohibited in (15.3.1) by large Hund's term, so it is not necessary to add Hubbard repulsion term $U \sum_i n_{i\uparrow} n_{i\downarrow}$ to Hamiltonian (15.3.1). In fact we can insert projection operators in the kinetic term in (15.3.1) and write $-t \sum_{<ij>\sigma} P c^+_{i\sigma} c_{j\sigma} P$ .

### 15.3.1. Homogeneous canting for small densities.

In classical physics conductivity electron can hop freely in FM surrounding of local spins, but cannot hop in AFM surrounding. Thus the only possibility for conductivity electron is to cant the local spins (belonging to different sublattices) when it hops from one site to a neighboring one (see Fig. 15.9).

Note that for large Hund's coupling $J_H > W$ (*W* is a bandwidth for conductivity electron) the term $-J_H \sum_i \vec{S}_i \vec{\sigma}_i$ corresponds to minimal energy for $\vec{S}_i \parallel \vec{\sigma}_i$. Correspondingly the total spin on site *i*, $|\vec{S}_{tot}| = S + \frac{1}{2}$ .

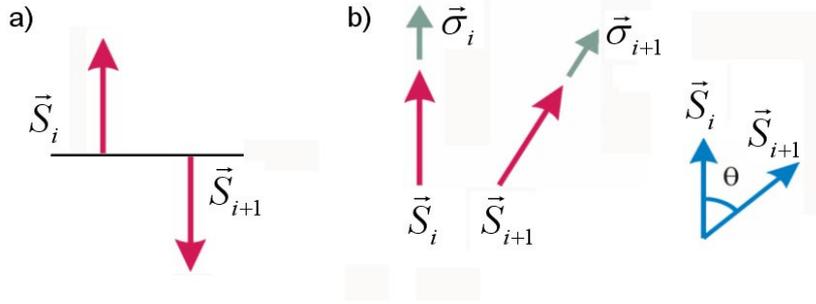

Fig. 15.9. Ideal AFM-lattice of local spins a) with local spins $\vec{S}_i$ and $\vec{S}_{i+1}$ on neighboring sites belonging to different sublattices. b) The canted structure with the angle θ between neighboring local spins $\vec{S}_i$ and $\vec{S}_{i+1}$, which created by conductivity electron hopping from site $i$, to a neighboring site $i + 1$.

An important point is that Ψ-function of conduction electrons has a spinor character. That is why electron hopping results in an effective rotation of the components of the Ψ-function on the angle θ/2 (where θ is the canting angle between sublattices). Thus an effective hopping integral for the electron in the surrounding of canted local spins reads [15.36, 15.38]:

$$t_{eff} = t\cos(\theta/2).\qquad(15.3.3)$$

### 15.3.2. Canted state instability.

An energy of a classical canted state of de Gennes reads:

$$E = -zt\cos(\theta/2)n + zJS^2\cos\theta\qquad(15.3.4)$$

where $z$ is the number of nearest-neighbors, $n$ is a concentration of conductivity electrons (for La$_{1-x}$Ca$_x$MnO$_3$: $n = 1-x$, where $x$ is hole doping). In (15.3.4) first term describes the decrease of kinetic energy and second one increase in the interaction energy between the local spins. If we minimize this energy with respect to $\cos(\theta/2)$ we get:

$$\frac{dE}{d\cos(\theta/2)} = 0 \;\; \text{and} \;\; \cos(\theta/2) = \frac{t}{4JS^2}n,\qquad(15.3.5)$$

thus an optimal $\cos(\theta/2)$ depends linearly upon doping. Correspondingly an optimal energy of a homogeneous canted state:

$$E = -\frac{zt^2n^2}{8JS^2} - zJS^2\qquad(15.3.6)$$

However it is easy to note that a canted state has a negative compressibility [15.29]:

$$\kappa^{-1} = \frac{d^2E}{dn^2} = -\frac{zt^2}{4JS^2} < 0.\qquad(15.3.7)$$

Negative compressibility reflects an instability of a homogeneous canted state towards phase-separation.

### 15.3.3. Small FM-polarons inside AFM-matrix.

The most energetically beneficial type of phase-separation corresponds to the creation of small FM-polarons (ferrons) inside AFM-matrix [15.14-15.16]. The energy of this phase-separated state reads for spherical magnetic polaron in 3D:

$$E_{pol} = -tn(z - \frac{\pi^2 d^2}{R^2}) + zJS^2 \frac{4}{3}\pi\left(\frac{R}{d}\right)^3 n - zJS^2\left[1 - \frac{4}{3}\pi\left(\frac{R}{d}\right)^3 n\right], \quad (15.3.8)$$

where $\Omega = \frac{4}{3}\pi\left(\frac{R}{d}\right)^3$ is a volume of a spherical polaron (spherical FM-droplet) in isotropic 3D case, $-tzn$ is a bottom of the band in an infinite FM-cluster (where electron can hop freely), $tn\frac{\pi^2 d^2}{R^2}$ is a delocalization kinetic energy of conductivity electron in a spherical potential well of the radius $R$ (see Fig. 15.10).

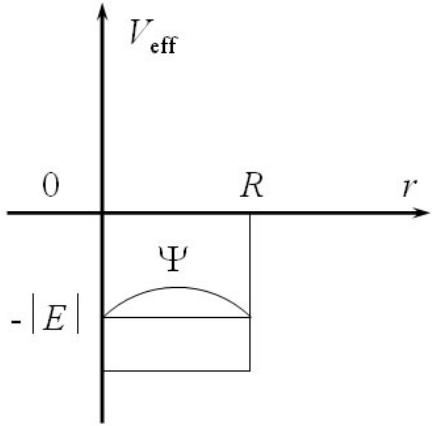

Fig. 15.10. The spatial extension of the conductivity electron $\Psi$-function corresponding to the deepest bound state in a spherical potential well with a width equal to the radius of the polaron $R$.

It can be obtained from boundary condition $\Psi(r = R) = 0$ for the $\Psi$-function $\Psi \sim \sin(kr)$ which corresponds to the deepest bound state of a conductivity electron in a spherical potential well of the width $R$ (self-trapping of conductivity electron in FM-droplet with a radius $R$).

The second term in (15.3.8) corresponds to the loss in AFM exchange interaction energy of local spins inside the ferrons, while the third term describes AFM exchange interaction energy between the local spins in the region which are free from FM-polarons (see Fig. 15.11).

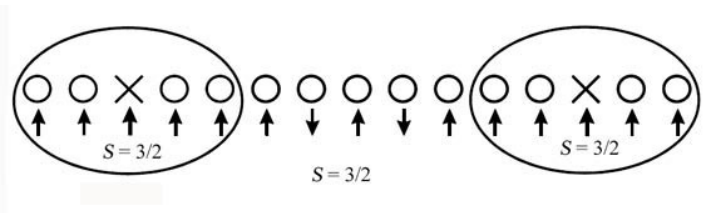

Fig. 15.11. Self-trapping of conductivity electron in a spherical droplet of radius $R$ for small density $n$ of conductivity electrons. The local spins $S = 3/2$ inside the droplets are oriented ferromagnetically. Outside the droplets there are no conductivity electrons (charge carriers) while the local spins $S = 3/2$ are oriented antiferromagnetically.

Minimization of a polaron energy (15.3.8) with respect to the polaronic radius $R$ yields in a 3D case:

$$\frac{dE_{pol}}{dR} = 0 \text{ and } \frac{R_{pol}}{d} = \left(\frac{\pi t}{2zJS^2}\right)^{1/5} \quad (15.3.9)$$

Correspondingly the optimal polaronic energy reads:

$$E_{pol} = -tzn + \frac{5}{3}\pi\, zn(\pi t)^{3/5}\left(2zJS^2\right)^{2/5} - zJS^2. \qquad (15.3.10)$$

The canting angle $\theta = 0$ inside the ferron (FM-region) and $\theta = \pi$ (AFM-region) outside the ferron. The boundary of Mott-Nagaev-Kasuya ferrons [15.14-15.16] is rigid: an angle $\theta$ changes from 0 to $\pi$ on a distance of the order of $d$. Note that for typical values of $t \sim 0.3$ eV and $JS^2 \sim 0.001$ eV the polaron radius $R \sim (2 \div 3)d \sim 10$ Å (has nanoscale dimensions) and the number of local spins inside the ferrons is $\frac{4}{3}\pi\left(\frac{R}{d}\right)^3 \sim 4\cdot(2^3 \div 3^3) \sim (30 \div 100)$. Thus a FM-polaron is a bound state of one conductivity electron and $(30 \div 100)$ local spins. Such an object is very difficult to describe diagrammatically, so the only possible description is a variational one presented in this Chapter.

Note also that we described nanoscale phase-separation at small electron density $n << 1$. In the opposite case of large electron density $n \to 1$ or correspondingly small hole doping $x = 1 - n << 1$ we have again the same nanoscale phase-separation on FM-droplets inside AFM-insulating matrix, but now the role of conductivity electron plays a hole. Thus effectively inside the ferrons we have one local spin $S = 3/2$ surrounding by total spins $S_{tot} = \left|S + \frac{1}{2}\right|$ oriented ferromagnetically, while outside the ferron we have antiferromagnetically oriented total spins $S_{tot} = \left|S + \frac{1}{2}\right|$ (see Fig. 15.12). This structure is totally equivalent to that in the t-J model [15.63] where local Kondo-singlets or Zhang-Rice singlets are formed by AFM Kondo-interaction $J_K$ between spin of conductivity electron $\sigma = 1/2$ and local spin $S = 1/2$. Here we have FM interaction $J_H$ between local spin $S = 3/2$ and conduction electron spin $\sigma = 1/2$ which leads to the formation of $S_{tot} = 2$ in the strong-coupling case.

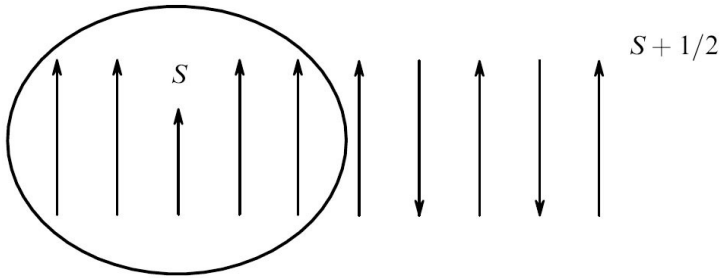

Fig. 15.12. Structure of phase-separated state at the electron densities $n \to 1$ (at the small doping level $x = 1 - n << 1$) [15.6].

### 15.3.4. Quantum canting.

In subsection 15.3.2 we considered classical canted state of de Gennes and proved its instability with respect to nanoscale phase-separation on small FM-polarons inside AFM-matrix (subsection 15.3.3).

In a more sophisticated quantum canting approach [15.14, 15.29] there are two bands, instead of one, corresponding respectively to $S_{tot} = \left|S + \frac{1}{2}\right|$, but $S_{tot}^z = S \pm \frac{1}{2}$.

Their spectra read (see Nagaev [15.14]):

$$t_{\pm}(\theta) = \frac{t}{2S+1}\left[\sqrt{2S+1+S^2\cos^2\frac{\theta}{2}} \pm S\cos\frac{\theta}{2}\right] \qquad (15.3.10)$$

where $t_+(\theta)$ corresponds to $S_{tot}^z = S + \frac{1}{2}$ and $t_-(\theta)$ - to $S_{tot}^z = S - \frac{1}{2}$.

For $\theta \to 0$ (FM-case) $t_+ = t$, $t_- = \dfrac{t}{2S+1}$. For $\theta \to \pi$ (AFM-case) $t_+ = t_- = \dfrac{t}{\sqrt{2S+1}}$. That is an interesting result. While in classical picture a conductivity electron cannot hop in AFM-background, in quantum approach its effective hopping integral is proportional to $\dfrac{t}{\sqrt{S}}$. Moreover for $\theta = \pi$ there is a "string"-type of motion for conductivity electron between the two bands: on site $i$ it forms $S_{tot} = S_{tot}^z = S + \frac{1}{2}$ with local spin $S$, while on neighboring site $i + 1$: $S_{tot} =\mid S + \dfrac{1}{2}\mid$ but $S_{tot}^z = S - \frac{1}{2}$ and so on with alternating values of $S_{tot}^z$ [15.40, 15.41]. Note that the classical one band canting of de Gennes corresponds for $S >> 1$ to the region of small angles where $S^2 \cos^2(\theta/2) >> 2S + 1$ and $t_+ \approx t \cos(\theta/2)$ while $t_- \to 0$.

## 15.3.5. Compromise between quantum canting and formation of FM-polarons.

An energy functional with the normalization condition for the $\Psi$-function of a polaronic state in the case of quantum canting reads for continuum model [15.39, 15.42, 15.43]:

$$F = -\int dV \left[ t(\theta)\left( z|\Psi|^2 + \Psi^* \Delta\Psi \right) - zJS^2 \cos^2 \frac{\theta}{2} - \beta|\Psi|^2 \right] \qquad (15.3.11)$$

with $t(\theta)$ given by (15.3.10) and $\beta$ is Legendre multiplier $\left( \int dV|\Psi|^2 = 1 \right)$, $\Delta$ is the Laplace operator, $z$ is the number of the nearest neighbors. The minimization of $F$ with respect to $\Psi^*$ yields:

$$\frac{\delta F}{\delta \Psi^*} = 0: \; t(\theta)(2z\Psi + \Delta\Psi) + \Delta[t(\theta)\Psi] - 2\beta\Psi = 0. \qquad (15.3.12)$$

In the same time the minimization of $F$ with respect to canting angle $\theta$ reads:

$$\frac{\delta F}{\delta\theta} = 0: \left[ \left( z|\Psi|^2 + \Psi^*\Delta\Psi \right)\frac{d\,t(\theta)}{d\cos\theta/2} - 2zJS^2 \cos\frac{\theta}{2} \right]\sin\frac{\theta}{2} = 0 \qquad (15.3.13)$$

<u>Electronic wave function and the dependence of the canting angle from the radius.</u>

The normalized electronic wave function $\Psi(r)$ and the dependence of the canting angle from the radius $r$ (function $\theta(r)$) were calculated in [15.42, 15.43] by the same numerical method which was proposed originally in [15.39] for 1D ferrons. The numerical results are presented on Fig. 15.13 for parameter $\alpha = t/JS^2 = 100$. Note that as we showed in subsection 15.3.3 the polaronic radius $R_{pol}$ given by (15.3.9) can be expressed in terms of $\alpha$ as $R_{pol} = d\left( \dfrac{\pi\alpha}{2z} \right)^{1/5}$ in 3D.

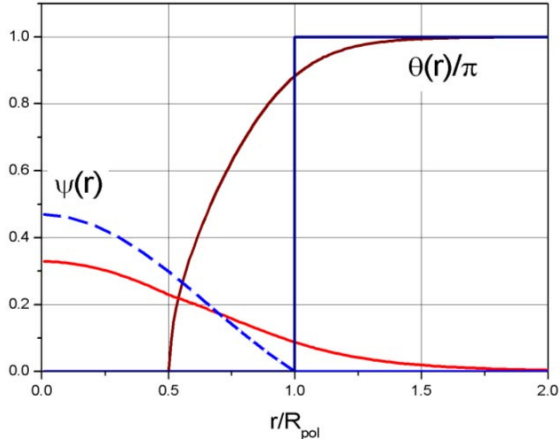

Fig. 15.13. Dependence of the electronic Ψ-function and the canting angle θ from dimensionless variable $r/R_{pol}$ for $\alpha = t/JS^2 = 100$. Solid line for $\Psi(r)$ corresponds to quantum two-band canting, while the dashed line to classical one band canting[15.42,15.43].

We see from Fig. 15.13 that a canting angle is essentially θ = 0 for $r < 0.5R_{pol}$ and then rather rapidly goes to π on the interval $0.5R_{pol} \div 1.5R_{pol}$.

The Ψ-function is zero outside ferrons (for $r > R_{pol}$) in the case of classical one band canting. In quantum two-band canting it has rather rapidly decreasing tail for $r > R_{pol}$.

Comparison of Nagaev-Mott and exact solution.

According to (15.3.10) the optimal polaronic energy in terms of parameter α reads for Nagaev-Mott FM-polarons:

$$\frac{E_{N-M}}{t} = -z + \frac{5}{3}\pi^2\left(\frac{2z}{\pi\alpha}\right)^{2/5} - \frac{z}{\alpha}.$$     (15.3.14)

In the same time the energy of AFM-string:

$$\frac{E_{AFM}}{t} = -\frac{z}{\sqrt{2S+1}}.$$     (15.3.15)

On Fig. 15.14 we present the comparison for the energy of Nagaev-Mott and exact numerical solution as functions of the parameter α.

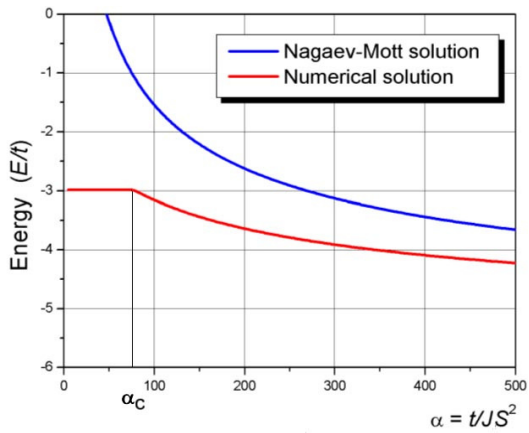

Fig. 15.14. Comparison of Nagaev-Mott solution for the FM-polarons and exact numerical solution for the reduced energy $E/t$ as functions of the parameter $\alpha = t/JS^2$. Thick solid line corresponds to Nagaev-Mott solution, while thin solid line to exact numerical solution[15.42,15.43].

From Fig. 15.14 we see that self-trapped FM-polaron state is beneficial for $\alpha > \alpha_C \approx 75$ (for $S = 3/2$). A bound state disappears and conductivity electron moves freely (a string type of motion) through AFM-media for $\alpha > \alpha_C$. Here the energy does not depend upon $\alpha$ in agreement with (15.3.15). Note that for most manganite families $\alpha > \alpha_C$ and FM-polarons are essentially beneficial even in a picture of the two-band quantum canting.

Note that in cuprates $t/J \sim 3$ and $S = \frac{1}{2}$ so parameter $\alpha << \alpha_C$. Hence the FM-polarons are usually not stabilized in high-$T_C$ compounds. Instead of them we have AFM-string (string-oscillators) of Bulaevskii, Nagaev, Khomskii, Brinkman, Rice [15.18, 15.19]. They represent the linear trace of frustrated spins which accompany the hole when it moves in AFM-background. Thus here we have a physics of spin-charge confinement considered in Chapter 13 analogous to the confinement physics in quark-gluon plasma in quantum chromodynamics (QCD) [15.44, 15.45].

## 15.4. Temperature ferrons. FM-polarons in a layered case.

In this section we will consider two more types of FM-polarons namely temperature ferrons and ferrons in a layered case.

### 15.4.1. Temperature ferrons.

At optimal concentration $x \sim x_{opt}$ and low temperatures $T < T_C$ ($T_C$ is Curie temperature), as we already mentioned, we have homogeneous state of FM-metal. However even in this range of concentrations at high temperatures $T > T_C$ there is again a phase-separation on FM-droplets but now inside paramagnetic (PM) insulating matrix.

The radius of the droplet for the hierarchy of the parameters $zJS^2 \leq T_C < T < t$ can be defined from the minimization of the free-energy [15.17]:

$$\Delta F = -tn\left(z - \frac{\pi^2 d^2}{R^2}\right) + T\ln(2S+1)\frac{4}{3}\pi\left(\frac{R}{d}\right)^3 n, \qquad (15.4.1)$$

where the second term describes the reduction of the spin-entropy inside the ferron.
As a result after minimization of $\Delta F$ with respect to polaron radius $R$ we get again in 3D case [15.17]:

$$R_{pol}^T \sim d\left(\frac{\pi a}{2T\ln(2S+1)}\right)^{1/5}. \qquad (15.4.2)$$

Note that effectively polaronic radius of temperature ferrons is given by the same expression as the radius of Nagaev-Mott-Kasuya FM-polaron at $T = 0$ with the substitution of $zJS^2$ by $T\ln(2S+1)$.

### 15.4.2. Polarons in a layered case.

Besides cubic (perovskite type) of manganites such as $Ln_{1-x}Ca_xMnO_3$ there is another family which is typified as layered manganites with the general chemical formula $(La, Ca)_{n+1}Mn_nO_{3n+1}$ ($n \geq 1$, for $n = 1$: $La_{2-x}Ca_xMnO_4$). The layered manganites strongly resemble high-$T_C$ compounds. Electronic transport in them takes place in conductive MnO-layers which are separated by insulating LaO(CaO)-layers (layers of charge reservoirs) see Fig. 15.15.

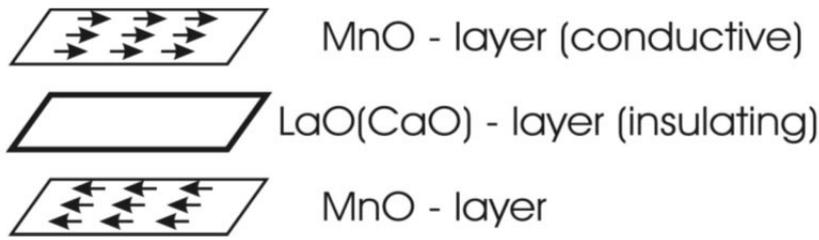

MnO - layer (conductive)

LaO(CaO) - layer (insulating)

MnO - layer

Fig. 15.15. The crystal and magnetic structure of layered manganites. There are conductive MnO-layers separated by insulating LaO(CaO)-layers in this family of manganites. Conductive layers are ferromagnetic but FM-moments of neighboring MnO layers are antiparallel (A1 magnetic structure)[15.6].

In the layered case we again have the tendency towards nanoscale phase-separation on FM-polarons inside AFM-matrix. The most favorable in a layered case is an ellipsoidal shape of a ferron elongated in the direction parallel to conductive layers (see Fig. 15.16 and [15.6]).

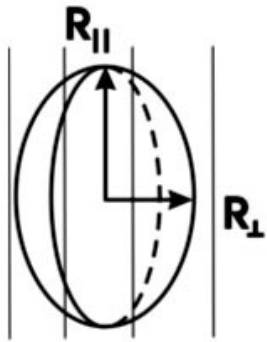

Fig. 15.16. FM-droplet of an ellipsoidal shape which is the most energetically beneficial in layered manganites. The droplet is elongated in the direction parallel to conductive FM layers of MnO ($R|| > R_\perp$)[15.6].

The volume of the ellipsoidal ferron is given by [15.6]:

$$\Omega = \frac{4\pi}{3}\left(\frac{R_\parallel}{d}\right)^2 \frac{R_\perp}{d}. \qquad (15.4.3)$$

Note that in the most general case of 3D anisotropic AFM – lattice the most beneficial shape of the ferrons again corresponds to the ellipsoid of rotation. Moreover its volume reads [15.33]:

$$\Omega = \frac{4\pi}{3}\left(\frac{\pi t_{eff}}{4\bar{J}S^2}\right)^{3/5}, \qquad (15.4.4)$$

where effective hopping integral $t_{eff}$ is given by geometrical average of the hopping integrals along main axis:

$$t_{eff} = (t_x \cdot t_y \cdot t_z)^{1/3}. \qquad (15.4.5)$$

In the same time an effective AFM exchange integral $\bar{J}$ is given by the algebraic average (by the sum) of Heisenberg exchanges along the main axis:

$$\bar{J} = (J_x + J_y + J_z) \qquad (15.4.6)$$

15.4.3. FM-polarons on a square-lattice.

On a square lattice in 2D the optimal shape of a ferron is a circle $\Omega = \pi (R/d)^2$ with a radius [15.33.15.6]:

$$\frac{R_{pol}}{d} \sim \left( \frac{t\alpha_o^2}{8\pi JS^2} \right)^{1/4}, \qquad (15.4.7)$$

where $\alpha_0 \approx 3\pi/4$ is a first zero of Bessel function, $J_0(kR) = 0$ for $kR_0 = \alpha_0$. Note that in 2D $R_{pol}/d$ $\sim (t/JS^2)^{1/4}$ instead of $(t/JS^2)^{1/5}$ in 3D-case [15.6].

For 2D anisotropic AFM – lattice the optimal shape of a ferron is an ellipse [15.33] with a volume:

$$\Omega = \pi \left( \frac{t_{eff}\alpha_0^2}{4\pi \bar{J}S^2} \right)^{1/2} \qquad (15.4.8)$$

where $t_{eff} = (t_x \cdot t_y)^{1/2}$ and $\bar{J} = (J_x + J_y)$.

Note that this type of nanoscale phase-separation is typical for a variety of quasi-2D (layered) cobaltates with low spin (a hole) in the center of a ferron surrounded (in analogy with the situation presented on Fig. 15.12) by high spin in case of hole-doping.

### 15.4.4. FM-polarons on a triangular lattice in 2D.

Finally in the end of this section let us consider frustrated lattices with AFM-interaction between neighboring local spins. On frustrated triangular lattice for planar (2D) spin configuration the most beneficial shape of a ferron corresponds again to the circle with a volume [15.31]:

$$\frac{\Omega_{triangle}}{\Omega_{square}} = \left( \frac{4}{3} \right)^{1/2} = \left( \frac{m_{triangle}}{m_{square}} \right)^{1/2} \qquad (15.4.9)$$

measured in terms of the volume of a ferron on a square lattice. It is possible to show [15.31] that the ratio of volumes is connected with the square-root of the effective mass ratio on triangular and square lattices in (15.4.9).

### 15.5. Free and bound magnetic polarons.

For very small doping concentration conductivity electrons are bound to Ca (Sr) impurity centers. Hence ferrons are also localized in this case [15.30]. An appearance of the conductivity in Mn – O subsystem and delocalization transition in the system of ferrons correspond to generalized Mott criterion [15.16, 15.42]. Having in mind that at low temperatures the system becomes metallic for $x_{met} \sim 16\%$ (at this concentration the FM-droplets start to overlap organizing an infinite metallic cluster), we conclude that ferrons are delocalized for $x_{Mott} < x < x_{met}$. For $x < x_{Mott}$ we have bound magnetic polarons [15.30]. We will show that, in contrast with free magnetic polarons, the bound polarons have a large intermediate region where a canting angle is changed gradually from almost 0 to almost $\pi$.

The existence of such ferrons (behaving effectively as magnetic impurities) was first assumed by de Gennes in 1960 in a seminal paper on the double exchange [15.36]. We can get them explicitly in a following simple model.

### 15.5.1. The minimal model for the bound magnetic polarons.

The bound magnetic polarons with extended coat of spin-distortions for $x << x_{Mott}$ are described by the Hamiltonian:

$$\hat{H} = -J_H \sum_i \bar{S}_i \vec{\sigma}_i - t \sum_{<n,m>\sigma} c_{n\sigma}^+ c_{m\sigma} + J \sum_{<n,m>} \bar{S}_n \bar{S}_m - V_{imp} \sum_{i\sigma} \frac{c_{n\sigma}^+ c_{n\sigma}}{|\vec{n} - \vec{n}_0|} - K \sum_n \left( S_{nx} \right)^2 \quad (15.5.1)$$

It is a double exchange model with Coulomb interaction between conductivity electrons and nonmagnetic Ca (Sr) donor impurity, as well as with one-site anisotropy. Ca (Sr) impurity is situated in the middle of some elementary lattice cells of the crystal (see Fig 15.17).

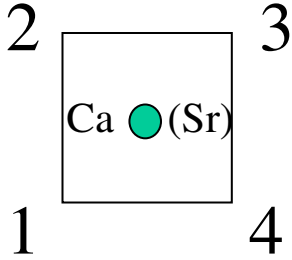

Fig 15.17. Nonmagnetic Ca (Sr) impurity is situated in the middle of some elementary cells of the crystal.

In Hamiltonian (15.5.1) we assume the following hierarchy of parameters [15.30]:
$$(J_H S \sim V_{imp}) >> t >> J S^2 >> K, \qquad (15.5.2)$$
where $V_{imp}$ is an amplitude of Coulomb attraction between impurity and conductivity electron, $K$ – is a constant of one-site anisotropy, $\vec{S}_n = S \vec{e}_n$ is a classical local spin, $\vec{n}$ is a unit vector, $\vec{\sigma} = 1/2 \, c^+_{n\mu} \vec{\sigma}_{\mu\nu} c_{n\nu}$ is a spin of conductivity electron, $\vec{\sigma}_{\mu\nu} = \{\sigma_x, \sigma_y, \sigma_z\}$ are Pauli matrices, $\vec{n}_0$ is a unit vector in the direction of impurity, $x$ is an easy magnetic axis.

15.5.2. The variational procedure.

The variational procedure is detaily described in [15.30, 15.31]. Here we present a brief sketch of the method. First of all note that:
$$\vec{S}_n = S(\cos\theta_n \sin\varphi_n, \sin\varphi_n \sin\theta_n, \cos\theta_n) - \qquad (15.5.3)$$
is parametrization of a classical spin ($S >> 1$) in 3D. In 2D case all $\theta_n = \pi/2$.
The $\Psi$-function of the conductivity electron in 3D reads:
$$|\Psi\rangle = \sum_n \Psi_n \left( \cos\frac{\theta_n}{2} c^+_{n\uparrow} + \sin\frac{\theta_n}{2} \exp(-i\varphi_n) c^+_{n\downarrow} \right) |0\rangle \, (15.5.4)$$
where $\sum_n |\Psi_n|^2 = 1$ is normalization condition.

The variational parameters of the problem are $\Psi_n$, $\theta_n$, $\varphi_n$. In the beginning we solve an electron problem (we minimize the energy of the system with respect to $\Psi_n$). The $\Psi$-function of the system in our approximation coincides with the $\Psi$-function of conductivity electron. Moreover $H_M |\Psi\rangle = E_M |\Psi\rangle$, where $E_M$ is magnetic energy: $E_M = J S^2 \sum_{<nm>} \cos\nu_{nm} - K S^2 \sum_n \sin^2\theta_n \cos^2\varphi_n$ and $\nu_{nm}$ is an angle between local spins $\vec{S}_n$ and $\vec{S}_m$ (see Fig 15.18) on the neighboring sites.

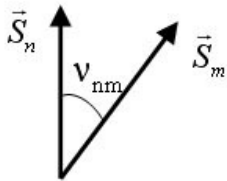

Fig 15.18. The angle $\nu_{nm}$ is an angle between local spins $\vec{S}_n$ and $\vec{S}_m$ on the neighboring sites.

Let us average the Hamiltonian (15.5.1) over $\Psi$-function $|\Psi\rangle$. Then we get:

$$\langle\Psi|H|\Psi\rangle = \langle\Psi|H_{el}|\Psi\rangle + \langle\Psi|H_m|\Psi\rangle = \langle\Psi|H_{el}|\Psi\rangle + E_M \qquad (15.5.5)$$

where

$$\langle\Psi|H_{el}|\Psi\rangle = -t\sum_{<nm>}(T_{nm}\Psi_n^*\Psi_m + c.c.) - \frac{J_H S}{2} - V_{imp}\sum_n \frac{|\Psi_n|^2}{|\vec{n}-\vec{n}_0|}, \qquad (15.5.6)$$

and

$$T_{nm} = \cos\frac{\nu_{nm}}{2}e^{-i\omega_{nm}} \qquad (15.5.7)$$

is classical de Gennes type canted state.

Note that in general $T_{nm}$ in (15.5.7) contains a phase-factor $\omega_{nm}$ [15.46] which is often called in the literature the topological Berry-phase [15.47]. However in our simple approach we will get the solutions with trivial topology both in 3D and 2D case. Formally

$$\cos\nu_{nm} = \cos\theta_n\cos\theta_m + \sin\theta_n\sin\theta_m\cos(\varphi_n - \varphi_m) \qquad (15.5.8)$$

and

$$\omega_{nm} = \arg\left[\cos\frac{\theta_n}{2}\cos\frac{\theta_m}{2} + \sin\frac{\theta_n}{2}\sin\frac{\theta_m}{2}e^{i(\varphi_n - \varphi_m)}\right]. \qquad (15.5.9)$$

At low doping the electron is bound by the impurity electrostatic potential. In the limit of strong electron-impurity coupling $V_{imp}\to\infty$, the electron wave-function $\Psi_n$ will be nonzero only on sites nearest to the impurity. Let us focus first on the 2D case. (The three-dimensional case can be considered in the similar way). Supposing that $\Psi_n \neq 0$ only for $\vec{n}_1 = (1,1)$, $\vec{n}_2 = (1,0)$, $\vec{n}_3 = (0,0)$ and $\vec{n}_4 = (0,1)$ for 2D square lattice (in total analogy it is nonzero only on 8 closest to impurity sites on 3D simple cubic lattice and on 3 closest to impurity sites on 2D frustrated triangular lattice). After minimization of $\langle\Psi|H|\Psi\rangle$ or equivalently of $\langle\Psi|H_{el}|\Psi\rangle$ we will get the system of 8 equations for 3D simple cubic lattice, 4 equations for 2D square lattice and 3 equations for 2D triangular lattice. From these equations it is possible to determine the energy of conductivity electron $E_{el}$. The most simple expression for $E_{el}$ we will have for 2D square lattice which reads:

$$E_{el} = -\frac{J_H S}{2} - V_{imp}\sqrt{2} - t\varepsilon(c_{ij}), \qquad (15.5.10)$$

where

$$\varepsilon(c_{ij}) = \frac{1}{\sqrt{2}}\left\{c_{12}^2 + c_{23}^2 + c_{34}^2 + c_{41}^2 + \sqrt{\left((c_{12}-c_{34})^2 + (c_{23}+c_{41})^2\right)\left((c_{12}+c_{34})^2 + (c_{23}-c_{41})^2\right)}\right\}^{1/2} \quad (15.5.11)$$

and we use the notation:

$$c_{ij} = \cos\frac{\nu_{n_i n_j}}{2} \qquad (15.5.12)$$

Note that in (15.5.11) we consider topologically trivial solution with $\Delta\omega = \omega_{12} + \omega_{23} + \omega_{34} + \omega_{41} = 0$ [15.30]. (For the nontrivial Berry-phase $\Delta\omega \neq 0$).

The electron energy $E_{el}$ in (15.5.10) has a minimum when all spins $\vec{S}_n$ are parallel to each other. Thus, we have a bound magnetic polaron state: a ferromagnetic core embedded in the antiferromagnetic matrix.

### 15.5.3. Magnetic structure of a bound ferron.

To find magnetic structure of a bound ferrons it is convenient to perform the transformation of the variational angles $\varphi_n \to \varphi_n + \pi$, $\theta_n \to \pi - \theta_n$ for one of the sublattices of cubic (quadratic) lattice. As a result, an AFM order becomes FM, and vice versa. Such a transformation allows us to work with continuously changing orientation of spins outside the ferron core. The total energy $E = E_{el} + E_M$ then reads (both in 3D cubic and 2D square lattices):

$$E = E_{el} + JS^2 \sum_{<nm>} (1 - \cos \nu_{nm}) - KS^2 \sum_n (\sin^2 \theta_n \cos^2 \varphi_n - 1). \qquad (15.5.13)$$

Here $E_M = 0$ for the state without a ferron.

If we minimize the total energy over the angles $\varphi_n$ and $\theta_n$ we get two types of the solutions. The first one corresponds to "bare" magnetic polaron. It is a bound magnetic polaron with completely polarized spins embedded in purely AFM-background. The total magnetic moment of such polaron is parallel to the easy axis. The ferron energy reads:

$$E_p^0 = -2t + 16 JS^2 \text{ for 2D square lattice} \qquad (15.5.14)$$

and

$$E_p^0 = -3t + 48 JS^2 \text{ for 3D simple cubic lattice.} \qquad (15.5.15)$$

The bound magnetic polaron (15.5.14), (15.5.15) corresponds to the trivial solution of the problem. However, there is also another solution corresponding to the magnetic polaron state with magnetic moment perpendicular to the easy axis. We call it a "coated" magnetic polaron. It creates long-range spin-distortions of AFM-matrix outside the region of electron localization. When such "coated" polaron is formed, we lose in gradient energy and in the energy of magnetic anisotropy, but we win in surface energy and in the magnetic exchange energy.

### 15.5.4. The "coated" ferrons on the 2D square lattice.

The easiest way to understand the structure of a "coated" ferron is to consider 2D square lattice. Here for planar configuration all the local spins lie in $xy$-plane (all $\theta_n = \pi/2$). The spin symmetry inside the ferron (after the lattice transformation) reads (see Fig. 15.19):

$$\varphi_1 = \varphi_3 = \varphi_0; \varphi_2 = \varphi_4 = -\varphi_0, \quad 0 < \varphi_0 < \pi/2 \qquad (15.5.16)$$

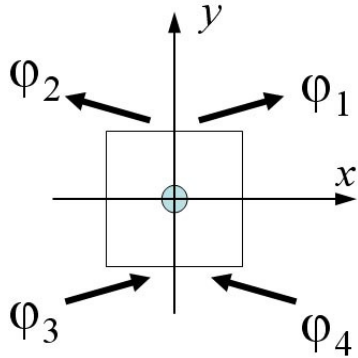

Fig. 15.19. The spin symmetry inside the ferron after the lattice transformation [15.30].

After minimization of the total energy with respect to $\varphi_n$ ($\nu_{nm} = \varphi_n - \varphi_n$) we get the following set of nonlinear equations:

$$\sum_{\Delta} \sin(\varphi_m - \varphi_n) - \frac{K_0}{2} \sin(2\varphi_n) = \frac{t}{2JS^2} \sum_i \delta_{n n_i} (-1)^i \cos^2 \varphi_{n_i}, \qquad (15.5.17)$$

where $K_0 = 2K/J$, $\delta_{nm}$ is the Kronecker symbol, and $|\vec{\Delta}| = |\vec{n} - \vec{m}|$ is the distance between neighboring sites on the square lattice. Thus $\vec{\Delta}$ equals to $(\pm 1, 0)$ and $(0, \pm 1)$.

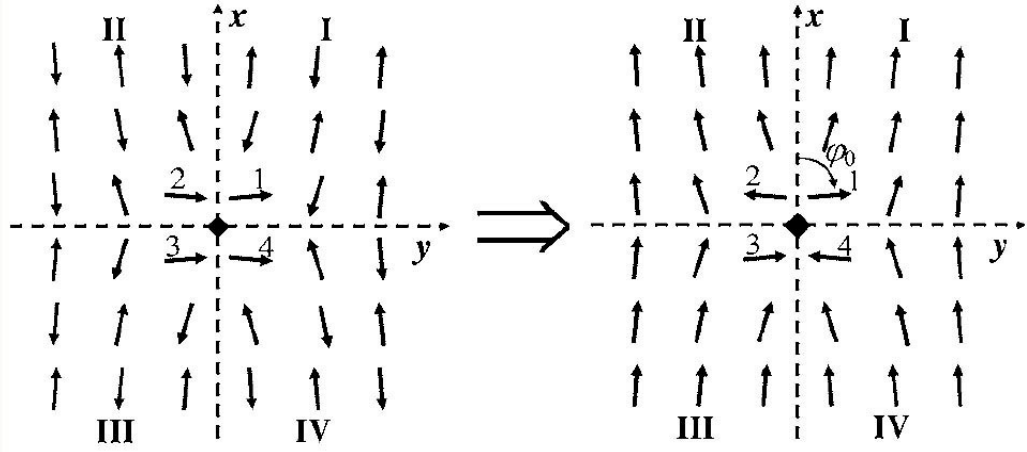

Fig. 15.20. "Coated" magnetic polaron before (left figure) and after (right figure) transformation of angles in one sublattice. The magnetic structure is calculated by solving (15.5.17) at $t/JS^2 = 50$ and $K_0 = 5 \cdot 10^{-3}$. At these values of parameters, $\varphi_0 \approx 85°$. $x$ is an easy axis [15.30].

The system of equations (15.5.17) with the boundary conditions (15.5.16) were solved numerically in [15.30] for the square cluster containing 40×40 sites. The further growth of the number of sites in the cluster does not change the obtained results. The initial angle $\varphi_0$ is also found. The calculated magnetic structure is presented on Fig. 15.20. The magnetization of the ferron core is smaller than the magnetization of the FM-saturation ($\varphi_0 < \pi/2$) for this solution. The coat of the ferrons has a magnetic moment which is antiparallel to the magnetic moment of the core. In the same tine both magnetic moments of the coat and of the core are perpendicular to the easy axis $x$. The angle $\varphi_0$ is close (but still smaller) than $\pi/2$. Note that on the left figure the core is almost ferromagnetic (ferrimagnetic) while the matrix corresponds to inhomogeneously distorted AFM-structure. In the same tine the on the right figure 15.20 we have a vice – a versa situation: the core is almost AFM, while the coat is inhomogeneous and ferromagnetic.

### 15.5.5. The "coated" ferrons in the continuum limit.

In order to get analytical estimations for the spatial distribution of the spin distortions, we find an approximate solution to (15.5.17) in the continuum limit. Namely, angles $\varphi_n$ are treated as values of continuous function $\varphi(\vec{r})$ at points $\vec{r} = \vec{n} - \vec{n}_0$ (further on in our calculation we sometimes put interatomic distance $d = 1$ but restore it in final expressions). Assuming that outside the magnetic polaron the following condition is met $|\varphi_{n+\Delta} - \varphi_n| \ll 1$, we can expand $\varphi(\vec{r} + \vec{\Delta})$ in the Taylor series up to the second order in $\Delta$:

$$\varphi(\vec{r} + \vec{\Delta}) \approx \varphi(\vec{r}) + \Delta^{\alpha}\partial_{\alpha}\varphi(\vec{r}) + \frac{1}{2}\Delta^{\alpha}\Delta^{\beta}\partial_{\alpha}\partial_{\beta}\varphi(\vec{r}) + ... \qquad (15.5.18)$$

Substituting this expansion into (15.5.17), we find that function $\varphi(\vec{r})$ outside the magnetic polaron should satisfy the 2D sine-Gordon equation [15.30]:

$$\Delta\varphi - \frac{\kappa_0}{2}\sin 2\varphi = 0 \qquad (15.5.19)$$

In the range of parameters under study, $K \ll J$, that is, $k_0 \ll 1$, we can linearize this equation. As a result, we obtain:

$$\Delta\varphi - \kappa_0\varphi = 0 \qquad (15.5.20)$$

15.5.6. The boundary conditions in the continuum limit.

This equation should be solved with the boundary conditions, which at infinity reads: $\varphi(\vec{r}) \to 0$ for $r \to \infty$, and with some boundary conditions at the surface of the magnetic polaron. We model the magnetic polaron by a circle of radius $R_{pol} = d/\sqrt{2}$ (see Fig. 15.21) and choose the Dirichlet boundary conditions as:

$$\varphi(\vec{r})\big|_{r=R_{pol}} = \tilde{\varphi}(\xi), \qquad (15.5.21)$$

where we introduce polar coordinates $(r, \xi)$ in the $xy$-plane. The function $\tilde{\varphi}(\xi)$ can be found in the following way. Note, that $\tilde{\varphi}(\xi)$ should satisfy the symmetry conditions (15.5.16) at points $\xi_i = \pi(2i+1)/4$:

$$\varphi(\xi_i) = \varphi_0, \quad i = 1, 2, 3, 4. \qquad (15.5.22)$$

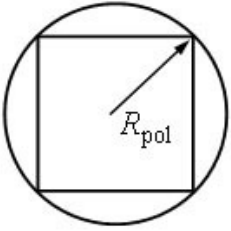

Fig. 15.21. The Dirichlet boundary condition on the surface of the magnetic polaron. We model the magnetic polaron by a circle of radius $R_{pol} = d/\sqrt{2}$ .

Since the function $\tilde{\varphi}(\xi)$ is a periodic one, it can be expanded in the Fourier series:

$$\tilde{\varphi}(\xi) = \sum_{m=0}^{\infty} (a_m \cos m\xi + b_m \sin m\xi). \qquad (15.5.23)$$

In (15.5.23) we neglect the terms with $m > 2$ which allows us to keep the minimum number of terms to satisfy the conditions (15.5.22). It follows from (15.5.22) that $a_0 = a_1 = a_2 = b_1 = 0$. Finally (having in mind that for $m = 0$ $\sin m\xi = 0$) we obtain:

$$\tilde{\varphi}(\xi) = \varphi_0 \sin 2\xi.$$

The solution to (15.5.20) with boundary condition (15.5.21) is:

$$\varphi(\vec{r}) = \frac{\varphi_0}{K_2(R_{pol}/r_0)} K_2(r/r_0) \sin 2\xi, \qquad (15.5.24)$$

where $r_0 = d/\sqrt{\kappa_0}$ and $K_2(x)$ is the Macdonald function [15.48]. In fact $r_0 = d/\sqrt{J/2\kappa} >> d$ plays the role of "coat" radius (the role of the relaxation length of spin distortions). Indeed, within the range $|\vec{r}| < r_0$ (more exactly for $R_{pol} < |\vec{r}| < r_0$) $\varphi(\vec{r})$ behaves as $R_{pol}^2/r^2$, whereas at large distances, it decreases exponentially $\varphi(\vec{r}) \sim \exp(-r/r_0)$. The function $\varphi(\vec{r})$ at $\xi = \pi/4$ and the numerical results for $\varphi_{n,n}$ are plotted on Fig. 15.22.

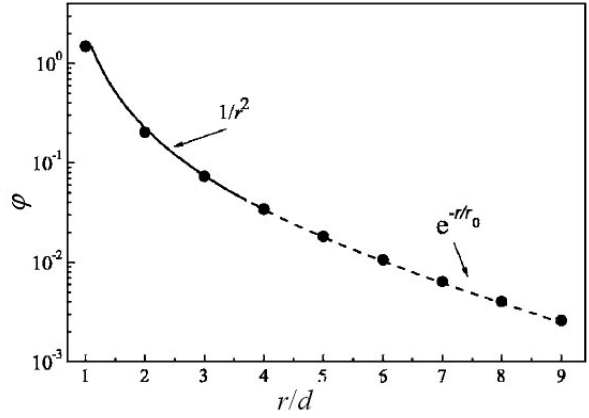

Fig. 15.22. A decrease with distance of an angle $\varphi(\vec{r})$ between distorted spin and easy axis ($x$-axis) for "coated" magnetic polaron on 2D square lattice. The circles correspond to exact numerical solution of the discrete problem on 40×40 square cluster at $t/JS^2 = 50$, $\kappa_0 = 5\cdot10^{-2}$ ($r_0 = 4.5\ d$). The solid curve corresponds to analytical solution (15.5.24) in the continuum limit. At small distances $\varphi(\vec{r})$ decreases proportionally to $1/r^2$ while at large distances $r >> r_0$ it decreases exponentially [15.30].

We see that for $t/JS^2 = 50$ and $\kappa_0 = 5\cdot10^{-2}$ an effective "coat" radius which separates slow power-law decrease of $\varphi(\vec{r})$ from a rapid exponential one is given by $r_0 = 4.5\ d$.

15.5.7. Energy of a "coated" ferron on quadratic lattice.

If we substitute the numerical solution of the discrete equations (15.5.17) into the total energy given by (15.5.10-15.5.12) and take into account that for planar spins configuration on 2D square lattice $\nu_{nm} = \varphi_n - \varphi_m$, than we get an exact numerical estimate for an energy of a "coated" magnetic polaron. However it is more simple to get an analytical estimate for the total energy in continuum limit. In this limit $E = E_{el} + E_M$ and correspondingly $E_{el}$ and $E_M$ are given by:

$$E_{el} = -\frac{J_H S}{2} - V_{imp}\sqrt{2} - 2t\sin\varphi_0, \qquad (15.5.25)$$

and

$$E_M = 8JS^2\left(1 + \frac{k_0}{4}\right)\sin^2\varphi_0 + \frac{JS^2}{2}\int\limits_{r \ge R_{pol}} d^2\vec{r}\left[(\nabla\varphi)^2 + k_0\varphi^2\right]. \qquad (15.5.26)$$

Substituting $\varphi(\vec{r})$ from (15.5.24) to (15.5.25) and performing the integration we get:

$$E_M = 8JS^2\left(1 + \frac{k_0}{4}\right)\sin^2\varphi_0 + \frac{JS^2\varphi_0^2}{2}I\left(\frac{R_{pol}}{r_0}\right), \qquad (15.5.27)$$

where

$$I(x) = 2\pi\left(1 + \frac{xK_1(x)}{2K_2(x)}\right). \qquad (15.5.28)$$

Since $R_{pol} << r_0$, the value of the function $I(x)$ for $x = R_{pol}/r_0 << 1$ is close to $2\pi$.

The optimal angle $\varphi_0$ is determined by the minimization of the ferron energy (15.5.25) and (15.5.27):

$$\cos\varphi_0 - \frac{4JS^2}{t}\left(1 + \frac{k_0}{4}\right)\sin(2\varphi_0) - \frac{JS^2\varphi_0}{2t}I\left(\frac{R_{pol}}{r_0}\right) = 0. \qquad (15.5.29)$$

For $R_{pol} << r_0$ from (15.5.29) it follows that:

$$\varphi_0 = \frac{\pi}{2}\left[1 - o\left(\frac{JS^2}{t}\right)\right]. \qquad (15.5.30)$$

We can compare now the energies of "bare" and "coated" magnetic polarons (see Fig. 15.23). We see that the energy difference $\Delta E = E_{pol} - E_{pol}^{(o)}$ between "coated" and "bare" magnetic polarons is negative at small values of anisotropy ($k_0 \ll 1$) thus stabilizing the "coated" ferron.

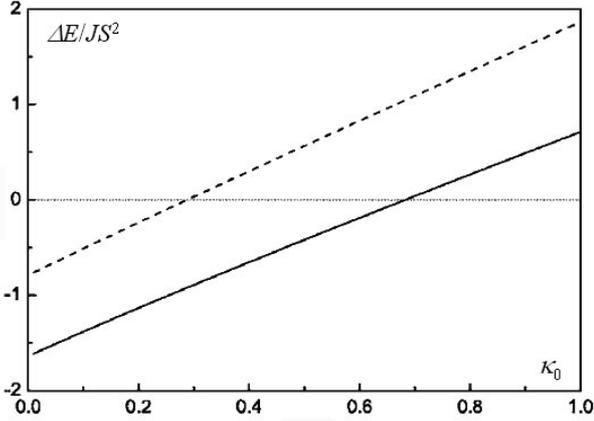

Fig. 15.23. The energy difference $\Delta E$ between "coated" and "bare" ferrons versus $\kappa_0 = 2K/J$ at $t/JS^2 = 50$. Solid curve corresponds to exact numerical solution, whereas dashed curve is calculated analytically in the continuum limit [15.30].

We see that in exact numerical solution the "coated" ferron is more energetically beneficial than a "bare" one till larger anisotropy ($k_0 \sim 0.6$ for $t/JS^2 = 50$) in exact solution than in approximate analytical solution in the continuum limit.

To coincide this section note that absolutely the same variational procedure for 3D simple cubic lattice yields in continuum approximation for $\varphi(\vec{r})$:

$$\varphi(r) \sim \frac{R_{pol}^4}{r^4} \text{ for } R_{\text{pol}} < r < r_0 \qquad (15.5.31)$$

and

$$\varphi(r) \sim \exp(-r/r_0) \text{ for } r > r_0 \qquad (15.5.32)$$

Finally for 2D frustrated triangular lattice we have very slowly decaying coat of spin-distortions [15.31]:

$$\varphi(r) \sim \frac{R_{pol}}{r} \text{ for } R_{\text{pol}} < r < r_0 \qquad (15.5.33)$$

and

$$\varphi(r) \sim \exp(-r/r_0) \text{ for } r > r_0. \qquad (15.5.34)$$

### 15.5.8. Generalized Mott criterion.

We consider that $x_{\text{Mott}}$ which governs the transition from bound to free magnetic polarons can be qualitatively defined via the radius $r_0$ of the coat-extension for bound magnetic polaron [15.31]. In 3D the coats overlap at:

$$x_{Mott} \sim \frac{3}{4\pi}\left(\frac{d}{r_0}\right)^3 \qquad (15.5.35)$$

So $x_{\text{Mott}} \sim (0.1 \div 0.2)\%$ for typical $r_0 \sim (5 \div 6)d$.

Note that for free magnetic polarons $R_{pol} \sim 2d$ in 3D cubic lattice. However, the structures of free and bound magnetic polarons are quite different. A free polaron has a saturated FM core of the size of $R_{pol}$ and very rapidly (exponentially) decaying spin-distortions outside it [15.14-15.16]. In the same time "coated" magnetic polaron bound by nonmagnetic donor impurity has a small and an almost saturated FM core of the radius $\sim d$ and extended coat of slowly-decaying spin-distortions at intermediate distances $d < r < r_0$. Their radius $r_0 >> d$. For $x << x_{Mott}$ the coats of the bound polarons do not overlap and we are effectively in an insulator (Mott limit) for strong Coulomb attraction $V_{imp} > V$. Note that if $V_{imp} < V_C$ (where $V_C \sim 2t$ see [15.31]), then according the preliminary estimates the FM core of the bound magnetic polaron starts to grow and could reach the typical value of the core of the free magnetic polaron. So, the free and bound magnetic polarons are very similar for $V_{imp} < V_C$. Note that in real manganites $V_{imp} \sim V_C$.

Finally for 2D square lattice:

$$x_{Mott} \sim \frac{1}{\pi}\left(\frac{d}{r_0}\right)^2 \sim (0.5 \div 1)\% . \qquad (15.5.36)$$

15.6. Phase separation in charge-ordered manganites.

In this Section we will consider the nanoscale phase separation and formation of small metallic droplets (polarons) in manganites and other systems with charge-ordering. Note that the problem of charge-ordering in magnetic oxides has attracted the attention of theorists since the discovery of the Verwey transition in magnetite $Fe_3O_4$ in the end of the 1930-s [15.49]. An early theoretical description of this phenomenon was given, e.g. in [15.50]. This problem was reexamined later in the numbers of papers in connection with the colossal magnetoresistance in manganites [15.51-15.52, 15.27]. The mechanisms stabilizing the charge ordered state can be different: the Coulomb repulsion of charge carriers (the energy minimization requires keeping the carriers as far away as possible, similarly to Wigner crystallization) or the electron-lattice interaction leading to the effective repulsion of electrons at the nearest-neighbor sites. In all cases, charge ordering can arise in mixed-valence systems if the electron bandwidth is sufficiently small. In the opposite case, the large electron kinetic energy stabilizes the homogeneous metallic state. In real materials, in contrast to the Wigner crystallization, the underlying lattice periodicity determines the preferential types of charge ordering (CO). Thus , in the simplest bipartite lattice, to which belongs the colossal magnetoresistance manganites of the type of $Ln_{1-x}A_xMnO_3$ (where Ln = La, Pr and A = Ca, Sr) or layered manganites $Ln_{2-x}A_xMnO_4$, $Ln_{2-2x}A_{1+2x}MnO_7$, the optimum conditions for the formation of the charge ordered state exist for the doping $x = \frac{1}{2}$ ($n = 1 - x = \frac{1}{2}$ - quarter filling). At this value of x the concentration of $Mn^{3+}$ and $Mn^{4+}$ are equal and the simple checkerboard arrangement is possible (see Fig. 15.24). The most remarkable fact here is that even at $x \neq \frac{1}{2}$ (in the underdoped manganites with $x < \frac{1}{2}$), only the simplest version of charge ordering is experimentally observed with the alternated checkerboard structure of the occupied and empty sites in the basal plane [15.53]. In other words, this structure corresponds to the doubling of the unit cell, whereas more complicated structures with a longer period (or even incommensurate structures) do not actually appear in this case.

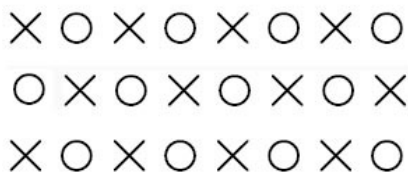

Fig. 15.24. Verwey charge ordering at $x = \frac{1}{2}$ with checkerboard distribution of $Mn^{3+}$ (crosses) and $Mn^{4+}$ (empty circles) ions, which on the figure play the role of electrons and holes respectively.

A natural question then arises as to how the extra or missing electrons can be re distributed for an arbitrary doping level such that the superstructure remains the same for $x = \frac{1}{2}$? To answer this question, the experimentalists introduced the concept of the incipient CO-state corresponding to the distortion of a long-range charge ordering by microscopic metallic clusters [15.54, 15.28]. In fact the existence of this state implies a kind of nanoscale phase separation [15.32] discussed in this Chapter. There is a growing evidence nowadays suggesting that an interplay between the charge ordering and the tendency toward phase separation plays the essential role in the CMR-materials [15.5, 15.6, 15.26, 15.57, 15.58].

In this section we consider a simple model allowing us to clarify the situation at an arbitrary doping. The model includes both the Coulomb repulsion of electrons on the neighboring sites and the magnetic interaction responsible for the magnetic ordering in manganites (in fact it is FM Kondo-lattice model with nn Coulomb repulsion). After demonstrating the instability of the system toward phase separation in certain doping ranges, we consider the simplest form of the phase separation, namely the formation of nanoscale metallic droplets in the insulating CO-matrix [15.32]. We estimate parameters of such droplets and complete the phase diagram of manganites illustrating the interplay between charge ordering, magnetic ordering and phase separation. We note that the CO mechanism considered below (Coulomb repulsion) is not the only one. The electron-lattice interaction can also play an important role (see e.g., [15.55]). In application to manganites, one must also take the orbital interactions into account [15.52, 15.55, 15.56]. Nanoscale phase separation in orbitally ordered matrix [15.34] will be considered in the next section. The orbital interactions explain the zig-zag structures [15.52] and may be also important in explaining the fact that the CO in half-doped perovskite manganites is a checkerboard one only in the basal plane, but it is "in-plane" along the $c$-direction. However the nature of this CO state is not clear yet. We also emphasize that the CO is often observed in manganites at higher temperatures then the magnetic ordering, and thus one must seek a model that does not heavily rely on magnetic interactions. In contrast to magnetic interactions, the Coulomb interaction is one of the important factors that is always present in the systems under consideration. Moreover it has a universal nature and does not critically depend on specific features of the particular system. Consequently, our treatment can be also applied to other CO systems such as magnetite $Fe_3O_4$ [15.49], cobaltites [15.59], nickelates [15.60] etc.

### 15.6.1. The simplest model for charge ordering.

As the starting point, we consider a simple lattice model for CO-state:

$$\hat{H} = -t \sum_{<i,j>} c_i^+ c_j + V \sum_{<i,j>} n_i n_j - \mu \sum_i n_i, \qquad (15.6.1)$$

where $t$ is the hopping integral, $V$ is the nearest-neighbor Coulomb interaction (a similar nn repulsion can be also obtained via the interaction with the breathing-type optical phonons), $\mu$ is the chemical potential, and $c_i^+$ and $c_j$ - are one-electron creation and annihilation operators, $n_i = c_i^+ c_i$ is the onesite density. The symbol $<i, j>$ denotes the summation over the nearest-neighbor sites. We omit the spin and orbital indices for simplicity in (15.6.1). As mentioned in the introduction to this section the spin and orbital effects play the important role in the formation of the real structure in specific compounds; in this section, however, we emphasize the most robust effects related to the nearest-neighbor Coulomb repulsion. The magnetic effects are discussed in the following subsections of the section 15.6. We also assume that the double occupancy does not occur in this model because of the strong onsite repulsion between electrons and thus we can insert projection operators in the kinetic term in (15.6.1). Effectively we consider extended Hubbard model with additional Coulomb repulsion on the neighboring sites under the condition: onsite Hubbard repulsion $U >> \{V, W -$ the bandwidth$\}$. This model is sometimes called Verwey model [15.49] or Shubin-Vonsovsky model [15.61]. Shubin-Vonsovsky model was considered in Chapter 11 with respect to superconductivity. Originally it was introduced for better description (in comparison with simple Hubbard model) of Mott's metal-dielectric transition. Note that: while the Mott-Hubbard localization (Mott-Hubbard dielectric state) corresponds to the density $n = 1$ ($x = 0$), Verwey localization (Verwey CO insulating state) corresponds to $n = \frac{1}{2}$ ($x = \frac{1}{2}$). Hamiltonian (15.6.1) explicitly accounts for the correlation effect that is most important for the formation of CO-state, namely, the electron repulsion on neighboring sites. The long-range part of the Coulomb interaction only leads to the renormalization of the bandwidth $W$ and does not significantly affect the properties of the uniform CO-state. However, it can produce a qualitative effect on the structure of phase-separated state (see the discussion in the beginning of the subsection 15.6.4).

The models of the type (15.6.1) with the nn repulsion responsible for the charge ordering are the most popular in describing this phenomenon, see e.g. [15.50, 15.51, 15.53, 15.62] and reference therein. Hamiltonian (15.6.1) captures the main physical effects. We will add some extra terms to it in the subsection 15.6.5 (where we will consider FM Kondo-lattice model with additional nn Coulomb repulsion).

In the main part of this section, we always speak about electrons. However, in application to real manganites we mostly have in mind less than half-doped (underdoped) systems of the type $Ln_{1-x}A_xMnO_3$ with $x < \frac{1}{2}$. For real system one must therefore substitute holes for our electrons. All the theoretical treatment in terms of hole density $x$ or electron density $n = (1 - x)$ remains qualitatively the same. There are some important differences, however, for the nanoscale phase-separated state for electron densities $n > \frac{1}{2}$ and $n < \frac{1}{2}$ connected e.g. with possible stripe formation. We will discuss these differences in the next subsections where we will study the nanoscale phase-separation in details. Note also that in principle there are hole-doped and electron-doped families of manganites in similarity with high-Tc materials, but the properties of electron-doped manganites are not so well understood yet.

In what follows, we consider the simplest case of square (2D) or cubic (3D) lattices, where the simple two-sublattice ordering occurs for $n = \frac{1}{2}$. As mentioned in the introduction to this section, this is the case in layered manganites, whereas in 3D perovskite manganites, this ordering occurs only in the basal plane (the ordering is "in-plane" along the $c$-direction). A more complicated model is apparently needed to account for this behavior (see [15.32] and discussion therein).

For $n = \frac{1}{2}$ a model (15.6.1) was analyzed in many papers; we follow the treatment of Khomsky in [15.50]. As mentioned above, the Coulomb repulsion (the second term in (15.6.1)) stabilizes the CO-state in the form of a checkerboard arrangement of the occupied and empty sites, whereas the first term (band energy) opposes this tendency. At arbitrary values of the electron density $n$ we first consider a homogeneous charge ordered solution and use the same ansatz as in [15.50], namely

$$n_i = n\left[1 + (-1)^i \tau\right] \qquad (15.6.2)$$

This expression implies the doubling of the lattice periodicity with the local densities:

$$n_1 = n(1+\tau); \quad n_2 = n(1-\tau) \qquad (15.6.3)$$

at the neighboring sites. We note that at $n = \frac{1}{2}$ for a general form of the electron dispersion without nesting, the charge ordered state exists only for a sufficiently strong repulsion $V > V_C = 2t$ [15.50] (or $zV > 2zt = W$). The order parameter is $\tau < 1$ for finite $V/2t$, and the ordering is not complete in general, i.e. an average electron density $n_i$ differs from zero or one even at $T = 0$. (The ideal CO-state described by Fig.15.24 with filled and empty sites is realized in the strong-coupling case $V \gg V_C$). Thus for Verwey localization (for insulating checkerboard CO-state) we need a hierarchy of parameters $U \gg V \gg W$ in Verwey (Shubin-Vonsovsky) model.

We use the coupled Green function approach as in [15.50] and [15.32], which yields:

$$\begin{cases} (E+\mu)G_1 - t_k G_2 - zVn(1-\tau)G_1 = \dfrac{1}{2\pi} \\ (E+\mu)G_2 - t_k G_1 - zVn(1+\tau)G_2 = 0, \end{cases} \qquad (15.6.4)$$

where $G_1$ and $G_2$ are the Fourier transforms of the normal lattice Green functions $G_{il} = <<c_i c_l^+>>$ for the sites $i$ and $l$ belonging respectively, to the same sublattice or to the different sublattices, $z$ is the number of nearest neighbors, and $t_k$ is the Fourier transform of the hopping matrix element. In deriving (15.6.4), we performed a mean-field decoupling and replaced the averages $<c_i^+ c_i>$ by the onsite densities $n_i$ in (15.6.4). The solution of (15.6.4) leads to the following spectrum:

$$E + \mu = Vnz \pm \sqrt{(Vn\tau z)^2 + t_k^2} = Vnz \pm \omega_k. \qquad (15.6.5)$$

The spectrum defined by (15.6.5) resembles the spectrum of superconductor (SC) and, hence, the first term under the square root is analogous to the superconducting gap squared. In other words, we can introduce the CO-gap by the formula: $\Delta = Vn\tau z$. It depends upon the density not only explicitly, but also via the density dependence of $\tau$. We thus obtain:

$$\omega_k = \sqrt{\Delta^2 + t_k^2} \qquad (15.6.6)$$

We note a substantial difference between the spectrum of CO-state (15.6.6) and SC-state, namely the chemical potential $\mu$ does not enter under the square root in (15.6.6) for $n \neq \frac{1}{2}$, which is in contrast to the spectrum of SC, where $\omega_k = \sqrt{\Delta^2 + (t_k - \mu)^2}$. We can then find the Green functions $G_1$ and $G_2$:

$$\begin{cases} G_1 = \dfrac{A_k}{E+\mu-Vnz-\omega_k - io} + \dfrac{B_k}{E+\mu-Vnz+\omega_k + io} \\ G_2 = \dfrac{t_k}{2\omega_k}\left[ \dfrac{1}{E+\mu-Vnz-\omega_k - io} + \dfrac{1}{E+\mu-Vnz+\omega_k + io} \right], \end{cases} \qquad (15.6.7)$$

where

$$A_k = \frac{1}{4\pi}\left(1 - \frac{\Delta}{\omega_k}\right); \quad B_k = \frac{1}{4\pi}\left(1 + \frac{\Delta}{\omega_k}\right). \qquad (15.6.8)$$

They have the two-pole structure, corresponding to the lower and upper Verwey bands. After the standard Wick transformation $E + io \rightarrow iE$ in the expression for $G_1$ we find the densities in the following form:

$$n_1 = n(1+\tau) = \int\left[\left(1 - \frac{\Delta}{\omega_k}\right)n_F(\omega_k - \mu + Vnz) + \left(1 + \frac{\Delta}{\omega_k}\right)n_F(-\omega_k - \mu + Vnz)\right]\frac{d^D\vec{k}}{2\Omega_{BZ}}$$

$$n_2 = n(1-\tau) = \int\left[\left(1 + \frac{\Delta}{\omega_k}\right)n_F(\omega_k - \mu + Vnz) + \left(1 - \frac{\Delta}{\omega_k}\right)n_F(-\omega_k + \mu + Vnz)\right]\frac{d^D\vec{k}}{2\Omega_{BZ}}, \qquad (15.6.9)$$

where $D = 3$ or $2$, $n_F(y) = \dfrac{1}{(e^{y/T}+1)}$ is the Fermi-Dirac distribution function and $\Omega_{BZ}$ is the volume of the first Brillouin zone.

Adding and subtracting the two equations for $n_1$ and $n_2$ we obtain the resulting systems of equations for $n$ and $\mu$:

$$1 = Vz\int\frac{1}{\omega_k}\left[n_F(-\omega_k - \mu + Vnz) - n_F(\omega_k - \mu + Vnz)\right]\frac{d^D\vec{k}}{2\Omega_{BZ}},$$

$$n = \int\left[n_F(-\omega_k - \mu + Vnz) + n_F(\omega_k - \mu + Vnz)\right]\frac{d^D\vec{k}}{2\Omega_{BZ}}. \qquad (15.6.10)$$

For low temperatures ($T \rightarrow 0$) and $n \leq \frac{1}{2}$, it is reasonable to assume that $\mu$ - $Vnz$ is negative. Therefore $n_F(\omega_k - \mu + Vnz) = 0$ and $n_F(-\omega_k - \mu + Vnz) = \theta(-\omega_k - \mu + Vnz)$ is the step function. It is easy to see that for $n = \frac{1}{2}$ the system of equations (15.6.10) yields identical results for all

$$-\Delta \leq \mu - Vnz \leq \Delta .$$  (15.6.11)

From this point of view $n = \frac{1}{2}$ is the indifferent equilibrium point. For infinitely small deviation from $n = \frac{1}{2}$, that is, for densities $n = \frac{1}{2}$ - 0, the chemical potential must be defined as:

$$\mu = -\Delta + \frac{Vz}{2} = \frac{Vz}{2}(1 - \tau).$$  (15.6.12)

If we consider the strong coupling case $V >> 2t$ ($zV >> W$) and assume a constant density of states inside the band, we have:

$$\tau = 1 - \frac{2W^2}{3V^2z^2}$$  (15.6.13)

for a simple cubic lattice and, therefore:

$$\mu = \frac{W^2}{3Vz}$$  (15.6.14)

We note that for the density $n = \frac{1}{2}$ the CO-gap $\Delta$ appears for an arbitrary interaction strength $V$. This is due to the existence of nesting in our simple model.

In the weak coupling case $V << 2t$ and with perfect nesting, we have:

$$\Delta \sim W \exp\left\{-\frac{W}{Vz}\right\}$$  (15.6.15)

and $\tau$ is exponentially small. For $zV >> W$ or, accordingly, for $V >> 2t$, it follows that $\Delta \approx Vz/2$ and $\tau \rightarrow 1$. As mentioned above, for a general form of the electron dispersion without nesting, the CO exists only if the interaction strength $V$ exceeds a certain critical value of the order of the bandwidth (see [15.50]). In what follows, we restrict ourselves to the physically more instructive strong-coupling case $V >> 2t$.

For the constant density of states (flat band) the integrals in (15.6.10) can be taken explicitly and the system of equations (15.6.10) can be easily solved for arbitrary $n$. We note, however, that in the strong coupling case $V >> 2t$ and the small density deviations from $\frac{1}{2}$ ($\delta << 1$), the results are not very sensitive to the form of the electron dispersion. That is why we do not need to solve the system of equations (15.6.10) exactly. We now consider the case where electron density $n = \frac{1}{2}$ - $\delta$ with $\delta << 1$ being the density deviation from $\frac{1}{2}$. In this case $\mu = \mu(\sigma, \tau)$ and we have two coupled equations for $\mu$ and $\tau$. As a result:

$$\mu(\delta) \approx Vnz(1 - \tau) - \frac{4W^2}{Vz}\delta^2 \approx \frac{W^2}{3Vz} + \frac{4W^2}{3Vz}\delta + o(\delta^2).$$  (15.6.16)

The energy of the CO-state (counted from the energy of a homogeneous metallic state $E_N = 0$) is therefore given by:

$$E_{CO}(\delta) = E_{CO}(0) - \frac{W^2}{3Vz}\delta - \frac{2W^2}{3Vz}\delta^2 + o(\delta^3),$$  (15.6.17)

where

$$E_{CO} = -\frac{W^2}{6Vz} -$$  (15.6.18)

is the energy precisely corresponding to the density $n = \frac{1}{2}$ and $|E_{CO}(0)| << W$ for $V >> 2t$. At the same time, the CO-gap $\Delta$ is given by:

$$\Delta \approx \frac{Vz}{2}\left[1 - 2\delta - \frac{2W^2}{3V^2z^2}(1 + 4\delta)\right].$$  (15.6.19)

The dependence of the chemical potential $\mu$ and the total energy $E$ on $\delta$ in (15.6.16) and (15.6.17) actually stems from this linear decrease of the energy gap $\Delta$ with the deviation from half-filling.

For $n > \frac{1}{2}$, the energy of the CO-state starts to increase rapidly due to a large contribution of the Coulomb interaction (the upper Verwey is partially filled for $n > \frac{1}{2}$ - that is the difference between $n > \frac{1}{2}$ and $n < \frac{1}{2}$ on the level of the Green functions in (15.6.7)). In other words, for $n > \frac{1}{2}$, contrary to the case where $n < \frac{1}{2}$, each extra electron put into the checkerboard CO-state, necessarily has occupied nearest-neighboring sites, increasing the total energy by $Vz|\delta|$. For $|\delta| = n - \frac{1}{2} > 0$, we then have:

$$E_{CO}(\delta) = E_{CO}(0) + \left( Vz - \frac{W^2}{3Vz} \right)|\delta| - \frac{2W^2}{3Vz}\delta^2 + o(\delta^3). \qquad (15.6.20)$$

Accordingly, the chemical potential is given by:

$$\mu(\delta) = Vz - \frac{W^2}{3Vz} - \frac{4W^2}{3Vz}|\delta| + o(\delta^2). \qquad (15.6.21)$$

It undergoes a jump equal to $Vz$ as $\tau \to 1$. We note that the CO-gap $\Delta$ is symmetric for $n > \frac{1}{2}$ and $n < \frac{1}{2}$ and is given by:

$$\Delta \approx \frac{Vz}{2}\left[ 1 - 2|\delta| - \frac{2W^2}{3V^2z^2}(1 + 4|\delta|) \right]. \qquad (15.6.22)$$

We could make the entire picture symmetric with respect to $n = \frac{1}{2}$ by shifting all the one-electron energy levels and the chemical potential by Hartree-Fock type shift $Vz/2$, i.e. defining:

$$\mu' = \mu - \frac{Vz}{2}. \qquad (15.6.23)$$

In terms of $\mu'$ expressions (15.6.16) and (15.6.21) can be written as:

$$\begin{cases} \mu' = -\dfrac{Vz}{2} + \dfrac{W^2}{3Vz} + \dfrac{4W^2}{3Vz}\delta; & n < 1/2, \\[2mm] \mu' = \dfrac{Vz}{2} - \dfrac{W^2}{3Vz} - \dfrac{4W^2}{3Vz}|\delta|; & n > 1/2. \end{cases} \qquad (15.6.24)$$

Similarly to the situation in semiconductors, we have $\mu' = 0$ precisely at the point $n = \frac{1}{2}$, which means that the chemical potential lies in the middle of the gap between lower and upper Verwey bands (see Fig. 15.25). At densities $n = \frac{1}{2} - 0$, the chemical potential $\mu' = -Vz/2$ coincides with the upper edge of the filled (lower) Verwey band.

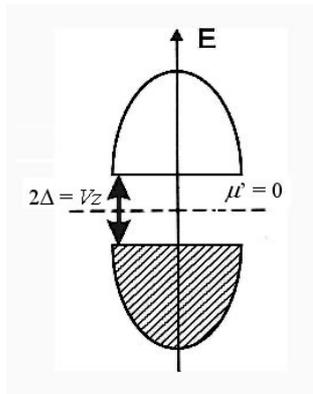

Fig. 15.25. Band structure of the Verwey (Shubin-Vonsovsky) model (15.6.1) at $n = \frac{1}{2}$. The lower Verwey band is completely filled. The upper Verwey band is empty. The chemical potential $\mu' = 0$ lies in the middle of the gap with the width $2\Delta$ [15.6,15.32].

### 15.6.2. The instability of the CO-state with respect to phase separation.

We now check the stability of the CO-state. At the densities close to $n = ½$, the dependence of its energy on the charge density has the form illustrated in Fig. 15.26. The figure clearly indicates a possible instability of the CO-state (the energy has a kink for $n \rightarrow ½$).

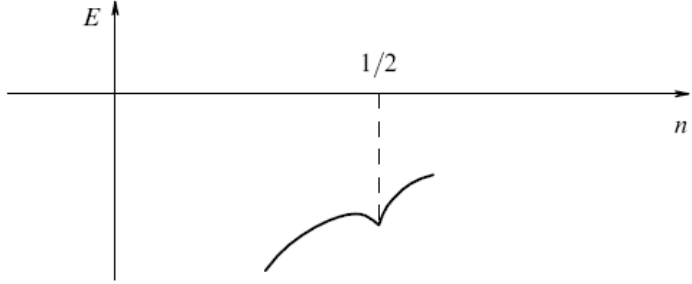

Fig. 15.26. The energy of the CO-state versus charge density for $n \rightarrow ½$. The energy has a kink for $n \rightarrow ½$ [15.32].

Indeed, the most remarkable implication of (15.6.16 - 15.6.22) is that the compressibility $\kappa$ of the homogeneous CO system is negative for the densities different from $½$,

$$\kappa^{-1} \sim \frac{d\mu}{dn} = -\frac{d\mu}{d\delta} = \frac{d^2E}{d\delta^2} = -\frac{4W^2}{3Vz} < 0, \qquad (15.6.25)$$

where $\delta = ½ - n$. This is a manifestation of the tendency toward the phase separation characteristic of the CO system with $\delta \neq 0$. The presence of a kink in $E_{CO}(\delta)$ (cf. (15.6.17) and (15.6.20)) implies that one of the states into which the system might separate would correspond to the checkerboard CO-state with $n = ½$, whereas the other would have a certain density $n'$ smaller or larger then $½$. This conclusion of [15.32] resembles that in [15.52] (see also [15.57] and [15.29]), although the detailed physical mechanism is different. The possibility of phase separation in Verwey model (15.6.1) away from $n = ½$ was also reported earlier in [15.58] for the infinite-dimensional case. In what follows, we focus our attention on the situation with $n < ½$ ($x > ½$ - overdoped manganites); the case where $n > ½$ in hole-doped manganites apparently has certain special properties – the existence of stripe phases, etc. [15.28], the detailed origin of which is not yet clear.

It is easy to understand the physics of the phase separation in our case. As follows from (15.6.22), the CO-gap decreases linearly with the density deviation from $n = ½$. Correspondingly, the energy of the homogeneous CO-state rapidly increases, and it is more favorable to "extract" extra holes from the CO-state, putting them into one part of the sample (for $n < ½$), while creating the "pure" checkerboard CO-state in the other part. The energy loss due to this redistribution of holes is overcompensated by the gain provided by a better charge order.

However the hole-rich regions would not be completely "empty", similarly to pores (clusters of vacancies) in crystals: we can gain extra energy by "dissolving" a certain amount of electrons there. In doing this, we decrease the band energy of the electrons due to their delocalization. Thus, this second phase would be a metallic one. The simplest state of this kind is a homogeneous metal with the electron concentration $n_m$. This concentration, as well as the relative volume of the metallic and CO-phases, can be easily calculated by minimizing the total energy of the system, The energy of the metallic part of the sample $E_m$ in 3D case is given by:

$$E_m = -tzn_m + c_1t(n_m)^{5/3} + V(n_m)^2, \qquad (15.6.26)$$

where $c_1$ is a constant.

Minimizing of (15.6.26) with respect to $n_m$, we find the equilibrium electron density in the metallic phase. For the strong coupling case $V > zt$, we obtain (neglecting a relatively small correction provided by the term with $(n_m)^{5/3}$):

$$n_{m0} \approx \frac{tz}{2V}.\qquad(15.6.27)$$

In accordance with this simple treatment, the system with $n_{m0} < n < \frac{1}{2}$ would therefore undergo the phase transition into the CO-phase with $n = \frac{1}{2}$ and the metallic phase with $n = n_{m0}$. For arbitrary $n$, the relative volumes $V_m$ and $V_{CO}$ of these phases can be found from Maxwell construction (see also Fig. 15.27):

$$\frac{V_m}{V_{CO}} = \frac{1/2 - n}{n - n_{m0}}.\qquad(15.6.28)$$

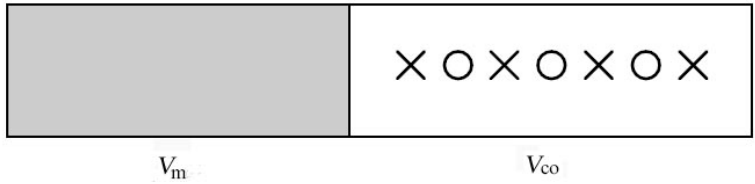

$V_m$ $\qquad\qquad$ $V_{CO}$

Fig. 15.27. Complete phase-separation into two large clusters (metallic and charge-ordered). $V_m$ and $V_{CO}$ are the volumes of the two clusters [15.6].

This implies that the metallic phase occupies the part $V_m$ of the total sample volume $V_S$ given by:

$$\frac{V_m}{V_S} = \frac{1/2 - n}{1/2 - n_{m0}}.\qquad(15.6.29)$$

The metallic phase would occupy the entire sample when the total density $n$ is less or equal to $n_{m0}$.

### 15.6.3. Nanoscale phase separation with metallic droplets inside CO- matrix.

As argued above, the system with a short-range repulsion described by (15.6.1) is unstable with respect to the phase separation for $n$ close but different from $\frac{1}{2}$. The long-range Coulomb forces would, however, prevent the full phase separation into large regions, containing all extra holes, and the pure $n = \frac{1}{2}$ CO-region. We can avoid this energy loss by formation (instead of one big metallic phase with many electrons) of the finite metallic clusters with smaller number of electrons. The limiting case would be a set of spherical droplets, each containing one electron. This state is similar to magnetic polarons ("ferrons") considered in the previous Sections of this Chapter.

Let us now estimate the characteristic parameters of these droplets. The main purpose of this treatment is to demonstrate that the energy of the state constructed in this way is lower then the energy of the homogeneous state, even if we treat these droplets rather crudely and do not optimize all their properties. In particular, we make the simplest assumption that the droplets have sharp boundaries and that the CO-state existing outside these droplets is not modified in their vicinity. This state can be treated as a variational one: optimizing the structure of the droplets boundary can only decrease its energy.

The energy (per unit volume) of the droplet state with the concentration of droplets $n_d$ can be written in total analogy with the ferron energy in the double-exchange model (see Section 15.3 and eq. (15.3.8)). This yields in 3D case:

$$E_{drop} = -t n_d \left( z - \frac{\pi^2 d^2}{R^2} \right) - \frac{W^2}{6Vz} \left[ 1 - n_d \frac{4}{3}\pi \left( \frac{R}{d} \right)^3 \right],\quad(15.6.30)$$

where $E_{CO}(\delta) = -\dfrac{W^2}{6Vz}$ is the energy of the CO-state for $n = \frac{1}{2}$ ($E_{\text{drop}} = E_{CO}(0)$ for $n_{\text{d}} = 0$), d – is the lattice constant and $R$ is the droplet (polaron) radius. The first term in (15.6.30) corresponds to the kinetic energy gain of the electron delocalization inside the metallic droplets and the second term describes the CO-energy in the remaining, insulating part of the sample. Minimization of the energy in (15.6.30) with respect to R yields for the optimal polaron radius:

$$\frac{R_{pol}}{d} \sim \left(\frac{V}{t}\right)^{1/5}. \qquad (15.6.31)$$

The optimal polaronic energy reads:

$$E_{pol} = -\frac{W^2}{6Vz} - tn_d z\left(1 - C_2\left(\frac{t}{V}\right)^{2/5}\right), \quad (15.6.32)$$

where $C_2$ is numerical constant.

The critical concentration $n_{\text{dC}}$ corresponds to the configuration where metallic droplets start to overlap, i.e., where the volume of the CO-phase (the second term in (15.6.30)) tends to zero. Hence,

$$n_{dC} = \frac{3}{4\pi}\left(\frac{d}{R_{pol}}\right)^3 \sim \left(\frac{t}{V}\right)^{3/5} \qquad (15.6.33)$$

Correspondingly in the 2D case $n_{dC} \sim \left(\dfrac{t}{V}\right)^{1/2}$ and $\dfrac{R_{pol}}{d} \sim \left(\dfrac{V}{t}\right)^{1/4}$ (we already used this formula for simple estimates in connection with phase separation in 2D Shubin-Vonsovsky (Verwey) model in Chapter 11).

Actually, one should include the surface energy contribution to the total energy of the droplet. The surface energy in 3D case should be of the order of $W^2R^2/V$. For large droplets, this contribution is small compared to the term $\sim R^3$ in (15.6.30), it would also be reduced for a "soft" droplet boundary. It is easy to show that even in the worst case of a small droplet (for the order of several lattice constants) with a sharp boundary, $R/d$ acquires the factor 1-0.2 $(t/V)^{1/5}$ related to the surface contribution. Thus, the corrections related to the surface would not exceed about 20% of the bulk value. That is why we ignore this term below.

Comparing (15.6.17) and (15.6.32) for one-electron droplets with $\delta = n_{\text{d}}$ we see that for the densities deviations from ½ ($\delta \le \delta_{\text{C}} = n_{\text{dC}}$) the energy of the phase separated state (15.6.32) is always lower that the energy of the homogeneous CO-state (for $n_{\text{d}} = 0$ and $\delta = 0$ the energies (15.6.32) and (15.6.17) coincide). The energy of a droplet state (15.6.32) is also lower than the energy of the fully separated state (15.6.26) obtained by Maxwell construction from the homogeneous metallic state. Correspondingly, the critical concentration $n_{\text{dC}}$ in (15.6.33) is larger then $n_{\text{mO}}$ in (15.6.27). There is no contradiction here: in the droplet (polaronic) state which we constructed the electrons are confined within the spheres of the radius $R$ in 3D and even when these droplets start to overlap at $n = n_{\text{dC}}$, occupying the entire sample, the electrons by construction, are still confined within their own spheres and avoid each other. In other words, a certain degree of CO-correlations is still present in our droplet state, decreasing the repulsion, and hence the total energy.

Thus the energy of the phase separated state with nanoscale metallic droplets inside insulating CO-matrix corresponds to the global minima of the energy for all $0 < \delta \le \delta_{\text{C}}$. This justifies our conclusion about the phase separation into CO-state with $n = \frac{1}{2}$ and a metallic state with small spherical droplets (see Fig. 15.28).

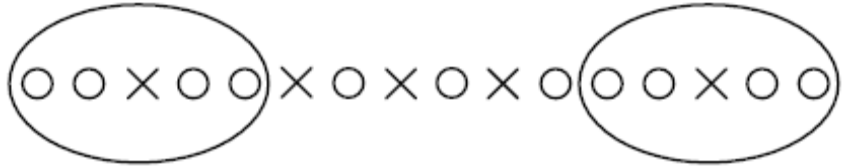

Fig. 15.28. Nanoscale phase separation on spherical metallic droplets with one conducting electron inside CO-insulating matrix [15.6].

Note that along with one-electron metallic droplets a nanoscale phase separation scenario of the kind shown on Fig. 15.29 can be also organized. Here a metallic droplet is formed by replacing one electron with a hole at the center of a droplet.

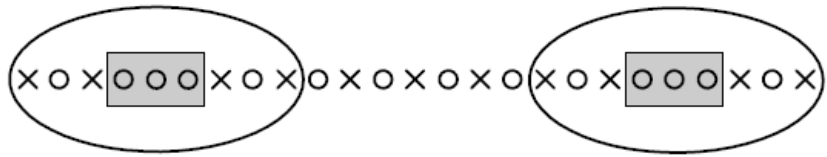

Fig. 15.29. Nanoscale phase separation scenario with an undermelted CO-state within a metallic droplet [15.6].

Note, however, that the energy of such undermelted CO-state (a "resonance-valence-bond" – RVB-state [15.67] for the Verwey model) is much more difficult to calculate than that for a one-electron metallic droplet, and this problem will not be considered in this Chapter. The nanoscale phase separation with one-electron droplet encountered here resembles that of a strongly interacting Hubbard model close to half-filling (for electron densities $n = 1 - \delta$). The CO-state corresponds to AFM-state in the Hubbard model and the role of nn Coulomb repulsion $V$ plays onsite Hubbard repulsion $U$. At $n = 1$ in the Hubbard model we have AFM-dielectric (insulating state) due to Mott- Hubbard localization scenario for $U >> W$ (we discussed it briefly in Chapter 11). Close to half-filling in the limit $U >> W$ we have phase separation on small metallic FM-droplets (containing one-hole) inside AFM insulating matrix due to Nagaoka theorem (which predicts FM-transition in the Hubbard model after addition of one hole to the half-filled AFM-structure in the limit $U \rightarrow \infty$) [15.65].

We would like to emphasize also that for electron density $n > \frac{1}{2}$, the compressibility of the CO-state is again negative $\kappa^{-1} = -\frac{4W^2}{6Vz} < 0$, and has the same value as for $n < \frac{1}{2}$. As a result it is again more favorable to create a phase-separated state for these densities. However, as already mentioned, the nature of the second phase with $n > \frac{1}{2}$ is not quite clear at present, and therefore, we do not consider this case here.

### 15.6.4. Phase separation in the extended double exchange model (with nn Coulomb interaction).

In the beginning of the subsection 15.3 we introduced FM Kondo lattice model (or double exchange model) as a minimal model to describe FM-polarons inside AFM-matrix at small densities $n << 1$ and in almost half-filled case for $x = 1 - n << 1$. To capture an additional possibility of the formation of metallic polarons (we will show that they are actually ferromagnetic also) inside CO-matrix in manganites we should add nn Coulomb repulsion between conductivity electrons to the double exchange model. Then the corresponding Hamiltonian of the extended model reads:

$$\hat{H} = -t \sum_{\langle i,j \rangle \sigma} c_{i\sigma}^{+} c_{j\sigma} + V \sum_{\langle i,j \rangle} n_i n_j - J_H \sum_i \vec{S}_i \vec{\sigma}_i + J \sum_{\langle i,j \rangle} \vec{S}_i \vec{S}_j \; . \qquad (15.6.34)$$

The discussion in Section 15.3 shows that we are working in the strong-coupling limit of FM Kondo lattice model $J_H S \gg W \gg JS^2$. In the same time as we discussed in Section 15.6 we are working in the strong-coupling limit of Verwey model. Thus it is reasonable to consider the following hierarchy of parameters in the extended model:

$$J_H S \gg V > W > JS^2 \; . \qquad (15.6.35)$$

We emphasize here once more that in manganites, strictly speaking, a correlative radius $r_s \geq 4$ due to large dielectric polarization $\varepsilon \sim (10 \div 20)$, so we are really on the border between weak-coupling perturbative approach (RPA-scheme with $V/t < 1$ which is valid for $r_s \leq 1$) and non-perturbative tight-binding case (for which $r_s \gg 1$ and it is reasonable to assume that $V > t$). In our view, however, the tight binding (strong-coupling) approximation $V > t$ is more preferable since it allows to consider all major phenomena in manganites on a simple qualitative level.

For small electron densities $n \ll 1$ the radius of FM-polaron embedded in the AFM-matrix in the absence of nn Coulomb interaction is given by $R_{pol}/d \sim \left(t/JS^2\right)^{1/5}$ and hence for densities which are far from percolation threshold $n \ll n_C \sim \left(JS^2/t\right)^{3/5}$ we have the following chain of inequalities (see Fig. 15.30):

$$d/n^{1/3} \gg R_{pol} > d \; . \qquad (15.6.36)$$

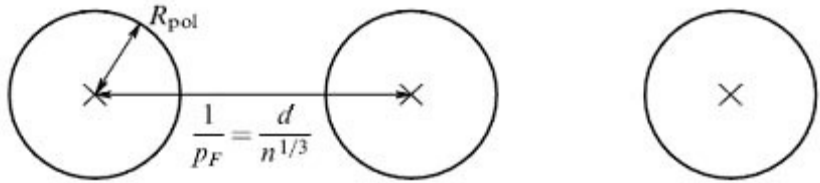

Fig. 15.30. Distribution of conduction electrons in a phase-separated state with FM polarons inside AFM matrix [15.6].

This chain of inequalities implies that the mean distance between the conduction electrons far exceeds the polaron radius, which in turn is larger than the intersite distance $d$.

Thus, for FM polarons (ferrons) with one conduction electron per ferron, even the allowance for the strong Coulomb interaction between electrons on neighboring sites does not lead to a charge redistribution. Therefore, upon including the Coulomb interaction, both the energy of the phase-separated state with FM polarons inside AFM matrix and that of a homogeneous state acquire only a Hartree-Fock correction term proportional to $(z/2)Vn^2$, so that the energy difference $E_{pol}$ - $E_{hom}$ between the polaron and the homogeneous state remains unchanged, and the global minimum for the energy of the system again corresponds to the phase separated FM polaron/AFM matrix state. The most important point to reemphasize here is that there is only one conduction electron in each ferron. Here lies the main difference between the small-scale phase separation and the large-scale separation with a large number of conduction electrons per FM cluster (droplet). Note, however, that within the simplest model (15.6.34) and in the case of large-scale separation, the electric charge within an AFM droplet may be made rarefied enough to avoid strong increase in the Coulomb interaction energy. This is likely to be a qualitative explanation of the experiments [15.23, 7.24, 15.28] showing FM droplets, 100 - 1000 Å across, with a large number of conduction electrons. Note also that the analysis of the large-scale phase separation probably requires considering the elastic energy of the lattice distortions caused by the formation of an inhomogeneous state. Such distortions may make it easier to change the electron density without violating the electroneutrality.

Phase separation for densities close to ½.

The energy of FM metallic droplet inside CO-matrix (which should be also AFM) for densities close to ½ is given by the combination of terms typical to FM-polarons inside AFM-matrix (15.3.8) and to metallic polarons inside CO-matrix (15.6.30):

$$E = -tn_d\left(z - \frac{\pi^2 d^2}{R^2}\right) + zJS^2\frac{4}{3}\pi\left(\frac{R}{d}\right)^3 n_d - zJS^2\left[1 - \frac{4}{3}\pi\left(\frac{R}{d}\right)^3 n_d\right] - \frac{W^2}{6Vz}\left(1 - \frac{4}{3}\pi\left(\frac{R}{d}\right)^3 n_d\right) \quad (15.6.37)$$

where $n_d$ is a droplet density. The first terms in (15.6.37) are identical to the magnetic polaron energy in the double exchange model (15.3.8) but with the electron density $n$ replaced by the droplet density $n_d$. At the same time the last term in (15.6.37) is identical to the second term in (15.6.30) corresponding to the energy of a homogeneous CO Verwey state. Minimization of the droplet energy (15.6.37) with respect to radius $R$ yields in the 3D case:

$$\frac{R_{pol}}{d} \sim \frac{1}{\left(t/V + JS^2/t\right)^{1/5}}. \quad (15.6.38)$$

Note that for $t/V \ll JS^2/t$, we obtain $R/d \sim \left(t/JS^2\right)^{1/5}$, and the double exchange result (15.3.9) is reproduced for the metallic droplet radius. In the opposite case $t/V \gg JS^2/t$, we have $R/d \sim \left(V/t\right)^{1/5}$, and we arrive to the Verwey model result (15.6.31). Accordingly, the critical concentration for the overlap of metallic droplets is:

$$n_{dC} \sim \left(\frac{t}{V} + \frac{JS^2}{t}\right)^{3/5}. \quad (15.6.39)$$

Physically, minimization of the total energy (15.6.37) with respect to the droplet radius implies that there is only one conduction electron inside the metallic droplet and that this electron is surrounded by FM-ordered local spins. At the same time, outside the droplets we have a CO (checkerboard) arrangement of conduction electrons surrounded by AFM-ordered local spins (Fig. 15.31).

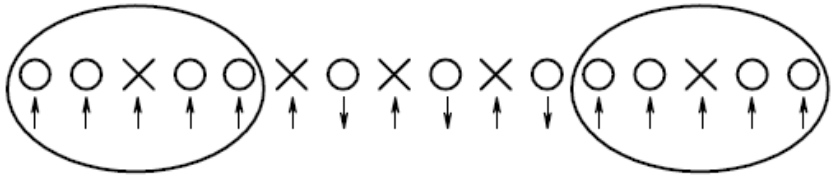

Fig. 15.31. Formation of metallic FM droplets of small radius within a charge-ordered AFM matrix[15.6].

This last result illustrates the main difference between the phase-separated states that are obtained in the extended model (15.6.34) at electron densities $n \to 0$ and $n \to$ ½. At low densities ($n \ll 1$), the conduction electron density outside FM polarons is zero and the entire electric charge is contained within metallic droplets. At the same time, at densities close to ½, most conduction electrons are localized in CO-regions outside the metallic droplets.

Finally, the phase diagram of the extended FM Kondo-lattice model with nn Coulomb interaction includes the following regions:

1) For $0 < n < (JS^2/t)^{3/5}$, the system separates into FM metallic droplets within an AFM insulating matrix.

Recollecting the results of Section 15.5 we should have in mind that if we include in the model (15.6.34) additional Coulomb attraction between a conduction electron and Ca(Sr) donor impurity $V_{imp}$ (see (15.5.1)) than the first interval $0 < n < (JS^2/t)^{3/5}$, actually is splited in two: for $0 < n < n_{Mott}$ we have bound magnetic polarons of de Gennes with extended coats of spin-distortions and for $n_{Mott} < n < (JS^2/t)^{3/5}$ we have free FM-polarons (Nagaev-Mott-Kasuya ferrons) with rigid boundary inside AFM-matrix.

2) For $(JS^2/t)^{3/5} < n < \frac{1}{2} - (t/V + JS^2/t)^{3/5} < \frac{1}{2}$ the system is a FM metal. Of course, we must have a certain "window" of parameters to satisfy this inequality. As we already discussed in real manganites we have $t/V \sim \frac{1}{2} \div \frac{1}{3}$ and $JS^2/t \sim 0.1$.

Therefore, the inequality $(JS^2/t)^{3/5} < \frac{1}{2} - (t/V + JS^2/t)^{3/5}$ is not necessarily met. Experimental evidence indicates that the desired parameter range exists for $La_{1-x}Ca_xMnO_3$, but definitely not for $Pr_{1-x}Ca_xMnO_3$.

3) Finally, for $\frac{1}{2} - (t/V + JS^2/t)^{3/5} < n < \frac{1}{2}$ we have phase separation into metallic FM droplets within an AFM charge-ordered matrix. Note that large Hund-rule exchange $J_H$ between a local spin $S$ and a conduction electron spin $\sigma$ can suppress an ideal AFM-structure inside CO- matrix and promote ferrimagnetism inside it (for total spins $S_{tot} = S + \frac{1}{2}$).

4) At $n = \frac{1}{2}$ a homogeneous CO-state is stabilized.

5) For $n > \frac{1}{2}$ metallic FM droplets can again appear inside charge-ordered AFM-matrix. Qualitatively, the only difference is that instead of one electron, one hole will be now localized inside the droplet (Fig. 15.32). However, in the electron-doped and hole-doped manganites the more complicated stripe phase can be stabilized both for $n > \frac{1}{2}$ and $n < \frac{1}{2}$

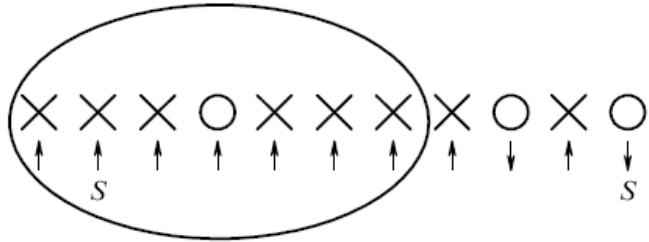

Fig. 15.32. Metallic FM droplet against the background of a charge- ordered AFM structure at densities $n > \frac{1}{2}$ [15.6].

6) Close to half-filling $1 - (JS_{tot}^2/t)^{3/5} < n = 1 - x < 1$ we have again phase separation on small FM polarons inside AFM matrix, but now inside a polaron we have a hole (a local spin $S$ in the surrounding of spins $S_{tot} = |S + \frac{1}{2}|$) (see Fig. 15.12).

## 15.7. Orbital ferrons.

In fact a two-band character of conducting $e_g$-orbitals (see Figs. 15.6 and 15.7) is important for conductivity electrons in manganites especially close to percolation threshold $x \sim 0.16$. Here the full model, which describes the manganites, is a very complicated Kondo-Hubbard model [15.67]. The beautiful physics of the phase separation and orbital ordering can be described, however, on the language of a reduced two-band Hubbard model, which neglects double exchange interactions and describe only the essential interactions inside the subspace of two conducting $e_g$-orbitals.

### 15.7.1. Two-band degenerate Hubbard model.

This model in a most general case was considered in Chapter 10. In manganites we deal with two-band degenerate Hubbard model describing by the Hamiltonian:

$$\hat{H}' = \sum_{<nm>\alpha\beta} t_{nm}^{\alpha\beta} c_{n\alpha\sigma}^+ c_{m\beta\sigma} + \varepsilon \sum_{m\sigma} n_{m\beta\sigma} - \mu \sum_{m\beta\sigma} n_{m\beta\sigma} +$$

$$+ \frac{1}{2} \sum_{\substack{m\beta \\ \sigma \neq \sigma'}} U^\beta n_{m\beta\sigma} \, n_{m\beta\sigma'} + \frac{1}{2} U' \sum_{\substack{m\sigma\sigma' \\ \alpha \neq \beta}} n_{m\alpha\sigma} \, n_{m\beta\sigma'}, \qquad (15.7.1)$$

where $\mu$ is the chemical potential, $\varepsilon$ is the energy-shift between the centers of the bands. In our case $\varepsilon = E_{JT}$ is connected with Jahn-Teller gap. $\{\alpha, \beta\}$ – are the two $e_g$-subbands, $<n, m>$ are neighboring sites, $\{\sigma, \sigma'\}$ are spin projections. Without loss of generality in the degenerate model we can assume $U_1 = U_2 = U' + 2J_H = U$, where strong on-site Hubbard repulsion $U \sim 10$ eV $>> J_H S \sim 1$ eV $>> t$. The total number of electrons (in both bands) per site $n_{tot} = 1 - x$, where we are interested in $x \leq x_C = 0.16$. In magnetic oxides $t_1 \sim t_2$. However in other materials such as uranium based HF-compounds (considered in Chapter 12) and possibly in overdoped cuprates they are different $t_1 >> t_2$.

### 15.7.2. Heisenberg-like orbital interaction.

The detailed analysis of the two-band degenerate Hubbard model is given in [15.67,15.68]. Here we present a brief sketch of the derivation of the effective model which allows to get an orbital ferron state. In the strong-coupling case for $U >> t$ and $x \leq 0.16$ we get on 2D square lattice with $e_g$-electrons on $|x^2 - y^2>$ and $|3z^2 - r^2>$-orbitals (see Figs. 15.7 and 15.8) the following effective Hamiltonian:

$$\hat{H}_{eff} = - \sum_{\substack{n,m \\ \alpha,\beta,\sigma}} P_{n\alpha\sigma} \, t_{nm}^{\alpha\beta} c_{n\alpha\sigma}^+ \, c_{m\beta\sigma} \, P_{m\beta\sigma} + J \sum_{<n,m>} \tau_n \, \tau_m \qquad (15.7.2)$$

It is an orbital t-J model. The effective model (15.7.2) is derived from the two-band Hubbard model in [15.68] by the canonical transformation similar to derivation of standard t-J model from one-band Hubbard model in [15.18]. Pseudospin operators in (15.7.2) $\vec{\tau}_n = \{\tau_n^x, \tau_n^z\}$ ,

$\tau_n^{x,z} = \frac{1}{2} \sum_{\alpha,\beta,\sigma} c_{n\alpha\sigma}^+ (\sigma_{\alpha\beta}^{x,z}) c_{n\beta\sigma}$ describe an orbital state, $\sigma_{\alpha\beta}^{x,z}$ are Pauli matrices, quadrupole interaction $J \sim t^2/U \sim (300 \div 400)$ K is analogous to superexchange interaction of AFM-type ($J > 0$) between two orbitals, $P_{m\alpha\beta}$ are projection operators, excluding double occupation of sites. The Hamiltonian (15.7.2) was firstly proposed by Kugel, Khomskii [15.35] to describe the orbital ordering in Jahn-Teller systems for n=1. The hopping integrals $t_{nm}^{\alpha\beta}$ are described by (2x2) matrix (see [15.34] for more detailed analysis) and read:

$$t_{nm}^{\alpha\beta} = \frac{t_0}{4} \begin{pmatrix} 3 & \mp\sqrt{3} \\ \mp\sqrt{3} & 1 \end{pmatrix}, \qquad (15.7.2)$$

where minus (plus) sign corresponds to $n$-$m$ bond parallel to $x$ ($y$) axis in [15.34].

### 15.7.3. Orbital ferrons in the orbital t-J model.

In the phase separated sate of the two-band Hubbard model (more precisely of the orbital t-J model) we deal with metallic orbital ferrons inside AFM orbital matrix. The derivation of the optimal radius of metallic orbital ferrons in [15.34] is exactly equivalent to the derivation of a FM-polaron inside AFM-matrix in 2D square lattice (see Subsection 15.4.3 and eq. (15.4.7)).

Characteristic radius of the droplet: $\dfrac{R_{pol}}{d} = \left(\dfrac{t_0 \alpha_0^2}{12 J S^2}\right)^{1/4}$ , where $\alpha_0 \approx 3\pi/4$ in the first zero of the

Bessel function (see Fig. 15.33). We can see that inside the circular ferrons we have the same orbitals (say $dx^2$-$y^2$), while outside the ferrons we have alternating order of $dx^2$-$y^2$ and $d3z^2 - r^2$ orbitals.

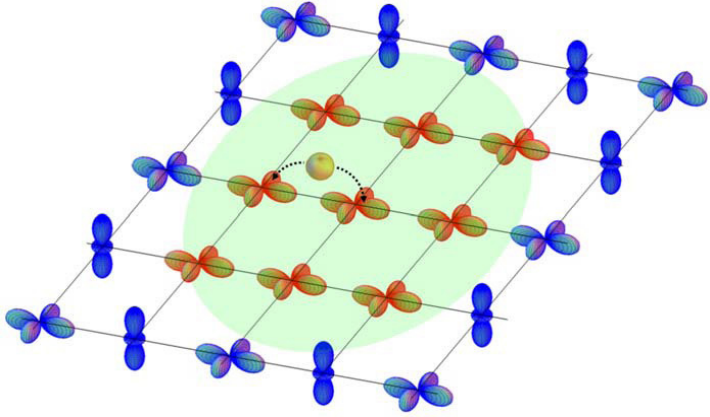

Fig. 15.33. Formation of metallic orbital ferrons inside AFM orbital matrix [15.34].

Concluding this section we can say that in Chapter 17 we considered free and bound magnetic polarons inside AFM and CO-matrices in 3D and layered manganites and other complex magnetic oxides (cobaltites, nickelates) as well as more exotic (but very beautiful) orbital ferrons inside AFM orbital matrix (inside the matrix with alternating $e_g$-orbitals in real space) (see the discussion of the gross phase-diagram on the Fig. 15.2 in subsection 15.2.1).

## 15.8 Experimental confirmation of the Gross phase-diagram and phase separation in manganites.

The phase diagram of manganites discussed in this Chapter is in good qualitative agreement both with experiments on nanoscale phase separation in manganites [15.69-15.74] and with numerical calculations by Dagotto et al., [15.55]. Note that real manganites are usually hole-doped, so that experimentally $x = 1 - n$ means the hole concentration. This does not matter much, however, because of the re-entrant character of the phase diagram for $n \to 0$ and $x = 1 - n \to 0$. Nevertheless, the phase diagram of real manganites differs considerably depending on whether it is electron- or hole-doped. The mechanisms of this asymmetry are not yet completely understood. In particular, the asymmetry can be related to the specific features of orbital ordering in manganites and on possible formation, as we have mentioned above, of inhomogeneous states other than droplet structure discussed here, charge- and orbital stripes, for example. To include all these aspects, however, a theory that goes beyond the content of this Chapter is needed.

### 15.8.1 Experimental confirmation of nano-scale phase separation.

Turning now to the experimental confirmation of our results, the beautiful nuclear magnetic resonance (NMR) experiments on $La_{1-x}Ca_xMnO_3$ [15.69] should be mentioned first. These experiments which employed $^{55}Mn$ nuclei, provided for the existence of two NMR frequencies in the sample (instead of only one as is typical for a homogeneous state), whose frequencies are naturally attributed to the FM and AFM-domains resulting from the phase separation in manganites. NMR measurements at La nuclei in La-Pr manganites led to similar conclusion [15.70].

Further experimental confirmation of phase separation in manganites comes from recent neutron scattering experiments [15.71, 15.74]. They showed that in the case of inelastic scattering there are two spin-wave modes, one of which has quadratic dispersion and corresponds to FM magnons, whereas the other has linear dispersion and corresponds to magnons in AFM phase.

Note that in elastic neutron scattering experiments the peak intensity $I(q)$ has a Lorentzian shape. The half-width of the peak at low densities $n \sim 0.05$ corresponds to the characteristic polaron radius $R_{pol} \sim 1/q_0 \sim 10$ Å [15.71]. At densities $n$ close to ½, the line half-width again corresponds to small-scale phase separation with a characteristic polaron size $R_{pol} \sim (10 \div 20)$ Å.

Note that similar measurements of the spin wave spectrum in a magnetic field using the AFM resonance technique are interpreted in [15.75] as favoring some nontrivial compromise between magnetic polaron formation and inhomogeneous spin canting. Probably these experimental results could be explained within a concept of bound magnetic polarons with extended coat of spin-distortions considered in Section 15.5.

The further experimental evidence in favor of the polaron picture was obtained by Babushkina et al., [15.72] who discovered a strongly non-linear current-voltage characteristic in La-Pr manganites close to the phase boundary between the FM and CO-states. This provides indirect evidence for percolative charge transfer [15.77,15.78] naturally activated by the phase separation process. The critical density for the overlap of polarons actually appears as the percolation threshold picture [15.77].

Finally, the experiments of Voloshin et al., [15.73] showed a shifted magnetization hysteresis in manganites with the center of the hysteresis loop shifting from the magnetic field $H$ = 0 to $H \sim 4 - 6$ T in the low-density state. The shift appears quite naturally within the polaron picture. To see this note that in a magnetic field the effective Heisenberg exchange is $J_{eff} S^2 = JS^2 - g\mu_B HS$ , where $g$ is the hyromagnetic ratio and $\mu_B$ is the Borh magneton. Therefore the polaron radius $\dfrac{R}{d} = \left( \dfrac{t}{J_{eff} S^2} \right)^{1/5}$ increases with the result that in strong magnetic fields the FM-polarons starts to overlap at lower densities $n_C(H) < n_C(0) \approx 0.16$.

### 15.8.2. Experiments on large scale phase separation. Formation of stripes.

Note that, recently, more direct experiments supporting the phase separation scenario have been carried out [15.23, 15.76]. In [15.76] metallic regions inside the insulating matrix were visualized via scanning tunneling microscopy in hole-doped manganites (STM experiments of Mydosh group with "catching" of FM metallic polaron by a needle of ST microscope for $x \sim$ 0.16). In [15.23] experiments on electron-diffraction performed by Cheong's group confirmed checkerboard CO-structure at $x = 0.5$ and also showed the coexistence of metallic FM domains and the insulating CO matrix for $x = 0.4$ in the experiments on dark image electron microscopy. However, both [15.76] and [15.23] actually report large scale phase separation, with metallic domains measuring $L \sim (100 \div 200)$ Å in size. Thus, the experiments reported in [15.76] and [15.23] neither contradict nor decisively verify the nanoscale polaron picture.

Finally, for $x > 0.5$ Mori et al., [15.28] showed the formation of stripes in experiments on electron diffraction (Fig. 15.34).

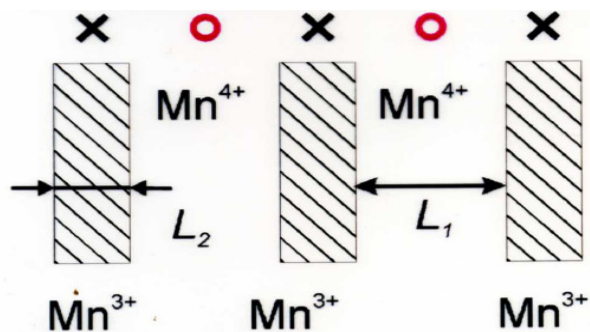

Fig. 15.34. Stripe structure measured by Mori et al., [15.28] in experiments on electron diffraction.

They reported incommensurate charge ordering with the distance between two stripes $L_1 \approx 3 L_2$, where $L_2$ is a stripe width. Note that the stripes are static in manganites due to Jahn-Teller effects in contrast with strongly fluctuating stripes in high-$T_C$ compounds.

Chapter 16. Mesoscopic transport properties in the phase-separated manganites.

The physics of mesoscopic transport phenomena [16.2, 16.3] is a very rapidly developing field in the modern condensed-matter physics which is very promising for the applications in small (nano-size) electron devices [16.4, 16.8] and very interesting for studying of the quantum interference effects [16.3, 16.5] and the quantum effects of localization in dirty or narrow-band (strongly-correlated) metals [16.6, 16.7] as well as classical percolation effects in doped semiconductors [16.9].

16.1. Mesoscopic transport properties in strongly-correlated electron systems.

In this Chapter we would like to build a bridge between the physics of strongly-correlated electrons [16.1] and the physics of mesoscopic transport phenomena studying the transport properties of the CMR-systems. Namely for the CMR-family of materials we will present the simple theory for the tunneling conductivity, magnetoresistance (MR) and 1/f – noise spectrum [16.17-16.19] in the low-doping case [16.10-16.15]. In this case as it was shown in Chapter 17, the CMR-systems are in the nano-scale phase-separated state. We will compare the theoretical predictions for transport properties of non-metallic phase-separated manganites with experimental observations in several families of 3D and layered CMR-materials [16.20-16.26].

16.1.1. Transport properties in non-metallic phase-separated manganites.

In Chapter 14 we considered the formation of nano-scale phase-separated state in different CMR-systems. In particular we studied the formation of small FM-polarons (FM droplets or ferrons) inside AFM, PM or CO-matrices. It is interesting to understand the nature of electron transport in the phase-separated state having in mind that small FM-polarons are metallic, while they are embedded in insulating AFM, PM or CO-matrices. We will consider here the transport mechanism connected with the electron tunneling from one FM-polaron to a neighboring one.

16.1.2. Tunneling conductivity in the phase-separated manganites.

Let us consider an insulating antiferromagnetic sample of volume $V_s$ in electric field $E$. The total number of magnetic polarons in the volume is $N$, and thus their spatial density is $n = N/V_s$. As mentioned in Chapter 17, the number of polarons is assumed to be equal to the number of charge carriers introduced by doping. Neglecting the conductivity of the insulating phase, we assume that charge carriers are only located within the droplets. The charge transfer can thus occurs either due to the motion of the droplets or due to electron tunneling. The former mechanism is less effective: indeed, the motion of a droplet is accompanied by a considerable rearrangement of the local magnetic structure, which results in the large effective mass of magnetic polarons. In addition, the droplets are expected to be easily pinned by crystal lattice defects. Thus, it is realistic to assume that the charge transport is essentially due to electron transitions between the droplets.

A magnetic polaron in the ground state contains one electron. As a result of the tunneling process, droplets with more than one electron are created, and some droplets become empty (the lifetime of such excitations is discussed in the end of this section). If the energy of an empty droplet $E(0)$ is taken to be zero, then the energy of a droplet with one electron can be estimated as $E(1) \sim t \ d^2/R^2$ (where $R$ is the ferron radius). This is essentially the kinetic energy of an electron localized in the sphere of radius $R$. In the same way, the energy of a two-electron magnetic polaron $E(2) \sim 2E(1) + U$, with $U$ the interaction energy of the two electrons. In all these estimates, we disregarded the surface energy, which as shown in Chapter 17 is expected to be small. Thus, $E(2) + E(0) > 2E(1)$, and the creation of two-electron droplets is associated with the energy barrier of the order of $A \equiv E(2) - 2E(1) \sim U$. It is clear that the interaction energy $U$ of two electrons in one droplet is determined mainly by the Coulomb repulsion of these electrons; hence $A \sim e^2/\varepsilon R$, where $\varepsilon$ is the static dielectric constant, which in real manganites can be rather large ($\varepsilon \sim 20$). We assume below that the mean distance between the droplets is $n^{-1/3} >> R$ (the droplets do not overlap and we are in non-metallic state far from the percolation threshold [16.9]). Then, $A$ is larger than the average Coulomb energy $e^2 n^{1/3}/\varepsilon$. Since the characteristic value of the droplet radius $R$ is of the order of 10 Å we have $A/k_B \sim 1000$ K and $A > k_B T$ in the case under study. In the following, we assume that the temperature is low, $A >> k_B T$, and we do not consider a possibility of the formation of the droplets with three or more electrons. Even in the case when these excitations are stable, it can be shown that far from the percolation threshold the strong Coulomb interaction suppresses their contribution to the conductivity giving rise only to the next-order terms with respect to $\exp(-A/k_B T)$.

Let us denote the numbers of single-electron, two-electron, and empty droplets as $N_1$, $N_2$, and $N_3$, respectively. According to our model, $N_2 = N_3$ (number of empty and two-electron droplets coincide), $N = N_1 + 2N_2$, and $N$ is constant. Before turning to conductivity, we evaluate the thermal averages of $N_1$ and $N_2$. To this end, we note that the number $P_N^m$ of possible states with m two-electron droplets and $m$ empty droplets equals $C_N^m C_{N-m}^m$, with $C_N^m$ being the binomial coefficients. Since the created pairs of droplets are independent, we write the partition function [16.27] in the form:

$$Z = \sum_{m=0}^{N/2} P_N^m \exp(-m\beta), \quad \beta = \frac{A}{k_B T}. \tag{16.1.1}$$

Though the sum can be evaluated exactly and expressed in terms of the Legendre polynomials for arbitrary $N$, it is more convenient to use the Stirling formula for the factorials [16.28] and the condition that the sample is macroscopic, $N >> 1$. Approximating the sum by an integral,

$$Z = \int_0^{N/2} dm \exp\left\{-m\beta - N\ln\left(1 - \frac{2m}{N}\right) + 2m\ln\left(\frac{N}{m} - 2\right)\right\}, \tag{16.1.2}$$

calculating it in the saddle-point approximation, and subsequently evaluating in the same way the statistical average of $N_2$,

$$\overline{N}_2 = Z^{-1} \sum_{m=0}^{N/2} P_N^m \exp(-m\beta) = -\frac{\partial}{\partial \beta} \ln Z, \tag{16.1.3}$$

we easily obtain:

$$\overline{N}_2 = N \exp\left(-\frac{A}{2k_B T}\right), \tag{16.1.4}$$

$\overline{N}_1 = N - 2\overline{N}_2 = N\left[1 - 2\exp\left(-\frac{A}{2k_B T}\right)\right] = \overline{N}_{1\sigma} + \overline{N}_{1-\sigma}$, where $\sigma$ is spin-projection of conductivity electron inside the ferron.

Now we calculate the conductivity. Within the framework of the proposed model the electron tunneling occurs via one of the four following processes illustrated in Fig. 16.1.

(i) In the initial state we have two droplets in the ground state (with spin-projections of the conductivity electrons $\sigma_1$ and $\sigma_2$ in them correspondingly), and after tunneling in the final state we have an empty droplet and a droplet with two electrons.

(ii) An empty droplet and a two-electron droplet in the initial state transform into two droplets in the ground state (two droplets with one electron, in the first one spin-projection of electron is $\sigma_1$, in the second one - $\sigma_2$).

(iii) A two-electron droplet and a single-electron droplet exchange their positions by transferring an electron from one droplet to the other.

(iv) An empty droplet and a single-electron droplet exchange their positions by transferring an electron from one droplet to the other.

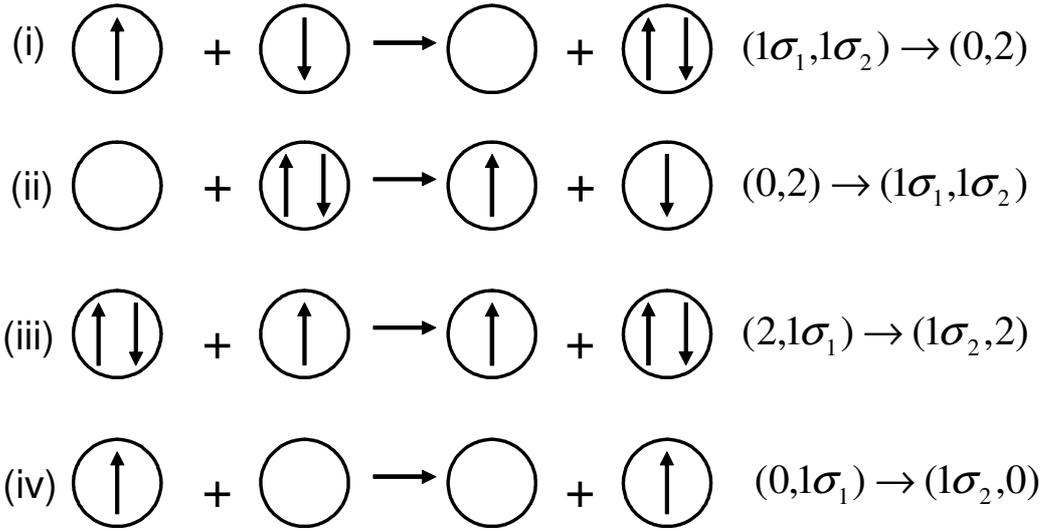

Fig. 16.1. Elementary tunneling processes.

Note that in the last two cases (iii) and (iv) $\sigma_1$, generally speaking, is not equal to $\sigma_2$. In the linear regime, all these processes contribute to the current density $j$ independently, $j = j_1 + j_2 + j_3 + j_4$. The contributions of the first two processes read:

$$j_{1,2} = e n_{1,2} \left\langle \sum_i v_{1,2}^i \right\rangle \qquad (16.1.5)$$

where $n_{1,2} = N_{1,2}/V_s$ are the densities of the single- and two-electron droplets, and <...> stands for statistical and time averages. The appearance of the factors $n_{1,2}$ reflects the fact that the electron tunnels from a single-electron droplet [process (i)] or two-electron (ii) droplet. The summation in (16.1.5) is performed over all magnetic polarons the electron can tunnel to - one-electron droplets for process (i) and empty droplets for process (ii). Finally, the components of the average electron velocity <$v_{1,2}^i$> along the direction of the electric field are obviously found as [16.29]:

$$\left\langle \sum_i v_{1,2}^i \right\rangle = \left\langle \sum_i \frac{r^i \cos\theta^i}{\tau_{1,2}(r^i,\theta^i)} \right\rangle, \qquad (16.1.6)$$

where $r^i$ and $\theta^i$ are the electron tunneling length (the distance between the droplets) and the angle between the electric field and the direction of motion, respectively, and $\tau_{1,2}(r^i,\theta^i)$ are characteristic times associated with the tunneling processes. The relation between $\tau_1(r,\theta)$ and $\tau_2(r,\theta)$ can be found from the following considerations. Near the equilibrium, the number of two-electron droplets, excited per unit time, equals to the number of the decaying two-electron droplets. We thus have the detailed balance relation,

$$\frac{\overline{N_1^2}}{\tau_1(r,\theta)} = \frac{\overline{N_2^2}}{\tau_2(r,\theta)}, \qquad (16.1.7)$$

where we have taken into account that the probability of the formation of a two-electron droplet is proportional to the total number $N_1$ of the single-electron states multiplied by the number of available hopping destinations, which also equals $N_1$. Similarly, the probability of decay of a two-electron droplet is proportional to $N_2 N_3 = N_2^2$. Equation (16.1.7) implies $\tau_2(r,\theta) = \tau_1(r,\theta)\exp(-A/k_B T)$. We write then the conventional expression for the tunneling times [16.29] in the following form:

$$\tau_{1,2}(r,\theta) = \omega_0^{-1}\exp\left\{\frac{r}{l} \pm \frac{A}{2k_B T} - \frac{eEr\cos\theta}{k_B T}\right\}, \qquad (16.1.8)$$

where $l$ and $\omega_0$ are the characteristic tunneling length and magnon frequency, and we have taken into account the contribution of the external electric field to the tunneling probability.

To perform the averaging, we assume that the centers of the magnetic polarons are randomly positioned in space and the average distance $n^{-1/3}$ between them is much larger than the droplet radius $R$. Both assumptions seem to be perfectly justified far below the percolation threshold. Then the averaged sum in (16.1.5) is essentially the space average of $v$, multiplied by the number of droplets available for hopping ($N_1$ for the process (i) and $N_2$ for process (ii)). Expanding in $eEl/k_B T \ll 1$, we obtain:

$$\left\langle \sum_i v_{1,2}^i \right\rangle = \frac{eE\omega_0}{k_B T} N_{1,2}\exp\left\{\mp\frac{A}{2k_B T}\right\}\langle r^2(\cos^2\theta)e^{-r/l}\rangle_V, \qquad (16.1.9)$$

where

$$\langle ... \rangle_V = \frac{1}{V_s}\int ...d^3\vec{r}. \qquad (16.1.10)$$

In (16.1.9) the electric field is outside the averaging. Rigorously speaking, this means that the characteristic hopping length $l$ is larger than the interdroplet distance $n^{-1/3}$ and our approach is valid only when the droplet concentration is not too small. Substituting (16.1.9) into (16.1.5) and performing the integration, we find:

$$j_{1,2} = \frac{32\pi e^2 E\omega_0 l^5 n_{1,2}^2}{k_B T}\exp\left\{\mp\frac{A}{2k_B T}\right\}. \qquad (16.1.11)$$

In processes (iii) and (iv) the free energy of the system is not changed after the tunneling, and we write the characteristic times as:

$$\tau_{3,4} = \omega_0^{-1}\exp\left\{\frac{r}{l} - \frac{eEr\cos\theta}{k_B T}\right\}. \qquad (16.1.12)$$

The contribution of these two processes to the current is calculated similarly to that of (i) and (ii). For process (iii) the number of magnetic polarons from which the electron may tunnel is $N_2$, whereas the number of accepting droplets is $N_1$. In the same way, for process (iv) these numbers are $N_1$ and $N_3 = N_2$, respectively. Consequently, the factors $n_{1,2}^2$ in (16.1.11) are replaced by $n_1 n_2$,

$$j_{3,4} = \frac{32\pi e^2 E\omega_0 l^5 n_1 n_2}{k_B T}. \qquad (16.1.13)$$

From (16.1.11) and (16.1.13) we now obtain the dc conductivity $\sigma = j/E$:

$$\sigma = \frac{32\pi e^2 \omega_0 l^5}{k_B T}\left(2n_1 n_2 + n_1^2 e^{-A/2k_B T} + n_2^2 e^{A/2k_B T}\right) \qquad (16.1.14)$$

In this subsection we are only interested in the average conductivity; fluctuations lead to the appearance of noise and are considered in subsection 16.1.5. Using (16.1.4), we find that all four processes illustrated in Fig.16.1 give identical contributions to the conductivity; for $A \gg k_B T$ the average conductivity (for which we retain the notation $\sigma$) reads:

$$\sigma = \frac{128\pi e^2 n^2 \omega_0 l^5}{k_B T} \exp(-A/2k_B T). \qquad (16.1.15)$$

We see that the conductivity increases with temperature as $\sigma(T) \sim T^{-1}\exp(-A/2k_B T)$, which is typical for tunneling systems (see, e.g., [16.29]).

At this point, let us discuss the applicability range of our model. The essence of our picture is the existence of different types of droplets. Only single-electron droplets are stable. Obviously, an empty droplet decays during the time of the order of $1/\omega_0$. On the other hand, following the above discussion, the empty droplet should acquire an electron from neighboring one-electron or two-electron droplets during the characteristic time $\tau_0$, which can be easily calculated based on the following considerations. The probability $P$ per unit time for an empty droplet to acquire one electron can be written as:

$$P = 4\pi\omega_0 \int_0^\infty e^{-r/l}\left[n_1 + n_2 \exp(A/k_B T)\right]r^2 dr, \qquad (16.1.16)$$

where the terms with $n_1$ and $n_2$ correspond to the electron transfer from single- and two-electron droplets, respectively. Performing integration in (16.1.16) and using (16.1.4), we find

$$\tau = \frac{1}{P} = \frac{\exp(-A/2k_B T)}{8\pi\omega_0 l^3 n}. \qquad (16.1.17)$$

Just the same estimate can be obtained for the characteristic time of electrons leaving two-electron droplets. For our picture with empty and two-electron droplets to be valid, the following condition must be met: $\tau_0 << \omega_0^{-1}$. Thus, our approach is valid at sufficiently low temperatures, $k_B T << A$, and for a not too small droplet density $n$.

The applicability of our approach also implies that $l > R$, $n^{-1/3}$. It is of interest to consider also the case of $l \sim R$ and/or low droplet concentrations. In this situation, in usual hopping systems, the conductivity strongly depends on the geometry of current paths [16.9]. This causes an exponential dependence of conductivity on the carrier concentration. However, our system turns out to be more complicated than those commonly invoked for hopping conductivity. It involves different types of hopping centers giving rise to an unusual geometry of current paths. Therefore, the conventional approaches used for hopping cannot be applied straightforwardly to the analysis of our model at low droplet concentration or at $l \sim R$. Despite these complications, we believe that the expression for the conductivity in the case $l \leq n^{-1/3}$ includes the percolation-related factor $\exp\{-\gamma/(n^{1/3} R)\}$, with $\gamma$ of the order one [16.9], though currently we have no rigorous proof of this statement.

The results below for magnetoresistance and noise are insensitive to this factor, and therefore we expect them to be valid in a general case.

### 16.1.3. Tunneling magnetoresistance in the phase-separated manganites.

As we already discussed, below the percolation threshold when the volume fraction of droplets $n < n_C$, a typical value of $A/k_B$ is mainly determined by Coulomb interaction between two electrons inside the droplet $A \sim e^2/\varepsilon R$ and has a typical value of 1000 K. Now we can use this estimate to analyze the magnetoresistance in non-metallic phase-separated manganites. To do that, we use the expression for the radius of the magnetic polaron, obtained in the Chapter 17 $R \sim d(t/JS^2)^{1/5}$. Recall once more that here $J \sim 100$ K is an AFM Heisenberg exchange between the local spins $S = 3/2$. It is natural to conclude that in the magnetic field the Heisenberg exchange integral $J$ decreases according to the formula $J(H)S^2 = J(0)S'^2 - g\mu_B HS$, where $\mu_B$ and $g$ are the Bohr magneton and the gyromagnetic ratio, respectively. Consequently, the value of $A$ is decreasing linearly in the experimentally accessible range of magnetic fields, and for the excitation energy we obtain

$$A(H) = A(0)\left[1 - bH\right]; \quad b = \frac{1}{5}\frac{g\mu_B}{J(0)S} \qquad (16.1.18)$$

It follows now from (16.1.15) that the magnetoresistance is negative and for temperatures $T < A/k_B$ reads:

$$|MR| = \frac{\rho(0) - \rho(H)}{\rho(H)} = \exp\left(\frac{A(0) - A(H)}{2k_B T}\right) - 1 = \exp\left(\frac{bHA}{2k_B T}\right) - 1. \qquad (16.1.19)$$

For low magnetic fields and not very small temperatures the absolute value of the magnetoresistance (MR) is small, $|MR| = bHA/2k_B T << 1$. In higher fields (but still $bH << 1$) the absolute value of magnetoresistance eventually exceeds 1 and behaves in exponential fashion, $|MR| = \exp\{bHA/2k_B T\}$. Note that for temperatures $T \leq A/k_B$ and for typical gyromagnetic ratios $g \sim 10$ the magnetoresistance in our region of doping becomes larger than 1 by absolute value only in relatively high magnetic fields $H \sim 10$ T.

To describe the actual experimental situation (especially at lower fields) in more detail, we need to take into account other important physical mechanisms, in particular, spin-dependent tunneling [16.11, 16.13, 16.15, 16.19, 16.20].

Effects of spin-assistant tunneling yields nontrivial preexponential factor for the magnetoresistance. The details of the calculations which are rather straightforward you can find in [16.11]. Here we present the brief sketch of the evaluation of MR(see [16.15]). The most important is that probability of tunneling depends on the mutual orientation of the electron spin and the magnetic moment of the droplet (see Fig.16.2). Orientation of the ferromagnetically regions in the external magnetic field $H$ leads to an increase in the transition probability and, hence, to a decrease in resistance with increasing field strength - in agreements with experiment. The conductivity of the system can be represented as $\sigma(H) = \sigma(0)<\Sigma(H)>$, where $\Sigma(H)$ is the "spin" contribution to the probability of electron tunneling. For this definition, MR = $<\Sigma(H)>$ - 1.

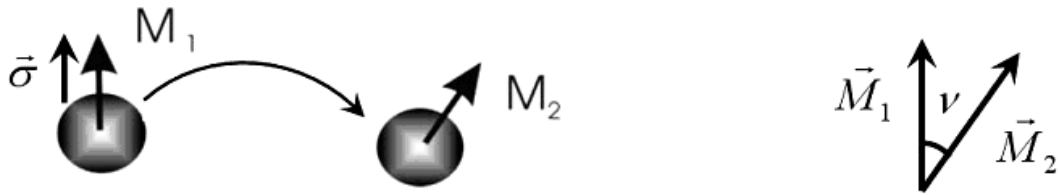

Fig.16.2. Spin-assistant tunneling of electron between two FM-droplets with magnetic moments $\vec{M}_1$ and $\vec{M}_2$. Angle $\nu$ is between $\vec{M}_1$ and $\vec{M}_2$.

Denoting the effective magnetic moment of the droplet by $M = \mu_B\, g N_{\text{eff}}\, S$ ($N_{\text{eff}}$ is a number of local spins in the droplet) and assuming the interaction between the droplets to be negligibly small, we write the Free-energy of the droplet in magnetic field in the following form [16.30, 16.31]:

$$U(H) = U(0) - M\,(H\cos\theta + H_a\cos^2\psi), \qquad (16.1.20)$$

where $\theta$ is the angle between the applied field $\vec{H}$ and the magnetic moment $\vec{M}$, $H_a$ is the anisotropy field, and $\psi$ is the angle between the anisotropy axis and the direction of the magnetic moment (for the sake of simplicity we consider the case of uniaxial anisotropy). Let $\vec{H}$ be parallel to $z$-axis, and let the anisotropy axis lie in the $(xz)$ plane and make the angle $\beta$ with vector $\vec{H}$. In this configuration (see Fig.16.3):

$$\cos\psi = \sin\theta\sin\beta\cos\varphi + \cos\theta\cos\beta, \qquad (16.1.21)$$

where $\varphi$ is the angle between the $x$-axis and the projection of $\vec{M}$ into $(xy)$ plane (see Fig.16.3).

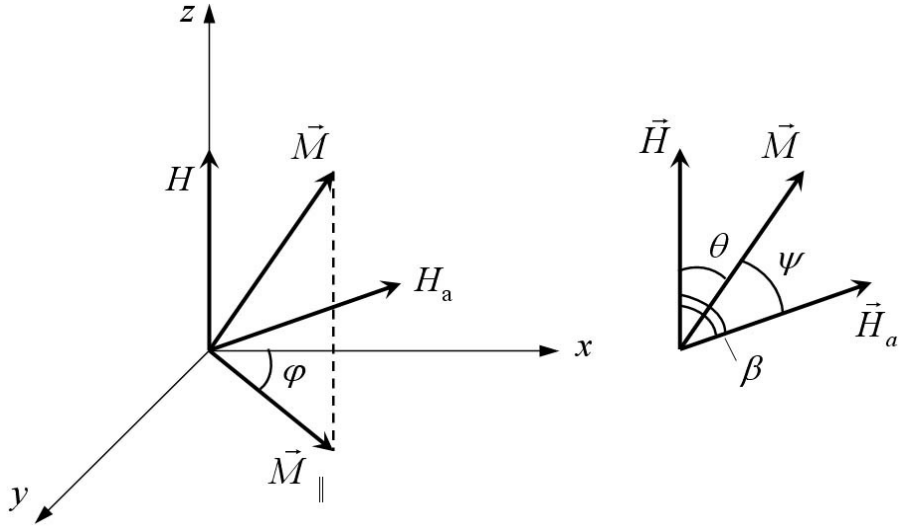

Fig.16.3. The mutual orientation of the magnetic field $\vec{H}$, anisotropy field $\vec{H}_a$ and magnetic moment of a droplet $\vec{M}$. Angle $\theta$ is between $\vec{H}$ and $\vec{M}$, angle $\psi$ is between $\vec{M}$ and $\vec{H}_a$, angle $\beta$ is between $\vec{H}$ and $\vec{H}_a$, angle $\varphi$ is between $x$-axis and projection of $\vec{M}$ on $(xy)$ plane. Magnetic field $\vec{H}$ is parallel to $z$-axis. Anisotropy field $\vec{H}_a$ lies in the $(xz)$ plane.

In the classical limit, a given orientation of vector $\vec{M}$ corresponds to the probability [16.27, 16.30]:

$$P(H, \theta, \varphi) = B(H)\exp\left[\frac{M\left(H\cos\theta + H_a\cos^2\psi(\theta,\varphi)\right)}{k_B T}\right], \qquad (16.1.22)$$

where $B(H)$ is the normalization factor. The eigenstates of an electron correspond to conservation of the spin-projection $\sigma = \pm \frac{1}{2}$ onto the effective field direction in a ferromagnetically correlated region. Let an electron interact with $Z$ magnetic moments of local spins $S$ in the droplet. The energy of this interaction is $E_\sigma = -J_{FM}SZ\sigma$ (where $J_{FM}$ is an effective ferromagnetic exchange interaction). Since the product $J_{FM}SZ$ is of the order of the Curie temperature ($T_C \sim 100$ K see the subsection 16.3.4), $E_\sigma$ is much greater than the energy of interaction between the electron spin and the magnetic field, provided that $H \ll 100$ Tesla. In this case, the effective field direction coincides with the direction of vector $\vec{M}$ and the probability for the electron spin projection to be $\sigma$ can be written as:

$$P_\sigma = \frac{\exp(-E_\sigma/k_B T)}{2ch(E_\sigma/k_B T)} \qquad (16.1.23)$$

Upon transfer from droplet 1 to droplet 2, an electron occurs in an effective field making an angle $\nu$ (see Fig.16.2) with that in the initial state, for which:

$$\cos\nu = \cos\theta_1\cos\theta_2 + \sin\theta_1\sin\theta_2\cos(\varphi_1 - \varphi_1) \qquad (16.1.24)$$

(indices 1 and 2 refer to the droplet number). Then, the work performed for the electron transfer from droplet 1 to droplet 2 is $\Delta E_\sigma = E_\sigma(1 - \cos\nu)$. Accordingly, the probability of this transfer is proportional to $\exp(-\Delta E_\sigma/k_B T)$. Taking into account all the probability factors introduced above, the final expression can be written as:

$$\langle \Sigma(H) \rangle = \int_0^{2\pi} d\varphi_1 \int_0^{2\pi} d\varphi_2 \int_0^{\pi} \sin\theta_1 d\theta_1 \int_0^{\pi} \sin\theta_2 d\theta_2 P(\theta_1\varphi_1)P(\theta_2\varphi_2) \sum_{\sigma=\pm 1/2} P_\sigma \exp\left(-\frac{\Delta E_\sigma}{k_B T}\right) \quad (16.1.25)$$

In the high-temperature range, where $k_B T$ is much greater compared to the Zeeman energy $\mu_B g S N_{eff} H$ ($N_{eff}$ is a number of local spins in the droplet) and the magnetic anisotropy energy $\mu_B gSN_{eff}H_a$, relations (16.1.22)–(16.1.25) yield in the strongly anisotropic case and low fields ($H \ll 10$ T):

$$|MR| = \frac{H^2 H_a}{T^5} \qquad (16.1.26)$$

and in the absence of anisotropy (for $H_a = 0$):

$$|MR| = \frac{H^2}{T^2} \qquad (16.1.27)$$

At higher fields magnetoresistance behaves as:

$$|MR| = \frac{H^2}{T} \qquad (16.1.28)$$

then goes on a plateau.

Thus for $H \leq 10$ T an absolute value of magnetoresistance behaves quadratic in field but temperature dependence of $|MR| = H^2/T^n$ ($n$ running from 1 to 5) is highly nontrivial. Finally in high fields $H \geq 10$ T it grows in exponential fashion with field (see Fig.16.4) and (16.1.19).

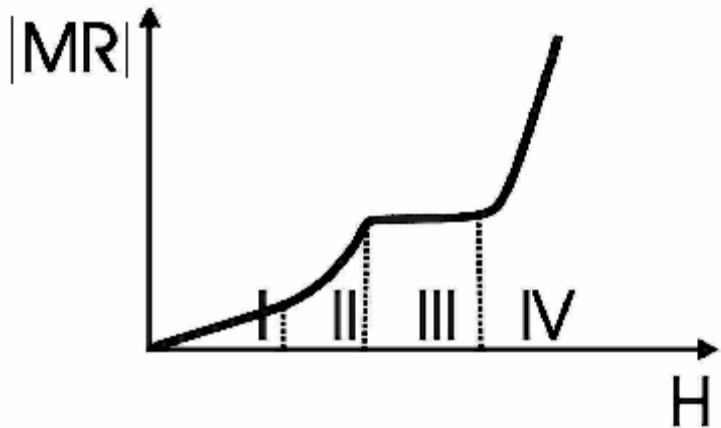

Fig.16.4. Absolute value of magnetoresistance |MR| in magnetic fields. The different regions I, II, III correspond to quadratic in field behavior $|MR| = H^2/T^n$ with $n = 2$ or $5$ in region I, $n = 1$ in region II, $n = 0$ (plateau) in region III. In region IV |MR| grows in exponential fashion with field.

The low-field regime I for |MR| (see (16.1.26) and Fig. 16.4) is confirmed by experimental observations of several groups for 6 families of 3D (cubic) and layered manganites [16.21-16.24]. We will discuss these results more detaily in the section 16.3.

16.2. $1/f$ – noise power spectrum.

Almost ten years ago Podzorov et al. [16.26] reported the observation of giant $1/f$ noise in perovskite manganites in the phase-separated regime. Generally, systems with distributed hopping lengths are standard objects that exhibit $1/f$ noise (for review, see [16.17, 16.18]). The purpose of this section is to study low-frequency noise within the framework of the model used to calculate the conductivity in subsection 16.1.3 and show that it has, indeed, $1/f$ form.

Starting from the Ohm's law $U = IL/\sigma S$ [16.1] (where $L$ and $S$ are the sample length and the cross-section, respectively) and assuming that the measuring circuit is stabilized ($I = const$), we can present the voltage noise at the frequency $\omega$, $<\delta U^2>_\omega$, in the following way:

$$\left\langle \delta U^2 \right\rangle_\omega = U_{dc}^2 \frac{\left\langle \delta \sigma^2 \right\rangle_\omega}{\sigma^2}, \qquad (16.2.1)$$

where $U_{dc}$ is the time-averaged voltage and $<\delta \sigma^2>_\omega$ is the noise spectrum of the fluctuations of the conductivity.

If we disregard possible fluctuations of temperature in the system, the only source of the fluctuations in our model is those of the occupation numbers $n_1$ and $n_2$. Using the conservation law $n_1 + 2n_2 = n$, we find from (16.1.14):

$$\delta\sigma = \sigma \frac{\delta n_2}{\bar{n}_2}\left[1 - 2\exp\left(-A/2k_B T\right)\right].$$  (16.2.2)

We thus need to find the fluctuation spectrum $<\delta n_2{}^2>_\omega$. Following the general prescription (see [16.30]), we recollect that the two-electron droplets decay via process (ii) on Fig.16.1 and the relaxation equation has the form:

$$\delta\dot{n}_2 = -\frac{\delta n_2}{\tau(r)}; \; \tau(r) = \omega_0^{-1}\exp\left(r/l - A/2k_B T\right).$$  (16.2.3)

where we have neglected the effect of the electric field. The fluctuation spectrum then reads [16.30]:

$$\left\langle \delta n_2{}^2 \right\rangle_\omega = \left\langle \delta n_2{}^2 \right\rangle_T \left\langle \sum_i \frac{2\tau(r^i)}{1 + \omega^2\tau^2(r^i)} \right\rangle,$$  (16.2.4)

where $<\delta n_2{}^2>_T$ is the thermal average of the variation of $n_2$ squared, and the summation is performed over the "empty droplet – two-electron droplet" pairs, with $r^i$ being the distance between the sites in a pair. Since all such pairs contribute to the noise, the average in (16.2.4) is essentially a spatial integral, with the main contribution coming from short distances,

$$\left\langle \delta n_2{}^2 \right\rangle_\omega = 8\pi\bar{n}_2 \left\langle \delta n_2{}^2 \right\rangle_T \int_0^\infty \frac{\tau(r)}{1 + \omega^2\tau^2(r)}r^2 dr.$$  (16.2.5)

Note that (16.2.5) is valid for an arbitrary relation between $l$ and $R$, not necessarily for $R \ll l$.

We are interested below in the frequency range

$$\tilde{\omega}_0 \exp\left(-L_s/l\right) \ll \omega \ll \tilde{\omega}_0, \quad \tilde{\omega}_0 = \omega_0\exp\left(A/2k_B T\right),$$  (16.2.6)

where $L_s$ is the smallest of the sample sizes. In this case, with the logarithmic accuracy we obtain for $A \gg k_B T$,

$$\left\langle \delta U^2 \right\rangle_\omega = U_{dc}^2 \frac{\left\langle \delta n_2{}^2 \right\rangle_T}{\bar{n}_2} \frac{4\pi^2 l^3}{\omega}\ln^2\frac{\tilde{\omega}_0}{\omega}.$$  (16.2.7)

Thus, in the wide range of sufficiently low frequencies (16.2.6), the noise power spectrum for our system has almost a $1/f$ form.

The variation $\left\langle \delta n_2{}^2 \right\rangle_T = V_s^{-2}\left(\overline{N_2^2} - \left(\overline{N_2}\right)^2\right)$ is easily found in the same way as (16.1.4),

$$\left\langle \delta n_2{}^2 \right\rangle_T = \frac{\bar{n}_2}{2V_s}.$$  (16.2.8)

Combining this with (16.2.7) we write the final expression for the spectral density of noise for $A \gg k_B T$ in the form:

$$\left\langle \delta U^2 \right\rangle_\omega = U_{dc}^2 \frac{2\pi^2 l^3}{V_s\omega}\ln^2\left(\frac{\omega_0 e^{A/2k_B T}}{\omega}\right).$$  (16.2.9)

Discussion. Comparison with experiments.

For the further discussion, it is convenient to rewrite (16.2.9) in the form:

$$\alpha_H = \frac{\left\langle \delta U^2 \right\rangle_\omega V_s \omega}{U_{dc}^2} = 2\pi^2 l^3 \ln^2\left(\frac{\tilde{\omega}_0}{\omega}\right) \quad -$$  (16.2.10)

is a Hooge constant [16.17,16.18].

It is remarkable that the noise spectrum in our model has a $1/f$ form up to very low frequencies. This is due to fluctuations in occupation numbers of droplets, associated with the creation and annihilation of extra electron-hole pairs. This mechanism of $1/f$ noise is specific for our model and is not present in standard hopping conduction.

Let us estimate the numerical value of the parameter $\alpha_H$, which is the standard measure of the strength of $1/f$ noise. This parameter is proportional to the third power of $l$. Simple estimates (analogous to that presented in the Chapter 17 for ferron radius $R$) reveal that, in general, $l$ is of the order or larger than $R$. Assuming again that the excitation energy is of the order of the Coulomb energy $A \sim e^2/\varepsilon R$, taking $\omega_0$ to be of the order of the Fermi-energy inside droplets (which means $\hbar\omega_0 \sim 300$ K for $n < n_c$), and estimating the tunneling length $l$ as being $l \geq 2 R = 20$ Å, we arrive at the conclusion that the parameter $\alpha_H$ is of the order $\alpha_H \sim (10^{-17} \div 10^{-16})$ cm$^3$ for $T < A/k_B$ ($T \sim 100 \div 200$ K) and $\omega \sim 1$ Hz $\div 1$ MHz. This value of $\alpha_H$ is by several orders of magnitude higher than that in the usual semiconducting materials [16.18]. Such a large magnitude of the noise can be attributed to the relatively low height of the potential barrier $A$ and to the relatively large tunneling length $l$. Formally, it is also related to the large value of the logarithm squared in (16.2.10).

According to (16.2.9) and (16.2.10), the noise power and the noise parameter (a Hooge constant) $\alpha_H$ are independent of the volume fraction occupied by the droplets. This result is valid in the intermediate range of $n$, when the droplet density is not too high and not too low. First, we assumed that the droplets are isolated point objects and that the tunneling between the two droplets is not affected by a third polaron. This is only valid provided the droplet density is far from the percolation threshold, $n << n_c$. On the other hand, the droplet density must not be too low since the conditions $N, N_1, N_2 >> 1$ are assumed to be met. Moreover, we neglected the possibilities of the disappearance of a droplet without an electron, the formation of a new droplet due to the electron tunneling, and the decay of two-electron droplets. Thus, the characteristic times of these processes should be longer than the characteristic tunneling time, and the average tunneling distance cannot be too high (see (16.1.17) and the discussion below it).

The above speculations imply that the following set of inequalities should be met, $R << n^{-1/3} << l$, for formula (16.1.15) for the conductivity to be valid. In general, the tunneling length should not be much larger than the droplet radius since just the same physical parameters determine these two characteristic distances. So, these inequalities could not be valid for real physical systems, and it is of interest to consider the situation where $R, l << n^{-1/3}$, which is beyond the scope of our model. However, some definite conclusions concerning the magnetoresistance and the noise power can be made at present.

First, the factor $\exp(A/2k_BT)$ in the temperature dependence of the conductivity is related to the number of carriers and appears due to the strong Coulomb repulsion of electrons in the droplet. It seems rather obvious that such a factor appears in the formula for the conductivity below the percolation threshold for an arbitrary relation between $R$ and $l$. On the other hand, in contrast to common hopping systems, a strong $1/f$-noise in our model results from fluctuations of state occupation numbers. Actually, our result for parameter $\alpha_H$ (16.2.10), only relies on the fact that $\delta\sigma/\sigma \sim \delta n_2/\bar{n}_2$. As we have mentioned previously, (16.2.5), which determines the spectral density of fluctuations of $n_2$, applies for an arbitrary relationship between $R$ and $l$. It follows then that the value of the parameter $\alpha_H$ for $1/f$-noise remains approximately the same under the (experimentally relevant) conditions $R \sim l$.

Another important point is that we disregard the direct Coulomb interaction between the droplets in comparison with the energy $A$. This is justified if the gas of the droplets is diluted, $n^{-1/3} >> R$. In this respect, we recollect that in standard hopping conduction systems (doped semiconductors) the main mechanism of low-frequency noise is an exchange of electrons between the infinite cluster and nearby finite clusters. In the absence of interactions it leads to the noise power proportional to $\omega^{-\delta}$, with the exponent $\delta$ being considerably below 1 [16.32] To explain 1/*f* noise in these systems, models involving Coulomb interactions were proposed [16.33, 16.34]. These sources of low-frequency noise are thus beyond our discussion. We also did not consider sources of noise different from resistance fluctuations. At least two other types of noise are inevitably present in the system: Nyquist-Johnson (thermal) noise [16.2], which is a consequence of fluctuation-dissipation theorem, and shot noise due to the discrete nature of electron charge (see [16.17] and [16.35] for review). Both these noises are frequency independent (white) at low frequencies. The magnitudes of Nyquist-Johnson, shot, and 1/*f* noises are governed by absolutely different parameters, and we do not attempt to compare them here, noting only that at low frequencies 1/*f* noise must dominate.

In our model, we assumed that the number of droplets *N* is fixed and strictly equal to the number of extra electrons. In actual systems, *N* can also fluctuate, and this can be an additional source of noise and of 1/*f* noise, in particular. However, this contribution depends critically on the heights of corresponding energy barriers and can vary for different systems.

As we have already mentioned, the main motivation of our work was the experimental study [16.26], which observed high 1/*f* noise power at high temperatures far from the metal-insulator transition. In the same experiment, the noise dropped to much lower levels at low temperatures in the metallic phase. This behavior of the noise power is consistent with the present model since in the metallic phase the electron tunneling contribution to the total conductivity is negligible. In the vicinity of the percolation transition the noise power increases drastically [16.26]. In this Chapter we do not attempt to describe the system of magnetic polarons close to the percolation threshold. However, we argue that the amplitude of 1/*f* noise is already large in the phase-separated regime even far from the percolation threshold.

### 16.3. Experimental confirmation of the theoretical predictions for tunneling conductivity.

In subsection 16.1.3 we present a simple model for tunneling conductivity. We consider droplets containing one or two electrons together with empty droplets. The direct generalization of (16.1.15) for the case of the droplets with $k$, $k + 1$ and $k - 1$ carriers [16.20] yields for the resistivity:

$$\rho = \frac{k_B T \exp\left(A/2k_B T\right)}{128\pi e^2 \, \omega_0 l^5 k n^2},\qquad(16.3.1)$$

$n$ is the concentration of ferromagnetic droplets. Electrical resistivity (16.3.1) exhibit a thermoactive behavior where the activation energy is equal to one half of Coulomb energy $A$. Expression (16.3.1) provides a fairly good description for the temperature dependence of the electrical resistivity for various manganites. As an illustration, in Figs. 16.5-16.8 we present experimental $\rho(T)$ curves for six different materials. Experimental data are plotted for samples reported in [16.21] by Babushkina et al., [16.22] by Fisher et al., [16.23] by Zhao et al., [16.24] by Wagner et al. The authors of these papers kindly provided us by the detailed numerical data on their measurements. As it could be seen from the figures and their captions, the examined samples differ in their chemical composition, type of crystal structure, magnitude of electrical resistivity (at fixed temperature, the latter varies for different samples by more than two orders of magnitude), and also by their low temperature behavior (which is metallic for some samples and insulating for the others). On the other hand, in the high-temperature range (above the point of ferromagnetic phase transition), $\rho(T)$ exhibits a similar behavior for all the samples, which is well fitted by the relationship $\rho(T) \sim T \exp(A/2k_B T)$ (solid lines in the Figs. 16.5-16.8).

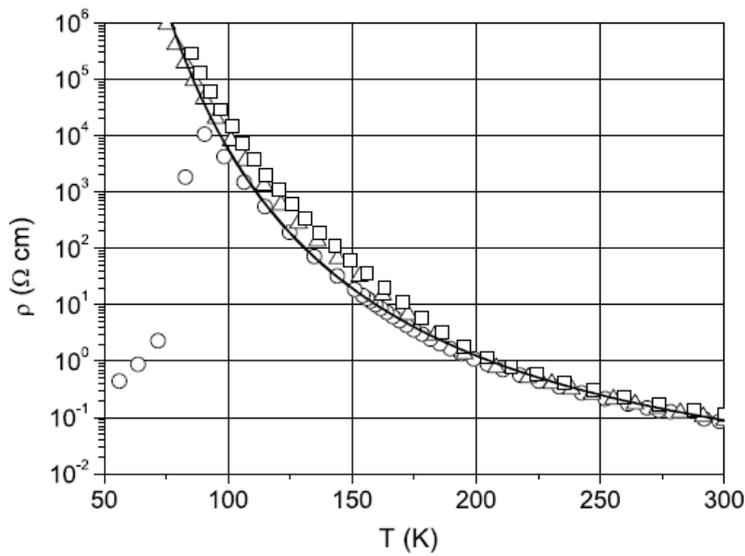

Fig. 16.5. Temperature dependence of the resistivity for $(La_{1-y}Pr_y)_{0.7}Ca_{0.3}MnO_3$ samples (Babushkina et al., 2003 [16.21]). Squares, triangles, and circles correspond to $y = 1$ (with $^{16}O \rightarrow$ $^{18}O$ isotope substitution), $y = 0.75$ (with $^{16}O \rightarrow$ $^{18}O$ isotope substitution), and $y = 0.75$ (with $^{16}O$), respectively. Solid line is the fit based on (16.3.1).

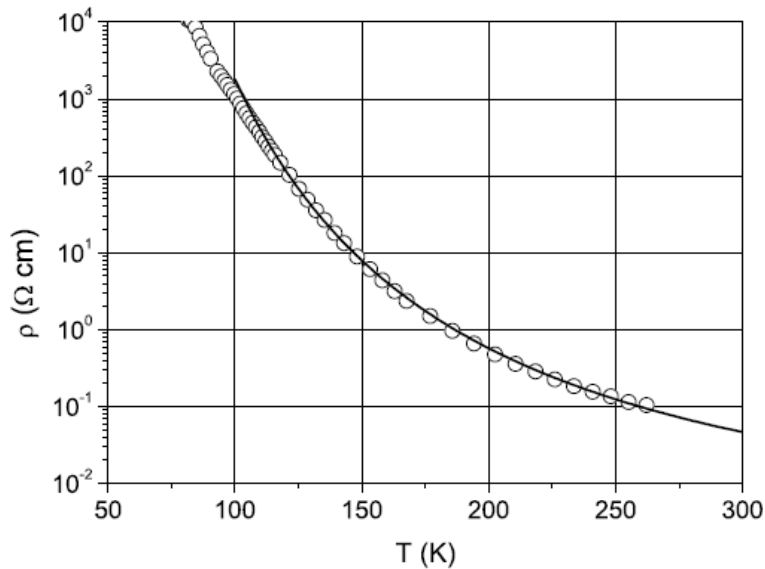

Fig. 16.6. Temperature dependence of the resistivity for $Pr_{0.71}Ca_{0.29}MnO_3$ sample (Fisher et al., 2003 [16.22]): experimental data (circles) and theoretical curve (solid line) based on (16.3.1).

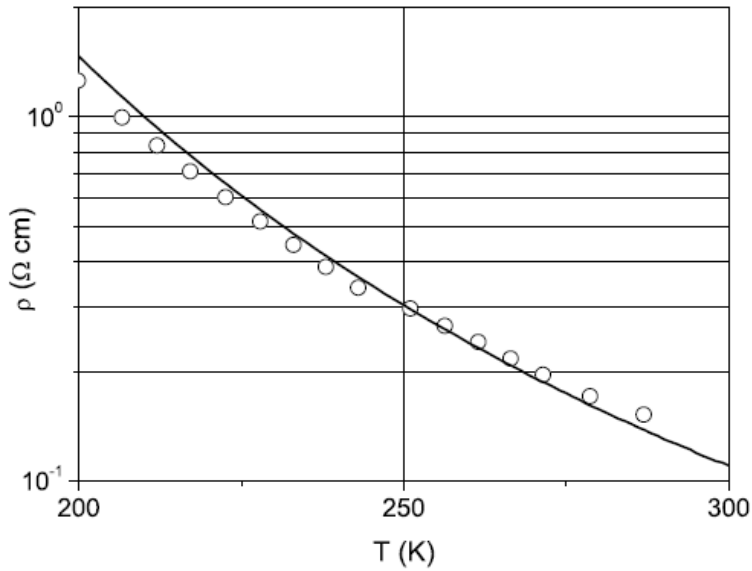

Fig. 16.7. Temperature dependence of the resistivity for a layered manganite $(La_{0.4}Pr_{0.6})_{1.2}Sr_{1.8}Mn_2O_7$ (Wagner et al., 2002 [16.24]): experimental data (circles) and theoretical curve (solid line) based on (16.3.1).

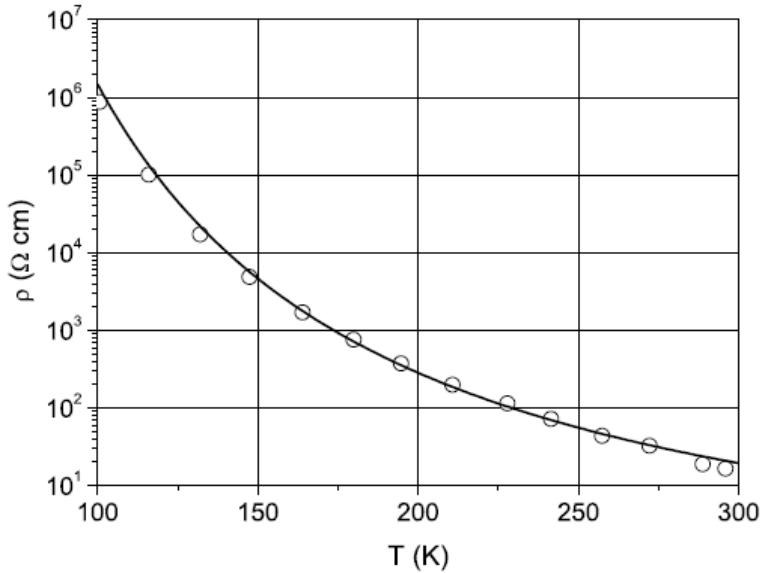

Fig. 16.8. Temperature dependence of the resistivity for $La_{0.8}Mg_{0.2}MnO_3$ sample (Zhao et al., 2001 [16.23]): experimental data (circles) and theoretical curve (solid line) based on (16.3.1).

Based on (16.3.1) and experimental data, one can deduce some quantitative characteristics of the phase-separated state. In particular, the analysis carried out in [16.23] demonstrated that an accurate estimate for the value of Coulomb energy $A$ can be found by fitting experimental data and using (16.3.1). The data represented in Figs. 16.5-16.8 suggest that the Coulomb barrier $A$ can be determined with an accuracy of 2-3% and its value lies in the narrow range from 3500 to 3700K (see Table 16.1). As it was mentioned in [16.10, 16.23] the characteristic frequency $\omega_0$ in (1) can also vary in a restricted range of $10^{13} \div 10^{14}$ Hz. This estimate might be derived, for example, from the uncertainty principle: $\hbar\omega_0 \sim \hbar^2/2mR^2$, where $R$ is a characteristic droplet size, and $m$ is the electron mass. Assuming $R \sim 1 - 2$ nm, one obtains the latter estimate. Note also that these values of a droplet size allow us to find an estimate for the barrier energy $A$, which is accurate within the order of magnitude. This energy is of the order of $e^2/\varepsilon R$, and substituting permittivity $\varepsilon \sim 10$, we get a value of $A$ consistent with the experimental data.

| Samples | $A$, K | $\rho$ (200K), $\Omega\cdot$cm | $l^5 n^2 k$, cm$^{-1}$ | Data source |
|---|---|---|---|---|
| $(La_{1-y}Pr_y)_{0.7}Ca_{0.3}MnO_3$ | 3650 | 1.25 | $2\cdot10^5$ | Fig. 16.5 (Babushkina et al., 2003) |
| $Pr_{0.71}Ca_{0.29}MnO_3$ | 3500 | 0.57 | $3\cdot10^5$ | Fig. 16.6 (Fisher et al., 2003) |
| $(La_{0.4}Pr_{0.6})_{1.2}Sr_{1.8}Mn_2O_7$ | 3600 | 1.5 | $1.5\cdot10^5$ | Fig. 16.7 (Wagner et al., 2002) |
| $La_{0.8}Mg_{0.2}MnO_3$ | 3700 | 283 | $1\cdot10^3$ | Fig. 16.8 (Zhao et al., 2001) |

Table 16.1.

It is rather difficult to estimate the tunneling length $l$. However, we can say that in the domain of the applicability of relationship (16.3.1), the length $l$ cannot be much smaller than an interdroplet spacing [16.10]. In another situation, the behavior of the resistivity would be different. In the quasiclassical approximation, the tunneling length is of the order of the characteristic size for the wave function provided the barrier height is comparable with the depth of the potential well. In our case, the size of the electron wave function is of the order of a ferron size, while the height of the barrier practically coincides with the depth of the potential well. The latter naturally follows from the model of ferron formation considered in Chapter 15. Therefore, it seems reasonable to assume the tunneling length to be of the same order as a ferron size (few nanometers), though, generally speaking, it can substantially differ from $R$.

It is rather nontrivial task to estimate the concentration $n$ of ferrons. In fact, following [16.23] concentration $n$ could be determined by the dopant concentration $x$ as $n \approx x/d^3$. Yet this approach would bring at least two contradictions. First, even under the moderate concentration of divalent element $x = 0.1 - 0.2$ the droplets should overlap giving rise to the continuous metallic and ferromagnetic cluster. However, the material could be insulating even at larger concentrations ($x = 0.5 \div 0.6$), at least, in a high-temperature range. Second, as it can be seen from the experimental data, the relation between a dopant concentration and the conductivity of manganites is relatively complicated - for some materials changing $x$ by a factor of two can change resistivity by two orders of magnitude [16.23], for other materials $\rho(x)$ exhibits even a nonmonotonic behavior in certain concentration ranges. Note that these discrepancies are essential not only for our model of phase-separation but also for other models dealing with the properties of manganites (e.g., polaronic models [16.36, 16.37]). Unfortunately, the authors of [16.23] do not take into account these considerations when analyzing their results from the standpoint of the existing theories of the conductivity in manganites. The natural conclusion is that the number of carriers, which contribute to the charge transfer processes does not coincide with the concentration of the divalent dopant $x$. This is particularly obvious in the case of charge ordering when some part of the carriers introduced by doping becomes localized and forms a regular structure.

Therefore, using expression (16.3.1) and experimental data, we are able to obtain also the value of the combination $l^5 n^2 k$. In Table 16.1, the values of Coulomb energy $A$, resistivity $\rho$ at 200 K and, combination $l^5 n^2 k$ are presented. All estimations were made based on (16.3.1) and the experimental data of Figs. 16.5 - 16.8. Note that whereas the accuracy of the estimate for $A$ is about $\pm$ 50 K, the combination $l^5 n^2 k$ could be estimated only by the order of magnitude (at least, due to the uncertainty in the values of frequency $\omega_0$).

16.3.1. Experiments on tunneling magnetoresistance.

In [16.21, 16.11, 16.12, 16.15] it was demonstrated that the model of phase separation considered here results in a rather specific dependence of the magnetoresistance MR on temperature and magnetic field. At relatively high temperatures and not very strong magnetic fields, the expression for the magnetoresistance reads [16.11, 16.12, 16.15](see also 16.1.26 and Fig. 16.4):

$$|MR| \approx 5 \cdot 10^{-3} \mu_B^3 S^5 N_{eff}^3 Z^2 g^3 J_{FM}^{2} \frac{H_a H^2}{(k_B T)^5},\qquad(16.3.2)$$

where $\mu_B$ is the Bohr magneton, $S$ is the average spin of a manganese ion, $N_{eff}$ is the number of manganese atoms in a droplet, $Z$ is the number of nearest neighbors of a manganese ion, $g$ is the Lande factor, $J_{FM}$ is an effective exchange integral of the ferromagnetic interaction, and $H_a$ is the effective field of magnetic anisotropy of a droplet. The $|MR| \sim H^2/T^5$ dependence was observed in the experiments for a number of manganites in the region of their non-metallic behavior (see [16.21, 16.22]). The same high-temperature behavior of the magnetoresistance can be obtained by processing the experimental data reported in [16.23, 16.24] (see Figs. 16.9 - 16.12).

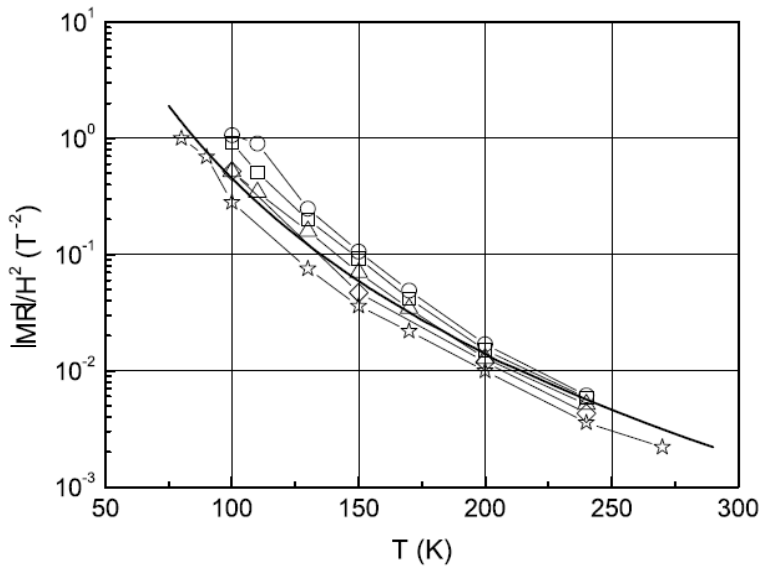

Fig. 16.9. Temperature dependence of $|MR|/H^2$ ratio for $(La_{1-y}Pr_y)_{0.7}Ca_{0.3}MnO_3$ samples (Babushkina et al., 2003 [16.21]). Squares, triangles, circles, diamonds, and asterisks correspond to $y = 0.75$, $y = 0.75$ (with 30% of $^{18}O$), $y = 0.75$ (with $^{16}O \rightarrow {}^{18}O$ isotope substitution), $y = 1$, and $y = 1$ (with $^{16}O \rightarrow {}^{18}O$ isotope substitution), respectively. Solid line is the fit based on (16.3.2) ($|MR| \sim 1/T^5$).

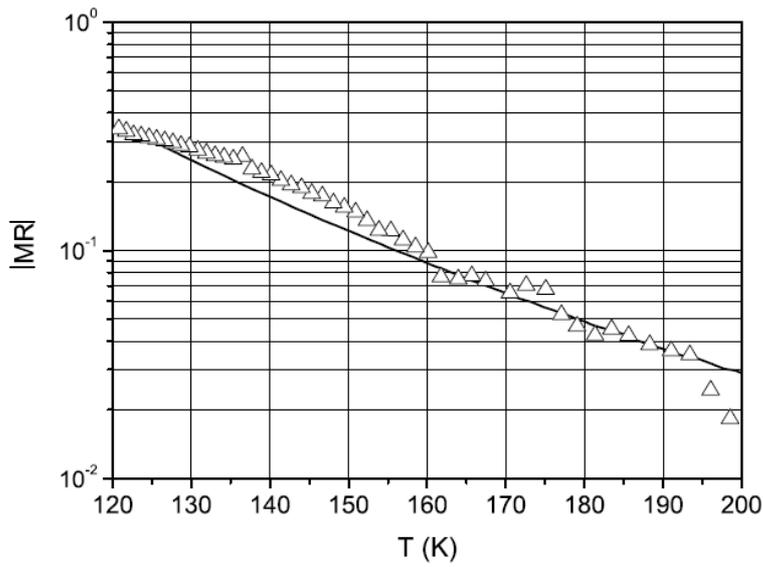

Fig. 16.10. Temperature dependence of the magnetoresistance for $Pr_{0.71}Ca_{0.29}MnO_3$ sample at $H$ = 2 T: experimental data (triangles) (Fisher et al., 2003 [16.22]) and theoretical curve (solid line) based on (16.3.2).

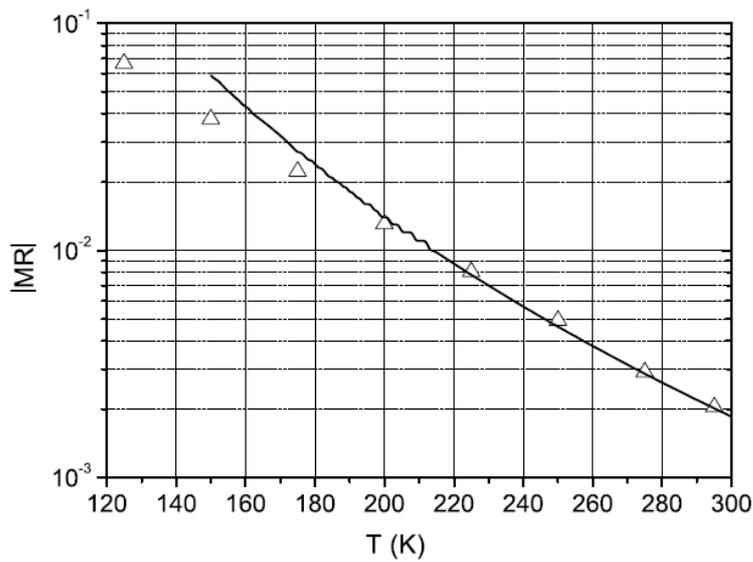

Fig. 16.11. Temperature dependence of the magnetoresistance for $(La_{0.4}Pr_{0.6})_{1.2}Sr_{1.8}Mn_2O_7$ sample (layered manganite) at $H$ = 1 T: experimental data (triangles) (Wagner et al., 2002 [16.24]) and theoretical curve (solid line) based on (16.3.2).

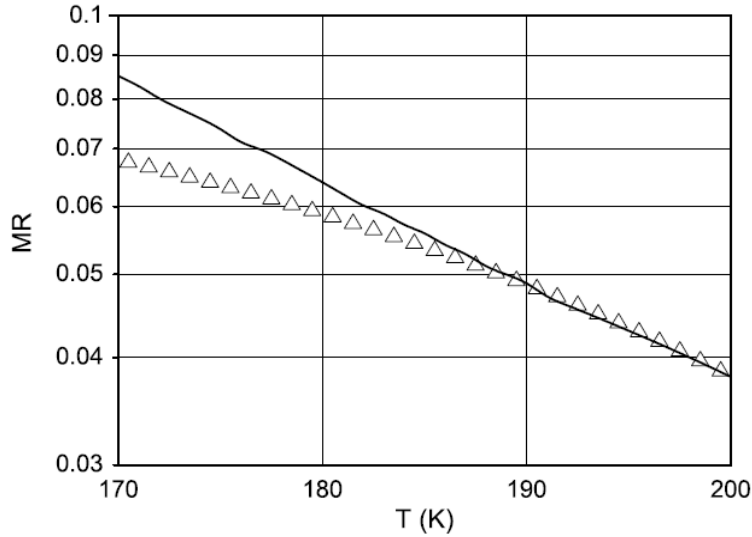

Fig. 16.12. Temperature dependence of the magnetoresistance for $La_{0.8}Mg_{0.2}MnO_3$ sample at $H = 1.5$ T: experimental data (triangles) (Zhao et al., 2001 [16.23]) and theoretical curve (solid line) based on (16.3.2).

The value of $S$ depends on the relative content of a trivalent and a tetravalent manganese ions and ranges from 3/2 to 2. Below it is assumed that $S = 2$ for all the estimations. Parameter $Z$ in (16.1.30) is, in fact, the number of manganese ions interacting with a conduction electron placed in a droplet. It is reasonable to assume that $Z$ is of the order of the number of nearest-neighbor sites around a manganese ion, i.e. $Z \approx 6$. The Lande factor g is determined from the experimental data. For manganese, $g$ is usually assumed to be close to its spin value 2. The exchange integral $J_{FM}$ characterizes the magnetic interaction between a conduction electron and the molecular field generated by ferromagnetically correlated spins in a droplet. It is this molecular field that produces a ferromagnetic state at low temperatures. Therefore, we can use a well-known relationship $S(S+1) Z J_{FM}/3 = k_B T_C$ of the molecular field theory [16.31] to evaluate the exchange integral (here $T_C$ is the Curie temperature). The value of $T_C$ is determined from the experiment (based on neutron diffraction or magnetization measurements). For example, in La-Pr-Ca manganites, it is about $100 - 120$ K [16.38].

The magnetic anisotropy of manganites related to crystal structure of these compounds is usually not too high. This implies that the main contribution to the effective field of a magnetic anisotropy $H_a$ stems from the shape anisotropy of a droplet and can be evaluated as $H_a = \pi(1 - 3\overline{N})M_S$, where $\overline{N}$ is the demagnetization factor [16.30] of the droplet (along the main axis), $M_S$ is the magnetic moment per unit volume ($M_S = M/V_S$) of the droplet. Below we assume a droplet to be sufficiently elongated $(\overline{N} \ll 1)$ and $M_S = Sg\mu_B/d^3$. Then $H_a \approx 2$ kOe.

The value of $N_{eff}$ is determined by the size of a droplet and it could be found from the neutron diffraction experiments. However, we are unaware of such measurements performed for the systems under discussion in a wide temperature range. Therefore, $N_{eff}$ is treated here as a fitting parameter. Hence, using (16.1.30) and the above estimates, we can determine the value of $N_{eff}$ from the experimental data on the magnetoresistance (in the range of parameters corresponding to $|MR| \sim H^2/T^5$). The results are summarized in Table 16.2. In Figs. 16.9 – 16.12, solid curves correspond to the fitting procedure based on (16.3.2). The value of $T_C$ was chosen to be equal to 120 K.

| Samples | $N_{eff}$ | $x$ | $k$ | Data source |
|---|---|---|---|---|
| $(La_{1-y}Pr_y)_{0.7}Ca_{0.3}MnO_3$ | 250 | 0.3 | 75 | Fig. 16.9 (Babushkina et al., 2003) |
| $Pr_{0.71}Ca_{0.29}MnO_3$ | 200 | 0.29 | 58 | Fig. 16.10 (Fisher et |
| | | | | al., 2003) |
| $(La_{0.4}Pr_{0.6})_{1.2}Sr_{1.8}Mn_2O_7$ | 250 | 0.4 | 100 | Fig. 16.11 (Wagner et al., 2002) |
| $La_{0.8}Mg_{0.2}MnO_3$ | 265 | 0.2 | 53 | Fig. 16.12 (Zhao et al., 2001) |

Table 16.2.

As a result, the size of the ferromagnetically correlated regions turns out to be nearly the same at temperatures about 200-300 K for all compositions under discussion. The volume of these regions is approximately equal to that of a ball with 7-8 lattice constants in diameter. It is natural to assume that within a droplet the number of charge carriers contributing to tunneling processes equals to the number of dopant atoms. Hence, we can write that entering in (16.1.29) $k = N_{eff} x$, where $x$ is the atomic percentage of dopants. The values of $x$ and $k$ are presented in Table 16.2.

16.3.2. Magnetic susceptibility.

The concentration of droplets can be evaluated based on the magnetic susceptibility data, if we assume that the dominant contribution to the susceptibility comes from the ferromagnetically correlated regions. At high temperatures ($k_B T \gg \mu_B g S N_{eff} H$, $\mu_B g S N_{eff} H_a$), susceptibility $\chi(T)$ can be written as

$$\chi(T) = \frac{n\left(\mu_B g S N_{eff}\right)^2}{3k_B\left(T - \theta\right)}, \qquad (16.3.3)$$

where $\theta$ is the Curie-Weiss constant [16.30, 16.31]. The results of the processing of the experimental data are presented in Table 16.3. In Figs. 16.13 - 16.16, the solid curves correspond to the fitting procedure based on (16.3.3). Using these results, we can also estimate the concentration of ferromagnetic phase as $p = nN_{eff}d^3$. For all the samples, the value of the lattice constant $d$ was taken to be equal to 3.9 Å. Based on the data of Tables 16.1 – 16.3, it is also possible to find an estimate for the tunneling length $l$.

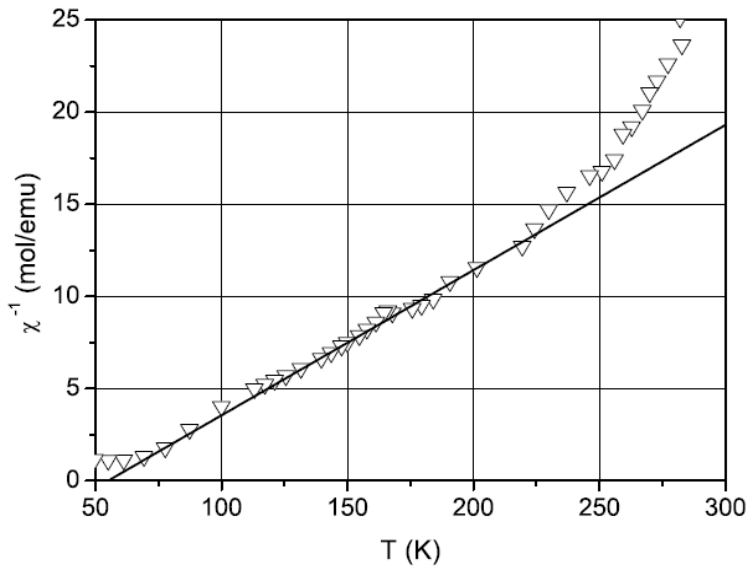

Fig. 16.13. Temperature dependence of the inverse magnetic susceptibility for $(La_{1-y}Pr_y)_{0.7}Ca_{0.3}MnO_3$ sample at $y = 1$: experimental data (triangles) (Babushkina et al., 2003 [16.21]) and theoretical curve (solid line) based on (16.3.3). For the other samples of this group,

the behavior of $\chi(T)$ at high temperatures is rather similar to that illustrated in this figure (see [16.21]).

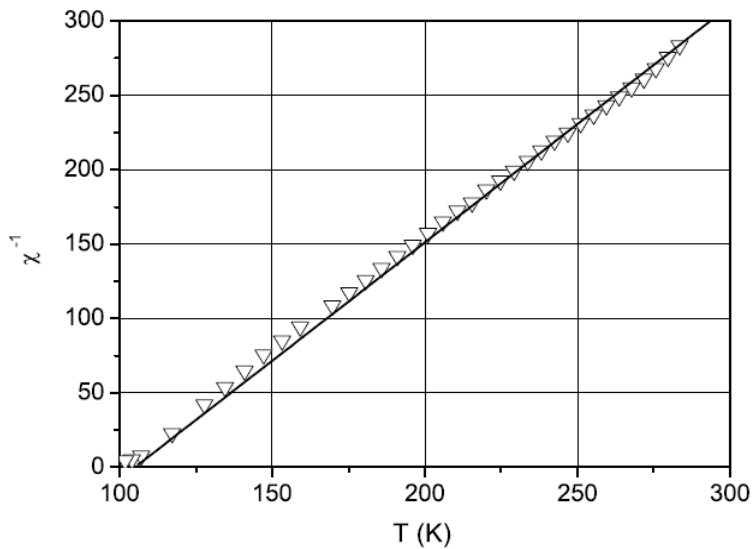

Fig. 16.14. Temperature dependence of the inverse magnetic susceptibility for $Pr_{0.71}Ca_{0.29}MnO_3$ sample: experimental data (triangles) (Fisher et al., 2003 [16.22]) and theoretical curve (solid line) based on (16.3.3). The sample was porous, its density was assumed to differ by a factor of 0.7 from the theoretical value.

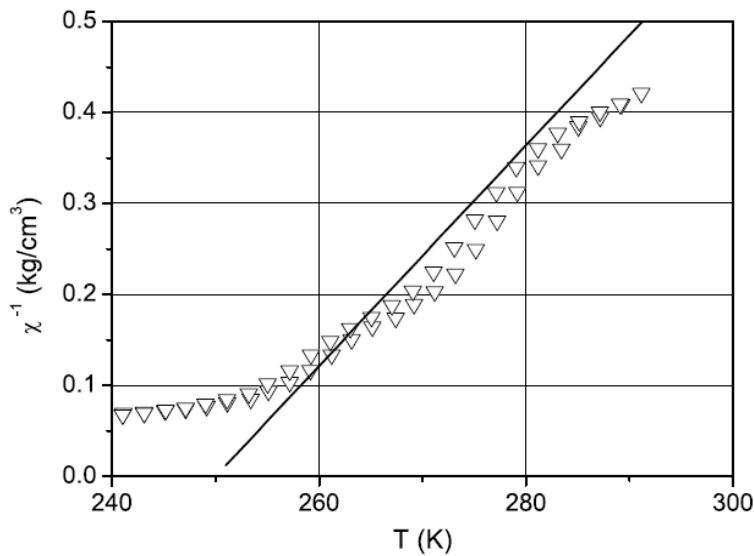

Fig. 16.15. Temperature dependence of the inverse magnetic susceptibility for the sample of $(La_{0.4}Pr_{0.6})_{1.2}Sr_{1.8}Mn_2O_7$ layered manganite: experimental data (triangles) (Wagner et al., 2002 [16.24]) and theoretical curve (solid line) based on (16.3.3).

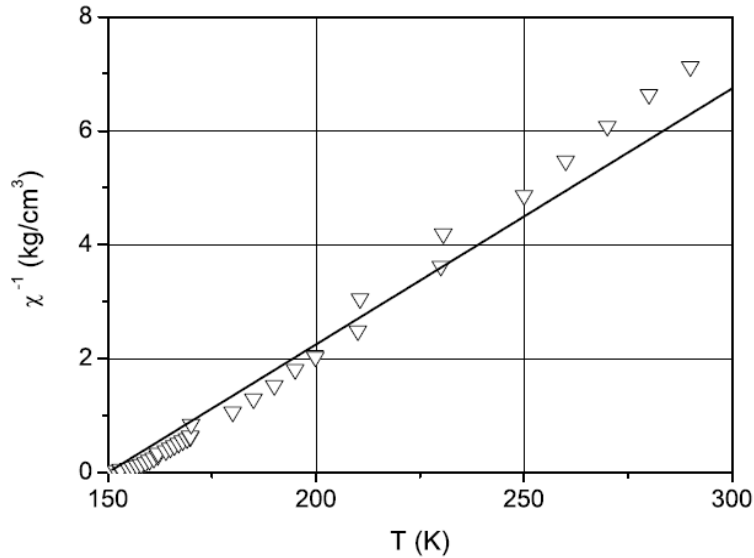

Fig. 16.16. Temperature dependence of the inverse magnetic susceptibility for $La_{0.8}Mg_{0.2}MnO_3$ sample: experimental data (triangles) (Zhao et al., 2001 [16.23]) and theoretical curve (solid line) based on (16.3.3).

| Samples | $\theta$, K | $n$, cm$^{-3}$ | $p$ | $l$, Å | Data source |
|---|---|---|---|---|---|
| $(La_{1-y}Pr_y)_{0.7}Ca_{0.3}MnO_3$ | 55 | $1.8 \cdot 10^{18}$ | 0.03 | 24 | Fig. 16.13 (Babushkina et al., 2003) |
| $Pr_{0.71}Ca_{0.29}MnO_3$ | 105 | $6.0 \cdot 10^{18}$ | 0.07 | 17 | Fig. 16.14 (Fisher et al., 2003) |
| $(La_{0.4}Pr_{0.6})_{1.2}Sr_{1.8}Mn_2O_7$ | 255 | $2.5 \cdot 10^{18}$ | 0.04 | 19 | Fig. 16.15 (Wagner et al., 2002) |
| $La_{0.8}Mg_{0.2}MnO_3$ | 150 | $0.6 \cdot 10^{18}$ | 0.01 | 14 | Fig. 16.16 (Zhao et al., 2001) |

Table 16.3.

16.3.3. Discussion. The triple point in manganites. Unresolved questions.

To sum up, the analysis performed in the previous sections demonstrates that a simple model of the electron tunneling between the ferromagnetically correlated regions (FM droplets) provides a possibility to describe the conductivity and the magnetoresistance data for a wide class of manganites. The comparison of the theoretical predictions with the experimental data on the temperature dependence of the resistivity, magnetoresistance, and magnetic susceptibility enables us to reveal various characteristics of the phase-separated state such as the size of FM droplets, their density, the number of electrons in a droplet and also to estimate the characteristic tunneling length of the charge carriers. The determined values of parameters appear to be rather reasonable. Indeed, the characteristic tunneling length turns out to be of the order of FM droplet size, the concentration of the ferromagnetic phase in the high-temperature range is substantially smaller than the percolation threshold and varies from about 1% to 7%.

Note also that the droplets contain 50-100 charge carriers, whereas parameter $A$ deduced from the experimental data is equal by the order of magnitude to the energy of Coulomb repulsion in a metallic ball of $(7 \div 8)d$ in diameter. The obtained numerical values for the droplet parameters (characteristic tunneling barrier, size, and tunneling length) are close for manganites with drastically different transport properties.

The large magnitude of the $1/f$ noise in the temperature range corresponding to the insulating state is another characteristic feature of the phase-separated manganites (see experimental results obtained in [16.26]. In the framework of the model of phase-separation discussed here we get the large value for the Hooge constant $\alpha_H$ in (16.2.10). At temperatures 100 – 200 K and frequencies 1 – 1000 s$^{-1}$ we get $\alpha_H = 10^{-16}$ cm$^3$. This value of $\alpha_H$ is by 3 – 5 orders of magnitude higher than the corresponding value for semiconductors.

Thus, we have a rather consistent scheme describing the transport properties of manganites under condition that the ferromagnetically correlated regions do not form a percolation cluster. Moreover, the presented approach proves to be valid for a fairly wide range of the dopant concentrations. However, as it was mentioned above, the relation between the concentration of ferromagnetic droplets and the doping level is far from being well understood. If the picture of the phase separation is believed to be applicable, it becomes obvious that not all electrons or holes introduced by doping participate in the transport processes. Below we try to present some qualitative arguments illustrating the possible difference in the effective concentration of charge carriers below and above the transition from paramagnetic to magnetically ordered state.

In the phase diagram of a typical manganite (see [16.39, 16.40] and Chapter 12 for a review), one would have the AFM state with FM-phase inclusions in the low-temperature range and at a low doping level. The transition from AFM to FM phase occurs upon doping. At high temperatures, manganites are in the paramagnetic (PM) state. When the temperature decreases, we observe the transition from PM to AFM or FM state depending on the doping level.

Let us consider the behavior of such a system in the vicinity of a triple point. In the AFM phase, radius $R$ of a region which one electron converts into FM state can be estimated as $R = d(\pi t/4JS^2Z)^{1/5}$ (see [16.40] and Chapter 17), where $J$ is an AFM exchange interaction between the local spins. For high-temperature PM phase, a radius $R_T$ of a region that one electron converts into FM state corresponds to the size of the so-called temperature ferron (see Chapter 17) and equals to $R_T = d(\pi t/4k_B T \ln(2S + 1))^{1/5}$. The critical concentration $x_C \approx 0.15$ of the overlapping of low-temperature ferrons can be derived from the estimate $x_C \approx 3/4\pi \, (d/R)^3$, while for the high-temperature ferrons it follows from the estimate $\delta_C \approx 3/4\pi \, (d/R_T)^3$. Substituting the expressions for the radii of the high- and the low-temperature ferrons to the ratio $x_C/\delta_C$, we obtain the following estimate for this ratio in the vicinity of the triple point corresponding to the coexistence of FM, AFM, and PM phases:

$$\frac{x_C}{\delta_C} \sim \left[\frac{T \ln(2S + 1)}{zJS^2}\right]^{3/5} \sim \left[\frac{T_C \ln(2S + 1)}{T_N}\right]^{3/5} \qquad (16.3.4)$$

where $T_C$ and $T_N$ are the Curie and the Neel temperatures [16.30], respectively. For the manganites under discussion, we have $T_C \sim T_N \sim (120 \div 150)$ K and $\ln(2S + 1) \sim 1.6$ for $S = 2$, hence $\delta_C \leq x_C$. The sign of this inequality is in agreement with experimental data which imply $\delta \sim (1 \div 7)\%$. Thus, we do not have a clear explanation of the charge disbalance in paramagnetic region in spite of the fact that the trend is correctly caught by our simple estimates. Probably, at $x > x_C$ (in real experiments the concentration $x$ can be as high as 50%), the residual charge is localized in the paramagnetic matrix outside the temperature ferrons. The detailed study of this problem will be a subject of the future investigations.